\documentclass[12pt]{article}

\usepackage{comment}

\usepackage{varwidth}
\usepackage{lipsum}
\usepackage{threeparttable}
\usepackage{makecell}
\usepackage{amsfonts}
\usepackage{amsmath}
\usepackage{mathrsfs}
\usepackage{fancyhdr}
\usepackage{epsfig,color,morefloats}
\usepackage{xcolor}
\usepackage[authoryear]{natbib}
\usepackage{booktabs,multirow,multicol,longtable}
\usepackage{amsthm}
\usepackage{amssymb}
\usepackage{color}
\usepackage{lscape}
\usepackage{rotating}
\usepackage{caption}
\usepackage{graphicx}
\graphicspath{{FDR_linear_model/}{FDR_linear_model/old figures/}}
\usepackage[breaklinks,colorlinks,linkcolor=blue,citecolor=blue,urlcolor=blue]{hyperref}
\usepackage{lineno}
\usepackage{cancel}
\usepackage{multirow}

\usepackage{setspace}

\usepackage{amsmath, amssymb, amsthm, latexsym,color, natbib}
\usepackage{fancyhdr}
\usepackage{enumitem}
\usepackage{geometry}
\usepackage{pdflscape}

\usepackage[resetlabels]{multibib}
\allowdisplaybreaks
\newcites{S}{References} 

\usepackage{authblk}

\usepackage{url}
\usepackage[utf8]{inputenc}
\usepackage{amsmath, amssymb, amsfonts, amsthm, color, bm, mathrsfs}
\usepackage[normalem]{ulem}
\usepackage{algorithm,algorithmic}
\allowdisplaybreaks[4]

\usepackage{float}
\usepackage{subfig}
\usepackage{overpic}

\newtheorem{theorem}{Theorem}
\newtheorem{lemma}{Lemma}

\newtheorem{remark}{Remark}
\newtheorem{assumption}{Condition}
\newtheorem{definition}{Definition}

\usepackage{enumitem}
\setenumerate[1]{label={[\arabic*]},leftmargin=1cm,itemsep=-3pt}

\allowdisplaybreaks
\usepackage{setspace}

\newcommand{\bA}{{\mathbf A}}
\newcommand{\bB}{{\mathbf B}}

\newcommand{\bX}{{\mathbf X}}

\newcommand{\bZ}{{\mathbf Z}}

\newcommand{\bI}{{\mathbf I}}
\newcommand{\bz}{{\mathbf z}}

\newcommand{\bV}{{\mathbf V}}

\newcommand{\bQ}{{\mathbf Q}}

\newcommand{\bD}{{\mathbf D}}
\newcommand{\bE}{{\mathbf E}}

\newcommand{\bu}{{\mathbf u}}
\newcommand{\bv}{{\mathbf v}}
\newcommand{\bw}{{\mathbf w}}

\newcommand{\bs}{{\mathbf s}}
\newcommand{\ba}{{\mathbf a}}
\newcommand{\bb}{{\mathbf b}}

\newcommand{\be}{{\mathbf e}}

\newcommand{\bt}{{\mathbf t}}

\newcommand{\by}{{\mathbf y}}
\newcommand{\bUpsilon}{{\mathbf \Upsilon}}

\newcommand{\bx}{\mathbf{x}}

\newcommand{\bbeta}  {\boldsymbol{\beta}}

\newcommand{\bdelta} {\boldsymbol{\delta}}

\newcommand{\bOmega}{\boldsymbol{\Omega}}
\newcommand{\bomega}{\boldsymbol{\omega}}
\newcommand{\bSigma}{\boldsymbol{\Sigma}}
\newcommand{\bDelta}{\boldsymbol{\Delta}}
\newcommand{\bgamma}{\boldsymbol{\gamma}}

\newcommand{\bTheta} {\boldsymbol{\Theta}}

\newcommand{\bPsi} {\boldsymbol{\Psi}}
\newcommand{\btheta} {\boldsymbol{\theta}}
\newcommand{\bxi} {\boldsymbol{\xi}}

\newcommand{\bzeta} {\boldsymbol{\zeta}}
\newcommand{\bGamma} {\boldsymbol{\Gamma}}
\newcommand{\bLambda} {\boldsymbol{\Lambda}}

\newcommand{\bzero}{{\mathbf 0}}

\newcommand{\bXi}{\boldsymbol{\Xi}}

\newcommand{\beps}{\boldsymbol{\varepsilon}}

\newcommand{\bzo}{{\mathbf 0}}
\newcommand{\E}{\mathbb{E}}
\newcommand{\bT}{\mathbf{T}}
\newcommand{\PP}{\mathbb{P}}

\theoremstyle{definition}

\def\spacingset#1{\renewcommand{\baselinestretch}%
	{#1}\small\normalsize} \spacingset{1}

\newcommand{\blind}{0}

\def\singlespace{\def\baselinestretch{1}\@normalsize}

\usepackage{microtype}   % character protrusion + font expansion -> tighter, cleaner line breaks
\makeatletter
\renewcommand\section{\@startsection{section}{1}{\z@}%
  {-2.8ex \@plus -.8ex \@minus -.2ex}{1.4ex \@plus .2ex}%
  {\renewcommand{\baselinestretch}{1}\normalfont\Large\bfseries}}
\renewcommand\subsection{\@startsection{subsection}{2}{\z@}%
  {-2.3ex \@plus -.6ex \@minus -.2ex}{1.0ex \@plus .2ex}%
  {\renewcommand{\baselinestretch}{1}\normalfont\large\bfseries}}
\renewcommand\subsubsection{\@startsection{subsubsection}{3}{\z@}%
  {-2.0ex \@plus -.6ex \@minus -.2ex}{0.9ex \@plus .2ex}%
  {\renewcommand{\baselinestretch}{1}\normalfont\normalsize\bfseries}}
\def\thm@space@setup{\thm@preskip=5pt plus 2pt minus 2pt \thm@postskip=\thm@preskip}
\makeatother
\begin{document}

\def\spacingset#1{\renewcommand{\baselinestretch}%
{#1}\small\normalsize} \spacingset{1}

%%%%%%%%%%%%%%%%%%%%%%%%%%%%%%%%%%%%%%%%%%%%%%%%%%%%%%%%%%%%%%%%%%%%%%%%%%%%%%

\if0\blind
{
  \spacingset{1.1}
  \title{\large \bf Controlling the False Discovery Rate in High-Dimensional Linear Models Using Model-X Knockoffs and 
$p$-values }
%  \author[1,2,3]{Jinyuan Chang}
  \author[1,2]{Jinyuan Chang}
  \author[1]{Chenlong Li}
  \author[3]{Cheng Yong Tang%\footnote{Address for correspondence: C. Y. Tang, Department of Statistics, Operations, and Data Science, Temple University, 1810 Liacouras Walk, Philadelphia, PA 19122 USA.  Email: yongtang@temple.edu  }
  }
  \author[4]{Zhengtian Zhu}

   \affil[1]{\it \small Joint Laboratory of Data Science and Business Intelligence, Institute of Statistical Interdisciplinary Research, Southwestern University of Finance and Economics, Chengdu, China}
   \affil[2]{\it \small State Key Laboratory of Mathematical Sciences, Academy of Mathematics and Systems Science, Chinese Academy of Sciences, Beijing, China}
%   \affil[3]{\it \small 
%China Center for Economic Research, Peking University, Beijing, China
%}
   \affil[3]{\it \small Department of Statistics, Operations, and Data Science, Temple University, Philadelphia, PA, USA.}
   \affil[4]{\it \small School of Mathematical Sciences, Key Laboratory of Intelligent Computing and Applications (Ministry of Education), Tongji University, Shanghai, China}

  \setcounter{Maxaffil}{0}
  \renewcommand\Affilfont{\itshape\small}
  \date{\vspace{-5ex}}
  
     \maketitle
} \fi

\if1\blind
{
  \bigskip
  \bigskip
  \bigskip
  \begin{center}
    {\Large \bf Controlling the False Discovery Rate in High-Dimensional Linear Models Using Model-X Knockoffs and 
$p$-values}
\end{center}
  \medskip
} \fi

\spacingset{1.3}
\begin{abstract}

We propose a novel multiple testing methodology for controlling the false discovery rate (FDR) in high-dimensional linear models that integrates model-X knockoff techniques with debiased penalized regression estimators. At the foundation of our methodology, we construct and study two sets of naturally paired high-dimensional test statistics and the associated $p$-values for evaluating the same null hypotheses. The first set is shown to be asymptotically mutually independent, justifying the use of the Benjamini-Hochberg procedure. We further exploit the pairing structure through a two-step procedure aimed at improving power. Our theoretical results establish the key properties of the framework with respect to asymptotic FDR control and formally characterize the associated power gains of the two-step procedure. Importantly, our framework accommodates general dependence in the design matrix. Extensive simulations demonstrate that our methods outperform existing approaches -- particularly those relying on empirical FDP estimates -- in both power and FDR control accuracy, with notable gains in settings involving weaker signals, small sample sizes, or low target FDR levels.

\end{abstract}

\noindent
{\sl Keywords:}  Benjamini-Hochberg procedure, debiased estimator,   false discovery rate,   model-X knockoff,  multiple testing,  penalized regression, $p$-values.

\spacingset{1.69}
\setlength{\abovedisplayskip}{0.2\baselineskip}
\setlength{\belowdisplayskip}{0.2\baselineskip}
\setlength{\abovedisplayshortskip}{0.2\baselineskip}
\setlength{\belowdisplayshortskip}{0.2\baselineskip}

\section{Introduction} \label{sec:intro}

\subsection{Overview}

Multiple testing is a foundational yet challenging aspect of statistical analysis and applied research. 
Adjusting for the multiplicity of statistical tests is essential for developing reliable methodologies that support valid scientific discoveries; see \cite{Benjamini2010} for a comprehensive overview and insights. 
This problem has received considerable attention across a wide range of disciplines, including molecular biology, genetics, clinical research, epidemiology, cognitive neuroscience, and others. 
In broad practical scenarios, controlling the false discovery rate (FDR) has emerged as a central methodological objective. 
FDR control aims to ensure that the expected proportion of false positives among the rejected hypotheses does not exceed a pre-specified level. 
The Benjamini-Hochberg (BH) procedure \citep{Benjamini1995Controlling}, which ranks \( p \)-values and compares them to an increasing sequence of thresholds, is the most widely used approach for achieving this goal.

Investigating and understanding the impact of dependence among multiple test statistics has been a central research focus in the area of multiple testing. Ideally, a methodologically valid approach must control the FDR at its pre-specified level regardless of the dependence structure among the test statistics.  The BH procedure is known to control the FDR  under the assumption of independence among test statistics. 
 However, when test statistics are dependent, the situation becomes considerably more complex. Under the condition of positive regression dependence on subsets (PRDS), the BH procedure has been shown to control the FDR conservatively \citep{BY2001, Sarkar2002}. Recent studies have further explored this issue by incorporating information and structure from additional models for the test statistics; see, for example, \cite{SunCai2008}, \cite{Fanetal2012}, and \cite{LeiFithian2018}. \cite{Fithian2022Conditional} investigate calibrating the FDR control procedure using information from the conditional distribution to handle dependent test statistics. Despite these advances, ensuring valid FDR control under arbitrary and unknown dependence structures remains a fundamental focus in the current state of knowledge.

The knockoff method, introduced by \cite{Barber2015Controlling}, has emerged as an influential tool for controlling the FDR. 
The central methodological step is to construct ``knockoff'' variables that replicate the dependence structure of the original variables while being conditionally independent of the response variable, given the original variables. 
By fitting a model to an augmented design matrix that includes both the original variables and their knockoff counterparts, knockoff-based procedures are developed to identify relevant variables while maintaining control of the FDR. 

Notably, a key distinction between the class of knockoff methods and the conventional BH procedure is that knockoff methods do not rely on \( p \)-values. Instead, they estimate the empirical false discovery proportion (FDP) directly. This design enables valid FDR control without requiring explicit knowledge of the dependence structure among the test statistics. In line with this perspective, several recent developments have focused on empirically evaluating the FDP without relying on \( p \)-values; see \cite{Liu2020}, \cite{Xing2021Controlling}, \cite{Daietal_2023_2}, and \cite{Daietal_2023}.

In contrast to FDP-based approaches, within the same setting as \cite{Barber2015Controlling}, \cite{Sarkar2022Adjusting} propose constructing  \( p \)-values upon incorporating knockoff copies into classical low-dimensional linear models. 
They introduce a two-step testing procedure that first conducts a selection step using the Bonferroni method, followed by applying the BH procedure to the selected variables in the second step. 
Their method controls the FDR under arbitrary dependence in the design matrix and performs competitively when the target FDR is low or signals are weak \citep{Sarkar2022Adjusting}.

\subsection{High-dimensional multiple testing}

Transitioning to high-dimensional paradigms introduces fundamental new challenges for multiple testing -- particularly in the construction of test statistics and the adjustment for multiplicity -- rendering this area inadequately understood and underdeveloped.

For estimating high-dimensional parameters, penalized regression methods have been extensively developed  with their statistical properties well studied; see, among others, the monographs by \cite{BV_2011} and \cite{Fan2020}. These estimators are typically sparse, making the joint distribution of their zero and non-zero components analytically intractable -- a longstanding challenge in statistical theory. This complexity poses significant obstacles for inference based on penalized estimators. To meet this challenge, substantial progress has been made through debiasing techniques \citep{Zhang2013, vandeGeer2014on, Ning2017}, which enable valid component-wise statistical inference by yielding asymptotically normal distributions for individual parameters.

However, advancing from component-wise inference to addressing multiplicity in high-dimensional models presents substantial challenges. A central difficulty lies in the complex and unknown dependence structure among components of penalized estimators, compounded by the intractability of their joint distribution. These issues pose significant obstacles to the methodological development  and theoretical understanding of multiple testing procedures in high-dimensional settings.

 A commonly used framework in the existing literature accommodates this challenge by relying on structural assumptions about the correlation among test statistics -- particularly the strength of pairwise correlations and the sparsity in the number of strongly correlated variable pairs; see \cite{Liu2013}, \cite{CaiLiu2016}, and reviewed by \cite{Caietal2023}.  Within this framework, test statistics, their limiting distributions, and the associated \( p \)-values derived from debiased estimators have been employed to develop multiple testing procedures; see \cite{JJ2019EJS} and \cite{Caietal2023}. 
However, this framework may fall short of fully accommodating the joint limiting distribution, which often involves intricate and unknown dependence structures. Such limitations can hinder the effectiveness of multiple testing procedures in high-dimensional settings. To the best of our knowledge, no existing methodology systematically addresses this challenge by explicitly incorporating the full joint distribution of test statistics under general dependence.

For high-dimensional regression models, the model-X knockoff method introduced by \cite{candes2018panning} extends the original approach of \cite{Barber2015Controlling}. Within the model-X framework, knockoff variables are constructed by exploiting knowledge of the joint distribution of the variables in the design matrix. Building on this construction, \cite{candes2018panning} develop and analyze a methodology based on empirically estimating the FDP, demonstrating its ability to control the FDR accommodating unknown dependence structures in the design matrix.
In the same direction through empirical FDP estimation, \cite{Xing2021Controlling} propose a two-stage procedure that performs variable selection in the first stage and estimates the FDP in the second stage. \cite{Daietal_2023_2} introduce a related approach that combines debiasing and data-splitting techniques to construct test statistics for FDP estimation.  While empirical FDP estimation-based  methods offer promising alternatives and avoid direct reliance on \( p \)-values, they may suffer from power loss in certain regimes due to the conservative nature and inherent variability of FDP estimation.

\subsection{Motivation, the framework,  and our contributions}

The current state of knowledge reveals a critical methodological dichotomy: while debiasing techniques provide component-wise asymptotic normality, they remain methodologically insufficient to properly account for unknown joint dependence structures required for reliable multiple testing inference; simultaneously, while knockoff methodologies offer robust FDR control, they are often constrained by a power deficit arising from the high variance inherent in empirical FDP estimation.

Our study seeks to bridge these parallel paradigms by integrating model-X knockoffs into the debiased Lasso framework.
We develop a new multiple-testing framework for high-dimensional linear models that controls the FDR under complex dependence structure. 
The method builds upon knockoff construction and debiasing techniques  to construct test statistics, derive their joint distribution, and obtain valid \( p \)-values. Our theory gives a rigorous foundation for inference in knockoff-augmented linear models, without relying on specific structural or sparsity assumptions about correlations.  
Unlike existing high-dimensional procedures -- e.g.\ \cite{JJ2019EJS}, \cite{Liu2013}, \cite{CaiLiu2016}, \cite{Caietal2023} -- which depend on such assumptions and therefore ignore much information from the dependence structures, our approach fully exploits the limiting joint distribution of the test statistics. By effectively incorporating informative dependence structures that are often overlooked by existing methods, the proposed framework offers potential for improved power while ensuring valid FDR control asymptotically.

At its core, our approach constructs two naturally paired sets of test statistics targeting the same underlying hypotheses. The methodological advancements of this framework are two-fold. First, and crucially, by establishing that one set is asymptotically mutually independent, we enable the direct and valid application of the BH procedure. Second, this paired architecture facilitates a two-step testing strategy -- sharing the structural logic of \cite{Sarkar2022Adjusting} -- which leverages a Bonferroni-based screening step to prune irrelevant hypotheses before applying the BH procedure in the second step. Our theoretical investigation characterizes the mechanism through which this two-step approach enhances discovery power, while empirical simulations confirm a robust overall performance with the most substantial advantages occurring in challenging regimes, such as those characterized by high sparsity or weak signals.

{Our framework represents a significant methodological and theoretical advancement over existing work, such as \cite{Sarkar2022Adjusting}. While their approach provides a valuable foundation, it is fundamentally restricted to low-dimensional settings due to its reliance on ordinary least squares, which becomes ill-posed in high-dimensional regimes. We address this critical limitation by developing what is, to our knowledge, the first systematic framework for FDR control explicitly designed for knockoff-augmented statistics in high dimensions. The key innovation lies in incorporating the limiting joint distribution of these high-dimensional statistics, which provides the theoretical and methodological grounding for valid high-dimensional inference.  Our analysis establishes asymptotic FDR control in these regimes and characterizes the power gains of the proposed two-step construction, showing how the use of paired test statistics can enhance discovery power. Empirically, we demonstrate substantial power improvements even in settings where \cite{Sarkar2022Adjusting} remains applicable, highlighting both the broader applicability and improved performance of our approach.}

Selective inference is a remarkable methodology for high-dimensional
statistical inference, designed for settings that involve a data-dependent
variable screening and/or model selection step. 
This framework relies on a selected model and then performs statistical
inference, such as constructing $p$-values, for the coefficients of that
model \citep{Lee2016Exact, Fithian2017Optimal}. High-dimensional multiple testing,
the aim of our study, targets a different problem: it tests all the hypotheses, one for each original
variable; see also Section~1.4 of \citet{candes2018panning}. In our approach, the validity of the FDR control is
justified by exploiting the asymptotic joint distribution of the proposed
test statistics. In contrast, the $p$-values from selective inference are
valid component-wise under the conditional distribution, and such
component-wise validity alone cannot support FDR control via the BH
procedure \citep{Dobriban2026BH, Lei2026BH}.  A detailed discussion of the connections
and distinctions, together with numerical comparisons against representative
selective-inference procedures, is provided in Section \ref{sec:selective inference} of the
Supplementary Material.

\subsection{Notation and organization}

For any positive integer $m$, we write $[m]=\{1,\ldots,m\}$. Denote the indicator function by $I(\cdot)$ and the $m$-dimensional identity matrix by $\bI_{m}$. For a vector $\bv=(v_{1},\ldots,v_{m})^{\top}\in\mathbb{R}^{m}$, $|\bv|_1 = \sum_{i=1}^{m} |v_i|$, $|\bv|_2 = (\sum_{i=1}^{m} |v_i|^2)^{1/2}$, and $|\bv|_0$ represents the number of nonzero entries of $\bv$. For two vectors $\bv, \bu\in\mathbb{R}^{m}$, we define the inner product $\langle \bv,\bu \rangle=\bv^{\top}\bu$. For a matrix $\bA=(a_{i,j})\in\mathbb{R}^{p\times q}$, we define the elementwise $\ell_{\infty}$-norm $|\bA|_{\infty}=\max_{i\in[p],j\in[q]}|a_{i,j}|$, the elementwise $\ell_{1}$-norm $|\bA|_{1}=\sum_{i\in[p],j\in[q]}|a_{i,j}|$, and the $\ell_{1}$-norm $\|\bA\|_{1}=\max_{j\in[q]}\sum_{i=1}^{p}|a_{i,j}|$. The maximum and the minimum singular values of $\bA$ are respectively denoted by $\sigma_{\max}(\bA)$ and $\sigma_{\min}(\bA)$. 
We use $\bA \succ 0$ to denote that $\bA$ is a positive definite matrix. We use $\lceil \cdot \rceil$ to denote the ceiling function. For a random variable $x$, we denote its sub-Gaussian norm 
$\|x\|_{\psi_2} = \sup_{q \ge 1}q^{-1/2}\{\mathbb E (\lvert x \rvert^q)\}^{1/q}$. For a random vector $\bx \in \mathbb R^m$, its sub-Gaussian norm is defined as $\|\bx\|_{\psi_2} = \sup_{\bu \in \mathbb{S}^{m-1}} \| \langle \bx, \bu \rangle \|_{\psi_2}$, where $\mathbb{S}^{m-1}$ denotes the unit sphere in $\mathbb R^m$. 

The rest of this paper is organized as follows. Section \ref{sec:method} outlines our methodology. Section \ref{sec:mainresults} presents the main results, including the theoretical properties of the proposed $p$-value-based multiple testing procedures. Section \ref{sec:sim} reports simulations, and Section \ref{sec:realdata} provides a demonstration with a real data example. Section \ref{sec:discussion} concludes the paper with a discussion. 

\section{Methodology} \label{sec:method}

\subsection{Setting and background}
We denote the design matrix by \( \bX = (\bx_1, \dots, \bx_n)^{\top} \), where \( \bx_i = (x_{i,1}, \ldots, x_{i,d})^{\top} \in \mathbb{R}^d \) for $i\in[n]$. In high-dimensional settings, the number of predictors \( d \) may exceed the sample size \( n \). For a random design, the rows \( \{\bx_i\}_{i=1}^n \) are treated as \( n \) independent copies of a \( d \)-dimensional random vector \( \bx \).
 Our investigation takes place in the context of the linear model:
\begin{align}\label{eq:lm}
\by = \bX\bbeta + \beps\,,
\end{align}
where \( \by= (y_1,\ldots,y_n)^\top \in \mathbb{R}^n \) is the response vector, \( \beps = (\varepsilon_1,\ldots,\varepsilon_n)^\top \in \mathbb{R}^n \) is the error term with $\mathbb{E}(\varepsilon_i)=0$ and ${\rm Var}(\varepsilon_i)=\sigma^2$, and \( \bbeta \in \mathbb{R}^d \) is the parameter vector.
The primary objective of interest is to simultaneously test the hypotheses:
\begin{align}\label{eq:mtest}
H_{0,j}: \beta_j = 0 \quad \text{versus} \quad H_{1,j}: \beta_j \neq 0\,, \quad j \in [d]\,.
\end{align}

Recently, the development of the so-called knockoff variables based approaches has been influential in addressing the multiple testing problem. In a seminal work, \cite{Barber2015Controlling} propose generating a new design matrix \( \tilde{\bX} = (\tilde{\bx}_1, \dots, \tilde{\bx}_n)^{\top} \), consisting of knockoff copies \( \{\tilde{\bx}_i\}_{i=1}^n \) of the variables \( \{\bx_i\}_{i=1}^n \) in \( \bX \). The new variables in \( \tilde{\bX} \) are conditionally independent of the response vector \( \by \) given the observed explanatory variables \( \bX \). As a remarkable property of \(\tilde{\bX}\), replacing any subset of the explanatory variables in the original design matrix \( \bX \) with their corresponding knockoff copies in \( \tilde{\bX} \) preserves the variance-covariance structure of \( \bX \) exactly.

By fitting a working linear model using the augmented design matrix $(\bX, \tilde{\bX})$:
\begin{align} \label{eq:lm2}
\by = (\bX, \tilde{\bX}) \bgamma + \beps\,,
\end{align}
we obtain a parameter estimate in $2d$ dimensions, denoted as 
$
\hat{\bgamma} = (\hat{\bbeta}_1^{\top}, \hat{\bbeta}_2^{\top})^{\top}$.
Due to the conditional independence of $\tilde{\bX}$ from $\by$ given $\bX$, the true parameter corresponding to $\hat{\bbeta}_2$ is $\mathbf{0}$. Specifically, if the true model is \eqref{eq:lm} with true parameter \(\bbeta_{0} :=(\beta^{0}_{1},\ldots,\beta^{0}_{d})^{\top}\), then the true value of \(\bgamma\) is \(\bgamma_0 = (\bbeta_0^\top, {\bf0}^\top)^\top\).

Based on the construction of knockoff variables, \cite{Barber2015Controlling} introduce a procedure that controls the FDR, regardless of the dependence structure in the design matrix. Their approach leverages the properties of the knockoff copies \( \tilde{\bX} \), using them in conjunction with the original data \((\bX, \by)\) to empirically estimate the FDP.
To this end, \cite{Barber2015Controlling} propose the construction of test statistics \( \{W_j\}_{j=1}^{d} \), derived from the procedure used to obtain the parameter estimates \( \hat{\bgamma} = (\hat{\bbeta}_1^{\top}, \hat{\bbeta}_2^{\top})^{\top} \in \mathbb{R}^{2d} \). These statistics are designed to quantify the evidence against each of the \( d \) null hypotheses \( H_{0,j}: \beta^0_j = 0 \), corresponding to the model \eqref{eq:lm}. A key requirement is that the statistics \(\{W_j\}_{j=1}^d\) are symmetric around zero under the null hypothesis.
The FDP is then estimated at a threshold \( t \) via
\begin{align} \label{eq:fdp}
\widehat{\text{FDP}}(t) = \frac{\#\{j : W_j < -t\}}{\#\{j : W_j > t\} \vee 1}\,.
\end{align}
 \cite{Barber2015Controlling} further propose to use a more conservative estimator:
\begin{align} \label{eq:fdp+}
\widehat{\text{FDP}}_+(t)= \frac{1 + \#\{j : W_j < -t\}}{\#\{j : W_j > t\} \vee 1}\,.
\end{align}
They show that selecting the threshold
$\tau = \min\{ t : \widehat{\text{FDP}}_+(t) \leq \alpha \}$ and rejecting all hypotheses for which \( W_j > \tau \) yields a multiple testing procedure that controls the FDR at the nominal level \( \alpha \). The additive adjustment in \eqref{eq:fdp+} is necessary, as noted by \cite{Barber2015Controlling}, to ensure both theoretical and empirical  control of the FDR.

Recently, \cite{Xing2021Controlling} propose an approach using a device called the Gaussian Mirror. For each $j\in[d]$, the \( j \)-th mirror statistic is constructed by replacing  the \( j \)-th column \( \bx_{\cdot,j} \) in the design matrix \( \bX \) with two perturbed versions:  
$\bx_{\cdot,j}^+ = \bx_{\cdot,j} + c_j \bxi_j$ and $\bx_{\cdot,j}^- = \bx_{\cdot,j} - c_j \bxi_j$, where \(\bxi_j \sim \mathcal{N}({\bf0}, \bI_n) \).   
A new set of parameter estimates is obtained by fitting the linear model \eqref{eq:lm} with \( \by \) and the new design matrix \( (\bx_{\cdot,j}^+, \bx_{\cdot,j}^-, \bX_{-j}) \), where \( \bX_{-j} \) denotes the matrix \( \bX \) without the \( j \)-th column. Let \( \hat{\beta}_j^+ \) and \( \hat{\beta}_j^- \) be the components of the estimator corresponding to \( \bx_{\cdot,j}^+ \) and \( \bx_{\cdot,j}^- \), respectively. Based on these estimates, the so-called mirror statistics can be constructed. An example of the mirror statistics is  
$
M_j = | \hat{\beta}_j^+ + \hat{\beta}_j^- | - | \hat{\beta}_j^+ - \hat{\beta}_j^- |$. 
In the same spirit as \eqref{eq:fdp}, \cite{Xing2021Controlling} propose to estimate the FDP as
\begin{align} \label{eq:fdp2}
\widehat{\text{FDP}}(t) = \frac{\#\{j : M_j < -t\}}{\#\{j : M_j > t\} \vee 1}\,.
\end{align}
A multiple testing procedure is then developed by finding
$
\tau = \min \{ t : \widehat{\text{FDP}}(t) \leq \alpha \}$,
and rejecting all null hypotheses for which \( M_j > \tau \).

With empirical FDP estimations such as those in \eqref{eq:fdp}--\eqref{eq:fdp2}, the resulting multiple testing procedures only require the test statistics, e.g., \( \{W_j\}_{j=1}^d \) or \( \{M_j\}_{j=1}^d \), without explicitly relying on the joint distributions of these statistics. As a result, these procedures become $p$-value-free, a key distinction from conventional methods rooted in \cite{Benjamini1995Controlling}.  
On the one hand, this offers significant practical convenience by eliminating the need for explicitly evaluating the joint distribution of the test statistics. On the other hand, it raises some critical concerns, particularly when addressing multiple testing problems in more challenging scenarios. Foremost, since FDP estimators such as \eqref{eq:fdp}--\eqref{eq:fdp2} are ratio estimators, they tend to exhibit higher variance when the numerator is small -- especially at low target FDR level \( \alpha \). This issue is further exacerbated when the sample size \( n \) is small, leading to greater variability in the FDP estimates and making the results of the multiple testing procedures unstable.  
In fact, the procedures of \cite{Barber2015Controlling} can become overly conservative in situations where the FDR level \( \alpha \) is low or the signal strength is weak; see \cite{Sarkar2022Adjusting}. 

In contrast, $p$-value-based multiple testing approaches attempt to fully incorporate data evidence in assessing the strength of evidence against the null hypotheses. Given the same model and data setup,  $p$-value-based methods, when appropriately designed, can be more powerful while still maintaining control of the FDR at a given nominal level.  
As an example, \cite{Sarkar2022Adjusting} propose a multiple testing procedure for settings with paired $p$-values for each null hypothesis. Their approach utilizes the properties of these paired $p$-values and follows a two-step procedure: first performing an initial selection using a Bonferroni-type method, and then applying the BH procedure to determine the final rejections. They demonstrate that the augmented design matrix containing the knockoff variables of \cite{Barber2015Controlling}, in the context of linear models, supports the development of paired $p$-values, and the resulting multiple testing procedure controls the FDR at the desired level.  
Empirically, \cite{Sarkar2022Adjusting} show that the $p$-value-based multiple testing procedure tends to be more powerful in challenging situations, such as when the sample size is small, the signal is weak, and the target FDR level is low.

When handling high-dimensional multiple testing problems -- e.g., when \( d \gg n \) -- substantial new challenges arise.  
For knockoff-based methods, the primary difficulty lies in constructing appropriate knockoff variables. In the high-dimensional linear model setting, \cite{candes2018panning} propose the model-X knockoff framework. Using model-X knockoff variables, an augmented design matrix can be constructed similarly to that in \cite{Barber2015Controlling}.
To estimate the high-dimensional linear model with the resulting \(n \times 2d\)  augmented design matrix, penalized regression methods are applicable; see \cite{BV_2011} and \cite{Fan2020}; a representative example is the Lasso estimator. To construct the statistics \( \{W_j\}_{j=1}^{d} \), one strategy is to use the solution path of the Lasso estimator for high-dimensional linear models.
Specifically, let
\begin{align} \label{eq:lasso}
    \hat{\bgamma} = \arg\min_{\bgamma \in \mathbb{R}^{2d}} \bigg\{ \frac{1}{n} | \by - (\bX, \tilde{\bX})\bgamma |_2^2 + \varrho_1 | \bgamma |_1 \bigg\}\,,
\end{align}
where \( \varrho_1>0 \) is a tuning parameter that controls the penalty on the \( \ell_1 \)-norm of \( \bgamma \). The Lasso estimator is convenient in practice due to the convex nature of the objective function in \eqref{eq:lasso}, as both the squared loss and the \( \ell_1 \)-penalty are convex. 
 As \(\varrho_1\) decreases, \(\hat{\boldsymbol{\gamma}}\) includes  selected variables in a sequential manner. 
Let \(\{\varrho_{1,j}\}_{j=1}^{2d}\) be the sequence of respective tuning parameter such that  \(\varrho_{1,j}\) corresponds to the \(j\)-th component  in the parameter firstly entering the solution path. 
Then the FDP can be estimated via \eqref{eq:fdp+} by letting \(W_j=\varrho_{1,j}-\varrho_{1,j+d}\).

In the Gaussian Mirror approach, \cite{Xing2021Controlling} develop a two-stage procedure. In the first stage, a variable selection stage is applied to obtain a low-dimensional estimated model, effectively reducing the dimensionality from \( d \) to \( s \), where \( s \ll d \). In the second stage, mirror statistics \( \{M_j\}_{j=1}^s \) are constructed based on the reduced model, and multiple testing is performed using FDP estimation as defined in \eqref{eq:fdp2}.

In the context of high-dimensional multiple testing, $p$-value-based methods face a fundamental difficulty: how to obtain reliable $p$-values associated with the test statistics for the high-dimensional model parameters.  Penalized regression methods with sparsity-inducing penalty functions are effective for estimation; however, the joint distribution of the resulting sparse parameter estimates is difficult to characterize; see \cite{Fan2020} and references therein. 
Without appropriate $p$-values, approaches such as that of \cite{Sarkar2022Adjusting} are not applicable for solving high-dimensional multiple testing problems.

\subsection{Multiple testing with high-dimensional $p$-values}\label{sec:fdrproc}
The primary objective of our study is to develop an effective \( p \)-value-based approach for high-dimensional multiple testing problems, based on the working model \eqref{eq:lm2} that incorporates the regenerated model matrix \( \tilde{\bX} \). 
Our methodology is designed to accommodate the dependence among test statistics arising in such settings.

In this study, we focus on the specific scenario with the model-X knockoffs elaborated extensively in \cite{candes2018panning}. In particular, in the context of the linear model \eqref{eq:lm} with a random design, the model-X knockoffs for a generic \( d \)-dimensional random vector \( \bx = (x_1, \ldots, x_d)^{\top} \) are a new family of random variables \( \tilde{\bx} = (\tilde{x}_1, \ldots, \tilde{x}_d)^{\top} \) constructed with the following two properties:
\begin{itemize}
    \item[(a)] \( (\bx^{\top}, \tilde{\bx}^{\top})^{\top}_{\text{swap}(S)} \overset{{\rm d}}{=} (\bx^{\top}, \tilde{\bx}^{\top})^{\top} \) for any \( S \subseteq[d] \), where \( \overset{{\rm d}}{=} \) denotes equality in distribution, and \( (\bx^{\top}, \tilde{\bx}^{\top})^{\top}_{\text{swap}(S)} \) is obtained from \( (\bx^{\top}, \tilde{\bx}^{\top})^{\top} \) by swapping the entries \( x_j \) and \( \tilde{x}_j \) for each \( j \in S \). 
    \item[(b)] \( \tilde{\bx} \perp \!\!\! \perp \by \mid \bx \).
\end{itemize}
Let \(\bSigma=\text{Var}(\bx)\). Then the properties of the model-X knockoff variables imply that the covariance matrix of \((\bx^{\top}, \tilde{\bx}^{\top})^{\top}\) is structured as
\begin{equation} \label{eq:kcov}
      \bGamma = \text{Var}\{(\bx^{\top}, \tilde{\bx}^{\top})^{\top}\}  =
    \begin{pmatrix}
        \bSigma & \bSigma - \bD \\
        \bSigma - \bD & \bSigma
    \end{pmatrix},
\end{equation}
where \( \bD = \text{diag}(\bs) \), and \( \bs \) is a hyperparameter chosen to ensure that the covariance matrix \(\bGamma\) is positive semidefinite. We refer to \cite{candes2018panning} for a comprehensive discussion of the model-X knockoff methodology and its constructions within high-dimensional linear models for practical implementations.

To address the high-dimensional estimation problem under model \eqref{eq:lm2}, the Lasso estimator \eqref{eq:lasso} offers a convenient solution. However, conducting statistical inference for high-dimensional model parameters remains challenging; see \cite{Hastieetal_2015}, \cite{Fan2020}, and references therein. In the context of multiple testing, 
a central challenge stems from the unknown distribution of the Lasso estimator, which involves intricate  dependence among its components.

To overcome these challenges, we propose utilizing a \textit{debiased estimator} \citep{Zhang2013, vandeGeer2014on} as the foundational element for constructing suitable test statistics and corresponding \( p \)-values. In brief, debiasing procedures begin with a sparse initial estimator, such as the Lasso estimator \eqref{eq:lasso}, and subsequently apply an additive correction to render the limiting joint distribution of the bias-corrected estimators tractable; see \cite{Ning2017}, \cite{Chang2020}, and \cite{Chang2023culling} for comprehensive expositions in broad context. Upon consistently estimating the limiting variances and covariances of the debiased estimator, one can construct the test statistics and associated \( p \)-values, thereby enabling valid statistical inference for high-dimensional model parameters.

In our setting with knockoff variables, let \( \bZ = (\bX, \tilde{\bX}) \), and define the debiased estimator as
\begin{equation}\label{de_est}
    \hat{\bgamma}^{(\rm bc)} = \hat{\bgamma} + n^{-1} \hat{\bTheta}^{\top} \bZ^{\top} (\by - \bZ \hat{\bgamma})\,.
\end{equation}
Here, \(\hat{\bgamma}\) denotes the Lasso estimator defined in \eqref{eq:lasso}, and \(\hat{\bTheta}\) is the so-called ``decorrelating'' matrix associated with \(\hat{\bGamma} = n^{-1}\bZ^{\top} \bZ\), satisfying \( |\hat{\bTheta}^{\top} \hat{\bGamma} -  \bI_{2d}|_{\infty} \leq \varrho_2 \) for some hyper-parameter \(\varrho_2>0\). A rigorous specification of \(\hat{\bTheta}\) is provided in Section \ref{sec:specselect}. 

Upon noting that the true parameter value in the working model \eqref{eq:lm2} is \(\bgamma_0\), we obtain
\begin{equation}\label{eq:est-debias}
	n^{1/2}\{\hat{\bgamma}^{(\rm bc)} - \bgamma_0\} = n^{-1/2} \hat{\bTheta}^{\top} \bZ^{\top} \beps + n^{1/2}(\hat{\bTheta}^{\top}\hat{\bGamma} -  \bI_{2d})(\bgamma_0 - \hat{\bgamma})\,.
\end{equation}
The properties of \(\hat{\bgamma}^{(\rm bc)}\), which incorporate the structure of the knockoff variables, are established in Theorem \ref{theo_random} in Section \ref{proof_thm0SM} of the Supplementary Material, forming the theoretical basis for the validity of our proposed multiple testing procedure. These technical developments are of independent interest, as they characterize the behavior of debiased estimators in the presence of knockoff variables, and are expected to be broadly useful for statistical inference in such settings.

Based on \( \hat{\bgamma}^{(\rm bc)} \) given in \eqref{de_est}, we propose constructing a transformed debiased estimator  
\begin{align}\label{eq:transform}
\bT \hat{\bgamma}^{(\rm bc)} = 
\begin{pmatrix}
\hat{\bbeta}_1^{(\rm bc)} \\
\hat{\bbeta}_2^{(\rm bc)}
\end{pmatrix}
~~\text{with}~~ 
\bT = 
\begin{pmatrix} 
\bI_d & \bI_d \\ 
\bI_d & -\bI_d 
\end{pmatrix}\,.
\end{align}
Since the true value of \( \bgamma \) is \( \bgamma_0 = (\bbeta_0^\top, {\bf0}^\top)^\top \), both \( \hat{\bbeta}_1^{(\rm bc)} \) and \( \hat{\bbeta}_2^{(\rm bc)} \in \mathbb{R}^d \) are valid estimators of \( \bbeta_0 \), provided that \( \hat{\bgamma}^{(\rm bc)} \) is consistent.  
Write \( \bTheta_0 = \bGamma^{-1} \). By the properties of the model-X knockoffs in \eqref{eq:kcov}, we have
\begin{equation*}
\bT \bTheta_0 \bT^\top =
\begin{pmatrix}
2(2\bSigma - \bD)^{-1} & {\bf0} \\
{\bf0} & 2\bD^{-1}
\end{pmatrix}\,.
\end{equation*}
As elaborated in Section \ref{proof_thm0SM} of the Supplementary Material, \( \bTheta_0 \) determines the limiting variance-covariance matrix of \( \hat{\bgamma}^{(\rm bc)} \), which also serves as the probability limit of the high-dimensional estimator \( \hat{\bTheta} \) under appropriate conditions.  
Intuitively, since \( \bT \bTheta_0 \bT^\top \) is block-diagonal, \( \hat{\bbeta}_1^{(\rm bc)} \) and \( \hat{\bbeta}_2^{(\rm bc)} \) are asymptotically independent. Additionally, because \( \bD \) is a diagonal matrix, the components of \( \hat{\bbeta}_2^{(\rm bc)} \) are asymptotically mutually independent. These intuitions will be rigorously confirmed by the theoretical results and the technical developments provided in Section \ref{proof_thm0SM} of the Supplementary Material.

Write \( \hat\bLambda = \bT \hat{\bTheta}^\top \hat{\bGamma} \hat{\bTheta} \bT^\top \), whose diagonal components are denoted by \( \{ \hat\Lambda_{j,j} \}_{j=1}^{2d} \). Let \( \hat{\sigma} \) be an estimator of the standard deviation \( \sigma \) of the model error \( \varepsilon_i \), which will be specified in Section \ref{sec:specselect}. Write \( \hat{\bbeta}_{1}^{(\rm bc)} = (\hat{\beta}_{1,1}^{(\rm bc)}, \ldots, \hat{\beta}_{1,d}^{(\rm bc)})^\top \) and \( \hat{\bbeta}_{2}^{(\rm bc)} = (\hat{\beta}_{2,1}^{(\rm bc)}, \ldots, \hat{\beta}_{2,d}^{(\rm bc)})^\top \). 
Based on \eqref{eq:est-debias} and \eqref{eq:transform}, for \( j \in [d] \), we define a set of paired test statistics \( (t_{1,j}, t_{2,j}) \) as follows: 
\begin{equation}\label{test1}
    t_{1,j} := \frac{n^{1/2} \hat{\beta}_{1,j}^{(\rm bc)}}{\hat{\sigma} \hat\Lambda_{j,j}^{1/2}} 
    ~~\text{and}~~ 
    t_{2,j} := \frac{n^{1/2} \hat{\beta}_{2,j}^{(\rm bc)}}{\hat{\sigma} \hat\Lambda_{j+d,j+d}^{1/2}}\,.
\end{equation} 
As rigorously established in our main theoretical analysis in Section \ref{proof_thm0SM} of the Supplementary Material, the two sets of test statistics, \( \{t_{1,j}\}_{j=1}^d \) and \( \{t_{2,j}\}_{j=1}^d \), are asymptotically independent of each other, and each set converges jointly in distribution to a multivariate normal distribution.
As a result of our rigorous theoretical development, the \( p \)-values associated with each \( t_{1,j} \) and \( t_{2,j} \) can be calculated with guaranteed validity, enabling a key component for developing a practically useful multiple testing method.  

The two sets of test statistics provide opportunities for developing high-dimensional multiple testing methodologies. 
In particular, we note that, since \( \bD \) is diagonal, the statistics \( \{t_{2,j}\}_{j=1}^d \) are asymptotically mutually independent. 
Consequently, the conventional BH procedure, as described in Algorithm \ref{method_1}, is applicable for solving \eqref{eq:mtest}. 
The construction of model-X knockoffs and the resulting variance structure in \eqref{eq:kcov} necessitate the positive definiteness constraint $2\bSigma - \bD \succ \bzero$. This requirement often results in small entries of the diagonal matrix $\bD$, and consequently, larger variance for certain entries of \( \hat{\bbeta}_{2}^{(\rm bc)} \). A BH procedure relying solely on $\{t_{2,j}\}_{j=1}^d$ may therefore exhibit insufficient discovery power, a limitation that motivates our integrated two-step approach. More importantly, the paired nature of the two sets of test statistics provides a unique opportunity to combine their strengths -- a structural advantage that facilitates power gains and motivates our integrated two-step approach. This paired $p$-value-based approach is detailed in Algorithm \ref{method_2} as a two-step framework: 
the first step performs an initial selection using the Bonferroni method, and the second step makes the final decision using the BH method with a set of adjusted \( p \)-values derived from the first step. 
Accordingly, we refer to the procedure in Algorithm \ref{method_2} as the Bonferroni-Benjamini-Hochberg method. The theoretical guarantees of asymptotic FDR control  for both Algorithms \ref{method_1} and \ref{method_2} are established in Section \ref{sec:theory}. See Theorems \ref{thm0} and \ref{thm1} for details. Furthermore, Theorem \ref{thm_power} in Section \ref{sec:theory} shows the power gains of Algorithm \ref{method_2} relative to Algorithm \ref{method_1}.

\floatname{algorithm}{Algorithm}
\begin{algorithm}
    \caption{Benjamini-Hochberg with test statistics $\{t_{2,j}\}_{j=1}^d$}
    \label{method_1}

    \begin{algorithmic}
        \STATE\textbf{Step 1}. Let $\tilde{P}_j=P_j^{(2)}=G(|t_{2, j}|)$ for each $j\in[d]$, where $G(t)=2\{1-\Phi(t)\}$ and $\Phi(t)$ is the cumulative distribution function of the standard normal distribution.
        \STATE\textbf{Step 2}. Given $\alpha \in(0,1)$, let $\tilde{P}_{(1)} \leq \cdots \leq \tilde{P}_{(d)}$ be the ordered versions of the $\tilde{P}_j$'s, find
        \begin{equation*}
            \tilde{R}=\max\bigg\{i\in[d]: \tilde{P}_{(i)} \leq \frac{i \alpha}{d}\bigg\}\,,
        \end{equation*}
        provided that the maximum exists; otherwise, let $\tilde{R}=0$.
        \STATE\textbf{Step 3}. Reject the null hypotheses corresponding to $\tilde{P}_{(j)}$ with $j \leq \tilde{R}$.
    \end{algorithmic}
\end{algorithm}

\floatname{algorithm}{Algorithm}
\begin{algorithm}
    \caption{Bonferonni-Benjamini-Hochberg with paired test statistics $\{(t_{1,j},t_{2,j})\}_{j=1}^d$ }
    \label{method_2}

    \begin{algorithmic}
        \STATE\textbf{Step 1}. Given $0<\sqrt{\alpha}<1$, for each $j\in[d]$, let
        \begin{align*}
            \tilde{P}_j= \left\{
        	\begin{aligned}
        		1\,, ~~~&\textrm{if}~P_j^{(1)}>\sqrt{\alpha} \,,\\
        		P_j^{(2)} \,, ~~&\textrm{if}~P_j^{(1)} \leq \sqrt{\alpha} \,,
        	\end{aligned}
        	\right.
        \end{align*}
            where $P_j^{(1)}=G(|t_{1, j}|)$ and $P_j^{(2)}=G(|t_{2, j}|)$.
        \STATE\textbf{Step 2}. Let $\tilde{P}_{(1)} \leq \cdots \leq \tilde{P}_{(d)}$ be the ordered versions of the $\tilde{P}_j$'s. Find
        \begin{equation*}
            \tilde{R}=\max\bigg\{i\in[d]: \tilde{P}_{(i)} \leq \frac{i \sqrt{\alpha}}{d}\bigg\}\,,
        \end{equation*}
        provided that the maximum exists; otherwise, let $\tilde{R}=0$.
        \STATE\textbf{Step 3}. Reject the null hypotheses corresponding to $\tilde{P}_{(j)}$ with $j \leq \tilde{R}$.
    \end{algorithmic}
\end{algorithm}

Here, Algorithm \ref{method_2} involves two pre-specified significance levels: one for the Bonferroni step and the other for the BH step. 
For simplicity in our presentation, we set both levels to \( \sqrt{\alpha} \) as a recommended balanced choice. 
More generally, in practice, one has the flexibility to use two different levels, denoted \( \alpha_1 \) and \( \alpha_2 \) for the Bonferroni and BH procedures, respectively, provided that \( \alpha_1 \alpha_2 = \alpha \), in order to maintain the pre-specified FDR level \( \alpha \); see the general form of Algorithm \ref{method_2} in Section \ref{proof_thm1} of the Supplementary Material.

While it is feasible within our framework to develop adaptive multiple testing procedures by incorporating an estimate of the proportion of true null hypotheses -- e.g., following the approach of \cite{Storey2002A} -- we omit this extension to maintain focus on our primary objective. 
Moreover, in high-dimensional multiple testing, the proportion of true nulls is typically high under reasonable scenarios, rendering such adjustments less critical in practice.

\subsection{Selection of $(\hat{\bTheta},\hat{\sigma})$}\label{sec:specselect}

We conclude this section with specific choices for estimating the intermediate parameters \( \hat{\bTheta} \) and \( \hat{\sigma} \) involved in the multiple testing procedures elaborated in Section~\ref{sec:fdrproc}. 

To ensure the effectiveness of the debiasing procedure, a suitable estimator \( \hat{\bTheta} \) is required in \eqref{de_est}. 
In this study, we adopt the CLIME approach proposed by \cite{cai2011constrained}. 
Specifically, the estimator \( \hat{\bTheta} \in \mathbb{R}^{2d \times 2d} \) is obtained by solving the following optimization problem:
\begin{equation}\label{CLIME}
    \min |\bTheta|_1 \quad \text{subject to} \quad |\hat{\bGamma} \bTheta - \bI_{2d}|_{\infty} \leq \varrho_2\,,
\end{equation}
where \( \varrho_2 > 0 \) is a tuning parameter.

For the estimator \( \hat{\sigma} \) of the standard deviation of the model error involved in \eqref{test1}, we employ the scaled Lasso estimator \citep{sun2012scaled}. 
This estimator is obtained by solving the following optimization problem:
\begin{align}\label{eq:vest}
    \min_{\bgamma \in \mathbb{R}^{2d},\, \sigma > 0} \bigg\{ \frac{1}{2\sigma n} | \by - \bZ \bgamma |_2^2 + \frac{\sigma}{2} + \varrho_3 |\bgamma|_1 \bigg\}\,,
\end{align}
where \( \varrho_3 > 0 \) is a tuning parameter. 
For the choice of \( \varrho_3 \), we follow the quantile-based penalty level introduced and analyzed by \cite{sun2013sparse}.

In summary, our work presents a novel multiple testing framework tailored for high-dimensional linear models. The core of our framework lies in a debiased Lasso methodology applied to an augmented design matrix \( (\bX, \tilde{\bX}) \) that incorporates knockoff variables, effectively addressing key inferential challenges in high-dimensional settings. A major innovation lies in the analytical properties of the transformed estimators, which enable the construction of a novel set of paired test statistics and the evaluation of the associated \( p \)-values. 
Our methodology, as elaborated in Algorithms \ref{method_1} and \ref{method_2}, facilitates not only the application of the conventional BH procedure but also a novel two-step Bonferroni-Benjamini-Hochberg procedure. 
Our theoretical results establish that both approaches guarantee control of the FDR, even under unknown dependence structures among the high-dimensional test statistics.

\section{Theoretical guarantees}\label{sec:mainresults}

\subsection{Overview}

Our investigation not only advances existing methodologies in practice, but also provides a theoretically sound solution to multiple testing in complex high-dimensional settings. 
The first objective of our comprehensive theoretical analysis is to establish that the transformed bias-corrected estimators defined in \eqref{eq:transform}, along with their associated \( p \)-values, possess the necessary asymptotic properties. 
Our main theoretical results confirm that the resulting high-dimensional test statistics in \eqref{test1} exhibit the required distributional behavior. 
As a second objective, we rigorously demonstrate that the proposed multiple testing procedures -- Algorithms \ref{method_1} and \ref{method_2} -- can control the FDR at the nominal level asymptotically.

Analyzing the theoretical properties of our proposed $p$-value-based multiple testing methods presents several significant challenges due to the intricate nature of high-dimensional models with knockoff variables. First, the linear model with the augmented matrix \( (\bX, \tilde{\bX}) \) requires dedicated analysis of the Lasso estimator \eqref{eq:lasso}, the debiased estimator \eqref{de_est}, and the transformation \eqref{eq:transform}. 
The inclusion of knockoff variables introduces complex dependence structures, making parameter estimation more challenging and requiring careful theoretical treatment to ensure proper control of these dependencies.
Further challenges arise from analyzing and ensuring the properties of the intermediate parameter estimates, including the high-dimensional precision matrix \( \hat{\bTheta} \) using the CLIME estimator \eqref{CLIME}, and the standard deviation estimate \( \hat{\sigma} \) through the scaled Lasso estimator \eqref{eq:vest}.

We present the relevant technical conditions. 

\begin{assumption}\label{ass:model_error}
The model error $\beps=(\varepsilon_1,\ldots,\varepsilon_n)^{\top}\sim\mathcal{N}({\bf0},\sigma^2\bI_{n})$ and is independent of $\bZ$.
\end{assumption}

\begin{assumption}\label{ass:random_Z}
The covariance matrix $\bGamma$ satisfies $C_{\min}\leq \sigma_{\min}(\bGamma) \leq \sigma_{\max}(\bGamma) \le C_{\max}$ for some constants $C_{\min}\in(0,\infty)$ and $C_{\max}\in[1,\infty)$, and $\bZ\bGamma^{-1/2}$ has independent sub-Gaussian rows with zero mean and sub-Gaussian norm $\|\bGamma^{-1/2}\bz\|_{\psi_2} = \kappa$ for some constant $\kappa \in (0,\infty)$, where $\bz=(\bx^{\top},\tilde{\bx}^{\top})^{\top}$.
\end{assumption}

\begin{assumption}\label{ass:CLIME}

Let
 $\mathcal{U}(M, q, s_d) = \{ \bTheta=(\theta_{i,j})_{2d\times 2d} : \bTheta \succ 0,\, \|\bTheta\|_1 \leq M,\, \max_{i \in [2d]} \sum_{j=1}^{2d} |\theta_{i,j}|^q \leq s_d \}
$ for some constants \( q \in[0,1) \) and $M\in(0,\infty)$.
Assume that the precision matrix \( \bTheta_0 = \bGamma^{-1} \in \mathcal{U}(M,q, s_d) \), and that the estimator \( \hat{\bTheta} \) is obtained from \eqref{CLIME} with tuning parameter \( \varrho_2 \ge a\{n^{-1}\log(2d)\}^{1/2} \) with \( a = \max\{C_0 M, 4e\kappa^2\sqrt{6(\tau+2)C_{\max}C_{\min}^{-1}}\} \), where \(\tau>0\) and \( C_0 = \eta^{-1}(2 + \tau + \eta^{-1}K_0^2) \) with  \( \eta = \min\{{1}/{8}, (4C_{\max}\kappa^{2}e)^{-1}\} \) and  \( K_{0} =(1 - 2C_{\max}\kappa^{2}e\eta)^{-1} \).
\end{assumption}

\begin{assumption}\label{ass0}\sloppy   
There exists $\mathcal{H}=\mathcal{H}_{n}\subseteq[d]$ such that $|\mathcal{H}|\geq\log\log d$ and $|\beta^0_{j}|\geq2\sqrt{2}C_{\min}^{-1/2}\sigma(n^{-1}\log d)^{1/2}$ for all $j\in\mathcal{H}$, where $C_{\min}$ is specified in Condition {\rm\ref{ass:random_Z}}.
\end{assumption}

Condition \ref{ass:model_error} is standard for analyzing linear models and has been widely assumed in many existing studies; see, for example,  \cite{vandeGeer2014on}. The independence between $\beps$ and $\tilde{\bX}$ follows directly from property (b) of the model-X knockoffs.
Condition \ref{ass:random_Z} imposes restrictions on the maximum and minimum singular values of $\bGamma$ and characterizes the tail behavior of $\bGamma^{-1/2}\bz$. This condition is essential for establishing that as $n \to \infty$, with high probability, the compatibility constant and the generalized coherence parameter associated with the random design matrix $\bZ$ remain bounded. These concepts play a crucial role in the theoretical analysis of high-dimensional Lasso estimators; see \cite{Hastieetal_2015} and references therein.
The conditions on $\bGamma$, $\bZ$, and $\bz$ in Condition \ref{ass:random_Z} are equivalent to imposing constraints on the singular values of $\bSigma$ and $\bD$, as well as the sub-Gaussian behavior of $\bx$. More detailed discussions on these equivalences can be found in Remarks \ref{remark_condition1} and \ref{prop_condition} below.
Condition \ref{ass:CLIME} pertains to the analysis of the CLIME estimator \eqref{CLIME}, following the same principles as in \cite{cai2011constrained}. It assumes that the true precision matrix $\bTheta_0$ belongs to a class of sparse matrices  and requires the tuning parameter $\varrho_2$ to be properly scaled to ensure the estimator converges to the true matrix at a desirable rate under the elementwise $\ell_{\infty}$-norm.
Finally, Condition \ref{ass0} concerns the minimal signal strength. It also imposes constraints on the number of false null hypotheses, ensuring it is not too small -- similar to requirement (2) in Theorem 2.1 of  \cite{liu2014phase}.

\begin{remark}\label{remark_condition1}
Condition {\rm\ref{ass:random_Z}} for $\bGamma$ is equivalent to imposing conditions on the singular values of $\bSigma$ and $\bD$, respectively. Notice that $\bD={\rm diag}(\bs)$ with $\bs\in\mathbb{R}^{d}$, and $\bGamma$ defined in \eqref{eq:kcov} is similar to ${\rm diag}(2\bSigma-\bD, \bD)$. On one hand, by Weyl inequality and Condition {\rm\ref{ass:random_Z}}, we have $C_{\min}\le \sigma_{\min}(\bD)\le 2\sigma_{\min}(\bSigma)-C_{\min}$, $2\sigma_{\max}(\bSigma)-C_{\max}\le \sigma_{\max}(\bD)\le C_{\max}$, and $C_{\min}\le\sigma_{\min}(\bSigma)\le\sigma_{\max}(\bSigma)\le C_{\max}$. On the other hand, letting $\tilde{C}_{\min}\le\sigma_{\min}(\bSigma)\le\sigma_{\max}(\bSigma)\le\tilde{C}_{\max}$ and $C_{\min}\le \sigma_{\min}(\bD)\le \sigma_{\max}(\bD)\le C_{\max}<2\tilde{C}_{\min}$ for some positive constants $\tilde{C}_{\min}$ and $\tilde{C}_{\max}$, then $\sigma_{\min}(\bGamma) \ge \min(2\tilde{C}_{\min}-C_{\max},C_{\min})>0$ and $\sigma_{\max}(\bGamma) \le \max(2\tilde{C}_{\max}-C_{\min},C_{\max})$.
\end{remark}

\begin{remark}\label{prop_condition}
If $\bX$ has independent sub-Gaussian rows, with zero mean and sub-Gaussian norm $\|\bx\|_{\psi_2} = \kappa_1$, then the model-X knockoff variate $\tilde{\bX}$ also has independent sub-Gaussian rows, with zero mean and sub-Gaussian norm $\|\tilde{\bx}\|_{\psi_2} = \kappa_1$. Furthermore, the augmented design matrix $\bZ$ has independent sub-Gaussian rows, with zero mean and sub-Gaussian norm $\kappa_1\le\|\bz\|_{\psi_2}\le 2\kappa_1$, and satisfies $\| \bz\|_{\psi_{2}}\sigma_{\max}^{-1}(\bGamma^{1/2})\le
\|\bGamma^{-1/2}\bz\|_{\psi_{2}}\le\| \bz\|_{\psi_{2}}\sigma_{\max}(\bGamma^{-1/2})$, where $\bz=(\bx^{\top},\tilde{\bx}^{\top})^{\top}$.
\end{remark}

\subsection{Main results}\label{sec:theory}

Let \( \mathcal{H}_0 = \{j \in [d] : \beta_{j}^{0} = 0\} \), and define
\begin{align*}
    {\rm FDR} = \mathbb{E}({\rm FDP})~~\textrm{with}~~{\rm FDP} = \frac{\sum_{j \in \mathcal{H}_0} I\{\tilde{P}_j \le \tilde{P}_{(\tilde{R})}\}}{\tilde{R} \vee 1}\,.
    \end{align*}
In our theoretical analysis, we focus on the challenging scenario where the number of predictors $d$ will diverge with the sample size $n$. Let $d_{0}$ be the number of true null hypotheses, and write $\pi_0 = d_0/d$. Based on the properties of \( \hat{\bgamma}^{(\rm bc)} \) given in \eqref{de_est} and \(\hat{\sigma}\) defined in \eqref{eq:vest} (see Theorem \ref{theo_random} and Lemma \ref{theo_scaled} in Section \ref{proof_thm0SM} of the Supplementary Material), we establish the following theorems, which guarantee control of the FDR under our proposed procedures. 

\begin{theorem}\label{thm0}
Let Conditions {\rm \ref{ass:model_error}--\rm \ref{ass0}} hold. Assume that $\tau>2$ and $|\bgamma_{0}|_{0}\le s_0<2d$ with $s_{0}\ll n^{1/2}(\log d)^{-3/2}$. Consider a suitable choice of the tuning parameters $\varrho_1\asymp\varrho_2\asymp\varrho_3\asymp\{n^{-1}\log(2d)\}^{1/2}$ in \eqref{eq:lasso}, \eqref{CLIME} and \eqref{eq:vest}, respectively. If $\log d \ll n^{1/(5+2\vartheta)}$ for some $\vartheta > 0$, then Algorithm {\rm \ref{method_1}} can control {\rm FDR} asymptotically at $\pi_0\alpha$, that is $\lim_{n\rightarrow\infty}\mathbb{P}({\rm FDP}\leq \pi_0 \alpha+\epsilon)=1$ 
for any $\epsilon>0$, and
$
\mathop{\lim\sup}_{n\rightarrow\infty}{\rm FDR}\leq \pi_0 \alpha$. 
\end{theorem}

\begin{theorem}\label{thm1}
Let the conditions of Theorem {\rm \ref{thm0}} hold. If $s_d^2\ll d$ and $s_{0}\lesssim n^{1/2}(\log d)^{-5/2-2\vartheta}$ for $\vartheta$ specified in Theorem {\rm\ref{thm0}}, then Algorithm {\rm \ref{method_2}} can control {\rm FDR} asymptotically at $\pi_0\alpha$, that is
$
\lim_{n\rightarrow\infty}\mathbb{P}({\rm FDP}\leq \pi_0 \alpha+\epsilon)=1$
for any $\epsilon>0$, and
	$\mathop{\lim\sup}_{n\rightarrow\infty}{\rm FDR}\leq \pi_0 \alpha$. 
\end{theorem}

The proofs of Theorems \ref{thm0} and \ref{thm1} are provided in Sections \ref{proof_thm0} and \ref{proof_thm1} of the Supplementary Material, respectively. Write 
$\mathcal{H}_1 = \{ j \in [d] : \beta_j^0 \neq 0 \}$.
The powers of Algorithms \ref{method_1} and \ref{method_2} can be, respectively, formulated as
\begin{align*}
    \Psi_1 
    = \mathbb{E}\Bigg[ 
    \frac{\sum_{j \in \mathcal{H}_1} 
    I\{ P_j^{(2)} \le \tilde{P}_{(\tilde{R}_1)} \}}
    {|\mathcal{H}_1|} 
    \Bigg]~~\text{and}~~\Psi_{\sqrt{\alpha}} 
    = \mathbb{E}\Bigg[ 
    \frac{\sum_{j \in \mathcal{H}_1} 
    I\{ \tilde{P}_j \le \tilde{P}_{(\tilde{R}_2)} \}}
    {|\mathcal{H}_1|} 
    \Bigg]\,,
\end{align*}
where $\tilde{R}_1$ and $\tilde{R}_2$ denote the rejection thresholds obtained in the second step of Algorithms \ref{method_1} and \ref{method_2}, respectively. 
For power comparison, we impose the following signal strength condition.

\begin{assumption}\label{Sass0}
For all $j \in \mathcal{H}_1$, it holds that
$|\beta_j^0| 
\ge 8\sqrt{2}\, C_{\min}^{-1/2}\sigma (n^{-1}\log d)^{1/2}$,
where $C_{\min}$ is specified in Condition {\rm\ref{ass:random_Z}}.
\end{assumption}

Theorem \ref{thm_power} shows that Algorithm \ref{method_2} dominates Algorithm \ref{method_1} in power, whose proof is provided in Section \ref{proof_power} of the Supplementary Material.

\begin{theorem}\label{thm_power}
Under Condition {\rm\ref{Sass0}} and the conditions of Theorem {\rm\ref{thm1}}, it holds that $\Psi_{\sqrt{\alpha}} \ge \Psi_1 - o(1)$ as $n \to \infty$. 
\end{theorem}

\begin{remark}
Theorem {\rm\ref{thm_power}} confirms the power advantage of Algorithm {\rm\ref{method_2}} over Algorithm {\rm\ref{method_1}}. As discussed earlier, although $\hat{\bbeta}_2^{(\rm bc)}$ satisfies the conditions required for applying the {\rm BH} procedure, it typically incurs a larger variance due to the constraint $2\bSigma - \bD \succ 0$. 
Algorithm {\rm\ref{method_2}} effectively exploits the pairing structure between $\hat{\bbeta}_1^{(\rm bc)}$ and $\hat{\bbeta}_2^{(\rm bc)}$. By incorporating this structure and performing the Bonferroni step, it achieves improved power while maintaining {\rm FDR} control. 
\end{remark}

\subsection{Further methodological and theoretical development} \label{sec:newtheory}

The properties of the knockoffs in \eqref{eq:kcov}, together with Conditions \ref{ass:model_error}--\ref{ass0}, ensure the theoretical validity of our framework as established in Theorems \ref{thm0}--\ref{thm_power}, including asymptotic FDR control and the power gain achieved by incorporating the pairing structure of the test statistics.

Extending the scope of the proposed framework is relevant in broader settings, particularly when (a) the knockoff construction procedure depends on additional estimated parameters, such as the variance--covariance matrix $\bSigma$ and/or its inverse; and (b) the model error distribution is more general, for example, sub-Gaussian.

These issues are addressed in our further investigations. In particular, in Section~\ref{F:F_0} of the Supplementary Material, we extend the methodological framework to accommodate estimated variance--covariance structures in the construction of knockoff variables, as well as more general error distributions, such as sub-Gaussian errors. In particular, we investigate a data-splitting strategy in the same spirit as \cite{Fan02012020}, where one part of the data is used to estimate the precision matrix of the design matrix, based on which the knockoff variables are constructed; the other part is then used to estimate the model parameters in \eqref{eq:lm2} and to construct the test statistics.
We show that this data-splitting strategy, combined with an estimated precision matrix in the knockoff construction, is capable of accommodating a broad class of design matrices under weaker assumptions on the distribution of the variables in the design matrix.  Our theoretical analysis shows that the proposed framework continues to control the FDR under these broader settings; see also its empirical performance presented in Section \ref{simu_ds} of the Supplementary Material.

\section{Simulations}\label{sec:sim}

\subsection{Setting}
In the simulations, data are generated from the linear model \eqref{eq:lm}, with model errors drawn independently from a standard normal distribution. The design matrix $\bX$ varies across three settings, reflecting different dependence structures between variables:

\begin{description}
    \item[Setting 1.] The rows of $\bX$ are generated from a multivariate normal distribution $\mathcal{N}({\bf 0},\bI_{d})$.
    \item[Setting 2.] The rows of $\bX$ are drawn from a multivariate normal distribution with an AR(1) dependence structure and correlation coefficient 0.4. This setting has been explored in the literature; see \cite{Barber2015Controlling}, \cite{candes2018panning}, and \cite{Sarkar2022Adjusting}.
    \item[Setting 3.] The rows of $\bX$ are generated from a multivariate normal distribution with block dependence. Specifically, samples are drawn from $\mathcal{N}({\bf0}, \bSigma)$, where $\bSigma = (\Sigma_{i,j})_{d\times d}$ with $\Sigma_{i,i}=1$ and $\Sigma_{i,j} = 0.2 \cdot I(\lceil i/20 \rceil = \lceil j/20 \rceil)$. This block structure has also been studied in prior work; see \cite{Fithian2022Conditional}. 
\end{description}

These designs allow us to investigate how different correlation structures in $\bX$ affect the performance of the statistical methods under consideration. 
The matrix containing the knockoff variables is constructed following the second-order model-X knockoff procedure of \cite{candes2018panning}. 
We generate knockoff variables by calling the {\texttt{R}}-function {\texttt{create.second\underline{~}order}} in the {\texttt{knockoff}} package. We use the {\texttt{R}}-function \texttt{cv.glmnet} in the \texttt{glmnet} package to obtain the Lasso estimator. To accelerate the computation of the precision matrix, we utilize the {\texttt{R}}-function {\texttt{Inverse}} provided by \cite{javanmard2014confidence}\footnote{\url{https://web.stanford.edu/~montanar/sslasso/home.html}}. The scaled Lasso estimator is implemented by the \texttt{R}-function \texttt{scalreg} in the \texttt{scalreg} package. 
We explore combinations of \( n \in \{200, 500\} \) and \( d \in \{n, 1.5n, 2n\} \). 
For each setting, we randomly assign \( k \) components of \( \bbeta_0 \) to be nonzero, with the remaining \( d - k \) components set to zero. 
The nonzero coefficients are assigned a common amplitude, which is varied over the interval \( [0.1, 0.5] \) using equally spaced values. 

The simulations consider FDR control levels \( \alpha \in \{0.05, 0.1, 0.2\} \).
Each simulation setting is repeated 100 times to ensure reliable performance estimates. 
In the simulations, we compare the proposed methods in Algorithms \ref{method_1} and \ref{method_2} with the FDR-controlling knockoff procedure of \cite{candes2018panning} and the Gaussian Mirror method of \cite{Xing2021Controlling}. The knockoff procedure is implemented by calling the \texttt{R}-function \texttt{knockoff.filter} in the \texttt{knockoff} package. We implement the Gaussian Mirror method using \texttt{R}-function \texttt{gm} provided by \cite{Xing2021Controlling}\footnote{\url{https://github.com/BioAlgs/GM}}.

\subsection{Summary of finding}

\subsubsection{FDR control}

Our results demonstrate that both proposed methods, as detailed in Algorithms \ref{method_1} and \ref{method_2}, 
control the FDR at the specified levels. 
In the absence of signal (i.e., when all null hypotheses are true), for all pre-specified FDR levels \( \alpha \in \{0.05, 0.1, 0.2\} \), the empirical FDRs achieved by our methods closely match the nominal levels. 
  Figure \ref{FDR-control-0.1} shows the empirical FDR across simulation replications at \(\alpha = 0.1\). Additional results for  FDR levels $\alpha = 0.05$ and $0.2$, with all true null hypotheses, are provided, respectively, in Figures \ref{FDR-control-0.05} and \ref{FDR-control-0.2} in Section \ref{SM1:figure} of the Supplementary Material.  
  These findings confirm the theoretical guarantees established in Section \ref{sec:mainresults} and provide strong empirical support for the validity of our procedures.

In contrast, the FDP-based knockoff approach using \eqref{eq:fdp+} tends to be overly conservative, often yielding no discoveries. This behavior is expected, and possible reasons for it are discussed in Sections \ref{sec:intro} and \ref{sec:method}. A notable observation is that the Gaussian Mirror method, when using \eqref{eq:fdp2} to estimate the empirical FDP, tends to inflate the empirical FDR when all null hypotheses are true. 
For instance, in the absence of signal, the empirical FDR of the Gaussian Mirror method often exceeds the nominal level by a substantial margin. 
As its FDR frequently surpasses 0.5 at \( \alpha = 0.1 \), we have omitted these results from Figure~\ref{FDR-control-0.1} for clarity.
A likely explanation lies in the behavior of the empirical FDP estimator \eqref{eq:fdp2}; see also the related discussion in \cite{Barber2015Controlling} regarding the estimator \eqref{eq:fdp}. 
To investigate this issue, we applied the adjustment strategy given in \eqref{eq:fdp+} -- in the spirit of the approach proposed by \cite{Barber2015Controlling} -- to estimate the empirical FDP for the Gaussian Mirror method. 
This adjustment allows the Gaussian Mirror method to control the FDR, albeit conservatively, similar to the knockoff procedure, but often at the cost of yielding no discoveries; see Figure~\ref{FDR-control-0.1}.

\subsubsection{Power comparisons}

Figure~\ref{Power-0.1} presents power results at \( \alpha = 0.1 \) for selected configurations of \( (n, d) \) under Setting 2 for generating the design matrix. 
Additional power results are provided in Section \ref{SM2:FIGURE} of the Supplementary Material; see Figures~\ref{Power-0.05} and~\ref{Power-0.2} for \( \alpha = 0.05 \) and \( 0.2 \), respectively.

With respect to power, our methods outperform the model-X knockoff approach in settings characterized by lower pre-specified FDR levels, weaker signals, and smaller sample sizes, demonstrating their effectiveness under more challenging conditions. 
A likely explanation is the variability introduced by the ratio estimators used in the empirical FDP calculations of knockoff-based methods \citep{candes2018panning}, which can substantially affect performance when signals are sparse and the total number of rejections is small. 
While the power of the knockoff method tends to improve with larger sample sizes and higher dimensionality, our methods remain competitive across all scenarios; see Figures~\ref{Power-large-0.05}--\ref{Power-large-0.2} in Section \ref{SM2:FIGURE} of the Supplementary Material for further details.

When the empirical FDP is estimated using \eqref{eq:fdp2}, the Gaussian Mirror method of \cite{Xing2021Controlling} tends to exhibit higher power. 
However, in these cases, the empirical FDR often exceeds the nominal level by relatively large margin, consistent with the observations when all null hypotheses are true. 
If the FDP estimation is adjusted using the FDP$+$ procedure in \eqref{eq:fdp+}, the empirical FDR is controlled at the nominal level, but the power of the Gaussian Mirror method decreases to levels comparable to those of the knockoff methods; see Figure~\ref{Power-0.1} and additional results in the Supplementary Material.

Notably, the Gaussian Mirror method employs a two-stage procedure (variable selection in the first stage and statistic evaluation in the second stage).
In this respect, we offer a cautionary observation: when the two-stage procedure of the Gaussian Mirror method fails to correctly identify the true contributing variables in the first stage, the approach becomes less competitive in terms of power. This limitation arises because once the true signals are missed during the initial variable selection stage, they cannot be recovered in the second stage. As demonstrated in the example provided in Section \ref{gm-extra} of the Supplementary Material, the Gaussian Mirror method struggles in settings with higher correlations among variables in the design matrix. See Figures \ref{Highly-correlated-0.05}--\ref{Highly-correlated-0.2} in the Supplementary Material for details. In such cases, Lasso-based variable selection in the first stage may fail, leading to relatively more frequent false exclusions of true contributing variables, particularly when signal strength is relatively weak. 
In these scenarios, the numerical results confirm that our methods outperform the Gaussian Mirror method, especially under more challenging scenarios, demonstrating better power.

\subsubsection{Further investigations of our methods}\label{sec:moresim}

\begin{figure}[htbp!]
	\centering
	\subfloat{\includegraphics[width=.27\columnwidth]{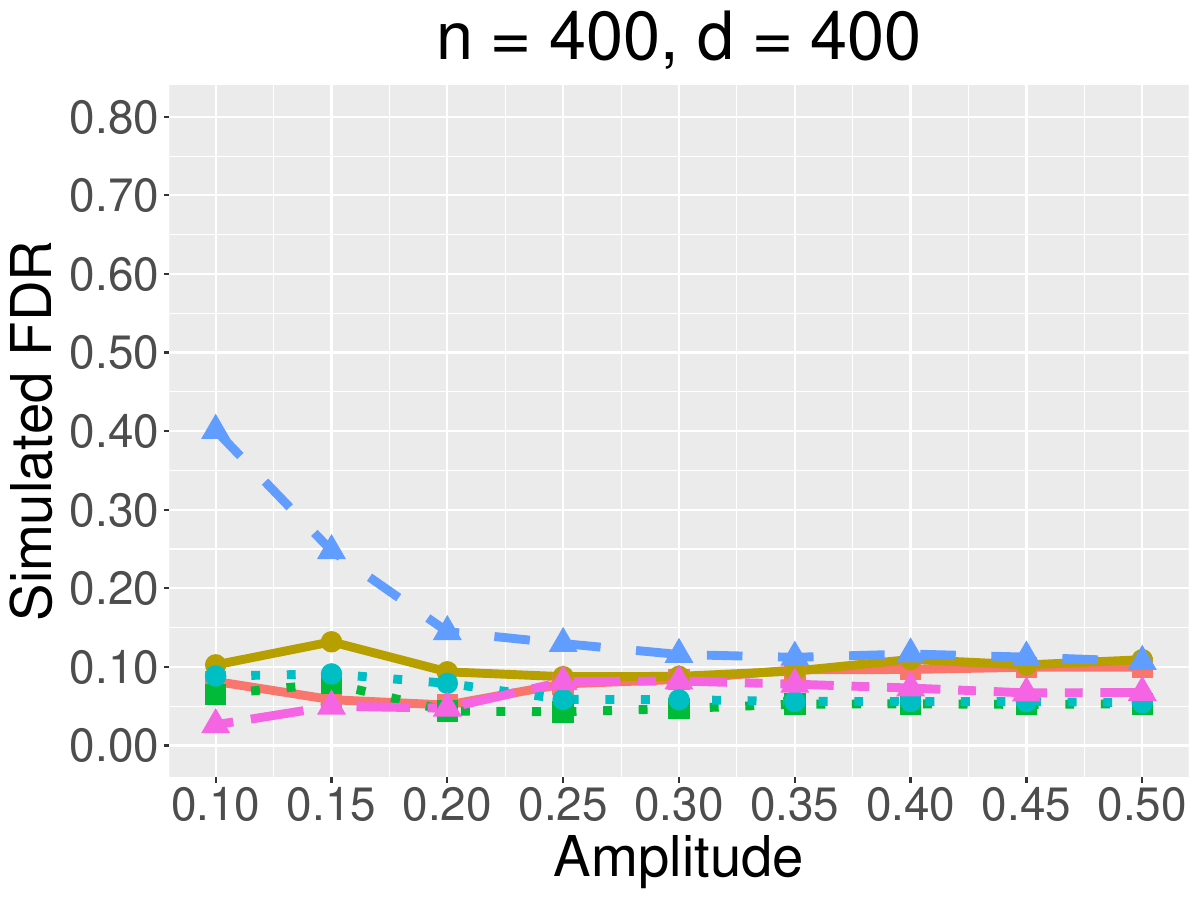}}\hspace{5pt}
	\subfloat{\includegraphics[width=.27\columnwidth]{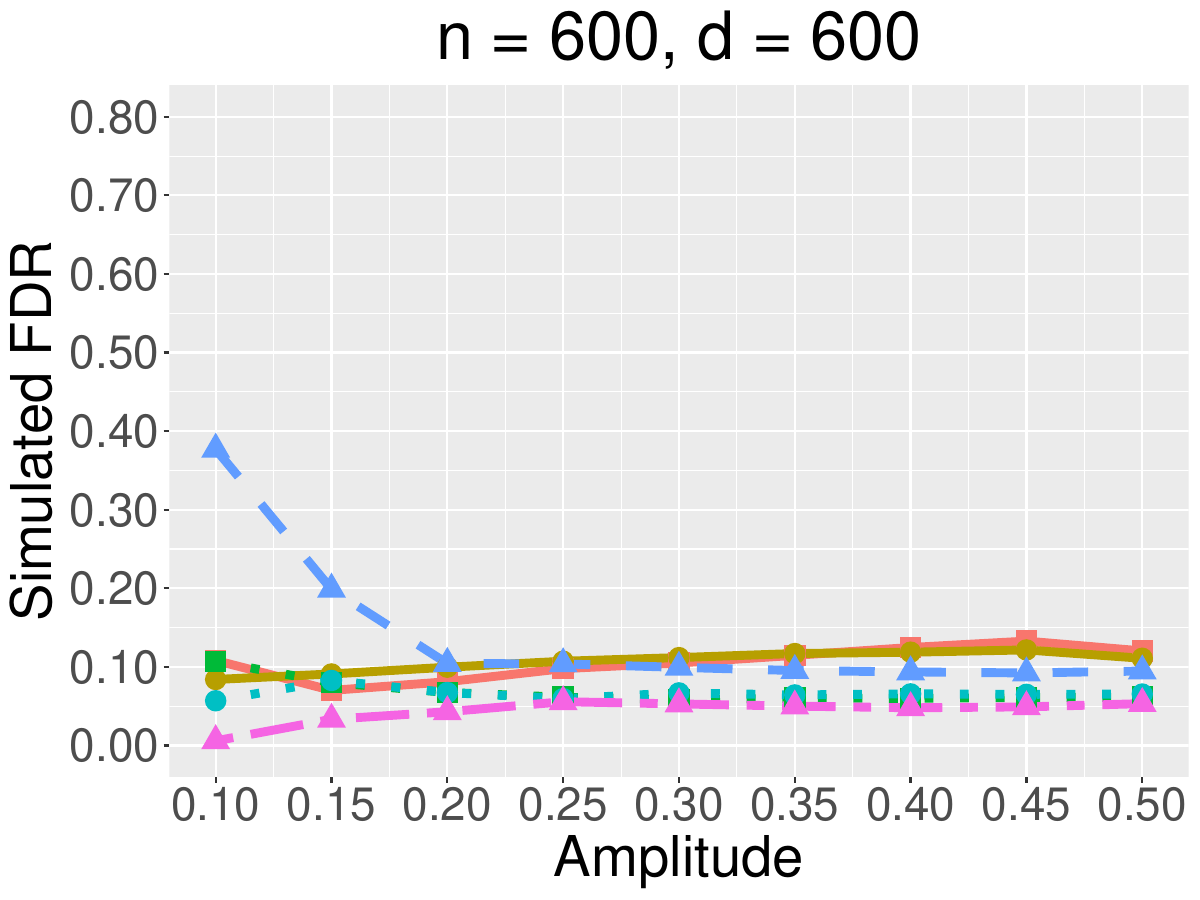}}\hspace{5pt}
	\subfloat{\includegraphics[width=.27\columnwidth]{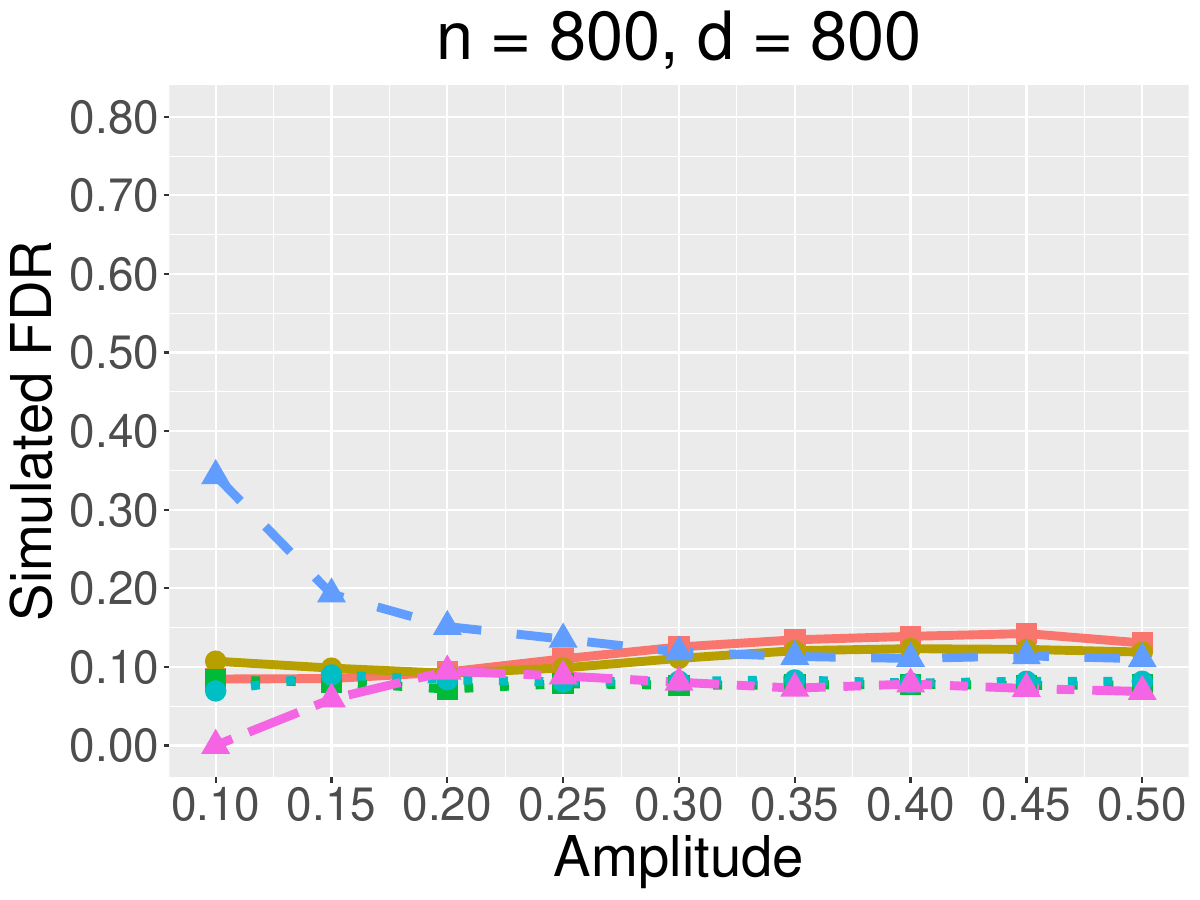}}\\
	\subfloat{\includegraphics[width=.27\columnwidth]{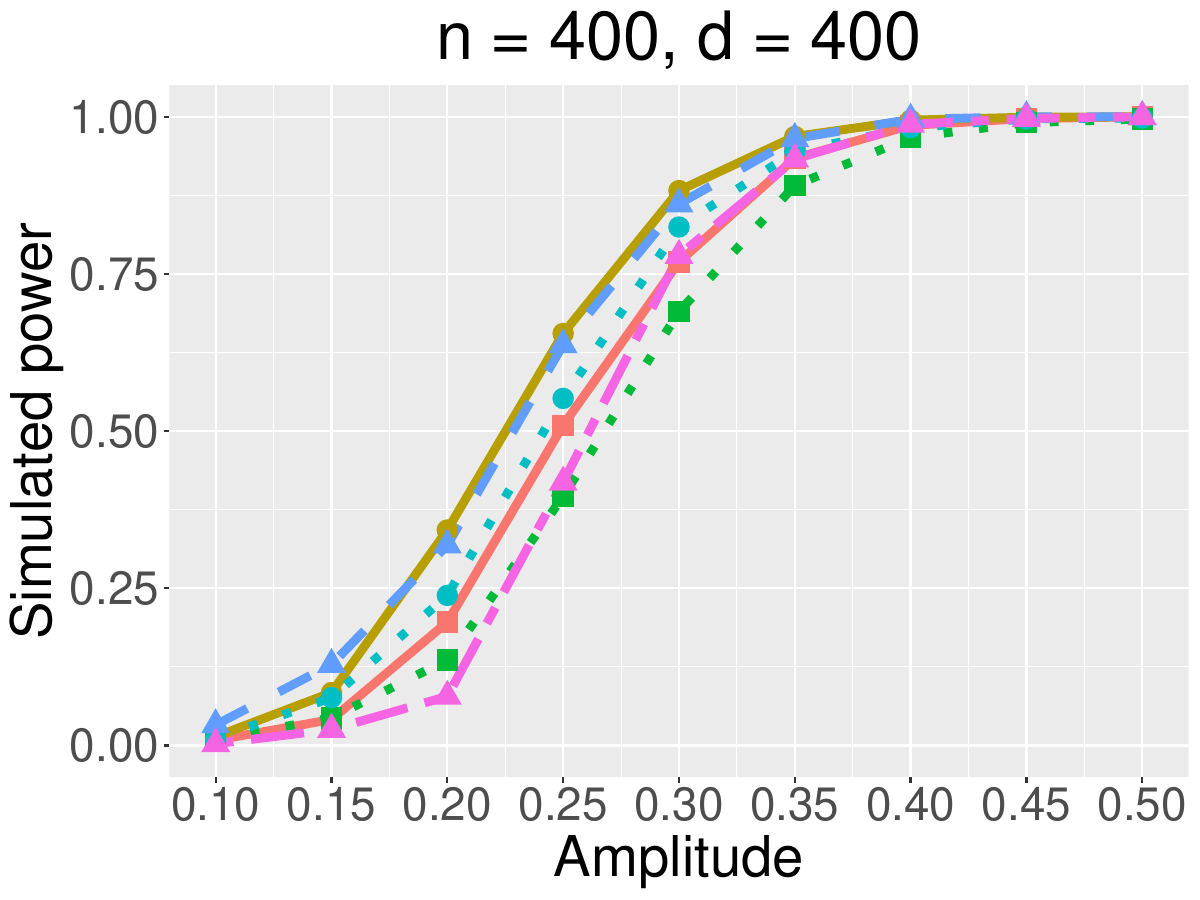}}\hspace{5pt}
    \subfloat{\includegraphics[width=.27\columnwidth]{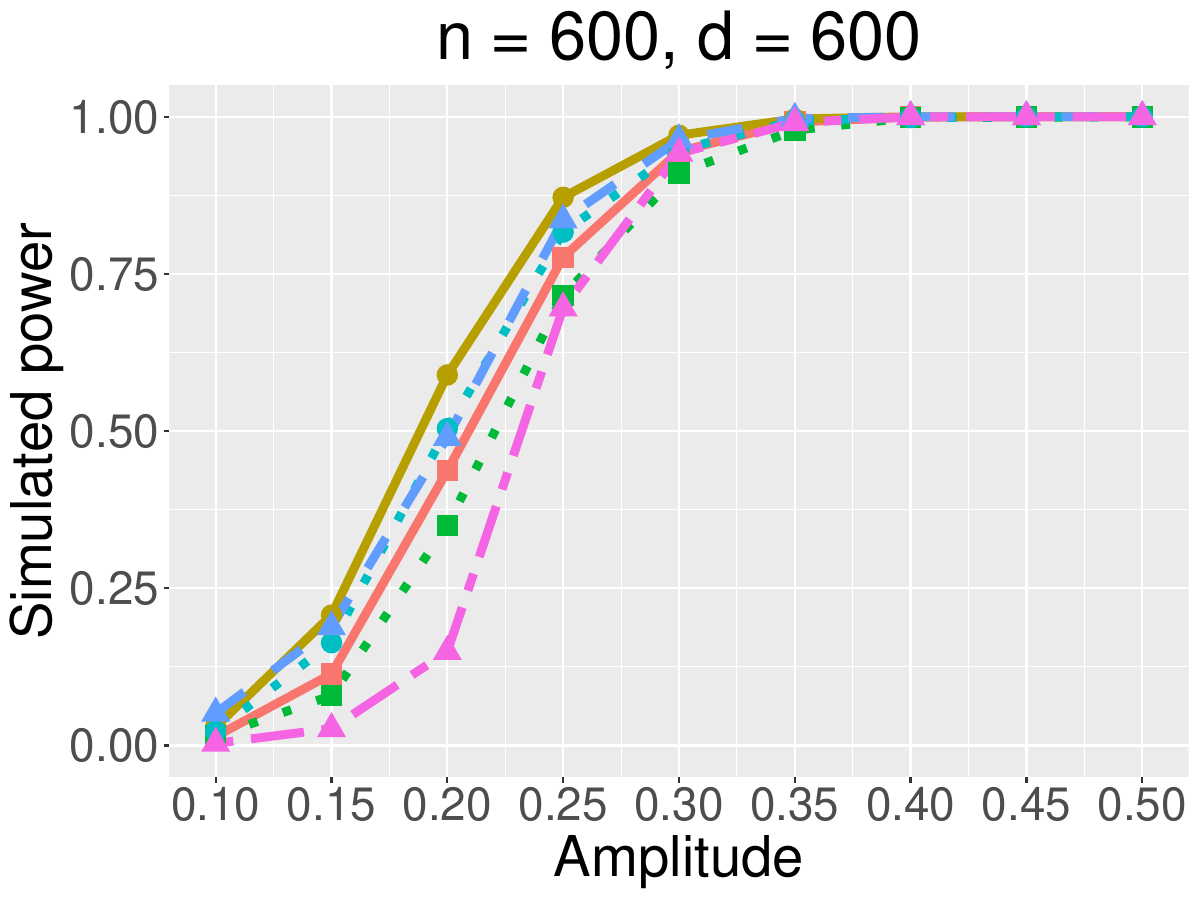}}\hspace{5pt}
    \subfloat{\includegraphics[width=.27\columnwidth]{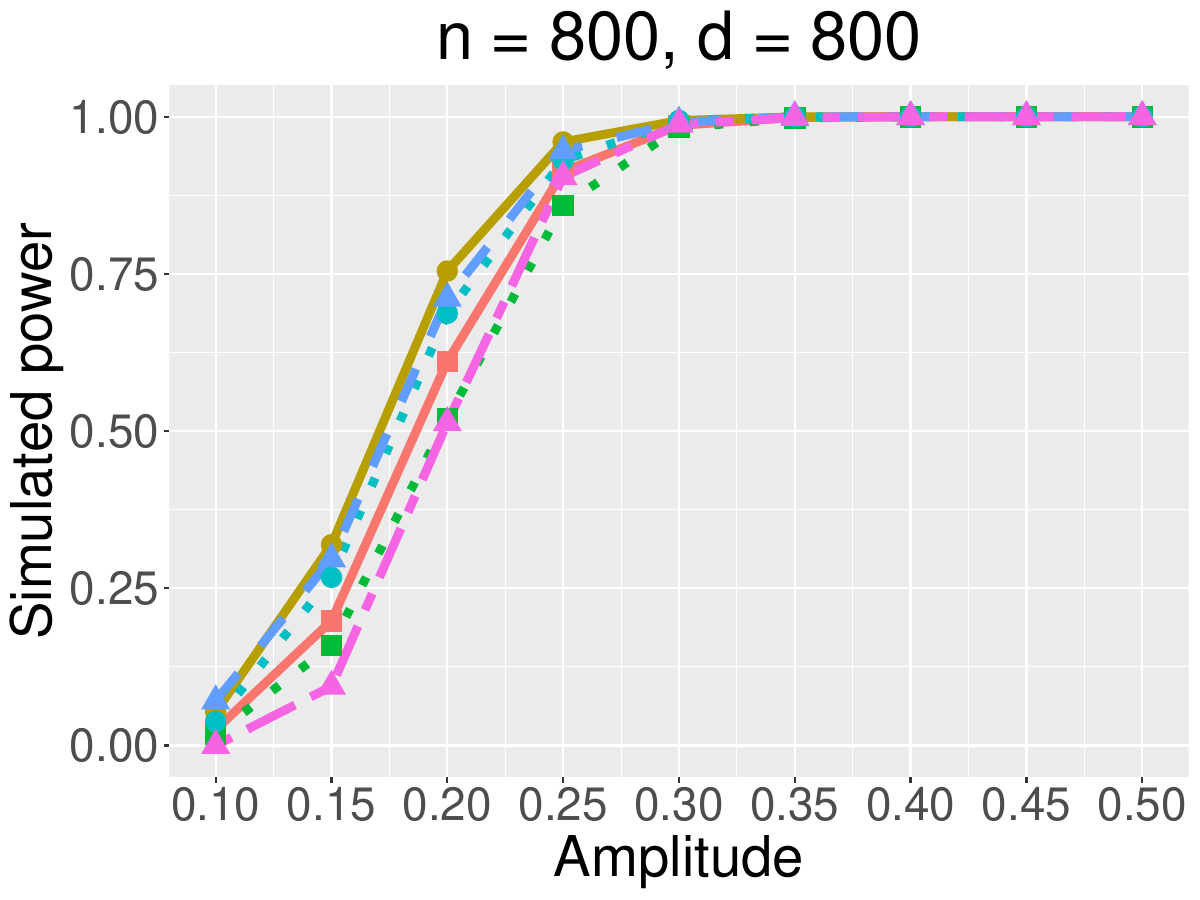}}
	\caption{\small Simulated FDR and power for the settings of $(n,d) = (400, 400)$, $(600, 600)$ and $(800, 800)$. The rows of the design matrix were generated from Setting 2. The sparsity level is $k = 15$ and the FDR level is $\alpha = 0.1$. The methods compared are Algorithm 1 (squares and green dotted line), two-stage Algorithm 1 (squares and red solid line), Algorithm 2 (circles and blue dotted line), two-stage Algorithm 2 (circles and yellow solid line), the Gaussian Mirror method of \cite{Xing2021Controlling} (triangles and blue dashed line), and the Gaussian Mirror method with FDP+ procedure (triangles and purple two-dashed line).}
    \label{Two-stage-0.1}
\end{figure}

Although our framework is formulated as a single-stage procedure based on debiasing techniques, it can be naturally extended to a two-stage approach that first performs greedy variable screening, followed by refined inference using our proposed methods. 
This extension may offer computational benefits and help reduce the effective dimensionality of the problem.
In the same spirit as the two-stage approach employed in the Gaussian Mirror method of \cite{Xing2021Controlling}, we implement a two-stage variant: we first apply variable selection using the same initial screening method, and then apply our \( p \)-value-based multiple testing procedures (Algorithms~\ref{method_1} and~\ref{method_2}) to the design matrix constructed from the selected variables. 
The results, presented in Figure~\ref{Two-stage-0.1}, highlight the competitive power of our methods and demonstrate strong potential for further enhancing their performance through this two-stage adaptation. Additional results are provided in Figures \ref{Two-stage-0.05} and \ref{Two-stage-0.2} in Section \ref{simu_twostage} of the Supplementary Material.

\begin{figure}[htbp!]
	\centering
	\subfloat{\includegraphics[width=.27\columnwidth]{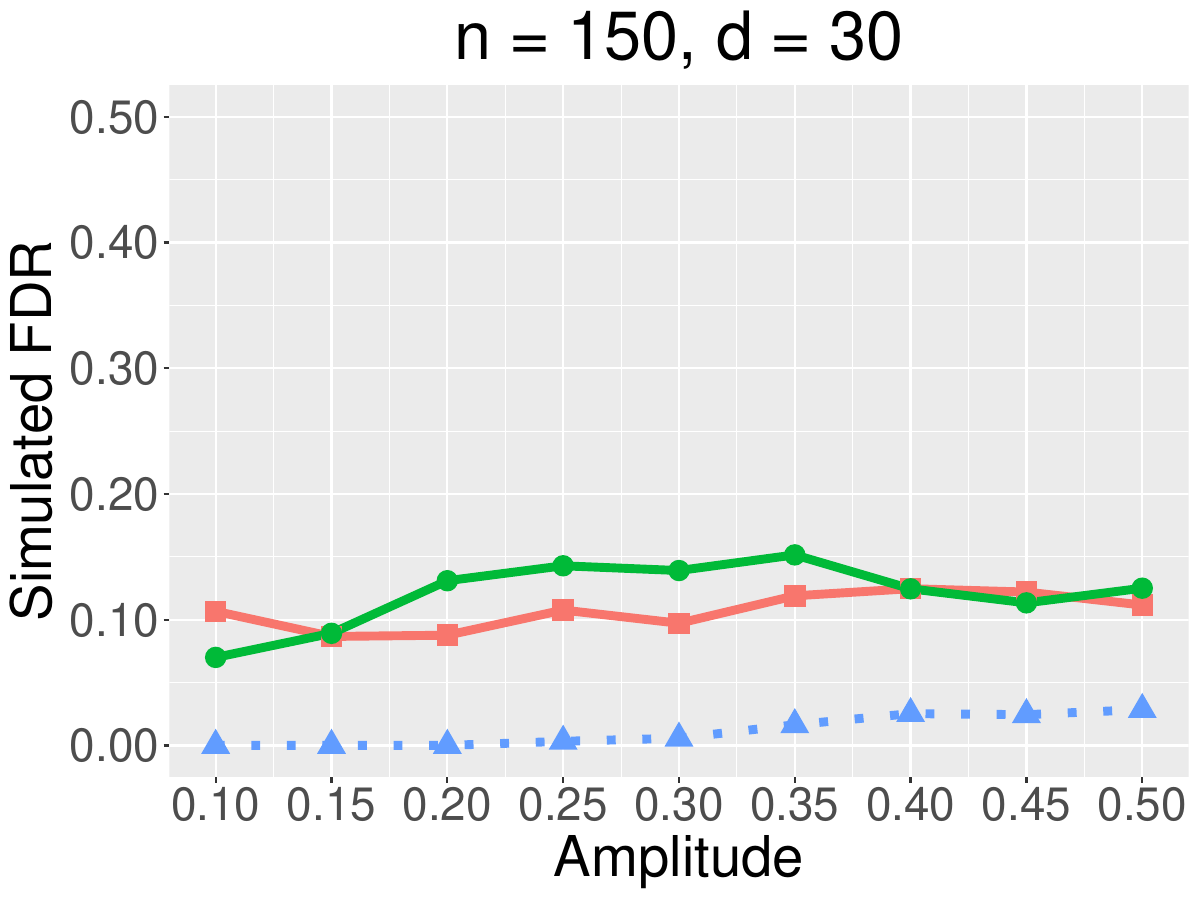}}\hspace{5pt}
	\subfloat{\includegraphics[width=.27\columnwidth]{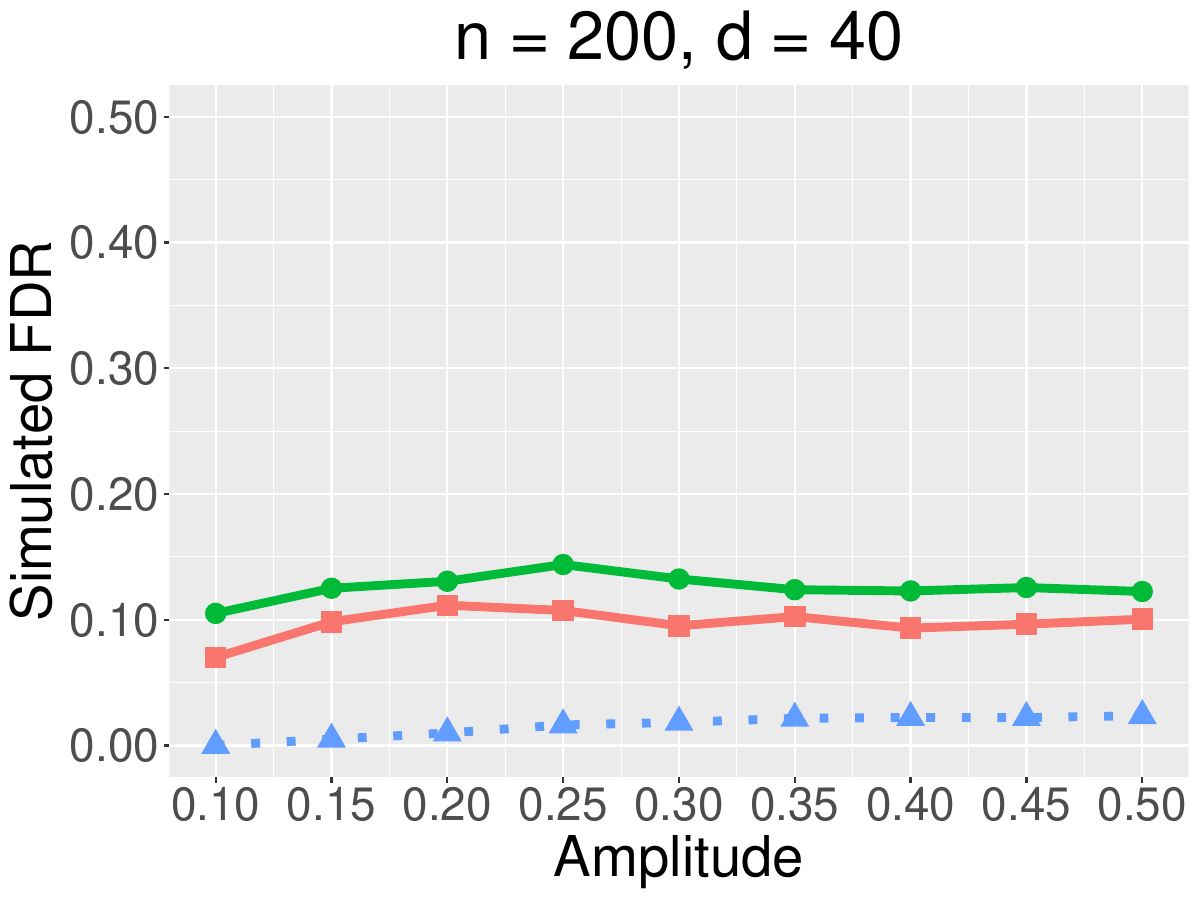}}\hspace{5pt}
	\subfloat{\includegraphics[width=.27\columnwidth]{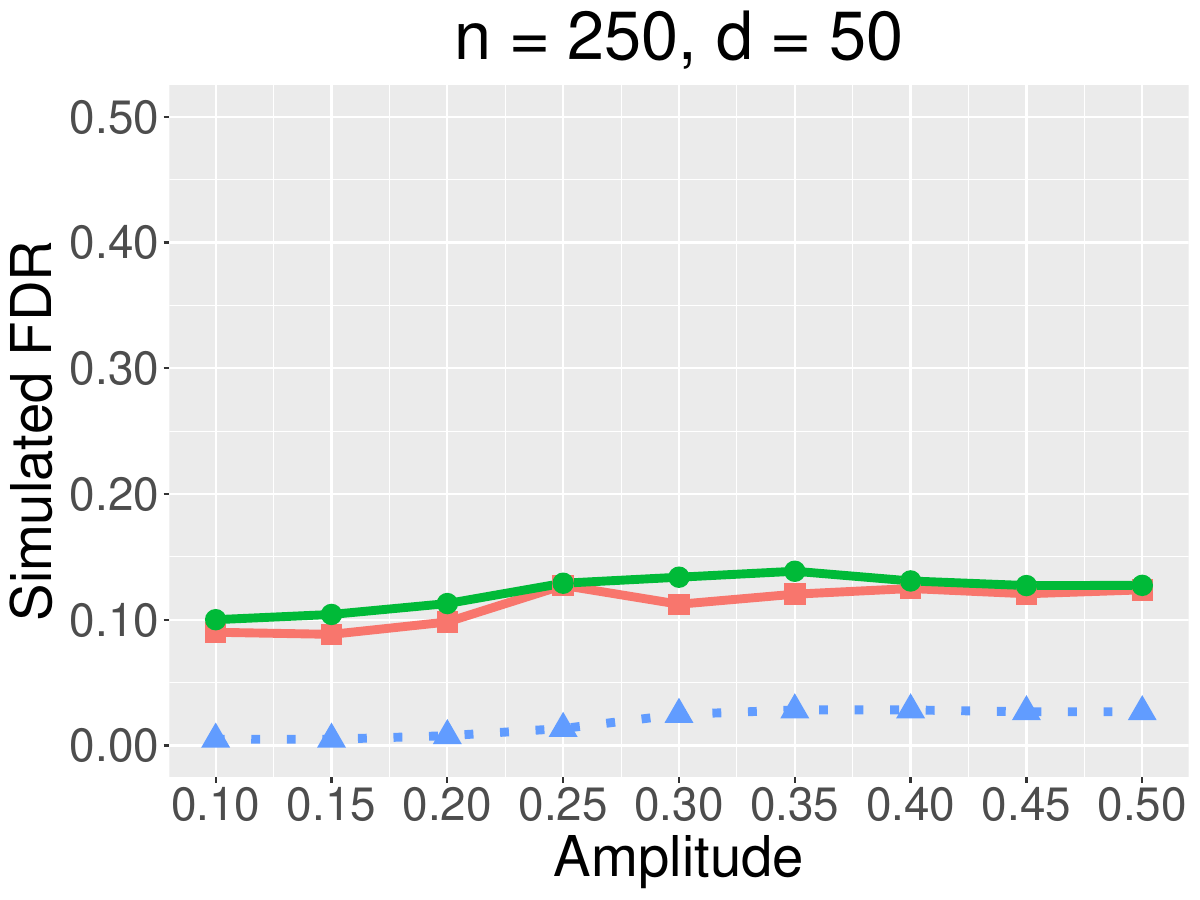}}\\
	\subfloat{\includegraphics[width=.27\columnwidth]{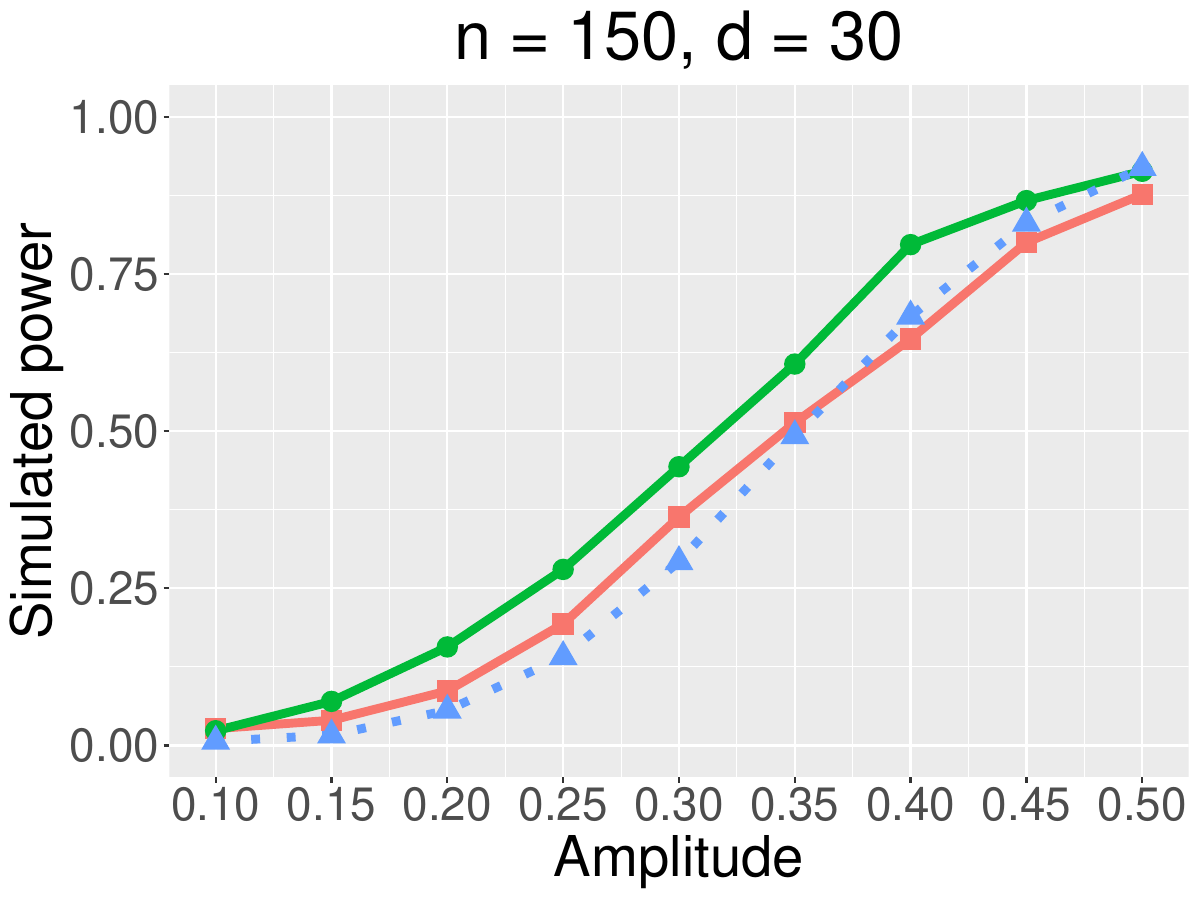}}\hspace{5pt}
    \subfloat{\includegraphics[width=.27\columnwidth]{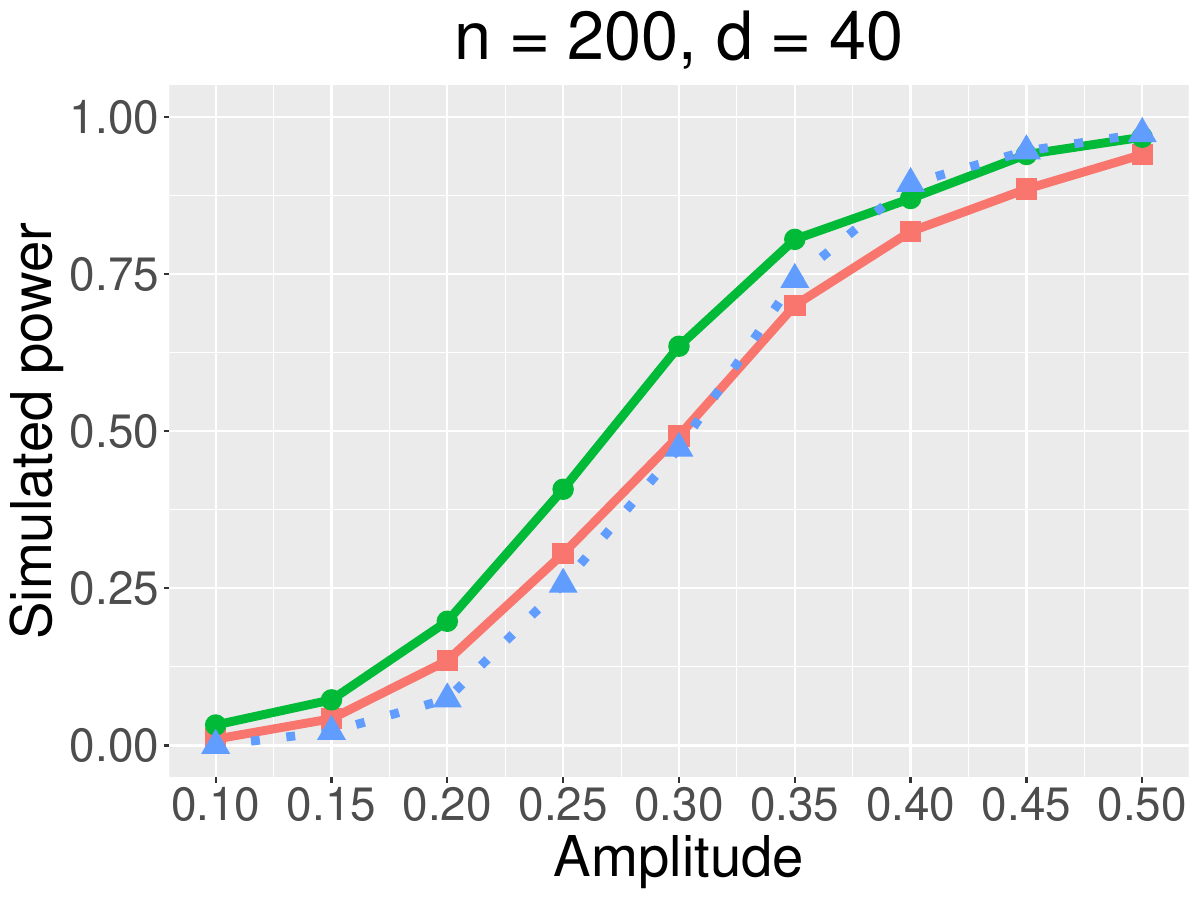}}\hspace{5pt}
    \subfloat{\includegraphics[width=.27\columnwidth]{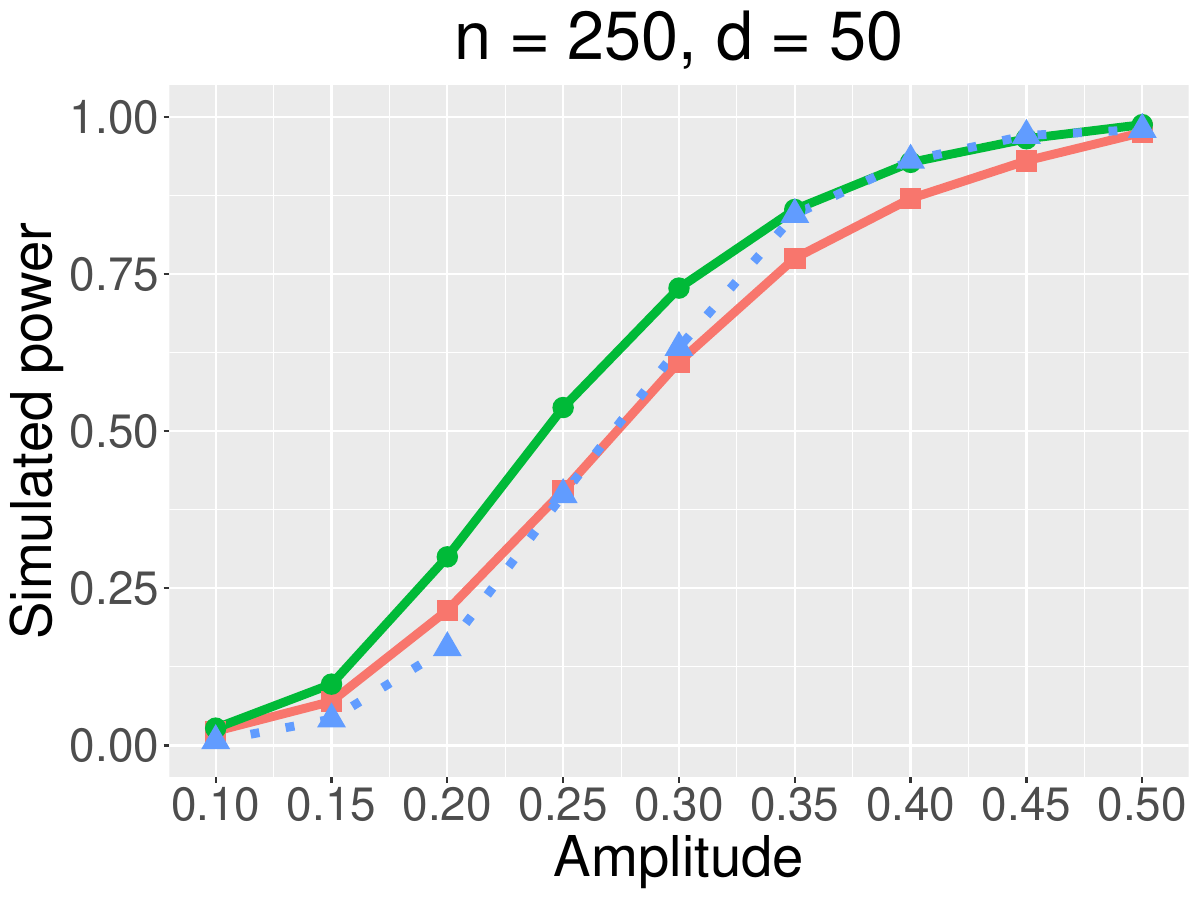}}
	\caption{\small Simulated FDR and power for the settings of $(n,d) = (150, 30)$, $(200,40)$ and $(250,50)$. The rows of the design matrix were generated from Setting 2. The sparsity level is $k = 0.1d$ and the FDR level is $\alpha = 0.1$. The methods compared are Algorithm 1 (squares and red solid line), Algorithm 2 (circles and green solid line), and the Bonferroni-Benjamini-Hochberg method of \cite{Sarkar2022Adjusting} (triangles and blue dotted line).}
    \label{Low-dimension-0.1}
\end{figure}

We also conduct additional numerical studies to demonstrate the benefits of using the debiased estimator, particularly in settings where low-dimensional approaches, such as that of \cite{Sarkar2022Adjusting}, are applicable. 
Figure~\ref{Low-dimension-0.1} presents results highlighting these advantages in cases where both our proposed methods and the method in \cite{Sarkar2022Adjusting} are applicable, specifically when \( n \geq 2d \). 
Additional results are provided in Figures~\ref{Low-dimension-0.05} and~\ref{Low-dimension-0.2} in
Section~\ref{lowd} of the Supplementary Material. While the method of \cite{Sarkar2022Adjusting} performs competitively in conventional low-dimensional linear models, it struggles in moderate-dimensional problems due to the limitations of the OLS approach. 
In contrast, the debiased method is crucial for generating valid \( p \)-values in high-dimensional settings. 
Unlike \cite{Sarkar2022Adjusting}, which does not incorporate a penalized estimator or a debiasing step, our methods leverage a parsimonious model structure and exploit the properties of high-dimensional estimators. 
High-dimensional variable selection promotes sparsity, while the debiasing step yields more accurate variance estimation. 
The OLS-based method suffers from inflated variance when the number of predictors is moderately large, rendering it inapplicable when \( n < 2d \).

 In summary, we find from the numerical results that our proposed methods effectively control the FDR while maintaining competitive power across various simulation settings. Overall, these findings establish the efficacy and practicality of our proposed approaches in high-dimensional multiple testing scenarios. Between our two proposed procedures, Algorithm~\ref{method_2} consistently demonstrates superior power compared to Algorithm~\ref{method_1}, confirming the theoretical advantage of the two-step multiple testing procedure established in Theorem \ref{thm_power}.

 In the simulation results reported above, the knockoff variables are generated using the second-order model-X knockoff procedure of \cite{candes2018panning}, where the variance-covariance matrix is estimated from the whole data. We have also conducted simulation studies using an estimated precision matrix within a data-splitting strategy, as described in Section~\ref{sec:newtheory}, where the validity of the procedure is established and shown to accommodate a broad class of design matrices under weaker distributional assumptions; see Section~\ref{F:F_0} of the Supplementary Material. Qualitatively, we find that the performance with data-splitting and an estimated covariance matrix is compelling and very comparable to the results reported above without data-splitting. Section~\ref{simu_ds} of the Supplementary Material provides some simulation results comparing our methods with and without data-splitting.

\subsection{Additional numerical studies}
To illustrate the benefit of using the augmented design matrix in our framework, 
Section~\ref{simu_pvalue} of the Supplementary Material compares the finite-sample performance of our method with the BH procedure applied to $p$-values constructed from the debiased Lasso using only the original design matrix $\bX$. 
The results indicate that this debiased-Lasso-based BH procedure without knockoff variables is highly sensitive to feature dependence and may fail to control the FDR under strong correlations. 
See Figures~\ref{FDR-control-origin-0.05}--\ref{Highly-correlated-origin-0.2} in the Supplementary Material for details.

We have also included additional simulation studies with non-Gaussian errors in Section~\ref{simu_nonGaussian} of the Supplementary Material. The results are broadly consistent with those obtained under normal errors, indicating that our method remains valid for non-Gaussian error distributions. See Figures \ref{FDR-control-nongaussian-0.1}--\ref{Two-stage-nongaussian-0.1} in the Supplementary Material for details.

\begin{figure}[htbp!]
	\centering
	\subfloat{\includegraphics[width=.27\columnwidth]{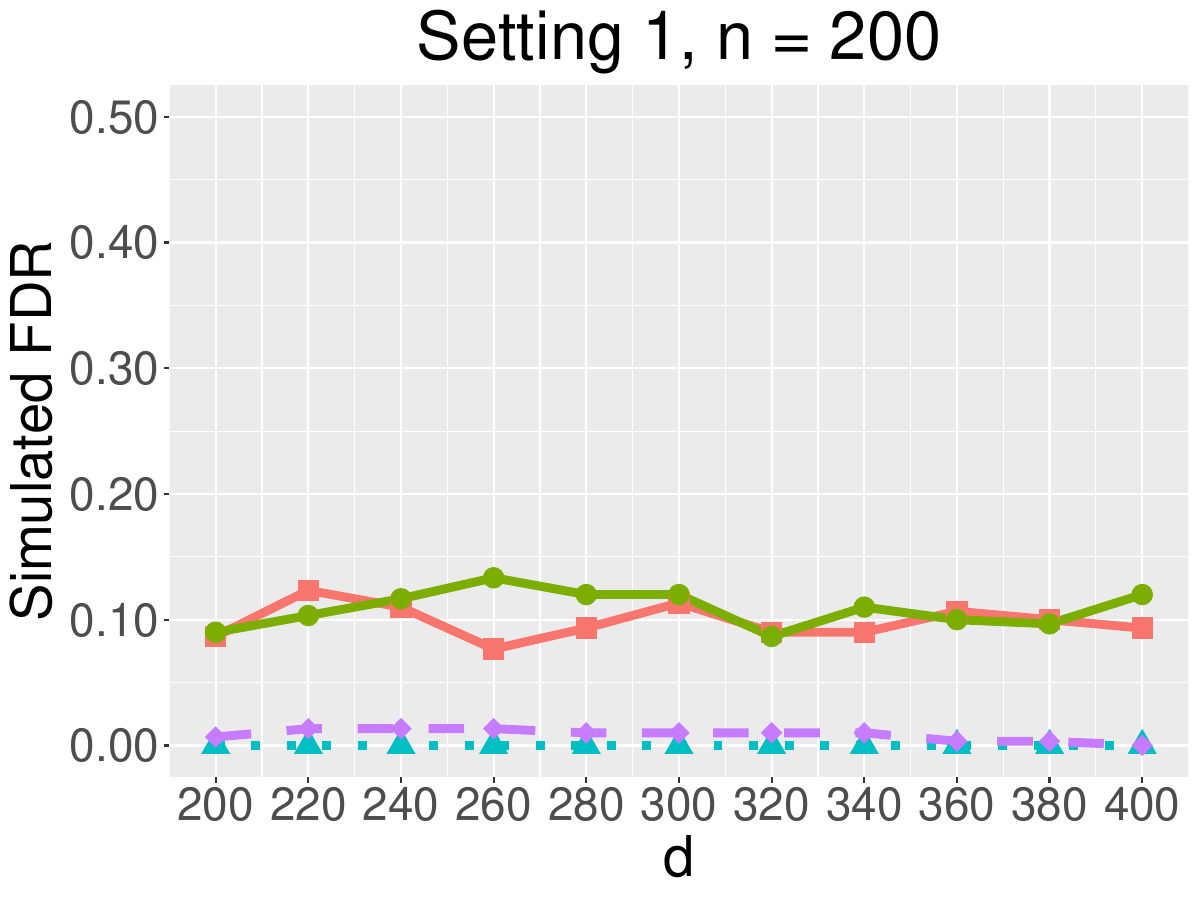}}\hspace{5pt}
	\subfloat{\includegraphics[width=.27\columnwidth]{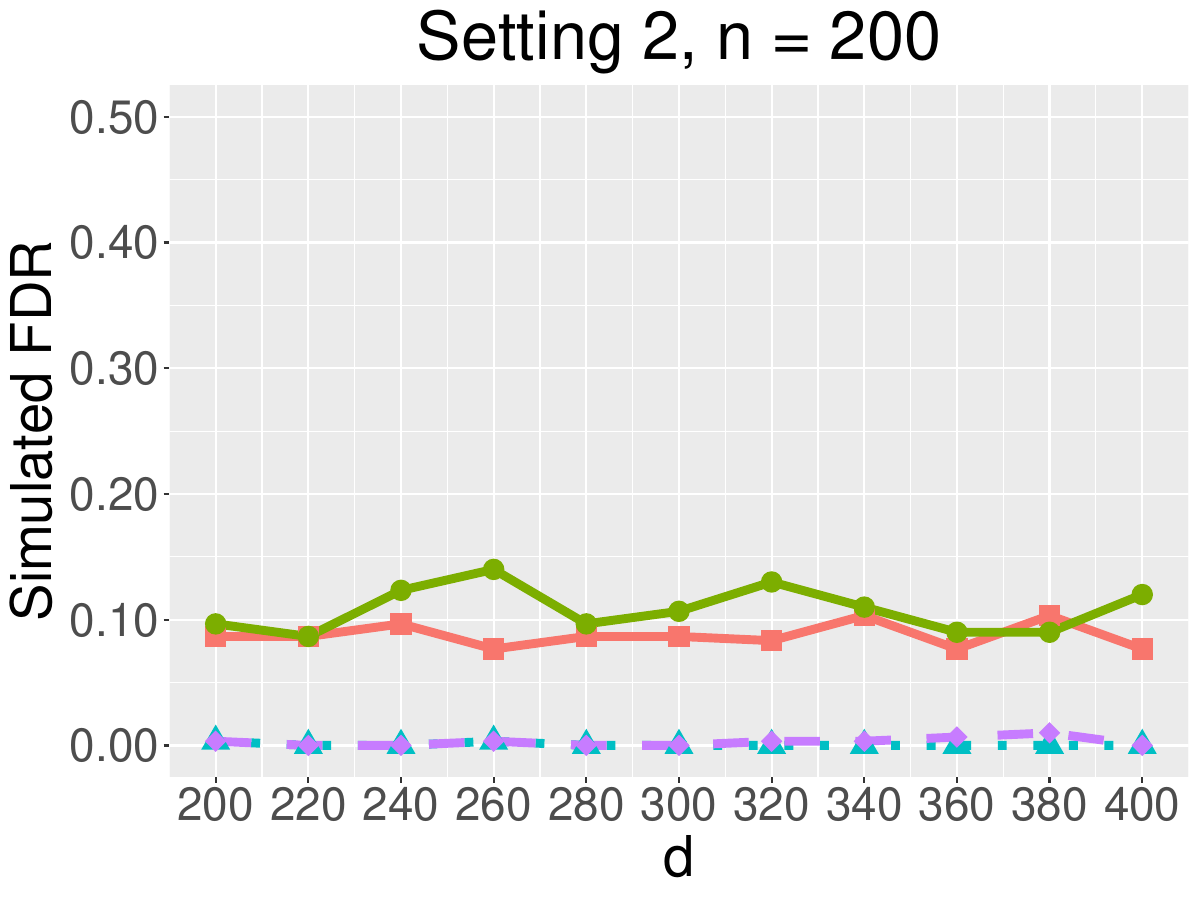}}\hspace{5pt}
	\subfloat{\includegraphics[width=.27\columnwidth]{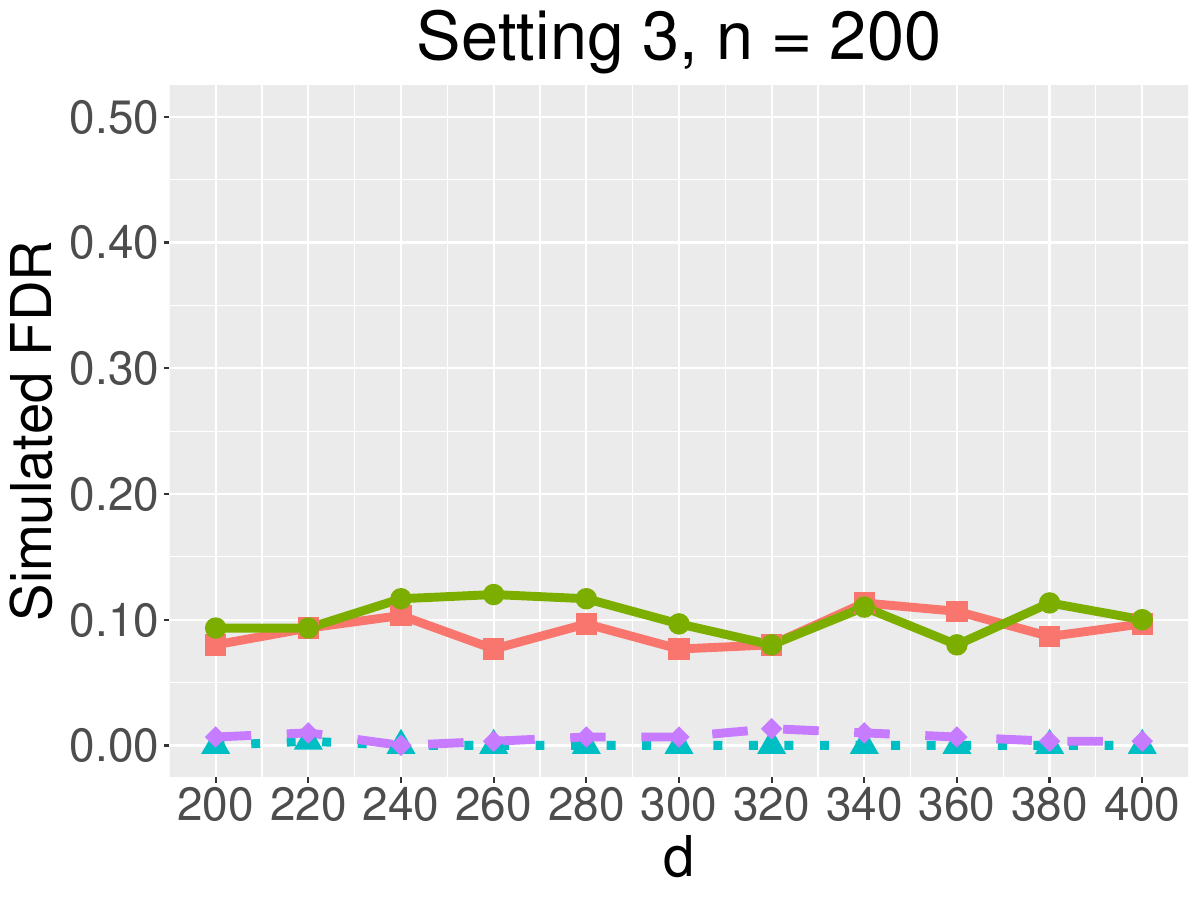}}\\
	\subfloat{\includegraphics[width=.27\columnwidth]{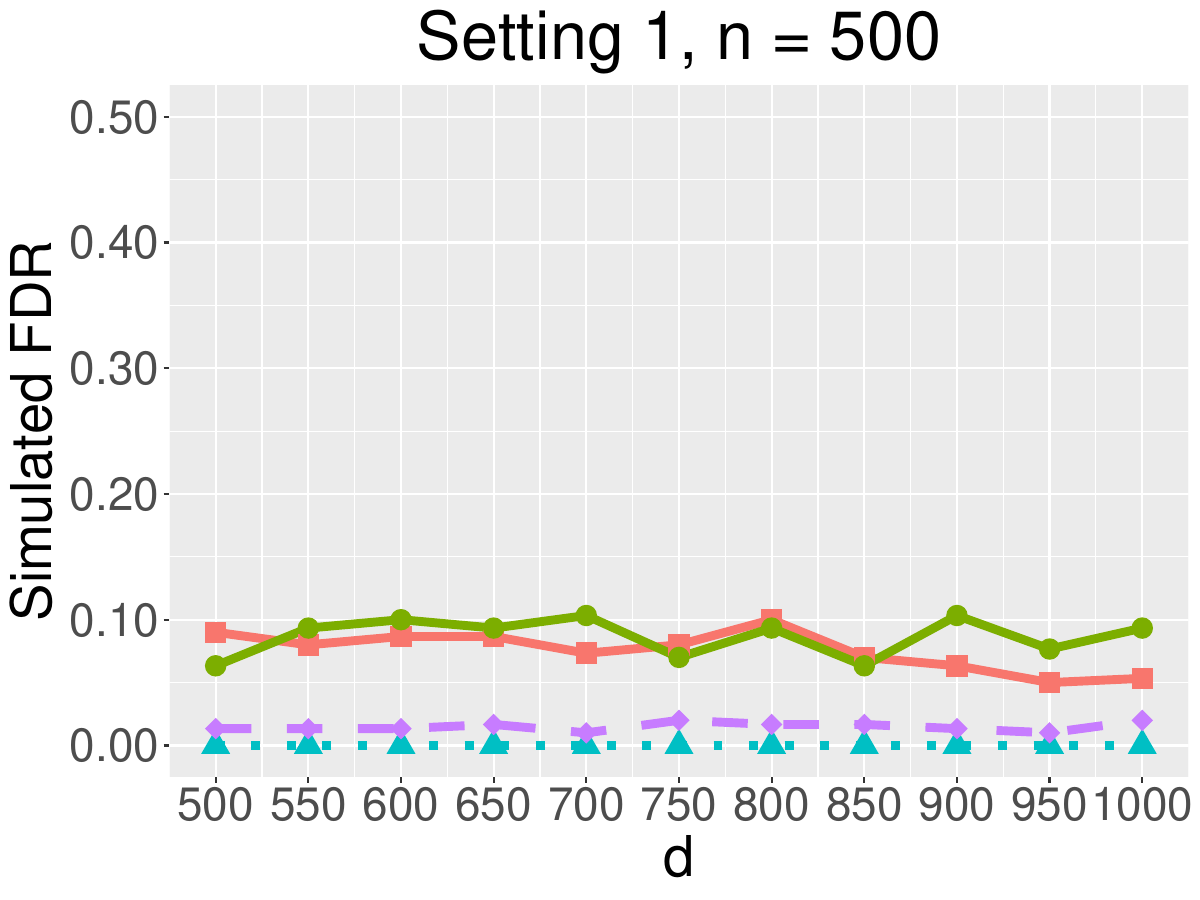}}\hspace{5pt}
    \subfloat{\includegraphics[width=.27\columnwidth]{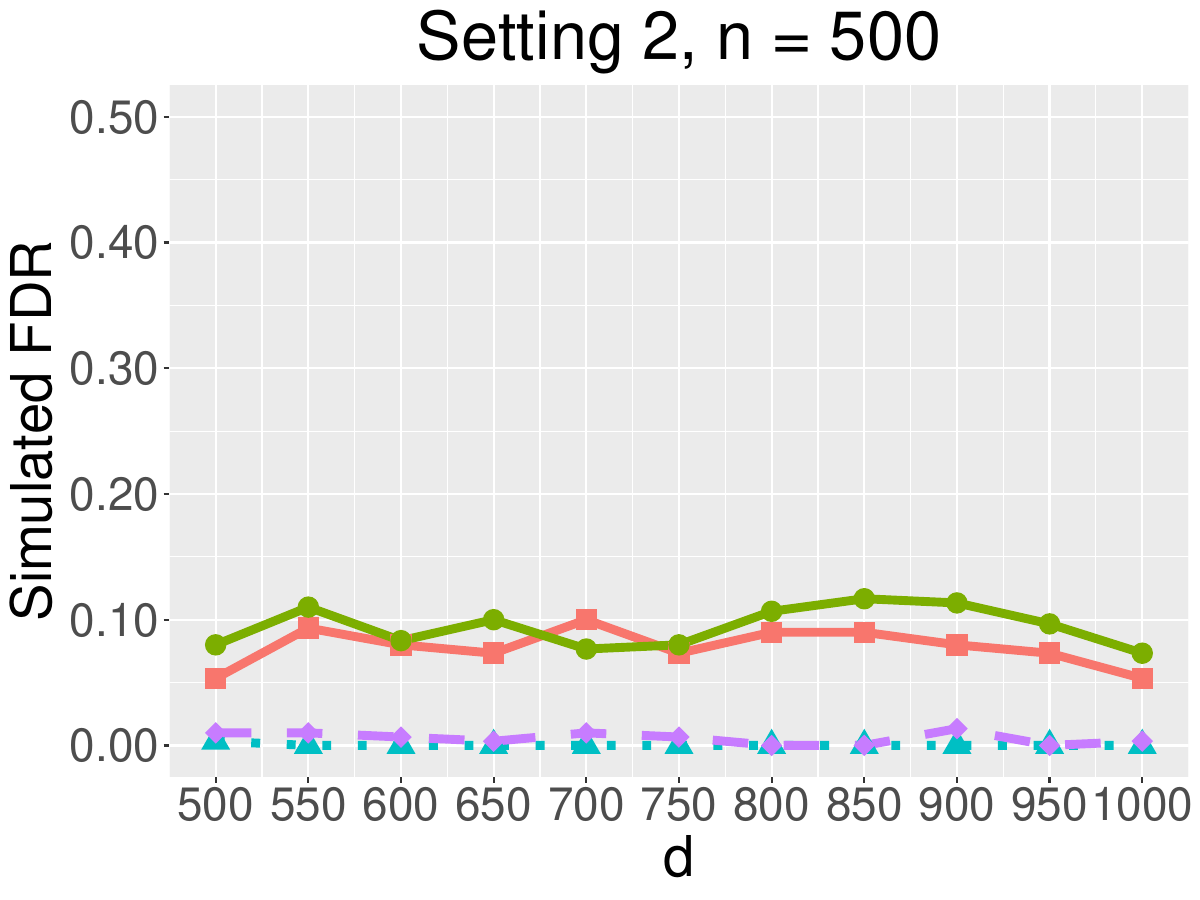}}\hspace{5pt}
    \subfloat{\includegraphics[width=.27\columnwidth]{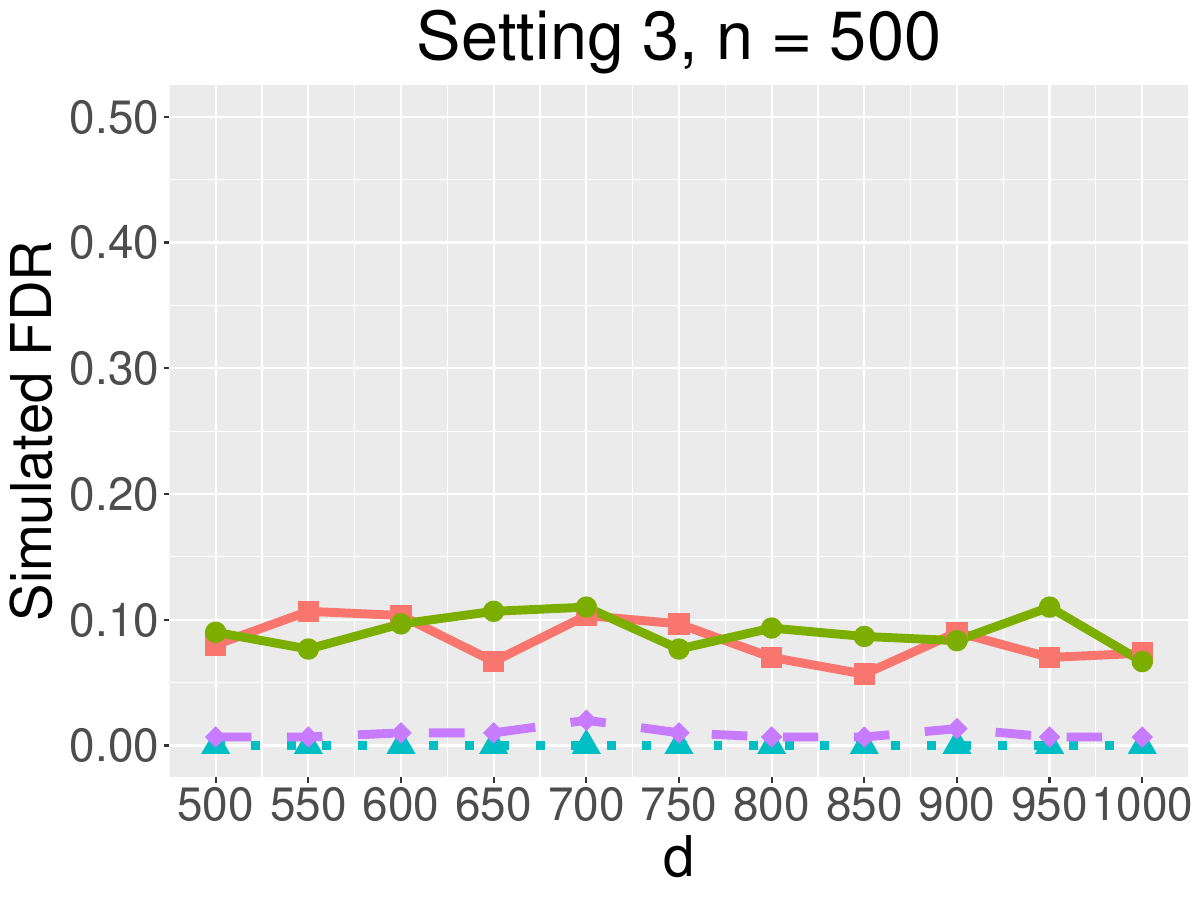}}
	\caption{\small Simulated false discovery rate (FDR) when all null hypotheses are true, for Setting 1, Setting 2, and Setting 3. The sample sizes of top row and bottom row are $n = 200$ and $n = 500$, respectively. The FDR level is $\alpha = 0.1$. The methods compared are Algorithm 1 (squares and red solid line), Algorithm 2 (circles and green solid line), the knockoff-based method of \cite{candes2018panning} (triangles and blue dotted line), and the Gaussian Mirror method of \cite{Xing2021Controlling} with FDP+ procedure (diamonds and purple dashed line).}
    \label{FDR-control-0.1}
\end{figure}

\begin{figure}[htbp!]
	\centering
	\subfloat{\includegraphics[width=.27\columnwidth]{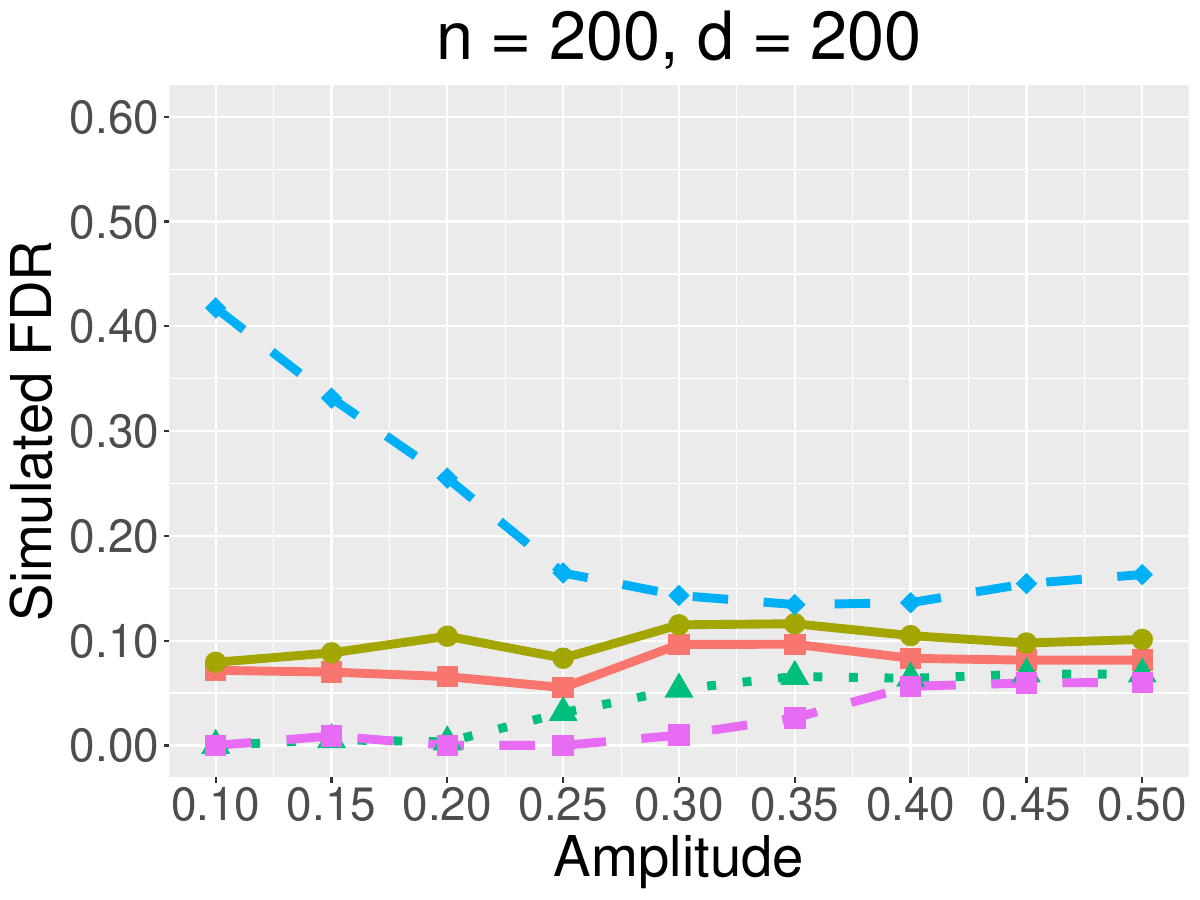}}\hspace{5pt}
	\subfloat{\includegraphics[width=.27\columnwidth]{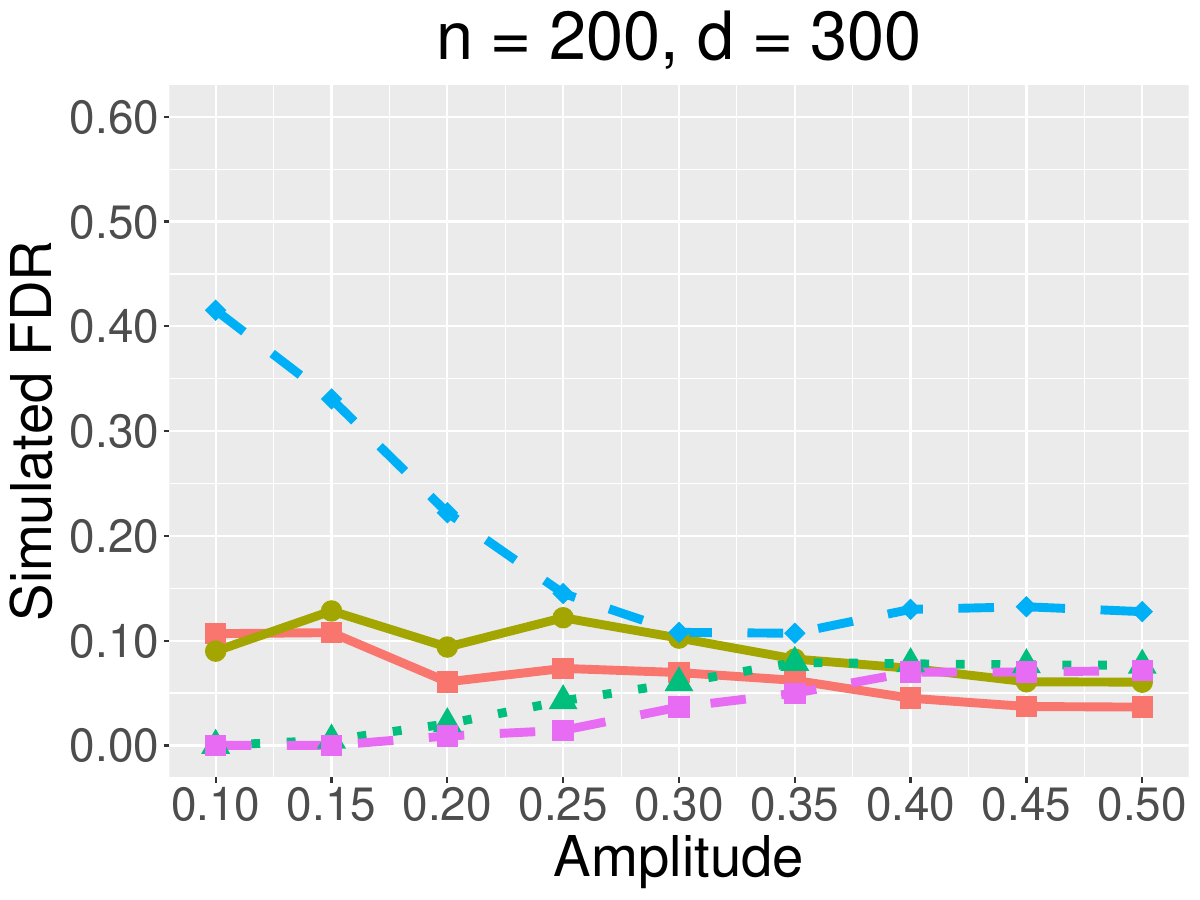}}\hspace{5pt}
	\subfloat{\includegraphics[width=.27\columnwidth]{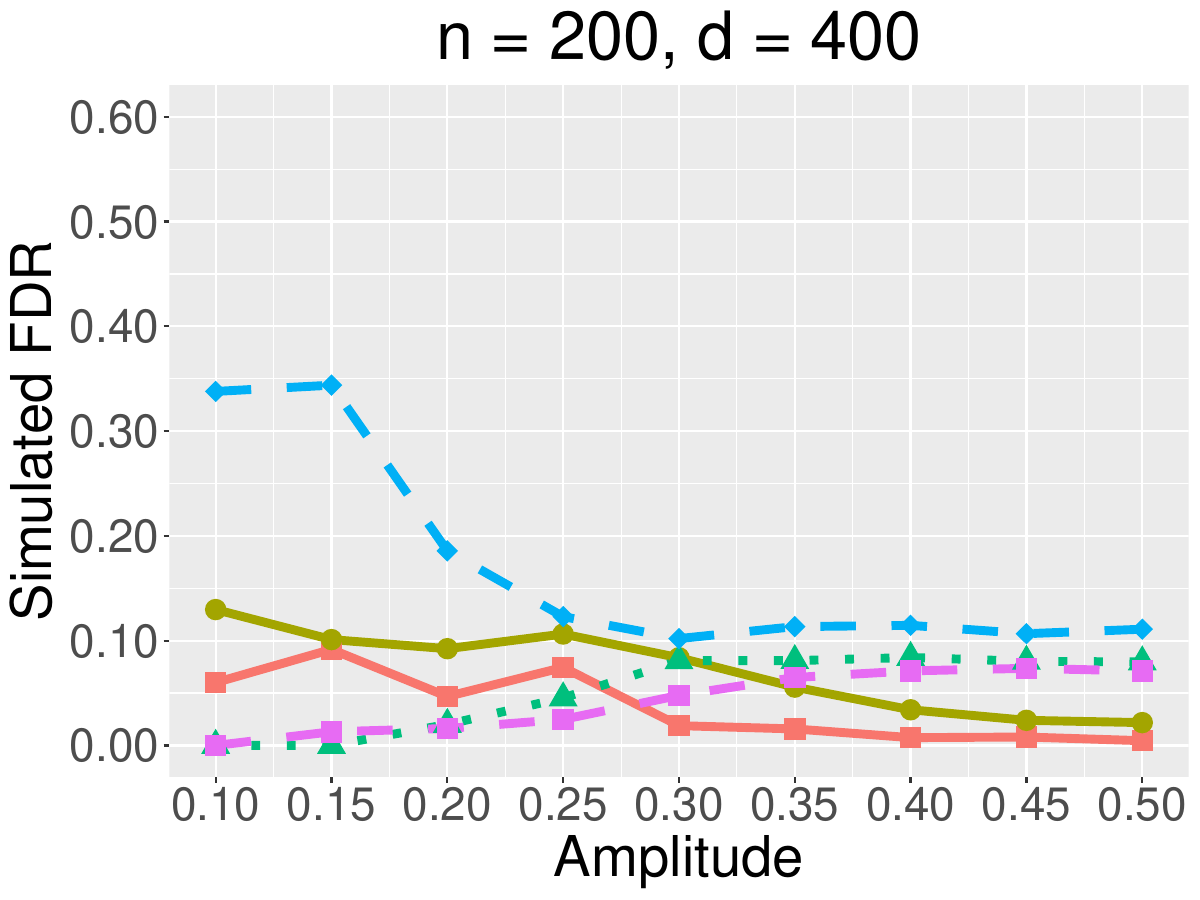}}\\
	\subfloat{\includegraphics[width=.27\columnwidth]{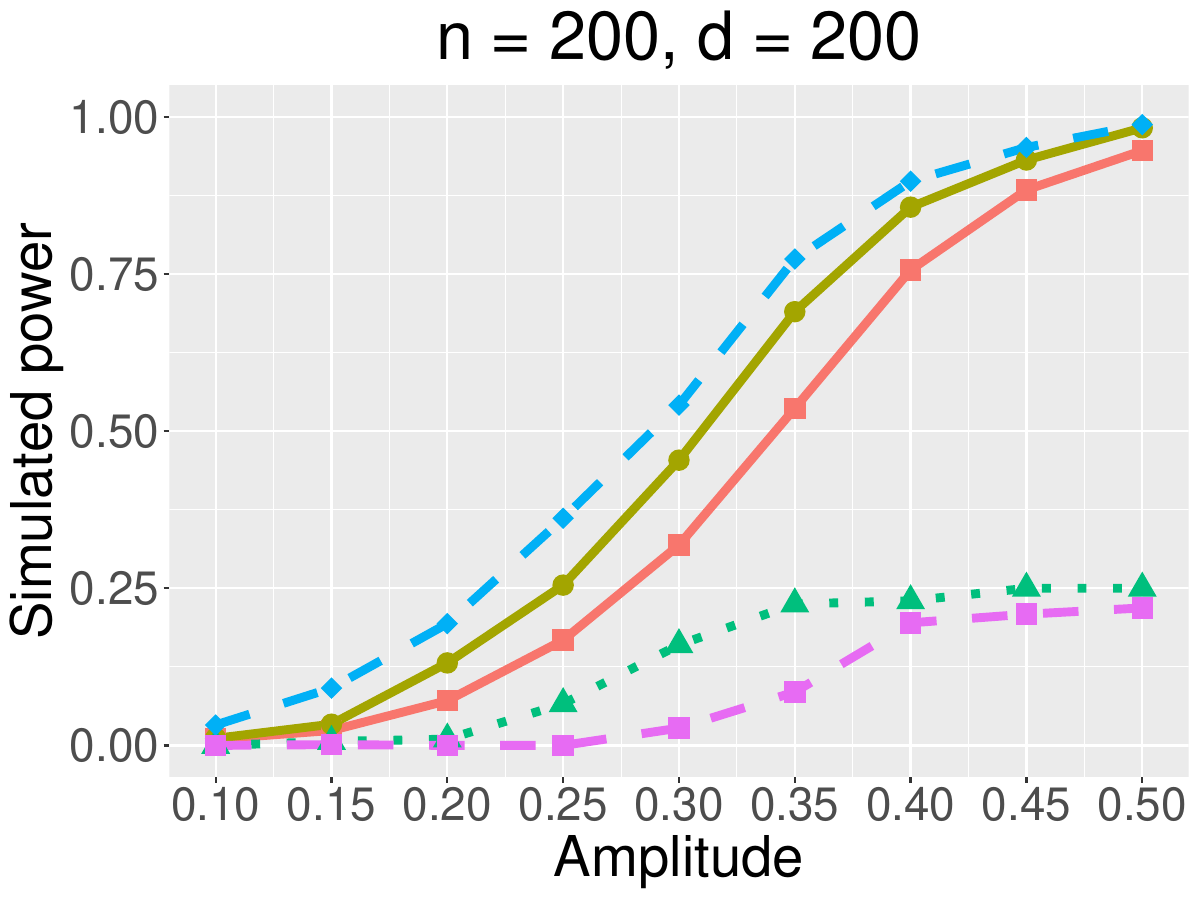}}\hspace{5pt}
    \subfloat{\includegraphics[width=.27\columnwidth]{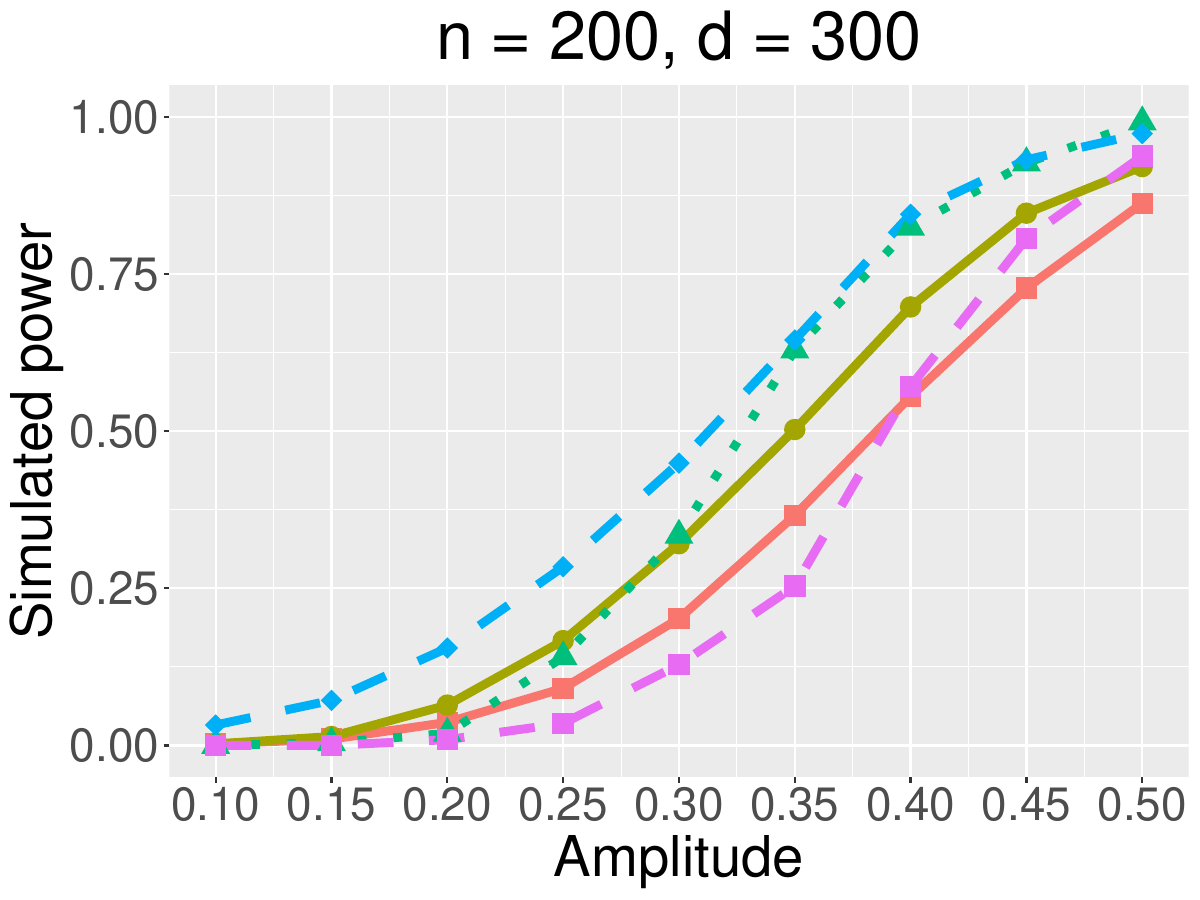}}\hspace{5pt}
    \subfloat{\includegraphics[width=.27\columnwidth]{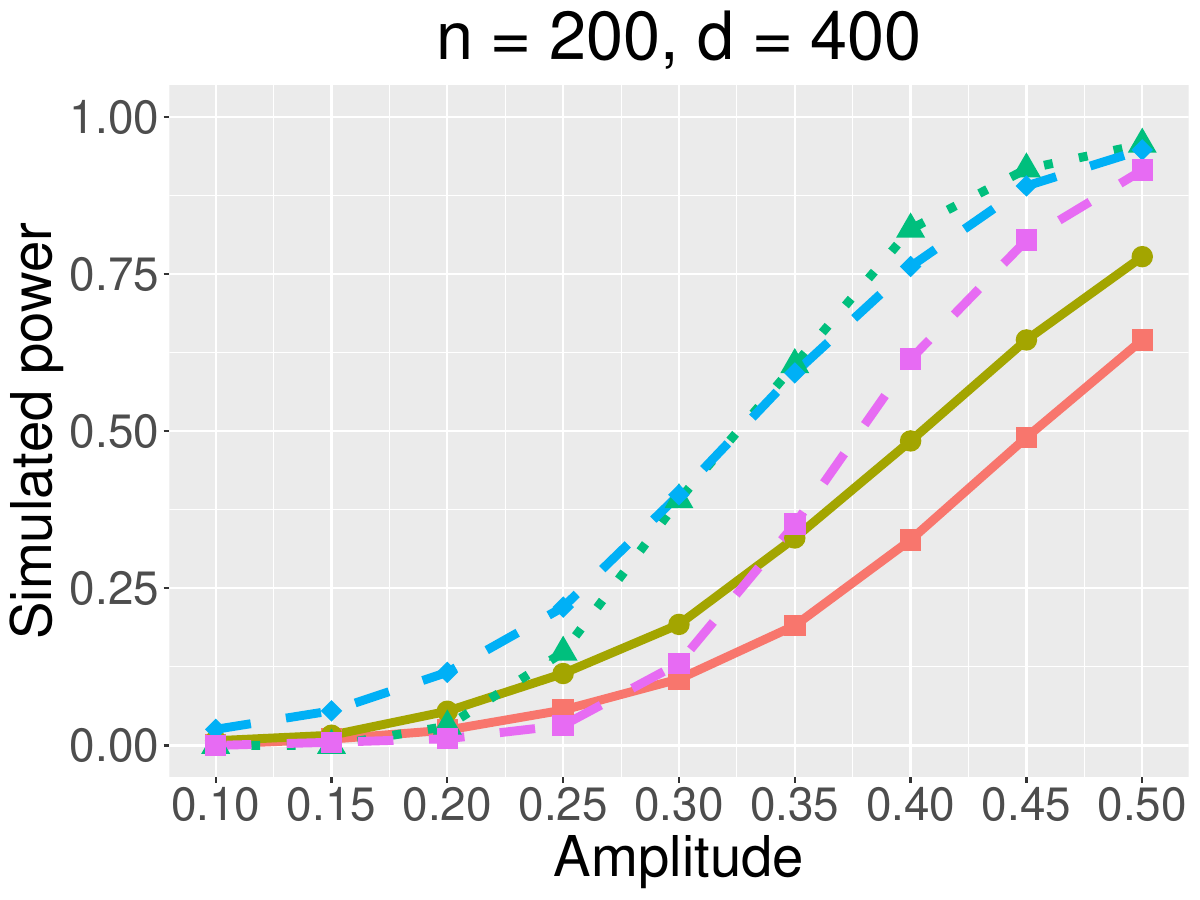}}
	\caption{\small Simulated FDR and power for the settings of $(n,d) = (200, 200)$, $ (200, 300)$ and $(200, 400)$. The rows of the design matrix were generated from Setting 2. The sparsity level is $k = 0.04d$ and the FDR level is $\alpha = 0.1$. The methods compared are Algorithm 1 (squares and red solid line), Algorithm 2 (circles and yellow solid line), the knockoff-based method of \cite{candes2018panning} (triangles and green dotted line), the Gaussian Mirror method of \cite{Xing2021Controlling} (diamonds and blue dashed line), and the Gaussian Mirror method with FDP+ procedure (squares and purple dashed line).}
    \label{Power-0.1}
\end{figure}

\section{Real data analysis} \label{sec:realdata}

We apply our methods to identify mutations in Human Immunodeficiency Virus Type 1 (HIV-1) associated with drug resistance, comparing them with competing approaches. The dataset analyzed, as described in \cite{Rhee2006genotypic}, comprises HIV-1 subtype B sequences from individuals with prior antiretroviral treatment. It includes mutations at protease and reverse transcriptase (RT) positions of HIV-1 subtype B sequences, conferring resistance to Protease Inhibitors (PIs), nucleoside reverse transcriptase inhibitors (NRTIs), and non-nucleoside RT inhibitors (NNRTIs).

In preprocessing the dataset, we follow the steps outlined in \cite{Barber2015Controlling}. The design matrix $\bX \in \{0,1\}^{n \times d}$ is constructed such that $x_{i,j} = 1$ if the $i$-th sample contains the $j$-th mutation, and $x_{i,j} = 0$ otherwise. For a specific drug, the $i$-th entry of the response vector $y_i$ represents the logarithm of the increase in resistance to that drug in the $i$-th patient. Using linear models to capture these effects, we apply our proposed methods to identify mutations in HIV-1 associated with resistance to Protease Inhibitors (PIs). Specifically, we focus on six of the seven drugs: APV, ATV, IDV, LPV, NFV, and RTV.

As an example with comparable $n$ and $d$, analyzing this real dataset provides valuable insights into the advantages of our $p$-value-based multiple testing methods. In particular, the methods of \cite{Barber2015Controlling} and \cite{Sarkar2022Adjusting} remain applicable, enabling a more extensive comparison to highlight the benefits of the proposed methodology. We compare our proposed methods with the knockoff method \citep{Barber2015Controlling}, the Gaussian Mirror method \citep{Xing2021Controlling}, and the Bonferroni-Benjamini-Hochberg method introduced by \cite{Sarkar2022Adjusting}.

Although there is no definitive ground truth in the real data, we validate our discoveries by comparing them against the treatment-selected mutation (TSM) panels provided in \cite{Rhee2005hiv}. These panels identify mutations observed more frequently in virus samples from patients treated with each drug compared to those never treated with that drug. Since these panels are independently derived from the dataset we analyze, they serve as a benchmark for validating the discoveries made by respective methods.

We investigate three different levels of FDR control: $\alpha = 0.05$, $0.1$, and $0.2$. Representative results are included in Figure \ref{fig_realdata}, with additional figures in Section \ref{SM3:figure} of the Supplementary Material. Summarizing the results, our methods demonstrate consistent performance across all cases, producing a reasonable number of findings with stable ``false discoveries", defined here as those not in the TSM panels. In contrast, the knockoff method \citep{Barber2015Controlling} yields no discoveries at $\alpha = 0.05$ for five out of the six drugs examined, as shown in Figure \ref{fig_realdata_0.05} of the Supplementary Material. This limited detection is consistent with our observation from the simulation studies. 
The Gaussian Mirror method \citep{Xing2021Controlling}, with empirical FDP evaluated by \eqref{eq:fdp2},  appears to identify the most discoveries across all scenarios, but its proportion of false discoveries also tends to exceed the target FDR level in most cases; see Figure \ref{fig_realdata}, and Figures \ref{fig_realdata_0.05} and \ref{fig_realdata_0.1} in the Supplementary Material. This aligns with our observations in simulations, where the Gaussian Mirror method,  with empirical FDP evaluated by \eqref{eq:fdp2},  tends to have a higher level of false discoveries when the FDR is low and/or the number of signals is few.

When $n \geq 2d$, where the method of \cite{Sarkar2022Adjusting} is applicable, our proposed methods demonstrate superior performance. Specifically, at $\alpha = 0.2$, as shown in Figure \ref{fig_realdata}, our method consistently identifies more findings, while the method of \cite{Sarkar2022Adjusting} makes fewer discoveries or even fails to detect any. This outcome is likely due to the higher variance of the estimators used in their approach, especially when $n$ is not sufficiently large and/or $d$ is relatively large. 
Qualitatively similar observations are reported in Figures \ref{fig_realdata_0.05} and \ref{fig_realdata_0.1} in the Supplementary Material. These comparisons highlight the advantages of using Lasso-penalized regression and a debiased approach in constructing the test statistics, which effectively address the limitations of conventional OLS methods in high-dimensional settings.

In summary, our proposed $p$-value-based multiple testing methods demonstrate competitive performance in addressing multiple testing in practical settings. We recommend our approach, particularly in cases where the target FDR level is lower or the signals from contributing variables are weaker, allowing for further investigation and comparisons.

\begin{figure}[htbp!]     
	\centering
	\subfloat[$n = 767, \ d = 201$]{\includegraphics[width=.31\columnwidth]{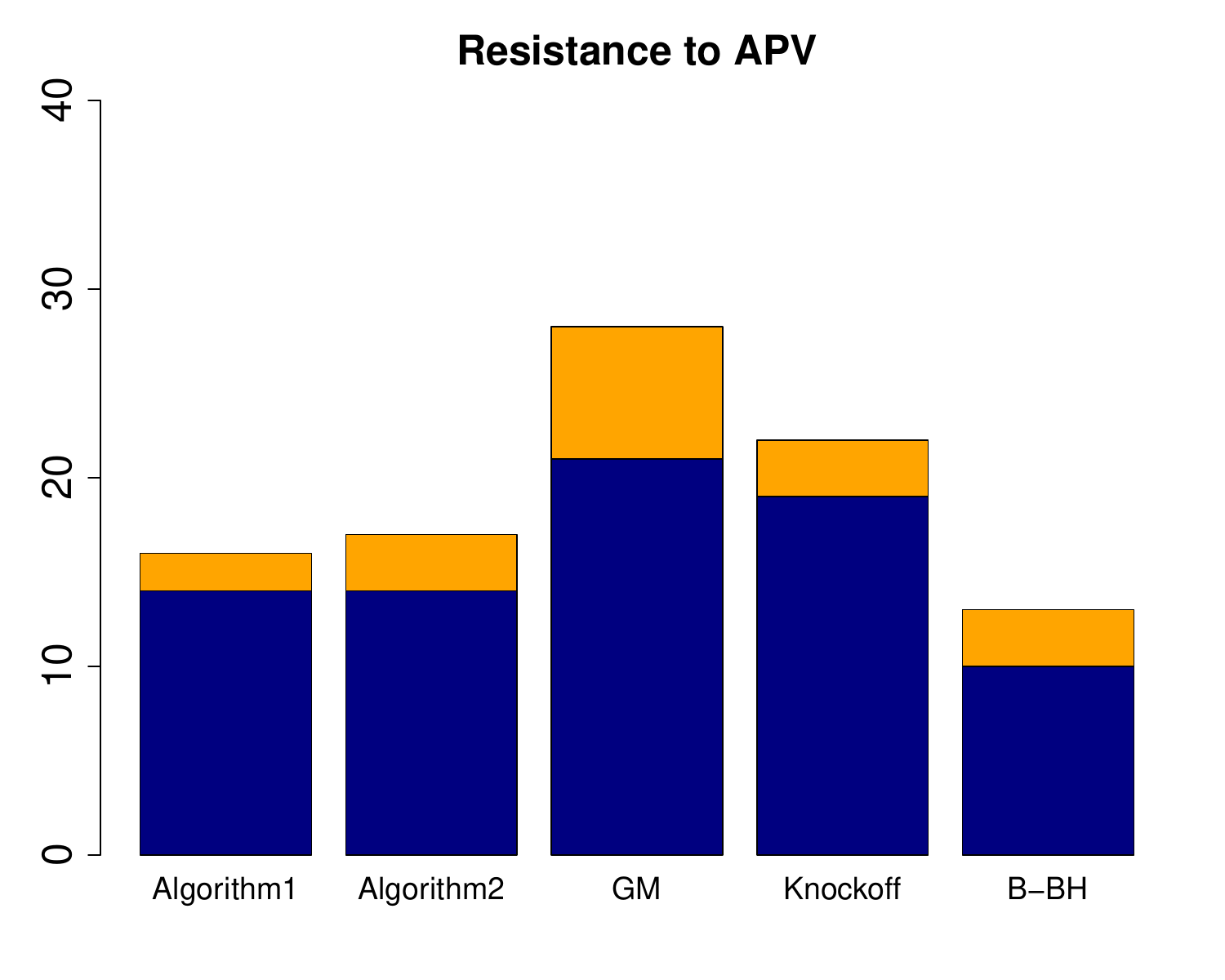}}\hspace{5pt}
	\subfloat[$n = 328, \ d = 147$]{\includegraphics[width=.31\columnwidth]{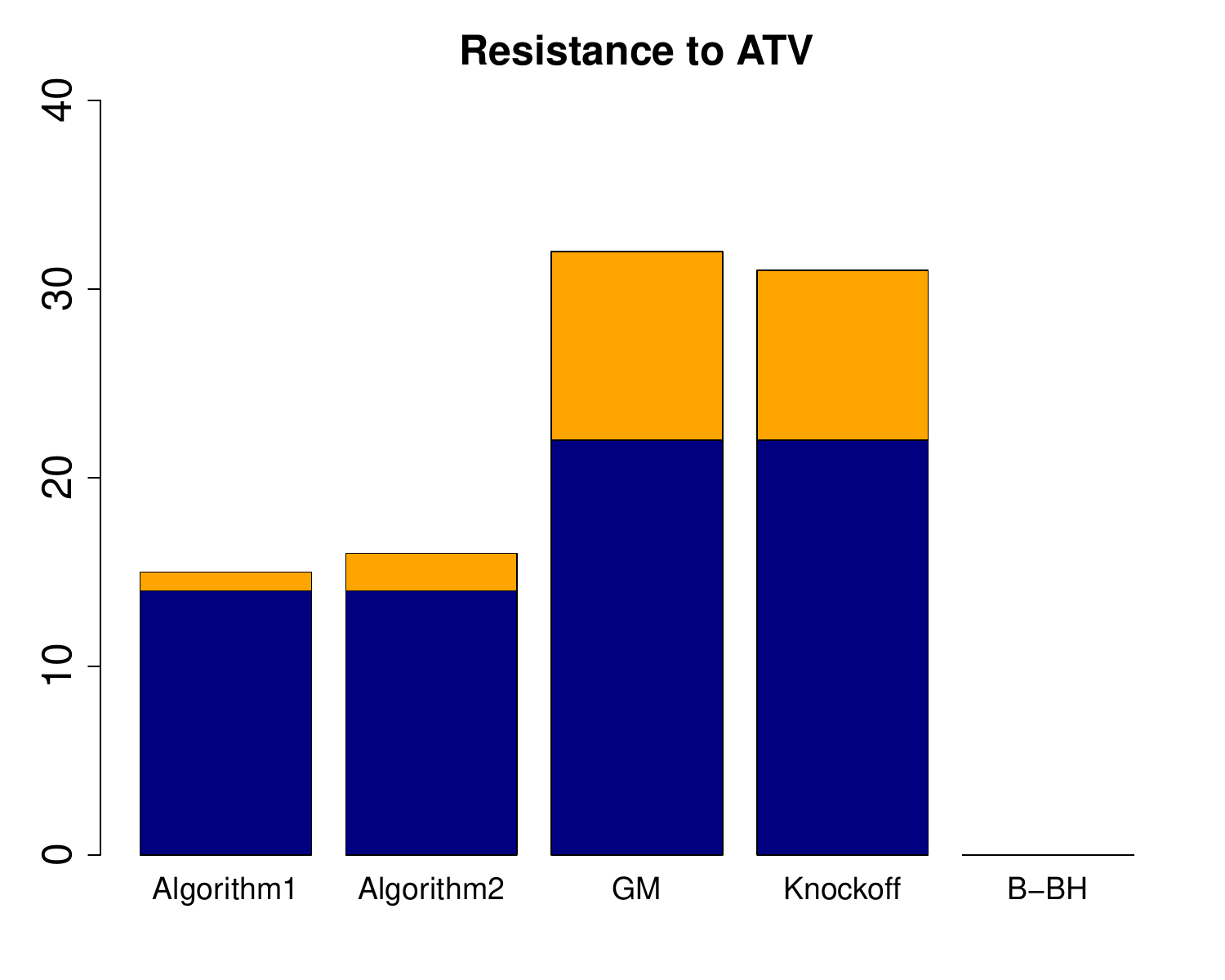}}\hspace{5pt}
	\subfloat[$n = 825, \ d = 206$]{\includegraphics[width=.31\columnwidth]{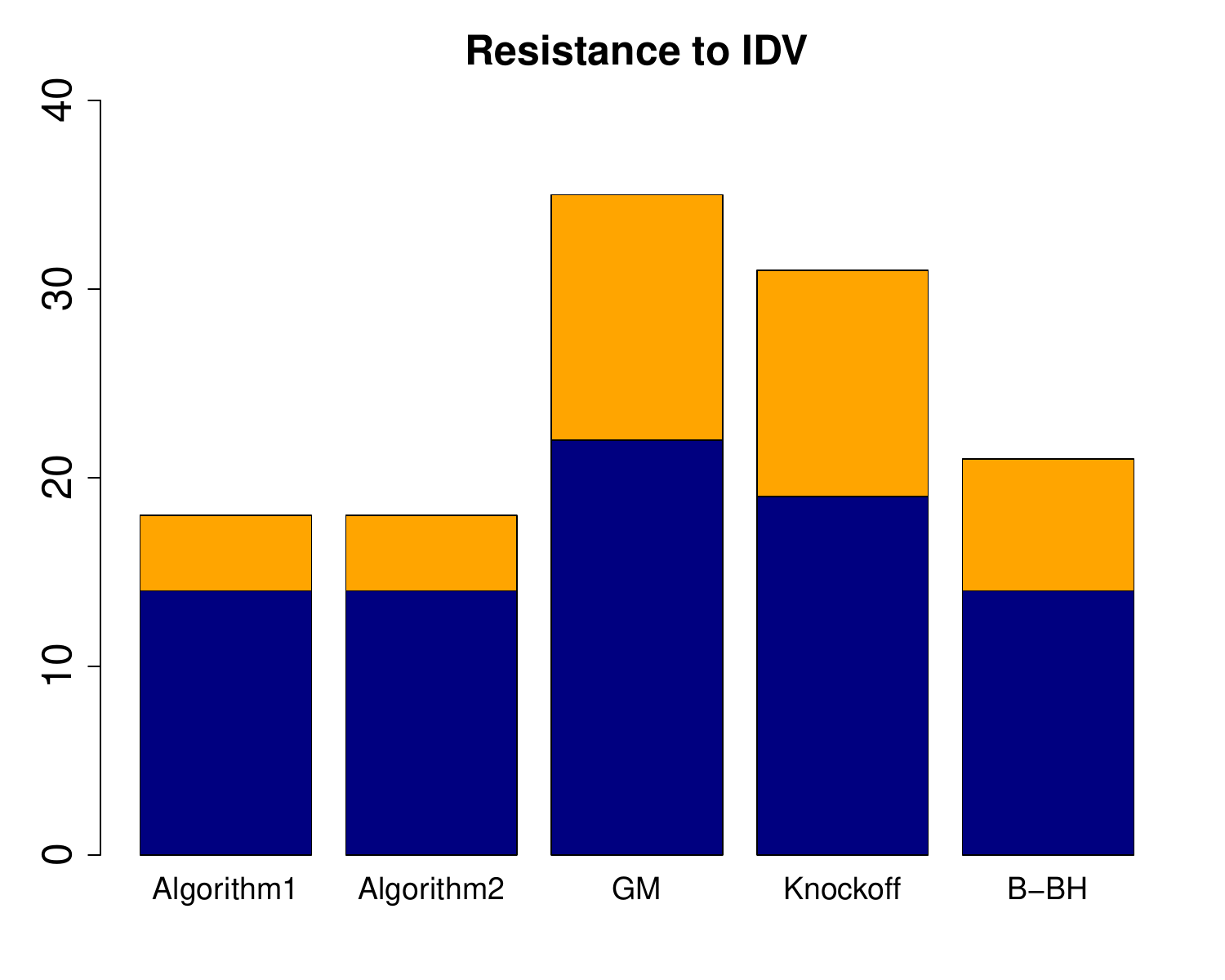}}\\
	\subfloat[$n = 515, \ d = 184$]{\includegraphics[width=.31\columnwidth]{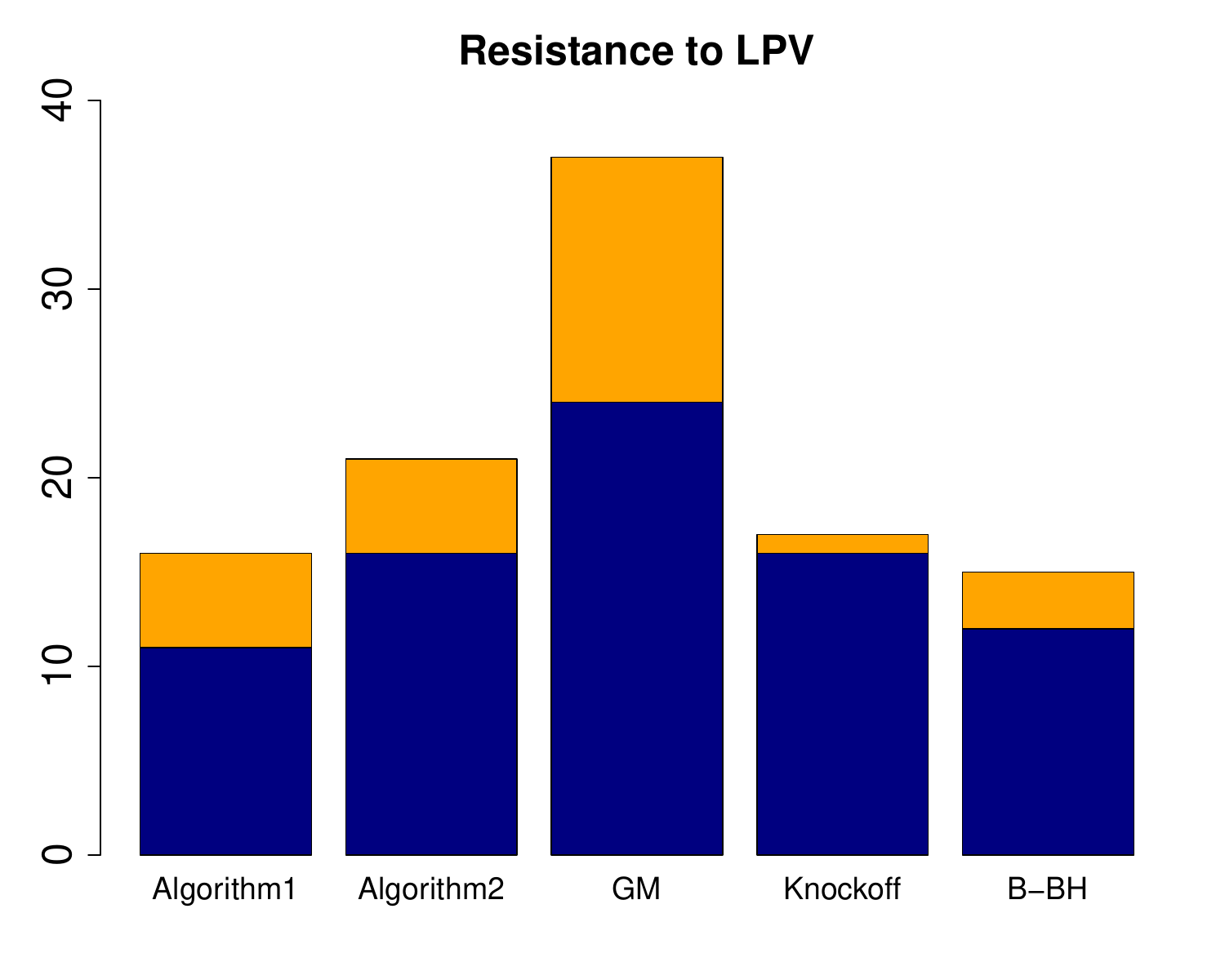}}\hspace{5pt}
    \subfloat[$n = 842, \ d = 207$]{\includegraphics[width=.31\columnwidth]{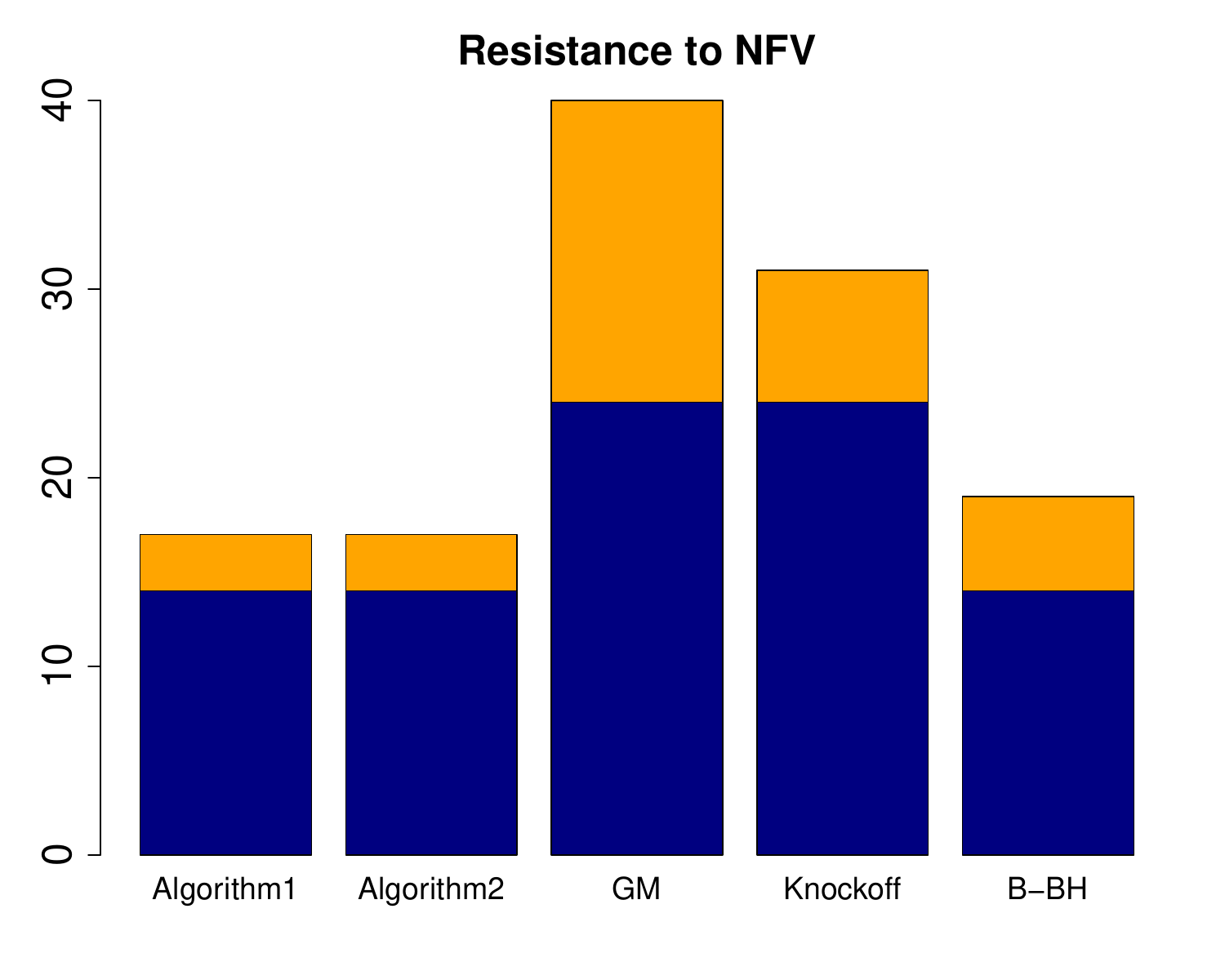}}\hspace{5pt}
    \subfloat[$n = 793, \ d = 205$]{\includegraphics[width=.31\columnwidth]{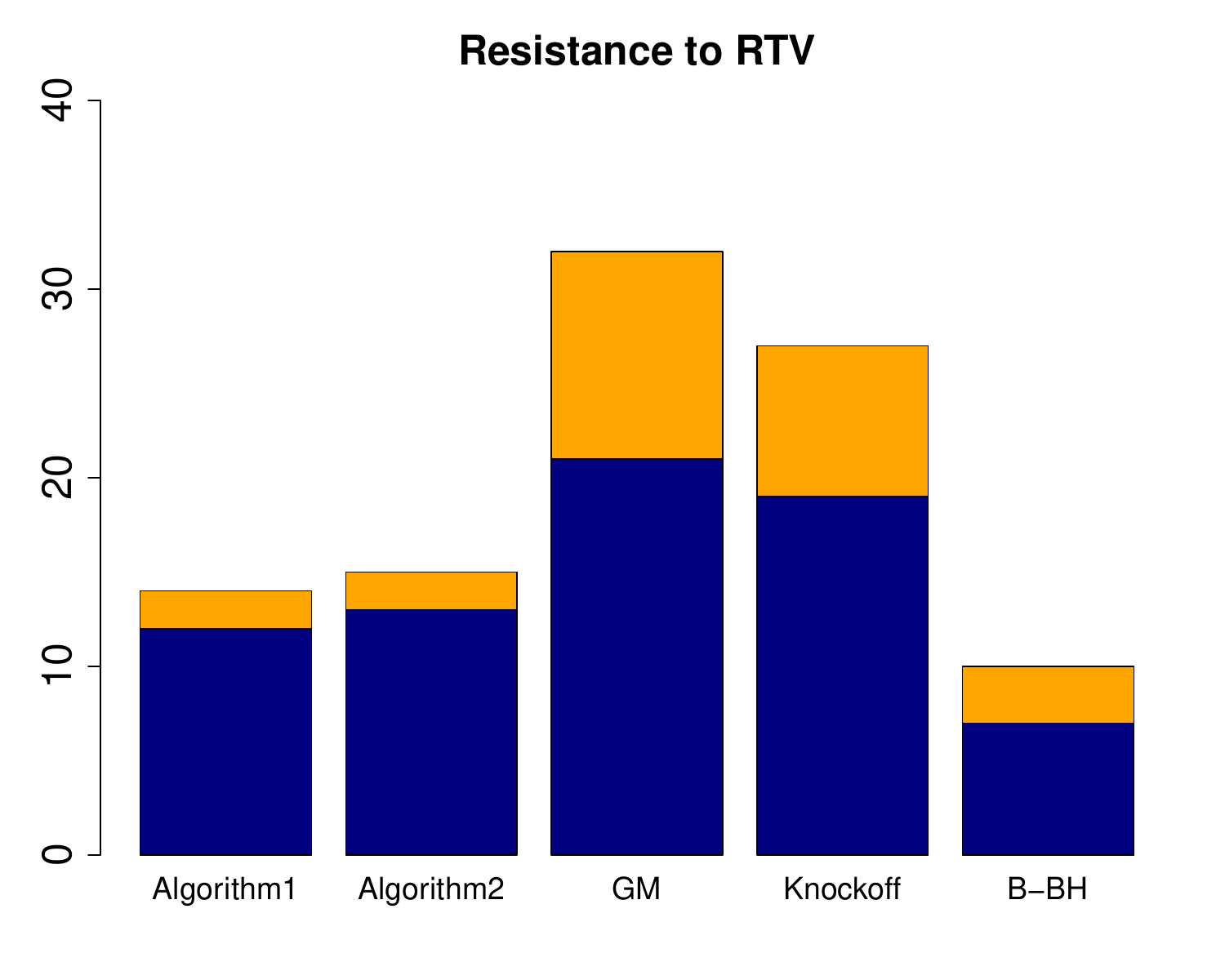}}
	\caption{\small Results of the real data example for $\alpha = 0.2$. Blue represents the number of discoveries that are in the treatment-selected mutation panels list, and yellow represents the number of discoveries not in the treatment-selected mutation panels list. The total number of HIV-1 protease positions in the treatment-selected mutation panels list is $34$. The methods compared are the proposed Algorithm 1 (Algorithm1), the proposed Algorithm 2 (Algorithm2), the Gaussian Mirror method of \cite{Xing2021Controlling} (GM), the knockoff-based method of \cite{Barber2015Controlling} (Knockoff), and the Bonferroni-Benjamini-Hochberg method of \cite{Sarkar2022Adjusting} (B-BH).
 }
\label{fig_realdata}
\end{figure}

\section{Discussion} \label{sec:discussion}

In conclusion, our study proposes and analyzes innovative $p$-value-based methodologies for high-dimensional multiple testing that effectively control the FDR asymptotically while enhancing power in challenging scenarios. The empirical results demonstrate the promising performance of our methods, particularly those utilizing $p$-values derived from the debiased estimator and paired test statistics from model-X knockoffs. By combining theoretical insights with methodological applications, our work represents a significant advancement in statistical methodology and theory, providing valuable tools for practical multiple testing applications.

High-dimensional multiple testing problems are inherently challenging, with numerous avenues for future development. For instance, constructing valid and useful test statistics, along with corresponding $p$-values, remains a difficult open problem for more general models beyond linear regression. Generalized linear models, such as Poisson and logistic regression, are particularly valuable tools. Although penalized likelihood approaches have been developed for model estimation \citep{Fan2020}, multiple testing methods for high-dimensional generalized linear models remain underexplored. The primary challenge lies in constructing valid test statistics and associated $p$-values, compounded by the complex dependence structures involved, which requires further dedicated effort. We plan to address these challenges in future projects.

\bibliographystyle{jasa}
\spacingset{0.95}\selectfont
\bibliography{mybib}

\clearpage

\appendix
\spacingset{1.7}\selectfont
\setlength{\abovedisplayskip}{0.2\baselineskip}
\setlength{\belowdisplayskip}{0.2\baselineskip}
\setlength{\abovedisplayshortskip}{0.2\baselineskip}
\setlength{\belowdisplayshortskip}{0.2\baselineskip}
\setcounter{equation}{0}

\begin{center}
	{\noindent\bf  \Large
		Supplementary Material for ``Controlling the False Discovery Rate in High-Dimensional Linear Models Using Model-X Knockoffs and $p$-values" 
        %by Jinyuan Chang, Chenlong Li, Cheng Yong Tang and Zhengtian Zhu 
        }  \\
% {\large \bf  (For online publication only)}
\end{center}

\renewcommand\thesection{\Alph{section}}
\renewcommand{\thepage}{S\arabic{page}}
\renewcommand{\thefigure}{F\arabic{figure}}
\renewcommand{\thelemma}{L\arabic{lemma}}
\renewcommand{\theassumption}{C\arabic{assumption}}
\renewcommand{\thetheorem}{T\arabic{theorem}}
\renewcommand{\thealgorithm}{A\arabic{algorithm}}
\renewcommand{\theremark}{R\arabic{remark}}

\setcounter{section}{0}
\setcounter{figure}{0}
\setcounter{page}{1}
\setcounter{theorem}{0}
\setcounter{lemma}{0}
\setcounter{assumption}{0}
\setcounter{algorithm}{0}
\setcounter{remark}{0}

\renewcommand{\theHfigure}{F\arabic{figure}}
\renewcommand{\theHlemma}{L\arabic{lemma}}
\renewcommand{\theHassumption}{C\arabic{assumption}}
\renewcommand{\theHtheorem}{T\arabic{theorem}}
\renewcommand{\theHalgorithm}{A\arabic{algorithm}}
\renewcommand{\theHremark}{R\arabic{remark}}

\renewcommand{\theequation}{S.\arabic{equation}}

The Supplementary Material is organized into Sections A--J, covering theoretical proofs, additional numerical experiments, supplementary empirical results, and a comparison with selective inference. Sections A--F present all theoretical proofs.

Section A contains the proofs of Theorems 1--3, all built upon the asymptotic representation of the debiased Lasso estimator on the knockoff-augmented design, established as Theorem T1. The proof of Theorem T1 in Section A.1 extends the debiased Lasso theory of \citeS{javanmard2014confidence_app} to the knockoff-augmented design. It relies on three auxiliary results: the bounds for the compatibility constant and generalized coherence parameter of \(\mathbf{Z}\) from Lemma L1, the \(\ell_1\)-consistency of the debiased Lasso estimator on the knockoff-augmented design from Lemma L2, and the elementwise convergence of the CLIME  estimator \citepS{cai2011constrained_app} from Lemma L3; these lemmas are proved in Section B. Together they yield a sharp bias bound, which ensures the asymptotic normality of the debiased estimator and, owing to the diagonal structure of the knockoff-augmented covariance matrix, the asymptotic independence properties of the transformed test statistics.

With this foundation, we apply the \textit{Many Approximate Means} framework of \citeS{belloni2018high_app} to prove Theorems 1 and 2. The proof of Theorem 1 in Section A.2 establishes FDR control for Algorithm 1; its general proof framework builds upon the results of \citeS{liu2014phase_app} and \citeS{belloni2018high_app}. The proof of Theorem 2 in Section A.3 incorporates the Bonferroni step, and the proof of Theorem 3 in Section A.4 verifies the power dominance of Algorithm 2 by comparing its rejection region with that of Algorithm 1 under the signal strength condition. Additional supporting results for the proofs of Theorems 1 and 2 are collected in Lemmas L4--L12 and proved in Sections B and C. 

Sections D--F extend the theory to the setting with sub-Gaussian errors and estimated, misspecified feature distributions. The proof of Theorem T2 in Section E.1 generalizes the bias bounds by additionally controlling the estimation errors of the covariance matrix and its inverse, using Lemmas L13--L16, proved in Section F, which characterize covariance matrix estimation accuracy, compatibility constant and generalized coherence parameter bounds, Lasso \(\ell_1\)-error, and CLIME consistency. The proofs of Theorems T3 and T4, given in Sections E.2 and E.3, then adapt the FDR arguments to the settings with misspecified feature distributions and sub-Gaussian errors. Additional supporting results for the proof of Theorems T3 and T4 are collected in Lemmas L17--L22 and proved in Section F.

Throughout the Supplementary Material, let $C$ be a generic positive constant, which may be different in different cases, and $\mathbb{S}^{2d-1}$ denote the unit sphere in $\mathbb R^{2d}$. 
Let $\be_{i}$ be a standard unit vector in $\mathbb{R}^{2d}$ with $1$ in the $i$-th coordinate and $0$ in all other coordinates. For a vector $\bv=(v_{1},\ldots,v_{m})^{\top}\in\mathbb{R}^{m}$ and a constant $a\in\mathbb{R}$, $\bv\le a$ means that $v_{i}\le a$ for all $i\in[m]$, $\bv_S$ is the subvector of $\bv$ with elements indexed in $S$, 
$\text{supp}(\bv)$ represents the positions of nonzero entries of $\bv$.
For a matrix $\bA=(a_{i,j})\in\mathbb{R}^{p\times q}$, 
we define the spectral norm $\|\bA\|_{2}=\sup_{|\bu|_2\le1}|\bA\bu|_{2}$, the $\ell_{\infty}$-norm $\|\bA\|_{\infty}=\max_{i\in[p]}\sum_{j=1}^{q}|a_{i,j}|$. 
We use $\lfloor \cdot \rfloor$ to denote the floor function.
For a random variable $x$, 
we denote its sub-exponential norm $\|x\|_{\psi_1} = \sup_{q \ge 1} q^{-1}\{\mathbb E (\lvert x \rvert^q)\}^{1/q}$. For a random vector $\bx \in \mathbb R^m$, its 
sub-exponential norm is defined as 
$\|\bx\|_{\psi_1} = \sup_{\bu \in \mathbb{S}^{m-1}} \| \langle \bx, \bu \rangle \|_{\psi_1}$,  
where $\mathbb{S}^{m-1}$ denotes the unit sphere in $\mathbb R^m$. 
Let $\mathcal A^{\rm c}$ denote the complement of event $\mathcal A$.

%%%%%%%%%%%%%%%%%%%%%%%%%%%%%%%%%%%%%%%%%%%%%%%%%%%%%%%%%%%%%%%%%

\section{Proofs of Theorems \ref{thm0}--\ref{thm_power}}\label{proof_thm0SM}

Recall that \( \bZ=(\bX,\tilde \bX) \in \mathbb{R}^{n \times 2d} \) is the augmented design matrix, and \(\hat{\bgamma}^{(\rm bc)}\) in \eqref{de_est} is the debiased estimator \eqref{de_est} with the estimator \(\hat{\bTheta} \in \mathbb{R}^{2d \times 2d}\) in \eqref{CLIME}. By letting \( \bw_{n} = n^{-1/2}\hat{\bTheta}^{\top} \bZ^\top \beps \), we have
\begin{equation} \label{eq:exp1}
    n^{1/2}\{\hat{\bgamma}^{(\rm bc)} - \bgamma_{0}\} = \bw_{n} + \bdelta_{n}\,,
\end{equation}
where \( \bw_{n}\,|\,\bZ \sim \mathcal{N}({\bf0}, \sigma^2 \hat{\bTheta}^{\top} \hat{\bGamma} \hat{\bTheta}) \), and \( \bdelta_{n} = n^{1/2}(\hat{\bTheta}^{\top} \hat{\bGamma} - \bI_{2d})(\bgamma_{0} - \hat{\bgamma}) \). To prove Theorems \ref{thm0}--\ref{thm_power}, we need Theorem \ref{theo_random} whose proof is given in Section \ref{proof_theo_random}.

\begin{theorem}\label{theo_random}
Let Conditions {\rm \ref{ass:model_error}--\ref{ass:CLIME}} hold and $|\bgamma_{0}|_{0}\le s_0$ for some integer $1\le s_0<2d$. For any given $\tau>0$ specified in Condition {\rm\ref{ass:CLIME}}, let $\hat{\bgamma}$ be the Lasso estimator given in \eqref{eq:lasso} with $\varrho_1$ satisfying \( \varrho_1 \ge 4\sigma\{3C_{\max}(\tau+1)n^{-1}\log(2d)\}^{1/2} \). Write $v_0 = 240000 c_*C_{\max}C_{\min}^{-1}\kappa^4$, $c_1 =(4c_*\kappa^4)^{-1}$ and $c_2 =a^2C_{\min}/(96e^2\kappa^4C_{\max}) - 2$ with $c_* = 32000e^{2}$. The following two assertions hold:
\begin{itemize}
\item[\rm (i)] If $n \ge \max\{v_0s_0\log(240eds_0^{-1}), 5000(\tau+1)\kappa^4\log(2d), 12.5(\tau+1)\log(2d), 6(c_{2}+2)\log(2d)\}$, then 
	\begin{equation*}
		\mathbb P\bigg(|\bdelta_{n}|_{\infty} > \frac{16\varrho_1\varrho_2s_0n^{1/2}}{C_{\min}}\bigg) \le 2e^{-c_1n} + 12d^{-\tau}\,.
	\end{equation*}

\item[\rm (ii)] If $n \ge \max\{6(c_{2}+2)\log (2d), \eta^{-1}\log(2d)\}$, then
\begin{align*}
\mathbb{P}\big(|\hat\bTheta^\top \hat{\bGamma} \hat\bTheta - \bTheta_0|_{\infty} \le 5M\varrho_{2}\big)\ge1 - 10d^{-\tau}\,.
\end{align*}
\end{itemize}
\end{theorem}

\subsection{Proof of Theorem \ref{theo_random}}\label{proof_theo_random}

Denote the compatibility constant of a symmetric matrix $\bA \in \mathbb{R}^{2d \times 2d}$ and a set $S \subseteq [2d]$ by $$\phi^2(\bA, S) = \min_{\btheta \in \mathbb{R}^{2d}:\,|\btheta_{S^{\mathrm{c}}}|_1 \leq 3|\btheta_S|_1} \frac{|S| \langle \btheta, \bA\btheta \rangle}{|\btheta_S|_1^2}\,. $$ 
Also, denote the generalized coherence parameter of \( \bZ \in\mathbb{R}^{n\times 2d}\) with respect to \( \bTheta \) by
$$\mu_*(\bZ; \bTheta) := | \bTheta^{\top} \hat{\bGamma} - \bI_{2d} |_{\infty}\,,$$
where $\hat{\bGamma}=n^{-1}\bZ^{\top}\bZ$. Let $\{\hat{\Gamma}_{j,j}\}_{j=1}^{2d}$ denote the diagonal components of $\hat\bGamma$. To prove Theorem \ref{theo_random}, we need 
Lemmas \ref{theo_condition}--\ref{lemm_clime} whose proofs are given in Sections \ref{pflem1a}--\ref{proof_lemm_clime}, respectively. 

\begin{lemma}\label{theo_condition}
	Let Condition {\rm\ref{ass:random_Z}} hold. The following two assertions hold:
	\begin{enumerate}
		\item[\rm(i)] For any given $\phi_0>0$, $1\le s_{0}<2d$ and $K>0$, write 
		\begin{equation*}
			\mathcal E_n(\phi_0,s_0,K) := \bigg\{\bZ \in \mathbb R^{n\times 2d}: \min_{S\subseteq[2d]:\,\lvert S \rvert \le s_0} \phi(\hat{\bGamma},S) \ge \phi_0,~ \max_{j \in [2d]} \hat{\Gamma}_{j,j}\le K \bigg\}\,.
		\end{equation*}
	Let $v_0= 240000 c_*C_{\max}C_{\min}^{-1}\kappa^4$ and $c_1 =(4c_*\kappa^4)^{-1}$ with $c_* = 32000e^{2}$. For any given $c>0$, if $K \ge C_{\max}[1 + 50\kappa^2\{cn^{-1}\log(2d)\}^{1/2}]$, then 
	\begin{equation*}
		\mathbb P\bigg\{\bZ \in \mathcal E_n\bigg(\frac{C_{\min}^{1/2}}{2},s_0,K\bigg)\bigg\} \ge 1-2e^{-c_1n}-4^{1-c}d^{-2c+1}
	\end{equation*}
	for $n \ge \max\{v_0s_0\log(240eds_0^{-1}),25c\log(2d)\}$.
	\item[\rm(ii)] For any given $a>0$, write $c_2 =a^2C_{\min}/(96e^2\kappa^4C_{\max}) - 2$ and 
	\begin{equation*}
		\mathcal G_n(a) := \bigg\{\bZ \in \mathbb R^{n\times 2d}: \mu_{\min}(\bZ) < a\sqrt{\frac{\log(2d)}{n}} \bigg\}
	\end{equation*}
	 with \( \mu_{\min}(\bZ) = \min_{\bTheta \in \mathbb{R}^{2d \times 2d}} \mu_*(\bZ; \bTheta) \). It holds that
        \begin{equation*}
		\mathbb P\{\bZ \in \mathcal G_n(a)\} \ge 1-8d^{-c_2}
	\end{equation*}
    for $n \ge 6(c_{2}+2)\log (2d)$.
	\end{enumerate}
\end{lemma}

\begin{lemma}\label{theo_fix}
For any given $\tau>0$, let $\hat{\bgamma}$ be the Lasso estimator given in \eqref{eq:lasso} with $\varrho_1$ satisfying \( \varrho_1 \ge 4\sigma\{3C_{\max}(\tau+1)n^{-1}\log(2d)\}^{1/2} \). Assume that \( |\bgamma_{0}|_{0} \leq s_0 \) and Conditions {\rm\ref{ass:model_error}} and {\rm\ref{ass:random_Z}} hold. Then
\begin{align*}
    \mathbb{P}\bigg(|\hat{\bgamma} - \bgamma_{0}|_1 > \frac{16\varrho_1 s_0}{C_{\min}}\bigg)\le4d^{-\tau}+2e^{-c_{1}n}
\end{align*}
provided that $n \ge \max\{v_0s_0\log(240eds_0^{-1}), 5000(\tau+1)\kappa^4\log(2d), 12.5(\tau+1)\log(2d)\}$,
where $v_0$ and $c_1$ are specified in Lemma {\rm\ref{theo_condition}(i)}.
\end{lemma}

\begin{lemma}\label{lemm_clime}
 Let Conditions {\rm\ref{ass:random_Z}} and {\rm\ref{ass:CLIME}} hold. Then, if $n\ge \eta^{-1}\log(2d)$,
\begin{align*}
    \mathbb{P}\big(| \hat{\bTheta} - \bTheta_0 |_{\infty} \le 4M\varrho_{2}\big)\ge
    1 - 2^{1-\tau}d^{-\tau}\,.
\end{align*}
\end{lemma}

For any given $\bZ\in\mathcal G_n(a)$ with $a$ specified in Condition \ref{ass:CLIME}, based on the definition of $\mathcal{G}_n(a)$, we know there exits a $\bTheta$ such that $| \bTheta^{\top} \hat{\bGamma} - \bI_{2d} |_{\infty} < a\{n^{-1}\log(2d)\}^{1/2}$, which implies the optimization problem \eqref{CLIME} has a solution. Restricted on $\bZ\in\mathcal G_n(a)$, based on the definition of $\hat{\bTheta}$, we have $|\hat{\bTheta}^{\top} \hat{\bGamma} - \bI_{2d}|_{\infty}\le \varrho_{2}$. Recall $\bdelta_{n} = n^{1/2}(\hat{\bTheta}^{\top} \hat{\bGamma} - \bI_{2d})(\bgamma_{0} - \hat{\bgamma})$. Hence, given $\bZ\in\mathcal G_n(a)$,
$$|\bdelta_{n}|_{\infty}\le n^{1/2}|\hat{\bTheta}^{\top} \hat{\bGamma} - \bI_{2d}|_{\infty}|\bgamma_{0} - \hat{\bgamma}|_{1}\le n^{1/2}\varrho_{2}|\bgamma_{0} - \hat{\bgamma}|_{1}\,.$$
For any given $\tau>0$, if $\varrho_1 \ge 4\sigma\{3C_{\max}(\tau+1)n^{-1}\log(2d)\}^{1/2}$, by Lemmas \ref{theo_condition}(ii) and \ref{theo_fix},
\begin{align*}
		\mathbb{P}\bigg( |\bdelta_{n}|_{\infty} > \frac{16\varrho_1\varrho_2s_0n^{1/2}}{C_{\min}} \bigg) &\le
		\mathbb P\bigg\{\bZ\in\mathcal G_n(a),|\bdelta_{n}|_{\infty} > \frac{16\varrho_1\varrho_2s_0n^{1/2}}{C_{\min}}\bigg\} 
        + \mathbb P\{\bZ\in\mathcal G^{\rm c}_n(a)\} \\
		&\le  \mathbb P\bigg\{\bZ\in\mathcal G_n(a), n^{1/2}\varrho_{2}|\bgamma_{0} - \hat{\bgamma}|_{1}>\frac{16\varrho_1\varrho_2s_0n^{1/2}}{C_{\min}} \bigg\} + 8d^{-c_{2}}\\
        &\le\mathbb{P}\bigg(|\hat{\bgamma} - \bgamma_{0}|_1 > \frac{16\varrho_1 s_0}{C_{\min}}\bigg) + 8d^{-\tau}\le2e^{-c_1n} + 12d^{-\tau}
\end{align*}
provided that $n \ge \max\{v_0s_0\log(240eds_0^{-1}), 5000(\tau+1)\kappa^4\log(2d), 12.5(\tau+1)\log(2d), 6(c_{2}+2)\log (2d)\}$, where $v_0$, $c_{1}$ and $c_2$ are specified in Lemma \ref{theo_condition}. By triangle inequality, we have 
\begin{equation}\label{eq:thm1-1}
|\hat\bTheta^{\top} \hat{\bGamma} \hat\bTheta - \bTheta_0|_{\infty} 
\le |(\hat\bTheta^{\top} \hat{ \bGamma} - \bI_{2d}) \hat\bTheta|_{\infty} + |\hat\bTheta - \bTheta_0|_{\infty}
\le|\hat\bTheta^{\top} \hat{\bGamma} - \bI_{2d}|_{\infty}\|\hat\bTheta\|_1
+ |\hat\bTheta - \bTheta_0|_{\infty}\,.
\end{equation}
As shown in Equation (25) of \citeS{cai2011constrained_app}, if $|\bTheta_0 \hat{\bGamma} - \bI_{2d}|_{\infty}\le \varrho_{2}$, then $\|\hat\bTheta\|_1 \le \|\bTheta_0\|_1 \le M$ and $|\hat{\bTheta}^{\top} \hat{\bGamma} - \bI_{2d}|_{\infty}\le \varrho_{2}$. By the proof of Lemma \ref{theo_condition}(ii) in Section \ref{pflem1a3}, if $n \ge 6(c_{2}+2)\log (2d)$, we have
\begin{align*}
\mathbb P(|\hat\bTheta^{\top} \hat{\bGamma} - \bI_{2d}|_{\infty}\|\hat\bTheta\|_1\le M\varrho_{2})
&\ge\mathbb P(|\hat\bTheta^{\top} \hat{\bGamma} - \bI_{2d}|_{\infty}\|\hat\bTheta\|_1\le M\varrho_{2},|\bTheta_0 \hat{\bGamma} - \bI_{2d}|_{\infty}\le \varrho_{2})\\
&\ge\mathbb P(|\hat\bTheta^{\top} \hat{\bGamma} - \bI_{2d}|_{\infty} \le \varrho_{2},|\bTheta_0^{\top} \hat{\bGamma} - \bI_{2d}|_{\infty}\le \varrho_{2})\\
&=\mathbb P(|\bTheta_0 \hat{\bGamma} - \bI_{2d}|_{\infty}\le \varrho_{2})\ge1 - 8d^{-c_{2}}
\ge1 - 8d^{-\tau}\,.
\end{align*}
Therefore, by \eqref{eq:thm1-1} and Lemma \ref{lemm_clime}, we have $\mathbb{P}\big(|\hat{\bTheta}^\top \hat{\bGamma} \hat{\bTheta} - \bTheta_0|_{\infty} \le 5M\varrho_{2}\big)\ge1 - 10d^{-\tau}$ for $n \ge \max\{6(c_{2}+2)\log (2d), \eta^{-1}\log(2d)\}$. We complete the proof of Theorem \ref{theo_random}. $\hfill\qedsymbol$

\subsection{Proof of Theorem \ref{thm0}}\label{proof_thm0}
Recall $G(t)=2\{1-\Phi(t)\}$. Let $G^{-1}(\cdot)$ denote the inverse function of $G(\cdot)$. To prove Theorem \ref{thm0}, we need Lemmas \ref{theo_scaled} and \ref{theo_random_trans} whose proofs are given in Sections \ref{proof_theo_scaled} and \ref{proof_theo_random_trans}, respectively.

\begin{lemma}\label{theo_scaled}
Assume that Condition {\rm\ref{ass:model_error}} and the conditions of Lemma {\rm \ref{theo_condition}} hold. Let $\hat{\sigma}$ be the estimator given in \eqref{eq:vest} with $\varrho_3$ satisfying $\varrho_3 \ge 10C_{\max}^{1/2}\{n^{-1}\log(2d)\}^{1/2}$ and $\varrho_3 \asymp \{n^{-1}\log(2d)\}^{1/2}$. If $1\le s_0 \ll n/\log(2d)$ and $|\bgamma_{0}|_{0}\le s_{0}$, then
\begin{equation*}
\lim_{n\rightarrow \infty}\mathbb{P} \bigg(\bigg\lvert \frac{\hat{\sigma}}{\sigma} - 1 \bigg\rvert \ge 16\varrho_3\sqrt{\frac{s_0}{C_{\min}}}\bigg) = 0\,.
\end{equation*}
\end{lemma}

\begin{lemma}\label{theo_random_trans}
Let the conditions of Theorem {\rm\ref{theo_random}} hold. The following two assertions hold:

{\rm (i)} If $n \ge \max\{v_0s_0\log(240eds_0^{-1}), 5000(\tau+1)\kappa^4\log(2d), 12.5(\tau+1)\log(2d), 6(c_{2}+2)\log(2d)\}$, then 
	\begin{align*}
		\mathbb P\bigg(|\bT\bdelta_{n}|_{\infty} > \frac{32\varrho_1\varrho_2s_0n^{1/2}}{C_{\min}}\bigg) \le 2e^{-c_1n} + 12d^{-\tau}\,,
	\end{align*}
where $v_0$, $c_1$ and $c_{2}$ are specified in Theorem {\rm\ref{theo_random}}.

{\rm (ii)} If $n \ge \max\{6(c_{2}+2)\log (2d), \eta^{-1}\log(2d)\}$, then
\begin{align*}
\mathbb{P}\big(|\bT\hat\bTheta^\top \hat{\bGamma} \hat\bTheta\bT^{\top} - \bT\bTheta_0\bT^{\top}|_{\infty} \le 20M\varrho_{2}\big)\ge1 - 10d^{-\tau}\,.
\end{align*}
\end{lemma}

Let $\bar{t}_{2, j}=|t_{2, j}|$ and $\bar{t}_{2, (1)} \geq \cdots \geq \bar{t}_{2,(d)}$ be the ordered sequence of $\{\bar{t}_{2, j}\}_{j=1}^{d}$. Define $t_{2,(0)}=+\infty$. By the definition of $\tilde{R}$ specified in Step 2 of Algorithm \ref{method_1}, we have $\tilde{R}=\max\{j=0,1, \ldots, d: \tilde{P}_{(j)}=G(\bar{t}_{2, (j)}) \leq j \alpha/d\}$. Then $G(\bar{t}_{2,(\tilde{R})})\leq \alpha\tilde{R} /d$ and $G(\bar{t}_{2,(l)})> \alpha l/d$ for all $l>\tilde{R}$, which implies $\bar{t}_{2,(l)}< \bar{t}_{2,(\tilde{R})}$ for all $l>\tilde{R}$. Hence, $|\{j\in[d]:\bar{t}_{2, j} \geq \bar{t}_{2,(\tilde{R})}\}|=\tilde{R}$, and rejecting all $H_{0,j}$'s with $\bar{t}_{2, j} \geq \bar{t}_{2,(\tilde{R})}$ is equivalent to rejecting all $H_{0,j}$'s corresponding to the first $\tilde{R}$ largest $\bar{t}_{2, j}$'s among $\{\bar{t}_{2, j}\}_{j=1}^{d}$. Without loss of generality, we assume $|\mathcal{H}|\le\sqrt{\log d}$. If $|\mathcal{H}|>\sqrt{\log d}$, we can redefine $\mathcal{H}$ by keeping only the first $\lfloor\sqrt{\log d}\rfloor$ elements of it. Our proof of Theorem \ref{thm0} mainly includes the following four steps:

{\bf Step 1}.
To show that
\begin{equation}\label{thm0-1}
\lim_{n\rightarrow\infty}\mathbb{P}\bigg\{\sum_{j=1}^d I(\bar{t}_{2, j}>\sqrt{2 \log d}) \geq|\mathcal{H}|\bigg\} = 1\,.
\end{equation}

{\bf Step 2}. To show that $\{\bar{t}_{2, j} \geq \bar{t}_{2,(\tilde{R})}\}=\{\bar{t}_{2, j} \geq \hat{t}\}$ for any $j\in[d]$, where
\begin{equation}\label{thm0-8}
\hat{t}:=\inf\bigg[t>0: G(t)\leq\frac{\alpha }{d}\max\bigg\{\sum_{j=1}^{d}I(\bar{t}_{2, j}\geq t),1\bigg\}\bigg]\,.
\end{equation}

{\bf Step 3}. To show that $\hat{t}$ specified in \eqref{thm0-8} satisfies
\begin{align}
&~\frac{dG(\hat{t})}{\max\{\sum_{j=1}^{d}I(\bar{t}_{2, j}\geq \hat{t}),1\}}=\alpha\,,\label{thm0-10}\\
&\lim_{n\rightarrow\infty}\mathbb{P}\bigg\{\hat{t} \leq G^{-1}\bigg(\frac{\alpha |\mathcal{H}|}{d}\bigg)\bigg\} = 1\,.\label{thm0-11}
\end{align}

{\bf Step 4}. To show that
\begin{equation}\label{thm0-16}
\begin{aligned}
\sup_{0\leq t\leq G^{-1}(\alpha|\mathcal{H}|/d)}\frac{\sum_{j\in \mathcal{H}_{0}}I(\bar{t}_{2, j}\geq t)}{d G(t)}\leq \frac{d_{0}}{d}+o_{\rm p}(1)\,,
\end{aligned}
\end{equation}
where $d_{0}=d-d_{1}$ is the number of true null hypotheses.

The proofs of Steps 1--4 are given in Sections \ref{proof_thm2_step1}--\ref{proof_thm2_step4}, respectively. Due to $\{\bar{t}_{2, j} \geq \bar{t}_{2,(\tilde{R})}\}=\{\bar{t}_{2, j} \geq \hat{t}\}$ for any $j\in[d]$, by \eqref{thm0-10}, 
\begin{equation*}
\text{FDP}=\frac{\sum_{j\in \mathcal{H}_{0}}I(\bar{t}_{2, j}\geq \hat{t})}{\max\{\sum_{j=1}^{d}I(\bar{t}_{2, j}\geq \hat{t}),1\}}=\frac{\sum_{j\in \mathcal{H}_{0}}I(\bar{t}_{2, j}\geq \hat{t})}{d G(\hat{t})/\alpha}\,.
\end{equation*}
By \eqref{thm0-11} and \eqref{thm0-16}, for any $\epsilon>0$,
\begin{align*}
&\mathbb{P}\bigg\{\frac{\sum_{j\in \mathcal{H}_{0}}I(\bar{t}_{2, j}\geq \hat{t})}{d G(\hat{t})}\leq \epsilon+\frac{d_{0}}{d}\bigg\}\\
&~~~~~\geq\mathbb{P}\bigg\{\sup_{0\leq t\leq G^{-1}(\alpha|\mathcal{H}|/d)}\frac{\sum_{j\in \mathcal{H}_{0}}I(\bar{t}_{2, j}\geq t)}{d G(t)}\leq \epsilon+\frac{d_{0}}{d},\hat{t} \leq G^{-1}\bigg(\frac{\alpha |\mathcal{H}|}{d}\bigg)\bigg\}\\
&~~~~~\geq\mathbb{P}\bigg\{\sup_{0\leq t\leq G^{-1}(\alpha|\mathcal{H}|/d)}\frac{\sum_{j\in \mathcal{H}_{0}}I(\bar{t}_{2, j}\geq t)}{d G(t)}\leq \epsilon+\frac{d_{0}}{d}\bigg\}-\mathbb{P}\bigg\{\hat{t} > G^{-1}\bigg(\frac{\alpha |\mathcal{H}|}{d}\bigg)\bigg\}\\
&~~~~~=1-o(1)\,,
\end{align*}
which implies
\begin{align*}
\mathbb{P}\bigg(\text{FDP}\leq\frac{d_{0}}{d}\alpha+\epsilon\bigg)
=\mathbb{P}\bigg\{\frac{\sum_{j\in \mathcal{H}_{0}}I(\bar{t}_{2, j}\geq \hat{t})}{d G(\hat{t})/\alpha}\leq\frac{d_{0}}{d}\alpha+\epsilon\bigg\}
\geq1-o(1)\,.
\end{align*}
Since $\text{FDR}=\mathbb{E}(\text{FDP})$ and $0\le\text{FDP}\le1$, it holds that $\lim\sup_{n\rightarrow \infty}\text{FDR}\leq \alpha d_{0}/d$. We complete the proof of Theorem \ref{thm0}. $\hfill\qedsymbol$

\subsubsection{Proof of Step 1}\label{proof_thm2_step1}

Write $\bw_{n}=(w_{1},\ldots,w_{2d})^{\top}$, $\bbeta_{0}=(\beta^0_{1},\ldots,\beta^0_{d})^{\top}$ and $\hat\bgamma^{(\rm bc)}=(\hat{\gamma}^{(\rm bc)}_{1},\ldots,\hat{\gamma}^{(\rm bc)}_{2d})^{\top}$. Let $\hat\bLambda :=(\hat\Lambda_{i,j})_{2d\times 2d} = \bT \hat{\bTheta}^\top \hat{\bGamma} \hat{\bTheta} \bT^\top$, $\tilde{\bD}:=(\tilde{D}_{i,j})_{d\times d}=2\bD^{-1}$ and $\tilde{\bdelta}_{n}:=(\tilde{\delta}_{1},\ldots,\tilde{\delta}_{2d})^{\top}=\bT \bdelta_{n}$. Using \eqref{eq:exp1}, for any $j\in[d]$, we have
\begin{align}\label{eq:thm2-step1-1}
t_{2,j} =\frac{n^{1/2}\{\hat{\gamma}^{(\rm bc)}_{j} - \hat{\gamma}^{(\rm bc)}_{j+d}\}}{\hat\sigma \hat\Lambda_{j+d,j+d}^{1/2}}= \frac{n^{1/2}\beta^0_{j}}{\hat\sigma\hat\Lambda_{j+d,j+d}^{1/2}}
+\frac{\sigma}{\hat\sigma}\tilde{w}_{j+d}+
\frac{\tilde{\delta}_{j+d}}{\hat\sigma\hat\Lambda_{j+d,j+d}^{1/2}}\,,
\end{align}
where $\tilde{w}_{j+d}=\sigma^{-1}\hat\Lambda_{j+d,j+d}^{-1/2}(w_{j}-w_{j+d})$. Since $\sigma_{\min}(\bD)\ge\sigma_{\min}(\bGamma)\ge C_{\min}$ and $\sigma_{\max}(\bD)\le\sigma_{\max}(\bGamma)\le C_{\max}$, then $2C_{\max}^{-1}\le \tilde{D}_{j,j}\le 2C_{\min}^{-1}$. Consider the events 
\begin{align*}
\tilde{\mathcal E}_{1,n}:=\bigg\{\bigg|\frac{\hat\sigma}{\sigma}-1\bigg|\le\frac{1}{25}\bigg\}
~~\text{and}~~
\tilde{\mathcal E}_{2,n}:=\bigg\{\max_{j \in \mathcal{H}}\bigg|\frac{\hat\Lambda_{j+d,j+d}}{\tilde{D}_{j,j}}-1\bigg|\le 10C_{\max}M\varrho_{2}\bigg\}\,.
\end{align*}
Select $\varrho_{2}\le(100\sqrt{2}-141)(910M)^{-1}C_{\max}^{-1}$. Given the events $\tilde{\mathcal E}_{1,n}$ and $\tilde{\mathcal E}_{2,n}$, by Condition \ref{ass0},
\begin{equation*}
\bigg|\frac{n^{1/2}\beta^0_{j}}{\hat\sigma\hat\Lambda_{j+d,j+d}^{1/2}}\bigg|
\ge\bigg|\frac{25n^{1/2}\beta^0_{j}}
{26(1+10C_{\max}M\varrho_{2})\sigma \tilde{D}_{j,j}^{1/2}}\bigg|\ge\sqrt{2\log d}+\frac{\sqrt{\log d}}{2}
\end{equation*}
for each $j\in\mathcal{H}$, which implies
\begin{align}\label{thm0-3}
&\mathbb{P}\bigg\{\sum_{j=1}^d I(\bar{t}_{2, j}>\sqrt{2 \log d})\geq |\mathcal{H}|\bigg\}
\geq \mathbb{P}\bigg[\bigcap_{j \in \mathcal{H}}\{\bar{t}_{2, j}> \sqrt{2 \log d}\}\bigg]\notag\\
&~~~~~=\mathbb{P}\bigg[\bigcap_{j \in \mathcal{H}}\bigg\{\bigg|\frac{n^{1/2}\beta^0_{j}}{\hat\sigma\hat\Lambda_{j+d,j+d}^{1/2}}
+\frac{\sigma}{\hat\sigma}\tilde{w}_{j+d}+
\frac{\tilde{\delta}_{j+d}}{\hat\sigma\hat\Lambda_{j+d,j+d}^{1/2}}\bigg|
>\sqrt{2 \log d}\bigg\}\bigg]\notag\\
&~~~~~\ge\mathbb{P}\bigg[\bigcap_{j \in \mathcal{H}}\bigg\{\bigg|\frac{n^{1/2}\beta^0_{j}}{\hat\sigma\hat\Lambda_{j+d,j+d}^{1/2}}\bigg|
-\bigg|\frac{\sigma}{\hat\sigma}\tilde{w}_{j+d}+
\frac{\tilde{\delta}_{j+d}}{\hat\sigma\hat\Lambda_{j+d,j+d}^{1/2}}\bigg|
>\sqrt{2 \log d}\bigg\}\bigg]\\
&~~~~~\geq\mathbb{P}\bigg[\bigcap_{j \in \mathcal{H}}\bigg\{\bigg|\frac{n^{1/2}\beta^0_{j}}{\hat\sigma\hat\Lambda_{j+d,j+d}^{1/2}}\bigg|
-\bigg|\frac{\sigma}{\hat\sigma}\tilde{w}_{j+d}+
\frac{\tilde{\delta}_{j+d}}{\hat\sigma\hat\Lambda_{j+d,j+d}^{1/2}}\bigg|
>\sqrt{2 \log d}\bigg\},\tilde{\mathcal E}_{1,n}, \tilde{\mathcal E}_{2,n}\bigg]\notag\\
&~~~~~\geq\mathbb{P}\bigg[\bigcap_{j \in \mathcal{H}}\bigg\{\bigg|\frac{\sigma}{\hat\sigma}\tilde{w}_{j+d}+
\frac{\tilde{\delta}_{j+d}}{\hat\sigma\hat\Lambda_{j+d,j+d}^{1/2}}\bigg|
<\frac{\sqrt{\log d}}{2}\bigg\},\tilde{\mathcal E}_{1,n}, \tilde{\mathcal E}_{2,n}\bigg]\,.\notag
\end{align}
Since $|x/y-1|\leq|(x/y)^2-1|$ for all $x, y>0$, given the event $\tilde{\mathcal E}_{2,n}$, we have
\begin{align*}
\frac{1}{\hat\Lambda_{j+d,j+d}^{1/2}}=
\frac{\tilde{D}_{j,j}^{1/2}}{\tilde{D}_{j,j}^{1/2}\hat\Lambda_{j+d,j+d}^{1/2}}\le
\frac{C_{\max}^{1/2}}{\sqrt{2}(1-10C_{\max}M\varrho_{2})}\le\frac{91C_{\max}^{1/2}}{90\sqrt{2}}\,,
\end{align*}
which implies
\begin{align}\label{eq:thm2-1}
&\mathbb{P}\bigg[\bigcap_{j \in \mathcal{H}}\bigg\{\bigg|\frac{\sigma}{\hat\sigma}\tilde{w}_{j+d}+
\frac{\tilde{\delta}_{j+d}}{\hat\sigma\hat\Lambda_{j+d,j+d}^{1/2}}\bigg|
<\frac{\sqrt{\log d}}{2}\bigg\},\tilde{\mathcal E}_{1,n}, \tilde{\mathcal E}_{2,n}\bigg]\notag\\
&~~~~~\geq\mathbb{P}\bigg[\bigcap_{j \in \mathcal{H}}\bigg\{|\tilde{w}_{j+d}|
<\frac{\hat\sigma}{\sigma}\cdot\frac{\sqrt{\log d}}{2}-\frac{|\tilde{\bdelta}_{n}|_{\infty}}
{\sigma \hat\Lambda_{j+d,j+d}^{1/2}}\bigg\},\tilde{\mathcal E}_{1,n}, \tilde{\mathcal E}_{2,n}\bigg]\notag\\
&~~~~~\geq\mathbb{P}\bigg[\bigcap_{j \in \mathcal{H}}\bigg\{|\tilde{w}_{j+d}|
<\frac{12\sqrt{\log d}}{25}-\frac{91C_{\max}^{1/2}|\tilde{\bdelta}_{n}|_{\infty}}
{90\sqrt{2}\sigma}\bigg\},\tilde{\mathcal E}_{1,n}, \tilde{\mathcal E}_{2,n}\bigg]\\
&~~~~~\geq\mathbb{P}\bigg[\bigcap_{j \in \mathcal{H}}\bigg\{|\tilde{w}_{j+d}|
<\frac{12\sqrt{\log d}}{25}-\frac{91C_{\max}^{1/2}|\tilde{\bdelta}_{n}|_{\infty}}
{90\sqrt{2}\sigma}\bigg\},|\tilde{\bdelta}_{n}|_{\infty}<1,\tilde{\mathcal E}_{1,n}, \tilde{\mathcal E}_{2,n}\bigg]\notag\\
&~~~~~\geq\mathbb{P}\bigg[\bigcap_{j \in \mathcal{H}}\bigg\{|\tilde{w}_{j+d}|
\leq\frac{12\sqrt{\log d}}{25}-\frac{91C_{\max}^{1/2}}
{90\sqrt{2}\sigma}\bigg\},|\tilde{\bdelta}_{n}|_{\infty}<1,\tilde{\mathcal E}_{1,n}, \tilde{\mathcal E}_{2,n}\bigg]\notag\\
&~~~~~\geq\mathbb{P}\bigg[\bigcap_{j \in \mathcal{H}}\bigg\{|\tilde{w}_{j+d}|
\leq\frac{12\sqrt{\log d}}{25}-\frac{91C_{\max}^{1/2}}
{90\sqrt{2}\sigma}\bigg\}\bigg]
-\mathbb{P}(|\tilde{\bdelta}_{n}|_{\infty}\ge1)
-\mathbb{P}(\tilde{\mathcal E}_{1,n}^{\rm c})-\mathbb{P}(\tilde{\mathcal E}_{2,n}^{\rm c})\,.\notag
\end{align}
Due to $\bw_{n}\,|\,\bZ\sim\mathcal N({\bf 0}, \sigma^{2} \hat{\bTheta}^{\top} \hat{\bGamma} \hat{\bTheta})$, then $\sigma^{-1}\bT\bw_{n}\,|\,\bZ\sim\mathcal N({\bf 0},\hat\bLambda)$, which implies $\tilde{w}_{j+d}\,|\,\bZ\sim\mathcal N(0,1)$. Hence,
\begin{align}\label{eq:thm2-2}
&\mathbb{P}\bigg[\bigcap_{j \in \mathcal{H}}\bigg\{|\tilde{w}_{j+d}|
\leq\frac{12\sqrt{\log d}}{25}-\frac{91C_{\max}^{1/2}}
{90\sqrt{2}\sigma}\bigg\}\bigg]
\geq1-\sum_{j \in \mathcal{H}}\mathbb{P}\bigg(|\tilde{w}_{j+d}|
>\frac{12\sqrt{\log d}}{25}-\frac{91C_{\max}^{1/2}}
{90\sqrt{2}\sigma}\bigg)\notag\\
&~~~~~=1-\sum_{j \in \mathcal{H}}\mathbb{E}\bigg\{\mathbb{P}\bigg(|\tilde{w}_{j+d}|>\frac{12\sqrt{\log d}}{25}-\frac{91C_{\max}^{1/2}}
{90\sqrt{2}\sigma}\,\bigg|\,\bZ\bigg)\bigg\}\\
&~~~~~\ge1-\sqrt{\log d}\cdot G\bigg(\frac{12\sqrt{\log d}}{25}-\frac{91C_{\max}^{1/2}}
{90\sqrt{2}\sigma}\bigg)\rightarrow1\notag
\end{align}
as $n\rightarrow\infty$. Recall $\varrho_{1}\asymp\varrho_{2}\asymp\varrho_{3}\asymp\{n^{-1}\log(2d)\}^{1/2}$ and $s_{0}\ll n^{1/2}/\log(2d)$. By Lemmas \ref{theo_scaled} and \ref{theo_random_trans}, we have 
$
\mathbb{P}(|\tilde{\bdelta}_{n}|_{\infty}\ge1)
+\mathbb{P}(\tilde{\mathcal E}_{1,n}^{\rm c})+\mathbb{P}(\tilde{\mathcal E}_{2,n}^{\rm c})= o(1)$. 
Together with \eqref{thm0-3}--\eqref{eq:thm2-2}, we have \eqref{thm0-1}. $\hfill\qedsymbol$

\subsubsection{Proof of Step 2}\label{proof_thm2_step2}

To prove this assertion, we consider three cases separately: (i) $\tilde{R}=0$, (ii) $\tilde{R}=d$, and (iii) $1\leq\tilde{R}<d$. 

\underline{{\it Case} (i): $\tilde{R}=0$.} Denote $\bar{t}_{2, (d+1)}:=-\infty$ and $\mathcal L:=\{l\in[d]: \bar{t}_{2, (l+1)}\neq\bar{t}_{2, (l)}\}$. By the definition of $\tilde{R}$, we have $G(\bar{t}_{2, (l)})>\alpha ld^{-1}$ for any $l\in[d]$. Therefore, for all $t\leq \bar{t}_{2, (1)}$, there exists $l\in \mathcal L$ such that $\bar{t}_{2,(l+1)}<t\leq \bar{t}_{2, (l)}$ and
\begin{align*}
G(t)\geq G(\bar{t}_{2, (l)})>\frac{\alpha l}{d}=\frac{\alpha }{d}\max\bigg[\sum_{j=1}^{d}I\{\bar{t}_{2, j}\geq \bar{t}_{2, (l)}\},1\bigg]=\frac{\alpha }{d}\max\bigg\{\sum_{j=1}^{d}I(\bar{t}_{2, j}\geq t),1\bigg\}\,,
\end{align*}
which implies $\hat{t}\ge \bar{t}_{2, (1)}$. By the definition of $\hat{t}$, if $\hat{t}= \bar{t}_{2, (1)}$, there exists a positive sequence $\{\epsilon_{k}\}$ with $\epsilon_{k}\rightarrow0$ such that
\begin{align*}
G(\bar{t}_{2, (1)}+\epsilon_{k})\leq\frac{\alpha }{d}\max\bigg[\sum_{j=1}^{d}I\{\bar{t}_{2, j}\geq \bar{t}_{2, (1)}+\epsilon_{k}\},1\bigg]=\frac{\alpha}{d}\,.
\end{align*}
Letting $\epsilon_k\downarrow0$ in the first term in above inequality, we have $G(\bar{t}_{2, (1)})\leq \alpha/d$, which implies $\tilde{R}>0$. Hence, it holds that $\hat{t}> \bar{t}_{2, (1)}$, which implies $\{\bar{t}_{2, j}\geq\hat{t}\}=\emptyset$ for any $j\in[d]$. Due to $\bar{t}_{2, (0)}=+\infty$, we have $\{\bar{t}_{2, j} \ge \bar{t}_{2, (0)}\}=\emptyset$ for any $j\in[d]$. Then, $\{\bar{t}_{2, j} \geq \bar{t}_{2,(0)}\}=\{\bar{t}_{2, j} \geq \hat{t}\}$ for any $j\in[d]$ in Case (i).

\underline{{\it Case} (ii): $\tilde{R}=d$.} By the definition of $\tilde{R}$, we have $$G(\bar{t}_{2, (d)})\leq \alpha=\frac{\alpha }{d}\max\bigg[\sum_{j=1}^{d}I\{\bar{t}_{2, j}\geq \bar{t}_{2, (d)}\},1\bigg]\,,$$ 
which implies $\hat{t}\leq \bar{t}_{2, (d)}$. Then $\{\bar{t}_{2,j}\geq \hat{t}\}$ is a certain event for any $j\in[d]$. Notice that $\{\bar{t}_{2,j}\geq \bar{t}_{2, (d)}\}$ is a certain event for any $j\in[d]$. Therefore, $\{\bar{t}_{2,j}\geq \hat{t}\}=\{\bar{t}_{2,j}\geq \bar{t}_{2, (d)}\}$ for any $j\in[d]$ in Case (ii). 

\underline{{\it Case} (iii): $1\leq\tilde{R}<d$.} By the definition of $\tilde{R}$, for all $\tilde{R}<l\leq d$, we have 
\begin{align}\label{thm0-9}
G(\bar{t}_{2, (l)})>\frac{\alpha l}{d}\,.
\end{align}
Define $\mathcal L_1:=\{\tilde{R}+1\leq l\leq d: \bar{t}_{2, (l+1)}\neq\bar{t}_{2, (l)}\}$. For all $t\leq \bar{t}_{2, (\tilde{R}+1)}$, there exists $l\in\mathcal L_1$ such that $\bar{t}_{2,(l+1)}<t\leq \bar{t}_{2, (l)}$ and
\begin{equation*}
G(t)\geq G(\bar{t}_{2, (l)})>\frac{\alpha l}{d}=\frac{\alpha }{d}\max\bigg\{\sum_{j=1}^{d}I(\bar{t}_{2, j}\geq t),1\bigg\}\,,
\end{equation*}
which implies $\hat{t}\ge \bar{t}_{2, (\tilde{R}+1)}$. If $\hat{t}= \bar{t}_{2, (\tilde{R}+1)}$, similar to Case (i), we have $G\{\bar{t}_{2, (\tilde{R}+1)}\}\leq \alpha(\tilde{R}+1)/d$, which contradicts \eqref{thm0-9}. Then, $\hat{t}>\bar{t}_{2, (\tilde{R}+1)}$. By the definition of $\tilde{R}$, we know $\bar{t}_{2, (\tilde{R}+1)}<\bar{t}_{2, (\tilde{R})}$ and
\begin{align*}
G(\bar{t}_{2, (\tilde{R})})\leq \frac{\alpha\tilde{R}}{d}=\frac{\alpha }{d}\max\bigg[\sum_{j=1}^{d}I\{\bar{t}_{2, j}\geq \bar{t}_{2, (\tilde{R})}\},1\bigg]\,,
\end{align*}
which implies $\hat{t}\leq \bar{t}_{2, (\tilde{R})}$. Hence, $\bar{t}_{2, (\tilde{R}+1)}<\hat{t}\leq \bar{t}_{2, (\tilde{R})}$. Therefore, $\{\bar{t}_{2,j}\geq \hat{t}\}=\{\bar{t}_{2,j}\geq \bar{t}_{2, (\tilde{R})}\}$ for any $j\in[d]$ in Case (iii).  $\hfill\qedsymbol$

\subsubsection{Proof of Step 3}\label{proof_thm2_step3}
By the definition of $\hat{t}$, for all $0<t<\hat{t}$, it holds that \begin{equation*}
\widetilde{\text{FDP}}(t):=\frac{dG(t)}{ \max\{\sum_{j=1}^{d}I(\bar{t}_{2, j}\geq t),1\}}>\alpha\,.
\end{equation*}
Since $I(\bar{t}_{2, j}>t)\geq I(\bar{t}_{2, j}>\hat{t})$ for all $t<\hat{t}$, we have
\begin{equation*}
\frac{dG(t)}{\max\{\sum_{j=1}^{d}I(\bar{t}_{2, j}\geq \hat{t}),1\}}\geq\widetilde{\text{FDP}}(t)>\alpha\,.
\end{equation*}
Letting $t\uparrow \hat{t}$ in the numerator of the first term in above inequality, we have $\widetilde{\text{FDP}}(\hat{t})\geq \alpha$. By the definition of $\hat{t}$, there exists a sequence $\{u_{k}\}$ with $u_{k}\geq\hat{t}$ and $u_{k}\rightarrow\hat{t}$ such that $\widetilde{\text{FDP}}(u_k)\leq\alpha$. Since $I(\bar{t}_{2, j}\geq \hat{t})\geq I(\bar{t}_{2, j}\geq u_k)$, we have
\begin{equation*}
\frac{dG(u_k)}{ \max\{\sum_{j=1}^{d}I(\bar{t}_{2, j}\geq \hat{t}),1\}}\leq \frac{dG(u_k)}{ \max\{\sum_{j=1}^{d}I(\bar{t}_{2, j}\geq u_k),1\}}=\widetilde{\text{FDP}}(u_k)\leq\alpha\,.
\end{equation*}
Letting $u_k\downarrow\hat{t}$ in the numerator of the first term in above inequality, we have $\widetilde{\text{FDP}}(\hat{t})\leq \alpha$. Therefore, we have (\ref{thm0-10}). 

By Proposition 2.5 of \citeS{dudley2014uniform}, we know $G(u)\leq e^{-u^2/2}$ for any $u>0$. By Condition \ref{ass0}, if $\log\log d \ge 1/\alpha$, then
\begin{align*}
\frac{\alpha|\mathcal{H}|}{d} \geq \frac{1}{d}=\exp \left\{-\frac{(\sqrt{2 \log d})^2}{2}\right\} \geq G(\sqrt{2 \log d})\,,
\end{align*}
which implies
\begin{equation}\label{thm0-13}
\begin{aligned}
G^{-1}\bigg(\frac{\alpha |\mathcal{H}|}{d}\bigg)\leq \sqrt{2\log d}\,.
\end{aligned}
\end{equation}
Then, by the definition of $\hat{t}$, we have
\begin{align*}
\mathbb{P}\bigg\{\hat{t} \leq G^{-1}\bigg(\frac{\alpha |\mathcal{H}|}{d}\bigg)\bigg\}
&\geq\mathbb{P}\bigg(\frac{\alpha |\mathcal{H}|}{d}\leq \frac{\alpha }{d}\max\bigg[\sum_{j=1}^{d}I\bigg\{\bar{t}_{2, j}\geq G^{-1}\bigg(\frac{\alpha |\mathcal{H}|}{d}\bigg)\bigg\},1\bigg]\bigg) \\
&\ge \mathbb{P}\bigg[\sum_{j=1}^{d}I\bigg\{\bar{t}_{2, j}\geq G^{-1}\bigg(\frac{\alpha |\mathcal{H}|}{d}\bigg)\bigg\}\ge|\mathcal{H}|\bigg]\\
&\ge \mathbb{P}\bigg\{\sum_{j=1}^{d}I(\bar{t}_{2, j}\geq \sqrt{2\log d})\ge|\mathcal{H}|\bigg\}\,.
\end{align*}
Together with \eqref{thm0-1}, we have \eqref{thm0-11}. $\hfill\qedsymbol$

\subsubsection{Proof of Step 4}\label{proof_thm2_step4}

Let $\breve{t}_{2,j}:=n^{1/2}(\hat{\gamma}^{(\rm bc)}_{j} - \hat{\gamma}^{(\rm bc)}_{j+d}-\beta^0_{j})/(\hat\sigma \hat\Lambda_{j+d,j+d}^{1/2})$ for $j\in[d]$. Since $\bar{t}_{2, j}=|\breve{t}_{2,j}|$ for all $j\in\mathcal{H}_{0}$, then
\begin{align*}
\sup_{0\leq t\leq G^{-1}(\alpha|\mathcal{H}|/d)}\frac{\sum_{j\in \mathcal{H}_{0}}I(\bar{t}_{2, j}\geq t)}{d G(t)}
\leq\sup_{0\leq t\leq G^{-1}(\alpha|\mathcal{H}|/d)}\bigg|\frac{\sum_{j\in \mathcal{H}_{0}}I(|\breve{t}_{2, j}|\geq t)}{d G(t)}-\frac{d_{0}}{d}\bigg|+\frac{d_{0}}{d}\,.
\end{align*}
To prove \eqref{thm0-16}, it suffices to show
\begin{equation}\label{thm0-17}
\begin{aligned}
\sup_{0\leq t\leq G^{-1}(\alpha|\mathcal{H}|/d)}\bigg|\frac{\sum_{j\in \mathcal{H}_{0}}I(|\breve{t}_{2, j}|\geq t)}{d G(t)}-\frac{d_{0}}{d}\bigg|=o_{\rm p}(1)\,.
\end{aligned}
\end{equation}
To prove \eqref{thm0-17}, we need Lemmas \ref{lem0} and \ref{lem_normal}, whose proofs are given in Sections \ref{proof_lem0} and \ref{proof_lem_normal}, respectively.
\begin{lemma}\label{lem0}
Let the conditions of Theorem {\rm \ref{thm0}} hold. For any sequence of positive constants $\{b_{d}\}$ such that $b_{d}\rightarrow \infty$ and $b_{d}\ll d$,
\begin{align*}
\sup_{0\leq t\leq G^{-1}(b_{d}/d)}\bigg|\frac{\sum_{j\in \mathcal{H}_{0}}I(|\tilde{w}_{j+d}|\geq t)}{d G(t)}-\frac{d_{0}}{d}\bigg|=o_{\rm p}(1)\,.
\end{align*}
\end{lemma}

\begin{lemma}\label{lem_normal}
For any sequence of constants $\{\epsilon_{n}\}$ such that $\epsilon_{n}\rightarrow0$,
\begin{equation*}
\begin{aligned}
\sup_{0\leq t \leq \sqrt{2\log d}}\bigg|1-\frac{G(t+\epsilon_{n}/\sqrt{\log d})}{G(t)}\bigg|=o(1)\,.
\end{aligned}
\end{equation*}
\end{lemma}

If $|\mathcal{H}|\rightarrow\infty$ and $|\mathcal{H}|\ll d$, by Lemma \ref{lem0} with $b_{d}=\alpha|\mathcal{H}|\{1+o(1)\}$, we have
\begin{align}\label{eq:thm2-1-6}
\sup_{0\leq t\leq G^{-1}(\alpha|\mathcal{H}|\{1+o(1)\}/d)}\bigg|\frac{\sum_{j\in \mathcal{H}_{0}}I(|\tilde{w}_{j+d}|\geq t)}{d G(t)}-\frac{d_{0}}{d}\bigg|=o_{\rm p}(1)\,.
\end{align}
Similar to \eqref{thm0-13}, if $\log\log d \ge 1/[\alpha\{1+o(1)\}]$, then
\begin{align}\label{eq:thm2-1-7}
G^{-1}\bigg(\frac{\alpha |\mathcal{H}|}{d}\{1+o(1)\}\bigg)\leq \sqrt{2\log d}\,.
\end{align}
Let $\upsilon_{n}=s_{0}^{1/2}n^{-1/4}(\log d)^{3/4}$. Since $s_{0}\ll n^{1/2}(\log d)^{-3/2}$, then $\upsilon_{n}=o(1)$. By (\ref{thm0-13}) and Lemma \ref{lem_normal} with $\epsilon_{n}=\upsilon_{n}$, 
\begin{align*}
G\bigg(t+\frac{\upsilon_{n}}{\sqrt{\log d}}\bigg)= G(t)\{1+o(1)\}\geq\frac{\alpha|\mathcal{H}|}{d}\{1+o(1)\}
\end{align*}
for any $0\leq t\leq G^{-1}(\alpha|\mathcal{H}|/d)$. Then, if $\log\log d \ge 1/[\alpha\{1+o(1)\}]$,
\begin{align*}
t+\frac{\upsilon_{n}}{\sqrt{\log d}}\leq G^{-1}\bigg(\frac{\alpha|\mathcal{H}|}{d}\{1+o(1)\}\bigg)
\end{align*}
for any $0\leq t\leq G^{-1}(\alpha|\mathcal{H}|/d)$, where the term $o(1)$ holds uniformly over $t \in[0,  G^{-1}(\alpha|\mathcal{H}|/d)]$. Consider the event 
\begin{align*}
\mathcal{D}=\bigg\{\max_{j\in[d]}|\breve{t}_{2,j}-\tilde{w}_{j+d}|>\frac{\upsilon_{n}}{\sqrt{\log d}} \bigg\}\,.
\end{align*}
Due to $\breve{t}_{2,j}=t_{2,j}-n^{1/2}\beta^0_{j}/(\hat\sigma \hat\Lambda_{j+d,j+d}^{1/2})$ and $2C_{\max}^{-1}\le \tilde{D}_{j,j}\le 2C_{\min}^{-1}$, by \eqref{eq:thm2-step1-1}, we have
\begin{align*}
|\breve{t}_{2,j}-\tilde{w}_{j+d}|&=\bigg|\frac{\sigma}{\hat\sigma}\tilde{w}_{j+d}
+\frac{\tilde{\delta}_{j+d}}{\hat\sigma\hat\Lambda_{j+d,j+d}^{1/2}}
-\tilde{w}_{j+d}\bigg|\\
&\leq\bigg|\frac{\sigma}{\hat\sigma}-1\bigg|\cdot|\tilde{w}_{j+d}|
+\bigg|\frac{\tilde{\delta}_{j+d}}{\sigma \tilde{D}_{j,j}^{1/2}}\bigg|\cdot\bigg|\frac{\sigma}{\hat\sigma}\bigg|
\cdot\bigg|\frac{\tilde{D}_{j,j}^{1/2}}{\hat\Lambda_{j+d,j+d}^{1/2}}\bigg|\\
&\leq\bigg|\frac{\sigma}{\hat\sigma}-1\bigg|\cdot|\tilde{w}_{j+d}|
+\frac{|\tilde{\bdelta}_{n}|_{\infty}C_{\max}^{1/2}}{\sqrt{2}\sigma }\cdot\bigg|\frac{\sigma}{\hat\sigma}\bigg|
\cdot\bigg|\frac{\tilde{D}_{j,j}^{1/2}}{\hat\Lambda_{j+d,j+d}^{1/2}}\bigg|\,,
\end{align*}
which implies
\begin{align}\label{eq:thm2-1-1}
\mathbb{P}(\mathcal{D})
&\leq\mathbb{P}\bigg(\bigg|\frac{\sigma}{\hat\sigma}-1\bigg|\cdot\max_{j\in[d]}|\tilde{w}_{j+d}|
+\frac{|\tilde{\bdelta}_{n}|_{\infty}C_{\max}^{1/2}}{\sqrt{2}\sigma }\cdot\bigg|\frac{\sigma}{\hat\sigma}\bigg|\cdot
\max_{j\in[d]}\bigg|\frac{\tilde{D}_{j,j}^{1/2}}{\hat\Lambda_{j+d,j+d}^{1/2}}\bigg|
>\frac{\upsilon_{n}}{\sqrt{\log d}}\bigg)
\notag\\
&\leq\mathbb{P}\bigg(\bigg|\frac{\sigma}{\hat\sigma}-1\bigg|
\cdot\max_{j\in[d]}|\tilde{w}_{j+d}|
>\frac{\upsilon_{n}}{2\sqrt{\log d}}\bigg)\notag\\
&\quad+\mathbb{P}\bigg(\frac{|\tilde{\bdelta}_{n}|_{\infty}C_{\max}^{1/2}}{\sqrt{2}\sigma}\cdot\bigg|\frac{\sigma}{\hat\sigma}\bigg|\cdot
\max_{j\in[d]}\bigg|\frac{\tilde{D}_{j,j}^{1/2}}{\hat\Lambda_{j+d,j+d}^{1/2}}\bigg|
>\frac{\upsilon_{n}}{2\sqrt{\log d}}\bigg)\notag\\
&\leq\mathbb{P}\bigg(\max_{j\in[d]}|\tilde{w}_{j+d}|
>2\sqrt{\log d}\bigg)
+\mathbb{P}\bigg(\bigg|\frac{\sigma}{\hat\sigma}-1\bigg|
>\frac{\upsilon_{n}}{4\log d}\bigg)
+\mathbb{P}\bigg(\frac{|\tilde{\bdelta}_{n}|_{\infty}C_{\max}^{1/2}}{\sqrt{2}\sigma}>\frac{\upsilon_{n}}{8\sqrt{\log d}}\bigg)\notag\\
&\quad+\mathbb{P}\bigg(\bigg|\frac{\sigma}{\hat\sigma}\bigg|
>2\bigg)+\mathbb{P}\bigg(
\max_{j\in[d]}\bigg|\frac{\tilde{D}_{j,j}^{1/2}}{\hat\Lambda_{j+d,j+d}^{1/2}}\bigg|
>2\bigg)\,.
\end{align}
Since $\tilde{w}_{j+d}\,|\,\bZ\sim\mathcal N(0,1)$, then 
\begin{align}\label{eq:thm2-1-2}
\mathbb{P}\bigg(\max_{j\in[d]}|\tilde{w}_{j+d}|
>2\sqrt{\log d}\bigg)
&=\mathbb{E}\bigg\{\mathbb{P}\bigg(\max_{j\in[d]}|\tilde{w}_{j+d}|
>2\sqrt{\log d}\,\bigg|\,\bZ\bigg)\bigg\}\notag\\
&\le dG(2\sqrt{\log d})=o(1)\,.
\end{align}
Recall $\upsilon_{n}=s_{0}^{1/2}n^{-1/4}(\log d)^{3/4}$. Due to $\varrho_{1}\asymp\varrho_{2}\asymp\varrho_{3}\asymp\{n^{-1}\log(2d)\}^{1/2}$ and $1\le s_{0}\ll n^{1/2}(\log d)^{-3/2}$, we have $\varrho_{3}s_{0}^{1/2}\ll \upsilon_{n}(\log d)^{-1}$ and $\varrho_{1}\varrho_{2}s_{0}n^{1/2}\ll \upsilon_{n}(\log d)^{-1/2}$. Then, by Lemmas \ref{theo_scaled} and \ref{theo_random_trans},
\begin{align}\label{eq:thm2-1-3}
\mathbb{P}\bigg(\bigg|\frac{\sigma}{\hat\sigma}-1\bigg|
>\frac{\upsilon_{n}}{4\log d}\bigg)
+\mathbb{P}\bigg(\frac{|\tilde{\bdelta}_{n}|_{\infty}C_{\max}^{1/2}}{\sqrt{2}\sigma}>\frac{\upsilon_{n}}{8\sqrt{\log d}}\bigg)
+\mathbb{P}\bigg(\bigg|\frac{\sigma}{\hat\sigma}\bigg|
>2\bigg)=o(1)\,.
\end{align}
Recall $\bOmega=(\Omega_{i,j})_{2d\times2d}=\bT\bTheta_0 \bT^\top$ with $\bTheta_0=\bGamma^{-1}$ and $C_{\min}\le\sigma_{\min}(\bGamma)\le\sigma_{\max}(\bGamma)\le C_{\max}$, and $\hat\bLambda =(\hat\Lambda_{i,j})_{2d\times 2d} = \bT \hat{\bTheta}^\top \hat{\bGamma} \hat{\bTheta} \bT^\top$. Notice that $2^{-1/2}\bT$ is an orthogonal matrix. Then $2C_{\max}^{-1}\le\sigma_{\min}(\bOmega)\le\sigma_{\max}(\bOmega)\le 2C_{\min}^{-1}$. Since $\tilde{D}_{j,j}=\Omega_{j+d,j+d}$ and $|x/y-1|\leq|(x/y)^2-1|$ for all $x, y>0$, by Lemma \ref{theo_random_trans},
\begin{align*}
&\mathbb{P}\bigg(
\max_{j\in[d]}\bigg|\frac{\tilde{D}_{j,j}^{1/2}}{\hat\Lambda_{j+d,j+d}^{1/2}}\bigg|
>2\bigg)
\le\mathbb{P}\bigg(
\max_{j\in[d]}\bigg|\frac{\tilde{D}_{j,j}^{1/2}}{\hat\Lambda_{j+d,j+d}^{1/2}}-1\bigg|
>1\bigg)\\
&~~~~~\le\mathbb{P}\bigg(
\max_{j\in[d]}\bigg|\frac{\tilde{D}_{j,j}}{\hat\Lambda_{j+d,j+d}}-1\bigg|
>1\bigg)
\le\mathbb{P}\bigg(\frac{|\bOmega-\hat\bLambda|_{\infty}}{\min_{j\in[d]}\hat\Lambda_{j+d,j+d}}>1\bigg)\\
&~~~~~\le
\mathbb{P}\bigg(\frac{|\bOmega-\hat\bLambda|_{\infty}}{\min_{j\in[d]}\hat\Lambda_{j+d,j+d}}>1,|\bOmega-\hat\bLambda|_{\infty}<C_{\max}^{-1}\bigg)+\mathbb{P}(|\bOmega-\hat\bLambda|_{\infty}\ge C_{\max}^{-1})\\
&~~~~~\le 2\mathbb{P}(|\bOmega-\hat\bLambda|_{\infty}\ge C_{\max}^{-1})
=o(1)\,.
\end{align*}
Together with \eqref{eq:thm2-1-1}--\eqref{eq:thm2-1-3}, we have 
\begin{align}\label{eq:thm2-2-1}
\mathbb{P}(\mathcal{D})=o(1)\,.
\end{align}
Consider two events 
\begin{align*}
&\mathcal{D}_1=\bigg\{\inf_{0\leq t\leq G^{-1}(\alpha|\mathcal{H}|/d)} \bigg\{\frac{\sum_{j \in \mathcal{H}_0} I(|\breve{t}_{2,j}|\geq t)}{d G(t)}-\frac{\sum_{j\in \mathcal{H}_{0}}I(|\tilde{w}_{j+d}|\geq t+\upsilon_{n}/\sqrt{\log d})}{d G(t)}\bigg\}\geq 0 \bigg\}\,,\\
&\mathcal{D}_2=\bigg\{\sup_{0\leq t\leq G^{-1}(\alpha|\mathcal{H}|/d)} \bigg\{\frac{\sum_{j \in \mathcal{H}_0} I(|\breve{t}_{2,j}|\geq t)}{d G(t)}-\frac{\sum_{j\in \mathcal{H}_{0}}I(|\tilde{w}_{j+d}|\geq t-\upsilon_{n}/\sqrt{\log d})}{d G(t)}\bigg\}\leq 0 \bigg\}\,.
\end{align*}
Recall $\mathcal{D}=\{\max_{j\in[d]}|\breve{t}_{2,j}-\tilde{w}_{j+d}|> \upsilon_{n}/\sqrt{\log d} \}$. Given the event $\mathcal{D}^{\rm c}$, if $|\tilde{w}_{j+d}|\geq t+\upsilon_{n}/\sqrt{\log d}$ and $t\ge0$, then $|\breve{t}_{2,j}|\geq t$, which implies $\mathcal{D}_1\cap\mathcal{D}^{\rm c}=\mathcal{D}^{\rm c}$. Therefore,
\begin{align}\label{eq:thm2-1-4}
\mathbb{P}(\mathcal{D}_1)\ge\mathbb{P}(\mathcal{D}_1\cap\mathcal{D}^{\rm c})
=\mathbb{P}(\mathcal{D}^{\rm c})=1-o(1)\,.
\end{align}
Given the event $\mathcal{D}^{\rm c}$, if $|\breve{t}_{2,j}|\geq t$ and $t\ge0$, then $|\tilde{w}_{j+d}|\geq t-\upsilon_{n}/\sqrt{\log d}$, which implies $\mathcal{D}_2\cap\mathcal{D}^{\rm c}=\mathcal{D}^{\rm c}$. Therefore,
\begin{align}\label{eq:thm2-1-5}
&\mathbb{P}(\mathcal{D}_2)\ge\mathbb{P}(\mathcal{D}_2\cap\mathcal{D}^{\rm c})
=\mathbb{P}(\mathcal{D}^{\rm c})=1-o(1)\,.
\end{align}
Given the events $\mathcal{D}_1$ and $\mathcal{D}_2$, we have
\begin{align}\label{eq:thm2-1-8}
\bigg|\frac{\sum_{j \in \mathcal{H}_0} I(|\breve{t}_{2,j}|\geq t)}{d G(t)}-\frac{d_{0}}{d}\bigg|
&\leq\max\bigg\{\underbrace{\bigg|\frac{\sum_{j \in \mathcal{H}_0} I(|\tilde{w}_{j+d}|\geq t+\upsilon_{n}/\sqrt{\log d})}{d G(t)}-\frac{d_{0}}{d}\bigg|}_{{\rm I}(t)}, \notag\\
&~~~~~~~~~~~~~\underbrace{\bigg|\frac{\sum_{j \in \mathcal{H}_0} I(|\tilde{w}_{j+d}|\geq t-\upsilon_{n}/\sqrt{\log d})}{d G(t)}-\frac{d_{0}}{d}\bigg|}_{{\rm II}(t)}\bigg\}
\end{align}
for any $0\leq t\leq G^{-1}(\alpha|\mathcal{H}|/d)$. Due to $\upsilon_{n}=o(1)$, if $\log\log d \ge 1/[\alpha\{1+o(1)\}]$, by Lemma \ref{lem_normal},
\begin{align*}
{\rm I}(t)&\le\frac{G(t+\upsilon_{n}/\sqrt{\log d})}{G(t)}
\bigg|\frac{\sum_{j \in \mathcal{H}_0} I(|\tilde{w}_{j+d}|\geq t+\upsilon_{n}/\sqrt{\log d})}{d G(t+\upsilon_{n}/\sqrt{\log d})}
-\frac{d_{0}}{d}\bigg|\\
&~~~+\frac{d_{0}}{d}\bigg|\frac{G(t+\upsilon_{n}/\sqrt{\log d})}{G(t)}-1\bigg|\\
&\le\{1+o(1)\}\sup_{0\leq u\leq G^{-1}(\alpha|\mathcal{H}|\{1+o(1)\}/d)}
\bigg|\frac{\sum_{j \in \mathcal{H}_0} I(|\tilde{w}_{j+d}|\geq u)}{d G(u)}-\frac{d_{0}}{d}\bigg|+o(1)\,,\\
{\rm II}(t)
&\le\frac{G(t-\upsilon_{n}/\sqrt{\log d})}{G(t)}
\bigg|\frac{\sum_{j \in \mathcal{H}_0} I(|\tilde{w}_{j+d}|\geq t-\upsilon_{n}/\sqrt{\log d})}{d G(t-\upsilon_{n}/\sqrt{\log d})}
-\frac{d_{0}}{d}\bigg|\\
&~~~+\frac{d_{0}}{d}\bigg|\frac{G(t-\upsilon_{n}/\sqrt{\log d})}{G(t)}-1\bigg|\\
&\le\{1+o(1)\}\sup_{-\upsilon_{n}/\sqrt{\log d}\leq u\leq G^{-1}(\alpha|\mathcal{H}|\{1+o(1)\}/d)}
\bigg|\frac{\sum_{j \in \mathcal{H}_0} I(|\tilde{w}_{j+d}|\geq u)}{d G(u)}-\frac{d_{0}}{d}\bigg|+o(1)\\
&=\{1+o(1)\}\sup_{0\leq u\leq G^{-1}(\alpha|\mathcal{H}|\{1+o(1)\}/d)}
\bigg|\frac{\sum_{j \in \mathcal{H}_0} I(|\tilde{w}_{j+d}|\geq u)}{d G(u)}-\frac{d_{0}}{d}\bigg|+o(1)
\end{align*}
for any $0\leq t\leq G^{-1}(\alpha|\mathcal{H}|/d)$, where the terms $o(1)$ hold uniformly over $t \in[0,  G^{-1}(\alpha|\mathcal{H}|/d)]$. Together with \eqref{eq:thm2-1-8}, given the events $\mathcal{D}_1$ and $\mathcal{D}_2$, if $\log\log d \ge 1/[\alpha\{1+o(1)\}]$, we have
\begin{align}\label{eq:thm2-thm3-1}
&\sup_{0\leq t\leq G^{-1}(\alpha|\mathcal{H}|/d)}\bigg|\frac{\sum_{j \in \mathcal{H}_0} I(|\breve{t}_{2,j}|\geq t)}{d G(t)}-\frac{d_{0}}{d}\bigg|\notag\\
&~~~~~~~~\leq\{1+o(1)\}\sup_{0\leq t\leq G^{-1}(\alpha|\mathcal{H}|\{1+o(1)\}/d)}
\bigg|\frac{\sum_{j \in \mathcal{H}_0} I(|\tilde{w}_{j+d}|\geq t)}{d G(t)}-\frac{d_{0}}{d}\bigg|+o(1)\,.
\end{align}
Due to $s_{0}\ll n^{1/2}(\log d)^{-3/2}$, $|\mathcal{H}|\rightarrow\infty$ and $|\mathcal{H}|\le \sqrt{\log d}$, by \eqref{eq:thm2-1-6}, \eqref{eq:thm2-1-7}, \eqref{eq:thm2-1-4} and \eqref{eq:thm2-1-5},
\begin{align*}
&\mathbb{P}\bigg\{\sup_{0\leq t\leq G^{-1}(\alpha|\mathcal{H}|/d)}\bigg|\frac{\sum_{j \in \mathcal{H}_0} I(|\breve{t}_{2,j}|\geq t)}{d G(t)}-\frac{d_{0}}{d}\bigg|>\epsilon \bigg\}\\
&~~~~~\leq\mathbb{P}\bigg[\{1+o(1)\}\sup_{0\leq t\leq G^{-1}(\alpha|\mathcal{H}|\{1+o(1)\}/d)}
\bigg|\frac{\sum_{j \in \mathcal{H}_0} I(|\tilde{w}_{j+d}|\geq t)}{d G(t)}-\frac{d_{0}}{d}\bigg|+o(1)>\epsilon,\mathcal{D}_{1},\mathcal{D}_{2}\bigg]\\
&~~~~~~~~+\mathbb{P}(\mathcal{D}_{1}^{\rm c})+\mathbb{P}(\mathcal{D}_{2}^{\rm c})\\
&~~~~~\le\mathbb{P}\bigg\{\sup_{0\leq t\leq G^{-1}(\alpha|\mathcal{H}|\{1+o(1)\}/d)}
\bigg|\frac{\sum_{j \in \mathcal{H}_0} I(|\tilde{w}_{j+d}|\geq t)}{d G(t)}-\frac{d_{0}}{d}\bigg|>\frac{\epsilon-o(1)}{1+o(1)}\bigg\}+o(1)=o(1)
\end{align*}
for any $\epsilon>0$. Hence, \eqref{thm0-17} holds. $\hfill\qedsymbol$

\subsection{Proof of Theorem \ref{thm1}}\label{proof_thm1}
Recall $G(t)=2\{1-\Phi(t)\}$. Let $G^{-1}(\cdot)$ denote the inverse function of $G(\cdot)$. The proof below is considered in the context of a setting more general than what we have considered in the main text. We will show that under the conditions of Theorem \ref{thm1}, the following Algorithm \ref{method_3} can control {\rm FDR} at $\pi_0\alpha$. Algorithm \ref{method_2} is the special case of Algorithm \ref{method_3} with $\lambda=\sqrt{\alpha}$.

\floatname{algorithm}{Algorithm}
\begin{algorithm}
    \caption{ Algorithm \ref{method_2} in its general form}
    \label{method_3}

    \begin{algorithmic}
        \STATE\textbf{Step 1}. Given $0<\alpha<1$ and $\alpha<\lambda<1$, for each $j\in[d]$, let
            \begin{align*}
                \tilde{P}_j=\left\{
                \begin{aligned}
                    1\,, ~~~~& \text{if}~~P_j^{(1)}>\lambda\,, \\
                    P_j^{(2)}\,, ~~& \text{if}~~P_j^{(1)} \leq \lambda\,,
                \end{aligned}
                \right.
            \end{align*}
            where $P_j^{(1)}=G(|t_{1, j}|)$ and $P_j^{(2)}=G(|t_{2, j}|)$.
        \STATE\textbf{Step 2}. Let $\tilde{P}_{(1)} \leq \cdots \leq \tilde{P}_{(d)}$ be the ordered versions of the $\tilde{P}_j$'s. Find
        \begin{equation*}
            \tilde{R}_{\lambda}=\max\bigg\{i\in[d]: \tilde{P}_{(i)} \leq \frac{i \alpha}{\lambda d}\bigg\}\,,
        \end{equation*}
        provided the maximum exists; otherwise, let $\tilde{R}_{\lambda}=0$.
        \STATE\textbf{Step 3}. Reject the null hypotheses corresponding to $\tilde{P}_{(j)}$ with $j \leq \tilde{R}_{\lambda}$.
    \end{algorithmic}
\end{algorithm}

Let $\bar{t}_{i, j}=|t_{i, j}|$ and $\bar{t}_{i, (1)} \geq \cdots \geq \bar{t}_{i,(d)}$ be the ordered sequence of $\{\bar{t}_{i, j}\}_{j=1}^{d}$, where $i \in[2]$. Define $\tilde{t}_{2, j}=\bar{t}_{2, j}I\{\bar{t}_{1, j} \geq G^{-1}(\lambda)\}$ and $\tilde{t}_{2,(0)}=+\infty$. Let $\tilde{t}_{2,(1)} \geq \cdots \geq \tilde{t}_{2, (d)}$ be the ordered sequence of $\{\tilde{t}_{2, j}\}_{j=1}^{d}$. Recall $P_j^{(1)}=G(|t_{1, j}|)$. Then $\{\bar{t}_{1,j}\geq G^{-1}(\lambda)\}=\{P_{j}^{(1)}\le \lambda\}$, which implies $G(\tilde{t}_{2, j})=\tilde{P}_{j}$. By the definition of $\tilde{R}_{\lambda}$, we have $\tilde{R}_{\lambda}=\max\{j=0,1, \ldots, d: \tilde{P}_{(j)}=G(\tilde{t}_{2, (j)}) \leq j \alpha/(\lambda d)\}$. Then $G(\tilde{t}_{2,(\tilde{R}_{\lambda})})\leq \alpha\tilde{R}_{\lambda} /(\lambda d)$ and $G(\tilde{t}_{2,(l)})> \alpha l/(\lambda d)$ for all $l>\tilde{R}_{\lambda}$, which implies $\tilde{t}_{2,(l)}< \tilde{t}_{2,(\tilde{R}_{\lambda})}$ for all $l>\tilde{R}_{\lambda}$. Hence, $|\{j\in[d]:\tilde{t}_{2, j} \geq \tilde{t}_{2,(\tilde{R}_{\lambda})}\}|=\tilde{R}_{\lambda}$, and rejecting all $H_{0,j}$'s with $\tilde{t}_{2, j} \geq \tilde{t}_{2,(\tilde{R}_{\lambda})}$ is equivalent to rejecting all $H_{0,j}$'s corresponding to the first $\tilde{R}_{\lambda}$ largest $\tilde{t}_{2, j}$'s among $\{\tilde{t}_{2, j}\}_{j=1}^{d}$. Without loss of generality, we assume $|\mathcal{H}|\le\sqrt{\log d}$. If $|\mathcal{H}|>\sqrt{\log d}$, we can redefine $\mathcal{H}$ by keeping only the first $\lfloor\sqrt{\log d}\rfloor$ elements of it. Our proof of Theorem \ref{thm1} mainly includes the following four steps:

{\bf Step 1}. To show that
\begin{equation}\label{thm1-1}
	\mathbb{P}\bigg\{\sum_{j=1}^d I(\tilde{t}_{2, j}>\sqrt{2 \log d}) \geq|\mathcal{H}|\bigg\} \rightarrow 1\,.
\end{equation}

{\bf Step 2}. To show that $\{\tilde{t}_{2, j} \geq \tilde{t}_{2,(\tilde{R}_{\lambda})}\}=\{\tilde{t}_{2, j} \geq \hat{t}_{\lambda}\}$ for any $j\in[d]$, where
\begin{equation}\label{thm1-4}
	\hat{t}_{\lambda}:=\inf\bigg[t>0: G(t)\leq\frac{\alpha}{d\lambda} \max\bigg\{\sum_{j=1}^{d}I(\tilde{t}_{2, j}\geq t),1\bigg\}\bigg]\,.
\end{equation}

{\bf Step 3}. To show that $\hat{t}_{\lambda}$ specified in \eqref{thm1-4} satisfies
\begin{align}
&~\frac{dG(\hat{t}_{\lambda})}{\max\{\sum_{j=1}^{d}I(\tilde{t}_{2, j}\geq \hat{t}_{\lambda}),1\}}=\frac{\alpha}{\lambda}\,,\label{thm1-10}\\
&\lim_{n\rightarrow\infty}\mathbb{P}\bigg\{\hat{t}_{\lambda} \leq G^{-1}\bigg(\frac{\alpha |\mathcal{H}|}{d\lambda}\bigg)\bigg\} = 1\,.\label{thm1-11}
\end{align}

{\bf Step 4}. To show that
\begin{align}\label{thm1-16}
\sup_{0< t\leq G^{-1}(\alpha|\mathcal{H}|/(d\lambda))}\frac{\sum_{j\in \mathcal{H}_{0}}I(\tilde{t}_{2, j}\geq t)}{d\lambda G(t)}\leq \frac{d_{0}}{d}+o_{\rm p}(1)\,,
\end{align}
where $d_{0} = d - d_1$ is the number of true null hypotheses.

The proofs of Steps 1--4 are given in Sections \ref{proof_thm3_step1}--\ref{proof_thm3_step4}, respectively. Due to $\{\tilde{t}_{2, j} \geq \tilde{t}_{2,(\tilde{R}_{\lambda})}\}=\{\tilde{t}_{2, j} \geq \hat{t}_{\lambda}\}$ for any $j\in[d]$, by \eqref{thm1-10}, 
\begin{equation*}
\text{FDP}=\frac{\sum_{j\in \mathcal{H}_{0}}I(\tilde{t}_{2, j}\geq \hat{t}_{\lambda})}{\max\{\sum_{j=1}^{d}I(\tilde{t}_{2, j}\geq \hat{t}_{\lambda}),1\}}=\frac{\sum_{j\in \mathcal{H}_{0}}I(\tilde{t}_{2, j}\geq \hat{t}_{\lambda})}{d \lambda G(\hat{t}_{\lambda})/\alpha}\,.
\end{equation*}
By the definition of $\hat{t}_{\lambda}$, we have 
\begin{align*}
\bigg[t>0: G(t)\leq \frac{\alpha}{d\lambda}\max\bigg\{\sum_{j=1}^{d}I(\tilde{t}_{2, j}\geq t),1\bigg\}\bigg]\subseteq\bigg\{t>0: G(t)\leq\frac{\alpha}{\lambda}\bigg\}\,,
\end{align*}
which implies $\hat{t}_{\lambda}\ge G^{-1}(\alpha/\lambda)>0$. Then $\mathbb{P}(\hat{t}_{\lambda} =0)=0$. Hence, by \eqref{thm1-11} and \eqref{thm1-16}, for any $\epsilon>0$,
\begin{align*}
&\mathbb{P}\bigg\{\frac{\sum_{j\in \mathcal{H}_{0}}I(\tilde{t}_{2, j}\geq \hat{t}_{\lambda})}{d\lambda G(\hat{t}_{\lambda})}\leq \epsilon+\frac{d_{0}}{d}\bigg\}\\
&~~~~~\geq\mathbb{P}\bigg\{\sup_{0< t\leq G^{-1}(\alpha|\mathcal{H}|/(\lambda d))}\frac{\sum_{j\in \mathcal{H}_{0}}I(\tilde{t}_{2, j}\geq t)}{d\lambda G(t)}\leq \epsilon+\frac{d_{0}}{d},0<\hat{t}_{\lambda} \leq G^{-1}\bigg(\frac{\alpha |\mathcal{H}|}{d\lambda }\bigg)\bigg\}\\
&~~~~~\geq\mathbb{P}\bigg\{\sup_{0< t\leq G^{-1}(\alpha|\mathcal{H}|/(\lambda d))}\frac{\sum_{j\in \mathcal{H}_{0}}I(\tilde{t}_{2, j}\geq t)}{d\lambda G(t)}\leq \epsilon+\frac{d_{0}}{d}\bigg\}
-\mathbb{P}\bigg\{\hat{t}_{\lambda} > G^{-1}\bigg(\frac{\alpha |\mathcal{H}|}{d\lambda }\bigg)\bigg\}\\
&~~~~~=1-o(1)\,,
\end{align*}
which implies
\begin{align*}
\mathbb{P}\bigg(\text{FDP}\leq\frac{d_{0}}{d}\alpha+\epsilon\bigg)
=\mathbb{P}\bigg\{\frac{\sum_{j\in \mathcal{H}_{0}}I(\tilde{t}_{2, j}\geq \hat{t}_{\lambda})}{d\lambda G(\hat{t}_{\lambda})/\alpha}\leq\frac{d_{0}}{d}\alpha+\epsilon\bigg\}
\geq1-o(1)\,.
\end{align*}
Since $\text{FDR}=\mathbb{E}(\text{FDP})$ and $0\le\text{FDP}\le1$, it holds that $\lim\sup_{n\rightarrow \infty}\text{FDR}\leq \alpha d_{0}/d$. We complete the proof of Theorem \ref{thm1}. $\hfill\qedsymbol$

\subsubsection{Proof of Step 1}\label{proof_thm3_step1}
Given $\lambda \in(0,1)$, if $\sqrt{2\log d}>G^{-1}(\lambda)$, then
$\{\bar{t}_{1,j}>\sqrt{2 \log d}\}\subseteq\{\bar{t}_{1,j}\geq G^{-1}(\lambda)\}$, which implies
\begin{align}\label{thm1-3}
		&\mathbb{P}\bigg\{\sum_{j=1}^d I(\tilde{t}_{2, j}>\sqrt{2 \log d})\geq |\mathcal{H}|\bigg\}
		\geq \mathbb{P}\bigg[\bigcap_{j \in \mathcal{H}}\{\tilde{t}_{2, j}> \sqrt{2 \log d}\}\bigg]\notag\\
		&~~~~~= \mathbb{P}\bigg[\bigcap_{j \in \mathcal{H}}\{\bar{t}_{2, j}>\sqrt{2 \log d}, \bar{t}_{1, j}\ge G^{-1}(\lambda)\}\bigg] \\
		&~~~~~\geq  \mathbb{P}\bigg[\bigcap_{j \in \mathcal{H}}\{\bar{t}_{2, j}>\sqrt{2 \log d}, \bar{t}_{1, j}>\sqrt{2 \log d}\}\bigg]\notag \\
		&~~~~~\geq \mathbb{P}\bigg[\bigcap_{j \in \mathcal{H}}\{\bar{t}_{2, j}>\sqrt{2 \log d}\}\bigg]+\mathbb{P}\bigg[\bigcap_{j \in \mathcal{H}}\{\bar{t}_{1, j} > \sqrt{2\log d}\}\bigg]-1\,.\notag
\end{align}
By \eqref{thm0-3}--\eqref{eq:thm2-2}, we have $\mathbb{P}[\bigcap_{j \in \mathcal{H}}\{\bar{t}_{2, j}>\sqrt{2 \log d}\}] = 1 - o(1)$. Recall $\bw_{n}=(w_{1},\ldots,w_{2d})^{\top}$ and $\bbeta_{0}=(\beta^0_{1},\ldots,\beta^0_{d})^{\top}$. Let $\hat\bLambda :=(\hat\Lambda_{i,j})_{2d\times 2d} = \bT \hat{\bTheta}^\top \hat{\bGamma} \hat{\bTheta} \bT^\top$ and $\tilde{\bdelta}_{n}:=(\tilde{\delta}_{1},\ldots,\tilde{\delta}_{2d})^{\top}=\bT \bdelta_{n}$. Using \eqref{eq:exp1}, for any $j\in[d]$, we have
\begin{align*}
	t_{1,j} =\frac{n^{1/2}\{\hat{\gamma}^{(\rm bc)}_{j} + \hat{\gamma}^{(\rm bc)}_{j+d}\}}{\hat\sigma \hat\Lambda_{j,j}^{1/2}}= \frac{n^{1/2}\beta^0_{j}}{\hat\sigma\hat\Lambda_{j,j}^{1/2}}
	+\frac{\sigma}{\hat\sigma}\tilde{w}_{j}+
	\frac{\tilde{\delta}_{j}}{\hat\sigma\hat\Lambda_{j,j}^{1/2}}\,,
\end{align*}
where $\tilde{w}_{j}=\sigma^{-1}\hat\Lambda_{j,j}^{-1/2}(w_{j}+w_{j+d})$. Parallel to the arguments for \eqref{thm0-3}--\eqref{eq:thm2-2}, it holds that $\mathbb{P}[\bigcap_{j \in \mathcal{H}}\{\bar{t}_{1, j}>\sqrt{2 \log d}\}] = 1 - o(1)$. Therefore, by \eqref{thm1-3}, we have \eqref{thm1-1}. $\hfill\qedsymbol$

\subsubsection{Proof of Step 2}\label{proof_thm3_step2}
The proof is identical to that of Step 2 in the proof of Theorem \ref{thm0} given in Section \ref{proof_thm2_step2}. We only need to replace $\{\bar{t}_{2, (j)}\}_{j=1}^{d}$, $\tilde R$ and $\hat{t}$ used there, respectively, by $\{\tilde{t}_{2, (j)}\}_{j=1}^{d}$, $\tilde{R}_{\lambda}$ and $\hat{t}_{\lambda}$. $\hfill\qedsymbol$

\subsubsection{Proof of Step 3}\label{proof_thm3_step3}
The proof is identical to that of Step 3 in the proof of Theorem \ref{thm0} given in Section \ref{proof_thm2_step3}. We only need to replace $\{\bar{t}_{2, j}\}_{j=1}^{d}$ and $\hat{t}$ used there, respectively, by $\{\tilde{t}_{2, j}\}_{j=1}^{d}$ and $\hat{t}_{\lambda}$. $\hfill\qedsymbol$

\subsubsection{Proof of Step 4}\label{proof_thm3_step4}

By the definition of $\tilde{t}_{2,j}$, $I(\tilde{t}_{2, j}\geq t) = I\{\bar{t}_{2, j}\geq t, \bar{t}_{1,j}\geq G^{-1}(\lambda)\}$ for any $t>0$, which implies
\begin{equation*}
	\begin{aligned}
		 \sup_{0< t\leq G^{-1}(\alpha|\mathcal{H}|/(d\lambda))}\frac{\sum_{j\in \mathcal{H}_{0}}I(\tilde{t}_{2, j}\geq t)}{d\lambda G(t)}
		&= \sup_{0< t\leq G^{-1}(\alpha|\mathcal{H}|/(d\lambda))}\frac{\sum_{j\in \mathcal{H}_{0}}I\{\bar{t}_{2, j}\geq t, \bar{t}_{1,j}\geq G^{-1}(\lambda)\}}{d\lambda G(t)}\\
        &\le  \sup_{0\le t\leq G^{-1}(\alpha|\mathcal{H}|/(d\lambda))}\frac{\sum_{j\in \mathcal{H}_{0}}I\{\bar{t}_{2, j}\geq t, \bar{t}_{1,j}\geq G^{-1}(\lambda)\}}{d\lambda G(t)}\,.
	\end{aligned}
\end{equation*}
Let $\breve{t}_{1,j}:=n^{1/2}\{\hat{\gamma}^{(\rm bc)}_{j} + \hat{\gamma}^{(\rm bc)}_{j+d}-\beta^0_{j}\}/(\hat\sigma \hat\Lambda_{j,j}^{1/2})$ and $\breve{t}_{2,j}:=n^{1/2}\{\hat{\gamma}^{(\rm bc)}_{j} - \hat{\gamma}^{(\rm bc)}_{j+d}-\beta^0_{j}\}/(\hat\sigma \hat\Lambda_{j+d,j+d}^{1/2})$ for $j\in[d]$. Since $\bar{t}_{1, j}=|\breve{t}_{1,j}|$ and $\bar{t}_{2, j}=|\breve{t}_{2,j}|$ for all $j\in\mathcal{H}_{0}$, then
	\begin{align*}
		&\sup_{0\leq t\leq G^{-1}(\alpha|\mathcal{H}|/(d\lambda))}\frac{\sum_{j\in \mathcal{H}_{0}}I\{\bar{t}_{2, j}\geq t, \bar{t}_{1,j}\geq G^{-1}(\lambda)\}}{d\lambda G(t)}\notag\\
		&~~~~~~~~\leq\sup_{0\leq t\leq G^{-1}(\alpha|\mathcal{H}|/(d\lambda))}\bigg|\frac{\sum_{j\in \mathcal{H}_{0}}I\{|\breve{t}_{2, j}|\geq t, |\breve{t}_{1, j}|\geq G^{-1}(\lambda)\}}{d\lambda G(t)}-\frac{d_{0}}{d}\bigg|+\frac{d_{0}}{d}\,.
	\end{align*}
To prove \eqref{thm1-16}, it suffices to show
\begin{equation}\label{thm1-17}
	\begin{aligned}
		\sup_{0\leq t\leq G^{-1}(\alpha|\mathcal{H}|/(d\lambda))}\bigg|\frac{\sum_{j\in \mathcal{H}_{0}}I\{|\breve{t}_{2, j}|\geq t, |\breve{t}_{1, j}|\geq G^{-1}(\lambda)\}}{d\lambda G(t)}-\frac{d_{0}}{d}\bigg|=o_{\rm p}(1)\,.
	\end{aligned}
\end{equation}
To prove (\ref{thm1-17}), we need Lemma \ref{lem1}, whose proof is given in Section \ref{proof_lem1}.
\begin{lemma}\label{lem1}
Let the conditions of Theorem {\rm\ref{thm1}} hold. For any sequence of positive constants $\{b_{d}\}$ such that $b_{d}\rightarrow \infty$ and $b_{d}\ll d$,
\begin{align*}
\sup_{0\leq t\leq G^{-1}(b_{d}/d)}\bigg|\frac{\sum_{j\in \mathcal{H}_{0}}I\{|\tilde{w}_{j+d}|\geq t, |\tilde{w}_{j}|\geq G_{l,n}^{-1}(\lambda)\}}{d\lambda G(t)}-\frac{d_{0}}{d}\bigg|=o_{\rm p}(1)\,,
\end{align*}
where $\lambda\in(0,1)$ is a constant, $G_{l,n}^{-1}(\lambda)=G^{-1}(\lambda)-(-1)^{l}\upsilon_{n}/\sqrt{\log d}$, $\upsilon_{n}=s_{0}^{1/2}n^{-1/4}(\log d)^{3/4}$ and $l\in[2]$.
\end{lemma}

If $|\mathcal{H}|\rightarrow\infty$ and $|\mathcal{H}|\ll d$, by Lemma \ref{lem1} with $b_{d}=\lambda^{-1}\alpha|\mathcal{H}|\{1+o(1)\}$, we have
\begin{align}\label{thm1-19}
\sup_{0\leq t\leq G^{-1}(\alpha|\mathcal{H}|\{1+o(1)\}/(d\lambda))}\bigg|\frac{\sum_{j\in \mathcal{H}_{0}}I\{|\tilde{w}_{j+d}|\geq t, |\tilde{w}_{j}|\geq G_{l,n}^{-1}(\lambda)\}}{d\lambda G(t)}-\frac{d_{0}}{d}\bigg|=o_{\rm p}(1)
\end{align}
for $l\in[2]$. Similar to \eqref{thm0-13}, if $\log\log d \ge \lambda/[\alpha\{1+o(1)\}]$, then
\begin{align}\label{eq:thm3-1-1}
G^{-1}\bigg(\frac{\alpha |\mathcal{H}|}{d\lambda}\{1+o(1)\}\bigg)\leq \sqrt{2\log d}\,.
\end{align}
Recall $\upsilon_{n}=s_{0}^{1/2}n^{-1/4}(\log d)^{3/4}$. Since $s_{0}\lesssim n^{1/2}(\log d)^{-5/2-2\vartheta}$, we have $\upsilon_{n}=o(1)$. Due to $\alpha|\mathcal{H}|/(d\lambda)>\alpha|\mathcal{H}|/d$, if $\log\log d \ge 1/\alpha$, by \eqref{thm0-13}, we have $G^{-1}(\alpha|\mathcal{H}|/(d\lambda))\le(2\log d)^{1/2}$. Then, by Lemma \ref{lem_normal} with $\epsilon_{n}=\upsilon_{n}$, 
\begin{equation*}
G\bigg(t+\frac{\upsilon_{n}}{\sqrt{\log d}}\bigg)= G(t)\{1+o(1)\}\geq\frac{\alpha|\mathcal{H}|}{d\lambda}\{1+o(1)\}
\end{equation*}
for any $0\leq t\leq G^{-1}(\alpha|\mathcal{H}|/(d\lambda))$. Therefore, $t+\upsilon_{n}/\sqrt{\log d}\leq G^{-1}(\alpha|\mathcal{H}|\{1+o(1)\}/(d\lambda))$ for any $0\leq t\leq G^{-1}(\alpha|\mathcal{H}|/(d\lambda))$ if $\log\log d \ge \max(\lambda/[\alpha\{1+o(1)\}],1/\alpha)$, where the term $o(1)$ holds uniformly over $t \in[0,  G^{-1}(\alpha|\mathcal{H}|/(d\lambda))]$. Consider the events 
\begin{align*}
&\mathcal{D}_{1}=\bigg\{\max_{j\in[d]}|\breve{t}_{1,j}-\tilde{w}_{j}|>\frac{\upsilon_{n}}{\sqrt{\log d}} \bigg\}\,,~~~~\mathcal{D}_{2}=\bigg\{\max_{j\in[d]}|\breve{t}_{2,j}-\tilde{w}_{j+d}|>\frac{\upsilon_{n}}{\sqrt{\log d}} \bigg\}\,,\\
&\mathcal{D}_{3}=\bigg\{\inf_{0\leq t\leq G^{-1}(\alpha|\mathcal{H}|/(d\lambda))}\bigg[\frac{\sum_{j \in \mathcal{H}_0} I\{|\breve{t}_{2,j}|\geq t,|\breve{t}_{1, j}|\geq G^{-1}(\lambda)\}}{d\lambda G(t)}\\
&~~~~~~~~~~~~~~~~~~~~
-\frac{\sum_{j\in \mathcal{H}_{0}}I\{|\tilde{w}_{j+d}|\geq t+\upsilon_{n}/\sqrt{\log d},|\tilde{w}_j|\geq G_{1,n}^{-1}(\lambda)\}}{d\lambda G(t)}\bigg]\ge0\bigg\}\,,\\
&\mathcal{D}_{4}=\bigg\{\sup_{0\leq t\leq G^{-1}(\alpha|\mathcal{H}|/(d\lambda))}\bigg[\frac{\sum_{j \in \mathcal{H}_0} I\{|\breve{t}_{2,j}|\geq t,|\breve{t}_{1, j}|\geq G^{-1}(\lambda)\}}{d\lambda G(t)}\\
&~~~~~~~~~~~~~~~~~~~~
-\frac{\sum_{j\in \mathcal{H}_{0}}I\{|\tilde{w}_{j+d}|\geq t-\upsilon_{n}/\sqrt{\log d},|\tilde{w}_j|\geq G_{2,n}^{-1}(\lambda)\}}{d\lambda G(t)}\bigg]\le0\bigg\}\,.
\end{align*}
Here, the event $\mathcal{D}_{2}$ is identical to the event $\mathcal{D}$ specified in Section \ref{proof_thm2_step4}. As shown in \eqref{eq:thm2-2-1}, $\mathbb{P}(\mathcal{D}_{2})=o(1)$. Parallel to \eqref{eq:thm2-2-1}, we can also show $\mathbb{P}(\mathcal{D}_{1})=o(1)$. Given $\mathcal{D}_{1}^{\rm c}\cap\mathcal{D}_{2}^{\rm c}$, if $|\tilde{w}_{j+d}|\geq t+\upsilon_{n}/\sqrt{\log d}$ and $|\tilde{w}_{j}|\geq G^{-1}_{1,n}(\lambda)$, then $|\breve{t}_{1,j}|\geq G^{-1}(\lambda)$ and $|\breve{t}_{2,j}|\geq t$, which implies $\mathcal{D}_3\cap\mathcal{D}_{1}^{\rm c}\cap\mathcal{D}_{2}^{\rm c}=\mathcal{D}_{1}^{\rm c}\cap\mathcal{D}_{2}^{\rm c}$. Hence,
\begin{align}\label{eq:thm3-1-2}
\mathbb{P}(\mathcal{D}_3)\ge\mathbb{P}(\mathcal{D}_3\cap\mathcal{D}_{1}^{\rm c}\cap\mathcal{D}_{2}^{\rm c})
=\mathbb{P}(\mathcal{D}_{1}^{\rm c}\cap\mathcal{D}_{2}^{\rm c})=1-o(1)\,.
\end{align}
Given $\mathcal{D}_{1}^{\rm c}\cap\mathcal{D}_{2}^{\rm c}$, if $|\breve{t}_{1,j}|\geq G^{-1}(\lambda)$ and $|\breve{t}_{2,j}|\geq t$, then $|\tilde{w}_{j}|\geq G^{-1}_{2,n}(\lambda)$ and $|\tilde{w}_{j+d}|\geq t-\upsilon_{n}/\sqrt{\log d}$, which implies $\mathcal{D}_4\cap\mathcal{D}_{1}^{\rm c}\cap\mathcal{D}_{2}^{\rm c}=\mathcal{D}_{1}^{\rm c}\cap\mathcal{D}_{2}^{\rm c}$. Therefore, 
\begin{align}\label{eq:thm3-1-3}
\mathbb{P}(\mathcal{D}_4)\ge\mathbb{P}(\mathcal{D}_4\cap\mathcal{D}_{1}^{\rm c}\cap\mathcal{D}_{2}^{\rm c})
=\mathbb{P}(\mathcal{D}_{1}^{\rm c}\cap\mathcal{D}_{2}^{\rm c})=1-o(1)\,.
\end{align}
Given $\mathcal{D}_3\cap\mathcal{D}_4$, it holds that
\begin{align*}
&\bigg|\frac{\sum_{j \in \mathcal{H}_0} I\{|\breve{t}_{2,j}|\geq t,|\breve{t}_{1,j}|\geq G^{-1}(\lambda)\}}{d\lambda G(t)}-\frac{d_{0}}{d}\bigg|\notag\\
&~~~~~\leq\max\bigg\{\bigg|\frac{\sum_{j \in \mathcal{H}_0} I\{|\tilde{w}_{j+d}|\geq t+\upsilon_{n}/\sqrt{\log d},|\tilde{w}_{j}|\geq G_{1,n}^{-1}(\lambda)\}}{d\lambda G(t)}-\frac{d_{0}}{d}\bigg|,\\
&~~~~~~~~~~~~~~~~~~\bigg|\frac{\sum_{j \in \mathcal{H}_0} I\{|\tilde{w}_{j+d}|\geq t-\upsilon_{n}/\sqrt{\log d},|\tilde{w}_{j}|\geq G_{2,n}^{-1}(\lambda)\}}{d\lambda G(t)}-\frac{d_{0}}{d}\bigg|\bigg\}
\end{align*}
for any $0\leq t\leq G^{-1}(\alpha|\mathcal{H}|/(d\lambda))$. Parallel to \eqref{eq:thm2-thm3-1}, given $\mathcal{D}_3\cap\mathcal{D}_4$, if $\log\log d \ge \max(\lambda/[\alpha\{1+o(1)\}],1/\alpha)$, we have
\begin{align*}
&\sup_{0\leq t\leq G^{-1}(\alpha|\mathcal{H}|/(d\lambda))}
\bigg|\frac{\sum_{j \in \mathcal{H}_0} I\{|\breve{t}_{2,j}|\geq t,|\breve{t}_{1,j}|\geq G^{-1}(\lambda)\}}{d\lambda G(t)}-\frac{d_{0}}{d}\bigg|\\
&~~~~~\leq\{1+o(1)\}\max_{l\in[2]}\sup_{0\leq t\leq G^{-1}(\alpha|\mathcal{H}|\{1+o(1)\}/(d\lambda))}
\bigg|\frac{\sum_{j \in \mathcal{H}_0} I\{|\tilde{w}_{j+d}|\geq t,|\tilde{w}_{j}|\geq G_{l,n}^{-1}(\lambda)\}}{d\lambda G(t)}-\frac{d_{0}}{d}\bigg|\\
&~~~~~~~~~+o(1)\,.
\end{align*}
Due to $s_{0}\lesssim n^{1/2}(\log d)^{-5/2-2\vartheta}$, $|\mathcal{H}|\rightarrow\infty$ and $|\mathcal{H}|\le \sqrt{\log d}$, by \eqref{thm1-19}--\eqref{eq:thm3-1-3}, we have
\begin{align*}
&\mathbb{P}\bigg\{\sup_{0\leq t\leq G^{-1}(\alpha|\mathcal{H}|/(d\lambda))}\bigg|\frac{\sum_{j \in \mathcal{H}_0} I\{|\breve{t}_{2,j}|\geq t,|\breve{t}_{1,j}|\geq G^{-1}(\lambda)\}}{d\lambda G(t)}-\frac{d_{0}}{d}\bigg|>\epsilon \bigg\}\\
&~~~~~\leq\mathbb{P}\bigg[\{1+o(1)\}\max_{l\in[2]}\sup_{0\leq t\leq G^{-1}(\alpha|\mathcal{H}|\{1+o(1)\}/(d\lambda))}\bigg|\frac{\sum_{j \in \mathcal{H}_0} I\{|\tilde{w}_{j+d}|\geq t,|\tilde{w}_{j}|\geq G_{l,n}^{-1}(\lambda)\}}{d\lambda G(t)}-\frac{d_{0}}{d}\bigg|\\
&~~~~~~~~~~~~~~~~~~+o(1)>\epsilon,\mathcal{D}_{3},\mathcal{D}_{4}\bigg]
+\mathbb{P}(\mathcal{D}_{3}^{\rm c})+\mathbb{P}(\mathcal{D}_{4}^{\rm c})\\
&~~~~~\le\mathbb{P}\bigg\{\max_{l\in[2]}\sup_{0\leq t\leq G^{-1}(\alpha|\mathcal{H}|\{1+o(1)\}/(d\lambda))}\bigg|\frac{\sum_{j \in \mathcal{H}_0} I\{|\tilde{w}_{j+d}|\geq t,|\tilde{w}_{j}|\geq G_{l,n}^{-1}(\lambda)\}}{d\lambda G(t)}-\frac{d_{0}}{d}\bigg|\\
&~~~~~~~~~~~~~~~~~~~~~~~>\frac{\epsilon-o(1)}{1+o(1)}\bigg\}+o(1)\\
&~~~~~=o(1)
\end{align*}
for any $\epsilon>0$. Hence, \eqref{thm1-17} holds. $\hfill\qedsymbol$

\subsection{Proof of Theorem \ref{thm_power}}\label{proof_power}

We define the power of Algorithm \ref{method_3} as follows: 
\begin{align}\label{eq:power1}
\Psi_{\lambda}=\mathbb{E}\bigg[\frac{\sum_{j\in\mathcal{H}_1}
I\{\tilde{P}_j\le\tilde{P}_{(\tilde{R}_{\lambda})}\}}{|\mathcal{H}_1|}\bigg]\,,
\end{align}
where $\lambda\in(\alpha,1]$. By applying the definition in \eqref{eq:power1}, the powers of Algorithms \ref{method_1} and \ref{method_2} can be calculated as $\Psi_{1}$ and $\Psi_{\sqrt{\alpha}}$, respectively. To establish Theorem \ref{thm_power}, it suffices to show that for any $\alpha<\lambda_1<\lambda_2\le1$, we have $\Psi_{\lambda_1}+o(1)\ge\Psi_{\lambda_2}$ as $n\rightarrow\infty$.  
Recall $\hat\bLambda =(\hat\Lambda_{i,j})_{2d\times 2d} = \bT \hat{\bTheta}^\top \hat{\bGamma} \hat{\bTheta} \bT^\top$ and $\bOmega=(\Omega_{i,j})_{2d\times2d}=\bT\bTheta_0 \bT^\top$ with $\bTheta_0=\bGamma^{-1}$ and $C_{\min}\le\sigma_{\min}(\bGamma)\le\sigma_{\max}(\bGamma)\le C_{\max}$. Then $2C_{\max}^{-1}\le\sigma_{\min}(\bOmega)\le\sigma_{\max}(\bOmega)\le 2C_{\min}^{-1}$. Consider the events 
\begin{align*}
&\qquad\quad\mathcal{A}_{\lambda}:=\bigcap_{j \in \mathcal{H}_1}\{P^{(1)}_j\le\lambda\}\,,
\quad\mathcal{B}_{1}:=\bigcap_{j \in \mathcal{H}_1}\{\bar{t}_{2,j}>3\sqrt{2\log d}\}\,,\\
&\quad~\mathcal{B}_{2}:=\bigcap_{j \in \mathcal{H}_0}\{\bar{t}_{2,j}\le2\sqrt{2\log d}\}\,,
\quad\mathcal{B}_{3}:=\bigcap_{j \in \mathcal{H}_1}\{\bar{t}_{1,j}>3\sqrt{2\log d}\}\,,\\
&\tilde{\mathcal E}_{1,n}:=\bigg\{\bigg|\frac{\hat\sigma}{\sigma}-1\bigg|\le\frac{1}{25}\bigg\}\,,
\quad\breve{\mathcal E}_{2,n}:=\bigg\{\max_{j \in [2d]}\bigg|\frac{\hat\Lambda_{j,j}}{\Omega_{j,j}}-1\bigg|\le 10C_{\max}M\varrho_{2}\bigg\}\,,
\end{align*}
where $\alpha<\lambda\le1$, $\bar{t}_{i, j}=|t_{i, j}|$ with $t_{i, j}$ involved in Algorithm \ref{method_3}. In the sequel, we write $\tilde{P}_j$ specified in Algorithm \ref{method_3} as $\tilde{P}_{j,\lambda}$ to highlight its dependence on $\lambda$.  
Let $\tilde{P}_{(1),\lambda} \leq \cdots \leq \tilde{P}_{(d),\lambda}$ be the ordered versions of the $\tilde{P}_{j,\lambda}$'s. For each $j\in[d]$, we denote $i_{j,\lambda}$ as the order of $\tilde{P}_{j,\lambda}$ among the ordered $p$-values $\{\tilde{P}_{(k),\lambda}\}_{k=1}^{d}$. Given the event $\mathcal{A}_{\lambda}$, we have $P_j^{(1)} \leq \lambda$ for all $j\in\mathcal{H}_1$, which implies $\tilde{P}_{j,\lambda}=P_j^{(2)}$ for all $j\in\mathcal{H}_1$.  Recall $P_j^{(2)}=G(\bar{t}_{2, j})$. Restricted on the event $\mathcal{B}_{1}\cap\mathcal{B}_{2}$, we have $\bar{t}_{2, j}>2\sqrt{2\log d}\ge\bar{t}_{2, k}$ for any $j\in\mathcal{H}_1$ and $k\in\mathcal{H}_0$, which implies $\tilde{P}_{(i_{j,\lambda}),\lambda}<\tilde{P}_{(i_{k,\lambda}),\lambda}$ for any $j\in\mathcal{H}_1$ and $k\in\mathcal{H}_0$. Therefore, given the event $\mathcal{A}_{\lambda}\cap\mathcal{B}_{1}\cap\mathcal{B}_{2}$, for any $\alpha<\lambda\le1$, $j\in\mathcal{H}_1$ and $k\in\mathcal{H}_0$, we have
\begin{align}\label{eq:thm_power4}
1\le i_{j,\lambda}\le|\mathcal{H}_1|<i_{k,\lambda}\le d\,.
\end{align}
For any $\alpha<\lambda_1<\lambda_2\le1$, write $\mathcal{C}_{\lambda_1,\lambda_2}:=\mathcal{A}_{\lambda_1}\cap\mathcal{A}_{\lambda_2}
\cap\mathcal{B}_{1}\cap\mathcal{B}_{2}$. Given the event $\mathcal{C}_{\lambda_1,\lambda_2}$, it follows from \eqref{eq:thm_power4} that the identities
\begin{align}\label{eq:thm_power4_1}
i_{j,\lambda_1}= i_{j,\lambda_2} \quad{\text{and}}\quad \tilde{P}_{(i_{j,\lambda_1}),\lambda_1}
=\tilde{P}_{(i_{j,\lambda_2}),\lambda_2}=P_j^{(2)}
\end{align}
hold for all $j\in\mathcal{H}_1$, which implies $\tilde{P}_{(i_{j,\lambda_1}),\lambda_1} \leq i_{j,\lambda_1} \alpha/(\lambda_1 d)$ if $\tilde{P}_{(i_{j,\lambda_2}),\lambda_2} \leq i_{j,\lambda_2} \alpha/(\lambda_2 d)$ and $j\in\mathcal{H}_1$. For any $\alpha<\lambda_1<\lambda_2\le1$ and $d\ge2$, we will establish that, given the event $\mathcal{C}_{\lambda_1,\lambda_2}$ and $j\in\mathcal{H}_1$, if $H_{0,j}$ is rejected by Algorithm \ref{method_3} with $\lambda=\lambda_2$, then it will also be rejected by Algorithm \ref{method_3} with $\lambda=\lambda_1$.

\underline{{\it Case} (i): $1\le\tilde{R}_{\lambda_2}\le|\mathcal{H}_1|$.} Given $j\in\mathcal{H}_1$, by Step 3 of Algorithm \ref{method_3}, $H_{0,j}$ is rejected by Algorithm \ref{method_3} with $\lambda=\lambda_2$ if and only if $i_{j,\lambda_2} \leq \tilde{R}_{\lambda_2}$. Based on the definition of $\tilde{R}_{\lambda_2}$, given the event $\mathcal{C}_{\lambda_1,\lambda_2}$, by \eqref{eq:thm_power4}, we know there exists $l\in\mathcal{H}_1$ such that $i_{l,\lambda_2}=\tilde{R}_{\lambda_2}$ and $\tilde{P}_{(i_{l,\lambda_2}),\lambda_2}\le i_{l,\lambda_2} \alpha/(\lambda_2 d)$. Given the event $\mathcal{C}_{\lambda_1,\lambda_2}$, by \eqref{eq:thm_power4_1}, we have $i_{l,\lambda_1}=i_{l,\lambda_2}=\tilde{R}_{\lambda_2}$ and $\tilde{P}_{(i_{l,\lambda_1}),\lambda_1}\le i_{l,\lambda_1} \alpha/(\lambda_1 d)$, which implies $\tilde{R}_{\lambda_1}\ge i_{l,\lambda_1}=\tilde{R}_{\lambda_2}$. Due to $i_{j,\lambda_2} \leq \tilde{R}_{\lambda_2}$, then $i_{j,\lambda_1}=i_{j,\lambda_2} \leq \tilde{R}_{\lambda_1}$, which implies $H_{0,j}$ will be rejected by Algorithm \ref{method_3} with $\lambda=\lambda_1$.

\underline{{\it Case} (ii): $|\mathcal{H}_1|<\tilde{R}_{\lambda_2}\le d$.} Based on the definition of $\tilde{R}_{\lambda_2}$, given the event $\mathcal{C}_{\lambda_1,\lambda_2}$, \eqref{eq:thm_power4} implies that (a) all null hypotheses $H_{0,j}$ for $j\in\mathcal{H}_1$ are rejected by Algorithm \ref{method_3} with $\lambda=\lambda_2$, and (b) there exists $l\in\mathcal{H}_0$ such that $i_{l,\lambda_2}=\tilde{R}_{\lambda_2}$ and $\tilde{P}_{(i_{l,\lambda_2}),\lambda_2}\le i_{l,\lambda_2} \alpha/(\lambda_2 d)$. By Proposition 2.5 of \citeS{dudley2014uniform}, we have $G(u)\leq e^{-u^2/2}$ for any $u>0$, and $G(u)\geq u^{-1}(2\pi)^{-1/2}e^{-u^2/2}$ for any $u>1$. Due to $l\in\mathcal{H}_0$, given the event $\mathcal{C}_{\lambda_1,\lambda_2}$, we have
\begin{align*}
\frac{1}{4\sqrt{\pi\log d}}\cdot\frac{1}{d^4}\le G(2\sqrt{2\log d})\le G(\bar{t}_{2, l})=\tilde{P}_{(i_{l,\lambda_2}),\lambda_2}
\le \frac{i_{l,\lambda_2}\alpha}{\lambda_2 d}\,.
\end{align*}
Notice that $|\mathcal{H}_1|\ge1$, $i_{l,\lambda_2}\le d$, and $d^4\ge4\sqrt{\pi\log d}$ for any $d\ge2$. For $j\in\mathcal{H}_1$, given the event $\mathcal{C}_{\lambda_1,\lambda_2}$, we have
\begin{align*}
\tilde{P}_{(i_{j,\lambda_2}),\lambda_2}
=G(\bar{t}_{2, j})\le G(3\sqrt{2\log d})\le\frac{1}{d^9}\le\frac{d^4}{4\sqrt{\pi\log d}}\cdot\frac{1}{d^9}
\le \frac{i_{l,\lambda_2}\alpha}{\lambda_2 d^2}
\le\frac{|\mathcal{H}_1|\alpha}{\lambda_2 d}\,,
\end{align*}
which implies 
\begin{align*}
\tilde{P}_{(|\mathcal{H}_1|),\lambda_2}
\le\frac{|\mathcal{H}_1|\alpha}{\lambda_2 d}\,.
\end{align*}
Together with \eqref{eq:thm_power4} and \eqref{eq:thm_power4_1}, given the event $\mathcal{C}_{\lambda_1,\lambda_2}$,  
\begin{align*}
\tilde{P}_{(|\mathcal{H}_1|),\lambda_1}
=\tilde{P}_{(|\mathcal{H}_1|),\lambda_2}
\le\frac{|\mathcal{H}_1|\alpha}{\lambda_2 d}
\le\frac{|\mathcal{H}_1|\alpha}{\lambda_1 d}\,,
\end{align*}
which implies $\tilde{R}_{\lambda_1}\ge|\mathcal{H}_1|$. Given the event $\mathcal{C}_{\lambda_1,\lambda_2}$, by Step 3 of Algorithm \ref{method_3} and \eqref{eq:thm_power4}, all null hypotheses $H_{0,j}$ for $j\in\mathcal{H}_1$ will be rejected by Algorithm \ref{method_3} with $\lambda=\lambda_1$.

Hence, for any $\alpha<\lambda_1<\lambda_2\le1$, $j\in\mathcal{H}_1$ and $d\ge2$, we have
\begin{align*}
I\{\tilde{P}_{j,\lambda_1}\le\tilde{P}_{(\tilde{R}_{\lambda_1}),\lambda_1},\mathcal{C}_{\lambda_1,\lambda_2}\}
\ge
I\{\tilde{P}_{j,\lambda_2}\le\tilde{P}_{(\tilde{R}_{\lambda_2}),\lambda_2},\mathcal{C}_{\lambda_1,\lambda_2}\}\,,
\end{align*}
which implies
\begin{align}\label{eq:thm_power6}
\Psi_{\lambda_2}
&=\mathbb{E}\bigg[\frac{\sum_{j\in\mathcal{H}_1}
I\{\tilde{P}_{j,\lambda_2}\le\tilde{P}_{(\tilde{R}_{\lambda_2}),\lambda_2},
\mathcal{C}_{\lambda_1,\lambda_2}\}}{|\mathcal{H}_1|}\bigg]
+\mathbb{E}\bigg[\frac{\sum_{j\in\mathcal{H}_1}
I\{\tilde{P}_{j,\lambda_2}\le\tilde{P}_{(\tilde{R}_{\lambda_2}),\lambda_2},
\mathcal{C}_{\lambda_1,\lambda_2}^{\rm c}\}}{|\mathcal{H}_1|}\bigg]\notag\\
&\le\mathbb{E}\bigg[\frac{\sum_{j\in\mathcal{H}_1}
I\{\tilde{P}_{j,\lambda_1}\le\tilde{P}_{(\tilde{R}_{\lambda_1}),\lambda_1},
\mathcal{C}_{\lambda_1,\lambda_2}\}}{|\mathcal{H}_1|}\bigg]
+\mathbb{P}(\mathcal{C}_{\lambda_1,\lambda_2}^{\rm c})\\
&\le\mathbb{E}\bigg[\frac{\sum_{j\in\mathcal{H}_1}
I\{\tilde{P}_{j,\lambda_1}\le\tilde{P}_{(\tilde{R}_{\lambda_1}),\lambda_1}\}}{|\mathcal{H}_1|}\bigg]
+\mathbb{P}(\mathcal{C}_{\lambda_1,\lambda_2}^{\rm c})
=\Psi_{\lambda_1}+\mathbb{P}(\mathcal{C}_{\lambda_1,\lambda_2}^{\rm c})\notag\,.
\end{align}
Select $\varrho_{2}\le(100\sqrt{2}-141)(910M)^{-1}C_{\max}^{-1}$. Given the events $\tilde{\mathcal E}_{1,n}$ and $\breve{\mathcal E}_{2,n}$, by Condition \ref{Sass0},
\begin{equation*}
\bigg|\frac{n^{1/2}\beta^0_{j}}{\hat\sigma\hat\Lambda_{j+d,j+d}^{1/2}}\bigg|
\ge\bigg|\frac{25n^{1/2}\beta^0_{j}}
{26(1+10C_{\max}M\varrho_{2})\sigma \Omega_{j+d,j+d}^{1/2}}\bigg|\ge4\sqrt{2\log d}+2\sqrt{\log d}
\end{equation*}
for each $j\in\mathcal{H}_1$. Following the same arguments of \eqref{thm0-3}--\eqref{eq:thm2-2}, we have
\begin{align}\label{eq:thm_power1}
\mathbb{P}(\mathcal{B}_{1})
&\geq\mathbb{P}\bigg[\bigcap_{j \in \mathcal{H}_1}\bigg\{\bigg|\frac{\sigma}{\hat\sigma}\tilde{w}_{j+d}+
\frac{\tilde{\delta}_{j+d}}{\hat\sigma\hat\Lambda_{j+d,j+d}^{1/2}}\bigg|
<(\sqrt{2}+2)\sqrt{\log d}\bigg\},\tilde{\mathcal E}_{1,n}, \breve{\mathcal E}_{2,n}\bigg]\notag\\
&\geq\mathbb{P}\bigg[\bigcap_{j \in \mathcal{H}_1}\bigg\{|\tilde{w}_{j+d}|
\leq\frac{24(\sqrt{2}+2)}{25}\sqrt{\log d}-\frac{91C_{\max}^{1/2}}
{90\sqrt{2}\sigma}\bigg\}\bigg]-\mathbb{P}(|\tilde{\bdelta}_{n}|_{\infty}\ge1)\\
&\quad~-\mathbb{P}(\tilde{\mathcal E}_{1,n}^{\rm c})-\mathbb{P}(\breve{\mathcal E}_{2,n}^{\rm c})\notag\\
&\geq1-|\mathcal{H}_1|\cdot G\bigg\{\frac{24(\sqrt{2}+2)}{25}\sqrt{\log d}-\frac{91C_{\max}^{1/2}}{90\sqrt{2}\sigma}\bigg\}
-\mathbb{P}(|\tilde{\bdelta}_{n}|_{\infty}\ge1)-\mathbb{P}(\tilde{\mathcal E}_{1,n}^{\rm c})-\mathbb{P}(\breve{\mathcal E}_{2,n}^{\rm c})\notag\,,
\end{align}
where $\tilde{\bdelta}_{n}:=(\tilde{\delta}_{1},\ldots,\tilde{\delta}_{2d})^{\top}=\bT \bdelta_{n}$, $\tilde{w}_{j+d}=\sigma^{-1}\hat\Lambda_{j+d,j+d}^{-1/2}(w_{j}-w_{j+d})$ and $\bw_{n}=(w_{1},\ldots,w_{2d})^{\top}$. Similar to \eqref{eq:thm_power1}, it holds that
\begin{align}
\mathbb{P}(\mathcal{B}_2)
&\geq1-|\mathcal{H}_0|\cdot G\bigg(\frac{48\sqrt{2\log d}}{25}-\frac{91C_{\max}^{1/2}}
{90\sqrt{2}\sigma}\bigg)\notag\\
&\quad~-\mathbb{P}(|\tilde{\bdelta}_{n}|_{\infty}\ge1)-\mathbb{P}(\tilde{\mathcal E}_{1,n}^{\rm c})-\mathbb{P}(\breve{\mathcal E}_{2,n}^{\rm c})\,,\label{eq:thm_power1-1}\\
\mathbb{P}(\mathcal{B}_{3})
&\geq1-|\mathcal{H}_1|\cdot G\bigg\{\frac{24(\sqrt{2}+2)}{25}\sqrt{\log d}-\frac{91C_{\max}^{1/2}}{90\sqrt{2}\sigma}\bigg\}\notag\\
&\quad~
-\mathbb{P}(|\tilde{\bdelta}_{n}|_{\infty}\ge1)
-\mathbb{P}(\tilde{\mathcal E}_{1,n}^{\rm c})
-\mathbb{P}(\breve{\mathcal E}_{2,n}^{\rm c})\,.\label{eq:thm_power1-1_1}
\end{align}
By Proposition 2.5 of \citeS{dudley2014uniform}, we know $G(u)\leq e^{-u^2/2}$ for any $u>0$, which implies
\begin{align}\label{eq:thm_power2}
G\bigg\{\frac{24(\sqrt{2}+2)}{25}\sqrt{\log d}-\frac{91C_{\max}^{1/2}}{90\sqrt{2}\sigma}\bigg\}
\le G\bigg(\frac{48\sqrt{2\log d}}{25}-\frac{91C_{\max}^{1/2}}
{90\sqrt{2}\sigma}\bigg) =o(d^{-1})\,.
\end{align}
Recall $\varrho_{1}\asymp\varrho_{2}\asymp\varrho_{3}\asymp\{n^{-1}\log(2d)\}^{1/2}$ and $s_{0}\ll n^{1/2}/\log(2d)$. By Lemmas \ref{theo_scaled} and \ref{theo_random_trans}, we have $
\mathbb{P}(|\tilde{\bdelta}_{n}|_{\infty}\ge1)
+\mathbb{P}(\tilde{\mathcal E}_{1,n}^{\rm c})+\mathbb{P}(\breve{\mathcal E}_{2,n}^{\rm c})= o(1)$. Due to $|\mathcal{H}_1|\le d$, $|\mathcal{H}_0|\le d$, by \eqref{eq:thm_power1}--\eqref{eq:thm_power2},
\begin{align}\label{eq:thm_power3}
\lim_{n\rightarrow\infty}\mathbb{P}(\mathcal{B}_{1})
=\lim_{n\rightarrow\infty}\mathbb{P}(\mathcal{B}_{2})
=\lim_{n\rightarrow\infty}\mathbb{P}(\mathcal{B}_{3}) = 1\,.
\end{align}
Since $\{P^{(1)}_j\le\lambda\}=\{\bar{t}_{1, j}\ge G^{-1}(\lambda)\}$ for all $j\in[d]$, if \(d > \exp[18^{-1}\{G^{-1}(\alpha)\}^2]\), we have $\mathcal{B}_{3}\subset\mathcal{A}_{\lambda_1}\cap\mathcal{A}_{\lambda_2}$ for any $\alpha<\lambda_1<\lambda_2\le1$. Then, by \eqref{eq:thm_power3}, for any $\alpha<\lambda_1<\lambda_2\le1$, 
\begin{align*}
\lim_{n\rightarrow\infty}\mathbb{P}(\mathcal{C}_{\lambda_1,\lambda_2}) 
=\lim_{n\rightarrow\infty}\mathbb{P}(\mathcal{A}_{\lambda_1}\cap\mathcal{A}_{\lambda_2}
\cap\mathcal{B}_{1}\cap\mathcal{B}_{2})= 1\,.
\end{align*}
Together with \eqref{eq:thm_power6}, for any $\alpha<\lambda_1<\lambda_2\le1$, we have $\Psi_{\lambda_2}\le\Psi_{\lambda_1}+o(1)$ as $n\rightarrow\infty$. We complete the proof of Theorem \ref{thm_power}. $\hfill\qedsymbol$

\section{Proofs of Lemmas \ref{theo_condition}--\ref{lem1}}\label{proof_theo_condition}
\subsection{Proof of Lemma \ref{theo_condition}(i)}\label{pflem1a}
Write $\phi_{0}=C_{\min}^{1/2}/2$. By the definition of the event $\mathcal E_n(\phi_0,s_0,K)$, we have
\begin{equation*}
\begin{aligned}
\mathcal E_n^{\rm c} (\phi_0,s_0,K) &\:= \mathcal B_{1,n}(\phi_0,s_0)\cup \mathcal B_{2,n}(K)\,,
\end{aligned}
\end{equation*}
where
$\mathcal B_{1,n}(\phi_0,s_0) = \{\bZ \in \mathbb R^{n\times 2d}: \min_{S\subseteq[2d]:\,\lvert S \rvert\le s_0} \phi(\hat{\bGamma},S)<\phi_0\}$ and $
\mathcal B_{2,n}(K) = \{\bZ \in \mathbb R^{n\times 2d}: \max_{j \in [2d]} \hat{\Gamma}_{j,j} > K\}$. As we will show in Sections \ref{pflem1a1} and \ref{pflem1a2},
\begin{align*}
\mathbb P\{\bZ\in\mathcal B_{1,n}(\phi_0,s_0)\} \le 2e^{-c_1n}~~\text{and}~~
\mathbb P\{\bZ\in\mathcal B_{2,n}(K)\} \le 4^{1-c}d^{-2c+1}\,.
\end{align*}
Hence, Lemma \ref{theo_condition}(i) holds. $\hfill\qedsymbol$

\subsubsection{Upper bound of $\mathbb P\{\bZ\in\mathcal B_{1,n}(\phi_0,s_0)\}$}\label{pflem1a1}
We first state the concept of restricted eigenvalue, introduced by \citeS{bickel2009simultaneous}.
\begin{definition}\label{def3}
Given a symmetric matrix $\bQ \in \mathbb R^{2d\times 2d}$, an integer $s_0 \ge 1$, and $L > 0$, the restricted eigenvalue of $\bQ$ is defined as
\begin{equation*}
\phi^2_{\rm RE}(\bQ,s_0,L) := \min_{S\subseteq [2d]:\, \lvert S \rvert \le s_0}\min_{\btheta\in\mathbb R^{2d}:\,|\btheta_{S^{\rm c}}|_1 \le L|\btheta_S|_1} \frac{\langle \btheta, \bQ\btheta\rangle}{|\btheta_S|_2^2}\,.
\end{equation*}
\end{definition}

Due to $\lambda_{\min}(\bGamma) = \sigma_{\min}(\bGamma) \ge C_{\min}$, we have $\phi_{\rm RE}(\bGamma,s_0,9) \ge \lambda^{1/2}_{\min}(\bGamma)\ge C^{1/2}_{\min}$. 
Let $\bt_0=\bGamma^{-1/2} \bz$ for $\bz$ specified in Condition \ref{ass:random_Z}. Recall $\|\bt_0\|_{\psi_{2}}=\sup_{\bu\in \mathbb{S}^{2d-1}}\sup_{q\ge1}q^{-1/2}\{\mathbb{E}(|\langle \bt_0, \bu\rangle|^{q})\}^{1/q}=\kappa$ and $k!\ge(k/e)^{k}$ for any integer $k\ge1$. Then, for any $\bu\in\mathbb{R}^{2d}$ with $|\bu|_2>0$, 
\begin{equation*}
\begin{aligned}
  \mathbb{E}\bigg\{\exp\bigg(\frac{\langle \bt_0, \bu\rangle^{2}}{t^{2}}\bigg)\bigg\} & = 1+\sum_{k=1}^{\infty}\frac{\mathbb{E}(\langle \bt_0, \bu\rangle^{2k})}{t^{2k}k!}=1+\sum_{k=1}^{\infty}\frac{\mathbb{E}(\langle \bt_0, \bu|\bu|_{2}^{-1}\rangle^{2k})|\bu|_{2}^{2k}}{t^{2k}k!}  \\
  &\le 1+\sum_{k=1}^{\infty}\frac{(2k)^{k}\kappa^{2k}|\bu|_{2}^{2k}}{t^{2k}k!}\le 1+\sum_{k=1}^{\infty}\bigg(\frac{2e\kappa^{2}|\bu|_{2}^{2}}{t^{2}}\bigg)^{k}\,,
\end{aligned}
\end{equation*}
which implies $\mathbb{E}\{\exp(t^{-2}\langle \bt_0, \bu\rangle^{2})\}\le2$ for any $t\ge2\sqrt{e}\kappa|\bu|_{2}$. Hence, $$\inf\bigg[t>0:\mathbb{E}\bigg\{\exp\bigg(\frac{\langle \bt_0, \bu\rangle^{2}}{t^{2}}\bigg)\bigg\}\le2\bigg]\le2\sqrt{e}\kappa|\bu|_{2}$$
for all $\bu\in\mathbb{R}^{2d}$. Note that $\mathbb{E}(|\langle \bt_0, \bu\rangle|^{2})=|\bu|_{2}^{2}$ for every $\bu\in\mathbb{R}^{2d}$. By Definition 5 of \citeS{rudelson2013reconstruction}, $\bt_0$ is isotropic $\psi_{2}$ random vector in $\mathbb{R}^{2d}$ with constant $\alpha=2\sqrt{e}\kappa$. Write $\bPsi=\bZ\bGamma^{-1/2}$ and $\bA=\bGamma^{1/2}$. Notice that the row vectors of $\bPsi$ are independent and have the same distribution as $\bt_0$. Therefore, by Theorem 6 of \citeS{rudelson2013reconstruction}, we have with probability at least $1 - 2\exp\{-\delta^2 n/(c_*\kappa^4)\}$,
\begin{equation}\label{eq:phi}
\frac{1}{\phi_{\rm RE}(\hat{\bGamma},s_0,3)} \le \frac{1}{(1-\delta)\phi_{\rm RE}(\bGamma,s_0,3)}
\end{equation}
for $n \ge c_*m\kappa^4\delta^{-2}\log\{120ed/(m\delta)\}$, where $\delta \in (0,1)$, $c_* = 32000e^{2}$, and
\begin{equation}\label{eq:m}
m = \min\bigg\{2d,\,s_0 + \frac{12960s_{0}}{\delta^2\phi^2_{\rm RE}(\bGamma,s_0,9)}\max_{j\in[2d]}|\bGamma^{1/2} \be_j|_2^2\bigg\}\,.
\end{equation}
Recall $s_0<2d$. Since $\phi_{\rm RE}(\bGamma,s_0,9) \ge C^{1/2}_{\min}$ and $\max_{j\in[2d]}|\bGamma^{1/2} \be_j|_2^2
\leq C_{\max}$, taking $\delta=1/2$ in \eqref{eq:m}, then
\begin{equation*}
s_{0}< m\le s_0 + \frac{51840s_0}{\phi^2_{\rm RE}(\bGamma,s_0,9)}\max_{j\in[2d]}|\bGamma^{1/2} \be_j|_2^2 \le s_0+ \frac{51840s_0 C_{\max}}{C_{\min}}\le \frac{60000s_0C_{\max}}{C_{\min}}\,.
\end{equation*}
Write $v_{0}=240000c_{*}C_{\max}C_{\min}^{-1}\kappa^{4}$. Then we have   $v_{0}s_{0}\log(240ed/s_{0})\ge4c_*m\kappa^{4}\log(240ed/m)$. 
Due to $\phi_{\rm RE}(\bGamma,s_0,3) \ge \phi_{\rm RE}(\bGamma,s_0,9) \ge C^{1/2}_{\min}$, taking $\delta = 1/2$ in \eqref{eq:phi}, we have
\begin{equation*}
\mathbb P \bigg\{\phi_{\rm RE}(\hat{\bGamma},s_0,3) \ge \frac{C^{1/2}_{\min}}{2}\bigg\} \ge 1-2e^{-c_1n}
\end{equation*}
for $n\ge v_{0}s_{0}\log(240ed/s_{0})$, where $c_1 = (4c_*\kappa^4)^{-1}$.
Recall
$$\phi^2(\hat{\bGamma}, S) = \min_{\btheta \in \mathbb{R}^{2d}:\,|\btheta_{S^{\mathrm{c}}}|_1 \leq 3|\btheta_S|_1} \frac{|S| \langle \btheta, \hat{\bGamma}\btheta \rangle}{|\btheta_S|_1^2} $$
for $S\subseteq[2d]$. Let $\phi_0=C_{\min}^{1/2}/2$. Since $\lvert S \rvert  |\btheta_S|_2^2 \ge |\btheta_S|_1^2$, we know $\min_{S\subseteq[2d]:\,\lvert S \rvert \le s_0} \phi(\hat{\bGamma},S) \ge \phi_{\rm RE}(\hat{\bGamma},s_0,3)$,
which implies
\begin{equation*}
\mathbb P\{\bZ\in\mathcal B_{1,n}(\phi_0,s_0)\} \le \mathbb P \bigg\{ \phi_{\rm RE}(\hat{\bGamma},s_0,3) < \frac{C^{1/2}_{\min}}{2} \bigg\} \le2e^{-c_1n}
\end{equation*}
for $n\ge v_{0}s_{0}\log(240ed/s_{0})$. $\hfill\qedsymbol$

\subsubsection{Upper bound of $\mathbb P\{\bZ\in\mathcal B_{2,n}(K)\}$}\label{pflem1a2}
Write $\bZ=(\bz_1,\ldots,\bz_n)^{\top}$. Recall $\hat{\bGamma}=n^{-1}\bZ^{\top} \bZ$. Let $\{\hat{\Gamma}_{j,j}\}_{j=1}^{2d}$ and $\{\Gamma_{j,j}\}_{j=1}^{2d}$ denote the diagonal components of $\hat\bGamma$ and $\bGamma$, respectively. For any $j\in[2d]$, we have
\begin{equation*}
\hat{\Gamma}_{j,j} - \Gamma_{j,j} = \frac{1}{n}\sum_{l=1}^n(\langle \bz_l,\be_j \rangle^2 -\Gamma_{j,j}) =: \frac{1}{n}\sum_{l=1}^n u_l^{(j)}\,.
\end{equation*}
By Condition \ref{ass:random_Z},
\begin{align*}
\|\langle \bz_{l},\be_j \rangle\|^2_{\psi_2}& = \|\langle \bGamma^{-1/2} \bz_{l}, \bGamma^{1/2} \be_j \rangle\|^2_{\psi_2}\\
&=\|\langle \bGamma^{-1/2} \bz_{l}, \bGamma^{1/2} \be_j|\bGamma^{1/2} \be_j|_2^{-1} \rangle\|^2_{\psi_2}\cdot|\bGamma^{1/2} \be_j|_2^{2}\\
&\le \|\bGamma^{-1/2} \bz_{l}\|_{\psi_2}^2 \cdot|\bGamma^{1/2} \be_j|^2_{2}=\kappa^2\be_j^{\top} \bGamma \be_j \le C_{\max}\kappa^2\,.
\end{align*}
Notice that $u_1^{(j)},\ldots,u_{n}^{(j)}$ are independent centered random variables and $\|\langle \bz_{l},\be_j \rangle^2\|_{\psi_1} \le  2\|\langle \bz_{l},\be_j \rangle\|^2_{\psi_2}$. According to Remark 5.18 of \citeS{vershynin_2012_app}, $\|u_l^{(j)}\|_{\psi_1} \le 2\|\langle \bz_l,\be_j \rangle^2\|_{\psi_1}\le4C_{\max}\kappa^2$, which implies
$\{u_l^{(j)}\}_{l=1}^{n}$ are sub-exponential with sub-exponential norm upper bounded by $4C_{\max}\kappa^2$. To obtain the upper bound of $\mathbb P\{\bZ\in\mathcal B_{2,n}(K)\}$, we need Lemma \ref{lem:Berns_ineq} whose proof
is given in Section \ref{proof_lem:Berns_ineq}.
\begin{lemma}[Bernstein-type inequality]\label{lem:Berns_ineq}
Let $\zeta_1,\ldots,\zeta_n$ be independent sub-exponential random variables with mean zero and $\max_{l\in[n]}\|\zeta_l\|_{\psi_1}\le\kappa'$. For any $v>0$, it holds that
\begin{equation*}
\mathbb P\bigg(\frac{1}{n}\bigg| \sum_{l=1}^n \zeta_l \bigg| \ge v \bigg) \le 2\exp\bigg[ -\frac{n}{6} \min\bigg\{ \bigg(\frac{v}{e\kappa'}\bigg)^2,\,\frac{v}{e\kappa'}  \bigg\} \bigg]\,.
\end{equation*}
\end{lemma}
By Lemma \ref{lem:Berns_ineq} with $\kappa'=4C_{\max}\kappa^2$, for any $j\in[2d]$, we obtain
\begin{equation*}
\mathbb P\bigg\{\frac{1}{n}\bigg\lvert \sum_{l=1}^n u_l^{(j)} \bigg\rvert \ge v \bigg\} \le 2\exp\bigg[ -\frac{n}{6} \min\bigg\{ \bigg(\frac{v}{4eC_{\max}\kappa^2}\bigg)^2,\,\frac{v}{4eC_{\max}\kappa^2}  \bigg\} \bigg]
\end{equation*}
for any $v>0$. For any given $c>0$, if $n\ge12c\log(2d)$, we have
\begin{align*}
\mathbb P \bigg( \max_{j \in [2d]} \hat{\Gamma}_{j,j}
\ge C_{\max} + v \bigg)
&\le \mathbb P \bigg( \max_{j \in [2d]} \hat{\Gamma}_{j,j} \ge \max_{j \in [2d]}\Gamma_{j,j} + v \bigg)\\
&\le \mathbb P \bigg( \max_{j \in [2d]} |\hat{\Gamma}_{j,j}-\Gamma_{j,j}| \ge  v \bigg)=\mathbb P \bigg\{ \max_{j \in [2d]} \frac{1}{n}\bigg|\sum_{l=1}^n u_l^{(j)}\bigg| \ge  v \bigg\}\\
&\le  4d\exp\bigg(-\frac{nv^2}{96C_{\max}^{2}e^2\kappa^4}\bigg)
\le4d\exp\{-2c\log(2d)\}\\
&\le 4^{1-c}d^{-2c+1}
\end{align*}
for any $v$ satisfying $\{192cn^{-1}\log(2d)\}^{1/2}\le v/(C_{\max}e\kappa^2)\le 4$. Let $v_{*}=50C_{\max}\kappa^2 \{cn^{-1}\log(2d)\}^{1/2}$. If $n\ge25c\log(2d)$, we know
$\{192cn^{-1}\log(2d)\}^{1/2}\le v_{*}/(C_{\max}e\kappa^2)\le 4$. Therefore, for any $K\ge K':=C_{\max}[1+50\kappa^2 \{cn^{-1}\log(2d)\}^{1/2}]$, we have
$$\mathbb P\{\bZ\in\mathcal B_{2,n}(K)\}=\mathbb P\bigg(\max_{j \in [2d]} \hat{\Gamma}_{j,j} \ge K\bigg)\le\mathbb P\bigg(\max_{j \in [2d]} \hat{\Gamma}_{j,j} \ge K'\bigg)\le 4^{1-c}d^{-2c+1}$$
for $n\ge25c\log(2d)$. $\hfill\qedsymbol$

\subsection{Proof of Lemma \ref{theo_condition}(ii)}\label{pflem1a3}
By the definition of $\mu_{\min}(\bZ)$, we have $\mu_{\min}(\bZ) \le | \bGamma^{-1} \hat{\bGamma} - \bI_{2d}|_{\infty}$. Write $\bZ=(\bz_1,\ldots,\bz_n)^{\top}$ and $\tilde{\bz}_l = \bGamma^{-1/2} \bz_l$. Recall $\hat{\bGamma}=n^{-1}\bZ^{\top}\bZ$. Then
	\begin{equation*}
		\bV =(v_{i,j})_{2d\times2d}:= \bGamma^{-1} \hat{\bGamma} - \bI_{2d} = \frac{1}{n}\sum_{l=1}^n (\bGamma^{-1} \bz_l\bz_l^\top
- \bI_{2d} ) = \frac{1}{n}\sum_{l=1}^n
(\bGamma^{-1/2} \tilde{\bz}_l \tilde{\bz}_l^\top
\bGamma^{1/2} - \bI_{2d} )\,.
	\end{equation*}
Let $v_l^{(i,j)} = \langle \bGamma^{-1/2}\be_{i}, \tilde{\bz}_l \rangle \langle \bGamma^{1/2}\be_{j}, \tilde{\bz}_l \rangle - \delta_{i,j}$ with $\delta_{i,j} = I(i=j)$, which is the $(i,j)$-th element of $\bGamma^{-1/2}\tilde{\bz}_l\tilde{\bz}_l^{\top}\bGamma^{1/2}
- \bI_{2d}$. Since $\bGamma^{-1/2}\mathbb{E}(\tilde{\bz}_l\tilde{\bz}_l^{\top})\bGamma^{1/2}
= \bI_{2d}$, then $\mathbb E\{v_l^{(i,j)}\} = 0$. By Remark 5.18 of \citeS{vershynin_2012_app} and Condition \ref{ass:random_Z}, it holds that
\begin{align*}
		\|v_l^{(i,j)}\|_{\psi_1} &\le 2\|\langle \bGamma^{-1/2}\be_{i}, \tilde{\bz}_l \rangle \langle \bGamma^{1/2}\be_{j}, \tilde{\bz}_l \rangle\|_{\psi_1}\\
        &\le 4\|\langle \bGamma^{-1/2}\be_{i}, \tilde{\bz}_l \rangle\|_{\psi_2} \|\langle \bGamma^{1/2}\be_{j}, \tilde{\bz}_l \rangle\|_{\psi_2}\\        &=4|\bGamma^{-1/2}\be_{i}|_{2}|\bGamma^{1/2}\be_{j}|_{2}\bigg\|\bigg\langle \frac{\bGamma^{-1/2}\be_{i}}{|\bGamma^{-1/2}\be_{i}|_{2}}, \tilde{\bz}_l \bigg\rangle\bigg\|_{\psi_2} \bigg\|\bigg\langle \frac{\bGamma^{1/2}\be_{j}}{|\bGamma^{1/2}\be_{j}|_{2}}, \tilde{\bz}_l \bigg\rangle\bigg\|_{\psi_2}\\
		&\le 4|\bGamma^{-1/2}\be_{i}|_{2}|\bGamma^{1/2}\be_{j}|_{2}\|\tilde{\bz}_l\|_{\psi_2}^{2} =4|\bGamma^{-1/2}\be_{i}|_{2}|\bGamma^{1/2}\be_{j}|_{2}\kappa^2 \\
		&\le 4(C_{\max}C_{\min}^{-1})^{1/2}\kappa^2\,.
\end{align*}
By Lemma \ref{lem:Berns_ineq} with $\kappa'=4(C_{\max}C_{\min}^{-1})^{1/2}\kappa^2$, we have
\begin{equation*}
	\mathbb P \bigg\{ \frac{1}{n}\bigg\lvert \sum_{l=1}^n v_l^{(i,j)} \bigg\rvert \ge v \bigg\} \le 2\exp\bigg[ -\frac{n}{6} \min\bigg\{\bigg(\frac{vC_{\min}^{1/2}}{4eC_{\max}^{1/2}\kappa^2}\bigg)^2,\frac{vC_{\min}^{1/2}}{4eC_{\max}^{1/2}\kappa^2}  \bigg\} \bigg]
\end{equation*}
for any $v>0$. For given $a>0$, choose $v = a\{n^{-1}\log(2d)\}^{1/2}$. Then
\begin{equation*}
	\mathbb P \bigg\{ \frac{1}{n}\bigg\lvert \sum_{l=1}^n v_l^{(i,j)} \bigg\rvert \ge a\sqrt{\frac{\log(2d)}{n}}\bigg\} \le 2d^{-a^2C_{\min}/(96e^2\kappa^4C_{\max})}
\end{equation*}
for $n \ge a^{2}C_{\min}(16e^{2}C_{\max}\kappa^4)^{-1} \log(2d)$. Recall $\bV=(v_{i,j})_{2d\times2d}$ with $v_{i,j} = n^{-1}\sum_{l=1}^n v_l^{(i,j)}$. Therefore,
\begin{align*}
\mathbb P\{\bZ \in \mathcal G_n(a)\} &= \mathbb{P}\bigg\{\mu_{\min}(\bZ) < a\sqrt{\frac{\log(2d)}{n}} \bigg\}
\ge1-\mathbb P\bigg\{|\bV|_{\infty}\ge a\sqrt{\frac{\log(2d)}{n}}\bigg\}\\
&\ge 1 - (2d)^2\cdot 2d^{-a^2C_{\min}/(96e^2\kappa^4C_{\max})} = 1 - 8d^{-c_2}
\end{align*}
for $n \ge a^{2}C_{\min}(16e^2C_{\max}\kappa^4)^{-1}\log(2d)=6(c_{2}+2)\log(2d)$, where $c_2 = a^2C_{\min}/(96e^2\kappa^4C_{\max}) - 2$. We complete the proof of Lemma \ref{theo_condition}(ii). $\hfill\qedsymbol$

\subsection{Proof of Lemma \ref{theo_fix}}\label{proof_theo_fix}
Consider the event
\begin{equation*}
\mathcal T :=\bigg\{\max_{j\in[2d]} \frac{2\lvert\beps^\top \bz^{(j)}\rvert}{n} \le \varrho_0\bigg\}\,,
\end{equation*}
where $\bz^{(j)}$ denotes the $j$-th column of $\bZ$. 
By Theorem 6.1 of \citeS{BV_2011_app}, given the event $\mathcal{T}$ and \(\bZ \in \mathcal E_n(C_{\min}^{1/2}/2,s_0,3C_{\max}/2)\), if  $\varrho_1 \ge 2\varrho_0$, we have 
$
|\hat{\bgamma} - \bgamma_{0}|_1 \le 16\varrho_1 s_0C_{\min}^{-1}$. 
By Lemma \ref{theo_condition}(i) with $c=(\tau+1)/2$ and $K=3C_{\max}/2$, if $\varrho_1 \ge 2\varrho_0$, it holds that 
\begin{align}
\mathbb{P}\bigg(|\hat{\bgamma} - \bgamma_{0}|_1 > \frac{16\varrho_1 s_0}{C_{\min}}\bigg)&\le \mathbb{P}\bigg\{|\hat{\bgamma} - \bgamma_{0}|_1 > \frac{16\varrho_1 s_0}{C_{\min}},\bZ\in\mathcal E_n\bigg(\frac{C_{\min}^{1/2}}{2},s_0,\frac{3C_{\max}}{2}\bigg),\mathcal{T}^{\rm c}\bigg\}\notag\\
&\quad+\mathbb{P}\bigg\{\bZ\in\mathcal E_n^{\rm c}\bigg(\frac{C_{\min}^{1/2}}{2},s_0,\frac{3C_{\max}}{2}\bigg)\bigg\}\label{eq:lem2-1}\\
&\le\mathbb{P}\bigg\{\bZ\in\mathcal E_n\bigg(\frac{C_{\min}^{1/2}}{2},s_0,\frac{3C_{\max}}{2}\bigg),\mathcal{T}^{\rm c}\bigg\}
+2e^{-c_1n} + 2d^{-\tau}\notag\\
&=\mathbb{E}\bigg[\mathbb P(\mathcal{T}^{\rm c} \,|\, \bZ)I\bigg\{\bZ \in \mathcal E_n\bigg(\frac{C_{\min}^{1/2}}{2},s_0,\frac{3C_{\max}}{2}\bigg)\bigg\}\bigg]+2e^{-c_1n} + 2d^{-\tau}\,,\notag
\end{align}
provided that $n \ge \max\{v_0s_0\log(240eds_0^{-1}), 5000(\tau+1)\kappa^4\log(2d), 12.5(\tau+1)\log(2d)\}$. Write $\zeta_j = n^{-1/2}\sigma^{-1}\beps^\top \bz^{(j)}$. Recall $\hat{\bGamma}=n^{-1}\bZ^{\top}\bZ$. By Condition \ref{ass:model_error}, we know $\zeta_j\,|\,\bZ\sim \mathcal N(0,\hat{\Gamma}_{j,j})$. 
For given $\tau>0$, let $\varrho_0 = 2\sigma \sqrt{3C_{\max}(\tau+1)n^{-1}\log(2d)}$. Then 
\begin{align*}
\mathbb P(\mathcal{T}^{\rm c} \,|\, \bZ)
&=\mathbb P \bigg\{\max_{j\in[2d]} \lvert \zeta_j \rvert > \sqrt{3C_{\max}(\tau+1)\log(2d)}\,\bigg|\,\bZ\bigg\}\\
&\le4d \exp\bigg\{-\frac{3C_{\max}(\tau+1)\log (2d)}{2\hat{\Gamma}_{j,j}}\bigg\}\,,
\end{align*}
which implies 
\begin{align*}
&\mathbb{E}\bigg[\mathbb P(\mathcal{T}^{\rm c} \,|\, \bZ)I\bigg\{\bZ \in \mathcal E_n\bigg(\frac{C_{\min}^{1/2}}{2},s_0,\frac{3C_{\max}}{2}\bigg)\bigg\}\bigg]\\
&~~~~~\le
\mathbb{E}\bigg[4d \exp\bigg\{-\frac{3C_{\max}(\tau+1)\log (2d)}{2\hat{\Gamma}_{j,j}}\bigg\}I\bigg\{\bZ \in \mathcal E_n\bigg(\frac{C_{\min}^{1/2}}{2},s_0,\frac{3C_{\max}}{2}\bigg)\bigg\}\bigg]\\
&~~~~~\le4d \exp\bigg\{-\frac{3C_{\max}(\tau+1)\log (2d)}{3C_{\max}}\bigg\}
=4d\cdot(2d)^{-\tau-1}\le2d^{-\tau}\,.
\end{align*}
Together with \eqref{eq:lem2-1}, if $\varrho_1 \ge 4\sigma \sqrt{3C_{\max}(\tau+1)n^{-1}\log(2d)}$, we have
\begin{align*}
\mathbb{P}\bigg(|\hat{\bgamma} - \bgamma_{0}|_1 > \frac{16\varrho_1 s_0}{C_{\min}}\bigg)\le4d^{-\tau}+2e^{-c_{1}n}
\end{align*}
for 
$n \ge \max\{v_0s_0\log(240eds_0^{-1}), 5000(\tau+1)\kappa^4\log(2d), 12.5(\tau+1)\log(2d)\}$. We complete the proof of Lemma \ref{theo_fix}. $\hfill\qedsymbol$

\subsection{Proof of Lemma \ref{lemm_clime}}\label{proof_lemm_clime}
Write $\bz=(\bx^\top,\tilde\bx^\top)^\top=(z_1,\ldots,z_{2d})^{\top}$. By Condition \ref{ass:random_Z} and Remark \ref{prop_condition}, we have
$$\| z_{i}\|_{\psi_{2}}=\| \langle \bz,\be_{i}\rangle\|_{\psi_{2}}\le\| \bz\|_{\psi_{2}}\le\sigma_{\max}(\bGamma^{1/2})\|\bGamma^{-1/2} \bz\|_{\psi_{2}}\le C_{\max}^{1/2}\kappa$$ 
for $i\in[2d]$, which implies
\begin{align}\label{Seq:lemm_clime1}
  \mathbb E(e^{tz_{i}^{2}})\le\mathbb E(e^{\eta z_{i}^{2}})\le1+\sum_{k=1}^{\infty}(2C_{\max}
  \kappa^{2}e\eta)^{k}\le\frac{1}{1-2C_{\max}
  \kappa^{2}e\eta}
\end{align}
for $|t|\le\eta$ with $\eta=\min\{1/8,(4C_{\max}\kappa^{2}e)^{-1}\}$. Recall $\hat{\bGamma}=n^{-1}\bZ^{\top}\bZ$. Therefore, as shown in Equation (29) of \citeS{cai2011constrained_app}, if $n\ge \eta^{-1}\log(2d)$,
\begin{align*}
    \mathbb{P} \bigg\{|\hat\bGamma-\bGamma|_{\infty}\ge C_{0}\sqrt{\frac{\log(2d)}{n}}\bigg\}\le 2^{1-\tau}d^{-\tau}\,,
\end{align*}
where $C_0$ is specified in Condition \ref{ass:CLIME}. By Theorem 6 of \citeS{cai2011constrained_app}, if $\varrho_2\ge \|\bTheta_0\|_{1}|\hat\bGamma-\bGamma|_{\infty}$, we have $|\hat\bTheta-\bTheta_0|_{\infty}\le4M\varrho_2$. 
As given in Condition \ref{ass:CLIME}, $\varrho_{2}\ge C_{0}M\{n^{-1}\log(2d)\}^{1/2}$ and $\|\bTheta_{0}\|_{1}\le M$. Therefore, for any given $\tau>0$, if $n\ge \eta^{-1}\log(2d)$,
\begin{align*}
    \mathbb{P}(|\hat\bTheta-\bTheta_0|_{\infty}\le4M\varrho_2)
    &\ge\mathbb{P}(|\hat\bTheta-\bTheta_0|_{\infty}\le4M\varrho_2,\varrho_2\ge \|\bTheta_0\|_{1}|\hat\bGamma-\bGamma|_{\infty})\\
    &=\mathbb{P}(\varrho_2\ge \|\bTheta_0\|_{1}|\hat\bGamma-\bGamma|_{\infty})\\
    &\ge\mathbb{P}\bigg\{\varrho_2\ge \|\bTheta_0\|_{1}|\hat\bGamma-\bGamma|_{\infty},|\hat\bGamma-\bGamma|_{\infty}< C_{0}\sqrt{\frac{\log(2d)}{n}}\bigg\}\\
    &=\mathbb{P} \bigg\{|\hat\bGamma-\bGamma|_{\infty}< C_{0}\sqrt{\frac{\log(2d)}{n}}\bigg\}\\
    &\ge1- 2^{1-\tau}d^{-\tau}\,.
\end{align*}
We complete the proof of Lemma \ref{lemm_clime}. $\hfill\qedsymbol$

\subsection{Proof of Lemma \ref{theo_scaled}}\label{proof_theo_scaled}
For any given $\tau>0$, by Lemma \ref{theo_condition}(i) with $c=(\tau+1)/2$, we have
    \begin{align}\label{eq:lem4-1}
        \mathbb P \bigg(\bigg| \frac{\hat{\sigma}}{\sigma} - 1 \bigg| \ge 16\varrho_3\sqrt{\frac{s_0}{C_{\min}}} \bigg)
        &\le \mathbb P \bigg\{\bigg| \frac{\hat{\sigma}}{\sigma} - 1 \bigg| \ge 16\varrho_3\sqrt{\frac{s_0}{C_{\min}}}, \bZ\in\mathcal{E}_n\bigg(\frac{C_{\min}^{1/2}}{2},s_{0},\frac{3C_{\max}}{2}\bigg)\bigg\}\notag\\
        &\quad~+ \mathbb P\bigg\{\bZ \in\mathcal{E}_n^{\rm c}\bigg(\frac{C_{\min}^{1/2}}{2},s_{0},\frac{3C_{\max}}{2}\bigg)\bigg\} \\
        &\le \mathbb P \bigg\{\bigg| \frac{\hat{\sigma}}{\sigma} - 1 \bigg| \ge 16\varrho_3\sqrt{\frac{s_0}{C_{\min}}}, \bZ\in\mathcal{E}_n\bigg(\frac{C_{\min}^{1/2}}{2},s_{0},\frac{3C_{\max}}{2}\bigg)\bigg\}\notag\\
        &\quad~+ 2e^{-c_1n}+2d^{-\tau}\notag
    \end{align}
for $n \ge \max\{v_0s_0\log(240eds_0^{-1}),12.5(\tau+1)\log(2d),5000(\tau+1)\kappa^{4}\log(2d)\}$, where $v_0$, $c_1$ and $\mathcal{E}_n(\cdot,\cdot,\cdot)$ are specified in Lemma \ref{theo_condition}(i). Consider the event 
$$\mathcal{D}:=\bigg\{\bigg|\frac{\bZ^{\top}\beps}{n\sigma^*}\bigg|_{\infty} \le \frac{\varrho_3}{4}\bigg\}\,.$$
Write $T_0=\text{supp}(\bgamma_{0})$ and $\sigma^*=n^{-1/2}|\by-\bZ\bgamma_{0}|_{2}$. Let $\eta(\cdot,\cdot,\cdot,\cdot)$ and $\kappa(\cdot,\cdot)$ be defined as in Equations (10) and (11) of \citeS{sun2012scaled_app}, respectively. Given $\bZ\in\mathcal{E}_n(C_{\min}^{1/2}/2,s_{0},3C_{\max}/2)$, we have $\kappa^2(3,T_0)\ge C_{\min}/4$. Due to $|\bgamma_{0}|_{0}\le s_{0}$, then
\begin{align*}
\eta_*(\sigma^*\varrho_3,3) := &\,\inf_{\bomega\in\mathbb{R}^{2d},\,T\subseteq[2d]}\eta(\sigma^*\varrho_3,3,\bomega,T)\\
\le&\,\eta(\sigma^*\varrho_3,3,\bgamma_0,T_0)
= \frac{4\cdot3^2(\sigma^*)^{2}\varrho_3^{2}|\bgamma_{0}|_{0}}{4^2\kappa^2(3,T_0)}
\le \frac{16s_0(\sigma^*)^{2}\varrho_3^2}{C_{\min}}\,,
\end{align*}
which implies $\tau_0 := \eta_*^{1/2}(\sigma^*\varrho_3,3)/\sigma^* \le 4(s_0C_{\min}^{-1})^{1/2}\varrho_3$. Recall $\varrho_3 \asymp \{n^{-1}\log(2d)\}^{1/2}$ and $s_0 \ll n/\log(2d)$. Given the event $\mathcal{D}$, we have $|\bZ^{\top}\beps/(n\sigma^*)|_{\infty} \le \varrho_3(1-\tau_0)/2$ for sufficiently large $n$. Hence, given the event $\mathcal{D}$, by Theorem 1 of \citeS{sun2012scaled_app}, we have $\max\{1- \hat{\sigma}/\sigma^*,1- \sigma^*/\hat{\sigma}\} \le \tau_0$ for sufficiently large $n$, which implies
    \begin{align*}
        \bigg|\frac{\hat\sigma}{\sigma^*} - 1\bigg|
        \le\max\bigg\{\tau_{0}, \frac{\tau_{0}}{1-\tau_{0}}\bigg\}\le 2\tau_0\le\frac{8s_0^{1/2}\varrho_3}{C_{\min}^{1/2}}
    \end{align*}
for sufficiently large $n$. Due to
    \begin{align*}
        \bigg| \frac{\hat{\sigma}}{\sigma} - 1 \bigg| 
        \le \bigg|\frac{\hat{\sigma}}{\sigma^*} - 1\bigg| \cdot \bigg|\frac{\sigma^*}{\sigma} - 1\bigg| + \bigg| \frac{\hat{\sigma}}{\sigma^*} - 1 \bigg| + \bigg|\frac{\sigma^*}{\sigma} - 1 \bigg|\,,
    \end{align*}
then
\begin{align}\label{eq:lem4-2}
        &\mathbb P \bigg\{\bigg| \frac{\hat{\sigma}}{\sigma} - 1 \bigg| \ge 16\varrho_3\sqrt{\frac{s_0}{C_{\min}}}, \bZ\in\mathcal{E}_n\bigg(\frac{C_{\min}^{1/2}}{2},s_{0},\frac{3C_{\max}}{2}\bigg)\bigg\}\notag\\
        &~~~~~\le  \mathbb P \bigg\{\mathcal{D},
        \bigg| \frac{\hat{\sigma}}{\sigma} - 1 \bigg| \ge 16\varrho_3\sqrt{\frac{s_0}{C_{\min}}},
        \bZ\in\mathcal{E}_n\bigg(\frac{C_{\min}^{1/2}}{2},s_{0},\frac{3C_{\max}}{2}\bigg)  \bigg\}\notag\\
        &\quad~~~~~+ \mathbb P \bigg\{\mathcal{D}^{\rm c},
        \bZ\in\mathcal{E}_n\bigg(\frac{C_{\min}^{1/2}}{2},s_{0},\frac{3C_{\max}}{2}\bigg)  \bigg\}\notag\\
        &~~~~~\le  \mathbb P \bigg\{\bigg|\frac{\hat\sigma}{\sigma^*} - 1\bigg|\le 8\varrho_3\sqrt{\frac{s_0}{C_{\min}}},
        \bigg| \frac{\hat{\sigma}}{\sigma} - 1 \bigg| \ge 16\varrho_3\sqrt{\frac{s_0}{C_{\min}}},
        \bZ\in\mathcal{E}_n\bigg(\frac{C_{\min}^{1/2}}{2},s_{0},\frac{3C_{\max}}{2}\bigg)  \bigg\}\notag \\
        &\quad~~~~~+\mathbb P \bigg\{\mathcal{D}^{\rm c},
         \bigg| \frac{\sigma^*}{\sigma} - 1 \bigg| < v,
         \bZ\in\mathcal{E}_n\bigg(\frac{C_{\min}^{1/2}}{2},s_{0},\frac{3C_{\max}}{2}\bigg)  \bigg\}
         +\mathbb P \bigg(\bigg| \frac{\sigma^*}{\sigma} - 1 \bigg| \ge v \bigg)\\
        &~~~~~\le  \mathbb P \bigg(\bigg|\frac{\sigma^*}{\sigma} - 1\bigg|\ge \frac{8\varrho_3s_0^{1/2}}{C_{\min}^{1/2}+8\varrho_3s_0^{1/2}}\bigg)
        +\mathbb P \bigg(\bigg| \frac{\sigma^*}{\sigma} - 1 \bigg| \ge v \bigg)\notag\\
        &\quad~~~~~+ \mathbb P \bigg\{\bigg|\frac{\bZ^{\top}\beps}{n\sigma}\bigg|_{\infty} > \frac{(1-v)\varrho_3}{4},
         \bZ\in\mathcal{E}_n\bigg(\frac{C_{\min}^{1/2}}{2},s_{0},\frac{3C_{\max}}{2}\bigg)  \bigg\}\notag
\end{align}
for any $0<v<1$. By Condition \ref{ass:model_error}, we know $n(\sigma^*\sigma^{-1})^2\sim\chi_n^2$. Since $|x/y-1|\leq|(x/y)^2-1|$ for all $x, y>0$, by inequality (5) of \citeS{ghosh2021exponential}, we have
\begin{align*}
        \mathbb P \bigg(\bigg|\frac{\sigma^*}{\sigma} - 1 \bigg| \ge v\bigg)
        \le \mathbb P \bigg\{\bigg| n\bigg(\frac{\sigma^*}{\sigma}\bigg)^2 - n \bigg| \ge nv \bigg\}
        \le2\exp\bigg\{ -\frac{nv^2}{4(1+v)} \bigg\}
\end{align*}
for any $0<v<1$. Due to $\varrho_3 \asymp \{n^{-1}\log(2d)\}^{1/2}$ and $s_0 \ll n/\log(2d)$, then
\begin{align*}
\frac{8\varrho_3s_0^{1/2}}{C_{\min}^{1/2}+8\varrho_3s_0^{1/2}}\asymp \varrho_3s_0^{1/2}\gg n^{-1/2}\,,
\end{align*}
which implies
\begin{align*}
\lim_{n\rightarrow \infty}\mathbb P \bigg(\bigg|\frac{\sigma^*}{\sigma} - 1\bigg|\ge \frac{8\varrho_3s_0^{1/2}}{C_{\min}^{1/2}+8\varrho_3s_0^{1/2}}\bigg)=0\,.
\end{align*}
Selecting $v=n^{-1/4}$, then 
\begin{align*}
        \lim_{n\rightarrow \infty}\mathbb P \bigg(\bigg| \frac{\sigma^*}{\sigma} - 1 \bigg| \ge n^{-1/4} \bigg)=0\,.
\end{align*}
By Condition \ref{ass:model_error}, we know $\bZ^{\top}\beps/(n^{1/2}\sigma)\,|\,\bZ \sim \mathcal N(\bf{0}, \hat{\bGamma})$. Selecting $\varrho_3 \ge 10C_{\max}^{1/2}\{n^{-1}\log(2d)\}^{1/2}$, then
    \begin{align*}
        &\mathbb P \bigg\{\bigg|\frac{\bZ^{\top}\beps}{n\sigma}\bigg|_{\infty} > \frac{(1-n^{-1/4})\varrho_3}{4},
         \bZ\in\mathcal{E}_n\bigg(\frac{C_{\min}^{1/2}}{2},s_{0},\frac{3C_{\max}}{2}\bigg)  \bigg\}\\
         &~~~~~=\mathbb{E}\bigg[\mathbb{P} \bigg\{\bigg|\frac{\bZ^{\top}\beps}{n^{1/2}\sigma}\bigg|_{\infty} > \frac{n^{1/2}(1-n^{-1/4})\varrho_3}{4} \,\bigg|\,\bZ \bigg\}I\bigg\{\bZ\in\mathcal{E}_n\bigg(\frac{C_{\min}^{1/2}}{2},s_{0},\frac{3C_{\max}}{2}\bigg)\bigg\}\bigg]\\
         &~~~~~\le\mathbb{E}\bigg[\sum_{j=1}^{2d} 2\exp\bigg[-\frac{\{n^{1/2}(1-n^{-1/4})\varrho_3/4\}^2}{2\hat{\Gamma}_{j,j}}\bigg]I\bigg\{\bZ\in\mathcal{E}_n\bigg(\frac{C_{\min}^{1/2}}{2},s_{0},\frac{3C_{\max}}{2}\bigg)\bigg\}\bigg]\\
        &~~~~~\le  4d\exp\bigg[-\frac{\{n^{1/2}(1-n^{-1/4})\varrho_3/4\}^2}{3C_{\max}}\bigg]\ll d^{-1}\,.
    \end{align*}
Therefore, by \eqref{eq:lem4-2}, it holds that
\begin{align*}
        \lim_{n\rightarrow\infty}\mathbb P \bigg\{\bigg| \frac{\hat{\sigma}}{\sigma} - 1 \bigg| \ge 16\varrho_3\sqrt{\frac{s_0}{C_{\min}}}, \bZ\in\mathcal{E}_n\bigg(\frac{C_{\min}^{1/2}}{2},s_{0},\frac{3C_{\max}}{2}\bigg)\bigg\}=0\,.
\end{align*}
Together with \eqref{eq:lem4-1}, we have
\begin{equation*}
\lim_{n\rightarrow \infty}\mathbb{P} \bigg(\bigg\lvert \frac{\hat{\sigma}}{\sigma} - 1 \bigg\rvert \ge 16\varrho_3\sqrt{\frac{s_0}{C_{\min}}}\bigg) = 0\,.
\end{equation*}
We complete the proof of Lemma \ref{theo_scaled}. $\hfill\qedsymbol$

\subsection{Proof of Lemma \ref{theo_random_trans}}\label{proof_theo_random_trans}

By the definition of $\bT$, we have $\|\bT^{\top}\|_{1}=2$. Then, $|\bT \bdelta_{n}|_{\infty}\le|\bdelta_{n}|_{\infty}\|\bT^{\top}\|_{1}=2|\bdelta_{n}|_{\infty}$. Hence, by Theorem \ref{theo_random}, it holds that
\begin{align*}
		\mathbb{P}\bigg( |\bT \bdelta_{n}|_{\infty} > \frac{32\varrho_1\varrho_2s_0n^{1/2}}{C_{\min}} \bigg) 
        \le \mathbb{P}\bigg( |\bdelta_{n}|_{\infty} > \frac{16\varrho_1\varrho_2s_0n^{1/2}}{C_{\min}} \bigg)
        \le 2e^{-c_1n} + 12d^{-\tau}
\end{align*}
for $n \ge \max\{v_0s_0\log(240ed/s_0), 5000(\tau+1)\kappa^4\log(2d), 12.5(\tau+1)\log(2d), 6(c_{2}+2)\log(2d)\}$, where $v_{0}$, $c_{1}$ and $c_{2}$ are constants specified in Theorem \ref{theo_condition}. Due to 
\begin{align*}
|\bT \hat{\bTheta}^\top \hat{\bGamma} \hat{\bTheta} \bT^\top - \bT \bTheta_{0} \bT^\top|_{\infty} &\le |\bT \hat{\bTheta}^\top \hat{\bGamma} \hat{\bTheta} - \bT \bTheta_{0} |_{\infty} \|\bT^\top\|_1
= 2|\bT \hat{\bTheta}^\top \hat{\bGamma} \hat{\bTheta} - \bT \bTheta_{0} |_{\infty}\\
&\le2|\hat{\bTheta}^\top \hat{\bGamma} \hat{\bTheta}  - \bTheta_{0}|_{\infty} \|\bT^\top\|_1
= 4|\hat{\bTheta}^\top \hat{\bGamma} \hat{\bTheta} - \bTheta_{0}|_{\infty}\,,
\end{align*}
then, by Theorem \ref{theo_random}, 
\begin{align*}
\mathbb{P}\big(|\bT \hat{\bTheta}^\top \hat{\bGamma} \hat{\bTheta} \bT^\top - \bT \bTheta_{0} \bT^\top|_{\infty}\le 20M \varrho_{2}\big)
\ge \mathbb{P}\big(|\hat{\bTheta}^\top \hat{\bGamma} \hat{\bTheta} - \bTheta_{0}|_{\infty}\le 5M \varrho_{2}\big)\ge1-10d^{-\tau}
\end{align*}
for $n \ge \max\{6(c_{2}+2)\log (2d), \eta^{-1}\log(2d)\}$. We complete the proof of Lemma \ref{theo_random_trans}. $\hfill\qedsymbol$

\subsection{Proof of Lemma \ref{lem0}}\label{proof_lem0}
For any $b_d \rightarrow \infty$ and $b_d\ll d$, let $v_0<v_1<\cdots<v_{m_d} \leq 1$ and $u_i=G^{-1}(v_i)$, where $v_0=b_d / d$, $v_i=b_d / d+b_d^{2 / 3} e^{i^\delta} / d$, $m_d=\log^{1 / \delta}\{(d-b_d) / b_d^{2 / 3}\}$ and $0<\delta<1$. We will specify $\delta$ later. Due to $u_{0}>u_{1}>\cdots>u_{m_d}\geq0$ and $G(u_{i})=v_{i}$ for $i\in[m_{d}]\cup\{0\}$, then
\begin{align}\label{eq:lem6-0-1}		
&\sup_{i\in[m_d-1]}\bigg|\frac{G(u_i)}{G(u_{i+1})}-1\bigg|
		=\sup_{i\in[m_d-1]}\bigg|\frac{b_d / d+b_d^{2 / 3} e^{i^\delta} / d}{b_d / d+b_d^{2 / 3} e^{(i+1)^\delta} / d}-1\bigg|
		=\sup_{i\in[m_d-1]}\frac{1- e^{i^\delta-(i+1)^{\delta}}}{b_d^{1/3}e^{-(i+1)^{\delta}}+1}\notag\\
&~~~~~\le\max\bigg\{\sup_{1\le i\le(\log b_{d})^{1/(2\delta)}}\frac{1- e^{i^\delta-(i+1)^{\delta}}}{b_d^{1/3}e^{-(i+1)^{\delta}}+1},
\sup_{(\log b_{d})^{1/(2\delta)}<i\le m_d-1}\frac{1- e^{i^\delta-(i+1)^{\delta}}}{b_d^{1/3}e^{-(i+1)^{\delta}}+1}\bigg\}\\
&~~~~~\le\max\bigg[\sup_{1\le i\le(\log b_{d})^{1/(2\delta)}}\frac{e^{(i+1)^{\delta}}}{b_d^{1/3}},
\sup_{(\log b_{d})^{1/(2\delta)}<i\le m_d-1}\{1- e^{i^\delta-(i+1)^{\delta}}\}\bigg]\,.\notag
\end{align}
If $\log b_d \ge16$, for any $1\le i\le(\log b_{d})^{1/(2\delta)}$, we have
\begin{align}\label{eq:lem6-0-2}	
\frac{e^{(i+1)^{\delta}}}{b_d^{1/3}}\le\frac{e^{\{(\log b_{d})^{1/(2\delta)}+1\}^{\delta}}}{b_d^{1/3}}\le 
\frac{e^{(\log b_{d})^{1/2}+1}}{b_d^{1/3}}=\frac{e^{(\log b_{d})^{-1/2}\log b_{d}+1}}{b_d^{1/3}}\le eb_d^{-1/12}\,.
\end{align}
By the Taylor expansion, $(1+u)^{\delta}=u^{\delta}+\delta(1+\xi)^{\delta-1}$ for any $u\ge1$, where $\xi\in(u-1,u)$. Then $\delta (1+i)^{\delta-1}\le(1+i)^{\delta}-i^{\delta}\le\delta i^{\delta-1}$ for $0<\delta<1$, which implies 
\begin{align*}	
1>e^{i^\delta-(i+1)^{\delta}}\ge e^{-\delta i^{\delta-1}}> e^{-\delta (\log b_d)^{(\delta-1)/(2\delta)}}
\end{align*}
for any $i>(\log b_{d})^{1/(2\delta)}$. Together with \eqref{eq:lem6-0-1} and \eqref{eq:lem6-0-2}, we have $\sup_{i\in[m_d-1]}|G(u_i)/G(u_{i+1})-1|=o(1)$. Since $b_d \rightarrow \infty$, we have
	\begin{align*}
		\bigg|\frac{G(u_0)}{G(u_{1})}-1\bigg|
		=\bigg|\frac{v_0}{v_{1}}-1\bigg|
		=\bigg|\frac{b_d / d}{b_d / d+b_d^{2 / 3} e / d}-1\bigg|
		=\frac{e}{b_d^{1/3}+ e}=o(1)\,.
	\end{align*}
Then, we have
\begin{align}\label{eq:lem6-lem8-1}
	\max_{i\in[m_d-1]\cup\{0\}}\bigg|\frac{G(u_i)}{G(u_{i+1})}-1\bigg|=o(1)\,.
\end{align}
For any $t\in [u_{i},u_{i-1}]$, we have
\begin{equation*}
\frac{\sum_{j \in \mathcal{H}_0} I(|\tilde{w}_{j+d}| \geq u_{i})}{d G(u_i)}\frac{G(u_i)}{G(u_{i-1})}
\ge \frac{\sum_{j \in \mathcal{H}_0} I(|\tilde{w}_{j+d}| \geq t)}{d G(t)}
\ge\frac{\sum_{j \in \mathcal{H}_0} I(|\tilde{w}_{j+d}| \geq u_{i-1})}{d G(u_{i-1})}\frac{G(u_{i-1})}{G(u_i)}\,.
\end{equation*}
To prove Lemma \ref{lem0}, it suffices to show that
\begin{equation}\label{lem0-1}
	\begin{aligned}
		\max_{i\in[m_d]\cup\{0\}}\bigg|\frac{\sum_{j \in \mathcal{H}_0} I(|\tilde{w}_{j+d}| \geq u_{i})}{d G(u_i)}-\frac{d_{0}}{d}\bigg| = o_{\rm p}(1)\,.
	\end{aligned}
\end{equation}
Write $\bOmega:=(\Omega_{i,j})_{2d\times2d}=\bT\bTheta_0 \bT^\top$ and $\hat\bLambda=(\hat\Lambda_{i,j})_{2d\times2d}=\bT \hat\bTheta^\top \hat{\bGamma} \hat\bTheta \bT^\top$. Let $\bOmega^{0}=(\Omega^{0}_{i,j})_{2d\times2d}$ with $\Omega^{0}_{i,j}=\Omega_{i,j}\Omega^{-1/2}_{i,i}\Omega^{-1/2}_{j,j}$, and $\hat\bLambda^{0}=(\hat\Lambda^{0}_{i,j})_{2d\times2d}$ with $\hat\Lambda^{0}_{i,j}=\hat\Lambda_{i,j}\hat\Lambda^{-1/2}_{i,i}\hat\Lambda^{-1/2}_{j,j}$. Consider the event
$$\mathcal{G}_{n}=\big\{|\bOmega^{0}-\hat\bLambda^{0}|_{\infty}\le(\log d)^{-2-\vartheta}\big\}\,.$$
To prove \eqref{lem0-1}, we need Lemma 
\ref{lem0_condition}, whose proof is given in Section \ref{proof_lem0_condition}.
\begin{lemma}\label{lem0_condition}
Let the conditions of Lemma {\rm\ref{lem0}} hold. Then $\mathbb P(\mathcal{G}_{n}^{\rm c})\lesssim d^{-\tau}$, where $\tau>2$ is specified in Theorem {\rm\ref{thm0}}.
\end{lemma}

Write $\rho_{i,j}={\rm Corr}(\tilde{w}_{i+d}, \tilde{w}_{j+d}\,|\,\bZ)$ and $\tilde{\bw}_{n}=\sigma^{-1}{\rm diag}(\hat{\Lambda}_{1,1}^{-1/2},\ldots,\hat{\Lambda}_{2d,2d}^{-1/2})\bT\bw_{n}$. Recall $\tilde{w}_{j+d}=\sigma^{-1}\hat\Lambda_{j+d,j+d}^{-1/2}(w_{j}-w_{j+d})$. Then $\tilde{w}_{j+d}$ is the $(j+d)$-th component of $\tilde{\bw}_{n}$. Due to $\sigma^{-1}\bT\bw_{n}\,|\,\bZ\sim\mathcal N({\bf 0},\hat\bLambda)$, then $\tilde{\bw}_{n}\,|\,\bZ\sim\mathcal N({\bf 0},\hat\bLambda^{0})$, which implies $\rho_{i,j}=\hat\Lambda^{0}_{i+d,j+d}$. Notice that $\Omega^{0}_{i+d,j+d}=0$ for $i\neq j$. Given the event $\mathcal{G}_{n}$, we have $|\rho_{i,j}| \leq(\log d)^{-2-\vartheta}$ for $i\neq j$. By Theorem 2.1.e of \citeS{lin2010probability}, 
\begin{align}\label{eq:lem6-0}
\mathbb{P}\big(\tilde{w}_{i+d}>t, \tilde{w}_{j+d}>t\,|\,\bZ\big) \leq\left\{\begin{aligned}
4^{-1}G(t) G\bigg\{\frac{(1-\rho_{i,j}) t}{(1-\rho_{i,j}^2)^{1 / 2}}\bigg\}\,,\qquad& \text { if }-1<\rho_{i,j} \leq 0\,, \\
4^{-1}(1+\rho_{i,j}) G(t) G\bigg\{\frac{(1-\rho_{i,j}) t}{(1-\rho_{i,j}^2)^{1 / 2}}\bigg\}\,, & \text { if } 0 \leq \rho_{i,j}<1\,,
\end{aligned}\right.
\end{align}
for any $t>0$. When $-(\log d)^{-2-\vartheta} \leq \rho_{i,j} \leq 0$, due to $(1-\rho_{i,j}) /(1-\rho_{i,j}^2)^{1 / 2} \geq 1$, we have $G\{(1-\rho_{i,j}) t /(1-\rho_{i,j}^2)^{1 / 2}\} \leq G(t)$, which implies that 
\begin{align}\label{eq:lem6-1}
\max_{i,j\in[d]:\,i\neq j}\mathbb{P}\big(\tilde{w}_{i+d}>t, \tilde{w}_{j+d}>t\,|\,\bZ\big) 
\leq \frac{G^2(t)}{4} \leq \frac{G^2(t)}{4}\{1+(\log d)^{-1-\vartheta}\}\,.
\end{align}
When $0<\rho_{i,j} \leq(\log d)^{-2-\vartheta}$, by the Taylor expansion, it holds that
$$
G\bigg\{\frac{(1-\rho_{i,j}) t}{(1-\rho_{i,j}^2)^{1 / 2}}\bigg\}
=G(t)+2\phi(\tilde{t})\bigg\{t-\frac{(1-\rho_{i,j}) t}{(1-\rho_{i,j}^2)^{1 / 2}}\bigg\}
$$
for $(1-\rho_{i,j}) t /(1-\rho_{i,j}^2)^{1 / 2}<\tilde{t}<t$, where $\phi(\cdot)$ is the density function of the standard normal distribution $\mathcal{N}(0,1)$. Then, if $\rho_{i,j}>0$,
\begin{align*}
G\bigg\{\frac{(1-\rho_{i,j}) t}{(1-\rho_{i,j}^2)^{1 / 2}}\bigg\}\{G(t)\}^{-1}
&=1+\frac{2\phi(\tilde{t})}{G(t)}\bigg\{t-\frac{(1-\rho_{i,j}) t}{(1-\rho_{i,j}^2)^{1 / 2}}\bigg\}\\
& \leq 1+\frac{t \phi(\tilde{t})}{t \phi(t) /(1+t^2)}\bigg\{1-\frac{(1-\rho_{i,j})}{(1-\rho_{i,j}^2)^{1 / 2}}\bigg\} \\
& \leq 1+\frac{\phi(\tilde{t})}{\phi(t)} \cdot 2 \rho_{i,j}(1+t^2)
\end{align*}
for any $t>0$. For $0<t\le (2\log d)^{1/2}$ and $\rho_{i,j} \leq(\log d)^{-2-\vartheta}$, it holds that
\begin{align*}
\frac{\phi(\tilde{t})}{\phi(t)}& =\exp \bigg\{\frac{1}{2}(t-\tilde{t})(t+\tilde{t})\bigg\}\leq \exp \bigg\{2\bigg(1-\sqrt{\frac{1-\rho_{i,j}}{1+\rho_{i,j}}}\bigg) \log d\bigg\}\\
& \leq \exp (4\rho_{i,j} \log d) \leq \exp \left\{4(\log d)^{-1-\vartheta}\right\} \lesssim 1\,.
\end{align*}
Therefore, 
\begin{align*}
G\bigg\{\frac{(1-\rho_{i,j}) t}{(1-\rho_{i,j}^2)^{1 / 2}} \bigg\} \leq G(t)\big[1+O\{(\log d)^{-1-\vartheta}\}\big]
\end{align*}
for any $t \in[0, (2\log d)^{1/2}]$ if $0<\rho_{i,j} \leq(\log d)^{-2-\vartheta}$, where the term $O\{(\log d)^{-1-\vartheta}\}$ holds uniformly over $t \in[0, (2\log d)^{1/2}]$. Then 
\begin{align*}
(1+\rho_{i,j}) G(t) G\bigg\{\frac{(1-\rho_{i,j}) t}{(1-\rho_{i,j}^2)^{1 / 2}} \bigg\} \leq G^2(t)\big[1+O\{(\log d)^{-1-\vartheta}\}\big]
\end{align*}
for any $t \in[0, (2\log d)^{1/2}]$ if $0<\rho_{i,j}\le (\log d)^{-2-\vartheta}$. Together with \eqref{eq:lem6-0} and \eqref{eq:lem6-1}, if 
$|\rho_{i,j}| \leq (\log d)^{-2-\vartheta}$, we have
\begin{align*}
\max_{i,j\in[d]:\,i\neq j}\mathbb{P}\big(\tilde{w}_{i+d}>t, \tilde{w}_{j+d}>t\,|\,\bZ\big) 
\leq\frac{G^2(t)}{4} \big[1+O\{(\log d)^{-1-\vartheta}\}\big]
\end{align*}
for any $t \in[0, (2\log d)^{1/2}]$. Analogously, if 
$|\rho_{i,j}| \leq (\log d)^{-2-\vartheta}$, for any $t \in[0, (2\log d)^{1/2}]$, same bound also holds for $\max_{i,j\in[d]:\,i\neq j}\mathbb{P}(\tilde{w}_{i+d}<-t, \tilde{w}_{j+d}>t\,|\,\bZ)$, $\max_{i,j\in[d]:\,i\neq j}\mathbb{P}(\tilde{w}_{i+d}<-t, \tilde{w}_{j+d}<-t\,|\,\bZ)$ and $\max_{i,j\in[d]:\,i\neq j}\mathbb{P}(\tilde{w}_{i+d}>t, \tilde{w}_{j+d}<-t\,|\,\bZ)$. Hence
\begin{align}\label{eq:lem6-2}
\max_{i,j\in[d]:\,i\neq j}\mathbb{P}\big(|\tilde{w}_{i+d}|>t, |\tilde{w}_{j+d}|>t\,|\,\bZ\big) 
\leq G^2(t)\big[1+O\{(\log d)^{-1-\vartheta}\}\big]\,.
\end{align}
Recall $\mathcal{G}_{n}=\{|\bOmega^{0}-\hat\bLambda^{0}|_{\infty}\le(\log d)^{-2-\vartheta}\}$. Due to $\tilde{w}_{j+d}\,|\,\bZ \sim \mathcal{N}(0,1)$, by Lemma \ref{lem0_condition} and \eqref{eq:lem6-2},
\begin{align}\label{eq:lem6-3}
	&\mathbb{E}\bigg(\bigg[\sum_{j \in \mathcal{H}_0}\big\{I(|\tilde{w}_{j+d}| \geq t)-G(t)\big\}\bigg]^2\bigg)\notag\\
    &~~~~~=\mathbb{E}\bigg\{\mathbb{E}\bigg(\bigg[\sum_{j \in \mathcal{H}_0}\big\{I(|\tilde{w}_{j+d}| \geq t)-G(t)\big\}\bigg]^2\,\bigg|\,\bZ\bigg)\bigg\}\notag\\
    &~~~~~= \sum_{i \in \mathcal{H}_0}\sum_{j \in \mathcal{H}_0}\mathbb{E}\big[\big\{\mathbb{P}(|\tilde{w}_{i+d}| \geq t,|\tilde{w}_{j+d}| \geq t\,|\,\bZ)-G^{2}(t)\big\}\big]\notag\\
	&~~~~~=d_{0}G(t)\{1-G(t)\}+\sum_{i \in \mathcal{H}_0}\sum_{j \in \mathcal{H}_0\setminus \{i\}}\mathbb{E}\big[\big\{\mathbb{P}(|\tilde{w}_{i+d}| \geq t,|\tilde{w}_{j+d}| \geq t\,|\,\bZ)-G^{2}(t)\big\}\big]\\
	&~~~~~\le d_{0}G(t)+d_{0}^{2}\mathbb{P}(\mathcal{G}_{n}^{\rm c})+\sum_{i \in \mathcal{H}_0}\sum_{j \in \mathcal{H}_0\setminus \{i\}}\mathbb{E}\big[I(\mathcal{G}_{n})\big\{\mathbb{P}(|\tilde{w}_{i+d}| \geq t,|\tilde{w}_{j+d}| \geq t\,|\,\bZ)-G^{2}(t)\big\}\big]
	\notag\\
	&~~~~~\lesssim dG(t)+d^{2-\tau}+\sum_{i \in \mathcal{H}_0}\sum_{j \in \mathcal{H}_0\setminus \{i\}}(G^{2}(t)[1+O\{(\log d)^{-1-\vartheta}\}]-G^{2}(t))\notag\\
	&~~~~~\lesssim dG(t)+d^{2-\tau}+d^{2}(\log d)^{-1-\vartheta}G^{2}(t)\notag
\end{align}
for any $t \in[0, (2\log d)^{1/2}]$. By Proposition 2.5 of \citeS{dudley2014uniform}, we know $G(u)\leq e^{-u^2/2}$ for any $u>0$. Then
\begin{align*}
		G\bigg\{\sqrt{2\log\bigg(\frac{d}{b_d}\bigg)}\bigg\}
		\leq\exp\bigg\{-\log\bigg(\frac{d}{b_d}\bigg)\bigg\}
		=\frac{b_d}{d}\,,
\end{align*}
which implies
\begin{align}\label{eq:lem6-4}
u_{0}=G^{-1}\bigg(\frac{b_d}{d}\bigg)\le \sqrt{2\log\bigg(\frac{d}{b_d}\bigg)}\,.
\end{align}
Take $1/(1+\vartheta)<\delta<1$. Due to $\tau>2$, $b_d \rightarrow \infty$ and $b_d\ll d$, for any $\epsilon>0$, by \eqref{eq:lem6-3}, \eqref{eq:lem6-4} and the Chebyshev inequality,
\begin{align*}
		&\sum_{i=0}^{m_{d}}\mathbb{P}\bigg\{\bigg|\frac{\sum_{j \in \mathcal{H}_0} I(|\tilde{w}_{j+d}| \geq u_{i})}{d G(u_i)}-\frac{d_{0}}{d}\bigg|\geq \epsilon\bigg\}\\
		&~~~~~= \sum_{i=0}^{m_{d}}\mathbb{P}\bigg[\bigg|\frac{\sum_{j \in \mathcal{H}_0} \{I(|\tilde{w}_{j+d}| \geq u_{i})-G(u_i)\}}{d G(u_i)}\bigg|\geq \epsilon\bigg]\\
		&~~~~~\lesssim \frac{1}{\epsilon^{2}}\bigg\{m_{d}(\log d)^{-1-\vartheta}+d^{-1}
\sum_{i=0}^{m_{d}}\frac{1}{G(u_{i})}+\frac{m_{d}d^{-\tau}}{G^{2}(u_{0})}\bigg\}\\
		&~~~~~\lesssim \frac{1}{\epsilon^{2}}\bigg\{(\log d)^{-1-\vartheta+1/\delta}+b_d^{-1}+b_d^{-2/3}\sum_{i=1}^{m_{d}}e^{-i^{\delta}}+(\log d)^{1/\delta}d^{2-\tau}b_d^{-2}\bigg\}=o(1)\,,
\end{align*}
which implies \eqref{lem0-1}. We complete the proof of Lemma \ref{lem0}. $\hfill\qedsymbol$

\subsection{Proof of Lemma \ref{lem_normal}}\label{proof_lem_normal}
Recall $G(t)=2\{1-\Phi(t)\}$. Then we have
\begin{align*}
&\sup_{0\leq t \leq \sqrt{2\log d}}\bigg|1-\frac{G(t+\epsilon_{n}/\sqrt{\log d})}{G(t)}\bigg|\notag\\
&~~~~~\leq2\max\bigg\{\underbrace{\sup_{0\leq t \leq 1}\frac{|\Phi(t+\epsilon_{n}/\sqrt{\log d})-\Phi(t)|}{G(1)}}_{\rm I},\underbrace{\sup_{1<t\leq \sqrt{2\log d}}\frac{|\Phi(t+\epsilon_{n}/\sqrt{\log d})-\Phi(t)|}{G(t)}}_{\rm II}\bigg\}\,.
\end{align*}
It suffices to show ${\rm I}=o(1)$ and ${\rm II}=o(1)$. Due to $\epsilon_{n}\rightarrow0$, then
\begin{align*}
\sup_{0\leq t \leq 1}\bigg|\Phi\bigg(t+\frac{\epsilon_{n}}{\sqrt{\log d}}\bigg)-\Phi(t)\bigg|=o(1)\,,
\end{align*}
which implies ${\rm I}=o(1)$. In the sequel, we will show ${\rm II}=o(1)$. 
Let $\phi(t)$ be the density function of the standard normal distribution. By the Taylor expansion, 
\begin{align}\label{eq:lem7-3}
\bigg|\Phi\bigg(t+\frac{\epsilon_{n}}{\sqrt{\log d}}\bigg)-\Phi(t)\bigg|=\frac{|\epsilon_n|}{\sqrt{\log d}}\phi\bigg(t+\frac{\theta_{t} \epsilon_n}{\sqrt{\log d}}\bigg)
\end{align}
for some $0<\theta_{t}<1$. If $\epsilon_n\ge0$, then
\begin{align*}
\phi\bigg(t+\frac{\theta_{t} \epsilon_n}{\sqrt{\log d}}\bigg)=\phi(t)+\frac{\theta_{t} \epsilon_n}{\sqrt{\log d}}\phi^{\prime}(\bar{t})=\phi(t)-\frac{\theta_{t} \epsilon_n\bar{t}}{\sqrt{\log d}}\phi(\bar{t})\,,
\end{align*}
where $t<\bar{t}< t+\theta_{t} \epsilon_n/\sqrt{\log d}$. For any $1<t\le\sqrt{2\log d}$, since $\bar{t} < t+\theta_{t} \epsilon_n/\sqrt{\log d}\leq t+1 \leq 2t \leq 2\sqrt{2\log d}$ for $\epsilon_n\le\sqrt{\log d}$, then $\theta_{t} \epsilon_n \bar{t}/\sqrt{\log d}\leq 2\sqrt{2}\epsilon_n$. Due to $\phi(\bar{t})\leq \phi(t)$, then
\begin{align*}
\phi(t)\ge\phi\bigg(t+\frac{\theta_{t} \epsilon_n}{\sqrt{\log d}}\bigg)\ge \phi(t)-2\sqrt{2}\epsilon_n\phi(t)\,,
\end{align*}
which implies $0\ge \phi(t+\theta_{t} \epsilon_n/\sqrt{\log d})/\phi(t)-1\ge -2\sqrt{2}\epsilon_n$. If $\epsilon_n<0$, then 
\begin{align*}
\phi\bigg(t+\frac{\theta_{t} \epsilon_n}{\sqrt{\log d}}\bigg)=\phi(t)-\frac{\theta_{t} \epsilon_n\bar{t}}{\sqrt{\log d}}\phi(\bar{t})
\end{align*}
for some $t+\theta_{t} \epsilon_n/\sqrt{\log d}<\bar{t}< t$ and $0<\theta_{t}<1$. If $\sqrt{\log d}\ge |\epsilon_n|$, we have $t+\epsilon_n/\sqrt{\log d}>0$ for any $1<t\leq \sqrt{2\log d}$, which implies $0<\bar{t}< t \leq \sqrt{2\log d}$. Then $\theta_{t}|\epsilon_n|\bar{t}/\sqrt{\log d}\leq \sqrt{2}|\epsilon_n|$, which implies
\begin{align*}
\phi(t)\le\phi\bigg(t+\frac{\theta_{t} \epsilon_n}{\sqrt{\log d}}\bigg)\le \phi(t)-\sqrt{2}\epsilon_n\phi(\bar{t})
\end{align*}
for any $1<t\leq \sqrt{2\log d}$. Due to 
\begin{align*}
\frac{\phi(\bar{t})}{\phi(t)}&=\exp\bigg(-\frac{\bar{t}^2}{2}+\frac{t^2}{2}\bigg)
=\exp\bigg\{\frac{(t+\bar{t})}{2}(t-\bar{t})\bigg\}\\
&\leq \exp\{t(t-\bar{t})\}\leq\exp\bigg\{\frac{\theta_{t}|\epsilon_n|\sqrt{2\log d}}{\sqrt{\log d}}\bigg\}\leq \exp(\sqrt{2}|\epsilon_n|)\,,
\end{align*}
then $0\le \phi(t+\theta_{t} \epsilon_n/\sqrt{\log d})/\phi(t)-1\le \sqrt{2}|\epsilon_n|\phi(\bar{t})/\phi(t)\le\sqrt{2}|\epsilon_n|\exp(\sqrt{2}|\epsilon_n|)$ for any $1<t\leq \sqrt{2\log d}$. Hence, if $\sqrt{\log d}\ge |\epsilon_n|$,
\begin{align*}
\bigg|\frac{\phi(t+\theta_{t} \epsilon_n/\sqrt{\log d})}{\phi(t)}-1\bigg|
\leq \max\{\sqrt{2}|\epsilon_n|\exp(\sqrt{2}|\epsilon_n|),2\sqrt{2}|\epsilon_n|\}=:\tilde{\epsilon}_n
\end{align*}
for any $1<t\leq \sqrt{2\log d}$. By Proposition 2.5 in \citeS{dudley2014uniform}, we know $\phi(u)\leq uG(u)$ for any $u\geq1$. By \eqref{eq:lem7-3}, if $\sqrt{\log d}\ge |\epsilon_n|$,
\begin{align*}
\bigg|\Phi\bigg(t+\frac{\epsilon_{n}}{\sqrt{\log d}}\bigg)-\Phi(t)\bigg|\leq\frac{|\epsilon_n|(\tilde{\epsilon}_n+1)\phi(t)}{\sqrt{\log d}}\leq\frac{|\epsilon_n|(\tilde{\epsilon}_n+1)tG(t)}{\sqrt{\log d}}\leq \sqrt{2}|\epsilon_n|(\tilde{\epsilon}_n+1)G(t)
\end{align*}
for any $1<t\leq \sqrt{2\log d}$, which implies ${\rm II}\le \sqrt{2}|\epsilon_n|(\tilde{\epsilon}_n+1)\rightarrow0$ as $\epsilon_n\rightarrow0$. We complete the proof of Lemma \ref{lem_normal}. $\hfill\qedsymbol$

\subsection{Proof of Lemma \ref{lem1}}\label{proof_lem1}
For any $b_d \rightarrow \infty$ and $b_d\ll d$, let $v_0<v_1<\cdots<v_{m_d} \leq 1$ and $u_i=G^{-1}(v_i)$, where $v_0=b_d / d$, $v_i=b_d / d+b_d^{2 / 3} e^{i^\delta} / d$, $m_d=\log^{1 / \delta}\{(d-b_d) / b_d^{2 / 3}\}$ and $0<\delta<1$. We will specify $\delta$ later. Recall $G_{l,n}^{-1}(\lambda)=G^{-1}(\lambda)-(-1)^{l}\upsilon_{n}/\sqrt{\log d}$, where $\upsilon_{n}=o(1)$ and $l\in[2]$. For any $t\in [u_{i},u_{i-1}]$, we have
\begin{align*}
& \frac{\sum_{j \in \mathcal{H}_0} I\{|\tilde{w}_{j+d}| \geq t,|\tilde{w}_{j}|\geq G_{l,n}^{-1}(\lambda)\}}{d\lambda G(t)}
\ge\frac{\sum_{j \in \mathcal{H}_0} I\{|\tilde{w}_{j+d}| \geq u_{i-1},|\tilde{w}_{j}|\geq G_{l,n}^{-1}(\lambda)\}}{d\lambda G(u_{i-1})}\frac{G(u_{i-1})}{G(u_i)}\,,\\
& \frac{\sum_{j \in \mathcal{H}_0} I\{|\tilde{w}_{j+d}| \geq t,|\tilde{w}_{j}|\geq G_{l,n}^{-1}(\lambda)\}}{d\lambda G(t)}
\le\frac{\sum_{j \in \mathcal{H}_0} I\{|\tilde{w}_{j+d}| \geq u_{i},|\tilde{w}_{j}|\geq G_{l,n}^{-1}(\lambda)\}}{d\lambda G(u_i)}\frac{G(u_i)}{G(u_{i-1})}\,.
\end{align*}
By \eqref{eq:lem6-lem8-1}, to prove Lemma \ref{lem1}, it suffices to show that
	\begin{align}\label{lem8-1}
		\max_{i\in[m_d]\cup\{0\}}\bigg|\frac{\sum_{j \in \mathcal{H}_0} I\{|\tilde{w}_{j+d}| \geq u_{i},|\tilde{w}_{j}|\geq G_{l,n}^{-1}(\lambda)\}}{d\lambda G(u_i)}-\frac{d_{0}}{d}\bigg| = o_{\rm p}(1)
	\end{align}
for $l\in[2]$. Recall $\bOmega=(\Omega_{i,j})_{2d\times2d}=\bT\bTheta_0 \bT^\top$, $\hat\bLambda=(\hat\Lambda_{i,j})_{2d\times2d}=\bT \hat\bTheta^\top \hat{\bGamma} \hat\bTheta \bT^\top$, $\bOmega^{0}=(\Omega^{0}_{i,j})_{2d\times2d}$ with $\Omega^{0}_{i,j}=\Omega_{i,j}\Omega^{-1/2}_{i,i}\Omega^{-1/2}_{j,j}$, $\hat\bLambda^{0}=(\hat\Lambda^{0}_{i,j})_{2d\times2d}$ with $\hat\Lambda^{0}_{i,j}=\hat\Lambda_{i,j}\hat\Lambda^{-1/2}_{i,i}\hat\Lambda^{-1/2}_{j,j}$, and $\tilde{\bw}_{n}=(\tilde{w}_{1},\ldots,\tilde{w}_{2d})^{\top}=\sigma^{-1}{\rm diag}(\hat{\Lambda}_{1,1}^{-1/2},\ldots,\hat{\Lambda}_{2d,2d}^{-1/2})\bT\bw_{n}$ satisfies $\tilde{\bw}_{n}\,|\,\bZ\sim\mathcal{N}({\bf 0},\hat\bLambda^{0})$. Let $\hat\bXi_{i,j}={\rm Var}(\tilde{\bw}_{i,j}\,|\,\bZ)$ with $\tilde{\bw}_{i,j}=(\tilde{w}_{i},\tilde{w}_{j})^{\top}$, $\hat\bXi_{i,j,d}={\rm Cov}(\tilde{\bw}_{i,j},\tilde{\bw}_{i+d,j+d}\,|\,\bZ)$, $\tilde{\bw}_{i,j,d}=(\tilde{\bw}_{i,j}^{\top},\tilde{\bw}_{i+d,j+d}^{\top})^{\top}$ and
\begin{align*}
  \hat\bXi_{i+d,j+d}^{(i,j)}:=
  \begin{pmatrix}
  \hat\bXi_{i,j}&\hat\bXi_{i,j,d}\\
  \hat\bXi_{i,j,d}^{\top}&\hat\bXi_{i+d,j+d}
  \end{pmatrix}
  ={\rm Var}(\tilde{\bw}_{i,j,d}\,|\,\bZ)\,.
\end{align*}
Write
\begin{align*}
\mathcal{S}_{i}=\{j \in \mathcal{H}_0: |\Omega_{i,j}| \geq(\log d)^{-2-\vartheta}\} ~~\text{and}~~\mathcal{S}_{i}^{\rm c}=\mathcal{H}_0\setminus\mathcal{S}_{i}
\end{align*}
for $i\in\mathcal{H}_0$, where $\vartheta>0$ is specified in Theorem \ref{thm0}. To prove \eqref{lem8-1}, we need Lemmas \ref{lem2} and \ref{lem6}, whose proofs are given in Sections \ref{proof_lem2} and \ref{proof_lem6}, respectively.

\begin{lemma}\label{lem2}
If $|\bOmega^{0}-\hat\bLambda^{0}|_{\infty}\le(\log d)^{-2-\vartheta}$, then
\begin{align}
&\mathbb{P}(|\tilde{w}_{i}|\ge \nu_{1},|\tilde{w}_{j}|\ge \nu_{1},
|\tilde{w}_{i+d}|\ge \nu_{2},|\tilde{w}_{j+d}|\ge \nu_{2}\,|\,\bZ)\label{lem8-1_1}\\
&~~~~~\le\mathbb{P}(|\tilde{w}_{i}|\ge \nu_{1},|\tilde{w}_{j}|\ge \nu_{1}\,|\,\bZ)
\mathbb{P}(|\tilde{w}_{i+d}|\ge \nu_{2},|\tilde{w}_{j+d}|\ge \nu_{2}\,|\,\bZ)
\big[1+O\{(\log d)^{-1-\vartheta}\}\big]+4d^{-2}\,,\notag\\
&\mathbb{P}\{\nu_{1}\le|\tilde{w}_{i}|\le 2(\log d)^{1/2},\nu_{2}\le|\tilde{w}_{i+d}|\le 2(\log d)^{1/2}
\,|\,\bZ\}\label{lem8-1_2}\\
&~~~~~\ge\mathbb{P}\{\nu_{1}\le|\tilde{w}_{i}|\le 2(\log d)^{1/2}\,|\,\bZ\}
\mathbb{P}\{\nu_{2}\le|\tilde{w}_{i+d}|\le 2(\log d)^{1/2}\,|\,\bZ\}
\big[1-O\{(\log d)^{-1-\vartheta}\}\big]\,,\notag\\
&\mathbb{P}(|\tilde{w}_{i}|\ge \nu_{1},|\tilde{w}_{i+d}|\ge \nu_{2}\,|\,\bZ)\label{lem8-1_3}\\
&~~~~~\le\mathbb{P}(|\tilde{w}_{i}|\ge \nu_{1}\,|\,\bZ)
\mathbb{P}(|\tilde{w}_{i+d}|\ge \nu_{2}\,|\,\bZ)
\big[1+O\{(\log d)^{-1-\vartheta}\}\big]+2d^{-2}\notag
\end{align}
for any $0\le \nu_{1},\nu_{2}\le (2\log d)^{1/2}$, where the terms $O\{(\log d)^{-1-\vartheta}\}$ hold uniformly over $\nu_{1},\nu_{2} \in[0, (2\log d)^{1/2}]$ and $i,j\in[d]$, $i\neq j$.
\end{lemma}

\begin{lemma}\label{lem6}
Assume that $\bTheta_0 \in \mathcal U(M,q,s_d)$ and $1/2<\rho<1$. If $d>\max\{4s_d^{2},(\log d)^{(4+2\vartheta)/(2\rho-1)}\}$, then $|\mathcal{S}_{i}|\le d^{\rho}$ for all $i\in \mathcal{H}_0$.
\end{lemma}

Notice that
\begin{align}\label{eq:lem6-2-1}
&\mathbb{E}\bigg[\bigg(\sum_{j \in \mathcal{H}_0}[I\{|\tilde{w}_{j+d}| \geq t, |\tilde{w}_{j}| \geq G_{l,n}^{-1}(\lambda)\}-\lambda G(t)]\bigg)^2\bigg]\notag\\
&~~~~~=\mathbb{E}\bigg(\mathbb{E}\bigg[\bigg(\sum_{j \in \mathcal{H}_0}[I\{|\tilde{w}_{j+d}| \geq t, |\tilde{w}_{j}| \geq G_{l,n}^{-1}(\lambda)\}-\lambda G(t)]\bigg)^2\,\bigg|\,\bZ\bigg]\bigg)\notag\\
&~~~~~=\sum_{i \in \mathcal{H}_0}\sum_{j \in \mathcal{H}_0\setminus\{i\}}
\mathbb{E}\big[\mathbb{P}\{|\tilde{w}_{i+d}| \geq t, |\tilde{w}_{i}| \geq G_{l,n}^{-1}(\lambda),|\tilde{w}_{j+d}| \geq t, |\tilde{w}_{j}| \geq G_{l,n}^{-1}(\lambda)\,|\,\bZ\}\notag\\
&~~~~~~~~~~~~~~~~~~~~~~~~~~~~~~
-\lambda G(t)\mathbb{P}\{|\tilde{w}_{i+d}| \geq t, |\tilde{w}_{i}| \geq G_{l,n}^{-1}(\lambda)\,|\,\bZ\}\\
&~~~~~~~~~~~~~~~~~~~~~~~~\underbrace{~~~~~
-\lambda G(t)\mathbb{P}\{|\tilde{w}_{j+d}| \geq t, |\tilde{w}_{j}| \geq G_{l,n}^{-1}(\lambda)\,|\,\bZ\}+\lambda^{2} G^{2}(t)\big]~~~~~~}_{{\rm I}_{i,j}(t)}\notag\\
&~~~~~~~~+\sum_{i \in \mathcal{H}_0}
\mathbb{E}\big[\mathbb{P}\{|\tilde{w}_{i+d}| \geq t, |\tilde{w}_{i}| \geq G_{l,n}^{-1}(\lambda)\,|\,\bZ\}\notag\\
&~~~~~~~~~~~~~~~~~\underbrace{~~~~~~~~~~~~~
-2\lambda G(t)\mathbb{P}\{|\tilde{w}_{i+d}| \geq t, |\tilde{w}_{i}| \geq G_{l,n}^{-1}(\lambda)\,|\,\bZ\}+\lambda^{2}G^{2}(t)\big]}_{{\rm II}_{i}(t)}\notag\,.
\end{align}
Recall $\mathcal{G}_{n}=\{|\bOmega^{0}-\hat\bLambda^{0}|_{\infty}\le(\log d)^{-2-\vartheta}\}$. Due to $\Omega^{0}_{i,j+d}=0$ for $i, j\in[d]$, given the event $\mathcal{G}_{n}$, we have $\max_{i,j\in[d]}|\hat\bXi_{i,j,d}|_{\infty} \leq(\log d)^{-2-\vartheta}$. Since $\upsilon_{n}=s_{0}^{1/2}n^{-1/4}(\log d)^{3/4}$ and $s_{0}\lesssim n^{1/2}(\log d)^{-5/2-2\vartheta}$, by \eqref{eq:lem7-3}, we have $G\{G^{-1}_{l,n}(\lambda)\}=\lambda[1+O\{(\log d)^{-1-\vartheta}\}]$. Hence, for any $0\le t\le (2\log d)^{1/2}$, by Lemma \ref{lem2}, it holds that
\begin{align*}
{\rm I}_{i,j}(t)
&\le\mathbb{E}\big[I(\mathcal{G}_{n})\mathbb{P}\{|\tilde{w}_{i+d}| \geq t, |\tilde{w}_{i}| \geq G_{l,n}^{-1}(\lambda),|\tilde{w}_{j+d}| \geq t, |\tilde{w}_{j}| \geq G_{l,n}^{-1}(\lambda)\,|\,\bZ\}\big]+\mathbb{P}(\mathcal{G}_{n}^{\rm c})\notag\\
&~~~-\lambda G(t)\mathbb{E}\big[I(\mathcal{G}_{n})\mathbb{P}\{2(\log d)^{1/2}\ge|\tilde{w}_{i+d}| \geq t, 2(\log d)^{1/2}\ge|\tilde{w}_{i}| \geq G_{l,n}^{-1}(\lambda)\,|\,\bZ\}\big]\\
&~~~-\lambda G(t)\mathbb{E}\big[I(\mathcal{G}_{n})\mathbb{P}\{2(\log d)^{1/2}\ge|\tilde{w}_{j+d}| \geq t, 2(\log d)^{1/2}\ge|\tilde{w}_{j}| \geq G_{l,n}^{-1}(\lambda)\,|\,\bZ\}\big]+\lambda^{2} G^{2}(t)\\
&\le\underbrace{\mathbb{E}\big[I(\mathcal{G}_{n})\mathbb{P}\{|\tilde{w}_{i+d}| \geq t, |\tilde{w}_{i}| \geq G_{l,n}^{-1}(\lambda),|\tilde{w}_{j+d}| \geq t, |\tilde{w}_{j}| \geq G_{l,n}^{-1}(\lambda)\,|\,\bZ\}\big]}_{{\rm I}_{i,j,1}(t)}+\mathbb{P}(\mathcal{G}_{n}^{\rm c})\notag\\
&~~~-2\lambda G(t)\underbrace{\mathbb{E}\big[I(\mathcal{G}_{n})\mathbb{P}\{2(\log d)^{1/2}\ge|\tilde{w}_{i+d}| \geq t\,|\,\bZ\}\mathbb{P}\{2(\log d)^{1/2}\ge|\tilde{w}_{i}| \geq G_{l,n}^{-1}(\lambda)\,|\,\bZ\}]}_{{\rm I}_{i,2}(t)}\\
&~~~~~~~~~~\times\big[1-O\{(\log d)^{-1-\vartheta}\}\big]+\lambda^{2} G^{2}(t)\,.
\end{align*}
Recall $\mathcal{S}_{i}=\{j \in \mathcal{H}_0: |\Omega_{i,j}| \geq(\log d)^{-2-\vartheta}\}$, and $\mathcal{S}_{i}^{\rm c}=\mathcal{H}_0\setminus\mathcal{S}_{i}$ for $i\in\mathcal{H}_0$. Due to $\Omega^{0}_{i,j}=\Omega_{i,j}\Omega^{-1/2}_{i,i}\Omega^{-1/2}_{j,j}$ and $2C_{\max}^{-1}\le\lambda_{\min}(\bOmega)\le\lambda_{\max}(\bOmega)\le 2C_{\min}^{-1}$, if $i\in\mathcal{H}_{0}$ and $j\in\mathcal{S}_{i}^{\rm c}$, we have $|\Omega_{i,j}^{0}|\le|\Omega_{i,j}|\lambda_{\max}(\bOmega^{-1})<2^{-1}C_{\max}(\log d)^{-2-\vartheta}$. Then, given the event $\mathcal{G}_{n}$, we have $\max_{i\in\mathcal{H}_{0},\,j\in\mathcal{S}_{i}^{\rm c}}|\hat\bXi_{i,j}-\bI_2|_{\infty} \leq(1+2^{-1}C_{\max})(\log d)^{-2-\vartheta}$. Parallel to \eqref{eq:lem6-2}, if $\max_{i\in\mathcal{H}_{0},\,j\in\mathcal{S}_{i}^{\rm c}}|\hat\bXi_{i,j}-\bI_2|_{\infty} \leq(1+2^{-1}C_{\max})(\log d)^{-2-\vartheta}$, 
\begin{align}\label{eq:lem6-2-2}
\max_{i\in\mathcal{H}_{0},\,j\in\mathcal{S}_{i}^{\rm c}}\mathbb{P}\big(|\tilde{w}_{i}|>t, |\tilde{w}_{j}|>t\,|\,\bZ\big) 
\leq G^2(t)\big[1+O\{(\log d)^{-1-\vartheta}\}\big]
\end{align}
for any $t \in[0, (2\log d)^{1/2}]$. Due to $G\{G^{-1}_{l,n}(\lambda)\}=\lambda[1+O\{(\log d)^{-1-\vartheta}\}]$, and $\max_{i,j\in[d]}|\hat\bXi_{i,j,d}|_{\infty} \leq(\log d)^{-2-\vartheta}$ on $\mathcal{G}_{n}$, if $i\in\mathcal{H}_{0}$ and $j\in\mathcal{S}_{i}^{\rm c}$, by Lemma \ref{lem2}, \eqref{eq:lem6-2} and \eqref{eq:lem6-2-2}, 
\begin{align*}
{\rm I}_{i,j,1}(t)&\le\mathbb{E}\big[I(\mathcal{G}_{n})\mathbb{P}\{|\tilde{w}_{i+d}| \geq t ,|\tilde{w}_{j+d}| \geq t\,|\,\bZ\}\mathbb{P}\{|\tilde{w}_{i}| \geq G_{l,n}^{-1}(\lambda), |\tilde{w}_{j}| \geq G_{l,n}^{-1}(\lambda)\,|\,\bZ\}\big]\\
&~~~\times\big[1+O\{(\log d)^{-1-\vartheta}\}\big]+4d^{-2}\\
&\le\lambda^{2}G^{2}(t)\big[1+O\{(\log d)^{-1-\vartheta}\}\big]+4d^{-2}
\end{align*}
for any $0\le t\le (2\log d)^{1/2}$. If $i\in\mathcal{H}_{0}$ and $j\in\mathcal{S}_{i}\setminus\{i\}$, given the event $\mathcal{G}_{n}$, by \eqref{eq:lem6-2}, we have 
\begin{align*}
{\rm I}_{i,j,1}(t)\le
\mathbb{E}\big[I(\mathcal{G}_{n})\mathbb{P}\{|\tilde{w}_{i+d}| \geq t,|\tilde{w}_{j+d}| \geq t\,|\,\bZ\}\big]
\le G^{2}(t)\big[1+O\{(\log d)^{-1-\vartheta}\}\big]
\end{align*}
for any $0\le t\le (2\log d)^{1/2}$. Due to $\tilde{w}_{l}\,|\,\bZ \sim \mathcal{N}(0,1)$ for $l\in[2d]$, $G\{G^{-1}_{l,n}(\lambda)\}=\lambda[1+O\{(\log d)^{-1-\vartheta}\}]$, and $G(t)\le e^{-t^2/2}$ for $t>0$, by Lemma \ref{lem0_condition} with $\tau>2$, we have 
\begin{align*}
{\rm I}_{i,2}(t)
&=\mathbb{E}(I(\mathcal{G}_{n})[G(t)-G\{2(\log d)^{1/2}\}][G\{ G_{l,n}^{-1}(\lambda)\}-G\{2(\log d)^{1/2}\}])\\
&\ge\mathbb{P}(\mathcal{G}_{n})\{G(t)-d^{-2}\}(\lambda[1+O\{(\log d)^{-1-\vartheta}\}]-d^{-2})\\
&\ge\{1-\mathbb{P}(\mathcal{G}_{n}^{\rm c})\}(\lambda G(t)[1+O\{(\log d)^{-1-\vartheta}\}]-O(d^{-2}))\\
&\ge \lambda G(t)[1+O\{(\log d)^{-1-\vartheta}\}]-O(d^{-2})
\end{align*}
for any $0\le t\le (2\log d)^{1/2}$. Therefore, by Lemma \ref{lem0_condition} with $\tau>2$, for any $0\le t\le (2\log d)^{1/2}$, 
\begin{align*}
{\rm I}_{i,j}(t) \lesssim\left\{\begin{array}{cl}
G^{2}(t)(\log d)^{-1-\vartheta}+d^{-2}\,, & \text { if } i\in\mathcal{H}_0,\, j\in\mathcal{S}_{i}^{\rm c} \,;\\
G^{2}(t) +d^{-2}\,, & \text { if } i\in\mathcal{H}_0,\, j\in\mathcal{S}_{i}\setminus\{i\}\,.
\end{array}\right.
\end{align*}
By Lemma \ref{lem0_condition} with $\tau>2$, and Lemma \ref{lem2}, we can also show ${\rm II}_{i}(t)\lesssim G(t)+d^{-2}$ for any $0\le t\le (2\log d)^{1/2}$. Together with \eqref{eq:lem6-2-1}, if $\tau>2$, by Lemma \ref{lem6},
\begin{align*}
&\mathbb{E}\bigg[\bigg(\sum_{j \in \mathcal{H}_0}[I\{|\tilde{w}_{j+d}| \geq t, |\tilde{w}_{j}| \geq G_{l,n}^{-1}(\lambda)\}-\lambda G(t)]\bigg)^2\bigg]\\
&~~~~~\le\sum_{i \in \mathcal{H}_0}\sum_{j \in \mathcal{H}_0\setminus\{i\}}{\rm I}_{i,j}(t)
+\sum_{i \in \mathcal{H}_0}{\rm II}_{i}(t)\\
&~~~~~=\sum_{i \in \mathcal{H}_0}\sum_{j\in\mathcal{S}_{i}^{\rm c}}{\rm I}_{i,j}(t)
+\sum_{i \in \mathcal{H}_0}\sum_{j\in\mathcal{S}_{i}\setminus\{i\}}{\rm I}_{i,j}(t)
+\sum_{i \in \mathcal{H}_0}{\rm II}_{i}(t)\\
&~~~~~\lesssim G^{2}(t)\{d^{\rho+1}+d^2(\log d)^{-1-\vartheta}\}+dG(t)+1
\end{align*}
for any $0\le t\le (2\log d)^{1/2}$. Take $1/(1+\vartheta)<\delta<1$. Due to $1/2<\rho<1$, $\tau>2$, $b_d \rightarrow \infty$ and $b_d\ll d$, for any $\epsilon>0$, by the Chebyshev inequality,
\begin{align*}
&\sum_{i=0}^{m_d} \mathbb{P}\bigg[\bigg|\frac{\sum_{j \in \mathcal{H}_0} I\{|\tilde{w}_{j+d}| \geq u_i, |\tilde{w}_{j}| \geq G_{l,n}^{-1}(\lambda)\}}{d\lambda G(u_i)}-\frac{d_{0}}{d}\bigg|\geq \epsilon\bigg]\\
&~~~~~\lesssim \sum_{i=0}^{m_d}\frac{G^{2}(u_{i})\{d^{\rho+1}+d^2(\log d)^{-1-\vartheta}\}+dG(u_{i})+1}{\epsilon^{2}d^{2}\lambda^{2}G^{2}(u_{i})}\\
&~~~~~\lesssim \frac{1}{\epsilon^{2}}\bigg[m_{d}\{(\log d)^{-1-\vartheta}+d^{\rho-1}\}+d^{-1}\sum_{i=0}^{m_d}\frac{1}{G(u_{i})}
+d^{-2}\sum_{i=0}^{m_d}\frac{1}{G^{2}(u_{i})}\bigg]\\
&~~~~~\le \frac{1}{\epsilon^{2}}\bigg[m_{d}\{(\log d)^{-1-\vartheta}+d^{\rho-1}\}+d^{-1}\sum_{i=0}^{m_d}\frac{1}{G(u_{i})}
+d^{-2}\sum_{i=0}^{m_d}\frac{1}{G(u_{0})G(u_{i})}\bigg]\\
&~~~~~\lesssim \frac{1}{\epsilon^{2}}\bigg\{(\log d)^{-1-\vartheta+1/\delta}+b_{d}^{-1}+b_{d}^{-2/3}\sum_{i=1}^{m_d}e^{-i^{\delta}}\bigg\}=o(1)\,,
\end{align*}
which implies \eqref{lem8-1}. We complete the proof of Lemma \ref{lem1}. $\hfill\qedsymbol$

\section{Proofs of Lemmas \ref{lem:Berns_ineq}--\ref{lem6}}\label{proof_lemmas_condition}
\subsection{Proof of Lemma \ref{lem:Berns_ineq}}\label{proof_lem:Berns_ineq}
Due to $\max_{l\in[n]}\|\zeta_l\|_{\psi_1}\le\kappa'$, then $\mathbb{E}(|\zeta_l|^p)\le(\kappa'p)^p$ for any $l\in[n]$ and $p\ge1$. Since $\mathbb{E}(\zeta_l)=0$ and $p!\geq(p / e)^p$ for any $p\ge1$, we have
\begin{align*}
\mathbb{E}(e^{\lambda \zeta_l})&=\mathbb{E}\bigg\{1+\lambda \zeta_l+\sum_{p=2}^{\infty} \frac{(\lambda \zeta_l)^p}{p!}\bigg\}
\le\mathbb{E}\bigg\{1+\lambda \zeta_l+\sum_{p=2}^{\infty} \frac{(\lambda |\zeta_l|)^p}{p!}\bigg\}\\
&=1+\sum_{p=2}^{\infty} \frac{\lambda^p \mathbb{E}(|\zeta_l|^p)}{p!}
\leq 1+\sum_{p=2}^{\infty} \frac{(\lambda p\kappa')^p}{(p / e)^p}\\
&=1+\sum_{p=2}^{\infty}(e \lambda\kappa')^p=1+\frac{(e \lambda\kappa')^2}{1-e \lambda\kappa'}\le\exp\bigg\{\frac{(e\lambda\kappa')^2}{1-a}\bigg\}
\end{align*}
for any $|\lambda| \leq a/(e\kappa')$ with $a\in(0,1)$. Let $S_n=\sum_{l=1}^n \zeta_l$. By Markov's inequality, for any $t>0$,
\begin{align*}
\mathbb{P}(S_n \geq t) \leq e^{-\lambda t} \prod_{l=1}^n \mathbb{E} (e^{\lambda \zeta_l})\leq\exp\bigg\{\frac{(e\lambda\kappa')^2n}{1-a}-\lambda t\bigg\}=:\exp(-\lambda t+K\lambda^2)\,,
\end{align*}
where $K=(1-a)^{-1}(e\kappa')^2n$. Selecting
\begin{align*}
\lambda=\min \bigg(\frac{t}{2 K}, \frac{a}{e\kappa'}\bigg)\,,
\end{align*}
it holds that
\begin{align*}
\mathbb{P}(S_n \geq t) \leq \left\{
\begin{aligned}
\exp\bigg(-\frac{t^2}{4 K}\bigg)\,,~~~&\text{if}~~\frac{t}{2 K}<\frac{a}{e\kappa'}\,,\\
\exp\bigg(-\frac{at}{2e\kappa'}\bigg)\,,~~&\text{if}~~\frac{t}{2 K}\ge\frac{a}{e\kappa'}\,,
\end{aligned}
\right.
\end{align*}
which implies
\begin{align*}
\mathbb{P}(S_n \geq t) \leq \exp \bigg\{-\min \bigg(\frac{t^2}{4 K}, \frac{at}{2e\kappa'}\bigg)\bigg\}
=\exp \bigg[-\min \bigg\{\frac{(1-a)t^2}{4(e\kappa')^2n}, \frac{at}{2e\kappa'}\bigg\}\bigg]\,.
\end{align*}
Letting $a = 1/3$ and $t=nv$, we have
\begin{align*}
\mathbb{P}(S_n \geq nv) \leq \exp \bigg\{-\min \bigg(\frac{t^2}{4 K}, \frac{at}{2e\kappa'}\bigg)\bigg\}
=\exp \bigg[-\frac{n}{6}\min \bigg\{\bigg(\frac{v}{e\kappa'}\bigg)^2, \frac{v}{e\kappa'}\bigg\}\bigg]\,.
\end{align*}
Analogously, we can show the same bound also holds for $\mathbb{P}(S_n \le -nv)$. We complete the proof of Lemma \ref{lem:Berns_ineq}. $\hfill\qedsymbol$

\subsection{Proof of Lemma \ref{lem0_condition}}\label{proof_lem0_condition}

Recall $\bOmega=(\Omega_{i,j})_{2d\times2d}=\bT\bTheta_0 \bT^\top$ and $C_{\min}\le\sigma_{\min}(\bGamma)\le\sigma_{\max}(\bGamma)\le C_{\max}$. Then $2C_{\max}^{-1}\le\sigma_{\min}(\bOmega)\le\sigma_{\max}(\bOmega)\le 2C_{\min}^{-1}$. Write $\hat\bLambda=(\hat\Lambda_{i,j})_{2d\times2d}=\bT \hat\bTheta^\top \hat{\bGamma} \hat\bTheta \bT^\top$, $\hat\bLambda^{0}=(\hat\Lambda^{0}_{i,j})_{2d\times2d}$ with $\hat\Lambda^{0}_{i,j}=\hat\Lambda_{i,j}\hat\Lambda^{-1/2}_{i,i}\hat\Lambda^{-1/2}_{j,j}$ and $\bOmega^{0}=(\Omega^{0}_{i,j})_{2d\times2d}$ with $\Omega^{0}_{i,j}=\Omega_{i,j}\Omega^{-1/2}_{i,i}\Omega^{-1/2}_{j,j}$. Due to $\hat\Lambda_{i,j}^{2}\le\hat\Lambda_{i,i}\hat\Lambda_{j,j}$ for all $i,j\in[2d]$, then
\begin{align}\label{eq:lem10-1}
\mathbb{P}\{|\bOmega^{0}-\hat\bLambda^{0}|_{\infty}>(\log d)^{-2-\vartheta}\}
&\le \mathbb{P}\bigg\{\max_{i,j \in [2d]}\frac{1}{\Omega_{i,i}^{1/2}\Omega_{j,j}^{1/2}}|\hat\Lambda_{i,j}|\bigg|
1-\frac{\Omega_{i,i}^{1/2}\Omega_{j,j}^{1/2}}{\hat\Lambda_{i,i}^{1/2}\hat\Lambda_{j,j}^{1/2}}\bigg|>
\frac{1}{2}(\log d)^{-2-\vartheta}\bigg\}\notag\\
&~~~+\mathbb{P}\bigg\{\max _{i,j \in [2d]}\frac{|
\Omega_{i,j}-\hat\Lambda_{i,j}|}{\Omega_{i,i}^{1/2}\Omega_{j,j}^{1/2}}>
\frac{1}{2}(\log d)^{-2-\vartheta}\bigg\}\notag\\
&\le \underbrace{\mathbb{P}\bigg\{\max_{i,j \in [2d]}|\hat\Lambda_{i,i}^{1/2}\hat\Lambda_{j,j}^{1/2}-\Omega_{i,i}^{1/2}\Omega_{j,j}^{1/2}|>
\frac{(\log d)^{-2-\vartheta}}{C_{\max}}\bigg\}}_{\rm I}\\
&~~~+\mathbb{P}\bigg\{|\bOmega-\hat\bLambda|_{\infty}>
\frac{(\log d)^{-2-\vartheta}}{C_{\max}}\bigg\}\,.\notag
\end{align}
Since $|x/y-1|\leq|(x/y)^2-1|$ for all $x, y>0$, we have
\begin{align}\label{eq:lem10-2}
{\rm I}&\le \mathbb{P}\bigg\{\max_{i,j \in [2d]}|\hat\Lambda_{i,i}\hat\Lambda_{j,j}-\Omega_{i,i}\Omega_{j,j}|>
\frac{ 2C_{\min}}{C_{\max}^{3}}(\log d)^{-2-\vartheta}\bigg\}\notag\\
&\le  \mathbb{P}\bigg\{\max_{i,j \in [2d]}|\hat\Lambda_{j,j}|\cdot|\hat\Lambda_{i,i}-\Omega_{i,i}|>
\frac{ C_{\min}}{C_{\max}^{3}}(\log d)^{-2-\vartheta}\bigg\}\notag\\
&~~~+\mathbb{P}\bigg\{\max_{i,j \in [2d]}|\Omega_{i,i}|\cdot|\hat\Lambda_{j,j}-\Omega_{j,j}|>
\frac{ C_{\min}}{C_{\max}^{3}}(\log d)^{-2-\vartheta}\bigg\}\\
&\le \mathbb{P}\bigg\{\max_{i,j \in [2d]}|\hat\Lambda_{j,j}|\cdot|\hat\Lambda_{i,i}-\Omega_{i,i}|>
\frac{ C_{\min}}{C_{\max}^{3}}(\log d)^{-2-\vartheta}\bigg\}\notag\\
&~~~+\mathbb{P}\bigg\{|\bOmega-\hat\bLambda|_{\infty}>\frac{ C_{\min}^2}{2C_{\max}^{3}}(\log d)^{-2-\vartheta}\bigg\}\,.\notag
\end{align}
Due to $\varrho_{2}\asymp\{n^{-1}\log(2d)\}^{1/2}$ and $\log(2d) \ll n^{1/(5+2\vartheta)}$, we have $\varrho_{2}\ll \{\log(2d)\}^{-2-\vartheta}$. For any given $\tau>0$, by Lemma {\rm\ref{theo_random_trans}}, we have
\begin{align*}
&\mathbb{P}\bigg\{\max_{i,j \in [2d]}|\hat\Lambda_{j,j}|\cdot|\hat\Lambda_{i,i}-\Omega_{i,i}|>
\frac{ C_{\min}}{C_{\max}^{3}}(\log d)^{-2-\vartheta}\bigg\}
+\mathbb{P}\bigg\{|\bOmega-\hat\bLambda|_{\infty}>\frac{ C_{\min}^2}{2C_{\max}^{3}}(\log d)^{-2-\vartheta}\bigg\}\\
&~~~~~\le  \mathbb{P}\bigg\{\max_{i,j \in [2d]}|\hat\Lambda_{j,j}|\cdot|\hat\Lambda_{i,i}-\Omega_{i,i}|>
\frac{ C_{\min}}{C_{\max}^{3}}(\log d)^{-2-\vartheta},|\bOmega-\hat\bLambda|_{\infty}\le \frac{ C_{\min}^2}{2C_{\max}^{3}}(\log d)^{-2-\vartheta}\bigg\}\\
&~~~~~~~~+2\mathbb{P}\bigg\{|\bOmega-\hat\bLambda|_{\infty}>\frac{ C_{\min}^2}{2C_{\max}^{3}}(\log d)^{-2-\vartheta}\bigg\}\\
&~~~~~\le \mathbb{P}\bigg\{\max_{i,j \in [2d]}|\hat\Lambda_{i,i}-\Omega_{i,i}|>\frac{2C_{\min}^{2}(\log d)^{-2-\vartheta}}{4C_{\max}^{3}+C_{\min}^{3}(\log d)^{-2-\vartheta}}\bigg\}+2\mathbb{P}\bigg\{|\bOmega-\hat\bLambda|_{\infty}>\frac{ C_{\min}^{2}}{2C_{\max}^{3}}(\log d)^{-2-\vartheta}\bigg\}\\
&~~~~~\le  \mathbb{P}\bigg\{|\bOmega-\hat\bLambda|_{\infty}>\frac{2C_{\min}^{2}(\log d)^{-2-\vartheta}}{4C_{\max}^{3}+C_{\min}^{3}(\log d)^{-2-\vartheta}}\bigg\}
+2\mathbb{P}\bigg\{|\bOmega-\hat\bLambda|_{\infty}>\frac{ C_{\min}^{2}}{2C_{\max}^{3}}(\log d)^{-2-\vartheta}\bigg\}
\lesssim  d^{-\tau}\,.
\end{align*}
Together with \eqref{eq:lem10-1} and \eqref{eq:lem10-2}, we have Lemma \ref{lem0_condition}. $\hfill\qedsymbol$

\subsection{Proof of Lemma \ref{lem2}}\label{proof_lem2}

Recall $\bOmega=(\Omega_{i,j})_{2d\times2d}=\bT\bTheta_0 \bT^\top$ and $\bOmega^{0}=(\Omega^{0}_{i,j})_{2d\times2d}$ with $\Omega^{0}_{i,j}=\Omega_{i,j}\Omega^{-1/2}_{i,i}\Omega^{-1/2}_{j,j}$. Due to $2C_{\max}^{-1}\le\lambda_{\min}(\bOmega)\le\lambda_{\max}(\bOmega)\le 2C_{\min}^{-1}$, then $C_{\min}C_{\max}^{-1}\le\lambda_{\min}(\bOmega^{0})\le\lambda_{\max}(\bOmega^{0})\le C_{\max}C_{\min}^{-1}$. For any distinct $i,j\in[d]$, write 
\begin{align*}
  \bXi_{i+d,j+d}^{(i,j)}:=
  \begin{pmatrix}
  \bXi_{i,j}&\bXi_{i,j,d}\\
  \bXi_{i,j,d}^{\top}&\bXi_{i+d,j+d}
  \end{pmatrix}\,,
\end{align*}
where
\begin{align*}
  \bXi_{i,j}=
  \begin{pmatrix}
  \Omega^{0}_{i,i}&\Omega^{0}_{i,j}\\
  \Omega^{0}_{i,j}&\Omega^{0}_{j,j}
  \end{pmatrix}\,,
  ~~\bXi_{i+d,j+d}=
  \begin{pmatrix}
  \Omega^{0}_{i+d,i+d}&\Omega^{0}_{i+d,j+d}\\
  \Omega^{0}_{i+d,j+d}&\Omega^{0}_{j+d,j+d}
  \end{pmatrix}\,, 
  ~~\bXi_{i,j,d}=
 \begin{pmatrix}
  \Omega^{0}_{i,i+d}&\Omega^{0}_{i,j+d}\\
  \Omega^{0}_{j,i+d}&\Omega^{0}_{j,j+d}
  \end{pmatrix}\,.
\end{align*}
Due to $\Omega^{0}_{i,j+d}=0$ for $i, j\in[d]$ and $\Omega^{0}_{l,l}=1$ for $l\in[2d]$, then $\bXi_{i,j,d}={\bf 0}$ and $|\bXi_{i+d,j+d}|_{\infty}\le1$. Recall $\tilde{\bw}_{n}=(\tilde{w}_{1},\ldots,\tilde{w}_{2d})^{\top}$ with $\tilde{\bw}_{n}\,|\,\bZ\sim\mathcal{N}({\bf 0},\hat\bLambda^{0})$, $\hat\bXi_{i,j}={\rm Var}(\tilde{\bw}_{i,j}\,|\,\bZ)$ with $\tilde{\bw}_{i,j}=(\tilde{w}_{i},\tilde{w}_{j})^{\top}$, $\hat\bXi_{i,j,d}={\rm Cov}(\tilde{\bw}_{i,j},\tilde{\bw}_{i+d,j+d}\,|\,\bZ)$, and
\begin{align*}
  \hat\bXi_{i+d,j+d}^{(i,j)}:=
  \begin{pmatrix}
  \hat\bXi_{i,j}&\hat\bXi_{i,j,d}\\
  \hat\bXi_{i,j,d}^{\top}&\hat\bXi_{i+d,j+d}
  \end{pmatrix}\,.
\end{align*}
Then $|\bXi_{i+d,j+d}^{(i,j)}-\hat\bXi_{i+d,j+d}^{(i,j)}|_{\infty}
\le|\bOmega^{0}-\hat\bLambda^{0}|_{\infty}\le(\log d)^{-2-\vartheta}$. If $\log d\ge (8C_{\max}C_{\min}^{-1})^{1/(2+\vartheta)}$, we have
\begin{align*}
\|\bXi_{i+d,j+d}^{(i,j)}-\hat\bXi_{i+d,j+d}^{(i,j)}\|_{2}
\le4|\bXi_{i+d,j+d}^{(i,j)}-\hat\bXi_{i+d,j+d}^{(i,j)}|_{\infty}\le\frac{C_{\min}}{2C_{\max}}\,.
\end{align*}
Due to $\lambda_{\min}(\bOmega^{0})\le\lambda_{\min}\{\bXi_{i+d,j+d}^{(i,j)}\}\le\lambda_{\max}\{\bXi_{i+d,j+d}^{(i,j)}\}\le\lambda_{\max}(\bOmega^{0})$, then
\begin{align}\label{eq:lem11-1}
\frac{C_{\min}}{2C_{\max}}\le\lambda_{\min}\{\hat\bXi_{i+d,j+d}^{(i,j)}\}\le\lambda_{\max}\{\hat\bXi_{i+d,j+d}^{(i,j)}\}\le\frac{3C_{\max}}{2C_{\min}}\,.
\end{align}
Therefore,
\begin{align*}
  \{\hat\bXi_{i+d,j+d}^{(i,j)}\}^{-1}&=
  \begin{pmatrix}
  \hat\bXi_{i,j}^{-1}+\tilde{\bXi}_{1} &  \tilde{\bXi}_{2}\\
  \tilde{\bXi}_{2}^{\top} &  \tilde{\bXi}_{3}
  \end{pmatrix}\,,
\end{align*}
where $\tilde{\bXi}_{1}=\hat\bXi_{i,j}^{-1}\hat\bXi_{i,j,d}\tilde{\bXi}_{3}\hat\bXi_{i,j,d}^{\top}\hat\bXi_{i,j}^{-1}$, $\tilde{\bXi}_{2}=-\hat\bXi_{i,j}^{-1}\hat\bXi_{i,j,d}\tilde{\bXi}_{3}$ and $\tilde{\bXi}_{3}=(\hat\bXi_{i+d,j+d}-\hat\bXi_{i,j,d}^{\top}\hat\bXi_{i,j}^{-1}\hat\bXi_{i,j,d})^{-1}$. Recall $\tilde{\bw}_{i,j,d}=(\tilde{\bw}_{i,j}^{\top},\tilde{\bw}_{i+d,j+d}^{\top})^{\top}$ with $\tilde{\bw}_{i,j}=(\tilde{w}_{i},\tilde{w}_{j})^{\top}$ and $\tilde{\bw}_{i+d,j+d}=(\tilde{w}_{i+d},\tilde{w}_{j+d})^{\top}$. Let $\bt=(\bt_{1}^{\top},\bt_{2}^{\top})^{\top}$ with $\bt_{1}=(t_{1},t_{2})^{\top}\in\mathbb{R}^{2}$ and $\bt_{2}=(t_{3},t_{4})^{\top}\in\mathbb{R}^{2}$, and $f_{\tilde{\bw}_{i,j,d}\,|\,\bZ}(\cdot)$, $f_{\tilde{\bw}_{i,j}\,|\,\bZ}(\cdot)$ and $f_{\tilde{\bw}_{i+d,j+d}\,|\,\bZ}(\cdot)$ be the conditional density functions of $\tilde{\bw}_{i,j,d}$, $\tilde{\bw}_{i,j}$ and $\tilde{\bw}_{i+d,j+d}$ given $\bZ$, respectively. Let $|\bA|:=\det(\bA)$ denote the determinant of square matrix $\bA$. Since $\tilde{\bw}_{n}\,|\,\bZ\sim\mathcal{N}({\bf 0},\hat\bLambda^{0})$ and $|\hat\bXi_{i+d,j+d}^{(i,j)}|=|\hat\bXi_{i,j}||\hat\bXi_{i+d,j+d}-\hat\bXi_{i,j,d}^{\top}\hat\bXi_{i,j}^{-1}\hat\bXi_{i,j,d}|$, then
\begin{align}\label{eq:lem11-2}
  f_{\tilde{\bw}_{i,j,d}\,|\,\bZ}(\bt)
  &=\frac{1}{(2\pi)^{2}|\hat\bXi_{i+d,j+d}^{(i,j)}|^{1/2}}\exp\big[-2^{-1}\bt^{\top}\{\hat\bXi_{i+d,j+d}^{(i,j)}\}^{-1} \bt\big]\notag\\
  &=\frac{\exp\{-2^{-1}(\bt_{1}^{\top}\hat\bXi_{i,j}^{-1}\bt_{1}
  +\bt_{1}^{\top}\tilde\bXi_{1}\bt_{1}+2\bt_{1}^{\top}\tilde{\bXi}_{2}\bt_{2}
  +\bt_{2}^{\top}\tilde{\bXi}_{3}\bt_{2})\} }{(2\pi)^{2}|\hat\bXi_{i,j}|^{1/2}|\hat\bXi_{i+d,j+d}-\hat\bXi_{i,j,d}^{\top}\hat\bXi_{i,j}^{-1}\hat\bXi_{i,j,d}|^{1/2}}\\ 
  &=f_{\tilde{\bw}_{i,j}\,|\,\bZ}(\bt_{1})f_{\tilde{\bw}_{i+d,j+d}\,|\,\bZ}(\bt_{2})\bigg(\underbrace{\frac{|\hat\bXi_{i+d,j+d}|}{|\hat\bXi_{i+d,j+d}-\hat\bXi_{i,j,d}^{\top}\hat\bXi_{i,j}^{-1}\hat\bXi_{i,j,d}|}}_{\rm I}\bigg)^{1/2}\notag\\
    &~~~\times\exp\{\underbrace{-\bt_{1}^{\top}\tilde{\bXi}_{2}\bt_{2}
    -2^{-1}\bt_{1}^{\top}\tilde\bXi_{1}\bt_{1}-2^{-1}\bt_{2}^{\top}(\tilde{\bXi}_{3}-\hat\bXi_{i+d,j+d}^{-1})\bt_{2}}_{{\rm II}(\bt_1,\bt_2)}\}\notag\,.
\end{align}
Recall $\bXi_{i,j,d}={\bf 0}$, $|\bXi_{i+d,j+d}|_{\infty}\le1$ and $|\bXi_{i+d,j+d}^{(i,j)}-\hat\bXi_{i+d,j+d}^{(i,j)}|_{\infty}\le(\log d)^{-2-\vartheta}$. Then
$|\hat\bXi_{i,j,d}|_{\infty}\le(\log d)^{-2-\vartheta}$ and $|\hat\bXi_{i+d,j+d}|_{\infty}\le1+(\log d)^{-2-\vartheta}$. By \eqref{eq:lem11-1}, we have $|\hat\bXi_{i,j}^{-1}|_{\infty}\le2C_{\max}C_{\min}^{-1}$ and $|\hat\bXi_{i+d,j+d}^{-1}|_{\infty}\le2C_{\max}C_{\min}^{-1}$. Since $\bXi_{i+d,j+d}=\bI_{2}$ and the diagonal elements of $\hat\bXi_{i+d,j+d}$ are equal to $1$, then $|\hat\bXi_{i+d,j+d}|= 1+O\{(\log d)^{-2-\vartheta}\}$ and $|\hat\bXi_{i+d,j+d}-\hat\bXi_{i,j,d}^{\top}\hat\bXi_{i,j}^{-1}\hat\bXi_{i,j,d}|=1+O\{(\log d)^{-2-\vartheta}\}$, which implies
\begin{align}\label{eq:lem11-1-1}
{\rm I}=1+O\{(\log d)^{-2-\vartheta}\}\,.
\end{align}
Notice that $\|\bA\|_{\infty}\le 2|\bA|_{\infty}$ and $\|\bA\bE\|_{\infty}\le\|\bA\|_{\infty}\|\bE\|_{\infty}$ for any $\bA,\bE\in\mathbb{R}^{2\times 2}$. Due to $|\hat\bXi_{i,j}^{-1}|_{\infty}\le2C_{\max}C_{\min}^{-1}$, $|\hat\bXi_{i+d,j+d}^{-1}|_{\infty}\le2C_{\max}C_{\min}^{-1}$ and $|\hat\bXi_{i,j,d}|_{\infty}\le(\log d)^{-2-\vartheta}$, then 
\begin{align*}
|\hat\bXi_{i+d,j+d}^{-1}\hat\bXi_{i,j,d}^{\top}\hat\bXi_{i,j}^{-1}\hat\bXi_{i,j,d}|_{\infty}
&\le\|\hat\bXi_{i+d,j+d}^{-1}\hat\bXi_{i,j,d}^{\top}\hat\bXi_{i,j}^{-1}\hat\bXi_{i,j,d}\|_{\infty}\\
&\le\|\hat\bXi_{i+d,j+d}^{-1}\|_{\infty}\|\hat\bXi_{i,j,d}^{\top}\|_{\infty}
\|\hat\bXi_{i,j}^{-1}\|_{\infty}\|\hat\bXi_{i,j,d}\|_{\infty}\\
&\le16|\hat\bXi_{i+d,j+d}^{-1}|_{\infty}|\hat\bXi_{i,j,d}^{\top}|_{\infty}
|\hat\bXi_{i,j}^{-1}|_{\infty}|\hat\bXi_{i,j,d}|_{\infty}=O\{(\log d)^{-4-2\vartheta}\}\,,
\end{align*}
which implies $( \bI_{2}
-\hat\bXi_{i+d,j+d}^{-1}\hat\bXi_{i,j,d}^{\top}\hat\bXi_{i,j}^{-1}\hat\bXi_{i,j,d})^{-1}- \bI_{2}
=\bDelta$ with $|\bDelta|_{\infty}=O\{(\log d)^{-4-2\vartheta}\}$. Due to $|\hat\bXi_{i+d,j+d}^{-1}|_{\infty}\le2C_{\max}C_{\min}^{-1}$, we have
\begin{align}\label{eq:lem11-1-2}
\tilde{\bXi}_{3}
=\hat\bXi_{i+d,j+d}^{-1}+\{( \bI_{2}
-\hat\bXi_{i+d,j+d}^{-1}\hat\bXi_{i,j,d}^{\top}\hat\bXi_{i,j}^{-1}\hat\bXi_{i,j,d})^{-1}- \bI_{2}\}\hat\bXi_{i+d,j+d}^{-1}
=\hat\bXi_{i+d,j+d}^{-1}+\tilde{\bDelta}\,,
\end{align}
where $|\tilde{\bDelta}|_{\infty}=O\{(\log d)^{-4-2\vartheta}\}$. Recall $\tilde{\bXi}_{1}=\hat\bXi_{i,j}^{-1}\hat\bXi_{i,j,d}\tilde{\bXi}_{3}\hat\bXi_{i,j,d}^{\top}\hat\bXi_{i,j}^{-1}$ and $\tilde{\bXi}_{2}=-\hat\bXi_{i,j}^{-1}\hat\bXi_{i,j,d}\tilde{\bXi}_{3}$. Since $|\hat\bXi_{i,j}^{-1}|_{\infty}\le2C_{\max}C_{\min}^{-1}$, $|\hat\bXi_{i+d,j+d}^{-1}|_{\infty}\le2C_{\max}C_{\min}^{-1}$ and $|\hat\bXi_{i,j,d}|_{\infty}\le(\log d)^{-2-\vartheta}$, by \eqref{eq:lem11-1-2}, then $|\tilde{\bXi}_{1}|_{\infty}=O\{(\log d)^{-4-2\vartheta}\}$ and $|\tilde{\bXi}_{2}|_{\infty}=O\{(\log d)^{-2-\vartheta}\}$. Therefore, we have ${\rm II}(\bt_1,\bt_2)=O\{(\log d)^{-1-\vartheta}\}$ for any $|\bt|_{\infty}\le 2(\log d)^{1/2}$. Notice that $e^{O\{(\log d)^{-1-\vartheta}\}}=1+O\{(\log d)^{-1-\vartheta}\}$ and $[1+O\{(\log d)^{-2-\vartheta}\}]^{1/2}=1+O\{(\log d)^{-2-\vartheta}\}$. Therefore, by \eqref{eq:lem11-2} and \eqref{eq:lem11-1-1}, 
\begin{align}\label{eq:lem11-2-1}
f_{\tilde{\bw}_{i,j,d}\,|\,\bZ}(\bt)=f_{\tilde{\bw}_{i,j}\,|\,\bZ}(\bt_{1})f_{\tilde{\bw}_{i+d,j+d}\,|\,\bZ}(\bt_{2})\big[1+O\{(\log d)^{-1-\vartheta}\}\big]
\end{align}
for any $|\bt|_{\infty}\le 2(\log d)^{1/2}$, where the term $O\{(\log d)^{-1-\vartheta}\}$ holds uniformly over $|\bt|_{\infty}\le 2(\log d)^{1/2}$ and $i,j\in[d]$. Let $\bt_{3}=(t_{5},t_{6})^{\top}$, and  $f_{\tilde{w}_{i}\,|\,\bZ}(\cdot)$ be the conditional density function of $\tilde{w}_{i}$ given $\bZ$. Analogously, we have
\begin{align}\label{eq:lem11-3-1}
f_{\tilde{\bw}_{i,i+d}\,|\,\bZ}(\bt_{3})=f_{\tilde{w}_{i}\,|\,\bZ}(t_{5})f_{\tilde{w}_{i+d}\,|\,\bZ}(t_{6})\big[1+O\{(\log d)^{-1-\vartheta}\}\big]
\end{align}
for any $|\bt_{3}|_{\infty}\le 2(\log d)^{1/2}$, where the term $O\{(\log d)^{-1-\vartheta}\}$ holds uniformly over $|\bt_{3}|_{\infty}\le 2(\log d)^{1/2}$ and $i\in[d]$. Let $\mathcal{U}_{1}=\{t:\,\nu_{1}\le|t|\le2(\log d)^{1/2}\}$ and $\mathcal{U}_{2}=\{t:\,\nu_{2}\le|t|\le2(\log d)^{1/2}\}$ for $0\le \nu_{1},\nu_{2}\le (2\log d)^{1/2}$. Then
\begin{align}\label{eq:lem11-2-2}
&\mathbb{P}(|\tilde{w}_{i}|\ge \nu_{1},|\tilde{w}_{j}|\ge \nu_{1},
|\tilde{w}_{i+d}|\ge \nu_{2},|\tilde{w}_{j+d}|\ge \nu_{2}\,|\,\bZ)\notag\\
&~~~~~\le\mathbb{P}(\tilde{w}_{i}\in \mathcal{U}_{1},\tilde{w}_{j}\in \mathcal{U}_{1},
\tilde{w}_{i+d}\in \mathcal{U}_{2},\tilde{w}_{j+d}\in \mathcal{U}_{2}\,|\,\bZ)\\
&~~~~~~~~+\mathbb{P}\{|\tilde{w}_{i}|>2(\log d)^{1/2}\,|\,\bZ\}
+\mathbb{P}\{|\tilde{w}_{j}|>2(\log d)^{1/2}\,|\,\bZ\}\notag\\
&~~~~~~~~+\mathbb{P}\{|\tilde{w}_{i+d}|>2(\log d)^{1/2}\,|\,\bZ\}
+\mathbb{P}\{|\tilde{w}_{j+d}|>2(\log d)^{1/2}\,|\,\bZ\}\notag\,.
\end{align}
For any $0\le \nu_{1},\nu_{2}\le (2\log d)^{1/2}$, by \eqref{eq:lem11-2-1}, it holds that
\begin{align}\label{eq:lem11-2-3}
&\mathbb{P}(\tilde{w}_{i}\in \mathcal{U}_{1},\tilde{w}_{j}\in \mathcal{U}_{1},
\tilde{w}_{i+d}\in \mathcal{U}_{2},\tilde{w}_{j+d}\in \mathcal{U}_{2}\,|\,\bZ)\notag\\
&~~~~~=\int_{\mathcal{U}_{1}}\int_{\mathcal{U}_{1}}\int_{\mathcal{U}_{2}}\int_{\mathcal{U}_{2}}
f_{\tilde{\bw}_{i,j,d}\,|\,\bZ}(\bt)\,{\rm d}t_{4}\,
{\rm d}t_{3}\,{\rm d}t_{2}\,{\rm d}t_{1}\notag\\
&~~~~~=\int_{\mathcal{U}_{1}}\int_{\mathcal{U}_{1}}\int_{\mathcal{U}_{2}}\int_{\mathcal{U}_{2}}
f_{\tilde{\bw}_{i,j}\,|\,\bZ}(\bt_{1})f_{\tilde{\bw}_{i+d,j+d}\,|\,\bZ}(\bt_{2})\big[1+O\{(\log d)^{-1-\vartheta}\}\big]\,{\rm d}t_{4}\,
{\rm d}t_{3}\,{\rm d}t_{2}\,{\rm d}t_{1}\\
&~~~~~\le\mathbb{P}(\tilde{w}_{i}\in \mathcal{U}_{1},\tilde{w}_{j}\in \mathcal{U}_{1}\,|\,\bZ)
\mathbb{P}(\tilde{w}_{i+d}\in \mathcal{U}_{2},\tilde{w}_{j+d}\in \mathcal{U}_{2}\,|\,\bZ)
\big[1+O\{(\log d)^{-1-\vartheta}\}\big]\notag\\
&~~~~~\le\mathbb{P}(|\tilde{w}_{i}|\ge \nu_{1},|\tilde{w}_{j}|\ge \nu_{1}\,|\,\bZ)
\mathbb{P}(|\tilde{w}_{i+d}|\ge \nu_{2},|\tilde{w}_{j+d}|\ge \nu_{2}\,|\,\bZ)
\big[1+O\{(\log d)^{-1-\vartheta}\}\big]\notag\,.
\end{align}
Analogously, by \eqref{eq:lem11-3-1}, for any $0\le \nu_{1},\nu_{2}\le (2\log d)^{1/2}$, we have
\begin{align*}
&\mathbb{P}\{\nu_{1}\le|\tilde{w}_{i}|\le 2(\log d)^{1/2},\nu_{2}\le|\tilde{w}_{i+d}|\le 2(\log d)^{1/2}\,|\,\bZ\}\\
&~~~~~\ge\mathbb{P}\{\nu_{1}\le|\tilde{w}_{i}|\le 2(\log d)^{1/2}\,|\,\bZ\}
\mathbb{P}\{\nu_{2}\le|\tilde{w}_{i+d}|\le 2(\log d)^{1/2}\,|\,\bZ\}
\big[1-O\{(\log d)^{-1-\vartheta}\}\big]\,.
\end{align*}
Due to $\tilde{w}_{l}\,|\,\bZ\sim\mathcal{N}(0,1)$ for $l\in[2d]$ and $G(x)\le e^{-x^2/2}$ for any $x>0$, then
\begin{align*}
\mathbb{P}\{|\tilde{w}_{i}|>2(\log d)^{1/2}\,|\,\bZ\}
=G\{2(\log d)^{1/2}\}\le d^{-2}\,.
\end{align*}
Analogously, same bound also holds for $\mathbb{P}\{|\tilde{w}_{j}|>2(\log d)^{1/2}\,|\,\bZ\}$, $\mathbb{P}\{|\tilde{w}_{i+d}|>2(\log d)^{1/2}\,|\,\bZ\}$ and $\mathbb{P}\{|\tilde{w}_{j+d}|>2(\log d)^{1/2}\,|\,\bZ\}$. Together with \eqref{eq:lem11-2-2} and \eqref{eq:lem11-2-3}, we have
\begin{align*}
&\mathbb{P}(|\tilde{w}_{i}|\ge \nu_{1},|\tilde{w}_{j}|\ge \nu_{1},
|\tilde{w}_{i+d}|\ge \nu_{2},|\tilde{w}_{j+d}|\ge \nu_{2}\,|\,\bZ)\\
&~~~~~\le\mathbb{P}(|\tilde{w}_{i}|\ge \nu_{1},|\tilde{w}_{j}|\ge \nu_{1}\,|\,\bZ)
\mathbb{P}(|\tilde{w}_{i+d}|\ge \nu_{2},|\tilde{w}_{j+d}|\ge \nu_{2}\,|\,\bZ)
\big[1+O\{(\log d)^{-1-\vartheta}\}\big]+4d^{-2}\,.
\end{align*}
Analogously, by \eqref{eq:lem11-3-1}, for any $0\le \nu_{1},\nu_{2}\le (2\log d)^{1/2}$, we have
\begin{align*}
\mathbb{P}(|\tilde{w}_{i}|\ge \nu_{1},|\tilde{w}_{i+d}|\ge \nu_{2}\,|\,\bZ)
\le\mathbb{P}(|\tilde{w}_{i}|\ge \nu_{1}\,|\,\bZ)
\mathbb{P}(|\tilde{w}_{i+d}|\ge \nu_{2}\,|\,\bZ)
\big[1+O\{(\log d)^{-1-\vartheta}\}\big]+2d^{-2}\,.
\end{align*}
We complete the proof of Lemma \ref{lem2}. $\hfill\qedsymbol$

\subsection{Proof of Lemma \ref{lem6}}\label{proof_lem6}
Recall $\bD={\rm diag}(\nu_{1},\ldots,\nu_{d})$ and 
\begin{align*}
\bGamma=\begin{pmatrix}\bSigma & \bSigma - \bD\\
\bSigma - \bD & \bSigma\end{pmatrix}\,.
\end{align*}
Write $\bA:=(a_{k,l})_{d\times d}=(2\bD-\bD\bSigma^{-1}\bD)^{-1}$, $\bB:=(b_{k,l})_{d\times d}= -(2\bD-\bD\bSigma^{-1}\bD)^{-1}(\bI_{d}-\bD\bSigma^{-1})$ and $\tilde{\bT}=2^{-1/2}\bT$. Recall $\bTheta_0=\bGamma^{-1}$. Then
  \begin{align}\label{eq:thm3-lem6-1}
    \bTheta_0:=(\theta_{i,j}^0)_{2d\times 2d}=\tilde{\bT}^{\top}
    \begin{pmatrix}(2\bSigma-\bD)^{-1}&{\bf0}\\ {\bf0}&\bD^{-1}\end{pmatrix}\tilde{\bT}
    =\begin{pmatrix}\bA&\bB\\ \bB&\bA\end{pmatrix}\,,
  \end{align}
which implies
  \begin{align}\label{eq:thm3-lem6-2}
    \bOmega=(\Omega_{i,j})_{2d\times2d}=\bT^{\top}\bTheta_0 \bT=
    \begin{pmatrix}2\bA+2\bB&{\bf0}\\{\bf0}&2\bA-2\bB\end{pmatrix}.
  \end{align}
Recall
\begin{equation*}
	\begin{aligned}
		\mathcal U(M,q,s_d) = \bigg\{ \bTheta=(\theta_{i,j})_{2d\times2d}:\bTheta \succ {\bf0}, \|\bTheta\|_1 \le M, \max_{i\in[2d]} \sum_{j=1}^{2d} \lvert \theta_{i,j}\rvert^q \le s_d \bigg\}
	\end{aligned}
\end{equation*}
for $0\le q < 1$. By \eqref{eq:thm3-lem6-2}, we have $|\Omega_{k,l}|^{q}\le2^{q}(|a_{k,l}|+|b_{k,l}|)^{q}\le 2(|a_{k,l}|^{q}+|b_{k,l}|^{q})$. Since $\bTheta_0\in \mathcal{U}(M,q,s_d)$, by \eqref{eq:thm3-lem6-1}, it holds that
\begin{align}\label{eq:thm3-lem6-3}
\max_{k\in[d]} \sum_{l=1}^{d} \lvert \Omega_{k,l}\rvert^q\le2\max_{k\in[d]} \sum_{l=1}^{d} (\lvert a_{k,l}\rvert^q+\lvert b_{k,l}\rvert^q)=2\max_{i\in[d]} \sum_{j=1}^{2d} \lvert \theta_{i,j}^0\rvert^q \le 2s_d\,.
\end{align}
Recall $\vartheta>0$, $1/2<\rho<1$ and $\mathcal{S}_{i}=\{j \in \mathcal{H}_0: |\Omega_{i,j}| \geq(\log d)^{-2-\vartheta}\}$. When $d>\max\{4s_d^{2},(\log d)^{(4+2\vartheta)/(2\rho-1)}\}$, if there exists some $k_{1}\in\mathcal{H}_0$ such that $|\mathcal{S}_{k_1}|> d^{\rho}$, then
\begin{align*}
\max_{k\in[d]} \sum_{l=1}^{d} \lvert \Omega_{k,l}\rvert^q
\ge\sum_{l\in\mathcal{H}_0} \lvert \Omega_{k_{1},l}\rvert^q
> d^{\rho}(\log d)^{-(2+\vartheta)q}\ge d^{\rho}(\log d)^{-2-\vartheta} > d^{1/2}> 2s_d\,,
\end{align*}
which contradicts \eqref{eq:thm3-lem6-3}. Then, if $d>\max\{4s_d^{2},(\log d)^{(4+2\vartheta)/(2\rho-1)}\}$, it holds that $|\mathcal{S}_{i}|\le d^{\rho}$ for all $i\in \mathcal{H}_0$. We complete the proof of Lemma \ref{lem6}. $\hfill\qedsymbol$

\section{Extension of theoretical results to estimated and misspecified feature distribution}\label{F:F_0}

\subsection{Second-order knockoffs using estimated moments}\label{F:F_1}

Assume $\bx$ is a $d$-dimensional random vector with $\mathbb{E}(\bx)=\bzero$ and ${\rm Var}(\bx)=\bSigma$. Let $\{\bx_i\}_{i=1}^{n}$ be the $n$ independent observations of $\bx$, where $\bx_i = (x_{i,1}, \ldots, x_{i,d})^{\top} \in \mathbb{R}^d$ for $i\in[n]$. Consider the linear model
\begin{align*}
\by = \bX\bbeta + \beps\,,
\end{align*}
where \( \bX = (\bx_1, \dots, \bx_n)^{\top} \in \mathbb{R}^{n\times d}\) is the design matrix, \( \by= (y_1,\ldots,y_{n})^\top \in \mathbb{R}^{n} \) is the response vector, \( \beps = (\varepsilon_1,\ldots,\varepsilon_{n})^\top \in \mathbb{R}^{n} \) is the error term with $\mathbb{E}(\varepsilon_i)=0$ and ${\rm Var}(\varepsilon_i)=\sigma^2$, and \( \bbeta \in \mathbb{R}^d \) is the parameter vector with the true value \(\bbeta_{0} :=(\beta^{0}_{1},\ldots,\beta^{0}_{d})^{\top}\). The core idea of model-X knockoffs is to construct the knockoff variables $\tilde\bx\in\mathbb{R}^d$ that satisfy two conditions: (i) exchangeable in distribution with the original covariates $\bx\in\mathbb{R}^d$, and (ii) independent of response conditional on the original covariates. The Gaussian knockoffs generator pretends that $\bx$ followed Gaussian distribution $\mathcal{N}(\bzo,\bSigma)$, and generates knockoff variables as
\begin{align}\label{Seq:gau_ture}
    \tilde\bx = (\bI_{d}-\bD\bSigma^{-1})\bx + (2\bD-\bD\bSigma^{-1}\bD)^{1/2}\bxi\,,
\end{align}
where $\bxi\sim\mathcal{N}(\bzo,\bI_{d})$ is independent of covariates $\bx$ and error term $\beps$, and $\bD = \text{diag}(\bs)$ with $\bs\in\mathbb{R}^d$ being a hyperparameter chosen to ensure that the covariance matrix $\bGamma := \text{Var}\{(\bx^{\top}, \tilde{\bx}^{\top})^{\top}\} $ is positive semidefinite. Notice that
\begin{align} \label{Seq:kcov}
      \bGamma = 
    \begin{pmatrix}
        \bSigma & \bSigma - \bD \\
        \bSigma - \bD & \bSigma
    \end{pmatrix}\,.
\end{align}

The Gaussian knockoffs generator is constructed such that its first two moments match those of $\bx$. In the scenarios considered in \citeS{candes2018panning_app}, the second moment of $\bx$ is assumed to be known. However, in numerous practical scenarios, the second moment of $\bx$ remains unknown. This necessitates constructing knockoffs using the estimated second moment of $\bx$. The precision matrix $\bOmega_0:=\bSigma^{-1}$ can be estimated either via unlabeled covariate data \citepS{Barber2018Robust_app} or via data-splitting on the design matrix \citepS{Fan02012020_app}. In practice, we can use $\{(\bx_i,y_i)\}_{i=1}^{n_1}$ to estimate the parameter vector $\bbeta$ and use $\{\bx_i\}_{i=1+n_1}^{n}$ to estimate the precision matrix $\bOmega_0$. 

To estimate $\bbeta$ based on $\{(\bx_i,y_i)\}_{i=1}^{n_1}$, in order to economize notation and without causing much confusion, we still use $\bX$, \( \by\) and \(\beps\) to denote, respectively, the design matrix, the response vector, and the error term, i.e.,
\begin{align}\label{Seq:lm_1}
\by = \bX\bbeta + \beps\,,
\end{align}
where $\bX = (\bx_1, \dots, \bx_{n_1})^{\top}\in\mathbb{R}^{n_1\times d}$, \( \by= (y_1,\ldots,y_{n_1})^\top \in \mathbb{R}^{n_1} \), and \( \beps = (\varepsilon_1,\ldots,\varepsilon_{n_1})^\top \in \mathbb{R}^{n_1} \).
We adopt the CLIME approach proposed by \citeS{cai2011constrained_app} to estimate the precision matrix $\bOmega_0$. Let $\{\hat{\bOmega}_1\}$ be the solution set of the following optimization problem:
\begin{align}\label{Seq:lm_111}
    \min |\bOmega|_1 \quad \text{subject to} \quad |\hat{\bSigma}\bOmega - \bI_{d}|_{\infty} \leq \varrho_4\,,
\end{align}
where $\hat{\bSigma}=n_2^{-1}\sum_{k=1}^{n_2} (\bx_{n_1+k}-\bar{\bx})(\bx_{n_1+k}-\bar{\bx})^\top$ with $\bar{\bx}=n_2^{-1}\sum_{k=1}^{n_2}\bx_{n_1+k}$, and $\varrho_4 > 0$ is a tuning parameter. 
Using Equation (2) in \citeS{cai2011constrained_app}, we symmetrize $\hat{\bOmega}_1$ to get $\hat{\bOmega}$, which we employ as $\bOmega_0$’s estimator.  
\citeS{candes2018panning_app} provide several construction methods for $\bD$ that render \eqref{Seq:kcov} positive definite. Here, we consider $\bD=\tilde{C}\lambda_{\max}^{-1}(\bOmega_0)\bI_{d}$ along with its estimator $\hat{\bD}=\tilde{C}\lambda_{\max}^{-1}(\hat{\bOmega})\bI_{d}$, where $\tilde{C}\in(0,2)$ is a constant. 
Notice that $2\hat{\bD}-\hat{\bD}\hat{\bOmega}\hat{\bD} \succ 0$.

Let $\mathcal{X}_{n_2}$ be the $\sigma$-algebra generated by $\{\bx_{n_1+k}\}_{k=1}^{n_2}$ with $n_2=n-n_1$. 
Assume $\bx$ follows a sub-Gaussian distribution (not necessarily Gaussian), 
and consider fully sample-based knockoff variables constructed as follows:
\begin{align}\label{Seq:gau_approx}
    \hat{\bx} = (\bI_{d}-\hat{\bD}\hat{\bOmega})\bx + (2\hat{\bD}-\hat{\bD}\hat{\bOmega}\hat{\bD})^{1/2}\bxi\,,
\end{align}
where $\bx$ and $\bxi$ are independent of $\mathcal{X}_{n_2}$. Then, it holds that
\begin{align}\label{Seq:con_cov}
    \bGamma_{n_2} &:= \text{Var}\{(\bx^{\top}, \hat{\bx}^{\top})^{\top}\,|\,\mathcal{X}_{n_2}\}\notag\\
    &~=
    \begin{pmatrix}
        \bSigma & \bSigma - \bSigma\hat{\bOmega}\hat{\bD} \\
        \bSigma - \hat{\bD}\hat{\bOmega}\bSigma & \bSigma-\bSigma\hat{\bOmega}\hat{\bD}-\hat{\bD}\hat{\bOmega}\bSigma+
        \hat{\bD}\hat{\bOmega}\bSigma\hat{\bOmega}\hat{\bD}+2\hat{\bD}-
        \hat{\bD}\hat{\bOmega}\hat{\bD}
    \end{pmatrix}\,.
\end{align}
Let $\{\bxi_i\}_{i=1}^{n_1}$ be $n_1$ independent copies of $\bxi$. Write the sample-based knockoffs data matrix as 
\begin{align}\label{Seq:gau_approx_mat}
    \hat{\bX} = \bX(\bI_{d}-\hat{\bOmega}\hat{\bD}) + \bUpsilon(2\hat{\bD}-\hat{\bD}\hat{\bOmega}\hat{\bD})^{1/2}\,,
\end{align}
where $\bUpsilon=(\bxi_1,\ldots,\bxi_{n_1})^\top$, and $\bX$ is specified in \eqref{Seq:lm_1}. Let \( \bZ = (\bX, \hat{\bX}) \) and $\hat{\bGamma} = n_1^{-1}\bZ^{\top} \bZ$. Consider the debiased estimator
\begin{equation}\label{Sde_est}
    \hat{\bgamma}^{(\rm bc)} = \hat{\bgamma} + n_1^{-1} \hat{\bTheta}^{\top} \bZ^{\top} (\by - \bZ \hat{\bgamma})\,,
\end{equation}
where $\by$ is specified in \eqref{Seq:lm_1}, the Lasso estimator \(\hat{\bgamma}\) is defined as
\begin{align} \label{Seq:lasso}
    \hat{\bgamma}=\arg\min_{\bgamma \in \mathbb{R}^{2d}} \bigg\{ \frac{1}{n_1} | \by - \bZ\bgamma |_2^2 + \varrho_1 | \bgamma |_1 \bigg\}\,,
\end{align}
and \( \hat{\bTheta} \in \mathbb{R}^{2d \times 2d} \) is obtained by solving the following optimization problem
\begin{equation}\label{SCLIME}
    \min |\bTheta|_1 \quad \text{subject to} \quad |\hat{\bGamma} \bTheta - \bI_{2d}|_{\infty} \leq \varrho_2\,.
\end{equation}
In \eqref{Seq:lasso} and \eqref{SCLIME}, $\varrho_1>0$ and $\varrho_2>0$ are two tuning parameters. Besides, the estimator \( \hat{\sigma} \) of the standard deviation of \( \varepsilon_i \) is obtained by solving the following optimization problem
\begin{align}\label{Seq:vest}
    \min_{\bgamma \in \mathbb{R}^{2d},\, \sigma > 0} \bigg\{ \frac{1}{2\sigma n_1} | \by - \bZ \bgamma |_2^2 + \frac{\sigma}{2} + \varrho_3 |\bgamma|_1 \bigg\}\,,
\end{align}
where \( \varrho_3 > 0 \) is a tuning parameter. 

Based on \( \hat{\bgamma}^{(\rm bc)} \) given in \eqref{Sde_est}, we propose constructing a transformed debiased estimator  
\begin{align}\label{Seq:transform}
\bT \hat{\bgamma}^{(\rm bc)} = 
\begin{pmatrix}
\hat{\bbeta}_1^{(\rm bc)} \\
\hat{\bbeta}_2^{(\rm bc)}
\end{pmatrix}
~~\text{with}~~ 
\bT = 
\begin{pmatrix} 
\bI_d & \bI_d \\ 
\bI_d & -\bI_d 
\end{pmatrix}\,,
\end{align}
where \( \hat{\bbeta}_{1}^{(\rm bc)} = (\hat{\beta}_{1,1}^{(\rm bc)}, \ldots, \hat{\beta}_{1,d}^{(\rm bc)})^\top \) and \( \hat{\bbeta}_{2}^{(\rm bc)} = (\hat{\beta}_{2,1}^{(\rm bc)}, \ldots, \hat{\beta}_{2,d}^{(\rm bc)})^\top \). Write \( \hat\bLambda = \bT \hat{\bTheta}^\top \hat{\bGamma} \hat{\bTheta} \bT^\top \), whose diagonal components are denoted by \( \{ \hat\Lambda_{j,j} \}_{j=1}^{2d} \).  
For \( j \in [d] \), we define a set of paired test statistics \( (t_{1,j}, t_{2,j}) \) as follows: 
\begin{equation}\label{Stest1}
    t_{1,j} := \frac{n_1^{1/2} \hat{\beta}_{1,j}^{(\rm bc)}}{\hat{\sigma} \hat\Lambda_{j,j}^{1/2}} 
    ~~\text{and}~~ 
    t_{2,j} := \frac{n_1^{1/2} \hat{\beta}_{2,j}^{(\rm bc)}}{\hat{\sigma} \hat\Lambda_{j+d,j+d}^{1/2}}\,.
\end{equation} 
Following the multiple testing framework established in Section \ref{sec:fdrproc}, we can apply either Algorithm \ref{method_1} to the 
marginal statistics 
\( \{t_{2,j}\}_{j=1}^d \), or Algorithm \ref{method_2} to the paired statistics \( \{(t_{1,j}, t_{2,j})\}_{j=1}^d \) to solve the multiple testing problem formulated in \eqref{eq:mtest}. Section \ref{Ssec:theory} shows that these two methods can ensure asymptotic FDR control.

\subsection{Main results}\label{Ssec:theory}
To establish the theoretical validity of the proposed methods, we require Conditions \ref{Sass:model_error}--\ref{Sass0:ass0} below. Condition \ref{Sass:model_error} is weaker than Condition \ref{ass:model_error}, which relaxes the Gaussian assumption of the model error to the sub-Gaussian assumption. Following the same principles as in \cite{cai2011constrained} for the CLIME estimator of the precision matrix, Conditions \ref{SSass_CLIME} and \ref{Sass:CLIME} are, respectively, used for the theoretical analysis of  $\hat{\bOmega}$ specified below \eqref{Seq:lm_1} and  $\hat\bTheta$ specified in \eqref{SCLIME}. Condition \ref{Sass0:ass0} is obtained by replacing $n$ in Condition \ref{ass0} with $n_1$.

\begin{assumption}\label{Sass:model_error}
The model error $\beps$ specified in \eqref{Seq:lm_1} has independent sub-Gaussian components with zero mean, covariance matrix ${\rm Var}(\beps)=\sigma^2\bI_{n_1}$, sub-Gaussian norm $\|\beps\|_{\psi_2} = \kappa_1$ for some constant $\kappa_1 \in (0,\infty)$, and is independent of $\bZ$.
\end{assumption}

\begin{assumption}\label{Sass:random_Z}
The covariance matrix $\bSigma$ satisfies $K_{\min}\leq \sigma_{\min}(\bSigma) \leq \sigma_{\max}(\bSigma) \le K_{\max}$ for some constants $K_{\min}\in(0,\infty)$ and $K_{\max}\in[1,\infty)$, and $\bx$ has sub-Gaussian norm $\|\bx\|_{\psi_2} = \kappa_2$ for some constant $\kappa_2 \in (0,\infty)$.
\end{assumption}

\begin{assumption}\label{SSass_CLIME}
Let
\begin{equation*}
	\begin{aligned}
		\mathcal{U}_1(M_1, q_1, s_{d,1}) = \bigg\{ \bA=(a_{i,j})_{d\times d} : \bA \succ 0,\, \|\bA\|_1 \leq M_1,\, \max_{i \in [d]} \sum_{j=1}^{d} |a_{i,j}|^{q_1} \leq s_{d,1} \bigg\}
	\end{aligned}
\end{equation*}
for some constants \( q_1 \in[0,1) \) and $M_1\in(0,\infty)$.
Assume that  \( \bOmega_0 = \bSigma^{-1} \in \mathcal{U}_1(M_1, q_1, s_{d,1}) \), \(n_2\ge\eta_1^{-1}\log d\) and the tuning parameter \( \varrho_4 = C_1 M_1(n_2^{-1}\log d)^{1/2} \) in \eqref{Seq:lm_111} with \( C_1 = 2\eta_1^{-2}(2 + \tau_1 + \eta_1^{-1}e^2K_1^2)^2 \), where \(\tau_1>0\), \( \eta_1 = \min\{{1}/{8}, (4\kappa_2^{2}e)^{-1}\} \) and  \( K_{1} =(1 - 2\kappa_2^{2}e\eta_1)^{-1} \) with $\kappa_2$ specified in Condition {\rm\ref{Sass:random_Z}}.
\end{assumption}

For $(K_{\min},K_{\max},\kappa_2)$ specified in Condition \ref{Sass:random_Z} and $(q_1,C_1,M_1)$ specified in Condition \ref{SSass_CLIME}, we write
\begin{align*}
&C_{\max}=\max\{2K_{\max}-\tilde CK_{\min},\tilde CK_{\max}\}\,,\quad C_{\min}=K_{\min}\cdot\min\{2-\tilde C,\tilde C\}\,,\\
&~~~~~~\tilde\kappa=\sqrt{2}C_3(\kappa_2+2)C_{\min}^{-1/2}\,,\quad\tilde C_{\max}=1.5C_{\max}\,,
\quad\tilde C_{\min}=0.5C_{\min}\,,\\
&~~~~~~~~~~~\tilde{C}_1=2(1+2^{1-q_1}+3^{1-q_1})4^{1-q_1}C_1^{1-q_1}\,,\quad C_2=2\tilde{C}C_{\max}^2\,,\\
&~~~~~~~\tilde{C}_2=2C_2
+4C_{\max}^2(\tilde C+1)+C_{\max}^3(\tilde C+1)^{2}(C_{\min}^{-1}+2^{-1}C_{\max}^{-1})\,,\\
&~~~\breve{C}_2=C_2
+3C_{\max}^2(\tilde C+1)+C_{\max}^3M_1(\tilde C+1)^{2}\,,\quad
\tilde C_{3}=C_3(\kappa_2+2)\,,\\
&~~~~~~~C_3=2+C_{\max}(\tilde C+1)(C_{\min}^{-1}+2^{-1}C_{\max}^{-1})\\
&~~~~~~~~~~~~~~+\{2C_{\max}(\tilde C+1)
+C_{\max}^2(\tilde C+1)^2(C_{\min}^{-1}+2^{-1}C_{\max}^{-1})\}^{1/2}\,.
\end{align*}

\begin{assumption}\label{Sass:CLIME}
Let
\begin{align*}
\mathcal{U}_{2}(M_{2}, q_{2}, s_{d,2}) = \bigg\{ \bTheta=(\theta_{i,j})_{2d\times 2d} : \bTheta \succ 0,\, \|\bTheta\|_1 \leq M_{2},\, \max_{i \in [2d]} \sum_{j=1}^{2d} |\theta_{i,j}|^{q_2} \leq s_{d,2} \bigg\}
\end{align*}
for some constants \( q_{2} \in[0,1) \) and $M_{2}\in(0,\infty)$.
Assume that \( \bTheta_0 = \bGamma^{-1} \in \mathcal{U}_{2}(M_{2},q_{2}, s_{d,2}) \), \(n_1\ge\eta_2^{-1}\log(2d)\) and \( \varrho_2 \ge a\{n_1^{-1}\log(2d)\}^{1/2}+ \breve{C}_2\tilde{C}_{1}M_{2}M_1^{2-2q_1}
s_{d,1}(n_2^{-1}\log d)^{(1-q_1)/2} \) in \eqref{SCLIME} with \( a = \max\{C_0 M_{2}, 4\sqrt{6}e\tilde\kappa^2
(\tau_2+2)^{1/2}\tilde C_{\max}^{1/2}\tilde C_{\min}^{-1/2}\} \), where \(\tau_2>0\) and \( C_0 = \eta_2^{-1}(2 + \tau_2 + \eta_2^{-1}K_2^2) \) with  \( \eta_2 = \min\{{1}/{8}, (4\tilde C_{3}^{2}e)^{-1}\} \) and  \( K_{2} =(1 - 2\tilde C_{3}^{2}e\eta_2)^{-1} \).
\end{assumption}

\begin{assumption}\label{Sass0:ass0}
There exists $\mathcal{H}\subseteq[d]$ such that $|\mathcal{H}|\geq\log\log d$ and $|\beta^0_{j}|\geq2\sqrt{2}C_{\min}^{-1/2}\sigma(n_1^{-1}\log d)^{1/2}$ for all $j\in\mathcal{H}$.
\end{assumption}

\begin{remark}\label{Sremark1}
Notice that $\bGamma$ defined in \eqref{Seq:kcov} is similar to ${\rm diag}(2\bSigma-\bD, \bD)$. Due to $\bD=\tilde{C}\lambda_{\max}^{-1}(\bOmega_0)\bI_{d}$ with $\bOmega_0=\bSigma^{-1}$, we have $\sigma_{\min}(\bD)=\sigma_{\max}(\bD)=\tilde{C}\lambda_{\min}(\bSigma)$, which implies 
\begin{align*}
&~~~~\sigma_{\min}(\bGamma)=\min\{\sigma_{\min}(2\bSigma-\bD), \sigma_{\min}(\bD)\}
=\min\{(2-\tilde C)\lambda_{\min}(\bSigma), \tilde C\lambda_{\min}(\bSigma)\}\,,\\
&\sigma_{\max}(\bGamma)=\max\{\sigma_{\max}(2\bSigma-\bD), \sigma_{\max}(\bD)\}
=\max\{2\lambda_{\max}(\bSigma)-\tilde C\lambda_{\min}(\bSigma), \tilde C\lambda_{\min}(\bSigma)\}\,.
\end{align*}
Together with Condition {\rm\ref{Sass:random_Z}}, we have $C_{\min}\le\sigma_{\min}(\bGamma)\le\sigma_{\max}(\bGamma)\le C_{\max}$.
\end{remark}

Recall \( \bZ=(\bX,\hat \bX) \in \mathbb{R}^{n_1 \times 2d} \), and let \(\bgamma_0 = (\bbeta_0^\top, {\bf0}^\top)^\top\) and \( \bw_{n_1} = n_1^{-1/2}\hat{\bTheta}^{\top} \bZ^\top \beps \) with \(\hat{\bTheta} \in \mathbb{R}^{2d \times 2d}\) specified in \eqref{SCLIME}. For \(\hat{\bgamma}^{(\rm bc)}\) specified in \eqref{Sde_est}, we have
\begin{equation} \label{Seq:exp1}
    n_1^{1/2}\{\hat{\bgamma}^{(\rm bc)} - \bgamma_{0}\} = \bw_{n_1} + \bdelta_{n_1}\,,
\end{equation}
where \( \E(\bw_{n_1}\,|\,\bZ)= {\bzo}\), \({\rm Var}(\bw_{n_1}\,|\,\bZ)=\sigma^2 \hat{\bTheta}^{\top} \hat{\bGamma} \hat{\bTheta} \) and \( \bdelta_{n_1} = n_1^{1/2}(\hat{\bTheta}^{\top} \hat{\bGamma} - \bI_{2d})(\bgamma_{0} - \hat{\bgamma}) \). To show the proposed methods can control FDR asymptotically, we need Theorem \ref{Stheo_random}, whose proof is given in Section \ref{Sproof_Stheo_random}.

\begin{theorem}\label{Stheo_random}
Let Conditions {\rm \ref{Sass:model_error}--\ref{Sass:CLIME}} hold, $|\bgamma_{0}|_{0}\le s_0$ for some integer $1\le s_0<2d$. For any given $\tau_2>0$ specified in Condition {\rm\ref{Sass:CLIME}}, let $\hat{\bgamma}$ be the Lasso estimator given in \eqref{Seq:lasso} with $\varrho_1$ satisfying \( \varrho_1 \ge 2\sqrt{6}\kappa_1\{C_4^{-1}\tilde C_{\max}(\tau_2+1)n_1^{-1}\log(2d)\}^{1/2} \), where $C_4>0$ denotes the absolute constant appeared in Proposition {\rm5.10} of {\rm\citeS{vershynin_2012_app}}.  Select $\tau_1$ specified in Condition {\rm\ref{SSass_CLIME}} satisfying $\tau_1\ge\tau_2$. Write $v_0 = 240000 c_*\tilde C_{\max}\tilde C_{\min}^{-1}\tilde\kappa^4$, $c_1 =(4c_*\tilde\kappa^4)^{-1}$ and $c_2 =a^2\tilde C_{\min}/(96e^2\tilde\kappa^4\tilde C_{\max}) - 2$ with $c_* = 32000e^{2}$. The following two assertions hold:
\begin{itemize}
\item[\rm(i)] If $n_1 \ge \max\{v_0s_0\log(240eds_0^{-1}), 5000(\tau_2+1)\tilde\kappa^4\log(2d), 12.5(\tau_2+1)\log(2d), 6(c_{2}+2)\log(2d)\}$ and $n_2
\ge(2\tilde{C}_2C_{\min}^{-1}\tilde{C}_1s_{d,1})^{2/(1-q_1)}M_1^4\log d$, then 
	\begin{equation*}
		\mathbb P\bigg(|\bdelta_{n_1}|_{\infty} > \frac{16\varrho_1\varrho_2s_0n_1^{1/2}}{\tilde C_{\min}}\bigg) \le 2e^{-c_1n_1} + 37d^{-\tau_2}\,.
	\end{equation*}

\item[\rm(ii)] If $n_1 \ge 6(c_{2}+2)\log (2d)$ and $n_2\ge(2\tilde{C}_2C_{\min}^{-1}\tilde{C}_1s_{d,1})^{2/(1-q_1)}M_1^4\log d$, then
\begin{align*}
\mathbb{P}\big(|\hat\bTheta^\top \hat{\bGamma} \hat\bTheta - \bTheta_0|_{\infty} \le 5M_{2}\varrho_{2}\big)
\ge1 - 36d^{-\tau_2}\,.
\end{align*}
\end{itemize}
\end{theorem}

Based on Theorem \ref{Stheo_random}, we can establish the asymptotic FDR control of our proposed methods as stated in Theorems \ref{Sthm0} and \ref{Sthm1}, whose proofs are given in Sections \ref{Sproof_thm0} and \ref{Sproof_thm1}, respectively.

\begin{theorem}\label{Sthm0}
Let Conditions {\rm \ref{Sass:model_error}--\ref{Sass0:ass0}} hold. Assume that $\tau_1\ge\tau_2>2$, $s_{d,1}^{2/(1-q_1)}\log d\ll n_2$ and $|\bgamma_{0}|_{0}\le s_0<2d$ with $s_{0}\ll n_1^{1/2}(\log d)^{-3/2}$. Consider a suitable choice of the tuning parameters $\varrho_1\asymp\varrho_2\asymp\varrho_3\asymp\{n_1^{-1}\log(2d)\}^{1/2}$ in \eqref{Seq:lasso}, \eqref{SCLIME} and \eqref{Seq:vest}, respectively. If $\log d\ll n_1^{1/14}$, then Algorithm {\rm \ref{method_1}} can control {\rm FDR} asymptotically at $\pi_0\alpha$, that is
\begin{equation*}
\lim_{n_1\rightarrow\infty}\mathbb{P}({\rm FDP}\leq \pi_0 \alpha+\epsilon)=1
\end{equation*}
for any $\epsilon>0$, and
\begin{equation*}
\mathop{\lim\sup}_{n_1\rightarrow\infty}{\rm FDR}\leq \pi_0 \alpha\,,
\end{equation*}
where $\pi_0 = d_0/d$, and $d_{0}$ is the number of true null hypotheses.
\end{theorem}

\begin{theorem}\label{Sthm1}
Let the conditions of Theorem {\rm \ref{Sthm0}} hold. If $s_{d,2}^2\ll d$ and $s_{0}\lesssim n_1^{1/2}(\log d)^{-9/2}$, then Algorithm {\rm \ref{method_2}} can control {\rm FDR} asymptotically at $\pi_0\alpha$, that is
\begin{equation*}
\lim_{n_1\rightarrow\infty}\mathbb{P}({\rm FDP}\leq \pi_0 \alpha+\epsilon)=1
\end{equation*}
for any $\epsilon>0$, and
	\begin{equation*}
		\mathop{\lim\sup}_{n_1\rightarrow\infty}{\rm FDR}\leq \pi_0 \alpha\,,
	\end{equation*}
	where $\pi_0 = d_0/d$, and $d_{0}$ is the number of true null hypotheses.
\end{theorem}

\section{Proofs of Theorems \ref{Stheo_random}--\ref{Sthm1}}

\subsection{Proof of Theorem \ref{Stheo_random}}\label{Sproof_Stheo_random}
Recall
$$\phi^2(\bA, S) := \min_{\btheta \in \mathbb{R}^{2d}:\,|\btheta_{S^{\mathrm{c}}}|_1 \leq 3|\btheta_S|_1} \frac{|S| \langle \btheta, \bA\btheta \rangle}{|\btheta_S|_1^2}\quad\text{and}\quad
\mu_*(\bZ; \bTheta) := | \bTheta^{\top} \hat{\bGamma} - \bI_{2d} |_{\infty}\,,
$$
where $\bA \in \mathbb{R}^{2d \times 2d}$ is a symmetric matrix, $S \subseteq [2d]$ is a set, \( \bZ \in\mathbb{R}^{n_1\times 2d}\) and $\hat{\bGamma}=n_1^{-1}\bZ^{\top}\bZ$. Let $\{\hat{\Gamma}_{j,j}\}_{j=1}^{2d}$ denote the diagonal components of $\hat\bGamma$. 
To prove Theorem \ref{Stheo_random}, we need Lemmas \ref{Stheo_con}--\ref{Slemm_clime} whose proofs are given in Sections \ref{Sproof_lem_cond}--\ref{Sproof_lemm_clime}, respectively.

\begin{lemma}\label{Stheo_con}
Let Conditions {\rm\ref{Sass:random_Z}} and {\rm\ref{SSass_CLIME}} hold. For $\bGamma_{n_2}$ specified in \eqref{Seq:con_cov}, consider two events 
\begin{gather*}
\mathcal{A}_{n_2}=\{\tilde C_{\min}\le\sigma_{\min}(\bGamma_{n_2})
\le\sigma_{\max}(\bGamma_{n_2})\le\tilde C_{\max}\}\,,\\
\mathcal{B}_{n_2}=\bigg\{|\bGamma_{n_2}-\bGamma|_{\infty}\le\breve{C}_2\tilde{C}_{1}M_1^{2-2q_1}
s_{d,1}\bigg(\frac{\log d}{n_2}\bigg)^{(1-q_1)/2}\bigg\}\,.
\end{gather*}
If $n_2\ge(2\tilde{C}_2C_{\min}^{-1}\tilde{C}_1s_{d,1})^{2/(1-q_1)}M_1^4\log d$, we have
\begin{align*}
\mathbb{P}(\mathcal{A}_{n_2})\ge 1-4d^{-\tau_1}\quad{\text{and}}\quad\mathbb{P}(\mathcal{B}_{n_2})\ge 1-8d^{-\tau_1}\,,
\end{align*}
where $\tau_1>0$ is specified in Condition {\rm\ref{SSass_CLIME}}.
\end{lemma}

\begin{lemma}\label{Stheo_condition}
	Let Conditions {\rm\ref{Sass:random_Z}} and {\rm\ref{SSass_CLIME}} hold. The following two assertions hold:
	\begin{enumerate}
		\item[\rm(i)] For any given $\phi_0>0$, $1\le s_{0}<2d$ and $K>0$, write 
		\begin{equation*}
			\mathcal E_{n_1}(\phi_0,s_0,K) := \bigg\{\bZ \in \mathbb R^{n_1\times 2d}: \min_{S\subseteq[2d]:\,\lvert S \rvert \le s_0} \phi(\hat{\bGamma},S) \ge \phi_0,~ \max_{j \in [2d]} \hat{\Gamma}_{j,j}\le K \bigg\}\,.
		\end{equation*}
	Let $v_0= 240000 c_*\tilde C_{\max}\tilde C_{\min}^{-1}\tilde\kappa^4$ and $c_1 =(4c_*\tilde \kappa^4)^{-1}$ with $c_* = 32000e^{2}$. For any given $c>0$, if $K \ge \tilde C_{\max}[1 + 50\tilde\kappa^2\{cn_1^{-1}\log(2d)\}^{1/2}]$, then 
	\begin{equation*}
		\mathbb P\bigg\{\bZ \in \mathcal E_{n_1}\bigg(\frac{\tilde C_{\min}^{1/2}}{2},s_0,K\bigg)\bigg\} \ge 1-2e^{-c_1n_1}-4^{1-c}d^{-2c+1}-16d^{-\tau_1}
	\end{equation*}	
    for $n_1 \ge \max\{v_0s_0\log(240eds_0^{-1}),25c\log(2d)\}$ and $n_2\ge(2\tilde{C}_2C_{\min}^{-1}\tilde{C}_1s_{d,1})^{2/(1-q_1)} M_1^4\log d$.
	\item[\rm(ii)] For any given $a>0$, write $c_2 =a^2\tilde C_{\min}/(96e^2\tilde\kappa^4\tilde C_{\max}) - 2$ and
	\begin{equation*}
		\mathcal G_{n_1}(a) := \bigg\{\bZ \in \mathbb R^{n_1\times 2d}: \mu_{\min}(\bZ) < a\sqrt{\frac{\log(2d)}{n_1}} \bigg\}
	\end{equation*}
   with \( \mu_{\min}(\bZ) = \min_{\bTheta \in \mathbb{R}^{2d \times 2d}} \mu_*(\bZ; \bTheta) \). It holds that
        \begin{equation*}
		\mathbb P\{\bZ \in \mathcal G_{n_1}(a)\} \ge 1-8d^{-c_2}-8d^{-\tau_1}
	\end{equation*}
    for $n_1 \ge 6(c_{2}+2)\log (2d)$ and $n_2\ge(2\tilde{C}_2C_{\min}^{-1}\tilde{C}_1s_{d,1})^{2/(1-q_1)}M_1^4\log d$.
	\end{enumerate}
\end{lemma}

\begin{lemma}\label{Stheo_fix}
For any given $\tau_2>0$, let $\hat{\bgamma}$ be the Lasso estimator given in \eqref{Seq:lasso} with $\varrho_1$ satisfying \( \varrho_1 \ge 2\sqrt{6}\kappa_1\{C_4^{-1}\tilde C_{\max}(\tau_2+1)n_1^{-1}\log(2d)\}^{1/2} \), where $\kappa_1$ is specified in Condition {\rm\ref{Sass:model_error}}, and $C_4>0$ denotes the absolute constant appeared in Proposition {\rm5.10} of {\rm\citeS{vershynin_2012_app}}. Assume that \( |\bgamma_{0}|_{0} \leq s_0 \) and Conditions {\rm\ref{Sass:model_error}--\ref{SSass_CLIME}} hold. Then
\begin{align*}
    \mathbb{P}\bigg(|\hat{\bgamma} - \bgamma_{0}|_1 > \frac{16\varrho_1 s_0}{\tilde C_{\min}}\bigg)\le 5d^{-\tau_2}+2e^{-c_{1}n_1}+16d^{-\tau_1}
\end{align*}
provided that $n_1 \ge \max\{v_0s_0\log(240eds_0^{-1}), 5000(\tau_2+1)\tilde\kappa^4\log(2d), 12.5(\tau_2+1)\log(2d)\}$ and $n_2
\ge(2\tilde{C}_2C_{\min}^{-1}\tilde{C}_1s_{d,1})^{2/(1-q_1)}M_1^4\log d$,
where $v_0$ and $c_1$ are specified in Lemma {\rm\ref{Stheo_condition}(i)}.
\end{lemma}

\begin{lemma}\label{Slemm_clime}
 Let Conditions {\rm\ref{Sass:random_Z}--\rm\ref{Sass:CLIME}} hold. If $n_2\ge(2\tilde{C}_2C_{\min}^{-1}\tilde{C}_1s_{d,1})^{2/(1-q_1)} M_1^4\log d$, then
\begin{align*}
    \mathbb{P}\big(| \hat{\bTheta} - \bTheta_0 |_{\infty} \le 4M_{2}\varrho_{2}\big)\ge
    1 - 2^{1-\tau_2}d^{-\tau_2}-16d^{-\tau_1}\,.
\end{align*}
\end{lemma}

Write $\varrho_2'=a\{n_1^{-1}\log(2d)\}^{1/2}+ \breve{C}_2\tilde{C}_{1}M_{2}M_1^{2-2q_1}
s_{d,1}(n_2^{-1}\log d)^{(1-q_1)/2}$ and $\tilde{\mathcal G}_{n_1}(a) := \{\bZ \in \mathbb R^{n_1\times 2d}: \mu_{\min}(\bZ) < \varrho_2'\}$.
For any given $\bZ\in\tilde{\mathcal G}_{n_1}(a)$ with $a$ specified in Condition \ref{Sass:CLIME}, based on the definition of $\tilde{\mathcal G}_{n_1}(a)$, we know there exits a $\bTheta$ such that $| \bTheta^{\top} \hat{\bGamma} - \bI_{2d} |_{\infty} < \varrho_2'\le\varrho_2$, which implies the optimization problem \eqref{SCLIME} has a solution. Restricted on $\bZ\in\tilde{\mathcal G}_{n_1}(a)$, based on the definition of $\hat{\bTheta}$, we have $|\hat{\bTheta}^{\top} \hat{\bGamma} - \bI_{2d}|_{\infty}\le \varrho_{2}$. Recall $\bdelta_{n_1} = n_1^{1/2}(\hat{\bTheta}^{\top} \hat{\bGamma} - \bI_{2d})(\bgamma_{0} - \hat{\bgamma})$. Hence, given $\bZ\in\tilde{\mathcal G}_{n_1}(a)$,
$$|\bdelta_{n_1}|_{\infty}\le n_1^{1/2}|\hat{\bTheta}^{\top} \hat{\bGamma} - \bI_{2d}|_{\infty}|\bgamma_{0} - \hat{\bgamma}|_{1}\le n_1^{1/2}\varrho_{2}|\bgamma_{0} - \hat{\bgamma}|_{1}\,.$$
Notice that $\mathcal G_{n_1}(a)\subset\tilde{\mathcal G}_{n_1}(a)$, $c_2\ge\tau_2$ and $\tau_1\ge\tau_2$. For any given $\tau_2>0$, if $\varrho_1 \ge 2\sqrt{6}\kappa_1\{C_4^{-1}\tilde C_{\max}(\tau_2+1)n_1^{-1}\log(2d)\}^{1/2}$, by Lemmas \ref{Stheo_condition}(ii) and \ref{Stheo_fix},
\begin{align*}
		&\mathbb{P}\bigg( |\bdelta_{n_1}|_{\infty} > \frac{16\varrho_1\varrho_2s_0n_1^{1/2}}{\tilde C_{\min}} \bigg) \le
		\mathbb P\bigg\{\bZ\in\tilde{\mathcal G}_{n_1}(a),|\bdelta_{n_1}|_{\infty} > \frac{16\varrho_1\varrho_2s_0n_1^{1/2}}{\tilde C_{\min}}\bigg\} 
        + \mathbb P\{\bZ\in\tilde{\mathcal G}_{n_1}^{\rm c}(a)\} \\
		&\qquad\le  \mathbb P\bigg\{\bZ\in\tilde{\mathcal G}_{n_1}(a), n_1^{1/2}\varrho_{2}|\bgamma_{0} - \hat{\bgamma}|_{1}>\frac{16\varrho_1\varrho_2s_0n_1^{1/2}}{\tilde C_{\min}} \bigg\}
 + 8d^{-c_{2}}+ 8d^{-\tau_1}\\
        &\qquad\le\mathbb{P}\bigg(|\hat{\bgamma} - \bgamma_{0}|_1 > \frac{16\varrho_1 s_0}{\tilde C_{\min}}\bigg) + 16d^{-\tau_2}\le2e^{-c_1n_1} + 37d^{-\tau_2}
\end{align*}
provided that $n_1 \ge \max\{v_0s_0\log(240eds_0^{-1}), 5000(\tau_2+1)\tilde\kappa^4\log(2d), 12.5(\tau_2+1)\log(2d), 6(c_{2}+2)\log (2d)\}$ and $n_2
\ge(2\tilde{C}_2C_{\min}^{-1}\tilde{C}_1s_{d,1})^{2/(1-q_1)}M_1^4\log d$.
As shown in Equation (25) of \citeS{cai2011constrained_app}, if $\varrho_2\ge \|\bTheta_0\|_{1}|\hat\bGamma-\bGamma|_{\infty}$, then $\|\hat\bTheta\|_1 \le \|\bTheta_0\|_1 \le M_{2}$ and $|\hat{\bTheta}^{\top} \hat{\bGamma} - \bI_{2d}|_{\infty}\le \varrho_{2}$. Recall $\tau_1\ge\tau_2$ and $c_2\ge\tau_2$. By \eqref{Seq:clime3} in the proof of Lemma \ref{Slemm_clime} in Section \ref{Sproof_lemm_clime}, if $n_1 \ge 6(c_{2}+2)\log (2d)$ and $n_2\ge(2\tilde{C}_2C_{\min}^{-1}\tilde{C}_1s_{d,1})^{2/(1-q_1)}M_1^4\log d$, we have
\begin{align}\label{Seq:thm2_1}
&\mathbb P(|\hat\bTheta^{\top} \hat{\bGamma} - \bI_{2d}|_{\infty}\|\hat\bTheta\|_1\le M_{2}\varrho_{2})
\ge\mathbb P(|\hat\bTheta^{\top} \hat{\bGamma} - \bI_{2d}|_{\infty}\|\hat\bTheta\|_1\le M_{2}\varrho_{2},\varrho_2\ge \|\bTheta_0\|_{1}|\hat\bGamma-\bGamma|_{\infty})\notag\\
&\qquad\ge\mathbb P(|\hat\bTheta^{\top} \hat{\bGamma} - \bI_{2d}|_{\infty} \le \varrho_{2},\varrho_2\ge \|\bTheta_0\|_{1}|\hat\bGamma-\bGamma|_{\infty})
=\mathbb P(\varrho_2\ge \|\bTheta_0\|_{1}|\hat\bGamma-\bGamma|_{\infty})\notag\\
&\qquad\ge1- 2^{1-\tau_2}d^{-\tau_2}-16d^{-\tau_1}
\ge1 - 18d^{-\tau_2}\,.
\end{align}
By triangle inequality, we have 
\begin{equation*}
|\hat\bTheta^{\top} \hat{\bGamma} \hat\bTheta - \bTheta_0|_{\infty} 
\le |(\hat\bTheta^{\top} \hat{ \bGamma} - \bI_{2d}) \hat\bTheta|_{\infty} + |\hat\bTheta - \bTheta_0|_{\infty}
\le|\hat\bTheta^{\top} \hat{\bGamma} - \bI_{2d}|_{\infty}\|\hat\bTheta\|_1
+ |\hat\bTheta - \bTheta_0|_{\infty}\,.
\end{equation*}
Together with \eqref{Seq:thm2_1} and Lemma \ref{Slemm_clime}, we have $\mathbb{P}\big(|\hat{\bTheta}^\top \hat{\bGamma} \hat{\bTheta} - \bTheta_0|_{\infty} \le 5M_{2}\varrho_{2}\big)\ge1 - 36d^{-\tau_2}$ for $n_1 \ge 6(c_{2}+2)\log (2d)$ and $n_2\ge(2\tilde{C}_2C_{\min}^{-1}\tilde{C}_1s_{d,1})^{2/(1-q_1)}M_1^4\log d$. We complete the proof of Theorem \ref{Stheo_random}. $\hfill\qedsymbol$

\subsection{Proof of Theorem \ref{Sthm0}}\label{Sproof_thm0}
Recall $G(t)=2\{1-\Phi(t)\}$. Let $G^{-1}(\cdot)$ be the inverse function of $G(\cdot)$. To prove Theorem \ref{Sthm0}, we need Lemmas \ref{Stheo_scaled} and \ref{Stheo_random_trans} whose proofs are given in Sections \ref{Sproof_theo_scaled} and \ref{Sproof_theo_random_trans}, respectively.

\begin{lemma}\label{Stheo_scaled}
Assume that Condition {\rm\ref{Sass:model_error}} and the conditions of Lemma {\rm \ref{Stheo_condition}} hold. Let $\hat{\sigma}$ be the estimator given in \eqref{Seq:vest} with $\varrho_3$ satisfying $\varrho_3 \ge 8C_4^{-1/2}\sigma^{-1}\kappa_1\tilde C_{\max}^{1/2}\{n_1^{-1}\log(2d)\}^{1/2}$ and $\varrho_3 \asymp \{n_1^{-1}\log(2d)\}^{1/2}$. If $1\le s_0 \ll n_1(\log d)^{-1}$, $|\bgamma_{0}|_{0}\le s_{0}$ and $s_{d,1}^{2/(1-q_1)}\log d\ll n_2$, then
\begin{equation*}
\lim_{d\rightarrow \infty}\mathbb{P} \bigg(\bigg\lvert \frac{\hat{\sigma}}{\sigma} - 1 \bigg\rvert \ge 16\varrho_3\sqrt{\frac{s_0}{\tilde C_{\min}}}\bigg) = 0\,.
\end{equation*}
\end{lemma}

\begin{lemma}\label{Stheo_random_trans}
Let the conditions of Theorem {\rm\ref{Stheo_random}} hold. The following two assertions hold:
\begin{itemize}
\item[\rm (i)] If $n_1 \ge \max\{v_0s_0\log(240eds_0^{-1}), 5000(\tau_2+1)\tilde\kappa^4\log(2d), 12.5(\tau_2+1)\log(2d), 6(c_{2}+2)\log(2d)\}$ and $n_2\ge(2\tilde{C}_2C_{\min}^{-1}\tilde{C}_1s_{d,1})^{2/(1-q_1)}
 M_1^4\log d$, then 
	\begin{align*}
		\mathbb P\bigg(|\bT\bdelta_{n_1}|_{\infty} > \frac{32\varrho_1\varrho_2s_0n_1^{1/2}}{\tilde C_{\min}}\bigg) \le 2e^{-c_1n_1} + 37d^{-\tau_2}\,.
	\end{align*}

\item[\rm (ii)] If $n_1 \ge 6(c_{2}+2)\log (2d)$ and $n_2\ge(2\tilde{C}_2C_{\min}^{-1}\tilde{C}_1s_{d,1})^{2/(1-q_1)}M_1^4\log d$, then
\begin{align*}
\mathbb{P}\big(|\bT\hat\bTheta^\top \hat{\bGamma} \hat\bTheta\bT^{\top} - \bT\bTheta_0\bT^{\top}|_{\infty} \le 20M_{2}\varrho_{2}\big)\ge1 - 36d^{-\tau_2}\,.
\end{align*}
\end{itemize}
\end{lemma}

Let $\bar{t}_{2, j}=|t_{2, j}|$ and $\bar{t}_{2, (1)} \geq \cdots \geq \bar{t}_{2,(d)}$ be the ordered sequence of $\{\bar{t}_{2, j}\}_{j=1}^{d}$. Define $t_{2,(0)}=+\infty$. By the definition of $\tilde{R}$ specified in Step 2 of Algorithm \ref{method_1}, we have $\tilde{R}=\max\{j=0,1, \ldots, d: \tilde{P}_{(j)}=G(\bar{t}_{2, (j)}) \leq j \alpha/d\}$. Then $G(\bar{t}_{2,(\tilde{R})})\leq \alpha\tilde{R} /d$ and $G(\bar{t}_{2,(l)})> \alpha l/d$ for all $l>\tilde{R}$, which implies $\bar{t}_{2,(l)}< \bar{t}_{2,(\tilde{R})}$ for all $l>\tilde{R}$. Hence, $|\{j\in[d]:\bar{t}_{2, j} \geq \bar{t}_{2,(\tilde{R})}\}|=\tilde{R}$, and rejecting all $H_{0,j}$'s with $\bar{t}_{2, j} \geq \bar{t}_{2,(\tilde{R})}$ is equivalent to rejecting all $H_{0,j}$'s corresponding to the first $\tilde{R}$ largest $\bar{t}_{2, j}$'s among $\{\bar{t}_{2, j}\}_{j=1}^{d}$. Without loss of generality, we assume $|\mathcal{H}|\le\sqrt{\log d}$. If $|\mathcal{H}|>\sqrt{\log d}$, we can redefine $\mathcal{H}$ by keeping only the first $\lfloor\sqrt{\log d}\rfloor$ elements of it. Our proof of Theorem \ref{Sthm0} mainly includes the following four steps:

{\bf Step 1}.
To show that
\begin{equation}\label{Sthm0-1}
\lim_{n_1\rightarrow\infty}\mathbb{P}\bigg\{\sum_{j=1}^d I(\bar{t}_{2, j}>\sqrt{2 \log d}) \geq|\mathcal{H}|\bigg\} = 1\,.
\end{equation}

{\bf Step 2}. To show that $\{\bar{t}_{2, j} \geq \bar{t}_{2,(\tilde{R})}\}=\{\bar{t}_{2, j} \geq \hat{t}\}$ for any $j\in[d]$, where
\begin{equation}\label{Sthm0-8}
\hat{t}:=\inf\bigg[t>0: G(t)\leq\frac{\alpha }{d}\max\bigg\{\sum_{j=1}^{d}I(\bar{t}_{2, j}\geq t),1\bigg\}\bigg]\,.
\end{equation}

{\bf Step 3}. To show that $\hat{t}$ specified in \eqref{Sthm0-8} satisfies
\begin{align}
&~\frac{dG(\hat{t})}{\max\{\sum_{j=1}^{d}I(\bar{t}_{2, j}\geq \hat{t}),1\}}=\alpha\,,\label{Sthm0-10}\\
&\lim_{n_1\rightarrow\infty}\mathbb{P}\bigg\{\hat{t} \leq G^{-1}\bigg(\frac{\alpha |\mathcal{H}|}{d}\bigg)\bigg\} = 1\,.\label{Sthm0-11}
\end{align}

{\bf Step 4}. To show that
\begin{equation}\label{Sthm0-16}
\begin{aligned}
\sup_{0\leq t\leq G^{-1}(\alpha|\mathcal{H}|/d)}\frac{\sum_{j\in \mathcal{H}_{0}}I(\bar{t}_{2, j}\geq t)}{d G(t)}\leq \frac{d_{0}}{d}+o_{\rm p}(1)\,,
\end{aligned}
\end{equation}
where $d_{0}=d-d_{1}$ is the number of true null hypotheses.

The proofs of Steps 1--4 are given in Sections \ref{Sproof_thm2_step1}--\ref{Sproof_thm2_step4}, respectively. Following the arguments of the proof of Theorem \ref{thm0} in Section \ref{proof_thm0}, by \eqref{Sthm0-1}--\eqref{Sthm0-16}, we have $\mathbb{P}(\text{FDP}\leq d^{-1}d_{0}\alpha+\epsilon)\geq1-o(1)$ and $\lim\sup_{n_1\rightarrow \infty}\text{FDR}\leq \alpha d_{0}/d$. We complete the proof of Theorem \ref{Sthm0}. $\hfill\qedsymbol$

\subsubsection{Proof of Step 1}\label{Sproof_thm2_step1}

Write $\bw_{n_1}=(w_{1},\ldots,w_{2d})^{\top}$, $\hat\bLambda :=(\hat\Lambda_{i,j})_{2d\times 2d} = \bT \hat{\bTheta}^\top \hat{\bGamma} \hat{\bTheta} \bT^\top$, $\tilde{\bD}:=(\tilde{D}_{i,j})_{d\times d}=2\bD^{-1}$ and $\tilde{\bdelta}_{n_1}:=(\tilde{\delta}_{1},\ldots,\tilde{\delta}_{2d})^{\top}=\bT \bdelta_{n_1}$. Consider the events 
\begin{align*}
\tilde{\mathcal E}_{1,n_1}:=\bigg\{\bigg|\frac{\hat\sigma}{\sigma}-1\bigg|\le\frac{1}{25}\bigg\}
~~\text{and}~~
\tilde{\mathcal E}_{2,n_1}:=\bigg\{\max_{j \in \mathcal{H}}\bigg|\frac{\hat\Lambda_{j+d,j+d}}{\tilde{D}_{j,j}}-1\bigg|\le 10C_{\max}M_{2}\varrho_{2}\bigg\}\,.
\end{align*}
Recall $\tau_1\ge\tau_2>2$, $\varrho_{1}\asymp\varrho_{2}\asymp\varrho_{3}\asymp\{n_1^{-1}\log(2d)\}^{1/2}$, $s_{0}\ll n_1^{1/2}/\log(2d)$ and $s_{d,1}^{2/(1-q_1)}\log d\ll n_2$. By Lemmas \ref{Stheo_scaled} and \ref{Stheo_random_trans}, we have 
$
\mathbb{P}(|\tilde{\bdelta}_{n_1}|_{\infty}\ge1)
+\mathbb{P}(\tilde{\mathcal E}_{1,n_1}^{\rm c})+\mathbb{P}(\tilde{\mathcal E}_{2,n_1}^{\rm c})= o(1)$. Notice that \eqref{thm0-3} and \eqref{eq:thm2-1} in Section \ref{proof_thm2_step1} also hold with replacing $\tilde{\mathcal E}_{1,n}$, $\tilde{\mathcal E}_{2,n}$, and $\tilde{\bdelta}_{n}$ used there, respectively, by $\tilde{\mathcal E}_{1,n_1}$, $\tilde{\mathcal E}_{2,n_1}$, and $\tilde{\bdelta}_{n_1}$. Thus,
\begin{align*}
\mathbb{P}\bigg\{\sum_{j=1}^d I(\bar{t}_{2, j}>\sqrt{2 \log d})\geq |\mathcal{H}|\bigg\}
\geq\mathbb{P}\bigg[\bigcap_{j \in \mathcal{H}}\bigg\{|\tilde{w}_{j+d}|
\leq\frac{12\sqrt{\log d}}{25}-\frac{91C_{\max}^{1/2}}
{90\sqrt{2}\sigma}\bigg\}\bigg]
-o(1)\,.
\end{align*}
Let $\bz^{(k)}$ denote the $k$-th column of $\bZ\hat{\bTheta}$ for $k\in[2d]$. Write $\tilde{w}_{j+d}=\sigma^{-1}\hat\Lambda_{j+d,j+d}^{-1/2}(w_{j}-w_{j+d})$ for $j\in[d]$. Since \( \bw_{n_1} = n_1^{-1/2}\hat{\bTheta}^{\top} \bZ^\top \beps \), we have $w_{k}=n_1^{-1/2}\beps^\top \bz^{(k)}$ for $k\in[2d]$, which implies $\tilde{w}_{j+d}=n_1^{-1/2}\sigma^{-1}\hat\Lambda_{j+d,j+d}^{-1/2}\{\bz^{(j)}-\bz^{(j+d)}\}^\top\beps$ and $\E(\tilde{w}_{j+d}\,|\,\bZ)=0$ for $j\in[d]$. Notice that $\tilde{w}_{j+d}$ is the $(j+d)$-th component of $\sigma^{-1}{\rm diag}(\hat{\Lambda}_{1,1}^{-1/2},\ldots,\hat{\Lambda}_{2d,2d}^{-1/2})\bT\bw_{n_1}$. Due to $\text{Var}(\sigma^{-1}\bT\bw_{n_1}\,|\,\bZ)=\hat\bLambda$, we have $\E(\tilde{w}_{j+d}^2\,|\,\bZ)=|n_1^{-1/2}\sigma^{-1}\hat\Lambda_{j+d,j+d}^{-1/2}\{\bz^{(j)}-\bz^{(j+d)}\}|_2^2\sigma^2=1$. By Condition \ref{Sass:model_error} and Proposition 5.10 of \citeS{vershynin_2012_app}, we have
\begin{align}\label{Seq:step1_1}
&\mathbb{P}\bigg[\bigcap_{j \in \mathcal{H}}\bigg\{|\tilde{w}_{j+d}|
\leq\frac{12\sqrt{\log d}}{25}-\frac{91C_{\max}^{1/2}}
{90\sqrt{2}\sigma}\bigg\}\bigg]
\geq1-\sum_{j \in \mathcal{H}}\mathbb{P}\bigg(|\tilde{w}_{j+d}|
>\frac{12\sqrt{\log d}}{25}-\frac{91C_{\max}^{1/2}}
{90\sqrt{2}\sigma}\bigg)\notag\\
&~~~~~=1-\sum_{j \in \mathcal{H}}\mathbb{E}\bigg\{\mathbb{P}\bigg(|\tilde{w}_{j+d}|>\frac{12\sqrt{\log d}}{25}-\frac{91C_{\max}^{1/2}}
{90\sqrt{2}\sigma}\,\bigg|\,\bZ\bigg)\bigg\}\\
&~~~~~\ge1-e\sqrt{\log d}\cdot\exp\bigg[-\frac{C_4\{12\sqrt{\log d}/25
-91C_{\max}^{1/2}/(90\sqrt{2}\sigma)\}^2}{\kappa_1^2\sigma^{-2}}\bigg]
\rightarrow1\notag
\end{align}
as $n_1\rightarrow\infty$, where $C_4$ denotes the absolute constant appeared in Proposition {\rm5.10} of \citeS{vershynin_2012_app}. Hence, \eqref{Sthm0-1} holds.  $\hfill\qedsymbol$

\subsubsection{Proof of Step 2}\label{Sproof_thm2_step2}
The proof is identical to that of Step 2 in the proof of Theorem \ref{thm0} given in Section \ref{proof_thm2_step2}.  $\hfill\qedsymbol$

\subsubsection{Proof of Step 3}\label{Sproof_thm2_step3}
The proof is identical to that of Step 3 in the proof of Theorem \ref{thm0} given in Section \ref{proof_thm2_step3}. $\hfill\qedsymbol$

\subsubsection{Proof of Step 4}\label{Sproof_thm2_step4}

To prove \eqref{Sthm0-16}, we need Lemma \ref{Slem0}, whose proof is given in Section \ref{Sproof_lem0}.

\begin{lemma}\label{Slem0}
Let the conditions of Theorem {\rm \ref{Sthm0}} hold. For any sequence of positive constants $\{b_{d}\}$ such that $b_{d}\rightarrow \infty$ and $b_{d}\ll d$,
\begin{align*}
\sup_{0\leq t\leq G^{-1}(b_{d}/d)}\bigg|\frac{\sum_{j\in \mathcal{H}_{0}}I(|\tilde{w}_{j+d}|\geq t)}{d G(t)}-\frac{d_{0}}{d}\bigg|=o_{\rm p}(1)\,.
\end{align*}
\end{lemma}

Based on \eqref{Seq:lem0_0}, \eqref{Seq:lem0_1_4} and \eqref{Seq:lem0_9} in the proof of Lemma \ref{Slem0} in Section \ref{Sproof_lem0}, we have 
\begin{align}\label{Seq:thm2-1-2}
\mathbb{P}\bigg(\max_{j\in[d]}|\tilde{w}_{j+d}|
>2\sqrt{\log d}\bigg)=o(1)
\end{align}
provided that $\log d\ll n_1^{1/14}$. Based on \eqref{Seq:thm2-1-2} and Lemmas \ref{lem_normal} and \ref{Stheo_scaled}--\ref{Slem0}, following the same arguments for the proof of Step 4 in Theorem \ref{thm0} in Section \ref{proof_thm2_step4}, we have \eqref{Sthm0-16}.  $\hfill\qedsymbol$

\subsection{Proof of Theorem \ref{Sthm1}}\label{Sproof_thm1}

Recall $G(t)=2\{1-\Phi(t)\}$. Let $G^{-1}(\cdot)$ denote the inverse function of $G(\cdot)$. We will show that under the conditions of Theorem \ref{Sthm1}, Algorithm \ref{method_3}, given in the proof of Theorem \ref{thm1} in Section \ref{proof_thm1}, can control {\rm FDR} at $\pi_0\alpha$. Notice that Algorithm \ref{method_2} is the special case of Algorithm \ref{method_3} with $\lambda=\sqrt{\alpha}$. Let $\bar{t}_{i, j}=|t_{i, j}|$ and $\bar{t}_{i, (1)} \geq \cdots \geq \bar{t}_{i,(d)}$ be the ordered sequence of $\{\bar{t}_{i, j}\}_{j=1}^{d}$, where $i \in[2]$. Define $\tilde{t}_{2, j}=\bar{t}_{2, j}I\{\bar{t}_{1, j} \geq G^{-1}(\lambda)\}$ and $\tilde{t}_{2,(0)}=+\infty$. Let $\tilde{t}_{2,(1)} \geq \cdots \geq \tilde{t}_{2, (d)}$ be the ordered sequence of $\{\tilde{t}_{2, j}\}_{j=1}^{d}$. Recall $P_j^{(1)}=G(|t_{1, j}|)$. Then $\{\bar{t}_{1,j}\geq G^{-1}(\lambda)\}=\{P_{j}^{(1)}\le \lambda\}$, which implies $G(\tilde{t}_{2, j})=\tilde{P}_{j}$ with $\tilde{P}_{j}$ specified in Algorithm \ref{method_3}. By the definition of $\tilde{R}_{\lambda}$ specified in Algorithm \ref{method_3}, we have $\tilde{R}_{\lambda}=\max\{j=0,1, \ldots, d: \tilde{P}_{(j)}=G(\tilde{t}_{2, (j)}) \leq j \alpha/(\lambda d)\}$. Then $G(\tilde{t}_{2,(\tilde{R}_{\lambda})})\leq \alpha\tilde{R}_{\lambda} /(\lambda d)$ and $G(\tilde{t}_{2,(l)})> \alpha l/(\lambda d)$ for all $l>\tilde{R}_{\lambda}$, which implies $\tilde{t}_{2,(l)}< \tilde{t}_{2,(\tilde{R}_{\lambda})}$ for all $l>\tilde{R}_{\lambda}$. Hence, $|\{j\in[d]:\tilde{t}_{2, j} \geq \tilde{t}_{2,(\tilde{R}_{\lambda})}\}|=\tilde{R}_{\lambda}$, and rejecting all $H_{0,j}$'s with $\tilde{t}_{2, j} \geq \tilde{t}_{2,(\tilde{R}_{\lambda})}$ is equivalent to rejecting all $H_{0,j}$'s corresponding to the first $\tilde{R}_{\lambda}$ largest $\tilde{t}_{2, j}$'s among $\{\tilde{t}_{2, j}\}_{j=1}^{d}$. Without loss of generality, we assume $|\mathcal{H}|\le\sqrt{\log d}$. If $|\mathcal{H}|>\sqrt{\log d}$, we can redefine $\mathcal{H}$ by keeping only the first $\lfloor\sqrt{\log d}\rfloor$ elements of it. Our proof of Theorem \ref{Sthm1} mainly includes the following four steps:

{\bf Step 1}. To show that
\begin{equation}\label{Sthm1-1}
	\mathbb{P}\bigg\{\sum_{j=1}^d I(\tilde{t}_{2, j}>\sqrt{2 \log d}) \geq|\mathcal{H}|\bigg\} \rightarrow 1\,.
\end{equation}

{\bf Step 2}. To show that $\{\tilde{t}_{2, j} \geq \tilde{t}_{2,(\tilde{R}_{\lambda})}\}=\{\tilde{t}_{2, j} \geq \hat{t}_{\lambda}\}$ for any $j\in[d]$, where
\begin{equation}\label{Sthm1-4}
	\hat{t}_{\lambda}:=\inf\bigg[t>0: G(t)\leq\frac{\alpha}{d\lambda} \max\bigg\{\sum_{j=1}^{d}I(\tilde{t}_{2, j}\geq t),1\bigg\}\bigg]\,.
\end{equation}

{\bf Step 3}. To show that $\hat{t}_{\lambda}$ specified in \eqref{Sthm1-4} satisfies
\begin{align}
&~\frac{dG(\hat{t}_{\lambda})}{\max\{\sum_{j=1}^{d}I(\tilde{t}_{2, j}\geq \hat{t}_{\lambda}),1\}}=\frac{\alpha}{\lambda}\,,\label{Sthm1-10}\\
&\lim_{n_1\rightarrow\infty}\mathbb{P}\bigg\{\hat{t}_{\lambda} \leq G^{-1}\bigg(\frac{\alpha |\mathcal{H}|}{d\lambda}\bigg)\bigg\} = 1\,.\label{Sthm1-11}
\end{align}

{\bf Step 4}. To show that
\begin{align}\label{Sthm1-16}
\sup_{0< t\leq G^{-1}(\alpha|\mathcal{H}|/(d\lambda))}\frac{\sum_{j\in \mathcal{H}_{0}}I(\tilde{t}_{2, j}\geq t)}{d\lambda G(t)}\leq \frac{d_{0}}{d}+o_{\rm p}(1)\,,
\end{align}
where $d_{0} = d - d_1$ is the number of true null hypotheses.

The proofs of Steps 1--4 are given in Sections \ref{Sproof_thm3_step1}--\ref{Sproof_thm3_step4}, respectively. Following the arguments for  the proof of Theorem \ref{thm1} in Section \ref{proof_thm1}, by \eqref{Sthm1-1}--\eqref{Sthm1-16}, we have $\mathbb{P}(\text{FDP}\leq d^{-1}d_{0}\alpha+\epsilon)\geq1-o(1)$ and $\lim\sup_{n_1\rightarrow \infty}\text{FDR}\leq \alpha d_{0}/d$. We complete the proof of Theorem \ref{Sthm1}. $\hfill\qedsymbol$

\subsubsection{Proof of Step 1}\label{Sproof_thm3_step1}
The proof is almost identical to that of Step 1 in the proof of Theorem \ref{thm1} given in Section \ref{proof_thm3_step1}. We only need to replace the events $\tilde{\mathcal E}_{1,n}$ and $\tilde{\mathcal E}_{2,n}$ used there, respectively, by the events $\tilde{\mathcal E}_{1,n_1}$ and $\tilde{\mathcal E}_{2,n_1}$ defined in Section \ref{Sproof_thm2_step1}, and also replace \eqref{eq:exp1} and \eqref{eq:thm2-2} used there by \eqref{Seq:exp1} and \eqref{Seq:step1_1}, respectively. $\hfill\qedsymbol$

\subsubsection{Proof of Step 2}\label{Sproof_thm3_step2}
The proof is identical to that of Step 2 in the proof of Theorem \ref{thm1} given in Section \ref{proof_thm3_step2}. $\hfill\qedsymbol$

\subsubsection{Proof of Step 3}\label{Sproof_thm3_step3}
The proof is identical to that of Step 3 in the proof of Theorem \ref{thm1} given in Section \ref{proof_thm3_step3}. $\hfill\qedsymbol$

\subsubsection{Proof of Step 4}\label{Sproof_thm3_step4}

To prove \eqref{Sthm1-16}, we employ the core proof argument almost identical to that for \eqref{thm1-16} in Section \ref{proof_thm3_step4}. Recall $\bw_{n_1}=(w_{1},\ldots,w_{2d})^{\top}= n_1^{-1/2}\hat{\bTheta}^{\top} \bZ^\top \beps$, $\bbeta_{0}=(\beta^0_{1},\ldots,\beta^0_{d})^{\top}$, $\hat\bLambda =(\hat\Lambda_{i,j})_{2d\times 2d} = \bT \hat{\bTheta}^\top \hat{\bGamma} \hat{\bTheta} \bT^\top$ and $\tilde{\bdelta}_{n_1}=(\tilde{\delta}_{1},\ldots,\tilde{\delta}_{2d})^{\top}=\bT \bdelta_{n_1}$. Write $\hat\bgamma^{(\rm bc)}=(\hat{\gamma}^{(\rm bc)}_{1},\ldots,\hat{\gamma}^{(\rm bc)}_{2d})^{\top}$, and let $\tilde{w}_{j}=\sigma^{-1}\hat\Lambda_{j,j}^{-1/2}(w_{j}+w_{j+d})$ and $\tilde{w}_{j+d}=\sigma^{-1}\hat\Lambda_{j+d,j+d}^{-1/2}(w_{j}-w_{j+d})$ for $j\in[d]$. To prove \eqref{Sthm1-16}, we need Lemma \ref{Slem1}, whose proof is given in Section \ref{Sproof_lem1}.
\begin{lemma}\label{Slem1}
Let the conditions of Theorem {\rm\ref{Sthm1}} hold. For any sequence of positive constants $\{b_{d}\}$ such that $b_{d}\rightarrow \infty$ and $b_{d}\ll d$,
\begin{align*}
\sup_{0\leq t\leq G^{-1}(b_{d}/d)}\bigg|\frac{\sum_{j\in \mathcal{H}_{0}}I\{|\tilde{w}_{j+d}|\geq t, |\tilde{w}_{j}|\geq G_{l,n_1}^{-1}(\lambda)\}}{d\lambda G(t)}-\frac{d_{0}}{d}\bigg|=o_{\rm p}(1)\,,
\end{align*}
where $\lambda\in(0,1)$ is a constant, $G_{l,n_1}^{-1}(\lambda)=G^{-1}(\lambda)-(-1)^{l}\upsilon_{n_1}/\sqrt{\log d}$, $\upsilon_{n_1}=s_{0}^{1/2}n_1^{-1/4}(\log d)^{3/4}$ and $l\in[2]$.
\end{lemma}

Lemma \ref{Slem1} is analogous to Lemma \ref{lem1} in Step 4 of the proof of Theorem \ref{thm1} in Section \ref{proof_thm3_step4}, with the only difference being that $G_{l,n}$ and $\upsilon_{n}$ employed in Lemma \ref{lem1} are replaced by $G_{l,n_1}$ and $\upsilon_{n_1}$ specified in Lemma \ref{Slem1}, respectively. By \eqref{Seq:transform}--\eqref{Seq:exp1}, for any $j\in[d]$, we have
\begin{align*}
	&~~~~~t_{1,j} =\frac{n_1^{1/2}\{\hat{\gamma}^{(\rm bc)}_{j} + \hat{\gamma}^{(\rm bc)}_{j+d}\}}{\hat\sigma \hat\Lambda_{j,j}^{1/2}}= \frac{n_1^{1/2}\beta^0_{j}}{\hat\sigma\hat\Lambda_{j,j}^{1/2}}
	+\frac{\sigma}{\hat\sigma}\tilde{w}_{j}+
	\frac{\tilde{\delta}_{j}}{\hat\sigma\hat\Lambda_{j,j}^{1/2}}\,,\\
   & t_{2,j} =\frac{n_1^{1/2}\{\hat{\gamma}^{(\rm bc)}_{j} - \hat{\gamma}^{(\rm bc)}_{j+d}\}}{\hat\sigma \hat\Lambda_{j+d,j+d}^{1/2}}= \frac{n_1^{1/2}\beta^0_{j}}{\hat\sigma\hat\Lambda_{j+d,j+d}^{1/2}}
+\frac{\sigma}{\hat\sigma}\tilde{w}_{j+d}+
\frac{\tilde{\delta}_{j+d}}{\hat\sigma\hat\Lambda_{j+d,j+d}^{1/2}}\,.
\end{align*}
Let $\breve{t}_{1,j}:=n_1^{1/2}\{\hat{\gamma}^{(\rm bc)}_{j} + \hat{\gamma}^{(\rm bc)}_{j+d}-\beta^0_{j}\}/(\hat\sigma \hat\Lambda_{j,j}^{1/2})$ and $\breve{t}_{2,j}:=n_1^{1/2}\{\hat{\gamma}^{(\rm bc)}_{j} - \hat{\gamma}^{(\rm bc)}_{j+d}-\beta^0_{j}\}/(\hat\sigma \hat\Lambda_{j+d,j+d}^{1/2})$ for $j\in[d]$.
Consider the events 
\begin{align*}
&\mathcal{D}_{1}=\bigg\{\max_{j\in[d]}|\breve{t}_{1,j}-\tilde{w}_{j}|>\frac{\upsilon_{n_1}}{\sqrt{\log d}} \bigg\}\,,~~~~\mathcal{D}_{2}=\bigg\{\max_{j\in[d]}|\breve{t}_{2,j}-\tilde{w}_{j+d}|>\frac{\upsilon_{n_1}}{\sqrt{\log d}} \bigg\}\,,\\
&\mathcal{D}_{3}=\bigg\{\inf_{0\leq t\leq G^{-1}(\alpha|\mathcal{H}|/(d\lambda))}\bigg[\frac{\sum_{j \in \mathcal{H}_0} I\{|\breve{t}_{2,j}|\geq t,|\breve{t}_{1, j}|\geq G^{-1}(\lambda)\}}{d\lambda G(t)}\\
&~~~~~~~~~~~~~~~~~~~~
-\frac{\sum_{j\in \mathcal{H}_{0}}I\{|\tilde{w}_{j+d}|\geq t+\upsilon_{n_1}/\sqrt{\log d},|\tilde{w}_j|\geq G_{1,n_1}^{-1}(\lambda)\}}{d\lambda G(t)}\bigg]\ge0\bigg\}\,,\\
&\mathcal{D}_{4}=\bigg\{\sup_{0\leq t\leq G^{-1}(\alpha|\mathcal{H}|/(d\lambda))}\bigg[\frac{\sum_{j \in \mathcal{H}_0} I\{|\breve{t}_{2,j}|\geq t,|\breve{t}_{1, j}|\geq G^{-1}(\lambda)\}}{d\lambda G(t)}\\
&~~~~~~~~~~~~~~~~~~~~
-\frac{\sum_{j\in \mathcal{H}_{0}}I\{|\tilde{w}_{j+d}|\geq t-\upsilon_{n_1}/\sqrt{\log d},|\tilde{w}_j|\geq G_{2,n_1}^{-1}(\lambda)\}}{d\lambda G(t)}\bigg]\le0\bigg\}\,.
\end{align*}
To establish \eqref{Sthm1-16}, we need 
\begin{align}
\mathbb{P}(\mathcal{D}_1)&=o(1)\,,\label{Seq:thmT4-2-1}\\
\mathbb{P}(\mathcal{D}_2)&=o(1)\,.\label{Seq:thmT4-2-2}
\end{align}
The proof of \eqref{Seq:thmT4-2-2} is almost identical to that of \eqref{eq:thm2-2-1}. We only need to replace \eqref{eq:thm2-1-2} used there by \eqref{Seq:thm2-1-2}, and also replace $n$ and $\upsilon_{n}$ used there by $n_1$ and $\upsilon_{n_1}$, respectively. Analogously, we can also prove \eqref{Seq:thmT4-2-1}. Following the same arguments for \eqref{eq:thm3-1-2} and \eqref{eq:thm3-1-3}, we can show 
\begin{align}\label{Seq:thmT4-2-3}
\mathbb{P}(\mathcal{D}_3)\ge1-o(1)\quad\text{and}\quad \mathbb{P}(\mathcal{D}_4)\ge1-o(1)\,.
\end{align}
Based on \eqref{Seq:thmT4-2-3} and Lemma \ref{Slem1}, following the same arguments for the proof of \eqref{thm1-16} in Section \ref{proof_thm3_step4}, with $s_d$ and $\vartheta$ used in that proof replaced by $s_{d,2}$ and $1$, respectively, we have \eqref{Sthm1-16}. $\hfill\qedsymbol$

\section{Proofs of Lemmas \ref{Stheo_con}--\ref{Slem6}}\label{Sproof_theo_condition}

\subsection{Proof of Lemma \ref{Stheo_con}}\label{Sproof_lem_cond}
Recall $\bx_i=(x_{i,1},\ldots,x_{i,d})^{\top}$. By Condition \ref{Sass:random_Z}, we have
$$\| x_{i,j}\|_{\psi_{2}}=\| \langle \bx_i,\be_{j}\rangle\|_{\psi_{2}}\le\| \bx_i\|_{\psi_{2}}=\kappa_2$$ 
for $j\in[d]$ and $i\in[n_1]$, which implies
\begin{align*}
  \max_{j\in[d]}\mathbb E(e^{tx_{i,j}^{2}})
  \le\max_{j\in[d]}\mathbb E(e^{\eta_1 x_{i,j}^{2}})
  \le1+\sum_{k=1}^{\infty}(2  \kappa_2^{2}e\eta_1)^{k}\le\frac{1}{1-2\kappa_2^{2}e\eta_1}
\end{align*}
for all $|t|\le\eta_1$ with $\eta_1=\min\{1/8,(4\kappa_2^{2}e)^{-1}\}$. By Theorem 1(a) of \citeS{cai2011constrained_app} and Condition \ref{SSass_CLIME}, it holds that
\begin{align}\label{Seq:lem_con1}
    \mathbb{P} \bigg\{\|\hat\bOmega-\bOmega_0\|_{2}\le \tilde{C}_{1}M_1^{2-2q_1}s_{d,1}\bigg(\frac{\log d}{n_2}\bigg)^{(1-q_1)/2}\bigg\}\ge 1-4d^{-\tau_1}\,,
\end{align}
where $\tilde{C}_1=2(1+2^{1-q_1}+3^{1-q_1})4^{1-q_1}C_1^{1-q_1}$, and $(q_1,\tau_1,C_1,M_1)$ is specified in Condition \ref{SSass_CLIME}. Recall $\hat{\bD}=\tilde{C}\lambda_{\max}^{-1}(\hat{\bOmega})\bI_{d}$ and $\bD=\tilde{C}\lambda_{\max}^{-1}(\bOmega_0)\bI_{d}$, where $\tilde{C}\in(0,2)$ is a constant. Then
$\|\hat{\bD}-\bD\|_{2}= \tilde{C}|\lambda_{\max}^{-1}(\hat\bOmega)-\lambda_{\max}^{-1}(\bOmega_0)|
\le \tilde{C}\|\hat\bOmega\|_2^{-1}\|\bOmega_0\|_2^{-1}\|\hat\bOmega-\bOmega_0\|_{2}$, which implies
\begin{align}\label{Seq:lem_con2}
\|\bGamma_{n_2}-\bGamma\|_2
&\le2\|\bD - \bSigma\hat{\bOmega}\hat{\bD}\|_2
+\|-\bSigma\hat{\bOmega}\hat{\bD}-\hat{\bD}\hat{\bOmega}\bSigma+
    \hat{\bD}\hat{\bOmega}\bSigma\hat{\bOmega}\hat{\bD}+2\hat{\bD}-
    \hat{\bD}\hat{\bOmega}\hat{\bD}\|_2\notag\\
&\le2\|\bD - \hat\bD\|_2
+4\|\bSigma(\bOmega_0-\hat{\bOmega})\hat{\bD}\|_2
+\|\hat{\bD}\hat{\bOmega}\bSigma(\hat{\bOmega}-\bOmega_0)\hat{\bD}\|_2\notag\\
&\le(2\tilde{C}\|\hat\bOmega\|_2^{-1}\|\bOmega_0\|_2^{-1}+4\|\bSigma\|_2\|\hat{\bD}\|_2
+\|\bSigma\|_2\|\hat{\bD}\|_2^2\|\hat{\bOmega}\|_2)\|\hat\bOmega-\bOmega_0\|_{2}\,.   
\end{align}
Recall $\mathcal{A}_{n_2}=\{0.5C_{\min}\le\sigma_{\min}(\bGamma_{n_2})\le\sigma_{\max}(\bGamma_{n_2})
\le1.5C_{\max}\}$ with $C_{\max}=\max\{2K_{\max}-\tilde CK_{\min},\tilde CK_{\max}\}$ and $C_{\min}=K_{\min}\cdot\min\{2-\tilde C,\tilde C\}$. By Condition \ref{Sass:random_Z} and Remark \ref{Sremark1}, we have $C_{\min}\le\sigma_{\min}(\bSigma)\le\sigma_{\max}(\bSigma)\le C_{\max}$, $C_{\min}\le\sigma_{\min}(\bGamma)\le\sigma_{\max}(\bGamma)\le C_{\max}$ and $C_{\min}\le\sigma_{\min}(\bD)\le\sigma_{\max}(\bD)\le C_{\max}$. Due to $\bOmega_0=\bSigma^{-1}$, we have $C_{\max}^{-1}\le\sigma_{\min}(\bOmega_0)\le\sigma_{\max}(\bOmega_0)\le C_{\min}^{-1}$. Write $\omega_{n_2}=\|\hat\bOmega-\bOmega_0\|_{2}$, $C_{n_2}=\tilde{C}C_{\max}|\|\bOmega_0\|_{2}-\omega_{n_2}|^{-1}$ and $C_2=2\tilde{C}C_{\max}^2$. If $\omega_{n_2}\le2^{-1}C_{\max}^{-1}$, then
\begin{align}\label{Seq:lem_con3}
&2\tilde{C}\|\hat\bOmega\|_2^{-1}\|\bOmega_0\|_2^{-1}+4\|\bSigma\|_2\|\hat{\bD}\|_2
+\|\bSigma\|_2\|\hat{\bD}\|_2^2\|\hat{\bOmega}\|_2\notag\\
&\quad\le 2C_{n_2}
+4C_{\max}(C_{\max}+C_{n_2}\omega_{n_2})+C_{\max}(C_{\max}+C_{n_2}\omega_{n_2})^{2}
(C_{\min}^{-1}+\omega_{n_2})\notag\\
&\quad\le 2C_2
+4C_{\max}^2(\tilde C+1)+C_{\max}^3(\tilde C+1)^{2}(C_{\min}^{-1}+2^{-1}C_{\max}^{-1})=:\tilde{C}_2\,.
\end{align}
Since $\tilde C_2>C_{\max}^2$, we have $\tilde C_2C_{\min}^{-1}>C_{\max}$. By \eqref{Seq:lem_con1}--\eqref{Seq:lem_con3} and Condition \ref{Sass:random_Z}, if $n_2\ge(2\tilde{C}_2C_{\min}^{-1}\tilde{C}_1
s_{d,1})^{2/(1-q_1)}M_1^4\log d$, then
\begin{align*}
\mathbb{P}(\mathcal{A}_{n_2})\ge&~\mathbb{P}\bigg(\|\bGamma_{n_2}-\bGamma\|_2\le\frac{C_{\min}}{2},\omega_{n_2}\le\frac{1}{2C_{\max}}\bigg)\\
\ge&~\mathbb{P}\bigg(\|\hat\bOmega-\bOmega_0\|_{2}\le\frac{C_{\min}}{2\tilde{C}_2}\bigg)\ge 1-4d^{-\tau_1}\,.
\end{align*}

Following the arguments of Equations (27) and (28) in \citeS{cai2011constrained_app}, we have
\begin{align}\label{Seq:lem_con5}
    \mathbb{P} \bigg\{\|\hat\bOmega-\bOmega_0\|_{1}\le \tilde{C}_{1}M_1^{2-2q_1}s_{d,1}\bigg(\frac{\log d}{n_2}\bigg)^{(1-q_1)/2}\bigg\}\ge 1-4d^{-\tau_1}\,.
\end{align}
By the definition of $\hat\bOmega$ and Equation (25) of \citeS{cai2011constrained_app}, if $|\bSigma-\hat\bSigma|_{\infty}\le\|\bOmega_0\|_1^{-1}\varrho_4$, then $\|\hat\bOmega\|_1\le M_1$. By Equation (28) of \citeS{cai2011constrained_app}, 
\begin{align*}
\PP\big(|\bSigma-\hat\bSigma|_{\infty}\le\|\bOmega_0\|_1^{-1}\varrho_4\big)
\ge\PP\big(|\bSigma-\hat\bSigma|_{\infty}\le M_1^{-1}\varrho_4\big)\ge1-4d^{-\tau_1}\,,
\end{align*}
which implies $\PP(\|\hat\bOmega\|_1\le M_1)\ge1-4d^{-\tau_1}$. Since $\bSigma$ is a covariance matrix, then $|\bSigma|_{\infty}\le\lambda_{\max}(\bSigma)\le C_{\max}$. Notice that $\|\hat{\bD}\|_1=\|\hat{\bD}\|_2$ and $|\hat{\bD}-\bD|_{\infty}= \tilde{C}|\lambda_{\max}^{-1}(\hat\bOmega)-\lambda_{\max}^{-1}(\bOmega_0)|
\le \tilde{C}\|\hat\bOmega\|_2^{-1}\|\bOmega_0\|_2^{-1}\omega_{n_2}$ with $\omega_{n_2}=\|\hat\bOmega-\bOmega_0\|_{2}$. Write $\tilde{\omega}_{n_2}=\|\hat\bOmega-\bOmega_0\|_{1}$. Due to $\omega_{n_2}\le\tilde{\omega}_{n_2}$, if $\|\hat\bOmega\|_1\le M_1$, it holds that 
\begin{align}\label{Seq:lem_con6}
|\bGamma_{n_2}-\bGamma|_{\infty}
&\le|\bD - \bSigma\hat{\bOmega}\hat{\bD}|_{\infty}
+|-\bSigma\hat{\bOmega}\hat{\bD}-\hat{\bD}\hat{\bOmega}\bSigma+
    \hat{\bD}\hat{\bOmega}\bSigma\hat{\bOmega}\hat{\bD}+2\hat{\bD}-
    \hat{\bD}\hat{\bOmega}\hat{\bD}|_{\infty}\notag\\
&\le|\bD - \hat\bD|_{\infty}
+3|\bSigma(\bOmega_0-\hat{\bOmega})\hat{\bD}|_{\infty}
+|\hat{\bD}\hat{\bOmega}\bSigma(\hat{\bOmega}-\bOmega_0)\hat{\bD}|_{\infty}\notag\\
&\le \tilde{C}\|\hat\bOmega\|_2^{-1}\|\bOmega_0\|_2^{-1}\omega_{n_2}
+3|\bSigma|_{\infty}\|\hat\bOmega-\bOmega_0\|_{1}\|\hat{\bD}\|_1
+|\bSigma|_{\infty}\|\hat{\bOmega}\|_{1}\|\hat{\bOmega}-\bOmega_0\|_1\|\hat{\bD}\|_1^2\notag\\   
&\le (\tilde{C}\|\hat\bOmega\|_2^{-1}\|\bOmega_0\|_2^{-1}
+3C_{\max}\|\hat{\bD}\|_2
+C_{\max}M_1\|\hat{\bD}\|_2^{2})\tilde{\omega}_{n_2}\,.
\end{align}
Similar to \eqref{Seq:lem_con3}, if $\omega_{n_2}\le2^{-1}C_{\max}^{-1}$, we have
\begin{align}\label{Seq:lem_con7}
&\tilde{C}\|\hat\bOmega\|_2^{-1}\|\bOmega_0\|_2^{-1}
+3C_{\max}\|\hat{\bD}\|_2+C_{\max}M_1\|\hat{\bD}\|_2^{2}\notag\\
&\quad\le C_{n_2}
+3C_{\max}(C_{\max}+C_{n_2}\omega_{n_2})+C_{\max}M_1(C_{\max}+C_{n_2}\omega_{n_2})^{2}\notag\\
&\quad\le C_2
+3C_{\max}^2(\tilde C+1)+C_{\max}^3M_1(\tilde C+1)^{2}=\breve{C}_2\,.
\end{align}
Recall $\mathcal{B}_{n_2}=\{|\bGamma_{n_2}-\bGamma|_{\infty}\le\breve{C}_2\tilde{C}_{1}M_1^{2-2q_1}
s_{d,1}(n_2^{-1}\log d)^{(1-q_1)/2}\}$, $\tilde C_2C_{\min}^{-1}>C_{\max}$ and $\tilde{\omega}_{n_2}\ge\omega_{n_2}$. By \eqref{Seq:lem_con5}--\eqref{Seq:lem_con7} and Condition \ref{Sass:random_Z}, if $n_2\ge(2\tilde{C}_2C_{\min}^{-1}\tilde{C}_1
s_{d,1})^{2/(1-q_1)}M_1^4\log d$, we have
\begin{align*}
\mathbb{P}(\mathcal{B}_{n_2})&\ge\mathbb{P} \bigg\{\tilde{\omega}_{n_2}\le \tilde{C}_{1}M_1^{2-2q_1}s_{d,1}\bigg(\frac{\log d}{n_2}\bigg)^{(1-q_1)/2},\omega_{n_2}\le\frac{1}{2C_{\max}},\|\hat\bOmega\|_1\le M_1\bigg\}\\
&\ge\mathbb{P} \bigg\{\tilde{\omega}_{n_2}\le \tilde{C}_{1}M_1^{2-2q_1}s_{d,1}\bigg(\frac{\log d}{n_2}\bigg)^{(1-q_1)/2},\tilde{\omega}_{n_2}\le\frac{C_{\min}}{2\tilde{C}_2},\|\hat\bOmega\|_1\le M_1\bigg\}\\
&=\mathbb{P} \bigg\{\tilde{\omega}_{n_2}\le \tilde{C}_{1}M_1^{2-2q_1}s_{d,1}\bigg(\frac{\log d}{n_2}\bigg)^{(1-q_1)/2},\|\hat\bOmega\|_1\le M_1\bigg\}
\ge 1-8d^{-\tau_1}\,.
\end{align*}
We complete the proof of Lemma \ref{Stheo_con}. $\hfill\qedsymbol$

\subsection{Proof of Lemma \ref{Stheo_condition}(i)}\label{Spflem1a}
Recall $\mathcal{X}_{n_2}=\sigma(\bx_{n_1+1},\ldots,\bx_{n})$. Define $\bar{\mathbb{E}}(\cdot)=\mathbb{E}(\cdot\,|\,\mathcal{X}_{n_2})$, $\bar{\mathbb{P}}(\cdot)=\mathbb{P}(\cdot\,|\,\mathcal{X}_{n_2})$ and $\|\bzeta\|_{\psi_{2}\,|\,\mathcal{X}_{n_2}}=\sup_{\bu\in \mathbb{S}^{m-1}}\sup_{q\ge1}q^{-1/2}\{\bar{\mathbb{E}}(|\langle \bzeta, \bu\rangle|^{q})\}^{1/q}$ for a random vector $\bzeta\in\mathbb{R}^m$. Write $\phi_{0}=\tilde C_{\min}^{1/2}/2$ with $\tilde C_{\min}$ defined above Condition \ref{Sass:CLIME}. By the definition of the event $\mathcal E_{n_1}(\phi_0,s_0,K)$, we have
\begin{equation*}
\begin{aligned}
\mathcal E_{n_1}^{\rm c} (\phi_0,s_0,K) &\:= \mathcal B_{1,n_1}(\phi_0,s_0)\cup \mathcal B_{2,n_1}(K)\,,
\end{aligned}
\end{equation*}
where
$\mathcal B_{1,n_1}(\phi_0,s_0) = \{\bZ \in \mathbb R^{n_1\times 2d}: \min_{S\subseteq[2d]:\,\lvert S \rvert\le s_0} \phi(\hat{\bGamma},S)<\phi_0\}$ and $
\mathcal B_{2,n_1}(K) = \{\bZ \in \mathbb R^{n_1\times 2d}: \max_{j \in [2d]} \hat{\Gamma}_{j,j} > K\}$. As we will show in Sections \ref{Spflem1a1} and \ref{Spflem1a2},
\begin{align*}
\mathbb P\{\bZ\in\mathcal B_{1,n_1}(\phi_0,s_0)\} \le 2e^{-c_1n_1}+8d^{-\tau_1}
\end{align*}
when $n_1\ge v_{0}s_{0}\log(240ed/s_{0})$ and $n_2\ge(2\tilde{C}_2C_{\min}^{-1}\tilde{C}_1s_{d,1})^{2/(1-q_1)}M_1^4\log d$, and for any given $c>0$ and $K\ge \tilde C_{\max}[1+50\tilde\kappa^2 \{cn_1^{-1}\log(2d)\}^{1/2}]$, 
\begin{align*}
\mathbb P\{\bZ\in\mathcal B_{2,n_1}(K)\} \le 4^{1-c}d^{-2c+1}+8d^{-\tau_1}
\end{align*}
when $n_1\ge25c\log(2d)$ and $n_2\ge(2\tilde{C}_2C_{\min}^{-1}\tilde{C}_1s_{d,1})^{2/(1-q_1)}M_1^4\log d$. Hence, Lemma \ref{Stheo_condition}(i) holds. $\hfill\qedsymbol$

\subsubsection{Upper bound of $\mathbb P\{\bZ\in\mathcal B_{1,n_1}(\phi_0,s_0)\}$}\label{Spflem1a1}
Recall $\mathcal{A}_{n_2}=\{\tilde{C}_{\min}\le\sigma_{\min}(\bGamma_{n_2}) 
\le\sigma_{\max}(\bGamma_{n_2})\le\tilde{C}_{\max}\}$. Given the event $\mathcal{A}_{n_2}$, for $\phi_{\rm RE}(\cdot,\cdot,\cdot)$ defined in Definition \ref{def3} in Section \ref{pflem1a1}, we have $\phi_{\rm RE}(\bGamma_{n_2},s_0,9) \ge \tilde{C}^{1/2}_{\min}$. Let $\bt_0=\bGamma_{n_2}^{-1/2} \hat\bz$ with $\hat\bz=(\bx^{\top},\hat\bx^{\top})^\top$. By \eqref{Seq:gau_approx}, it holds that
\begin{align}\label{Seq:norm1}
    \hat\bz = \begin{pmatrix}
                \bI_{d} & \bzero \\
                \bI_{d}-\hat{\bD}\hat{\bOmega} & (2\hat{\bD}-\hat{\bD}\hat{\bOmega}\hat{\bD})^{1/2} 
              \end{pmatrix}
                   \begin{pmatrix}
                     \bx \\
                     \bxi 
                   \end{pmatrix}
        =:\bA_{n_2}\bz_0\,,
\end{align}
where $\bxi\sim\mathcal{N}(\bzo,\bI_{d})$ is independent of $\bx$ and $\mathcal{X}_{n_2}$. By Condition \ref{Sass:random_Z}, we have $\|\bx\|_{\psi_{2}}=\kappa_2$. Let $\Gamma(\cdot)$ be the Gamma function. Notice that $\langle \bxi, \bu\rangle\sim\mathcal{N}(0,1)$ for any $\bu\in\mathbb{S}^{d-1}$. Due to $\Gamma(1/2)=\sqrt{\pi}$, $\Gamma(k/2)\le(k/2)^{k/2}$ and $k^{1/k}\le e^{1/e}$ for any $k\ge2$, by Equation (5.6) of \citeS{vershynin_2012_app}, 
\begin{align*}
\{\mathbb{E}(|\langle \bxi, \bu\rangle|^{q})\}^{1/q}
&=\sqrt{2}\bigg[\frac{\Gamma\{(1+q)/2\}}{\Gamma(1/2)}\bigg]^{1/q}
\le\sqrt{2}\bigg(\frac{1+q}{2}\bigg)^{(1+q)/(2q)}\le\sqrt{2q}q^{1/(2q)}
\le2\sqrt{q}
\end{align*}
for any $q\ge1$, which implies $\|\bxi\|_{\psi_{2}}\le 2$. Notice that 
\begin{align}\label{Seq:norm2}
\|\bA_{n_2}\|_2&\le\|\bI_{d}\|_2+\|\bI_{d}-\hat{\bD}\hat{\bOmega}\|_2
+\|(2\hat{\bD}-\hat{\bD}\hat{\bOmega}\hat{\bD})^{1/2}\|_2\notag\\
&\le2+\|\hat{\bD}\|_2\|\hat{\bOmega}\|_2+(2\|\hat{\bD}\|_2+\|\hat{\bD}\|_2^2\|\hat{\bOmega}\|_2)^{1/2}\,.
\end{align}
Following the arguments of \eqref{Seq:lem_con3}, we have $\|\hat{\bD}\|_2\le C_{\max}+C_{n_2}\omega_{n_2}$ and $\|\hat{\bOmega}\|_2\le C_{\min}^{-1}+\omega_{n_2}$, where $C_{n_2}=\tilde{C}C_{\max}|\|\bOmega_0\|_2-\omega_{n_2}|^{-1}$ and $\omega_{n_2}=\|\hat{\bOmega}-\bOmega_0\|_2$. Recall $C_2=2\tilde{C}C_{\max}^2$. If $\omega_{n_2}\le2^{-1}C_{\max}^{-1}$, by \eqref{Seq:norm2}, we have
\begin{align}\label{Seq:norm2_1}
\|\bA_{n_2}\|_2&\le2+C_{\max}(\tilde C+1)(C_{\min}^{-1}+2^{-1}C_{\max}^{-1})\notag\\
&\quad~+\{2C_{\max}(\tilde C+1)
+C_{\max}^2(\tilde C+1)^2(C_{\min}^{-1}+2^{-1}C_{\max}^{-1})\}^{1/2}=:C_3\,.
\end{align}
Define the event $\mathcal{C}_{n_2}=\{\omega_{n_2}\le2^{-1}C_{\max}^{-1}\}$. By \eqref{Seq:lem_con1}, if $n_2\ge(2C_{\max}\tilde{C}_1s_{d,1})^{2/(1-q_1)}M_1^4\log d$, then
\begin{align}\label{Seq:lem_con3_0}
\mathbb{P}(\mathcal{C}_{n_2})
=\mathbb{P}\bigg(\|\hat\bOmega-\bOmega_0\|_{2}\le\frac{1}{2C_{\max}}\bigg)\ge 1-4d^{-\tau_1}\,.
\end{align}
Due to $\bz_0=(\bx^{\top},\bxi^\top)^\top$, we have $\|\bz_0\|_{\psi_{2}}\le\|\bx\|_{\psi_{2}}+\|\bxi\|_{\psi_{2}}\le \kappa_2+2$. Given the event $\mathcal{C}_{n_2}$, we know $\bA_{n_2}$ is an invertible matrix. Therefore, given the event $\mathcal{C}_{n_2}$, by \eqref{Seq:norm1} and \eqref{Seq:norm2_1}, it holds that
\begin{align}\label{Seq:lem_con3_1}
\|\hat\bz\|_{\psi_{2}\,|\,\mathcal{X}_{n_2}}
\le\|\bz_0\|_{\psi_{2}}\sigma_{\max}(\bA_{n_2})\le C_3(\kappa_2+2)\,.
\end{align}
Due to $\bt_0=\bGamma_{n_2}^{-1/2} \hat\bz$, given the event $\mathcal{A}_{n_2}\cap~\mathcal{C}_{n_2}$, we have $\|\bt_0\|_{\psi_{2}\,|\,\mathcal{X}_{n_2}}
\le\|\hat\bz\|_{\psi_{2}\,|\,\mathcal{X}_{n_2}}\sigma_{\max}(\bGamma_{n_2}^{-1/2})\le \sqrt{2}C_3(\kappa_2+2)C_{\min}^{-1/2}=:\tilde{\kappa}$. Recall $k!\ge(k/e)^{k}$ for any integer $k\ge1$. Then, given the event $\mathcal{A}_{n_2}\cap\mathcal{C}_{n_2}$, for any $\bu\in\mathbb{R}^{2d}$ with $|\bu|_2>0$, 
\begin{equation*}
\begin{aligned}
  \bar{\mathbb{E}}\bigg\{\exp\bigg(\frac{\langle \bt_0, \bu\rangle^{2}}{t^{2}}\bigg)\bigg\} & = 1+\sum_{k=1}^{\infty}\frac{\bar{\mathbb{E}}(\langle \bt_0, \bu\rangle^{2k})}{t^{2k}k!}=1+\sum_{k=1}^{\infty}\frac{\bar{\mathbb{E}}(\langle \bt_0, \bu|\bu|_{2}^{-1}\rangle^{2k})|\bu|_{2}^{2k}}{t^{2k}k!}  \\
  &\le 1+\sum_{k=1}^{\infty}\frac{(2k)^{k}\tilde\kappa^{2k}|\bu|_{2}^{2k}}{t^{2k}k!}\le 1+\sum_{k=1}^{\infty}\bigg(\frac{2e\tilde\kappa^{2}|\bu|_{2}^{2}}{t^{2}}\bigg)^{k}\,,
\end{aligned}
\end{equation*}
which implies $\bar{\mathbb{E}}\{\exp(t^{-2}\langle \bt_0, \bu\rangle^{2})\}\le2$ for any $t\ge2\sqrt{e}\tilde\kappa|\bu|_{2}$. Hence, $$\inf\bigg[t>0:\bar{\mathbb{E}}\bigg\{\exp\bigg(\frac{\langle \bt_0, \bu\rangle^{2}}{t^{2}}\bigg)\bigg\}\le2\bigg]\le2\sqrt{e}\tilde\kappa|\bu|_{2}$$
for all $\bu\in\mathbb{R}^{2d}$. Notice that, if $\bGamma_{n_2}\succ 0$, $\bar{\mathbb{E}}(|\langle \bt_0, \bu\rangle|^{2})=|\bu|_{2}^{2}$ for every $\bu\in\mathbb{R}^{2d}$. Given the event $\mathcal{A}_{n_2}\cap\mathcal{C}_{n_2}$, by Definition 5 of \citeS{rudelson2013reconstruction}, $\bt_0$ is conditional isotropic $\psi_{2}$ random vector in $\mathbb{R}^{2d}$ with constant $\alpha=2\sqrt{e}\tilde\kappa$. Write $\bPsi=\bZ\bGamma_{n_2}^{-1/2}$ and $\bA=\bGamma_{n_2}^{1/2}$. Notice that the row vectors of $\bPsi$ are conditional independent and have the same conditional distribution as $\bt_0$. Hence, given the event $\mathcal{A}_{n_2}\cap\mathcal{C}_{n_2}$, by Theorem 6 of \citeS{rudelson2013reconstruction}, we have with conditional probability at least $1 - 2\exp\{-\delta^2 n_1/(c_*\tilde\kappa^4)\}$,
\begin{equation}\label{Seq:phi}
\frac{1}{\phi_{\rm RE}(\hat{\bGamma},s_0,3)} \le \frac{1}{(1-\delta)\phi_{\rm RE}(\bGamma_{n_2},s_0,3)}
\end{equation}
for $n_1 \ge c_*m\tilde\kappa^4\delta^{-2}\log\{120ed/(m\delta)\}$, where $\delta \in (0,1)$, $c_* = 32000e^{2}$, and
\begin{equation}\label{Seq:m}
m = \min\bigg\{2d,\,s_0 + \frac{12960s_{0}}{\delta^2\phi^2_{\rm RE}(\bGamma_{n_2},s_0,9)}\max_{j\in[2d]}|\bGamma_{n_2}^{1/2} \be_j|_2^2\bigg\}\,.
\end{equation}
Recall $s_0<2d$. Given the event $\mathcal{A}_{n_2}$, due to $\phi_{\rm RE}(\bGamma_{n_2},s_0,9) \ge \tilde{C}^{1/2}_{\min}$ and $\max_{j\in[2d]}|\bGamma_{n_2}^{1/2} \be_j|_2^2
\leq \tilde C_{\max}$, taking $\delta=1/2$ in \eqref{Seq:m}, then
\begin{equation*}
s_{0}< m\le s_0 + \frac{51840s_0}{\phi^2_{\rm RE}(\bGamma_{n_2},s_0,9)}\max_{j\in[2d]}|\bGamma_{n_2}^{1/2} \be_j|_2^2 \le s_0+ \frac{51840s_0 \tilde C_{\max}}{\tilde C_{\min}}\le \frac{60000s_0\tilde C_{\max}}{\tilde C_{\min}}\,.
\end{equation*}
Write $v_{0}=240000c_{*}\tilde C_{\max}\tilde C_{\min}^{-1}\tilde \kappa^{4}$. Then we have $v_{0}s_{0}\log(240ed/s_{0})\ge4c_*m\tilde\kappa^{4}\log(240ed/m)$. Given the event $\mathcal{A}_{n_2}\cap\mathcal{C}_{n_2}$, due to $\phi_{\rm RE}(\bGamma_{n_2},s_0,3) \ge \phi_{\rm RE}(\bGamma_{n_2},s_0,9) \ge \tilde{C}^{1/2}_{\min}$, taking $\delta = 1/2$ in \eqref{Seq:phi}, we have
\begin{equation*}
\bar{\mathbb P} \bigg\{\phi_{\rm RE}(\hat{\bGamma},s_0,3) \ge \frac{\tilde C^{1/2}_{\min}}{2}\bigg\} \ge 1-2e^{-c_1n_1}
\end{equation*}
for $n_1\ge v_{0}s_{0}\log(240ed/s_{0})$, where $c_1 = (4c_*\tilde\kappa^4)^{-1}$.
Recall
$$\phi^2(\hat{\bGamma}, S) = \min_{\btheta \in \mathbb{R}^{2d}:\,|\btheta_{S^{\mathrm{c}}}|_1 \leq 3|\btheta_S|_1} \frac{|S| \langle \btheta, \hat{\bGamma}\btheta \rangle}{|\btheta_S|_1^2} $$
for $S\subseteq[2d]$. Let $\phi_0=\tilde C_{\min}^{1/2}/2$. Since $\lvert S \rvert  |\btheta_S|_2^2 \ge |\btheta_S|_1^2$, we know $\min_{S\subseteq[2d]:\,\lvert S \rvert \le s_0} \phi(\hat{\bGamma},S) \ge \phi_{\rm RE}(\hat{\bGamma},s_0,3)$,
which implies
\begin{equation*}
I(\mathcal{A}_{n_2}\cap\mathcal{C}_{n_2})\bar{\mathbb P}\{\bZ\in\mathcal B_{1,n_1}(\phi_0,s_0)\} \le I(\mathcal{A}_{n_2}\cap\mathcal{C}_{n_2})\bar{\mathbb P}  \bigg\{ \phi_{\rm RE}(\hat{\bGamma},s_0,3) < \frac{\tilde C^{1/2}_{\min}}{2} \bigg\} \le2e^{-c_1n_1}
\end{equation*}
for $n_1\ge v_{0}s_{0}\log(240ed/s_{0})$. Recall $\tilde C_2C_{\min}^{-1}>C_{\max}$. By Lemma \ref{Stheo_con} and \eqref{Seq:lem_con3_0}, 
\begin{align}\label{Seq:lem_con4}
\mathbb P\{\bZ\in\mathcal B_{1,n_1}(\phi_0,s_0)\}
&\le\mathbb{E}[I(\mathcal{A}_{n_2}\cap\mathcal{C}_{n_2})\bar{\mathbb P}\{\bZ\in\mathcal B_{1,n_1}(\phi_0,s_0)\}] + \mathbb P(\mathcal{A}_{n_2}^{\rm c})+\mathbb P(\mathcal{C}_{n_2}^{\rm c})\notag\\
&\le2e^{-c_1n_1}+8d^{-\tau_1}
\end{align}
for $n_1\ge v_{0}s_{0}\log(240ed/s_{0})$ and $n_2\ge(2\tilde{C}_2C_{\min}^{-1}\tilde{C}_1s_{d,1})^{2/(1-q_1)}M_1^4\log d$. $\hfill\qedsymbol$

\subsubsection{Upper bound of $\mathbb P\{\bZ\in\mathcal B_{2,n_1}(K)\}$}\label{Spflem1a2}
Recall that, given the event $\mathcal{A}_{n_2}\cap \mathcal{C}_{n_2}$, we have $\|\bGamma_{n_2}^{-1/2} \hat\bz\|_{\psi_{2}\,|\,\mathcal{X}_{n_2}}\le
\tilde{\kappa}$ with $\hat\bz$ defined in \eqref{Seq:norm1} and $\tilde{\kappa}=\sqrt{2}C_3(\kappa_2+2)C_{\min}^{-1/2}$. Following the arguments of the proof in Section \ref{pflem1a2} and \eqref{Seq:lem_con4}, for any given $c>0$, if $K\ge \tilde C_{\max}[1+50\tilde\kappa^2 \{cn_1^{-1}\log(2d)\}^{1/2}]$, we have
\begin{align*}
\mathbb P\{\bZ\in\mathcal B_{2,n_1}(K)\}
&\le\mathbb{E}[I(\mathcal{A}_{n_2}\cap\mathcal{C}_{n_2})\bar{\mathbb P}\{\bZ\in\mathcal B_{2,n_1}(K)\}]+ \mathbb P(\mathcal{A}_{n_2}^{\rm c})+\mathbb P(\mathcal{C}_{n_2}^{\rm c})\\
&\le 4^{1-c}d^{-2c+1}+8d^{-\tau_1}
\end{align*}
for $n_1\ge25c\log(2d)$ and $n_2\ge(2\tilde{C}_2C_{\min}^{-1}\tilde{C}_1s_{d,1})^{2/(1-q_1)}M_1^4\log d$. $\hfill\qedsymbol$

\subsection{Proof of Lemma \ref{Stheo_condition}(ii)}

Following the arguments of the proof in Section \ref{pflem1a3} and \eqref{Seq:lem_con4}, it holds that
\begin{align*}
\mathbb P\{\bZ \in \mathcal G_{n_1}(a)\} &\ge 
\mathbb{E}[I(\mathcal{A}_{n_2}\cap\mathcal{C}_{n_2})\bar{\mathbb P}\{\bZ \in \mathcal G_{n_1}(a)\}]
\ge (1 - 8d^{-c_2})_{+}\mathbb{P}(\mathcal{A}_{n_2}\cap\mathcal{C}_{n_2})\\
&\ge (1-8d^{-c_2})_{+}(1-8d^{-\tau_1})\ge1-8d^{-c_2}-8d^{-\tau_1}
\end{align*}
for $n_1 \ge 6(c_{2}+2)\log(2d)$ and $n_2\ge(2\tilde{C}_2C_{\min}^{-1}\tilde{C}_1s_{d,1})^{2/(1-q_1)}M_1^4\log d$, where $\cdot_+=\max(\cdot,0)$ and
$$c_2 = a^2\tilde C_{\min}/(96e^2\tilde\kappa^4\tilde C_{\max}) - 2\,.$$ 
We complete the proof of Lemma \ref{Stheo_condition}(ii). $\hfill\qedsymbol$

\subsection{Proof of Lemma \ref{Stheo_fix}}\label{Sproof_theo_fix}
Consider the event
\begin{equation*}
\mathcal T :=\bigg\{\max_{j\in[2d]} \frac{2\lvert\beps^\top \bz^{(j)}\rvert}{n_1} \le \varrho_0\bigg\}\,,
\end{equation*}
where $\bz^{(j)}$ denotes the $j$-th column of $\bZ$. 
By Theorem 6.1 of \citeS{BV_2011_app}, given the events $\mathcal{T}$ and \(\bZ \in \mathcal E_{n_1}(\tilde C_{\min}^{1/2}/2,s_0,3\tilde C_{\max}/2)\), if  $\varrho_1 \ge 2\varrho_0$, we have 
$
|\hat{\bgamma} - \bgamma_{0}|_1 \le 16\varrho_1 s_0\tilde C_{\min}^{-1}$. 
By Lemma \ref{Stheo_condition}(i) with $c=(\tau_2+1)/2$ and $K=3\tilde C_{\max}/2$, if $\varrho_1 \ge 2\varrho_0$, then 
\begin{align}\label{Seq:lem2-1}
\mathbb{P}\bigg(|\hat{\bgamma} - \bgamma_{0}|_1 > \frac{16\varrho_1 s_0}{\tilde C_{\min}}\bigg)&\le \mathbb{P}\bigg\{|\hat{\bgamma} - \bgamma_{0}|_1 > \frac{16\varrho_1 s_0}{\tilde C_{\min}},\bZ\in\mathcal E_{n_1}\bigg(\frac{\tilde C_{\min}^{1/2}}{2},s_0,\frac{3\tilde C_{\max}}{2}\bigg),\mathcal{T}^{\rm c}\bigg\}\notag\\
&\quad+\mathbb{P}\bigg\{\bZ\in\mathcal E_{n_1}^{\rm c}\bigg(\frac{\tilde C_{\min}^{1/2}}{2},s_0,\frac{3\tilde C_{\max}}{2}\bigg)\bigg\}\\
&\le\mathbb{P}\bigg\{\bZ\in\mathcal E_{n_1}\bigg(\frac{\tilde C_{\min}^{1/2}}{2},s_0,\frac{3\tilde C_{\max}}{2}\bigg),\mathcal{T}^{\rm c}\bigg\}
+2e^{-c_1n_1} + 2d^{-\tau_2}+16d^{-\tau_1}\notag\\
&=\mathbb{E}\bigg[\mathbb P(\mathcal{T}^{\rm c} \,|\, \bZ)I\bigg\{\bZ \in \mathcal E_{n_1}\bigg(\frac{\tilde C_{\min}^{1/2}}{2},s_0,\frac{3\tilde C_{\max}}{2}\bigg)\bigg\}\bigg]\notag\\
&~~~+2e^{-c_1n_1} + 2d^{-\tau_2}+16d^{-\tau_1}\notag
\end{align}
provided that $n_1 \ge \max\{v_0s_0\log(240eds_0^{-1}), 5000(\tau_2+1)\tilde\kappa^4\log(2d), 12.5(\tau_2+1)\log(2d)\}$ and $n_2\ge(2\tilde{C}_2C_{\min}^{-1}
\tilde{C}_1s_{d,1})^{2/(1-q_1)}M_1^4\log d$. Write $\zeta_j = n_1^{-1/2}\sigma^{-1}\beps^\top \bz^{(j)}$. Recall $\hat{\bGamma}=n_1^{-1}\bZ^{\top}\bZ$. By Condition \ref{Sass:model_error}, we know $\E(\zeta_j\,|\,\bZ)=0$ and $\E(\zeta_j^2\,|\,\bZ)=|n_1^{-1/2}\sigma^{-1}\bz^{(j)}|_2^2\sigma^2=\hat{\Gamma}_{j,j}$, where $\hat{\Gamma}_{j,j}$ is the $j$-th diagonal component of $\hat\bGamma$. For given $\tau_2>0$, let $\varrho_0 = \sqrt{6}\kappa_1 \{C_4^{-1}\tilde C_{\max}(\tau_2+1)n_1^{-1}\log(2d)\}^{1/2}$, where $C_4>0$ denotes the absolute constant appeared in Proposition 5.10 of \citeS{vershynin_2012_app}. By Condition \ref{Sass:model_error} and Proposition 5.10 of \citeS{vershynin_2012_app}, we have
\begin{align*}
\mathbb P(\mathcal{T}^{\rm c} \,|\, \bZ)
&=\mathbb P \bigg\{\max_{j\in[2d]} \lvert \zeta_j \rvert > 2^{-1}\sigma^{-1}\kappa_1\sqrt{6C_4^{-1}\tilde C_{\max}(\tau_2+1)\log(2d)}\,\bigg|\,\bZ\bigg\}\\
&\le  \sum_{j=1}^{2d}e\cdot\exp\bigg\{-\frac{3\tilde C_{\max}(\tau_2+1)\log (2d)}{2\hat{\Gamma}_{j,j}}\bigg\}\,,
\end{align*}
which implies
\begin{align*}
&\mathbb{E}\bigg[\mathbb P(\mathcal{T}^{\rm c} \,|\, \bZ)I\bigg\{\bZ \in \mathcal E_{n_1}\bigg(\frac{\tilde C_{\min}^{1/2}}{2},s_0,\frac{3\tilde C_{\max}}{2}\bigg)\bigg\}\bigg]\\
&~~~~~\le
\mathbb{E}\bigg[ \sum_{j=1}^{2d}e\cdot\exp\bigg\{-\frac{3\tilde C_{\max}(\tau_2+1)\log (2d)}{2\hat{\Gamma}_{j,j}}\bigg\}I\bigg\{\bZ \in\mathcal E_{n_1}\bigg(\frac{\tilde C_{\min}^{1/2}}{2},s_0,\frac{3\tilde C_{\max}}{2}\bigg)\bigg\}\bigg]\\
&~~~~~\le 2ed\exp\bigg\{-\frac{3\tilde C_{\max}(\tau_2+1)\log (2d)}{3 \tilde C_{\max}}\bigg\}
=2ed\cdot(2d)^{-(\tau_2+1)}\le3d^{-\tau_2}\,.
\end{align*}
Together with \eqref{Seq:lem2-1}, if $\varrho_1 \ge 2\sqrt{6}\kappa_1\{C_4^{-1}\tilde C_{\max}(\tau_2+1)n_1^{-1}\log(2d)\}^{1/2}$, we have
\begin{align*}
\mathbb{P}\bigg(|\hat{\bgamma} - \bgamma_{0}|_1 > \frac{16\varrho_1 s_0}{\tilde C_{\min}}\bigg)\le
5d^{-\tau_2}+2e^{-c_{1}n_1}+16d^{-\tau_1}
\end{align*}
for 
$n_1 \ge \max\{v_0s_0\log(240eds_0^{-1}), 5000(\tau_2+1)\tilde\kappa^4\log(2d), 12.5(\tau_2+1)\log(2d)\}$ and $n_2\ge(2\tilde{C}_2C_{\min}^{-1}
\tilde{C}_1s_{d,1})^{2/(1-q_1)}M_1^4\log d$. We complete the proof of Lemma \ref{Stheo_fix}. $\hfill\qedsymbol$

\subsection{Proof of Lemma \ref{Slemm_clime}}\label{Sproof_lemm_clime}

Recall $\mathcal{X}_{n_2}=\sigma(\bx_{n_1+1},\ldots,\bx_{n})$, $\mathcal{A}_{n_2}=\{\tilde{C}_{\min}\le\sigma_{\min}(\bGamma_{n_2})
\le\sigma_{\max}(\bGamma_{n_2})\le\tilde{C}_{\max}\}$ and $\mathcal{C}_{n_2}=\{\omega_{n_2}\le2^{-1}C_{\max}^{-1}\}$. Write $\hat\bz=(\hat z_1,\ldots,\hat z_{2d})^{\top}$ with $\hat\bz$ defined in \eqref{Seq:norm1}, $\bar{\mathbb{E}}(\cdot)=\mathbb{E}(\cdot\,|\,\mathcal{X}_{n_2})$, $\bar{\mathbb{P}}(\cdot)=\mathbb{P}(\cdot\,|\,\mathcal{X}_{n_2})$ and $\|\bzeta\|_{\psi_{2}\,|\,\mathcal{X}_{n_2}}=\sup_{\bu\in \mathbb{S}^{m-1}}\sup_{q\ge1}q^{-1/2}\{\bar{\mathbb{E}}(|\langle \bzeta, \bu\rangle|^{q})\}^{1/q}$ for a random vector $\bzeta\in\mathbb{R}^m$. Given the event $\mathcal{C}_{n_2}$, by \eqref{Seq:lem_con3_1}, we have
$\| \hat z_{i}\|_{\psi_{2}\,|\,\mathcal{X}_{n_2}}\le\| \hat\bz\|_{\psi_{2}\,|\,\mathcal{X}_{n_2}}\le C_3(\kappa_2+2)=\tilde C_3$ 
for each $i\in[2d]$, which implies
\begin{align*}
  \bar\E(e^{t\hat z_{i}^{2}})\le\bar\E(e^{\eta_2 \hat z_{i}^{2}})\le1+\sum_{k=1}^{\infty}(2\tilde C_{3}^2e\eta_2)^{k}\le\frac{1}{1-2\tilde C_{3}^{2}e\eta_2}
\end{align*}
for $|t|\le\eta_2$ with $\eta_2=\min\{1/8,(4\tilde C_{3}^{2}e)^{-1}\}$. Recall $\hat{\bGamma}=n_1^{-1}\bZ^{\top}\bZ$. Therefore, given the event $\mathcal{A}_{n_2}\cap\mathcal{C}_{n_2}$, by Equation (29) of \citeS{cai2011constrained_app}, we have
\begin{align}\label{Seq:clime1}
    \bar{\mathbb{P}} \bigg\{|\hat\bGamma-\bGamma_{n_2}|_{\infty}\ge C_{0}\sqrt{\frac{\log(2d)}{n_1}}\bigg\}\le 2^{1-\tau_2}d^{-\tau_2}\,,
\end{align}
where $C_0>0$ is specified in Condition \ref{Sass:CLIME}. By Lemma \ref{Stheo_con}, we have 
\begin{align*}
\mathbb{P}\bigg\{|\bGamma_{n_2}-\bGamma|_{\infty}\le\breve{C}_2\tilde{C}_{1}M_1^{2-2q_1}
s_{d,1}\bigg(\frac{\log d}{n_2}\bigg)^{(1-q_1)/2}\bigg\}\ge 1-8d^{-\tau_1}
\end{align*}
for $n_2\ge(2\tilde{C}_2C_{\min}^{-1}\tilde{C}_1s_{d,1})^{2/(1-q_1)}M_1^4\log d$, where $\tau_1>0$ is specified in Condition \ref{SSass_CLIME}. Together with Lemma \ref{Stheo_con}, \eqref{Seq:lem_con3_0} and \eqref{Seq:clime1}, for any given $\tau_2>0$, we have
\begin{align}\label{Seq:clime2}
    &\mathbb{P} \bigg\{|\hat\bGamma-\bGamma|_{\infty}\ge C_{0}\sqrt{\frac{\log(2d)}{n_1}}
    +\breve{C}_2\tilde{C}_{1}M_1^{2-2q_1}s_{d,1}\bigg(\frac{\log d}{n_2}\bigg)^{(1-q_1)/2}\bigg\}\notag\\
    &\quad\le
    \mathbb{P} \bigg\{|\hat\bGamma-\bGamma_{n_2}|_{\infty}\ge C_{0}\sqrt{\frac{\log(2d)}{n_1}}\bigg\}
    +\mathbb{P}\bigg\{|\bGamma_{n_2}-\bGamma|_{\infty}>\breve{C}_2\tilde{C}_{1}M_1^{2-2q_1}
s_{d,1}\bigg(\frac{\log d}{n_2}\bigg)^{(1-q_1)/2}\bigg\}\notag\\
    &\quad\le    \E\bigg[I(\mathcal{A}_{n_2}\cap\mathcal{C}_{n_2})\bar{\mathbb{P}} \bigg\{|\hat\bGamma-\bGamma_{n_2}|_{\infty}\ge C_{0}\sqrt{\frac{\log(2d)}{n_1}}\bigg\}\bigg]
    +\PP(\mathcal{A}_{n_2}^{\rm c}\cup\mathcal{C}_{n_2}^{\rm c})+8d^{-\tau_1}\notag\\
    &\quad\le 2^{1-\tau_2}d^{-\tau_2}+16d^{-\tau_1}
\end{align}
for $n_2\ge(2\tilde{C}_2C_{\min}^{-1}\tilde{C}_1s_{d,1})^{2/(1-q_1)}
M_1^4\log d$. If $\varrho_2\ge \|\bTheta_0\|_{1}|\hat\bGamma-\bGamma|_{\infty}$, by Equations (24) and (25) of \citeS{cai2011constrained_app}, we have $|\bI_{2d}-\hat\bGamma\bTheta_0|_{\infty}\le\varrho_2$ and $\|\hat\bTheta\|_1\le\|\bTheta_0\|_1$, which implies $|\hat\bGamma(\hat\bTheta-\bTheta_0)|_{\infty}\le|\hat\bGamma\hat\bTheta-\bI_{2d}|_{\infty}
+|\bI_{2d}-\hat\bGamma\bTheta_0|_{\infty}\le2\varrho_2$. Recall $\|\bTheta_0\|_1\le M_{2}$. Therefore, if $\varrho_2\ge \|\bTheta_0\|_{1}|\hat\bGamma-\bGamma|_{\infty}$, we have
\begin{align*}
|\bGamma(\hat\bTheta-\bTheta_0)|_{\infty}&\le|\hat\bGamma(\hat\bTheta-\bTheta_0)|_{\infty}
+|(\hat\bGamma-\bGamma)(\hat\bTheta-\bTheta_0)|_{\infty}\\
&\le2\varrho_2+|\hat\bGamma-\bGamma|_{\infty}\|\hat\bTheta-\bTheta_0\|_{1}\\
&\le2\varrho_2+2\|\bTheta_0\|_{1}|\hat\bGamma-\bGamma|_{\infty}\le4\varrho_2\,,
\end{align*}
which implies $|\hat\bTheta-\bTheta_0|_{\infty}\le
\|\bTheta_0\|_1|\bGamma(\hat\bTheta-\bTheta_0)|_{\infty}\le4M_{2}\varrho_2$. 
By Condition \ref{Sass:CLIME}, we know $\varrho_{2}\ge C_{0}M_{2}\{n_1^{-1}\log(2d)\}^{1/2}
+\breve{C}_2\tilde{C}_{1}M_{2}M_1^{2-2q_1}
s_{d,1}(n_2^{-1}\log d)^{(1-q_1)/2}$. Then, for any given $\tau_2>0$, by \eqref{Seq:clime2}, if $n_2\ge(2\tilde{C}_2C_{\min}^{-1}\tilde{C}_1s_{d,1})^{2/(1-q_1)}
M_1^4\log d$, we have
\begin{align}\label{Seq:clime3}
    &\mathbb{P}(|\hat\bTheta-\bTheta_0|_{\infty}\le4M_{2}\varrho_2)
    \ge\mathbb{P}(\varrho_2\ge \|\bTheta_0\|_{1}|\hat\bGamma-\bGamma|_{\infty})\notag\\
    &\quad\ge\mathbb{P}\bigg\{\varrho_2\ge \|\bTheta_0\|_{1}|\hat\bGamma-\bGamma|_{\infty},|\hat\bGamma-\bGamma|_{\infty}< C_{0}\sqrt{\frac{\log(2d)}{n_1}}+\breve{C}_2\tilde{C}_{1}M_1^{2-2q_1}s_{d,1}
    \bigg(\frac{\log d}{n_2}\bigg)^{(1-q_1)/2}\bigg\}\notag\\
    &\quad=\mathbb{P} \bigg\{|\hat\bGamma-\bGamma|_{\infty}< C_{0}\sqrt{\frac{\log(2d)}{n_1}}
    +\breve{C}_2\tilde{C}_{1}M_1^{2-2q_1}s_{d,1}\bigg(\frac{\log d}{n_2}\bigg)^{(1-q_1)/2}\bigg\}\notag\\
    &\quad\ge1- 2^{1-\tau_2}d^{-\tau_2}-16d^{-\tau_1}\,.
\end{align}
We complete the proof of Lemma \ref{Slemm_clime}. $\hfill\qedsymbol$

\subsection{Proof of Lemma \ref{Stheo_scaled}}\label{Sproof_theo_scaled}

Recall $c=(\tau_2+1)/2$ and $\tau_1\ge\tau_2$. Write $\sigma^*=n_1^{-1/2}|\by-\bZ\bgamma_0|_2$. Following the arguments of \eqref{eq:lem4-1} and \eqref{eq:lem4-2}, if $n_1 \ge \max\{v_0s_0\log(240eds_0^{-1}),12.5(\tau_2+1)\log(2d),5000(\tau_2+1)\tilde\kappa^{4}\log(2d)\}$ is sufficiently large and $n_2\ge(2\tilde{C}_2C_{\min}^{-1}\tilde{C}_1s_{d,1})^{2/(1-q_1)}M_1^4\log d$, by Lemmas \ref{Stheo_condition}(i), we have
\begin{align}\label{Seq:lem4-1}
    \mathbb P \bigg(\bigg| \frac{\hat{\sigma}}{\sigma} - 1 \bigg| \ge 16\varrho_3\sqrt{\frac{s_0}{\tilde C_{\min}}} \bigg)
    &\le \mathbb P \bigg\{\bigg| \frac{\hat{\sigma}}{\sigma} - 1 \bigg| \ge 16\varrho_3\sqrt{\frac{s_0}{\tilde C_{\min}}}, \bZ\in\mathcal{E}_{n_1}\bigg(\frac{\tilde C_{\min}^{1/2}}{2},s_{0},\frac{3\tilde C_{\max}}{2}\bigg)\bigg\}\notag \\   
    &\quad+ 2e^{-c_1n_1}+18d^{-\tau_2}\notag\\
    &\le  \mathbb P \bigg(\bigg|\frac{\sigma^*}{\sigma} - 1\bigg|\ge \frac{8\varrho_3s_0^{1/2}}{\tilde C_{\min}^{1/2}+8\varrho_3s_0^{1/2}}\bigg)
        +\mathbb P \bigg(\bigg| \frac{\sigma^*}{\sigma} - 1 \bigg| \ge v \bigg)\notag\\
        &\quad+ \mathbb P \bigg\{\bigg|\frac{\bZ^{\top}\beps}{n_1\sigma}\bigg|_{\infty} > \frac{(1-v)\varrho_3}{4},
         \bZ\in\mathcal{E}_{n_1}\bigg(\frac{\tilde C_{\min}^{1/2}}{2},s_{0},\frac{3\tilde C_{\max}}{2}\bigg)  \bigg\}\notag\\
        &\quad + 2e^{-c_1n_1}+18d^{-\tau_2}
\end{align}
for any $0<v<1$. Due to $\sigma^*=n_1^{-1/2}|\by-\bZ\bgamma_0|_2$ and $\beps=(\varepsilon_1,\ldots,\varepsilon_{n_1})^{\top}$, we have 
 $n_1(\sigma^*\sigma^{-1})^2-n_1=\sigma^{-2}|\beps|_2^{2}-n_1
 =\sum_{i=1}^{n_1}(\sigma^{-2}\varepsilon_i^2-1)$. By Condition \ref{Sass:model_error}, Lemma 5.14 and Remark 5.18 of \citeS{vershynin_2012_app}, for each $i\in[n_1]$, we have
$$\|\sigma^{-2}\varepsilon_i^2-1\|_{\psi_1}\le2\|\sigma^{-2}\varepsilon_i^2\|_{\psi_1}
 \le4\|\sigma^{-1}\varepsilon_i\|_{\psi_2}^2=4\sigma^{-2}\|\varepsilon_i\|_{\psi_2}^2
 \le4\sigma^{-2}\kappa_1^2\,.$$
Since $|x/y-1|\leq|(x/y)^2-1|$ for all $x, y>0$, by Lemma \ref{lem:Berns_ineq} with $\kappa'=4\sigma^{-2}\kappa_1^2$, we have 
\begin{align*}
        \mathbb P \bigg(\bigg|\frac{\sigma^*}{\sigma} - 1 \bigg| \ge v\bigg)
        &\le \mathbb P \bigg\{\bigg| n_1\bigg(\frac{\sigma^*}{\sigma}\bigg)^2 - n_1 \bigg| \ge n_1v \bigg\}\\
        &\le2\exp\bigg[ -\frac{n_1}{6} \min\bigg\{ \bigg(\frac{v\sigma^{2}}{4e\kappa_1^2}\bigg)^2,\,\frac{v\sigma^{2}}{4e\kappa_1^2}  \bigg\} \bigg]
\end{align*}
for any $0<v<1$. Due to $\varrho_3 \asymp \{n_1^{-1}\log(2d)\}^{1/2}$ and $s_0 \ll n_1/\log(2d)$, then
\begin{align*}
1\gg\frac{8\varrho_3s_0^{1/2}}{\tilde C_{\min}^{1/2}+8\varrho_3s_0^{1/2}}\asymp \varrho_3s_0^{1/2}\gg n_1^{-1/2}\,,
\end{align*}
which implies
\begin{align}\label{Seq:lem4-2}
\lim_{n_1\rightarrow \infty}\mathbb P \bigg(\bigg|\frac{\sigma^*}{\sigma} - 1\bigg|\ge \frac{8\varrho_3s_0^{1/2}}{\tilde C_{\min}^{1/2}+8\varrho_3s_0^{1/2}}\bigg)=0\,.
\end{align}
Selecting $v=n_1^{-1/4}$, then 
\begin{align}\label{Seq:lem4-3}
        \lim_{n_1\rightarrow \infty}\mathbb P \bigg(\bigg| \frac{\sigma^*}{\sigma} - 1 \bigg| \ge n_1^{-1/4} \bigg)=0\,.
\end{align}
Let $\bz^{(j)}$ denote the $j$-th column of $\bZ$ and $\hat\Gamma_{j,j}$ be the $j$-th diagonal element of $\hat\bGamma$. Notice that $|n_1^{-1/2}\sigma^{-1}\bz^{(j)}|_2^2=\sigma^{-2}\hat\Gamma_{j,j}$. By Condition \ref{Sass:model_error} and Proposition 5.10 of \citeS{vershynin_2012_app}, 
\begin{align*}
        \mathbb{P} \bigg(\bigg|\frac{\bZ^{\top}\beps}{n_1^{1/2}\sigma}\bigg|_{\infty} > v \,\bigg|\,\bZ \bigg)
        \le\sum_{j=1}^{2d}\mathbb{P} \bigg\{\bigg|\frac{\bz^{(j),\top}\beps}{n_1^{1/2}\sigma}\bigg| > v \,\bigg|\,\bZ \bigg\}
        \le e\sum_{j=1}^{2d} \exp\bigg(-\frac{C_4\sigma^2v^2}{\kappa_1^2\hat{\Gamma}_{j,j}}\bigg)\,,
\end{align*}
where $C_4>0$ denotes the absolute constant appeared in Proposition 5.10 of \citeS{vershynin_2012_app}. Selecting $\varrho_3 \ge 8C_4^{-1/2}\sigma^{-1}\kappa_1
\tilde C_{\max}^{1/2}\{n_1^{-1}\log(2d)\}^{1/2}$, then
\begin{align*}
        &\mathbb P \bigg\{\bigg|\frac{\bZ^{\top}\beps}{n_1\sigma}\bigg|_{\infty} > \frac{(1-n_1^{-1/4})\varrho_3}{4},
         \bZ\in\mathcal{E}_{n_1}\bigg(\frac{\tilde C_{\min}^{1/2}}{2},s_{0},\frac{3\tilde C_{\max}}{2}\bigg)  \bigg\}\\
         &~~~~~=\mathbb{E}\bigg[\mathbb{P} \bigg\{\bigg|\frac{\bZ^{\top}\beps}{n_1^{1/2}\sigma}\bigg|_{\infty} > \frac{n_1^{1/2}(1-n_1^{-1/4})\varrho_3}{4} \,\bigg|\,\bZ \bigg\}I\bigg\{\bZ\in\mathcal{E}_{n_1}\bigg(\frac{\tilde C_{\min}^{1/2}}{2},s_{0},\frac{3\tilde C_{\max}}{2}\bigg)\bigg\}\bigg]\\
         &~~~~~\le e\cdot\mathbb{E}\bigg[\sum_{j=1}^{2d} \exp\bigg[-\frac{C_4\sigma^2\{n_1^{1/2}(1-n_1^{-1/4})\varrho_3/4\}^2}{\kappa_1^2\hat{\Gamma}_{j,j}}\bigg]
         I\bigg\{\bZ\in\mathcal{E}_{n_1}\bigg(\frac{\tilde C_{\min}^{1/2}}{2},s_{0},\frac{3\tilde C_{\max}}{2}\bigg)\bigg\}\bigg]\\
        &~~~~~\le  2ed\exp\bigg[-\frac{C_4\sigma^2\{n_1^{1/2}(1-n_1^{-1/4})\varrho_3/4\}^2}{3\kappa_1^2\tilde C_{\max}/2}\bigg]\ll d^{-1}\,.
\end{align*}
Due to $s_0\log(2d) \ll n_1$ and $s_{d,1}^{2/(1-q_1)}\log d\ll n_2$, by \eqref{Seq:lem4-1}--\eqref{Seq:lem4-3}, we have
\begin{equation*}
\lim_{d\rightarrow \infty}\mathbb{P} \bigg(\bigg\lvert \frac{\hat{\sigma}}{\sigma} - 1 \bigg\rvert \ge 16\varrho_3\sqrt{\frac{s_0}{\tilde C_{\min}}}\bigg) = 0\,.
\end{equation*}
We complete the proof of Lemma \ref{Stheo_scaled}. $\hfill\qedsymbol$

\subsection{Proof of Lemma \ref{Stheo_random_trans}}\label{Sproof_theo_random_trans}
Following the arguments of the proof of Lemma \ref{theo_random_trans} in Section \ref{proof_theo_random_trans}, by Theorem \ref{Stheo_random}, we know that Lemma \ref{Stheo_random_trans} holds.
$\hfill\qedsymbol$

\subsection{Proof of Lemma \ref{Slem0}}\label{Sproof_lem0}
Write $\bOmega:=(\Omega_{i,j})_{2d\times2d}=\bT\bTheta_0 \bT^\top$ and $\hat\bLambda=(\hat\Lambda_{i,j})_{2d\times2d}=\bT \hat\bTheta^\top \hat{\bGamma} \hat\bTheta \bT^\top$. Let $\bOmega^{0}=(\Omega^{0}_{i,j})_{2d\times2d}$ with $\Omega^{0}_{i,j}=\Omega_{i,j}\Omega^{-1/2}_{i,i}\Omega^{-1/2}_{j,j}$, and $\hat\bLambda^{0}=(\hat\Lambda^{0}_{i,j})_{2d\times2d}$ with $\hat\Lambda^{0}_{i,j}=\hat\Lambda_{i,j}\hat\Lambda^{-1/2}_{i,i}\hat\Lambda^{-1/2}_{j,j}$. Consider the event
$\mathcal{G}_{n_1}=\{|\bOmega^{0}-\hat\bLambda^{0}|_{\infty}\le(\log d)^{-3}\}$. Recall $\varrho_{2}\asymp\{n_1^{-1}\log(2d)\}^{1/2}$. Parallel to the proof of Lemma \ref{lem0_condition} in Section \ref{proof_lem0_condition}, by Lemma \ref{Stheo_random_trans}(ii), if $\log d\ll n_{1}^{1/7}$, we have 
\begin{align}\label{Seq:lem0_0}
\mathbb P(\mathcal{G}_{n_1}^{\rm c})\lesssim d^{-\tau_2}\,.
\end{align}
Write $\tilde{\bw}_{n_1}=\sigma^{-1}{\rm diag}(\hat{\Lambda}_{1,1}^{-1/2},\ldots,\hat{\Lambda}_{2d,2d}^{-1/2})\bT\bw_{n_1}$ with \( \bw_{n_1}=(w_{1},\ldots,w_{2d})^{\top}=n_1^{-1/2}\hat{\bTheta}^{\top} \bZ^\top \beps \). Then $\tilde{w}_{j+d}=\sigma^{-1}\hat\Lambda_{j+d,j+d}^{-1/2}(w_{j}-w_{j+d})$ is $(j+d)$-th element of $\tilde{\bw}_{n_1}$. Due to ${\rm Var}(\sigma^{-1}\bT\bw_{n_1}\,|\,\bZ)=\hat\bLambda$, we have ${\rm Var}(\tilde{\bw}_{n_1}\,|\,\bZ)=\hat\bLambda^{0}$, which implies ${\rm Cov}(\tilde{w}_{i+d},\tilde{w}_{j+d}\,|\,\bZ)=\hat\Lambda^{0}_{i+d,j+d}$. Write $\tilde\bomega_{i,j,d}=(\tilde{w}_{i+d},\tilde{w}_{j+d})^{\top}$. Notice that $\Omega^{0}_{i+d,j+d}=0$ for $i\neq j$ and $\hat\Lambda^{0}_{i+d,i+d}=\Omega^{0}_{i+d,i+d}=1$ for $i\in[d]$. Given the event $\mathcal{G}_{n_1}$, we have
\begin{align}\label{Seq:lem0_1}
\|{\rm Var}(\tilde\bomega_{i,j,d}\,|\,\bZ)-\bI_2\|_2\le\|{\rm Var}(\tilde\bomega_{i,j,d}\,|\,\bZ)-\bI_2\|_1\le(\log d)^{-3}\,.
\end{align}
Recall $\beps=(\varepsilon_1,\ldots,\varepsilon_{n_1})^{\top}$. Let $\bz^{(k)}=(z^{(k)}_{1},\ldots,z^{(k)}_{n_1})^{\top}$ denote the $k$-th column of $\bZ\hat{\bTheta}$ for $k\in[2d]$. Since \( \bw_{n_1}=(w_{1},\ldots,w_{2d})^{\top} = n_1^{-1/2}\hat{\bTheta}^{\top} \bZ^\top \beps \), we have $w_{k}=n_1^{-1/2}\beps^\top \bz^{(k)}$ for $k\in[2d]$, which implies $\tilde{w}_{j+d}=n_1^{-1/2}\sigma^{-1}\hat\Lambda_{j+d,j+d}^{-1/2}\beps^\top \{\bz^{(j)}-\bz^{(j+d)}\}$ and $\E(\tilde{w}_{j+d}\,|\,\bZ)=0$ for $j\in[d]$. Therefore, 
\begin{align}\label{Seq:lem0_1_1}
\tilde\bomega_{i,j,d}&=n_1^{-1/2}\sigma^{-1}
\begin{pmatrix}
  \hat\Lambda_{i+d,i+d}^{-1/2} & 0 \\
  0 & \hat\Lambda_{j+d,j+d}^{-1/2} 
\end{pmatrix}
(
  \bz^{(i)}-\bz^{(i+d)} , \bz^{(j)}-\bz^{(j+d)} 
)^{\top}
\beps\notag\\
&=n_1^{-1/2}\sigma^{-1}\sum_{k=1}^{n_1}
\begin{pmatrix}
  \hat\Lambda_{i+d,i+d}^{-1/2} & 0 \\
  0 & \hat\Lambda_{j+d,j+d}^{-1/2} 
\end{pmatrix}
\begin{pmatrix}
  z^{(i)}_k-z^{(i+d)}_k\\
  z^{(j)}_k-z^{(j+d)}_k 
\end{pmatrix}
\varepsilon_{k}\notag\\
&=:n_1^{-1/2}\sigma^{-1}\sum_{k=1}^{n_1}\ba_{i,j,d,k}\varepsilon_{k}
=:\sum_{k=1}^{n_1}\bzeta^{(i,j,d)}_{k}\,.
\end{align}
Conditional on $\bZ$, it follows from Condition \ref{Sass:model_error} that  $\{\bzeta^{(i,j,d)}_{k}\}_{k=1}^{n_1}$ are 2-dimensional independent random vectors with mean zero and covariance matrix ${\rm Var}\{\bzeta^{(i,j,d)}_{k}\,|\,\bZ\}=n_1^{-1}\ba_{i,j,d,k}\ba_{i,j,d,k}^\top$. Write $\bZ=(\hat\bz_1,\ldots,\hat\bz_{n_1})^\top$ and $\hat{\bTheta}=(\hat\btheta_1,\ldots,\hat\btheta_{2d})$. Consider the event 
$$\mathcal{D}=\bigg\{\min_{i\in[2d]}|\hat\Lambda_{i,i}|\ge C_{\max}^{-1},\|\hat{\bTheta}\|_1\le M_2,
\max_{k\in[n_1]}|\hat\bz_k|_{\infty}\le\log (dn_1)\bigg\}\,.$$ 
Since $\bOmega=\bT\bTheta_0 \bT^\top$ and $\sigma_{\min}(\bTheta_0)\ge C_{\max}^{-1}$, we have $\sigma_{\min}(\bOmega)\ge2C_{\max}^{-1}$, which implies $\Omega_{i,i}\ge2C_{\max}^{-1}$ for $i\in[2d]$. Due to $\varrho_{2}\asymp\{n_1^{-1}\log(2d)\}^{1/2}$, by Lemma \ref{Stheo_random_trans}(ii), we have
\begin{align}\label{Seq:lem0_1_2}
\mathbb{P}\bigg(\min_{i\in[2d]}|\hat\Lambda_{i,i}|< C_{\max}^{-1}\bigg)
\le\mathbb{P}(|\bOmega-\hat\bLambda|_{\infty}> C_{\max}^{-1})
\lesssim  d^{-\tau_2}
\end{align}
provided that $\log d\ll n_1$. As shown in Equation (25) of \citeS{cai2011constrained_app}, if $\varrho_2\ge \|\bTheta_0\|_{1}|\hat\bGamma-\bGamma|_{\infty}$, then $\|\hat\bTheta\|_1 \le M_{2}$. Due to $\tau_1\ge\tau_2$, by \eqref{Seq:clime3}, we have
\begin{align}\label{Seq:lem0_1_3}
    &\mathbb{P}(\|\hat{\bTheta}\|_1> M_2)
    \le\mathbb{P}(\varrho_2< \|\bTheta_0\|_{1}|\hat\bGamma-\bGamma|_{\infty})\lesssim  d^{-\tau_2}\,.
\end{align}
Recall $\mathcal{X}_{n_2}=\sigma(\bx_{n_1+1},\ldots,\bx_{n})$ and $\mathcal{C}_{n_2}=\{\omega_{n_2}\le2^{-1}C_{\max}^{-1}\}$ with $\omega_{n_2}=\|\hat{\bOmega}-\bOmega_0\|_2$. By \eqref{Seq:lem_con3_0}, we have $\PP(\mathcal{C}_{n_2}^{\rm c})\lesssim d^{-\tau_2}$. As shown in the proof of Lemma \ref{Slemm_clime} in Section \ref{Sproof_lemm_clime}, given the event $\mathcal{C}_{n_2}$, we have
$\max_{k\in[n_1]}\| \hat\bz_k\|_{\psi_{2}\,|\,\mathcal{X}_{n_2}}\le \tilde C_3$. Write $\hat\bz_k=(\hat z_{k,1},\ldots,\hat z_{k,2d})^\top$. Given the event $\mathcal{C}_{n_2}$, by Lemma 5.5 of \citeS{vershynin_2012_app}, we have
\begin{align*}
&\PP\bigg\{\max_{k\in[n_1]}|\hat\bz_k|_{\infty}> \log(dn_1)\,\bigg|\,\mathcal{X}_{n_2}\bigg\}
\le\PP\bigg\{\max_{k\in[n_1]}\max_{i\in[2d]}|\hat z_{k,i}|>\log(dn_1)\,\bigg|\,\mathcal{X}_{n_2}\bigg\}\\
&\qquad\le\sum_{k=1}^{n_1}\sum_{i=1}^{2d}\PP\big\{|\hat z_{k,i}|>\log(dn_1)\,|\,\mathcal{X}_{n_2}\big\}
\le 2e n_1 d\exp[-C_5\{\log(dn_1)\}^2]
\lesssim d^{-\tau_2}\,,
\end{align*}
where $C_5$ denotes a absolute constant appeared in Lemma 5.5 of \citeS{vershynin_2012_app}. Then
\begin{align*}
\PP\bigg\{\max_{k\in[n_1]}|\hat\bz_k|_{\infty}>\log(dn_1)\bigg\}
\le\E\bigg[I(\mathcal{C}_{n_2})\PP\bigg\{\max_{k\in[n_1]}|\hat\bz_k|_{\infty}>\log(dn_1)\,\bigg|\,\mathcal{X}_{n_2}\bigg\}\bigg]+\PP(\mathcal{C}_{n_2}^{\rm c})\lesssim d^{-\tau_2}\,.
\end{align*}
Together with \eqref{Seq:lem0_1_2} and \eqref{Seq:lem0_1_3}, we have
\begin{align}\label{Seq:lem0_1_4}
\PP(\mathcal{D}^{\rm c})\lesssim d^{-\tau_2}
\end{align}
provided that $\log d\ll n_1$. Notice that $z^{(j)}_{k}=\hat\bz_k^\top\hat\btheta_j$ for $j\in[2d]$ and $k\in[n_1]$. Given the event $\mathcal{D}$, we have 
\begin{align}\label{Seq:Slem0_1}
\max_{i,j\in[d]}|\ba_{i,j,d,k}|_{\infty}&\le \bigg(\min_{i\in[2d]}|\hat\Lambda_{i,i}|\bigg)^{-1/2}\max_{k\in[n_1]}\max_{j\in[d]}|\hat\bz_k^\top(\hat\btheta_j-\hat\btheta_{j+d})|\notag\\
&\le\bigg(\min_{i\in[2d]}|\hat\Lambda_{i,i}|\bigg)^{-1/2}\max_{k\in[n_1]}|\hat\bz_k|_{\infty}\max_{j\in[d]}\|\hat\btheta_j-\hat\btheta_{j+d}\|_1\\
&\le 2\bigg(\min_{i\in[2d]}|\hat\Lambda_{i,i}|\bigg)^{-1/2}
\|\hat{\bTheta}\|_1\max_{k\in[n_1]}|\hat\bz_k|_{\infty}
\le2 C_{\max}^{1/2}M_2\log(dn_1)\,.\notag
\end{align}
Let
\begin{align*}
\hat\bzeta^{(i,j,d)}_{k}=\bzeta^{(i,j,d)}_{k}I\{|\bzeta^{(i,j,d)}_{k}|_2\le (\log d)^{-4}\}
-\E\big[\bzeta^{(i,j,d)}_{k}I\{|\bzeta^{(i,j,d)}_{k}|_2\le (\log d)^{-4}\}\,|\,\bZ\big]
\end{align*}
and $\tilde\bzeta^{(i,j,d)}_{k}=\bzeta^{(i,j,d)}_{k}-\hat\bzeta^{(i,j,d)}_{k}$ for $k\in[n_1]$. Write $\bzeta^{(i,j,d)}_{k}=(\zeta^{(i,j,d)}_{k,1},\zeta^{(i,j,d)}_{k,2})^{\top}$ and $\hat\bzeta^{(i,j,d)}_{k}=(\hat\zeta^{(i,j,d)}_{k,1},\hat\zeta^{(i,j,d)}_{k,2})^{\top}$. Due to $|\sum_{k=1}^{n_1}\zeta^{(i,j,d)}_{k,l}|\le|\sum_{k=1}^{n_1}\hat\zeta^{(i,j,d)}_{k,l}|
+|\sum_{k=1}^{n_1}\tilde\bzeta^{(i,j,d)}_{k}|_2$ for $l\in[2]$, we have 
\begin{align}\label{Seq:Slem0_2}
&\mathbb{P}\bigg\{\bigg|\sum_{k=1}^{n_1}\zeta^{(i,j,d)}_{k,1}\bigg|\ge  t,\bigg|\sum_{k=1}^{n_1}\zeta^{(i,j,d)}_{k,2}\bigg|\ge  t\,\bigg|\,\bZ\bigg\}\notag\\
&\quad\le\mathbb{P}\bigg\{\bigg|\sum_{k=1}^{n_1}\hat\zeta^{(i,j,d)}_{k,1}\bigg|\ge  t-(\log d)^{-2},\bigg|\sum_{k=1}^{n_1}\hat\zeta^{(i,j,d)}_{k,2}\bigg|\ge t-(\log d)^{-2}\,\bigg|\,\bZ\bigg\}\notag\\
&\qquad+\PP\bigg\{\bigg|\sum_{k=1}^{n_1}\tilde\bzeta^{(i,j,d)}_{k}\bigg|_2\ge (\log d)^{-2}\,\bigg|\,\bZ\bigg\}\,.
\end{align}
Given the event $\mathcal{D}$, by \eqref{Seq:lem0_1_1} and \eqref{Seq:Slem0_1}, we have
\begin{align}\label{Seq:Slem0_3}
\Delta&:=\sum_{k=1}^{n_1}\E\big[|\bzeta^{(i,j,d)}_{k}|_2I\{|\bzeta^{(i,j,d)}_{k}|_2> (\log d)^{-4}\}\,|\,\bZ\big]\notag\\
&~=n_1^{-1/2}\sigma^{-1}
\sum_{k=1}^{n_1}\E\big[|\ba_{i,j,d,k}\varepsilon_{k}|_2I\{|\ba_{i,j,d,k}\varepsilon_{k}|_2> \sigma n_1^{1/2}(\log d)^{-4}\}\,|\,\bZ\big]\\
&~\lesssim n_1^{-1/2}\log(dn_1)
\sum_{k=1}^{n_1}\E\big[|\varepsilon_{k}|I\{2\sqrt{2} C_{\max}^{1/2}M_2\log(dn_1)|\varepsilon_{k}|> \sigma n_1^{1/2}(\log d)^{-4}\}\,|\,\bZ\big]\notag\\
&~\lesssim n_1^{-1/2}\log(dn_1)
\sum_{k=1}^{n_1}\E\big(|\varepsilon_{k}|I\big[|\varepsilon_{k}|> 4^{-1}\sqrt{2} C_{\max}^{-1/2}M_2^{-1}\sigma n_1^{1/2}\{\log(dn_1)\}^{-5}\big]\big)\,.\notag
\end{align}
Due to $\|\beps\|_{\psi_{2}}= \kappa_1$, for any given $m>0$, we have $\max_{k\in[n_1]}\mathbb E(|\varepsilon_{k}|^m)=O(1)$, which implies
\begin{align*}
&\sum_{k=1}^{n_1}\E\big(|\varepsilon_{k}|I\big[|\varepsilon_{k}|> 4^{-1}\sqrt{2} C_{\max}^{-1/2}M_2^{-1}\sigma n_1^{1/2}\{\log(dn_1)\}^{-5}\big]\big)\\
&\qquad\le\big[4^{-1}\sqrt{2} C_{\max}^{-1/2}M_2^{-1}\sigma n_1^{1/2}\{\log(dn_1)\}^{-5}\big]^{-5}\sum_{k=1}^{n_1}\E(|\varepsilon_{k}|^6)
\lesssim n_1^{-3/2}\{\log(dn_1)\}^{25}\,.
\end{align*}
Therefore, given the event $\mathcal{D}$, by \eqref{Seq:Slem0_3}, we have
\begin{align}\label{Seq:Slem0_4}
&\bigg|\sum_{k=1}^{n_1}\E\big[\bzeta^{(i,j,d)}_{k}I\{|\bzeta^{(i,j,d)}_{k}|_2> (\log d)^{-4}\}\,|\,\bZ\big]\bigg|_2
\le \sum_{k=1}^{n_1}\big|\E\big[\bzeta^{(i,j,d)}_{k}I\{|\bzeta^{(i,j,d)}_{k}|_2> (\log d)^{-4}\}\,|\,\bZ\big]\big|_2\notag\\
&\qquad\le\Delta\lesssim n_1^{-2}\{\log(dn_1)\}^{26}
=o\{(\log d)^{-2}\}
\end{align}
provided that $\log d\ll n_1^{1/14}$. Given the event $\mathcal{D}$, if $\Delta\le 2^{-1}(\log d)^{-2}$, by \eqref{Seq:Slem0_1}, we have
\begin{align}\label{Seq:Slem0_4_1}
&\PP\bigg\{\bigg|\sum_{k=1}^{n_1}\tilde\bzeta^{(i,j,d)}_{k}\bigg|_2\ge (\log d)^{-2}\,\bigg|\,\bZ\bigg\}\notag\\
&\qquad\le\PP\bigg\{\bigg|\sum_{k=1}^{n_1}\bzeta^{(i,j,d)}_{k}I\{|\bzeta^{(i,j,d)}_{k}|_2> (\log d)^{-4}\}\bigg|_2\ge 2^{-1}(\log d)^{-2}\,\bigg|\,\bZ\bigg\}\notag\\
&\qquad \le\sum_{k=1}^{n_1}\PP\big\{|\bzeta^{(i,j,d)}_{k}|_2> (\log d)^{-4}\,|\,\bZ\big\}
=\sum_{k=1}^{n_1}\PP\big\{n_1^{-1/2}\sigma^{-1}|\ba_{i,j,d,k}\varepsilon_{k}|_2> (\log d)^{-4}\,|\,\bZ\big\}\notag\\
&\qquad\le
\sum_{k=1}^{n_1}\PP\big[|\varepsilon_{k}|> 4^{-1}\sqrt{2}\sigma C_{\max}^{-1/2}M_2^{-1} n_1^{1/2}\{\log(dn_1)\}^{-5}\,|\,\bZ\big]\notag\\
&\qquad=
\sum_{k=1}^{n_1}\PP\big[|\varepsilon_{k}|> 4^{-1}\sqrt{2}\sigma C_{\max}^{-1/2}M_2^{-1} n_1^{1/2}\{\log(dn_1)\}^{-5}\big]
\end{align}
By Lemma 5.5 of \citeS{vershynin_2012_app}, we have
\begin{align*}
&\PP\big\{|\varepsilon_{k}|> C n_1^{1/2}\{\log(dn_1)\}^{-5}\big\}
\le e \exp\big[-C_5C^2 n_1\{\log(dn_1)\}^{-10}\big]
\end{align*}
for any $C>0$, where $C_5$ denotes a absolute constant appeared in Lemma 5.5 of \citeS{vershynin_2012_app}. Together with \eqref{Seq:Slem0_4} and \eqref{Seq:Slem0_4_1}, given the event $\mathcal{D}$, we have
\begin{align}\label{Seq:Slem0_3_1}
\PP\bigg\{\bigg|\sum_{k=1}^{n_1}\tilde\bzeta^{(i,j,d)}_{k}\bigg|_2\ge (\log d)^{-2}\,\bigg|\,\bZ\bigg\}
\le O\{(\log d)^{-3/2}G^2(t)\}
\end{align}
uniformly over $0\le t\le2(\log d)^{1/2}$, provided that $\log d\ll n_1^{1/14}$. Therefore, given the event $\mathcal{D}$, if $\log d\ll n_1^{1/14}$, by \eqref{Seq:Slem0_2}, we have
\begin{align}\label{Seq:Slem0_5}
&\mathbb{P}\bigg\{\bigg|\sum_{k=1}^{n_1}\zeta^{(i,j,d)}_{k,1}\bigg|\ge  t,\bigg|\sum_{k=1}^{n_1}\zeta^{(i,j,d)}_{k,2}\bigg|\ge  t\,\bigg|\,\bZ\bigg\}\notag\\
&\qquad\le\mathbb{P}\bigg\{\bigg|\sum_{k=1}^{n_1}\hat\zeta^{(i,j,d)}_{k,1}\bigg|\ge  t-(\log d)^{-2},\bigg|\sum_{k=1}^{n_1}\hat\zeta^{(i,j,d)}_{k,2}\bigg|\ge t-(\log d)^{-2}\,\bigg|\,\bZ\bigg\}\notag\\
&\qquad\quad+O\{(\log d)^{-3/2}G^2(t)\}
\end{align}
uniformly over $0\le t\le2(\log d)^{1/2}$. Similarly, given the event $\mathcal{D}$, if $\log d\ll n_1^{1/14}$, 
\begin{align}\label{Seq:Slem0_5_1}
&\mathbb{P}\bigg\{\bigg|\sum_{k=1}^{n_1}\zeta^{(i,j,d)}_{k,1}\bigg|\ge  t,\bigg|\sum_{k=1}^{n_1}\zeta^{(i,j,d)}_{k,2}\bigg|\ge  t\,\bigg|\,\bZ\bigg\}\notag\\
&\qquad\ge\mathbb{P}\bigg\{\bigg|\sum_{k=1}^{n_1}\hat\zeta^{(i,j,d)}_{k,1}\bigg|\ge  t+(\log d)^{-2},\bigg|\sum_{k=1}^{n_1}\hat\zeta^{(i,j,d)}_{k,2}\bigg|\ge t+(\log d)^{-2}\,\bigg|\,\bZ\bigg\}\notag\\
&\qquad\quad-O\{(\log d)^{-3/2}G^2(t)\}
\end{align}
uniformly over $0\le t\le2(\log d)^{1/2}$. Let $\bu=(u_1,u_2)^{\top}$ with $\bu \,|\,\bZ\sim\mathcal{N}({\bf0},{\rm Var}\{\sum_{k=1}^{n_1}\hat\bzeta^{(i,j,d)}_{k}\,|\,\bZ\})$. By Theorem 1 in \citeS{1987On}, we have
\begin{align}
&\mathbb{P}\bigg\{\bigg|\sum_{k=1}^{n_1}\hat\zeta^{(i,j,d)}_{k,1}\bigg|\ge  t-(\log d)^{-2},\bigg|\sum_{k=1}^{n_1}\hat\zeta^{(i,j,d)}_{k,2}\bigg|\ge t-(\log d)^{-2}\,\bigg|\,\bZ\bigg\}\notag\\
&\quad\le\mathbb{P}\big\{|u_1|\ge t-2(\log d)^{-2},|u_2|\ge t-2(\log d)^{-2}\,|\,\bZ\big\}+c_{1,1}\exp\{-c_{1,2}(\log d)^2\}\,,\label{Seq:Slem0_6}\\
&\mathbb{P}\bigg\{\bigg|\sum_{k=1}^{n_1}\hat\zeta^{(i,j,d)}_{k,1}\bigg|\ge  t+(\log d)^{-2},\bigg|\sum_{k=1}^{n_1}\hat\zeta^{(i,j,d)}_{k,2}\bigg|\ge t+(\log d)^{-2}\,\bigg|\,\bZ\bigg\}\notag\\
&\quad\ge\mathbb{P}\big\{|u_1|\ge t+2(\log d)^{-2},|u_2|\ge t+2(\log d)^{-2}\,|\,\bZ\big\}-c_{1,1}\exp\{-c_{1,2}(\log d)^2\}\,,\label{Seq:Slem0_7}
\end{align}
where $c_{1,1}>0$ and $c_{1,2}>0$ are two absolute constants. Recall $\hat\bzeta^{(i,j,d)}_{k}=\bzeta^{(i,j,d)}_{k}I\{|\bzeta^{(i,j,d)}_{k}|_2\le (\log d)^{-4}\}
-\E[\bzeta^{(i,j,d)}_{k}I\{|\bzeta^{(i,j,d)}_{k}|_2\le (\log d)^{-4}\}\,|\,\bZ]$ and $\E\{\bzeta^{(i,j,d)}_{k}\,|\,\bZ\}=0$. Given the event $\mathcal{D}$, we have
\begin{align*}
&\bigg\|{\rm Var}\bigg\{\sum_{k=1}^{n_1}\hat\bzeta^{(i,j,d)}_{k}\,\bigg|\,\bZ\bigg\}-{\rm Var}\bigg\{\sum_{k=1}^{n_1}\bzeta^{(i,j,d)}_{k}\,\bigg|\,\bZ\bigg\}\bigg\|_2\notag\\
&\quad=\bigg\|\sum_{k=1}^{n_1}\big[\E\{\bzeta^{(i,j,d)}_{k}\bzeta^{(i,j,d),\top}_{k}I\{|\bzeta^{(i,j,d)}_{k}|_2> (\log d)^{-4}\}\,|\,\bZ\}\notag\\
&\quad\qquad\qquad+\E[\bzeta^{(i,j,d)}_{k}I\{|\bzeta^{(i,j,d)}_{k}|_2> (\log d)^{-4}\}\,|\,\bZ]
(\E[\bzeta^{(i,j,d)}_{k}I\{|\bzeta^{(i,j,d)}_{k}|_2> (\log d)^{-4}\}\,|\,\bZ])^\top\big]\bigg\|_2\notag\\
&\quad\le2\sum_{k=1}^{n_1}\E\{|\bzeta^{(i,j,d)}_{k}|_2^2I\{|\bzeta^{(i,j,d)}_{k}|_2> (\log d)^{-4}\}\,|\,\bZ\}
=o\{(\log d)^{-3}\}
\end{align*}
provided that $\log d\ll n_1^{1/12}$, where the last step is based on the similar arguments of \eqref{Seq:Slem0_4}. Recall $\mathcal{G}_{n_1}=\{|\bOmega^{0}-\hat\bLambda^{0}|_{\infty}\le(\log d)^{-3}\}$. Given the event $\mathcal{G}_{n_1}\cap\mathcal{D}$, by \eqref{Seq:lem0_1} and \eqref{Seq:lem0_1_1}, it holds that
\begin{align}\label{Seq:Slem0_7_1}
\bigg\|{\rm Var}\bigg\{\sum_{k=1}^{n_1}\hat\bzeta^{(i,j,d)}_{k}\,\bigg|\,\bZ\bigg\}-\bI_2\bigg\|_2
=O\{(\log d)^{-3}\}
\end{align}
provided that $\log d\ll n_1^{1/12}$. Given the event $\mathcal{G}_{n_1}\cap\mathcal{D}$, based on \eqref{Seq:Slem0_7_1}, with the density of multivariate normal random variable, if $\log d\ll n_1^{1/12}$, it is easy to show that
\begin{align*}
\mathbb{P}\big\{|u_1|\ge t-2(\log d)^{-2},|u_2|\ge t-2(\log d)^{-2}\,|\,\bZ\big\}
\le\{1+O(\log d)^{-3/2}\}G^2(t)
\end{align*}
uniformly over $0\le t\le2(\log d)^{1/2}$. On the other hand, notice that
$c_{1,1}\exp\{-c_{1,2}(\log d)^2\}= O\{(\log d)^{-2}\}G^2(t)$ uniformly over $0\le t\le2(\log d)^{1/2}$. Therefore, given the event $\mathcal{G}_{n_1}\cap\mathcal{D}$, if $\log d\ll n_1^{1/12}$, by \eqref{Seq:Slem0_6}, we have
\begin{align*}
&\mathbb{P}\bigg\{\bigg|\sum_{k=1}^{n_1}\hat\zeta^{(i,j,d)}_{k,1}\bigg|\ge  t-(\log d)^{-2},\bigg|\sum_{k=1}^{n_1}\hat\zeta^{(i,j,d)}_{k,2}\bigg|\ge t-(\log d)^{-2}\,\bigg|\,\bZ\bigg\}\\
&\qquad\le\{1+O(\log d)^{-3/2}\}G^2(t)
\end{align*}
uniformly over $0\le t\le2(\log d)^{1/2}$. Similarly, given the event $\mathcal{G}_{n_1}\cap\mathcal{D}$, if $\log d\ll n_1^{1/12}$, by \eqref{Seq:Slem0_7}, we have
\begin{align}\label{Seq:lem0_7_1}
&\mathbb{P}\bigg\{\bigg|\sum_{k=1}^{n_1}\hat\zeta^{(i,j,d)}_{k,1}\bigg|\ge  t+(\log d)^{-2},\bigg|\sum_{k=1}^{n_1}\hat\zeta^{(i,j,d)}_{k,2}\bigg|\ge t+(\log d)^{-2}\,\bigg|\,\bZ\bigg\}\notag\\
&\qquad\ge\{1-O(\log d)^{-3/2}\}G^2(t)
\end{align}
uniformly over $0\le t\le2(\log d)^{1/2}$. Recall $\tilde{w}_{i+d}=\sum_{k=1}^{n_1}\zeta^{(i,j,d)}_{k,1}$ and $\tilde{w}_{j+d}=\sum_{k=1}^{n_1}\zeta^{(i,j,d)}_{k,2}$. Given the event $\mathcal{G}_{n_1}\cap\mathcal{D}$, together with \eqref{Seq:Slem0_5} and \eqref{Seq:Slem0_5_1}, we have
\begin{align}\label{Seq:lem0_8}
\max_{i,j\in[d]:\,i\neq j}\sup_{0\le t\le 2\sqrt{\log d}}\bigg|\frac{\mathbb{P}(|\tilde{w}_{i+d}|>t, |\tilde{w}_{j+d}|>t\,|\,\bZ)}{G^2(t)}-1\bigg|
\leq O\{(\log d)^{-3/2}\}
\end{align}
provided that $\log d\ll n_1^{1/14}$. Parallel to the arguments for \eqref{Seq:lem0_8}, given the event $\mathcal{G}_{n_1}\cap\mathcal{D}$, it also holds that
\begin{align}\label{Seq:lem0_9}
\max_{i\in[d]}\sup_{0\le t\le 2\sqrt{\log d}}\bigg|\frac{\mathbb{P}(|\tilde{w}_{i+d}|>t\,|\,\bZ)}{G(t)}-1\bigg|
\leq O\{(\log d)^{-3/2}\}
\end{align}
provided that $\log d\ll n_1^{1/14}$. By \eqref{Seq:lem0_0} and \eqref{Seq:lem0_1_4}, we have $\PP(\mathcal{G}_{n_1}^{\rm c})+\PP(\mathcal{D}^{\rm c})\lesssim d^{-\tau_2}$ provided that $\log d\ll n_{1}^{1/7}$. Parallel to the arguments for \eqref{lem0-1} and the proof of Lemma \ref{lem0} in Section \ref{proof_lem0}, by \eqref{Seq:lem0_8} and \eqref{Seq:lem0_9}, we know Lemma \ref{Slem0} holds. $\hfill\qedsymbol$

\subsection{Proof of Lemma \ref{Slem1}}\label{Sproof_lem1}

Recall $\bOmega=(\Omega_{i,j})_{2d\times2d}=\bT\bTheta_0 \bT^\top$, $\hat\bLambda=(\hat\Lambda_{i,j})_{2d\times2d}=\bT \hat\bTheta^\top \hat{\bGamma} \hat\bTheta \bT^\top$, $\bOmega^{0}=(\Omega^{0}_{i,j})_{2d\times2d}$ with $\Omega^{0}_{i,j}=\Omega_{i,j}\Omega^{-1/2}_{i,i}\Omega^{-1/2}_{j,j}$, $\hat\bLambda^{0}=(\hat\Lambda^{0}_{i,j})_{2d\times2d}$ with $\hat\Lambda^{0}_{i,j}=\hat\Lambda_{i,j}\hat\Lambda^{-1/2}_{i,i}\hat\Lambda^{-1/2}_{j,j}$, and $\tilde{\bw}_{n_1}=(\tilde{w}_{1},\ldots,\tilde{w}_{2d})^{\top}=\sigma^{-1}{\rm diag}(\hat{\Lambda}_{1,1}^{-1/2},\ldots,\hat{\Lambda}_{2d,2d}^{-1/2})\bT\bw_{n_1}$ satisfies ${\rm Var}(\tilde{\bw}_{n_1}\,|\,\bZ)=\hat\bLambda^{0}$. Let $\tilde{\bw}_{i,j,d}=(\tilde{w}_{i},\tilde{w}_{j},\tilde{w}_{i+d},\tilde{w}_{j+d})^{\top}$ for $i,j\in[d]$ with $i\neq j$, and write 
\begin{align}\label{Seq:Slem1_0}
  \bOmega^{0}_{i,j,d}:=
  \begin{pmatrix}
  \Omega^{0}_{i,i}&\Omega^{0}_{i,j}&\Omega^{0}_{i,i+d}&\Omega^{0}_{i,j+d}\\
  \Omega^{0}_{j,i}&\Omega^{0}_{j,j}&\Omega^{0}_{j,i+d}&\Omega^{0}_{j,j+d}\\
  \Omega^{0}_{i+d,i}&\Omega^{0}_{i+d,j}&\Omega^{0}_{i+d,i+d}&\Omega^{0}_{i+d,j+d}\\
  \Omega^{0}_{j+d,i}&\Omega^{0}_{j+d,j}&\Omega^{0}_{j+d,i+d}&\Omega^{0}_{j+d,j+d}
  \end{pmatrix}
  =  
  \begin{pmatrix}
  1&\Omega^{0}_{i,j}&0&0\\
  \Omega^{0}_{j,i}&1&0&0\\
  0&0&1&0\\
  0&0&0&1
  \end{pmatrix}\,.
\end{align}
Write $\bZ=(\hat\bz_1,\ldots,\hat\bz_{n_1})^\top$. Consider two events $\mathcal{G}_{n_1}=\{|\bOmega^{0}-\hat\bLambda^{0}|_{\infty}\le(\log d)^{-3}\}$ and 
$$\mathcal{D}=\bigg\{\min_{i\in[2d]}|\hat\Lambda_{i,i}|\ge C_{\max}^{-1},\|\hat{\bTheta}\|_1\le M_2,
\max_{k\in[n_1]}|\hat\bz_k|_{\infty}\le\log (dn_1)\bigg\}\,.$$ 
Given the event $\mathcal{G}_{n_1}$, we have
\begin{align}\label{Seq:lem1_1}
\|{\rm Var}(\tilde\bw_{i,j,d}\,|\,\bZ)-\bOmega^{0}_{i,j,d}\|_2\le\|{\rm Var}(\tilde\bw_{i,j,d}\,|\,\bZ)-\bOmega^{0}_{i,j,d}\|_1\le3(\log d)^{-3}\,.
\end{align}
Recall that $\bz^{(k)}=(z^{(k)}_{1},\ldots,z^{(k)}_{n_1})^{\top}$ is the $k$-th column of $\bZ\hat{\bTheta}$, and \( \bw_{n_1}=(w_{1},\ldots,w_{2d})^{\top} = n_1^{-1/2}\hat{\bTheta}^{\top} \bZ^\top \beps \) with $\beps=(\varepsilon_1,\ldots,\varepsilon_{n_1})^{\top}$. Parallel to the arguments for \eqref{Seq:lem0_1_1}, we have
\begin{align*}
\tilde\bw_{i,j,d}
&=\sum_{k=1}^{n_1}n_1^{-1/2}\sigma^{-1}
\begin{pmatrix}
  \hat\Lambda_{i,i}^{-1/2} & 0 & 0 & 0 \\
  0 & \hat\Lambda_{j,j}^{-1/2} & 0 & 0 \\
  0 & 0 & \hat\Lambda_{i+d,i+d}^{-1/2} & 0 \\
  0 & 0 & 0 & \hat\Lambda_{j+d,j+d}^{-1/2} 
\end{pmatrix}
\begin{pmatrix}
  z^{(i)}_k+z^{(i+d)}_k\\
  z^{(j)}_k+z^{(j+d)}_k\\ 
  z^{(i)}_k-z^{(i+d)}_k\\
  z^{(j)}_k-z^{(j+d)}_k 
\end{pmatrix}
\varepsilon_{k}\\
&=:n_1^{-1/2}\sigma^{-1}\sum_{k=1}^{n_1}\bb_{i,j,d,k}\varepsilon_{k}
=:\sum_{k=1}^{n_1}\bxi^{(i,j,d)}_{k}\,.\notag
\end{align*}
Conditional on $\bZ$, by Condition \ref{Sass:model_error}, $\{\bxi^{(i,j,d)}_{k}\}_{k=1}^{n_1}$ are independent 4-dimensional random vectors with mean zero and covariance matrix ${\rm Var}\{\bxi^{(i,j,d)}_{k}\,|\,\bZ\}=n_1^{-1}\bb_{i,j,d,k}\bb_{i,j,d,k}^\top$. Let
\begin{align*}
\hat\bxi^{(i,j,d)}_{k}=\bxi^{(i,j,d)}_{k}I\{|\bxi^{(i,j,d)}_{k}|_2\le (\log d)^{-4}\}
-\E\big[\bxi^{(i,j,d)}_{k}I\{|\bxi^{(i,j,d)}_{k}|_2\le (\log d)^{-4}\}\,|\,\bZ\big]
\end{align*}
and $\tilde\bxi^{(i,j,d)}_{k}=\bxi^{(i,j,d)}_{k}-\hat\bxi^{(i,j,d)}_{k}$ for $k\in[n_1]$. Parallel to  the arguments for \eqref{Seq:Slem0_7_1}, given the event $\mathcal{G}_{n_1}\cap\mathcal{D}$, by \eqref{Seq:lem1_1}, it holds that
\begin{align}\label{Seq:Slem1_2}
\bigg\|{\rm Var}\bigg\{\sum_{k=1}^{n_1}\hat\bxi^{(i,j,d)}_{k}\,\bigg|\,\bZ\bigg\}-\bOmega^{0}_{i,j,d}\bigg\|_2
=O\{(\log d)^{-3}\}
\end{align}
provided that $\log d\ll n_1^{1/12}$. Write $\bxi^{(i,j,d)}_{k}=(\xi^{(i,j,d)}_{k,1},\ldots,\xi^{(i,j,d)}_{k,4})^{\top}$, $\hat\bxi^{(i,j,d)}_{k}=(\hat\xi^{(i,j,d)}_{k,1},\ldots,\hat\xi^{(i,j,d)}_{k,4})^{\top}$, $\breve \nu_{l,\ell}=\nu_{l}-\ell(\log d)^{-2}$ for $l,\ell\in[2]$. Due to $|\sum_{k=1}^{n_1}\xi^{(i,j,d)}_{k,l}|\le|\sum_{k=1}^{n_1}\hat\xi^{(i,j,d)}_{k,l}|
+|\sum_{k=1}^{n_1}\tilde\bxi^{(i,j,d)}_{k}|_2$ for $l\in[4]$, we have 
\begin{align}\label{Seq:Slem1_3}
&\mathbb{P}\bigg\{\bigg|\sum_{k=1}^{n_1}\xi^{(i,j,d)}_{k,1}\bigg|\ge \nu_{1},\bigg|\sum_{k=1}^{n_1}\xi^{(i,j,d)}_{k,2}\bigg|\ge \nu_{1},
\bigg|\sum_{k=1}^{n_1}\xi^{(i,j,d)}_{k,3}\bigg|\ge \nu_{2},\bigg|\sum_{k=1}^{n_1}\xi^{(i,j,d)}_{k,4}\bigg|\ge \nu_{2}\,\bigg|\,\bZ\bigg\}\notag\\
&\quad\le\mathbb{P}\bigg\{\bigg|\sum_{k=1}^{n_1}\hat\xi^{(i,j,d)}_{k,1}\bigg|\ge \breve\nu_{1,1},\bigg|\sum_{k=1}^{n_1}\hat\xi^{(i,j,d)}_{k,2}\bigg|\ge \breve\nu_{1,1},
\bigg|\sum_{k=1}^{n_1}\hat\xi^{(i,j,d)}_{k,3}\bigg|\ge \breve\nu_{2,1},\bigg|\sum_{k=1}^{n_1}\hat\xi^{(i,j,d)}_{k,4}\bigg|\ge \breve\nu_{2,1}\,\bigg|\,\bZ\bigg\}\notag\\
&\qquad+\PP\bigg\{\bigg|\sum_{k=1}^{n_1}\tilde\bxi^{(i,j,d)}_{k}\bigg|_2\ge (\log d)^{-2}\,\bigg|\,\bZ\bigg\}\,.
\end{align}
Parallel to the arguments for \eqref{Seq:Slem0_3_1}, given the event $\mathcal{D}$, we have
\begin{align}\label{Seq:Slem1_4}
\PP\bigg\{\bigg|\sum_{k=1}^{n_1}\tilde\bxi^{(i,j,d)}_{k}\bigg|_2\ge (\log d)^{-2}\,\bigg|\,\bZ\bigg\}
 =o(d^{-4})
\end{align}
provided that $\log d\ll n_1^{1/14}$. Let $\bu_{i,j,d}=(u_i,u_j,u_{i+d},u_{j+d})^{\top}$ with $\bu_{i,j,d}\,|\,\bZ\sim\mathcal{N}({\bf0},\hat\bOmega^{0}_{i,j,d})$, where $\hat\bOmega^{0}_{i,j,d}={\rm Var}\{\sum_{k=1}^{n_1}\hat\bxi^{(i,j,d)}_{k}\,|\,\bZ\}$. By Theorem 1 in \citeS{1987On}, we have
\begin{align}\label{Seq:Slem1_5}
&\mathbb{P}\bigg\{\bigg|\sum_{k=1}^{n_1}\hat\xi^{(i,j,d)}_{k,1}\bigg|\ge \breve\nu_{1,1},\bigg|\sum_{k=1}^{n_1}\hat\xi^{(i,j,d)}_{k,2}\bigg|\ge \breve\nu_{1,1},
\bigg|\sum_{k=1}^{n_1}\hat\xi^{(i,j,d)}_{k,3}\bigg|\ge \breve\nu_{2,1},\bigg|\sum_{k=1}^{n_1}\hat\xi^{(i,j,d)}_{k,4}\bigg|\ge \breve\nu_{2,1}\,\bigg|\,\bZ\bigg\}\notag\\
&\quad\le\mathbb{P}(|u_i|\ge\breve\nu_{1,2},|u_j|\ge\breve\nu_{1,2},
|u_{i+d}|\ge\breve\nu_{2,2},|u_{j+d}|\ge\breve\nu_{2,2}\,|\,\bZ)+c_{1,4}\exp\{-c_{2,4}(\log d)^2\}\notag\\
&\quad\le\mathbb{P}(|u_i|\ge\breve\nu_{1,2},|u_j|\ge\breve\nu_{1,2},
|u_{i+d}|\ge\breve\nu_{2,2},|u_{j+d}|\ge\breve\nu_{2,2}\,|\,\bZ)+o(d^{-2})\,,
\end{align}
where $c_{1,4}>0$ and $c_{2,4}>0$ are two absolute constants. Let $\tilde\bu_{i,j,d}=(\tilde \bu_{i,j}^{\top},\tilde \bu_{i+d,j+d}^\top)^{\top}\sim\mathcal{N}({\bf0},\bOmega^{0}_{i,j,d})$ with $\tilde \bu_{i,j}=(\tilde u_{i},\tilde u_{j})^\top$ and $\tilde \bu_{i+d,j+d}=(\tilde u_{i+d},\tilde u_{j+d})^\top$. To prove Lemma \ref{Slem1}, we need Lemmas \ref{Slem2} and \ref{Slem6}, whose proofs are given in Sections \ref{Sproof_lem2} and \ref{Sproof_lem6}, respectively.

\begin{lemma}\label{Slem2}
If $|\bOmega^{0}_{i,j,d}-\hat\bOmega^{0}_{i,j,d}|_{\infty}\le C_6(\log d)^{-3}$ for some constant $C_6>0$, then
\begin{align*}
&\mathbb{P}(|u_{i}|\ge v_{1},|u_{j}|\ge v_{1},
|u_{i+d}|\ge v_{2},|u_{j+d}|\ge v_{2}\,|\,\bZ)\\
&~~~~~\le\mathbb{P}(|\tilde{u}_{i}|\ge v_{1},|\tilde{u}_{j}|\ge v_{1})
\mathbb{P}(|\tilde{u}_{i+d}|\ge v_{2},|\tilde{u}_{j+d}|\ge v_{2})
\big[1+O\{(\log d)^{-2}\}\big]+o(d^{-2})\,,\\
&\mathbb{P}\{v_{1}\le|u_{i}|\le 3(\log d)^{1/2},v_{2}\le|u_{i+d}|\le 3(\log d)^{1/2}\,|\,\bZ\}\\
&~~~~~\ge\mathbb{P}\{v_{1}\le|\tilde{u}_{i}|\le 3(\log d)^{1/2}\}
\mathbb{P}\{v_{2}\le|\tilde{u}_{i+d}|\le 3(\log d)^{1/2}\}
\big[1-O\{(\log d)^{-2}\}\big]\,,\\
&\mathbb{P}(|u_{i}|\ge v_{1},|u_{i+d}|\ge v_{2}\,|\,\bZ)\\
&~~~~~\le\mathbb{P}(|\tilde{u}_{i}|\ge v_{1})
\mathbb{P}(|\tilde{u}_{i+d}|\ge v_{2})
\big[1+O\{(\log d)^{-2}\}\big]+o(d^{-2})
\end{align*}
for any $-2(\log d)^{-2}\le v_{1},v_{2}\le (2\log d)^{1/2}+2(\log d)^{-2}$, where $i,j\in[d]$, $i\neq j$, the terms $O\{(\log d)^{-2}\}$ and $o(d^{-2})$ hold uniformly over $-2(\log d)^{-2}\le v_{1},v_{2}\le (2\log d)^{1/2}+2(\log d)^{-2}$. 
\end{lemma}

\begin{lemma}\label{Slem6}
For any $0\le t\le (2\log d)^{1/2}$, if $\log d>\sqrt{2}$, it holds that
\begin{align*}
\mathbb{P}\big\{|\tilde{u}_{i}|\ge t-2(\log d)^{-2},|\tilde{u}_{j}|\ge t-2(\log d)^{-2}\big\}
=\mathbb{P}(|\tilde{u}_{i}|\ge t,|\tilde{u}_{j}|\ge t)\big[1+O\{(\log d)^{-3/2}\}\big]\,,
\end{align*}
where the term $O\{(\log d)^{-3/2}\}$ holds uniformly over $0\le t\le (2\log d)^{1/2}$.
\end{lemma}

Recall $\tilde{w}_{i}=\sum_{k=1}^{n_1}\xi^{(i,j,d)}_{k,1}$, $\tilde{w}_{j}=\sum_{k=1}^{n_1}\xi^{(i,j,d)}_{k,2}$, $\tilde{w}_{i+d}=\sum_{k=1}^{n_1}\xi^{(i,j,d)}_{k,3}$ and $\tilde{w}_{j+d}=\sum_{k=1}^{n_1}\xi^{(i,j,d)}_{k,4}$. Notice that $\breve \nu_{l,2}=\nu_{l}-2(\log d)^{-2}\in[-2(\log d)^{-2},(2\log d)^{1/2}]$ for $0\le \nu_{l}\le (2\log d)^{1/2}$ and $l\in[2]$. Analogous to Lemma \ref{Slem6}, for any $0\le t\le (2\log d)^{1/2}$, if $\log d>\sqrt{2}$, it holds that
\begin{align}\label{Seq:Slem1_5_1}
&\mathbb{P}\big\{|\tilde{u}_{i+d}|\ge t-2(\log d)^{-2},|\tilde{u}_{j+d}|\ge t-2(\log d)^{-2}\big\}\notag\\
&\qquad=\mathbb{P}(|\tilde{u}_{i+d}|\ge t,|\tilde{u}_{j+d}|\ge t)\big[1+O\{(\log d)^{-3/2}\}\big]\,,
\end{align}
where the term $O\{(\log d)^{-3/2}\}$ holds uniformly over $0\le t\le (2\log d)^{1/2}$. Given the event $\mathcal{G}_{n_1}\cap\mathcal{D}$, if $\sqrt{2}\le\log d\ll n_1^{1/14}$, by \eqref{Seq:Slem1_2}--\eqref{Seq:Slem1_5_1} and Lemmas \ref{Slem2} and \ref{Slem6}, we have
 \begin{align}\label{Seq:Slem1_6}
&\mathbb{P}\big(|\tilde{w}_{i}|\ge \nu_{1},|\tilde{w}_{j}|\ge \nu_{1},
|\tilde{w}_{i+d}|\ge \nu_{2},|\tilde{w}_{j+d}|\ge \nu_{2}\,|\,\bZ\big)\notag\\
&\quad\le\mathbb{P}(|u_i|\ge\breve\nu_{1,2},|u_j|\ge\breve\nu_{1,2},
|u_{i+d}|\ge\breve\nu_{2,2},|u_{j+d}|\ge\breve\nu_{2,2}\,|\,\bZ)+o(d^{-2})\\
&\quad\le\mathbb{P}(|\tilde{u}_{i}|\ge\breve\nu_{1,2},|\tilde{u}_{j}|\ge\breve\nu_{1,2})
\mathbb{P}(|\tilde{u}_{i+d}|\ge\breve\nu_{2,2},|\tilde{u}_{j+d}|\ge\breve\nu_{2,2})
\big[1+O\{(\log d)^{-2}\}\big]+o(d^{-2})\notag\\
&\quad=\mathbb{P}(|\tilde{u}_{i}|\ge \nu_{1},|\tilde{u}_{j}|\ge \nu_{1})
\mathbb{P}(|\tilde{u}_{i+d}|\ge \nu_{2},|\tilde{u}_{j+d}|\ge \nu_{2})
\big[1+O\{(\log d)^{-3/2}\}\big]+o(d^{-2})\notag
\end{align}
for any $0\le \nu_{1},\nu_{2}\le (2\log d)^{1/2}$, where the terms $O\{(\log d)^{-3/2}\}$ and $o(d^{-2})$ hold uniformly over $0\le \nu_{1},\nu_{2}\le (2\log d)^{1/2}$. Let $\tilde \nu_{l,2}=\nu_{l}+2(\log d)^{-2}$ for $l\in[2]$. Parallel to the arguments for \eqref{Seq:Slem0_5}--\eqref{Seq:lem0_7_1}, given the event $\mathcal{D}$, if $2^{2/5}\le\log d\ll n_1^{1/14}$, by \eqref{Seq:Slem1_4}, we have
\begin{align}
&\mathbb{P}\big\{\nu_{1}\le|\tilde{w}_{i}|\le 4(\log d)^{1/2},\nu_{2}\le|\tilde{w}_{i+d}|\le 4(\log d)^{1/2}\,|\,\bZ\big\}\notag\\
&~~~~~\ge\mathbb{P}\big\{\tilde\nu_{1,2}\le|u_{i}|\le 4(\log d)^{1/2}-2(\log d)^{-2},\tilde\nu_{2,2}\le|{u}_{i+d}|\le 4(\log d)^{1/2}-2(\log d)^{-2}\,|\,\bZ\big\}\notag\\
&\qquad~~-o(d^{-2})\notag\\
&~~~~~\ge\mathbb{P}\big\{\tilde\nu_{1,2}\le|u_{i}|\le 3(\log d)^{1/2},\tilde\nu_{2,2}\le|{u}_{i+d}|\le 3(\log d)^{1/2}\,|\,\bZ\big\}-o(d^{-2})\,,\notag\\
&\mathbb{P}(|\tilde{w}_{i}|\ge \nu_{1},|\tilde{w}_{i+d}|\ge \nu_{2}\,|\,\bZ)\le\mathbb{P}(|{u}_{i}|\ge \breve\nu_{1,2},|{u}_{i+d}|\ge \breve\nu_{2,2}\,|\,\bZ)
+o(d^{-2})\notag
\end{align}
for any $0\le \nu_{1},\nu_{2}\le (2\log d)^{1/2}$, where the term $o(d^{-2})$ holds uniformly over $0\le \nu_{1},\nu_{2}\le (2\log d)^{1/2}$. Therefore, given the event $\mathcal{G}_{n_1}\cap\mathcal{D}$, if $2^{2/5}\le\log d\ll n_1^{1/14}$, by \eqref{Seq:Slem1_2} and Lemma \ref{Slem2}, we have
\begin{align}
&\mathbb{P}\big\{\nu_{1}\le|\tilde{w}_{i}|\le 4(\log d)^{1/2},\nu_{2}\le|\tilde{w}_{i+d}|\le 4(\log d)^{1/2}\,|\,\bZ\big\}\notag\\
&~~~~~\ge\mathbb{P}\big\{\tilde\nu_{1,2}\le|\tilde{u}_{i}|\le 3(\log d)^{1/2}\big\}
\mathbb{P}\big\{\tilde\nu_{2,2}\le|\tilde{u}_{i+d}|\le 3(\log d)^{1/2}\big\}\notag\\
&\qquad~~\times\big[1-O\{(\log d)^{-2}\}\big]-o(d^{-2})\,,\label{Seq:Slem1_7}\\
&\mathbb{P}(|\tilde{w}_{i}|\ge \nu_{1},|\tilde{w}_{i+d}|\ge \nu_{2}\,|\,\bZ)\notag\\
&~~~~~\le\mathbb{P}(|\tilde{u}_{i}|\ge \breve\nu_{1,2})
\mathbb{P}(|\tilde{u}_{i+d}|\ge \breve\nu_{2,2})
\big[1+O\{(\log d)^{-2}\}\big]+o(d^{-2})\label{Seq:Slem1_8}
\end{align}
for any $0\le \nu_{1},\nu_{2}\le (2\log d)^{1/2}$, where the terms $O\{(\log d)^{-2}\}$ and $o(d^{-2})$ hold uniformly over $0\le \nu_{1},\nu_{2}\le (2\log d)^{1/2}$.

To establish Lemma \ref{Slem1}, we employ the core proof argument almost identical to that for Lemma \ref{lem1} in Section \ref{proof_lem1}. To adapt to the present setting, we need to replace the parameters $\vartheta$ and $\tau$ used there, respectively, by the constant $1$ and the parameter $\tau_2$, and replace $G_{l,n}^{-1}(\lambda)=G^{-1}(\lambda)-(-1)^{l}\upsilon_{n}/\sqrt{\log d}$ with $\upsilon_{n}=s_{0}^{1/2}n^{-1/4}(\log d)^{3/4}$ and $l\in[2]$ used there by $G_{l,n_1}^{-1}(\lambda)=G^{-1}(\lambda)-(-1)^{l}\upsilon_{n_1}/\sqrt{\log d}$ with $\upsilon_{n_1}=s_{0}^{1/2}n_1^{-1/4}(\log d)^{3/4}$ and $l\in[2]$. We also need to replace the event $\mathcal{G}_n$ used there by the event $\mathcal{G}_{n_1}\cap\mathcal{D}$. Consequently, the upper bound on $\mathbb{P}(\mathcal{G}_n^{\rm c})$ provided in Lemma \ref{lem0_condition} is no longer applicable in the present setting. Instead, we use Equations \eqref{Seq:lem0_0} and \eqref{Seq:lem0_1_4} to establish the upper bound on $\mathbb{P}(\mathcal{G}_{n_1}^{\rm c}\cup\mathcal{D}^{\rm c})$. 

Moreover, Equations \eqref{Seq:Slem1_6}--\eqref{Seq:Slem1_8} in the present setting are the analogues of Equations \eqref{lem8-1_1}--\eqref{lem8-1_3} in Lemma \ref{lem2} which are used for the proof of Lemma \ref{lem1}. For the right-hand side of \eqref{lem8-1_1}, we need to replace $(\tilde w_{i},\tilde w_{j},\tilde w_{i+d},\tilde w_{j+d})$ and $(O\{(\log d)^{-1-\vartheta}\},4d^{-2})$ used there, respectively, by $(\tilde u_{i},\tilde u_{j},\tilde u_{i+d},\tilde u_{j+d})$ and $(O\{(\log d)^{-3/2}\},o(d^{-2}))$; for the right-hand side of \eqref{lem8-1_2}, we need to replace $(\tilde w_{i},\tilde w_{i+d})$ and $(\nu_1,\nu_2,2(\log d)^{1/2})$ used there, respectively, by $(\tilde u_{i},\tilde u_{i+d})$ and $(\tilde\nu_{1,2},\tilde\nu_{2,2},3(\log d)^{1/2})$, and additionally subtract $o(d^{-2})$; for the right-hand side of \eqref{lem8-1_3}, we need to replace $(\tilde w_{i},\tilde w_{i+d})$, $(\nu_1,\nu_2)$ and $2d^{-2}$ used there, respectively, by $(\tilde u_{i},\tilde u_{i+d})$, $(\breve\nu_{1,2},\breve\nu_{2,2})$ and $o(d^{-2})$. Since $|\tilde\nu_{l,2}-\nu_{l}|=|\breve\nu_{l,2}-\nu_{l}|=O\{(\log d)^{-2}\}$ for $l\in[2]$, to establish Lemma \ref{Slem1}, we need to show that for $\epsilon_d=O\{(\log d)^{-2}\}$, the relation
\begin{align}\label{Seq:Slem1_9}
G(t+\epsilon_d)=G(t)\big[1+O\{(\log d)^{-3/2}\}\big]
\end{align}
holds uniformly over $0\le t\le\sqrt{2\log d}$. Recall $G(t)=2\{1-\Phi(t)\}$. Parallel to the proof of Lemma \ref{lem_normal} in Section \ref{proof_lem_normal} for $\epsilon_d=O\{(\log d)^{-2}\}$, we know \eqref{Seq:Slem1_9} holds. Furthermore, analogous to Lemma \ref{lem6}, for any given $1/2<\rho<1$, Condition \ref{Sass:CLIME} implies that if $d>\max\{4s_{d,2}^{2},(\log d)^{6/(2\rho-1)}\}$, then $|\mathcal{S}_{i}|\le d^{\rho}$ for all $i\in \mathcal{H}_0$, where $\mathcal{S}_{i}=\{j \in \mathcal{H}_0: |\Omega_{i,j}| \geq(\log d)^{-3}\}$ and $\mathcal{S}_{i}^{\rm c}=\mathcal{H}_0\setminus\mathcal{S}_{i}$. 

Due to $s_{d,2}^2\ll d$ and $(\log d)^{6/(2\rho-1)}\ll  d$ for any given $1/2<\rho<1$, the conclusion of Lemma \ref{Slem1} follows directly from \eqref{Seq:lem0_0}, \eqref{Seq:lem0_1_4} and \eqref{Seq:Slem1_6}--\eqref{Seq:Slem1_9}. $\hfill\qedsymbol$

\subsection{Proof of Lemma \ref{Slem2}}\label{Sproof_lem2}

Recall $\bOmega=(\Omega_{i,j})_{2d\times2d}=\bT\bTheta_0 \bT^\top$ and $\bOmega^{0}=(\Omega^{0}_{i,j})_{2d\times2d}$ with $\Omega^{0}_{i,j}=\Omega_{i,j}\Omega^{-1/2}_{i,i}\Omega^{-1/2}_{j,j}$. Due to $2C_{\max}^{-1}\le\lambda_{\min}(\bOmega)\le\lambda_{\max}(\bOmega)\le 2C_{\min}^{-1}$, then 
\begin{align}\label{Seq:lem11-0}
\frac{C_{\min}}{C_{\max}}\le\lambda_{\min}(\bOmega^{0})\le\lambda_{\max}(\bOmega^{0})\le \frac{C_{\max}}{C_{\min}}\,.
\end{align}
For any distinct $i,j\in[d]$, write 
\begin{align*}
  \bOmega^{0}_{i,j,d}:=
  \begin{pmatrix}
  \bXi_{i,j}&\bXi_{i,j,d}\\
  \bXi_{i,j,d}^{\top}&\bXi_{i+d,j+d}
  \end{pmatrix}\,,
\end{align*}
where
\begin{align*}
  \bXi_{i,j}=
  \begin{pmatrix}
  \Omega^{0}_{i,i}&\Omega^{0}_{i,j}\\
  \Omega^{0}_{i,j}&\Omega^{0}_{j,j}
  \end{pmatrix}\,,
  ~~\bXi_{i+d,j+d}=
  \begin{pmatrix}
  \Omega^{0}_{i+d,i+d}&\Omega^{0}_{i+d,j+d}\\
  \Omega^{0}_{i+d,j+d}&\Omega^{0}_{j+d,j+d}
  \end{pmatrix}\,, 
  ~~\bXi_{i,j,d}=
 \begin{pmatrix}
  \Omega^{0}_{i,i+d}&\Omega^{0}_{i,j+d}\\
  \Omega^{0}_{j,i+d}&\Omega^{0}_{j,j+d}
  \end{pmatrix}\,.
\end{align*}
By \eqref{Seq:Slem1_0}, we have $\bXi_{i,j,d}={\bf 0}$ and $|\bXi_{i+d,j+d}|_{\infty}\le1$. Recall $\bu_{i,j,d}=(u_{i},u_{j},u_{i+d},u_{j+d})^{\top}$ with $\bu_{i,j,d}\,|\,\bZ\sim\mathcal{N}({\bf0},\hat\bOmega^{0}_{i,j,d})$, where $\hat\bOmega^{0}_{i,j,d}={\rm Var}\{\sum_{k=1}^{n_1}\hat\bxi^{(i,j,d)}_{k}\,|\,\bZ\}$. Due to $|\bOmega^{0}_{i,j,d}-\hat\bOmega^{0}_{i,j,d}|_{\infty}\le C_6(\log d)^{-3}$, if $\log d\ge (8C_{\max}C_{\min}^{-1}C_6)^{1/3}$, we have
\begin{align*}
\|\bOmega^{0}_{i,j,d}-\hat\bOmega^{0}_{i,j,d}\|_{2}
\le4|\bOmega^{0}_{i,j,d}-\hat\bOmega^{0}_{i,j,d}|_{\infty}\le\frac{C_{\min}}{2C_{\max}}\,.
\end{align*}
Due to $\lambda_{\min}(\bOmega^{0})\le\lambda_{\min}(\bOmega^{0}_{i,j,d})\le\lambda_{\max}(\bOmega^{0}_{i,j,d})\le\lambda_{\max}(\bOmega^{0})$, then
\begin{align}\label{Seq:lem11-1}
\frac{C_{\min}}{2C_{\max}}\le\lambda_{\min}(\hat\bOmega^{0}_{i,j,d})\le\lambda_{\max}(\hat\bOmega^{0}_{i,j,d})\le\frac{3C_{\max}}{2C_{\min}}\,.
\end{align}
Letting $\hat\bXi_{i,j}={\rm Var}(\bu_{i,j}\,|\,\bZ)$ with $\bu_{i,j}=(u_{i},u_{j})^{\top}$, $\hat\bXi_{i+d,j+d}={\rm Var}(\bu_{i+d,j+d}\,|\,\bZ)$ with $\bu_{i+d,j+d}=(u_{i+d},u_{j+d})^{\top}$, and $\hat\bXi_{i,j,d}={\rm Cov}(\bu_{i,j},\bu_{i+d,j+d}\,|\,\bZ)$, we have
\begin{align*}
  \{\hat\bOmega^{0}_{i,j,d}\}^{-1}&=
  \begin{pmatrix}
  \hat\bXi_{i,j}^{-1}+\tilde{\bXi}_{1} &  \tilde{\bXi}_{2}\\
  \tilde{\bXi}_{2}^{\top} &  \tilde{\bXi}_{3}
  \end{pmatrix}\,,
\end{align*}
where $\tilde{\bXi}_{1}=\hat\bXi_{i,j}^{-1}\hat\bXi_{i,j,d}\tilde{\bXi}_{3}\hat\bXi_{i,j,d}^{\top}\hat\bXi_{i,j}^{-1}$, $\tilde{\bXi}_{2}=-\hat\bXi_{i,j}^{-1}\hat\bXi_{i,j,d}\tilde{\bXi}_{3}$ and $\tilde{\bXi}_{3}=(\hat\bXi_{i+d,j+d}-\hat\bXi_{i,j,d}^{\top}\hat\bXi_{i,j}^{-1}\hat\bXi_{i,j,d})^{-1}$. Recall $\tilde\bu_{i,j,d}=(\tilde \bu_{i,j}^{\top},\tilde \bu_{i+d,j+d}^\top)^{\top}\sim\mathcal{N}({\bf0},\bOmega^{0}_{i,j,d})$ with $\tilde \bu_{i,j}=(\tilde u_{i},\tilde u_{j})^\top$ and $\tilde \bu_{i+d,j+d}=(\tilde u_{i+d},\tilde u_{j+d})^\top$. Write $\bt=(\bt_{1}^{\top},\bt_{2}^{\top})^{\top}$ with $\bt_{1}=(t_{1},t_{2})^{\top}\in\mathbb{R}^{2}$ and $\bt_{2}=(t_{3},t_{4})^{\top}\in\mathbb{R}^{2}$. Let $f_{\tilde{\bu}_{i,j}}(\cdot)$ and $f_{\tilde{\bu}_{i+d,j+d}}(\cdot)$ be the density functions of $\tilde{\bu}_{i,j}$ and $\tilde{\bu}_{i+d,j+d}$, respectively, and $f_{\bu_{i,j,d}\,|\,\bZ}(\cdot)$ be the conditional density function of $\bu_{i,j,d}$ given $\bZ$. Let $|\bA|:=\det(\bA)$ denote the determinant of square matrix $\bA$. 
Since $\bu_{i,j,d}\,|\,\bZ\sim\mathcal{N}({\bf0},\hat\bOmega^{0}_{i,j,d})$ and $|\hat\bOmega^{0}_{i,j,d}|=|\hat\bXi_{i,j}||\hat\bXi_{i+d,j+d}-\hat\bXi_{i,j,d}^{\top}\hat\bXi_{i,j}^{-1}\hat\bXi_{i,j,d}|$, then
\begin{align}\label{Seq:lem11-2}
  &f_{\bu_{i,j,d}\,|\,\bZ}(\bt)
  =\frac{1}{(2\pi)^{2}|\hat\bOmega^{0}_{i,j,d}|^{1/2}}\exp\big[-2^{-1}\bt^{\top}\{\hat\bOmega^{0}_{i,j,d}\}^{-1} \bt\big]\notag\\
  &\quad~=\frac{\exp\{-2^{-1}(\bt_{1}^{\top}\hat\bXi_{i,j}^{-1}\bt_{1}
  +\bt_{1}^{\top}\tilde\bXi_{1}\bt_{1}+2\bt_{1}^{\top}\tilde{\bXi}_{2}\bt_{2}
  +\bt_{2}^{\top}\tilde{\bXi}_{3}\bt_{2})\} }{(2\pi)^{2}|\hat\bXi_{i,j}|^{1/2}|\hat\bXi_{i+d,j+d}-\hat\bXi_{i,j,d}^{\top}\hat\bXi_{i,j}^{-1}\hat\bXi_{i,j,d}|^{1/2}}\\ 
  &\quad~=f_{\tilde{\bu}_{i,j}}(\bt_{1})f_{\tilde{\bu}_{i+d,j+d}}(\bt_{2})\bigg(\underbrace{\frac{|\bXi_{i,j}||\bXi_{i+d,j+d}|}
  {|\hat\bXi_{i,j}||\hat\bXi_{i+d,j+d}-\hat\bXi_{i,j,d}^{\top}\hat\bXi_{i,j}^{-1}\hat\bXi_{i,j,d}|}}_{\rm I}\bigg)^{1/2}\notag\\
  &\quad~~~~\times\exp\{\underbrace{-\bt_{1}^{\top}\tilde{\bXi}_{2}\bt_{2}
    -2^{-1}\bt_{1}^{\top}\tilde\bXi_{1}\bt_{1}-2^{-1}\bt_{2}^{\top}(\tilde{\bXi}_{3}-\bXi_{i+d,j+d}^{-1})\bt_{2}
    -2^{-1}\bt_{1}^{\top}(\hat\bXi_{i,j}^{-1}-\bXi_{i,j}^{-1})\bt_{1}}_{{\rm II}(\bt_1,\bt_2)}\}\notag\,.
\end{align}
Due to $\bXi_{i,j,d}={\bf 0}$ and $|\bOmega^{0}_{i,j,d}-\hat\bOmega^{0}_{i,j,d}|_{\infty}\le C_6(\log d)^{-3}$, we have $|\hat\bXi_{i,j,d}|_{\infty}\le C_6(\log d)^{-3}$, 
$|\hat\bXi_{i+d,j+d}-\bXi_{i+d,j+d}|_{\infty}\le C_6(\log d)^{-3}$ and $|\hat\bXi_{i,j}-\bXi_{i,j}|_{\infty}\le C_6(\log d)^{-3}$. By \eqref{Seq:lem11-0} and \eqref{Seq:lem11-1}, we have $|\bXi_{i,j}^{-1}|_{\infty}\le C_{\max}C_{\min}^{-1}$, $|\hat\bXi_{i,j}^{-1}|_{\infty}\le2C_{\max}C_{\min}^{-1}$ and $|\hat\bXi_{i+d,j+d}^{-1}|_{\infty}\le2C_{\max}C_{\min}^{-1}$. Notice that $\bXi_{i+d,j+d}=\bI_{2}$. Then $|\bXi_{i+d,j+d}|=1$, $|\bXi_{i,j}^{-1}\hat\bXi_{i,j}|= 1+O\{(\log d)^{-3}\}$ and $|\hat\bXi_{i+d,j+d}-\hat\bXi_{i,j,d}^{\top}\hat\bXi_{i,j}^{-1}\hat\bXi_{i,j,d}|=1+O\{(\log d)^{-3}\}$, which implies
\begin{align}\label{Seq:lem11-1-1}
{\rm I}=1+O\{(\log d)^{-3}\}\,.
\end{align}
Notice that $\|\bA\|_{\infty}\le 2|\bA|_{\infty}$ and $\|\bA\bE\|_{\infty}\le\|\bA\|_{\infty}\|\bE\|_{\infty}$ for any $\bA,\bE\in\mathbb{R}^{2\times 2}$. Due to $|\hat\bXi_{i,j}^{-1}|_{\infty}\le2C_{\max}C_{\min}^{-1}$, $|\hat\bXi_{i+d,j+d}^{-1}|_{\infty}\le2C_{\max}C_{\min}^{-1}$ and $|\hat\bXi_{i,j,d}|_{\infty}\le C_6(\log d)^{-3}$, then 
\begin{align*}
|\hat\bXi_{i+d,j+d}^{-1}\hat\bXi_{i,j,d}^{\top}\hat\bXi_{i,j}^{-1}\hat\bXi_{i,j,d}|_{\infty}
&\le\|\hat\bXi_{i+d,j+d}^{-1}\hat\bXi_{i,j,d}^{\top}\hat\bXi_{i,j}^{-1}\hat\bXi_{i,j,d}\|_{\infty}\\
&\le\|\hat\bXi_{i+d,j+d}^{-1}\|_{\infty}\|\hat\bXi_{i,j,d}^{\top}\|_{\infty}
\|\hat\bXi_{i,j}^{-1}\|_{\infty}\|\hat\bXi_{i,j,d}\|_{\infty}\\
&\le16|\hat\bXi_{i+d,j+d}^{-1}|_{\infty}|\hat\bXi_{i,j,d}^{\top}|_{\infty}
|\hat\bXi_{i,j}^{-1}|_{\infty}|\hat\bXi_{i,j,d}|_{\infty}=O\{(\log d)^{-6}\}\,,
\end{align*}  
which implies $( \bI_{2}
-\hat\bXi_{i+d,j+d}^{-1}\hat\bXi_{i,j,d}^{\top}\hat\bXi_{i,j}^{-1}\hat\bXi_{i,j,d})^{-1}- \bI_{2}
=\bDelta_1$ with $|\bDelta_1|_{\infty}=O\{(\log d)^{-6}\}$. Hence,
\begin{align}\label{Seq:lem11-1-2}
\tilde{\bXi}_{3}
=\hat\bXi_{i+d,j+d}^{-1}+\{( \bI_{2}
-\hat\bXi_{i+d,j+d}^{-1}\hat\bXi_{i,j,d}^{\top}\hat\bXi_{i,j}^{-1}\hat\bXi_{i,j,d})^{-1}- \bI_{2}\}\hat\bXi_{i+d,j+d}^{-1}
=\hat\bXi_{i+d,j+d}^{-1}+\bDelta_2\,,
\end{align}
where $|\bDelta_2|_{\infty}=O\{(\log d)^{-6}\}$. Due to 
$|\hat\bXi_{i+d,j+d}-\bXi_{i+d,j+d}|_{\infty}\le C_6(\log d)^{-3}$ and $|\hat\bXi_{i,j}-\bXi_{i,j}|_{\infty}\le C_6(\log d)^{-3}$, we have
\begin{align}\label{Seq:lem11-1-3}
\hat\bXi_{i,j}^{-1}=\bXi_{i,j}^{-1}+\bDelta_3\quad\text{and}\quad
\hat\bXi_{i+d,j+d}^{-1}=\bXi_{i+d,j+d}^{-1}+\bDelta_4\,,
\end{align}
where $|\bDelta_3|_{\infty}=O\{(\log d)^{-3}\}$ and $|\bDelta_4|_{\infty}=O\{(\log d)^{-3}\}$. Recall $\tilde{\bXi}_{1}=\hat\bXi_{i,j}^{-1}\hat\bXi_{i,j,d}\tilde{\bXi}_{3}\hat\bXi_{i,j,d}^{\top}\hat\bXi_{i,j}^{-1}$ and $\tilde{\bXi}_{2}=-\hat\bXi_{i,j}^{-1}\hat\bXi_{i,j,d}\tilde{\bXi}_{3}$. Since $|\hat\bXi_{i,j}^{-1}|_{\infty}\le2C_{\max}C_{\min}^{-1}$, $|\hat\bXi_{i+d,j+d}^{-1}|_{\infty}\le2C_{\max}C_{\min}^{-1}$ and $|\hat\bXi_{i,j,d}|_{\infty}\le C_6(\log d)^{-3}$, by \eqref{Seq:lem11-1-2}, we have $|\tilde{\bXi}_{1}|_{\infty}=O\{(\log d)^{-6}\}$ and $|\tilde{\bXi}_{2}|_{\infty}=O\{(\log d)^{-3}\}$. Therefore, ${\rm II}(\bt_1,\bt_2)=O\{(\log d)^{-2}\}$ for any $|\bt|_{\infty}\le 3(\log d)^{1/2}$. Due to $e^{O\{(\log d)^{-2}\}}=1+O\{(\log d)^{-2}\}$ and $[1+O\{(\log d)^{-3}\}]^{1/2}=1+O\{(\log d)^{-3}\}$, for any $|\bt|_{\infty}\le 3(\log d)^{1/2}$, by \eqref{Seq:lem11-2} and \eqref{Seq:lem11-1-1}, we have
\begin{align}\label{Seq:lem11-2-1}
f_{\bu_{i,j,d}\,|\,\bZ}(\bt)=f_{\tilde{\bu}_{i,j}}(\bt_{1})f_{\tilde{\bu}_{i+d,j+d}}(\bt_{2})\big[1+O\{(\log d)^{   -2}\}\big]\,,
\end{align}
where the term $O\{(\log d)^{-2}\}$ holds uniformly over $|\bt|_{\infty}\le 3(\log d)^{1/2}$ and $i,j\in[d]$, $i\neq j$. Let $f_{\tilde u_{i}}(\cdot)$ and $f_{\tilde u_{i+d}}(\cdot)$ be the density function of $\tilde u_{i}$ and $\tilde u_{i+d}$, respectively, and  $f_{\bu_{i,i+d}\,|\,\bZ}(\cdot)$ be the conditional density function of $\bu_{i,i+d}=(u_i,u_{i+d})^{\top}$ given $\bZ$. Analogous to \eqref{Seq:lem11-2-1}, we can also show \begin{align}\label{Seq:lem11-3-1}
f_{\bu_{i,i+d}\,|\,\bZ}(\bt_{3})=f_{\tilde u_{i}}(t_{5})f_{\tilde u_{i+d}}(t_{6})\big[1+O\{(\log d)^{-2}\}\big]
\end{align}
for any $|\bt_{3}|_{\infty}\le 3(\log d)^{1/2}$ with $\bt_{3}=(t_{5},t_{6})^{\top}$, where the term $O\{(\log d)^{-2}\}$ holds uniformly over $|\bt_{3}|_{\infty}\le 3(\log d)^{1/2}$ and $i\in[d]$. 

Let $\mathcal{U}_{1}=\{t:\,v_{1}\le|t|\le3(\log d)^{1/2}\}$ and $\mathcal{U}_{2}=\{t:\,v_{2}\le|t|\le3(\log d)^{1/2}\}$ for $-2(\log d)^{-2}\le v_{1},v_{2}\le (2\log d)^{1/2}+2(\log d)^{-2}$. Then
\begin{align}\label{Seq:lem11-2-2}
&\mathbb{P}(|u_{i}|\ge v_{1},|u_{j}|\ge v_{1},
|u_{i+d}|\ge v_{2},|u_{j+d}|\ge v_{2}\,|\,\bZ)\notag\\
&~~~~~\le\mathbb{P}(u_{i}\in \mathcal{U}_{1},u_{j}\in \mathcal{U}_{1},
u_{i+d}\in \mathcal{U}_{2},u_{j+d}\in \mathcal{U}_{2}\,|\,\bZ)\\
&~~~~~~~~+\mathbb{P}\{|u_{i}|>3(\log d)^{1/2}\,|\,\bZ\}
+\mathbb{P}\{|u_{j}|>3(\log d)^{1/2}\,|\,\bZ\}\notag\\
&~~~~~~~~+\mathbb{P}\{|u_{i+d}|>3(\log d)^{1/2}\,|\,\bZ\}
+\mathbb{P}\{|u_{j+d}|>3(\log d)^{1/2}\,|\,\bZ\}\notag\,.
\end{align}
For any $-2(\log d)^{-2}\le v_{1},v_{2}\le (2\log d)^{1/2}+2(\log d)^{-2}$, by \eqref{Seq:lem11-2-1}, it holds that
\begin{align}\label{Seq:lem11-2-3}
&\mathbb{P}(u_{i}\in \mathcal{U}_{1},u_{j}\in \mathcal{U}_{1},
u_{i+d}\in \mathcal{U}_{2},u_{j+d}\in \mathcal{U}_{2}\,|\,\bZ)\notag\\
&~~~~~=\int_{\mathcal{U}_{1}}\int_{\mathcal{U}_{1}}\int_{\mathcal{U}_{2}}\int_{\mathcal{U}_{2}}
f_{\bu_{i,j,d}\,|\,\bZ}(\bt)\,{\rm d}t_{4}\,
{\rm d}t_{3}\,{\rm d}t_{2}\,{\rm d}t_{1}\notag\\
&~~~~~=\int_{\mathcal{U}_{1}}\int_{\mathcal{U}_{1}}\int_{\mathcal{U}_{2}}\int_{\mathcal{U}_{2}}
f_{\tilde{\bu}_{i,j}}(\bt_{1})f_{\tilde{\bu}_{i+d,j+d}}(\bt_{2})\big[1+O\{(\log d)^{-2}\}\big]\,{\rm d}t_{4}\,
{\rm d}t_{3}\,{\rm d}t_{2}\,{\rm d}t_{1}\\
&~~~~~\le\mathbb{P}(\tilde{u}_{i}\in \mathcal{U}_{1},\tilde{u}_{j}\in \mathcal{U}_{1})
\mathbb{P}(\tilde{u}_{i+d}\in \mathcal{U}_{2},\tilde{u}_{j+d}\in \mathcal{U}_{2})
\big[1+O\{(\log d)^{-2}\}\big]\notag\\
&~~~~~\le\mathbb{P}(|\tilde{u}_{i}|\ge v_{1},|\tilde{u}_{j}|\ge v_{1})
\mathbb{P}(|\tilde{u}_{i+d}|\ge v_{2},|\tilde{u}_{j+d}|\ge v_{2})
\big[1+O\{(\log d)^{-2}\}\big]\notag\,.
\end{align}
Analogously, by \eqref{Seq:lem11-3-1}, for any $-2(\log d)^{-2}\le v_{1},v_{2}\le (2\log d)^{1/2}+2(\log d)^{-2}$, we have
\begin{align*}
&\mathbb{P}\{v_{1}\le|u_{i}|\le 3(\log d)^{1/2},v_{2}\le|u_{i+d}|\le 3(\log d)^{1/2}\,|\,\bZ\}\\
&~~~~~\ge\mathbb{P}\{v_{1}\le|\tilde{u}_{i}|\le 3(\log d)^{1/2}\}
\mathbb{P}\{v_{2}\le|\tilde{u}_{i+d}|\le 3(\log d)^{1/2}\}
\big[1-O\{(\log d)^{-2}\}\big]\,.
\end{align*}
Let $\breve\sigma_i^2={\rm Var}(u_{i}\,|\,\bZ)$. Due to $\bu_{i,j,d}\,|\,\bZ\sim\mathcal{N}({\bf0},\hat\bOmega^{0}_{i,j,d})$ and $|\bOmega^{0}_{i,j,d}-\hat\bOmega^{0}_{i,j,d}|_{\infty}\le C_6(\log d)^{-3}$, by \eqref{Seq:Slem1_0}, we have $\breve\sigma_i^2\le1+C_6(\log d)^{-3}$, which implies 
$\breve\sigma_i^{-1}\ge\{1+C_6(\log d)^{-3}\}^{-1/2}$. Notice that $G(x)\le e^{-x^2/2}$ for any $x>0$. Therefore, it holds that
\begin{align*}
\mathbb{P}\{|u_{i}|>3(\log d)^{1/2}\,|\,\bZ\}&\le\mathbb{P}\big[\breve\sigma_i^{-1}|u_{i}|>3\{1+C_6(\log d)^{-3}\}^{-1/2}(\log d)^{1/2}\,|\,\bZ\big]\\
&=G\big[3\{1+C_6(\log d)^{-3}\}^{-1/2}(\log d)^{1/2}\big]=o(d^{-2})\,.
\end{align*}
The same bound also holds for $\mathbb{P}\{|u_{j}|>3(\log d)^{1/2}\,|\,\bZ\}$, $\mathbb{P}\{|u_{i+d}|>3(\log d)^{1/2}\,|\,\bZ\}$ and $\mathbb{P}\{|u_{j+d}|>3(\log d)^{1/2}\,|\,\bZ\}$. Together with \eqref{Seq:lem11-2-2} and \eqref{Seq:lem11-2-3}, we have
\begin{align*}
&\mathbb{P}(|u_{i}|\ge v_{1},|u_{j}|\ge v_{1},
|u_{i+d}|\ge v_{2},|u_{j+d}|\ge v_{2}\,|\,\bZ)\\
&~~~~~\le\mathbb{P}(|\tilde{u}_{i}|\ge v_{1},|\tilde{u}_{j}|\ge v_{1})
\mathbb{P}(|\tilde{u}_{i+d}|\ge v_{2},|\tilde{u}_{j+d}|\ge v_{2})
\big[1+O\{(\log d)^{-2}\}\big]+o(d^{-2})
\end{align*}
for any $-2(\log d)^{-2}\le v_{1},v_{2}\le (2\log d)^{1/2}+2(\log d)^{-2}$. Analogously, by \eqref{Seq:lem11-3-1}, we have
\begin{align*}
\mathbb{P}(|u_{i}|\ge v_{1},|u_{i+d}|\ge v_{2}\,|\,\bZ)
\le\mathbb{P}(|\tilde{u}_{i}|\ge v_{1})
\mathbb{P}(|\tilde{u}_{i+d}|\ge v_{2})
\big[1+O\{(\log d)^{-2}\}\big]+o(d^{-2})
\end{align*}
for any $-2(\log d)^{-2}\le v_{1},v_{2}\le (2\log d)^{1/2}+2(\log d)^{-2}$. We complete the proof of Lemma \ref{Slem2}. $\hfill\qedsymbol$

\subsection{Proof of Lemma \ref{Slem6}}\label{Sproof_lem6}

Due to $\tilde\bu_{i,j,d}=(\tilde \bu_{i,j}^{\top},\tilde \bu_{i+d,j+d}^\top)^{\top}\sim\mathcal{N}({\bf0},\bOmega^{0}_{i,j,d})$ with $\tilde \bu_{i,j}=(\tilde u_{i},\tilde u_{j})^\top$ and $\tilde \bu_{i+d,j+d}=(\tilde u_{i+d},\tilde u_{j+d})^\top$, by \eqref{Seq:Slem1_0}, we have ${\rm Var}(\tilde u_i)={\rm Var}(\tilde u_j)=1$. Let $\rho_{i,j}={\rm Cov}(\tilde u_i,\tilde u_j)$. By \eqref{Seq:lem11-0}, we have
${\rm det}\{{\rm Var}(\tilde \bu_{i,j})\}=1-\rho_{i,j}^2\ge\lambda_{\min}^2\{{\rm Var}(\tilde \bu_{i,j})\}\ge C_{\min}^2C_{\max}^{-2}$, which implies $|\rho_{i,j}|\le(1-C_{\min}^2C_{\max}^{-2})^{1/2}=:\rho_0$. Let 
\begin{align*}
F(t):=\mathbb{P}(\tilde{u}_{i}\ge t,\tilde{u}_{j}\ge t)=\int_{t}^{\infty}\int_{t}^{\infty}\varphi(v_1,v_2)\,{\rm d}v_1\,{\rm d}v_2 
\end{align*}
with
\begin{align*}
\varphi(v_1,v_2)=\frac{1}{2\pi(1-\rho_{i,j}^2)^{1/2}}\exp\bigg\{-\frac{v_1^2+v_2^2-2\rho_{i,j} v_1v_2 }{2(1-\rho_{i,j}^2)}\bigg\}\,.
\end{align*}
Due to $1-\Phi(u)<u^{-1}e^{-u^2/2}$ for any $u>0$, it holds that 
\begin{align}\label{Seq:Slem2_1}
0\ge F'(t)&=-\int_{t}^{\infty}\varphi(v_1,t)\,{\rm d}v_1-\int_{t}^{\infty}\varphi(t,v_2)\,{\rm d}v_2\notag\\
&=-\frac{e^{-t^2/2}}{\pi}\int_{\tilde \rho_{i,j} t}^{\infty}e^{-u^2/2}\,{\rm d}u
=-\frac{\sqrt{2}e^{-t^2/2}}{\sqrt{\pi}}\{1-\Phi(\tilde \rho_{i,j} t)\}\\
&>-\frac{\sqrt{2}e^{-t^2/2}}{\sqrt{\pi}}\cdot\frac{e^{-(\tilde \rho_{i,j} t)^2/2}}{\tilde \rho_{i,j} t}
=-\frac{\sqrt{2}}{\sqrt{\pi}\tilde \rho_{i,j} t}e^{-t^2/(1+\rho_{i,j})}\notag
\end{align}
for any $t>0$, where $\tilde{\rho}_{i,j}=\{(1-\rho_{i,j})(1+\rho_{i,j})^{-1}\}^{1/2}$. Notice that $v_1^2+v_2^2-2\rho_{i,j} v_1v_2\le(1+|\rho_{i,j}|)(v_1^2+v_2^2)$ for any $v_1,v_2\in \mathbb{R}$. 

\underline{{\it Case} (i): $-\rho_0\le\rho_{i,j}\le0$.} 
Notice that $1\le\tilde\rho_{i,j}\le\{(1+\rho_0)(1-\rho_0)^{-1}\}^{1/2}=:\tilde\rho_0$. Due to $1-\Phi(u)>(2\pi)^{-1/2}u(1+u^2)^{-1}e^{-u^2/2}$ for any $u>0$, it holds that 
\begin{align}\label{Seq:Slem2_2}
F(t)&\ge\int_{t}^{\infty}\int_{t}^{\infty}\frac{1}{2\pi(1-\rho_{i,j}^2)^{1/2}}\exp\bigg\{-\frac{v_1^2+v_2^2}{2(1+\rho_{i,j})}\bigg\}\,{\rm d}v_1\,{\rm d}v_2\notag\\
&=\frac{1}{\tilde\rho_{i,j}}\bigg[1-\Phi\bigg\{\frac{t}{(1+\rho_{i,j})^{1/2}}\bigg\}\bigg]^2
>\frac{1}{2\pi\tilde\rho_{i,j}}\bigg\{\frac{(1+\rho_{i,j})^{-1/2}t}{1+(1+\rho_{i,j})^{-1}t^2}\bigg\}^2e^{-t^2/(1+\rho_{i,j})}\notag\\
&=\frac{(1+\rho_{i,j})t^2}{2\pi\tilde \rho_{i,j}(1+\rho_{i,j}+t^2)^2}e^{-t^2/(1+\rho_{i,j})}
>\frac{(1-\rho_0)t^2}{2\pi\tilde\rho_0(2+t^2)^2}e^{-t^2/(1+\rho_{i,j})}
\end{align}
for any $t>0$. 

\underline{{\it Case} (ii): $0<\rho_{i,j}\le\rho_0$.} Notice that $1>\tilde\rho_{i,j}\ge\{(1-\rho_0)(1+\rho_0)^{-1}\}^{1/2}=:\tilde\rho_1$. By Theorem 2.1.e of \citeS{lin2010probability}, we have
\begin{align}\label{Seq:Slem2_2_1}
F(t)&\ge\{1-\Phi(t)\}\big\{1-\Phi(\tilde \rho_{i,j}t)\big\}
>\frac{te^{-t^2/2}}{(2\pi)^{1/2}(1+t^2)}\cdot\frac{\tilde\rho_{i,j}te^{-(\tilde\rho_{i,j}t)^2/2}}{(2\pi)^{1/2}\{1+(\tilde\rho_{i,j}t)^2\}}\notag\\
&=\frac{\tilde\rho_{i,j}t^2}{2\pi(1+t^2)(1+\tilde\rho_{i,j}^2t^2)}e^{-t^2/(1+\rho_{i,j})}
>\frac{\tilde\rho_1t^2}{2\pi(1+t^2)^2}e^{-t^2/(1+\rho_{i,j})}
\end{align}
for any $t>0$. 

Notice that $|\rho_{i,j}|\le\rho_0$, $\tilde\rho_1\le\tilde\rho_{i,j}\le\tilde\rho_0$ and $t-2(\log d)^{-2}>2^{-1}t$ for $t>2$ and $\log d>\sqrt{2}$. For any $2<t\le(2\log d)^{1/2}$, by \eqref{Seq:Slem2_1}--\eqref{Seq:Slem2_2_1}, if $\log d>\sqrt{2}$ and $t-2(\log d)^{-2}\le u\le t$, then
\begin{align}\label{Seq:Slem2_3}
\bigg|\frac{F'(u)}{F(t)}\bigg|&<\max\bigg\{\frac{2\pi\tilde\rho_0(2+t^2)^2}{\tilde\rho_{i,j} u (1-\rho_0)t^2},\frac{2\pi(1+t^2)^2}{\tilde\rho_{i,j}u \tilde\rho_1t^2}\bigg\}
e^{-(u^2-t^2)/(1+\rho_{i,j})}\notag\\
&\le\max\bigg\{\frac{4\pi\tilde\rho_0}{\tilde\rho_1 (1-\rho_0)},\frac{4\pi}{ \tilde\rho_1^2}\bigg\}\frac{(2+t^2)^2}{t^3}e^{|u^2-t^2|/(1-\rho_0)}\notag\\
&\le\max\bigg\{\frac{4\pi\tilde\rho_0}{\tilde\rho_1 (1-\rho_0)},\frac{4\pi}{ \tilde\rho_1^2}\bigg\}\bigg(\frac{4}{t^{3}}+\frac{4}{t}+t\bigg)e^{4(\log d)^{-2}t/(1-\rho_0)}\\
&\le\max\bigg\{\frac{4\pi\tilde\rho_0}{\tilde\rho_1 (1-\rho_0)},\frac{4\pi}{ \tilde\rho_1^2}\bigg\}
\bigg\{\frac{5}{2}+(2\log d)^{1/2}\bigg\}e^{4\sqrt{2}(\log d)^{-3/2}/(1-\rho_0)}\notag\\
&=O\{(\log d)^{1/2}\}\,.\notag
\end{align}
By Taylor expansion, $|F\{t-2(\log d)^{-2}\}-F(t)|=2(\log d)^{-2}|F'\{t-2(\log d)^{-2}\theta_t\}|$
for some $0<\theta_t<1$. Therefore, for any $2<t\le(2\log d)^{1/2}$, by \eqref{Seq:Slem2_3}, if $\log d>\sqrt{2}$, we have
\begin{align}\label{Seq:Slem2_4}
\bigg|\frac{F\{t-2(\log d)^{-2}\}}{F(t)}-1\bigg|&=2(\log d)^{-2}\bigg|\frac{F'\{t-2(\log d)^{-2}\theta_t\}}{F(t)}\bigg|
\le O\{(\log d)^{-3/2}\}\,.
\end{align} 
For any $2(\log d)^{-2}< t\le2$ and $0<\theta_t<1$, by \eqref{Seq:Slem2_1}, we have $F(t)\ge F(2)$ and
\begin{align*}
|F'\{t-2(\log d)^{-2}\theta_t\}|&=\frac{e^{-\{t-2(\log d)^{-2}\theta_t\}^2/2}}{\pi}\int_{\tilde \rho_{i,j} \{t-2(\log d)^{-2}\theta_t\}}^{\infty}e^{-u^2/2}{\rm d}u\le1\,,
\end{align*}
which implies
\begin{align*}
\bigg|\frac{F\{t-2(\log d)^{-2}\}}{F(t)}-1\bigg|\le\frac{2(\log d)^{-2}}{ F(2)}= O\{(\log d)^{-2}\}\,.
\end{align*}
Recall $F(t)=\mathbb{P}(\tilde{u}_{i}\ge t,\tilde{u}_{j}\ge t)$. Together with \eqref{Seq:Slem2_4}, we have
\begin{align*}
\mathbb{P}\big\{\tilde{u}_{i}\ge t-2(\log d)^{-2},\tilde{u}_{j}\ge t-2(\log d)^{-2}\big\}
=\mathbb{P}(\tilde{u}_{i}\ge t,\tilde{u}_{j}\ge t)\big[1+O\{(\log d)^{-3/2}\}\big]
\end{align*}
for any $2(\log d)^{-2}<t\le(2\log d)^{1/2}$. The same result also holds for $\mathbb{P}(\tilde{u}_{i}\le -t,\tilde{u}_{j}\ge t)$, $\mathbb{P}(\tilde{u}_{i}\ge t,\tilde{u}_{j}\le -t)$ and $\mathbb{P}(\tilde{u}_{i}\le -t,\tilde{u}_{j}\le -t)$ when $2(\log d)^{-2}<t\le(2\log d)^{1/2}$. Hence,
\begin{align}\label{Seq:Slem2_5}
&\mathbb{P}\big\{|\tilde{u}_{i}|\ge t-2(\log d)^{-2},|\tilde{u}_{j}|\ge t-2(\log d)^{-2}\big\}\notag\\
&\qquad=\mathbb{P}(|\tilde{u}_{i}|\ge t,|\tilde{u}_{j}|\ge t)\big[1+O\{(\log d)^{-3/2}\}\big]
\end{align}
for any $2(\log d)^{-2}<t\le(2\log d)^{1/2}$. Notice that $G\{2(\log d)^{-2}\}=1-O\{(\log d)^{-2}\}$ with $G(t)=2\{1-\Phi(t)\}$. For any $0\le t\le2(\log d)^{-2}$, it holds that 
\begin{align*}
\mathbb{P}(|\tilde{u}_{i}|\ge t,|\tilde{u}_{j}|\ge t)
&\ge\mathbb{P}\{|\tilde{u}_{i}|\ge 2(\log d)^{-2},|\tilde{u}_{j}|\ge 2(\log d)^{-2}\}\\
&\ge\mathbb{P}\{|\tilde{u}_{i}|\ge 2(\log d)^{-2}\}+\mathbb{P}\{|\tilde{u}_{j}|\ge 2(\log d)^{-2}\}-1\\
&=2G\{2(\log d)^{-2}\}-1
=1-O\{(\log d)^{-2}\}\,,
\end{align*}
which implies 
\begin{align*}
\mathbb{P}(|\tilde{u}_{i}|\ge t,|\tilde{u}_{j}|\ge t)\cdot\underbrace{\frac{1}{1-O\{(\log d)^{-2}\}}}_{=1+O\{(\log d)^{-2}\}}\ge1\,.
\end{align*}
Due to $\mathbb{P}\{|\tilde{u}_{i}|\ge t-2(\log d)^{-2},|\tilde{u}_{j}|\ge t-2(\log d)^{-2}\}=1$ for any $t\in[0,2(\log d)^{-2}]$, we have
\begin{align*}
\mathbb{P}(|\tilde{u}_{i}|\ge t,|\tilde{u}_{j}|\ge t)\big[1+O\{(\log d)^{-2}\}\big]&\ge\mathbb{P}\{|\tilde{u}_{i}|\ge t-2(\log d)^{-2},|\tilde{u}_{j}|\ge t-2(\log d)^{-2}\}\\
&\ge\mathbb{P}(|\tilde{u}_{i}|\ge t,|\tilde{u}_{j}|\ge t)\big[1-O\{(\log d)^{-2}\}\big]\,,
\end{align*}
which implies 
\begin{align*}
\mathbb{P}\big\{|\tilde{u}_{i}|\ge t-2(\log d)^{-2},|\tilde{u}_{j}|\ge t-2(\log d)^{-2}\big\}
=\mathbb{P}(|\tilde{u}_{i}|\ge t,|\tilde{u}_{j}|\ge t)\big[1+O\{(\log d)^{-3/2}\}\big]
\end{align*}
for any $0\le t\le2(\log d)^{-2}$. Together with \eqref{Seq:Slem2_5}, we have Lemma \ref{Slem6}.  $\hfill\qedsymbol$

\section{Additional simulation results}
Additional results of simulations are reported here.

\subsection{Additional simulation results of FDR control}\label{SM1:figure}
\begin{figure}[htbp!]
	\centering
	\subfloat{\includegraphics[width=.27\columnwidth]{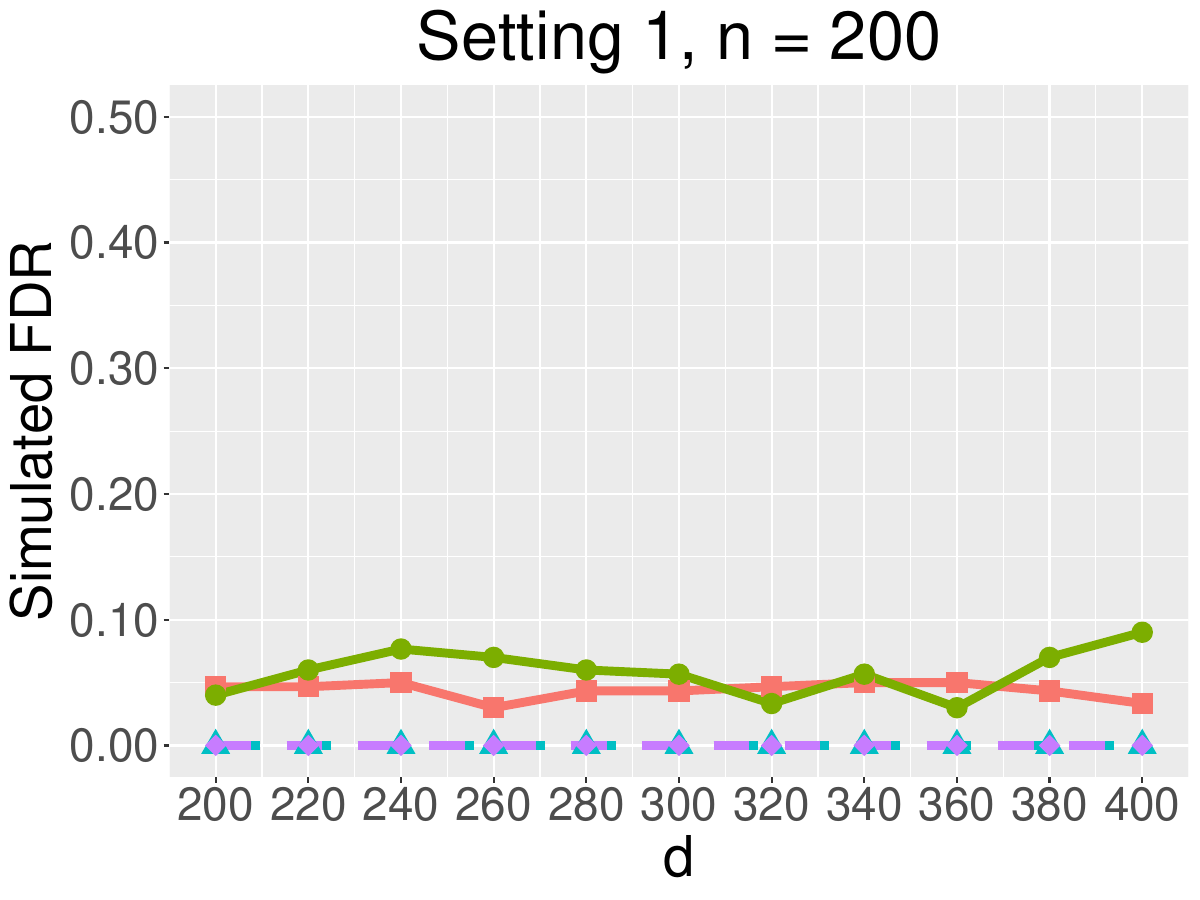}}\hspace{5pt}
	\subfloat{\includegraphics[width=.27\columnwidth]{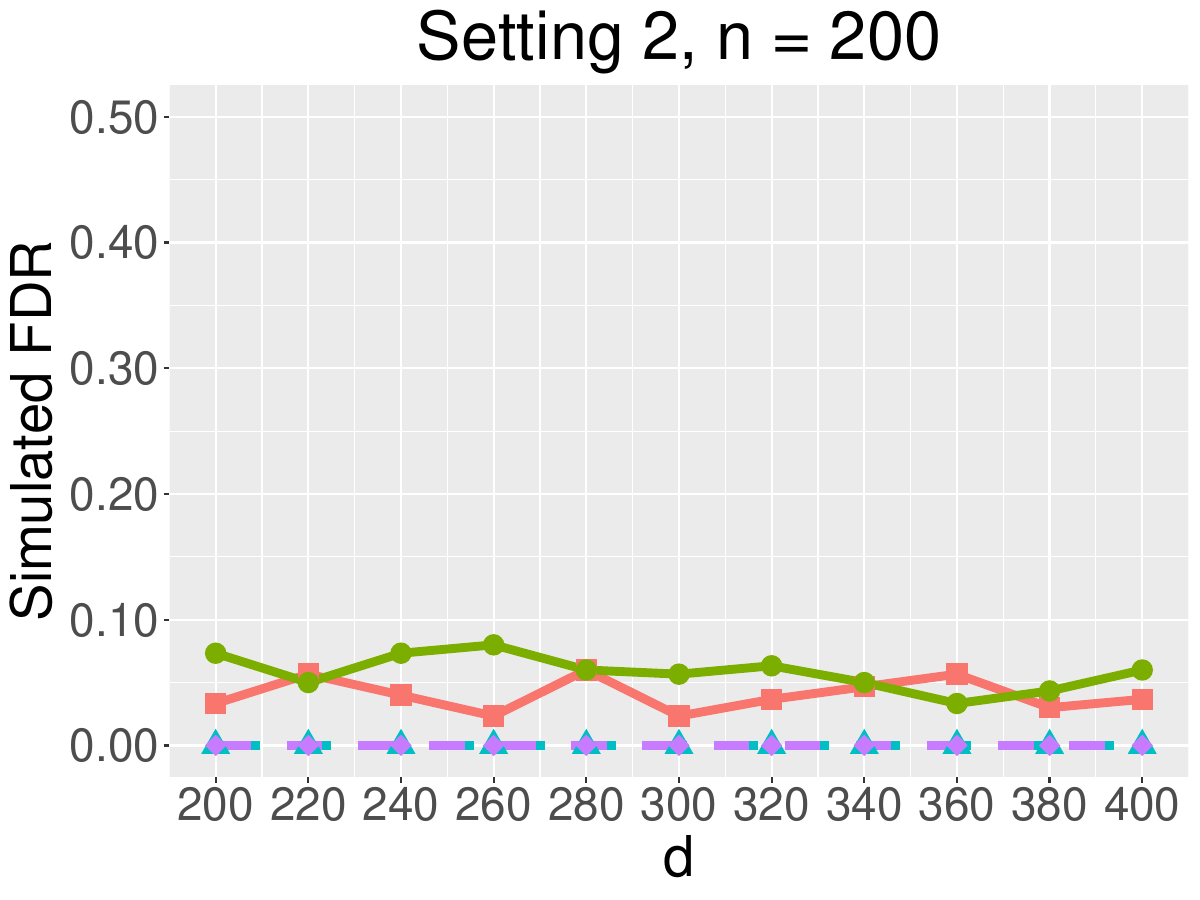}}\hspace{5pt}
	\subfloat{\includegraphics[width=.27\columnwidth]{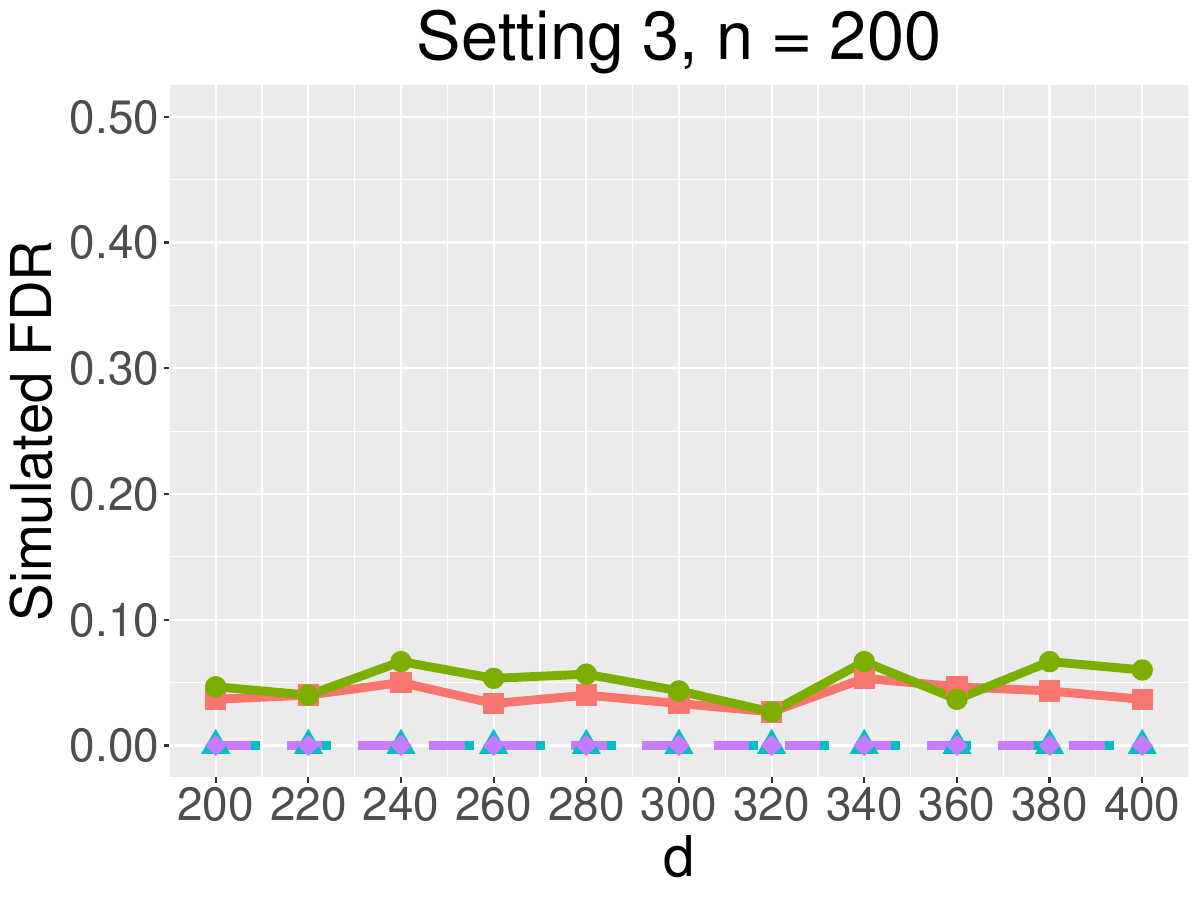}}\\
	\subfloat{\includegraphics[width=.27\columnwidth]{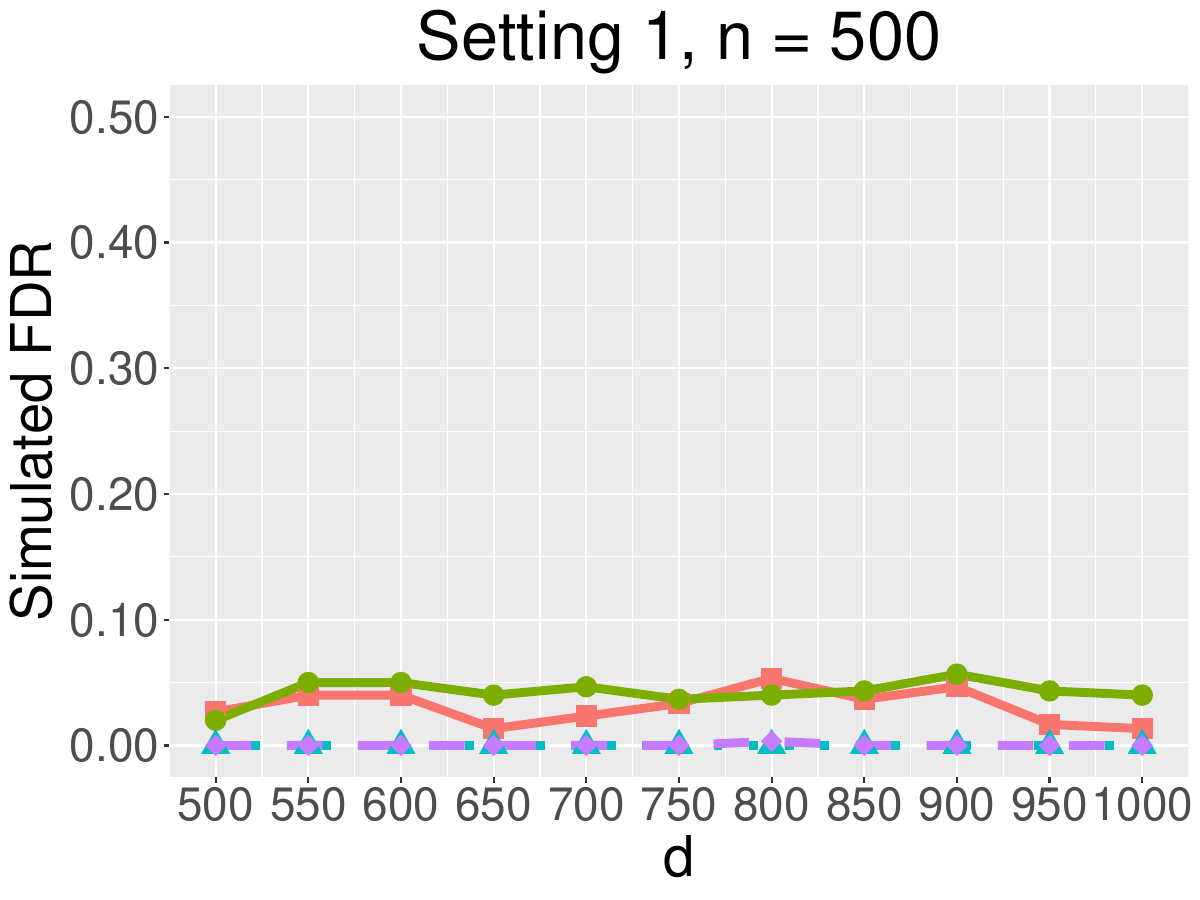}}\hspace{5pt}
    \subfloat{\includegraphics[width=.27\columnwidth]{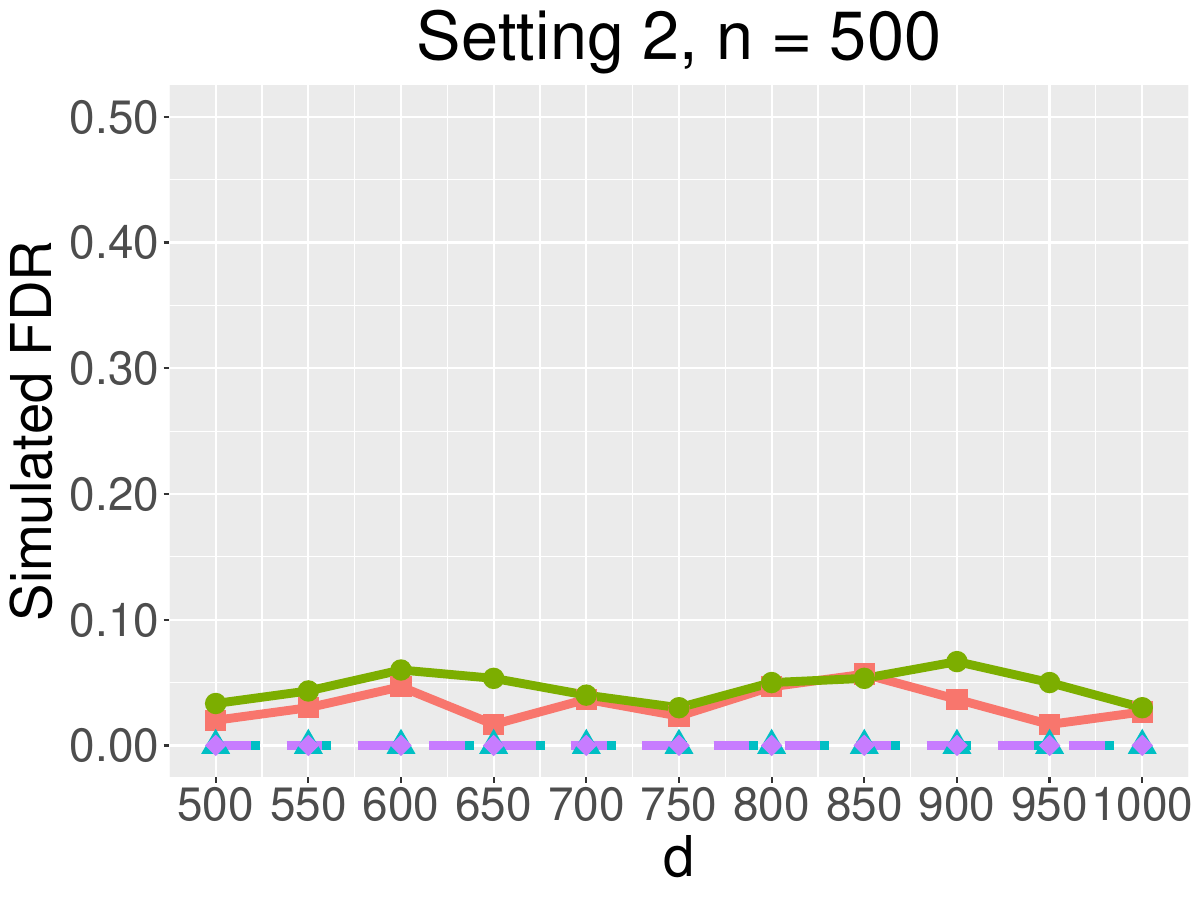}}\hspace{5pt}
    \subfloat{\includegraphics[width=.27\columnwidth]{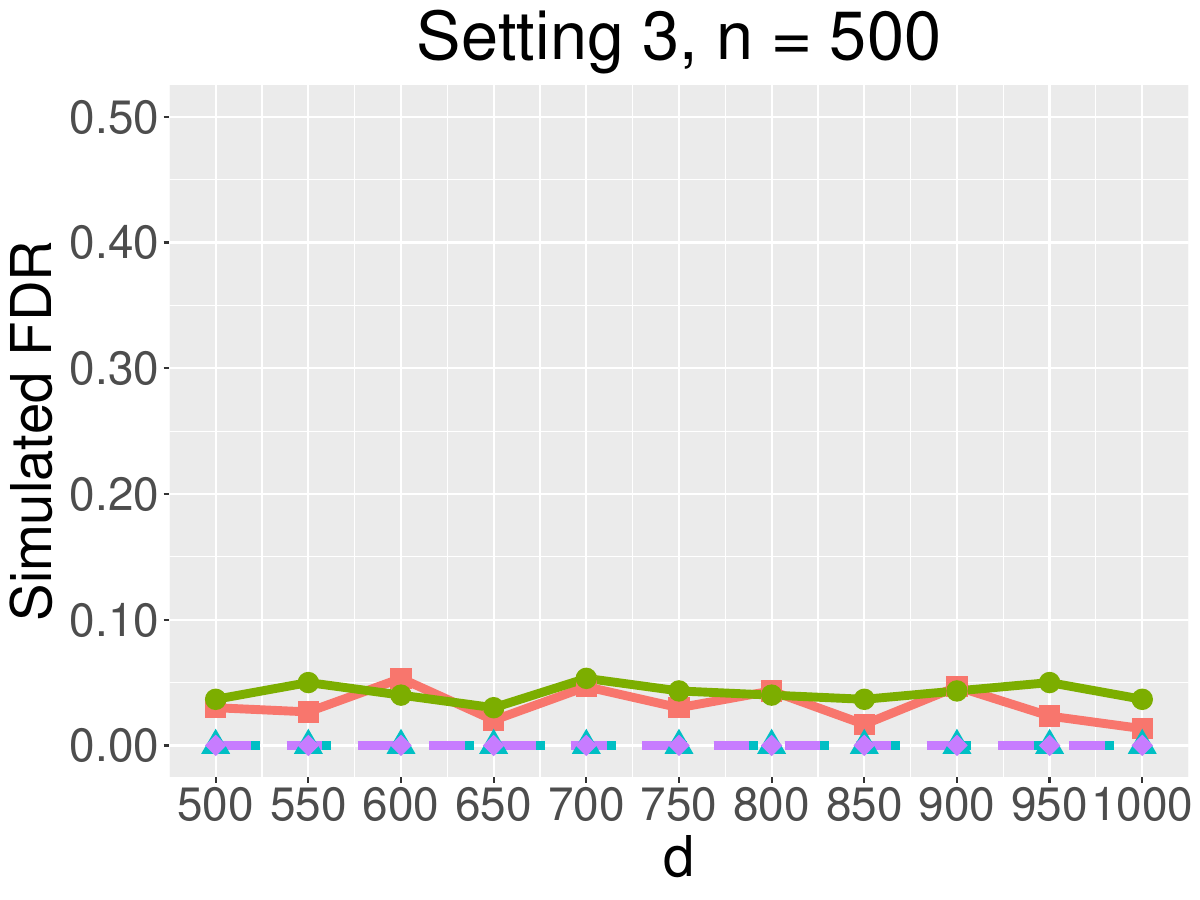}}
	\caption{\small Simulated FDR when all null hypotheses are true, for Setting 1, Setting 2, and Setting 3. The FDR level is $\alpha = 0.05$. The methods compared are Algorithm 1 (squares and red solid line), Algorithm 2 (circles and green solid line), the knockoff-based method of \cite{candes2018panning} (triangles and blue dotted line), and the Gaussian Mirror method of \cite{Xing2021Controlling} with FDP+ procedure (diamonds and purple dashed line).}
    \label{FDR-control-0.05}
\end{figure}

\begin{figure}[htbp!]
	\centering
	\subfloat{\includegraphics[width=.27\columnwidth]{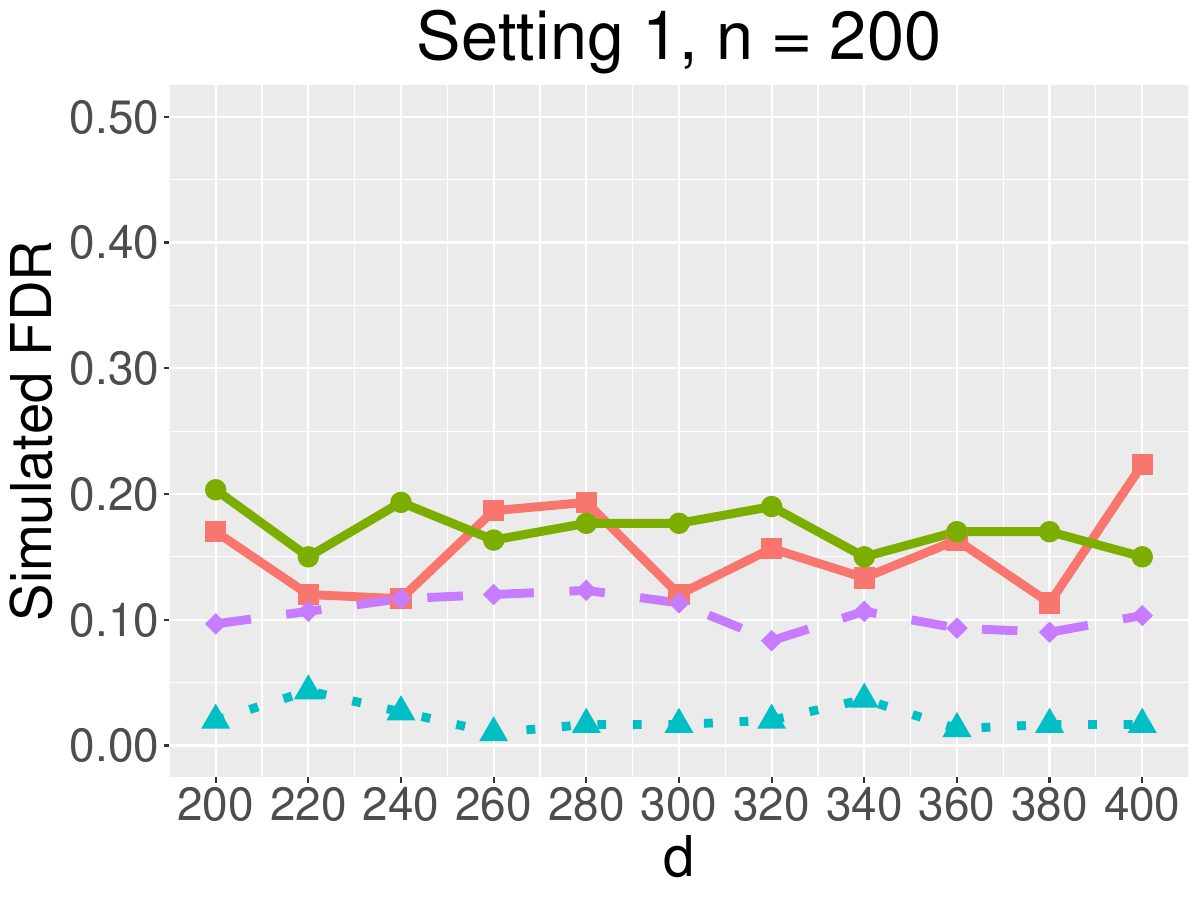}}\hspace{5pt}
	\subfloat{\includegraphics[width=.27\columnwidth]{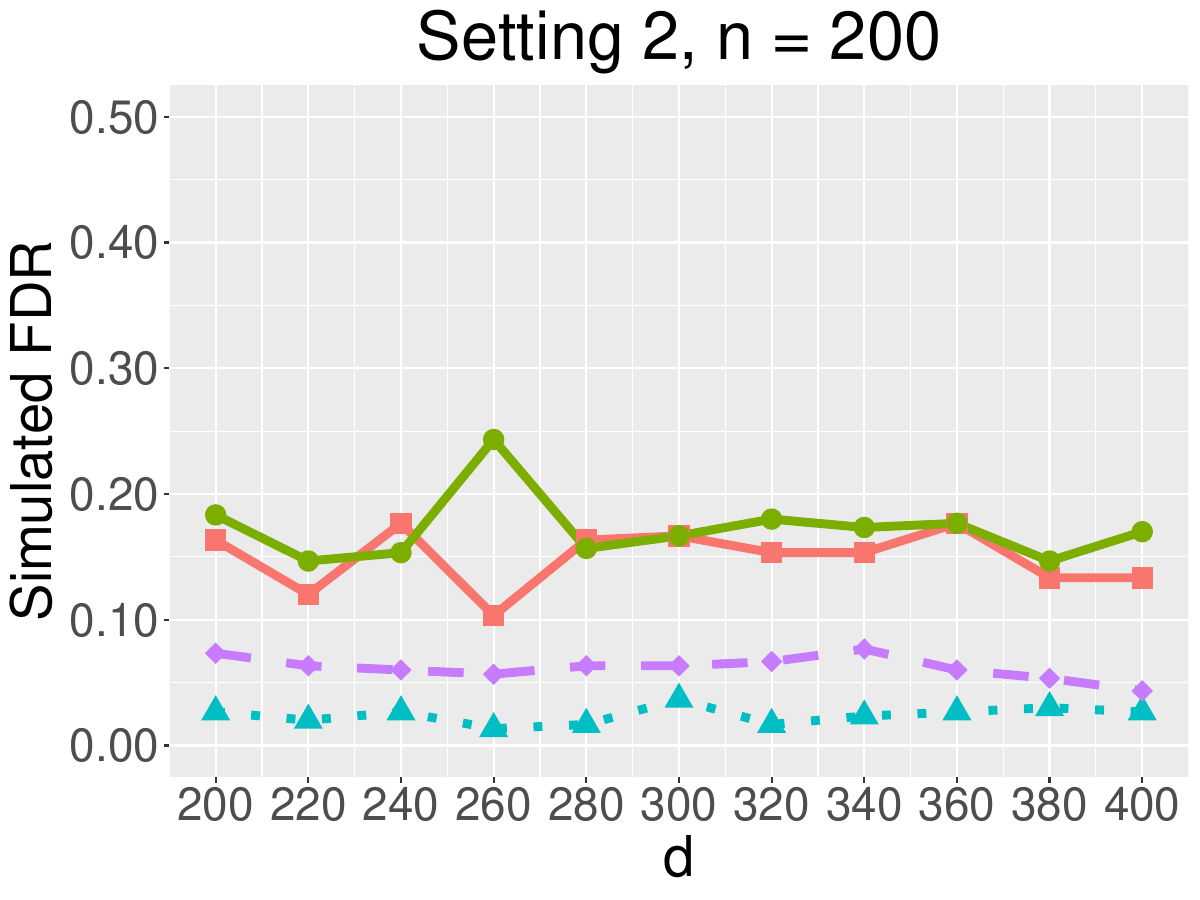}}\hspace{5pt}
	\subfloat{\includegraphics[width=.27\columnwidth]{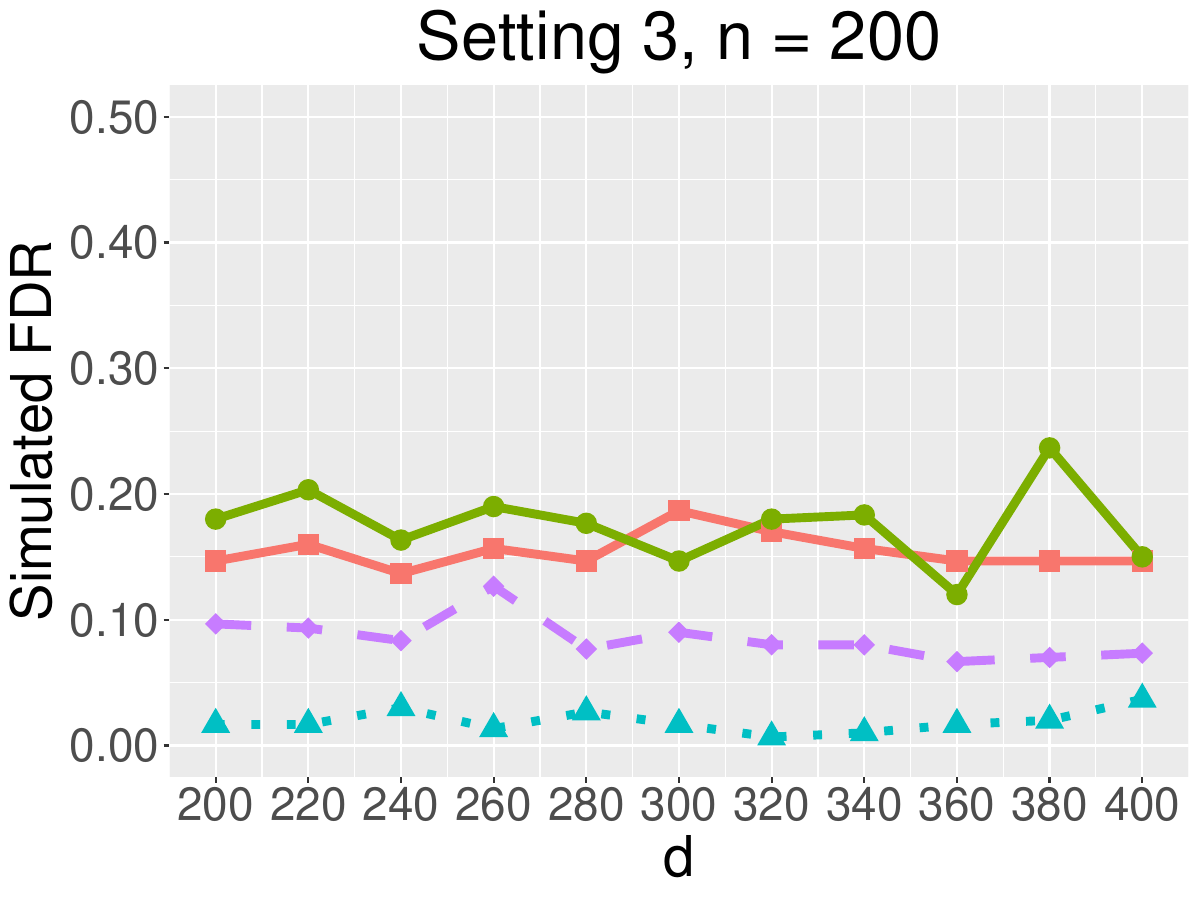}}\\
	\subfloat{\includegraphics[width=.27\columnwidth]{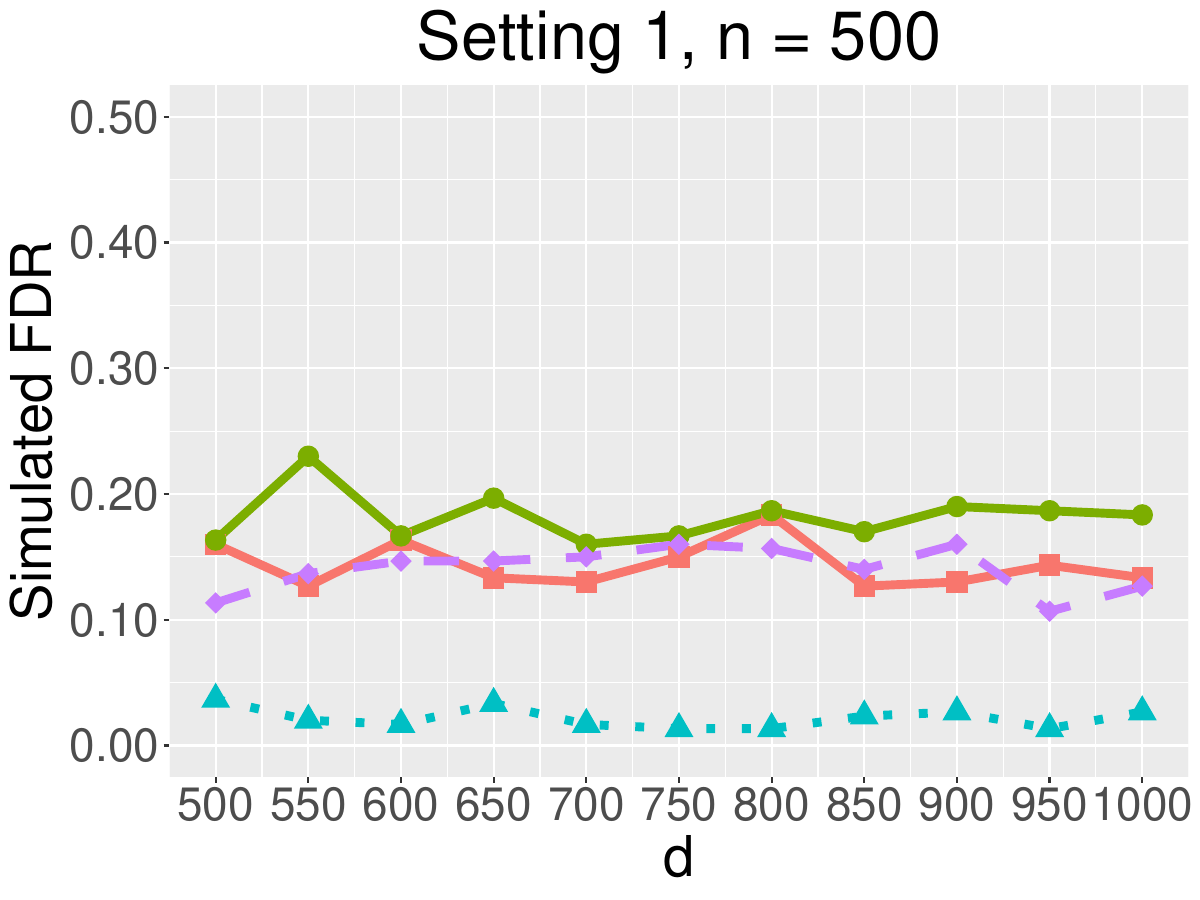}}\hspace{5pt}
    \subfloat{\includegraphics[width=.27\columnwidth]{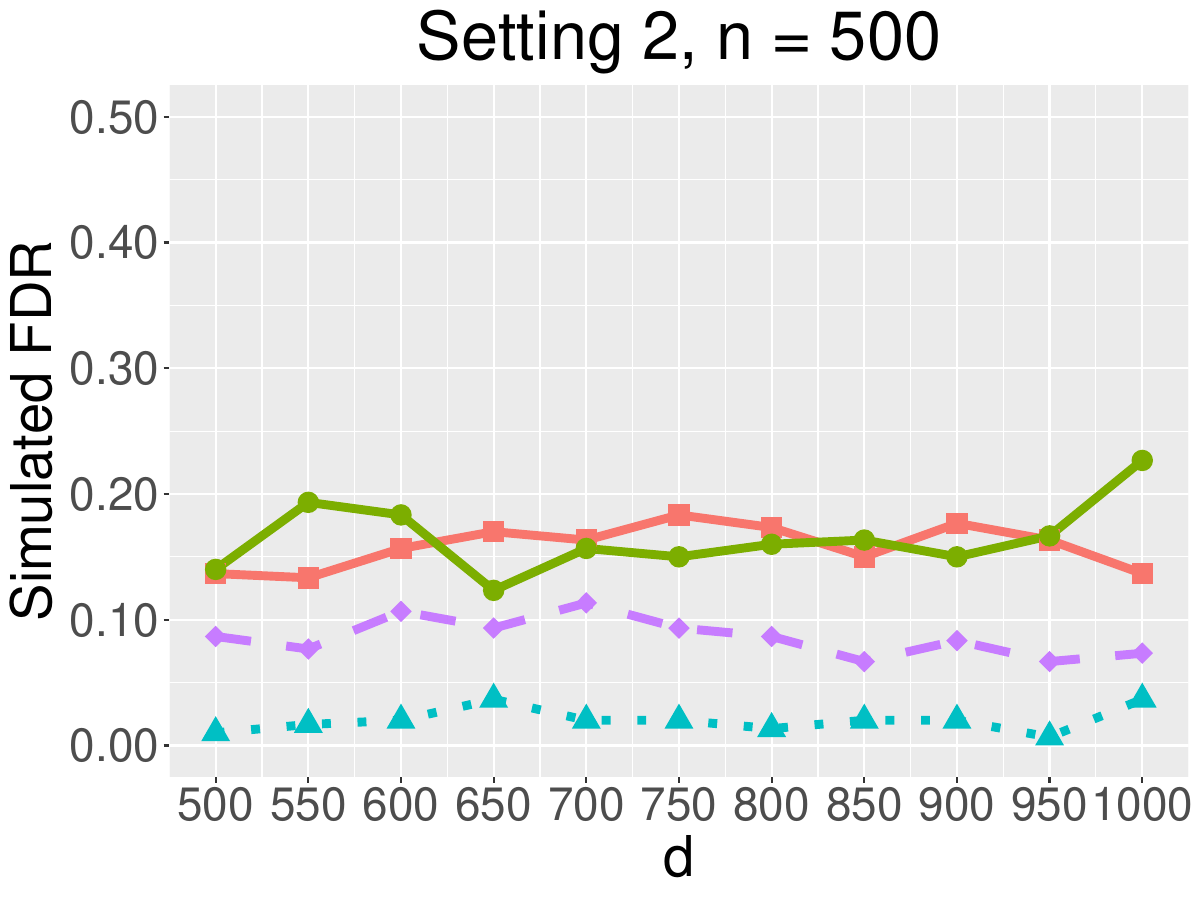}}\hspace{5pt}
    \subfloat{\includegraphics[width=.27\columnwidth]{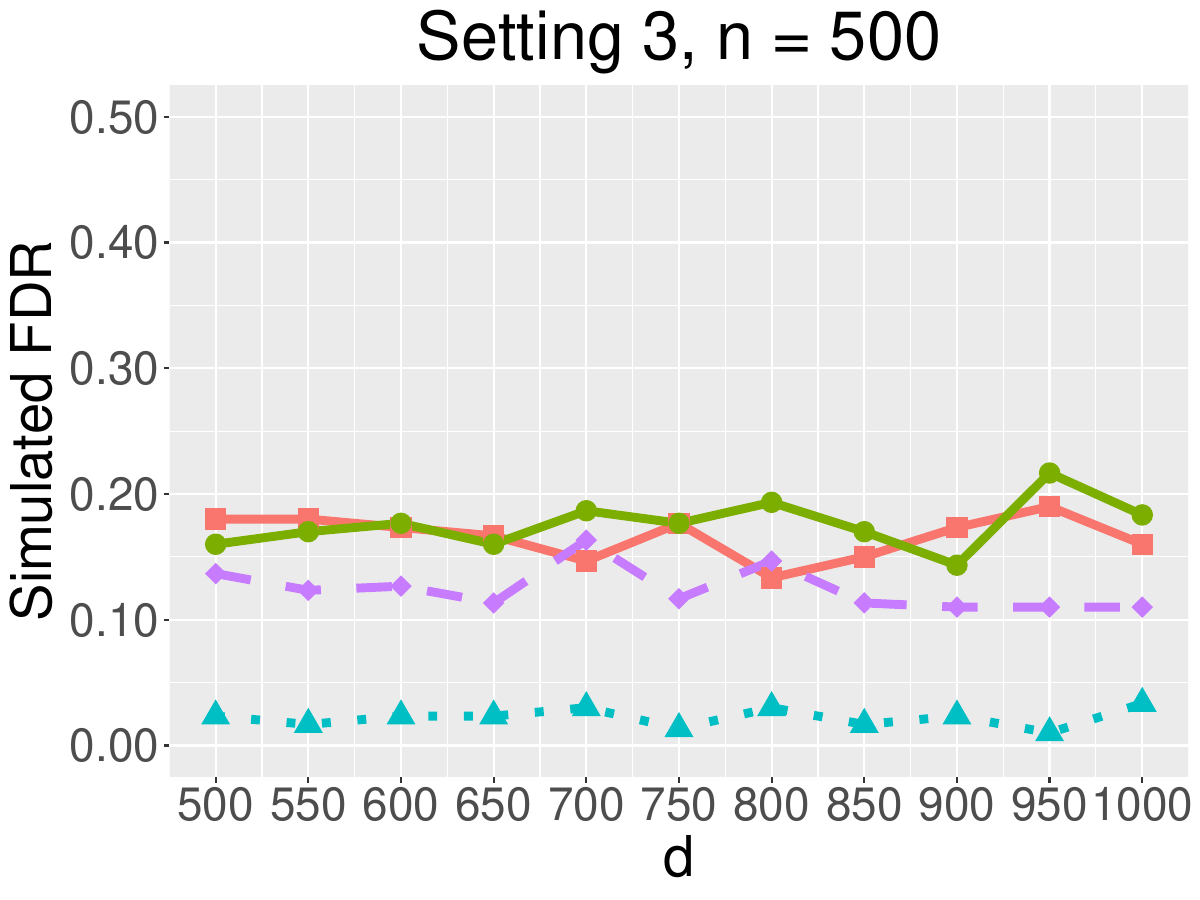}}
	\caption{\small Simulated FDR when all null hypotheses are true, for Setting 1, Setting 2, and Setting 3. The FDR level is $\alpha = 0.2$. The methods compared are Algorithm 1 (squares and red solid line), Algorithm 2 (circles and green solid line), the knockoff-based method of \cite{candes2018panning} (triangles and blue dotted line), and the Gaussian Mirror method of \cite{Xing2021Controlling} with FDP+ procedure (diamonds and purple dashed line).}
    \label{FDR-control-0.2}
\end{figure}

We first provide all additional simulation results when all null hypotheses are true. Figures \ref{FDR-control-0.05} and \ref{FDR-control-0.2} report the FDR control results under FDR levels $0.05$ and $0.2$, respectively.

\subsection{Additional simulation results of power analysis}\label{SM2:FIGURE}
\begin{figure}[htbp!]
	\centering
	\subfloat{\includegraphics[width=.27\columnwidth]{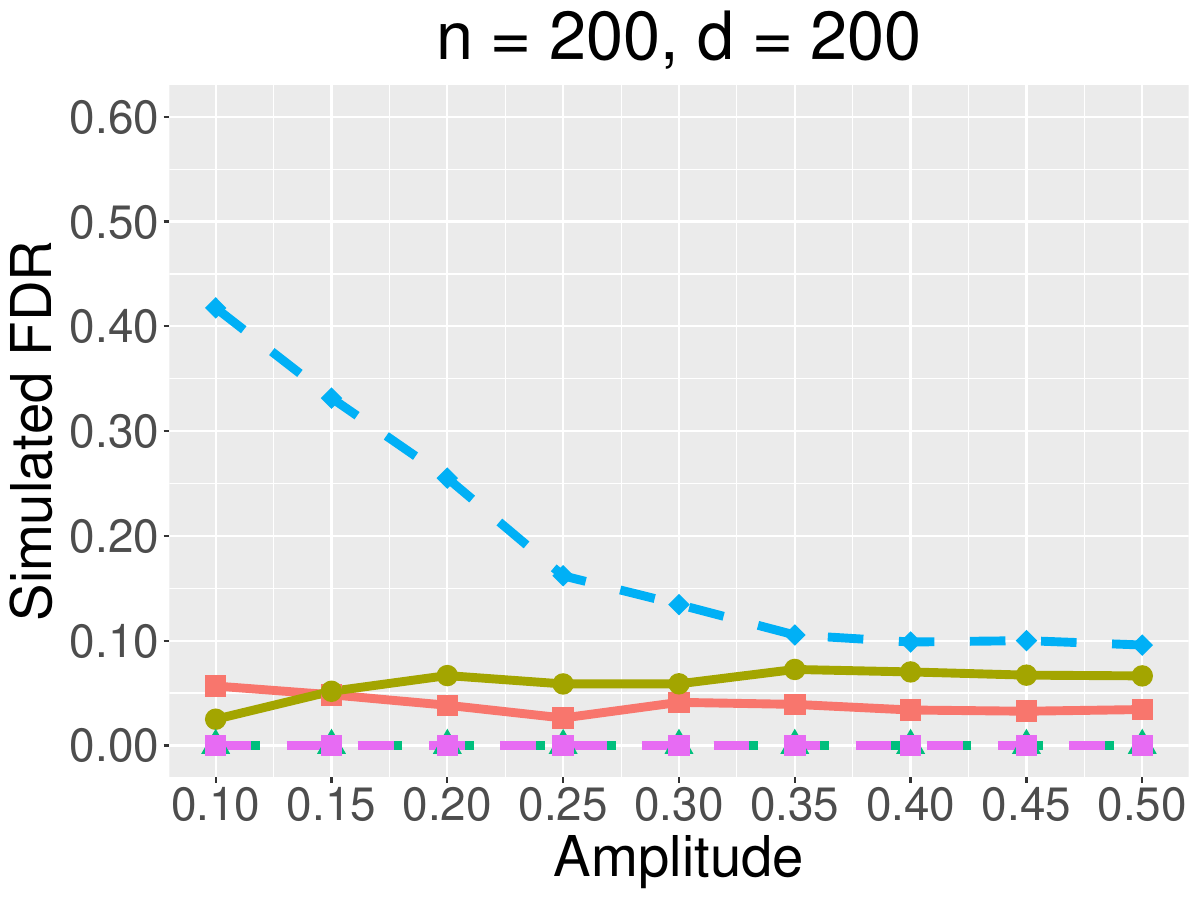}}\hspace{5pt}
	\subfloat{\includegraphics[width=.27\columnwidth]{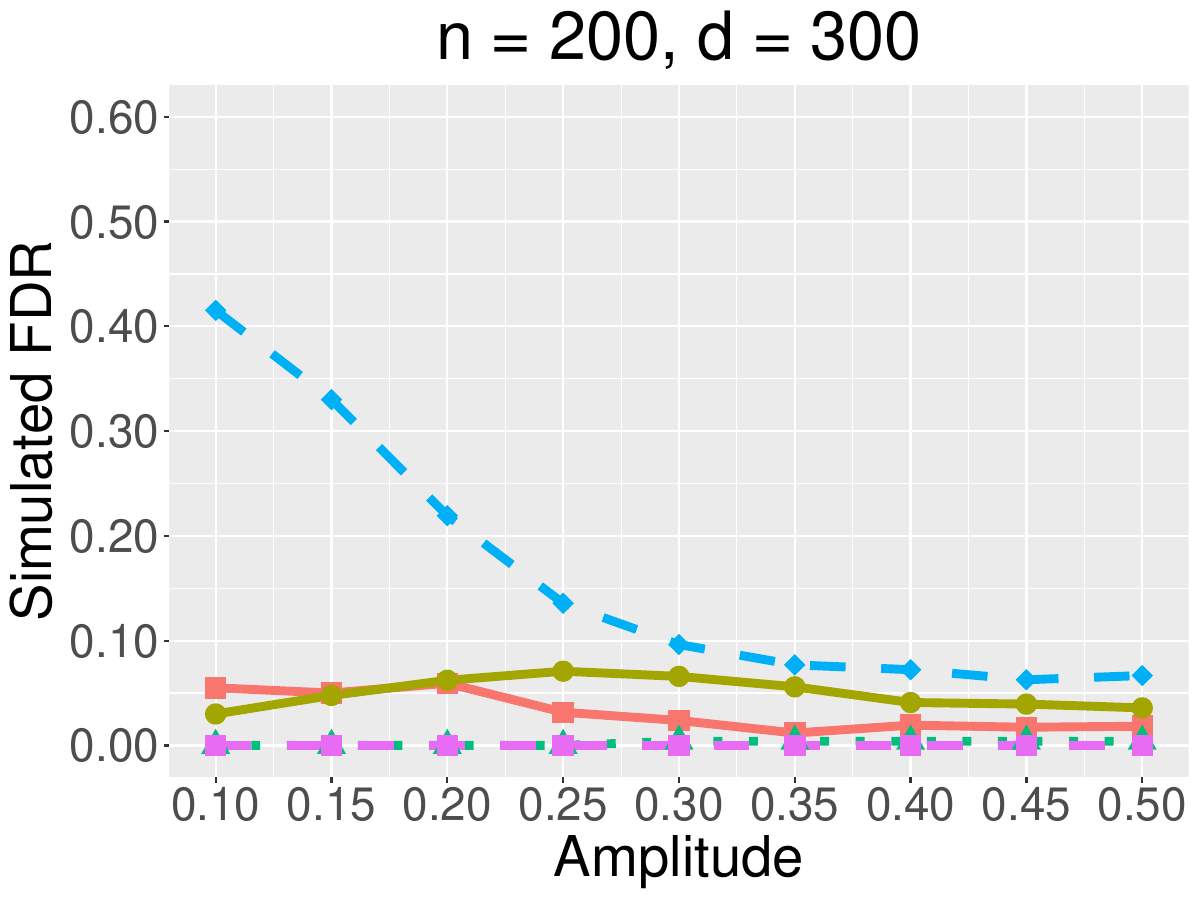}}\hspace{5pt}
	\subfloat{\includegraphics[width=.27\columnwidth]{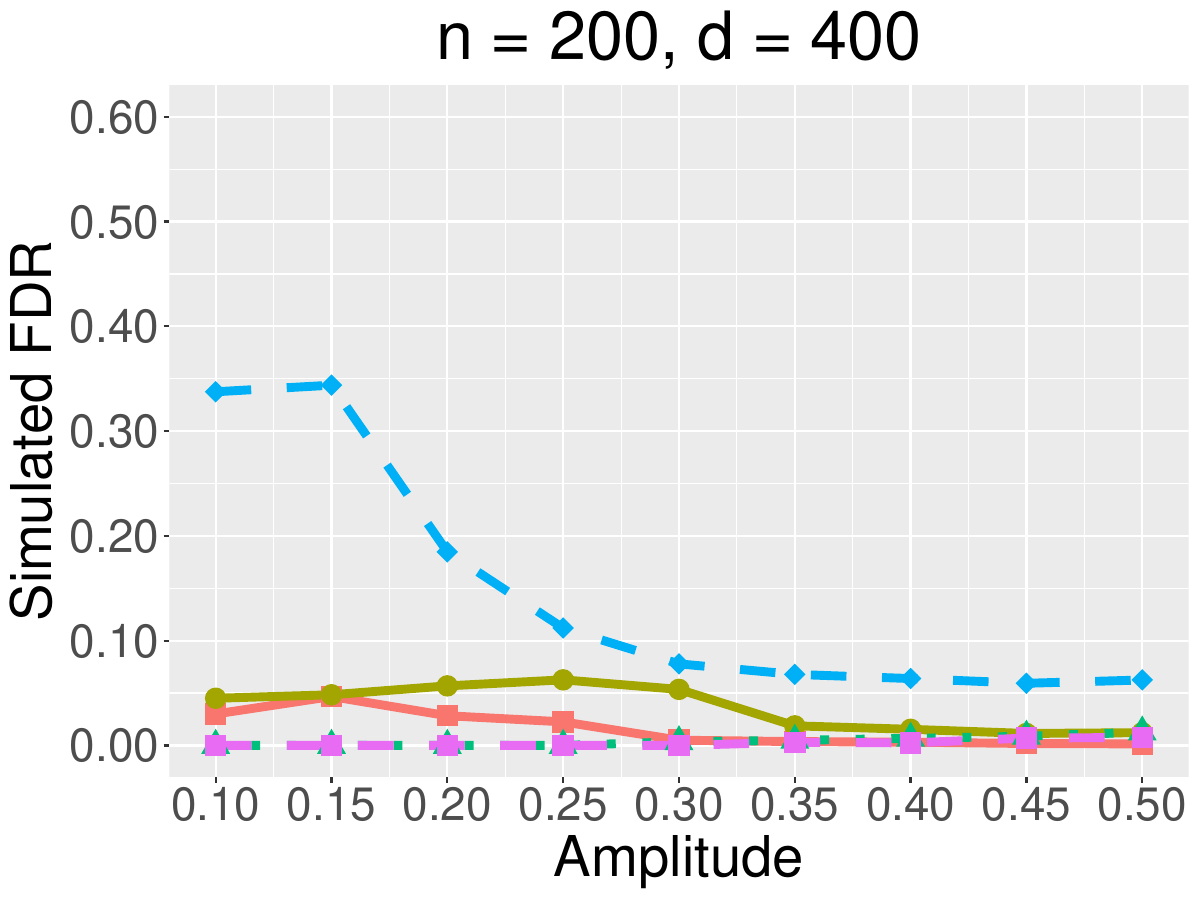}}\\
	\subfloat{\includegraphics[width=.27\columnwidth]{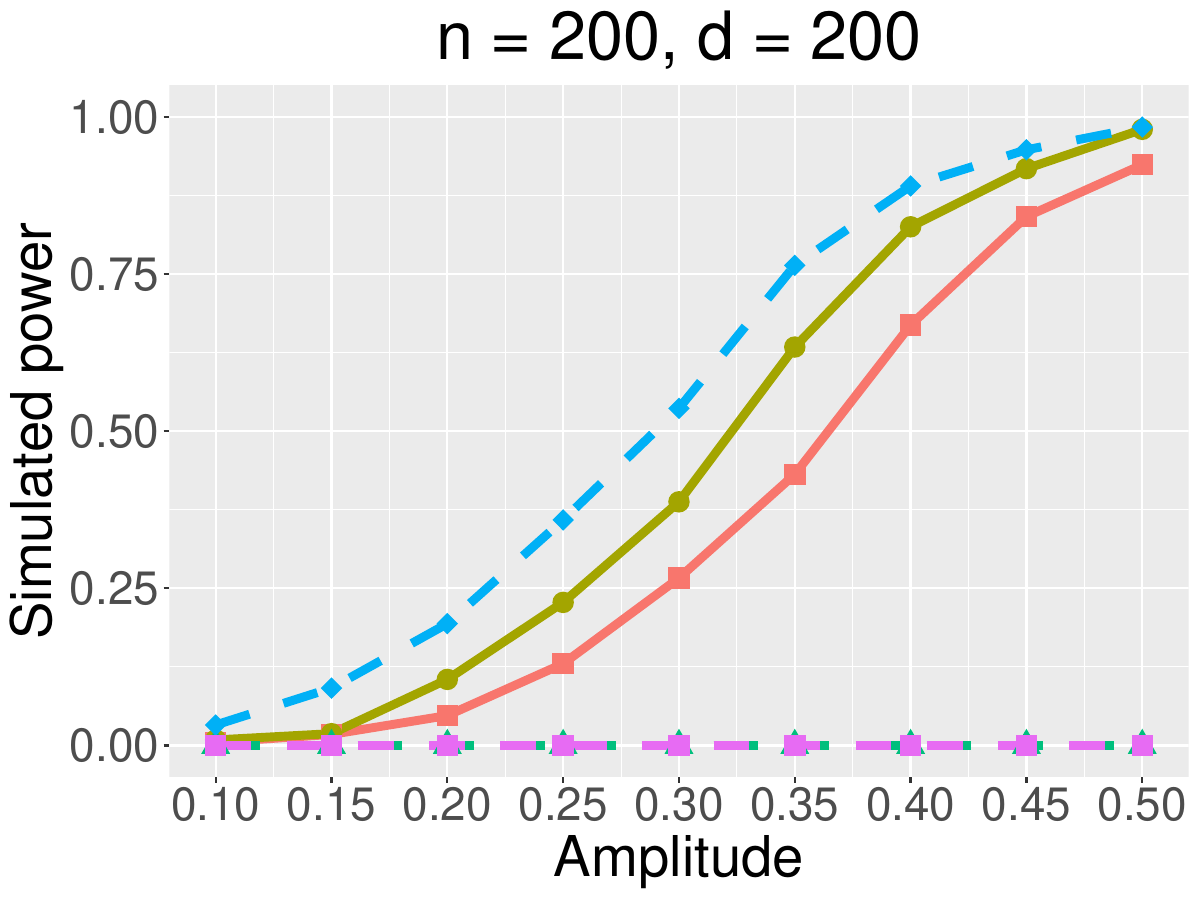}}\hspace{5pt}
    \subfloat{\includegraphics[width=.27\columnwidth]{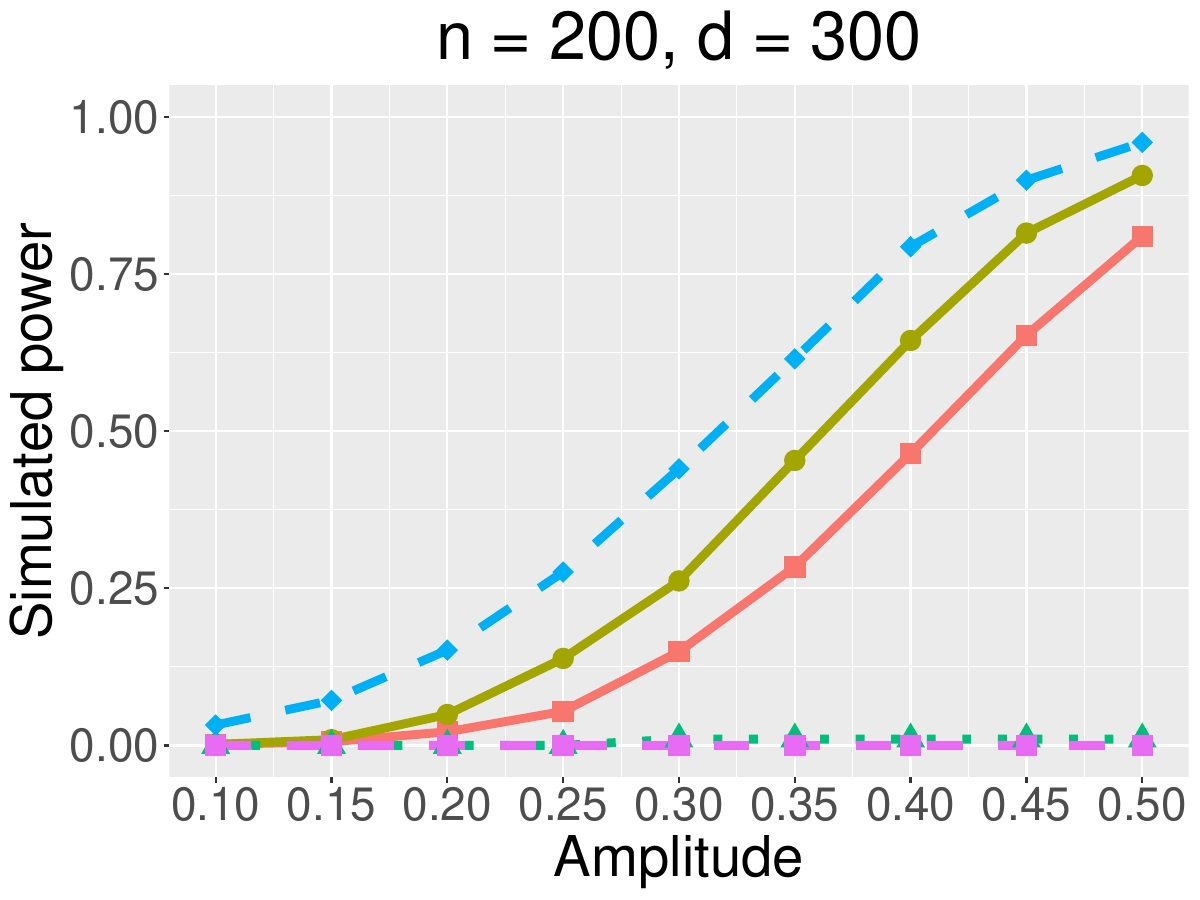}}\hspace{5pt}
    \subfloat{\includegraphics[width=.27\columnwidth]{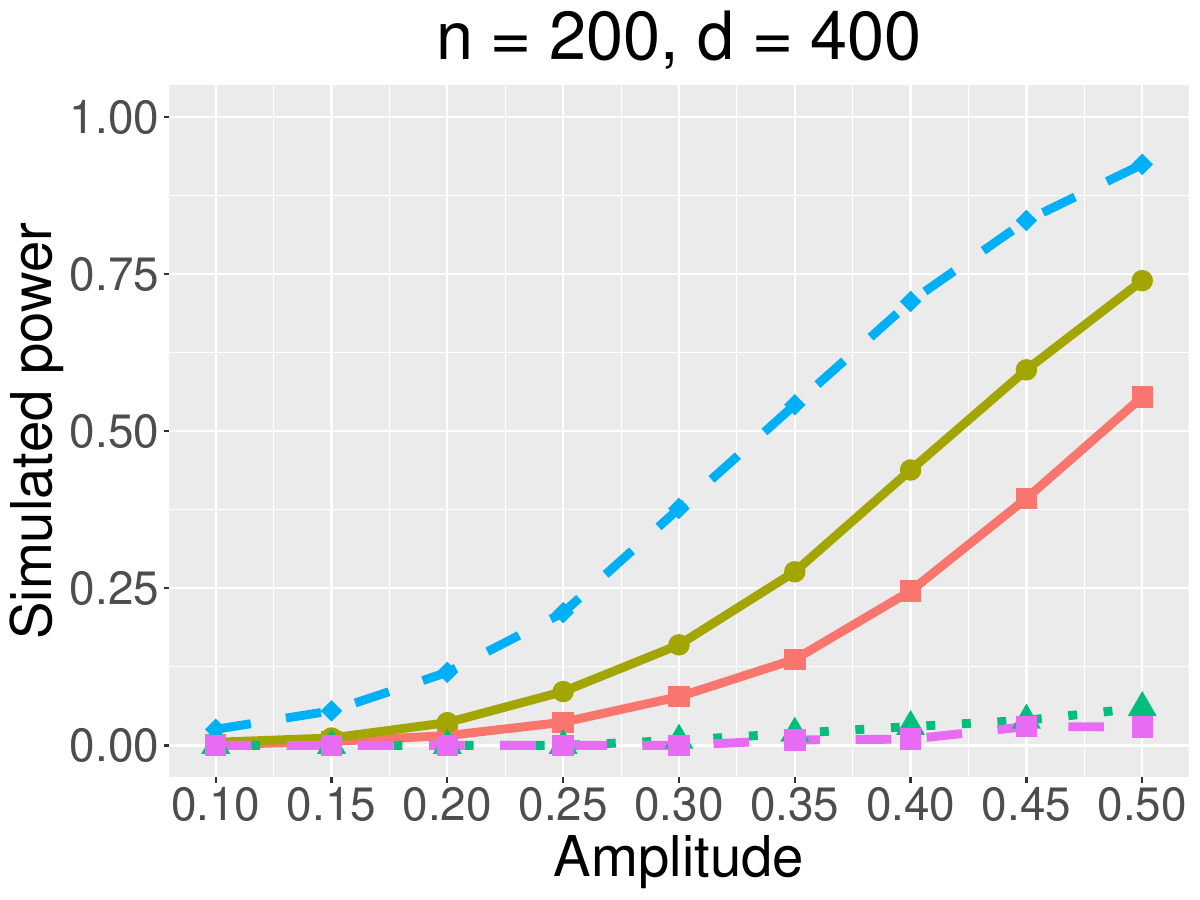}}
	\caption{\small Simulated FDR and power for different combinations of $(n,d)$. The rows of the design matrix were generated from Setting 2. The sparsity level is $k = 0.04d$ and the FDR level is $\alpha = 0.05$. The methods compared are Algorithm 1 (squares and red solid line), Algorithm 2 (circles and yellow solid line), the knockoff-based method of \cite{candes2018panning} (triangles and green dotted line), the Gaussian Mirror method of \cite{Xing2021Controlling} (diamonds and blue dashed line), and the Gaussian Mirror method with FDP+ procedure (squares and purple dashed line).}
    \label{Power-0.05}
\end{figure}

\begin{figure}[htbp!]
	\centering
	\subfloat{\includegraphics[width=.27\columnwidth]{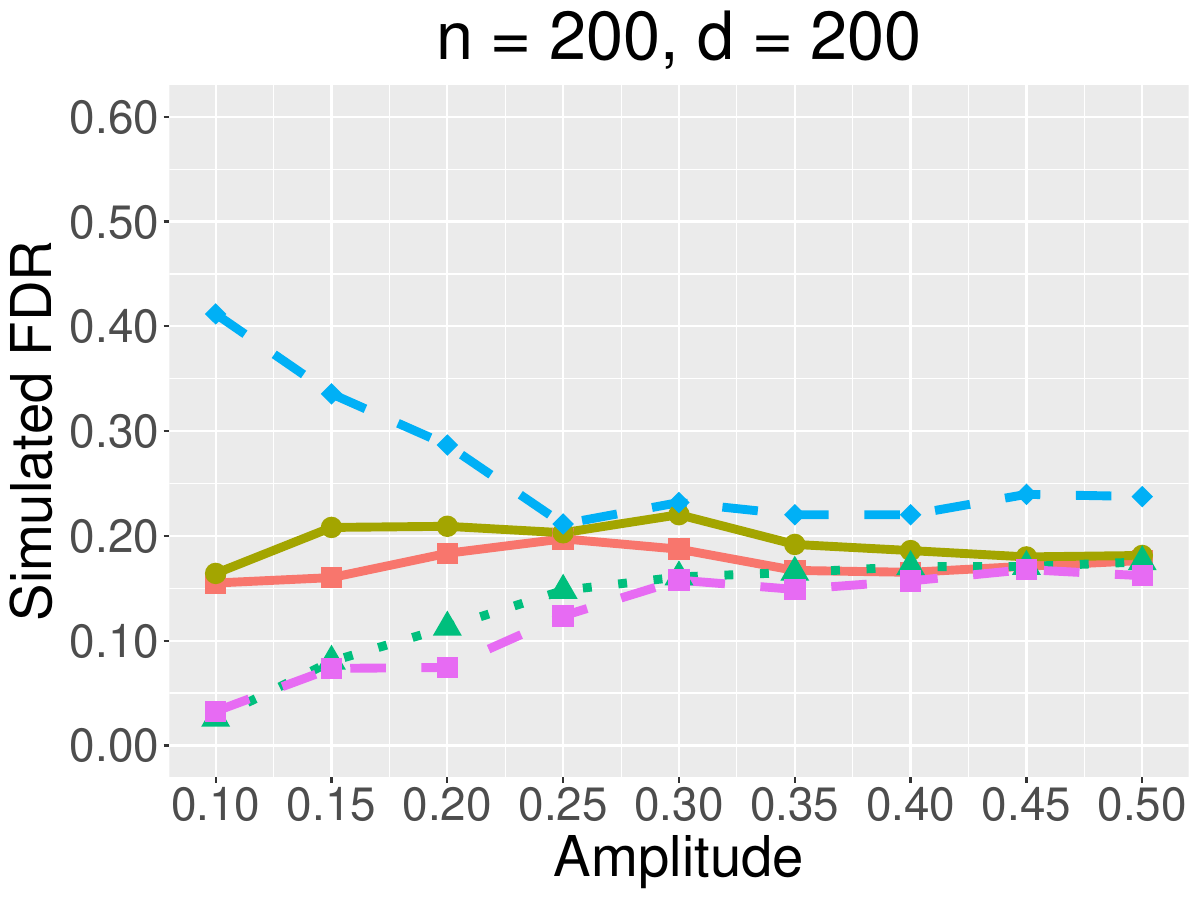}}\hspace{5pt}
	\subfloat{\includegraphics[width=.27\columnwidth]{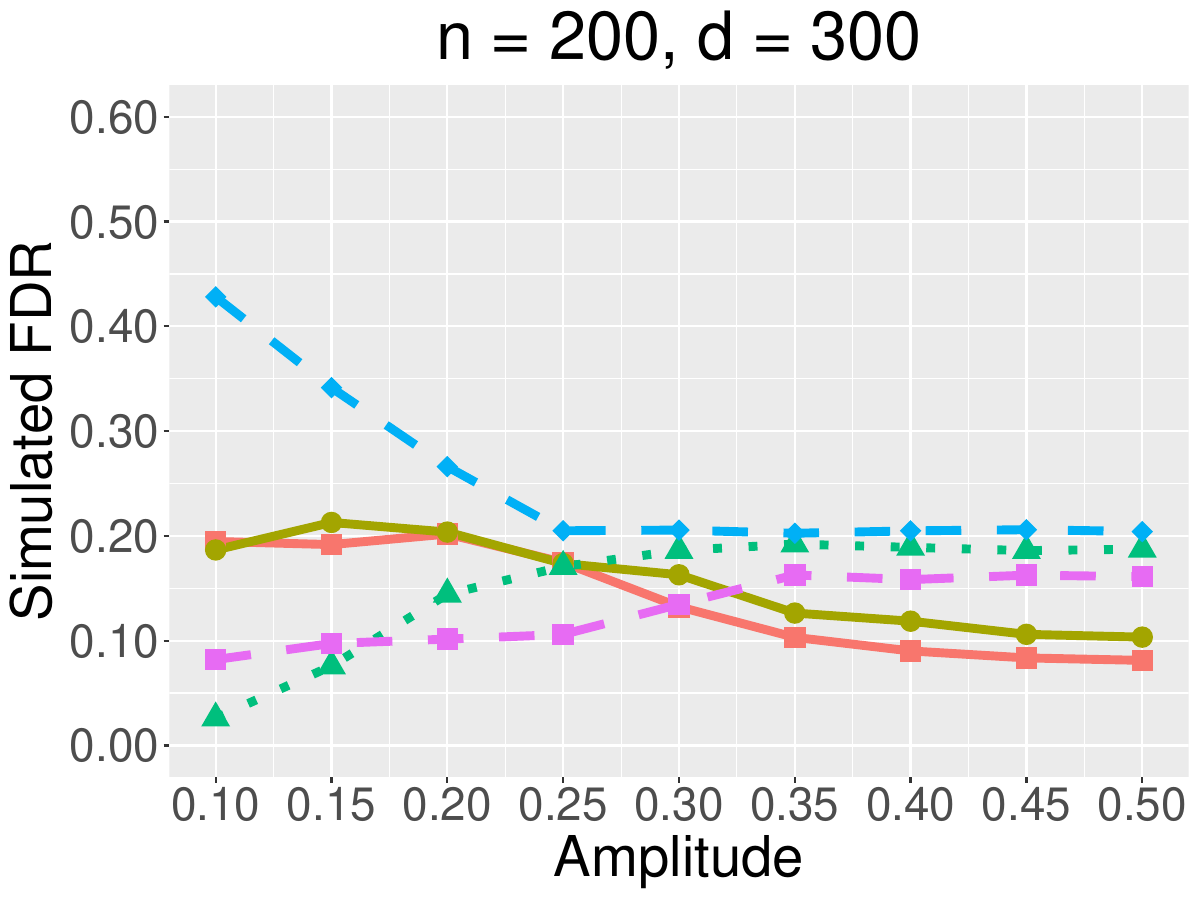}}\hspace{5pt}
	\subfloat{\includegraphics[width=.27\columnwidth]{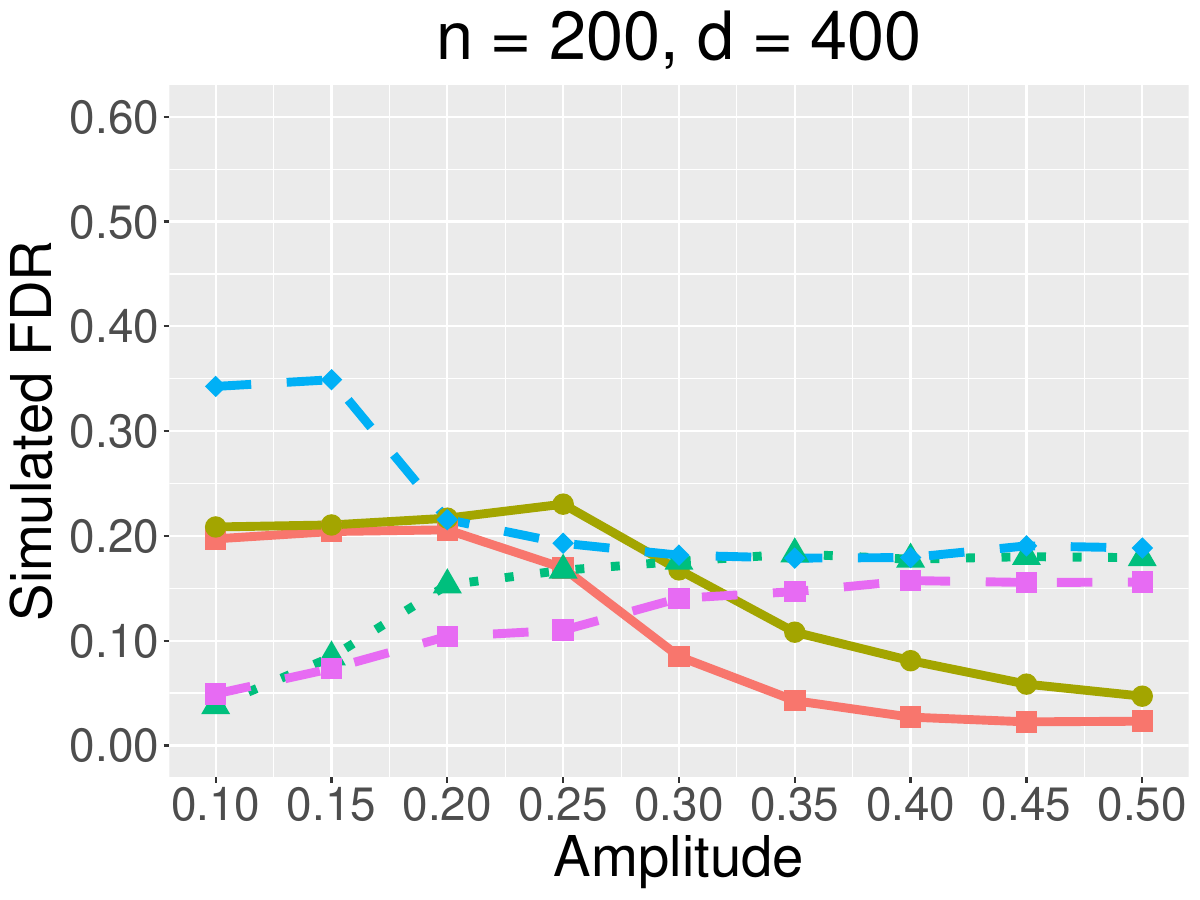}}\\
	\subfloat{\includegraphics[width=.27\columnwidth]{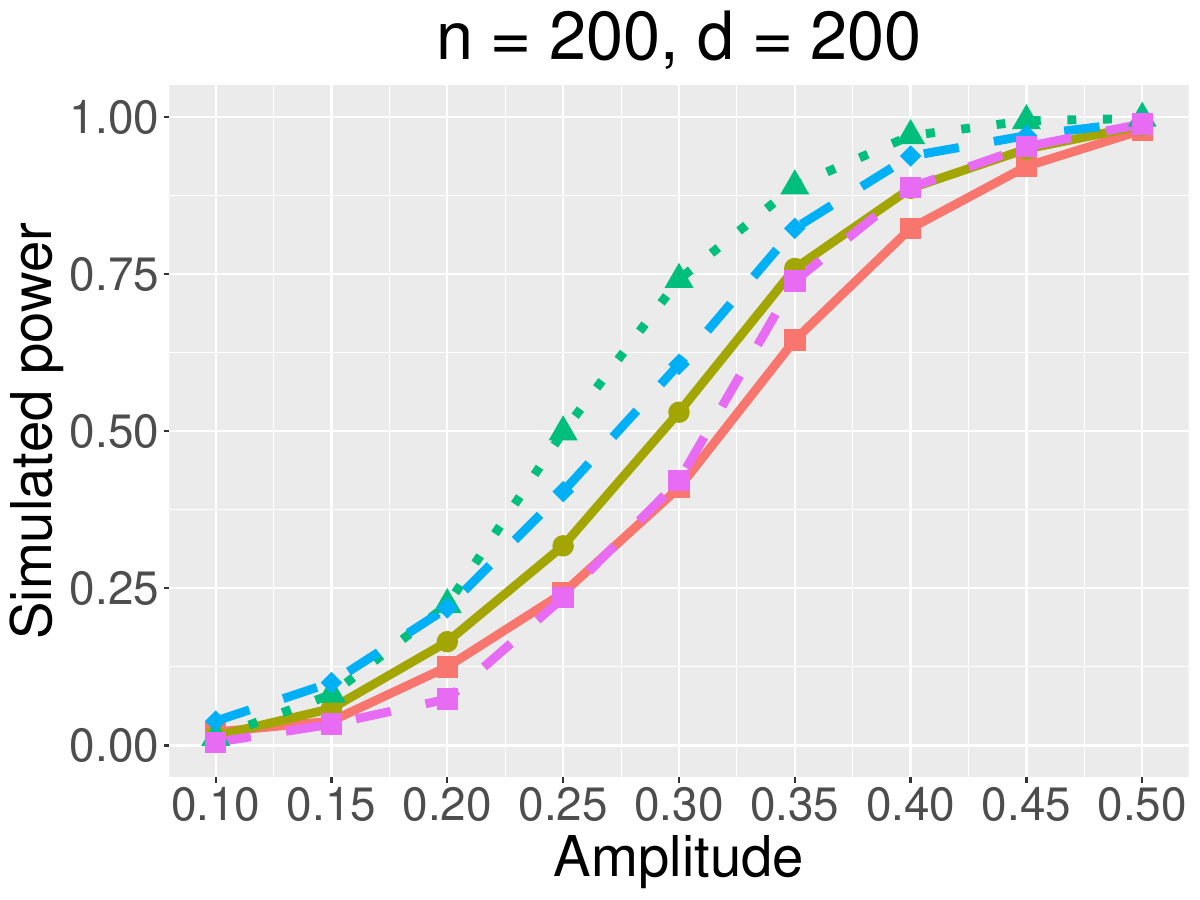}}\hspace{5pt}
    \subfloat{\includegraphics[width=.27\columnwidth]{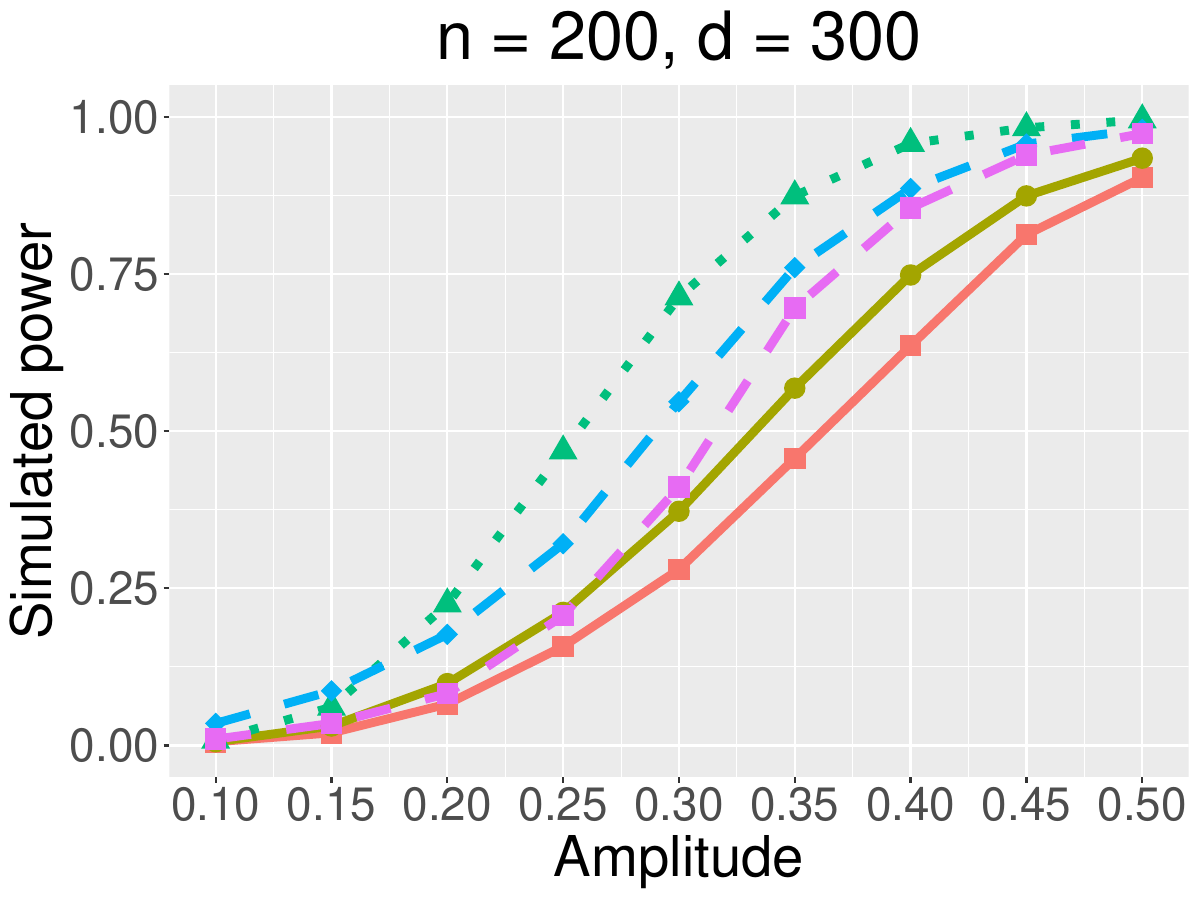}}\hspace{5pt}
    \subfloat{\includegraphics[width=.27\columnwidth]{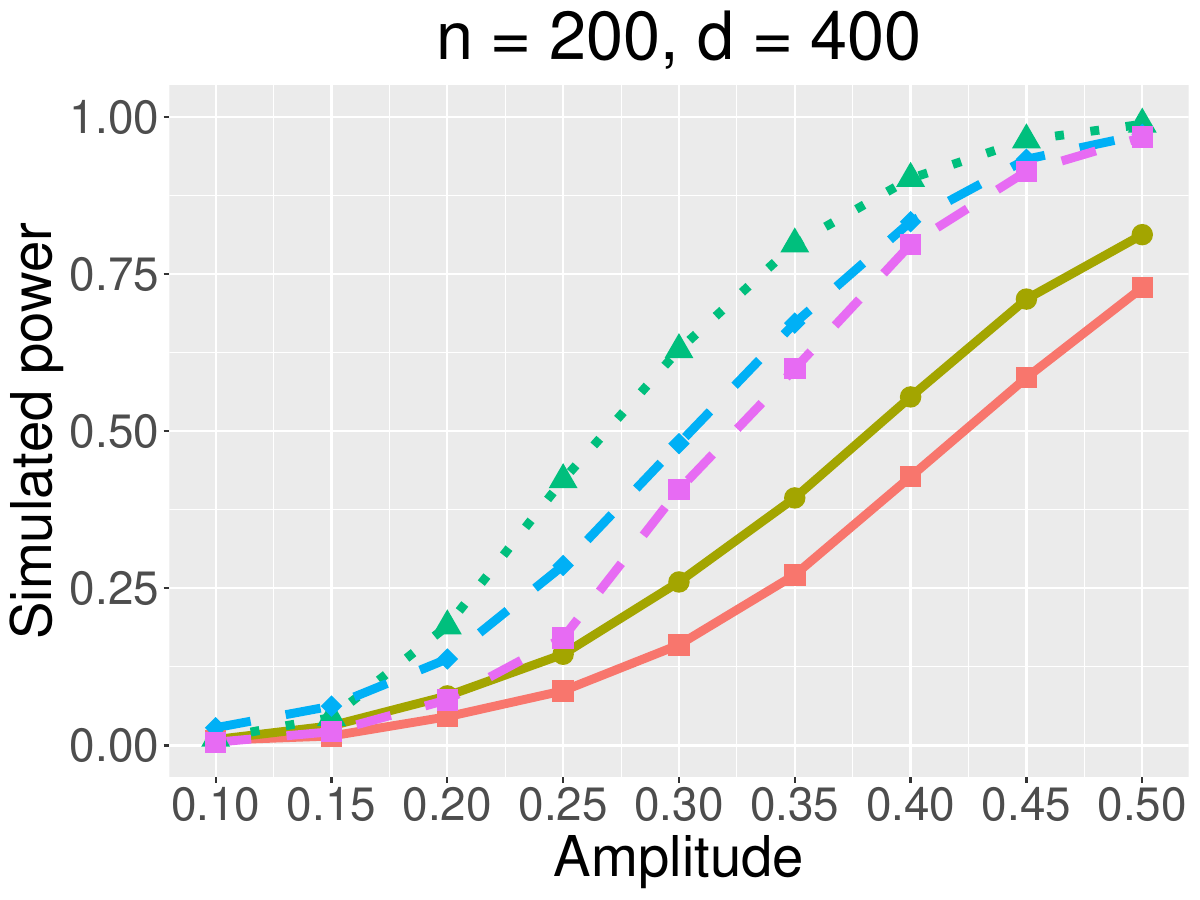}}
	\caption{\small Simulated FDR and power for different combinations of $(n,d)$. The rows of the design matrix were generated from Setting 2. The sparsity level is $k = 0.04d$ and the FDR level is $\alpha = 0.2$. The methods compared are Algorithm 1 (squares and red solid line), Algorithm 2 (circles and yellow solid line), the knockoff-based method of \cite{candes2018panning} (triangles and green dotted line), the Gaussian Mirror method of \cite{Xing2021Controlling} (diamonds and blue dashed line), and the Gaussian Mirror method with FDP+ procedure (squares and purple dashed line).}
    \label{Power-0.2}
\end{figure}

\begin{figure}[htbp!]
	\centering
	\subfloat{\includegraphics[width=.27\columnwidth]{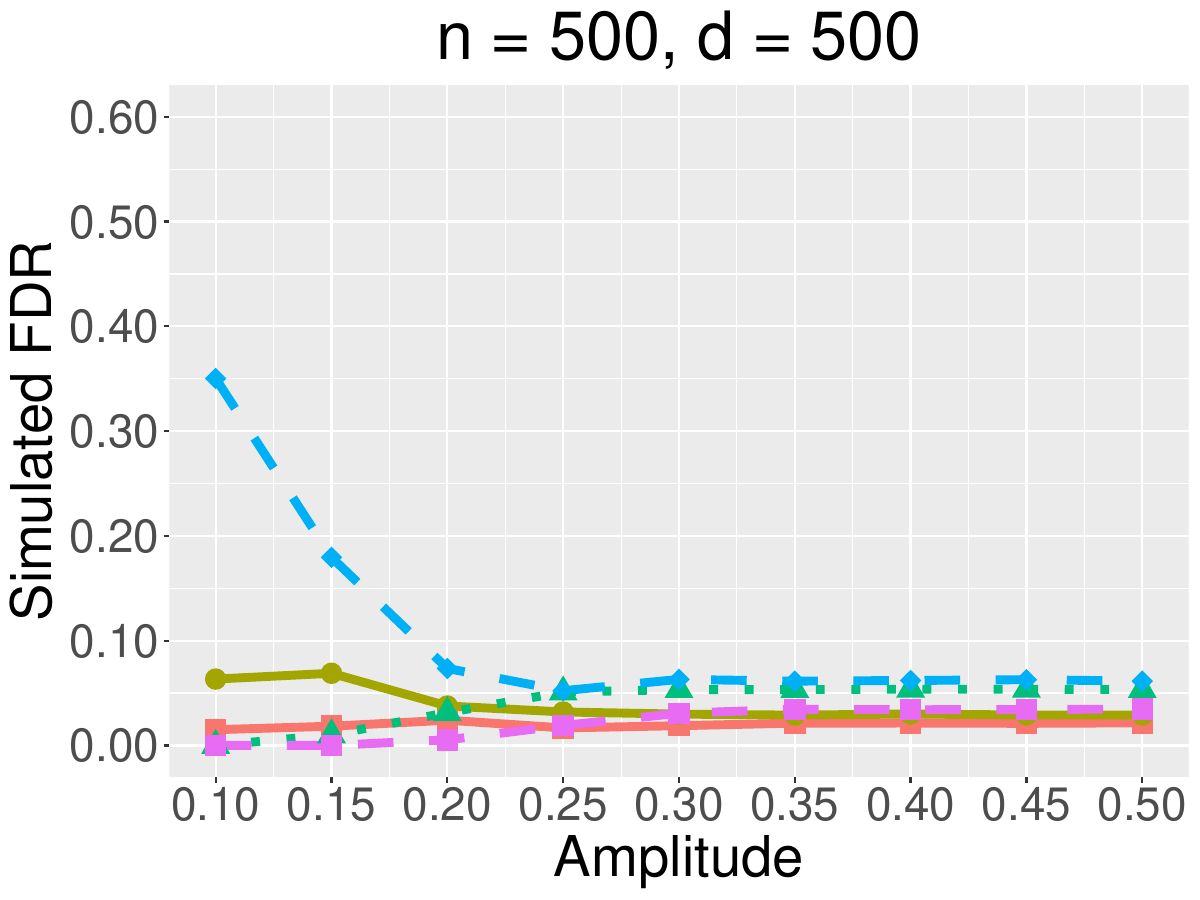}}\hspace{5pt}
	\subfloat{\includegraphics[width=.27\columnwidth]{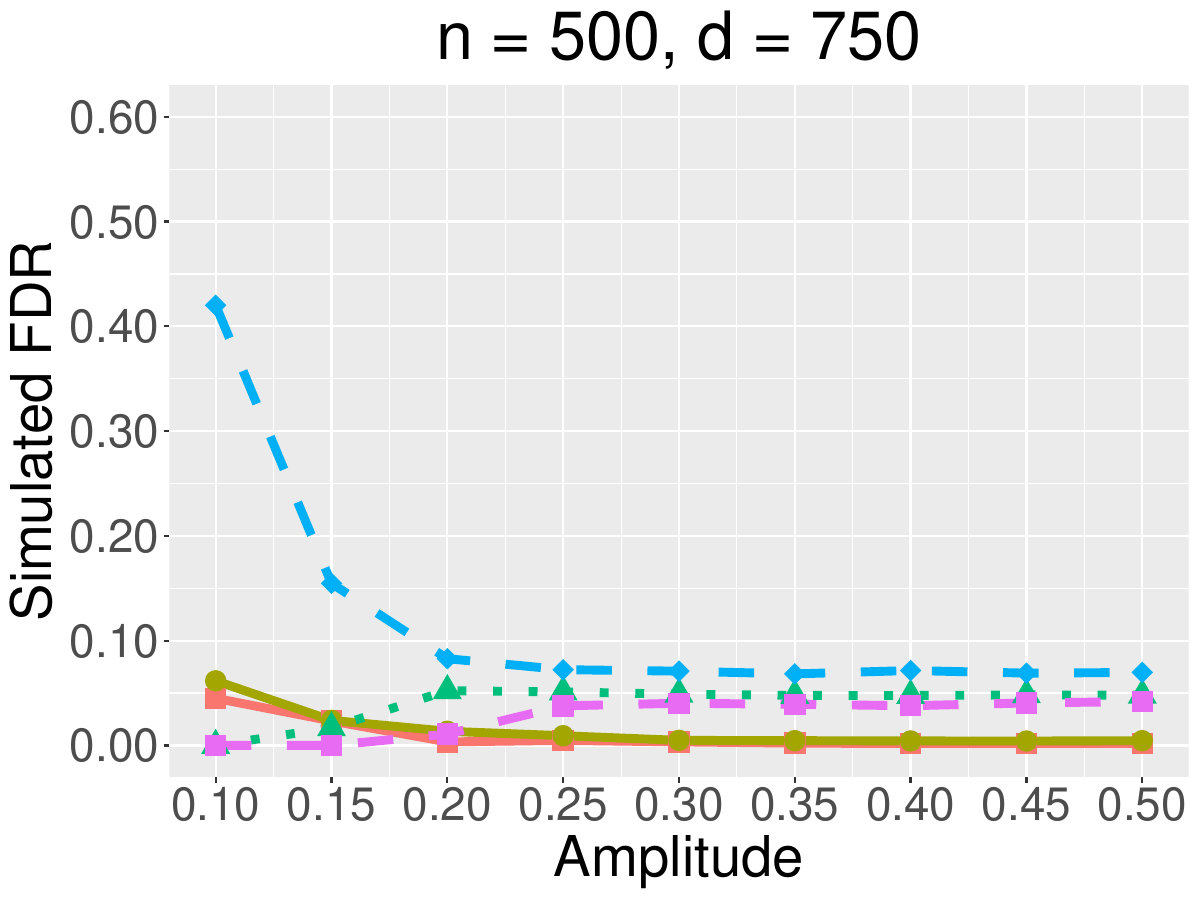}}\hspace{5pt}
	\subfloat{\includegraphics[width=.27\columnwidth]{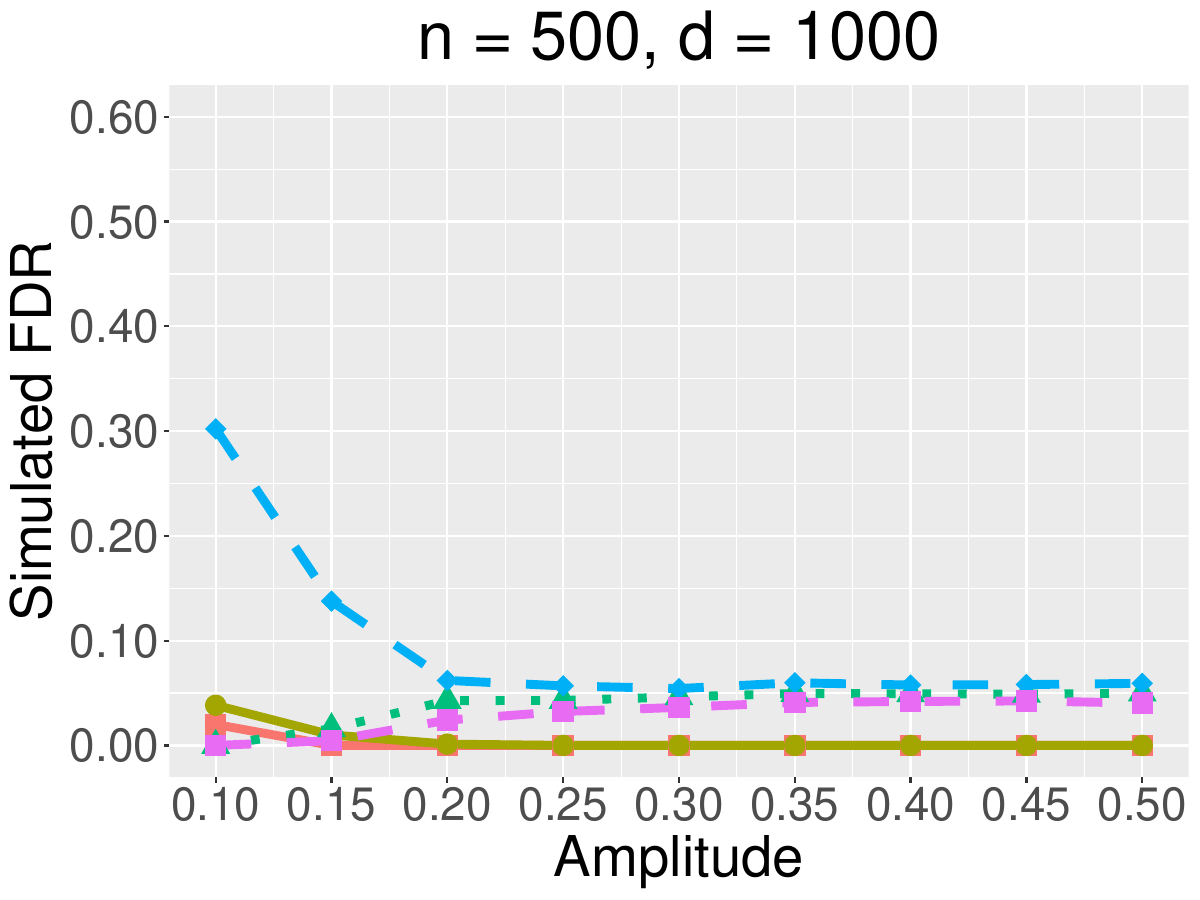}}\\
	\subfloat{\includegraphics[width=.27\columnwidth]{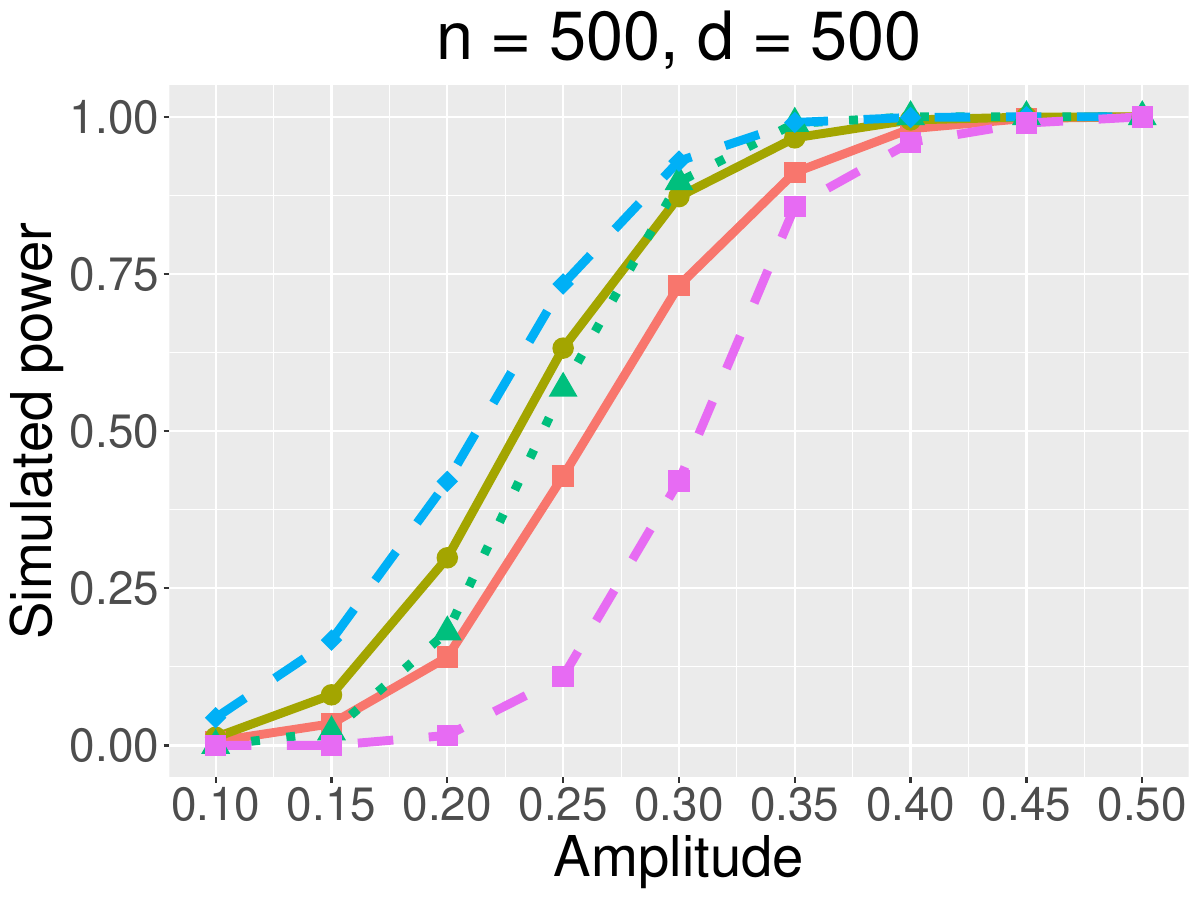}}\hspace{5pt}
    \subfloat{\includegraphics[width=.27\columnwidth]{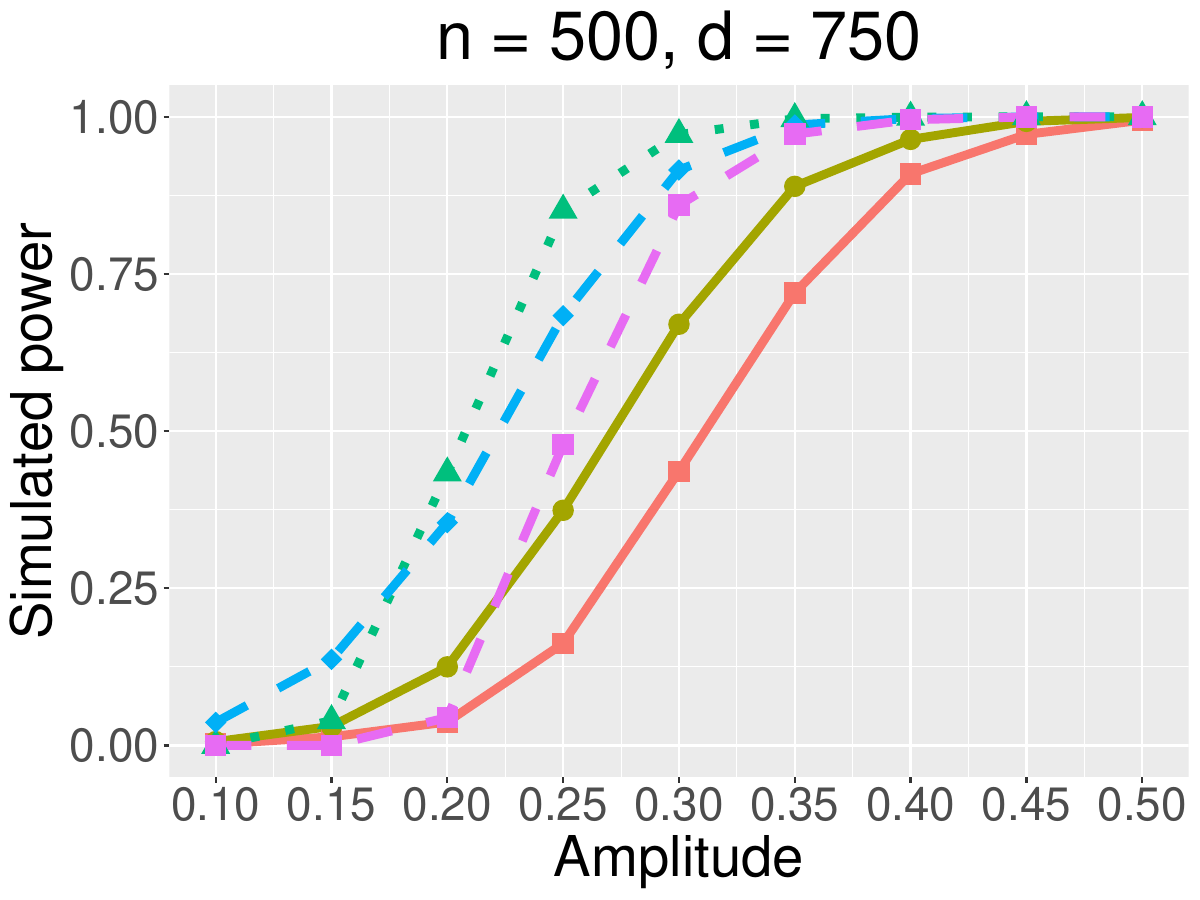}}\hspace{5pt}
    \subfloat{\includegraphics[width=.27\columnwidth]{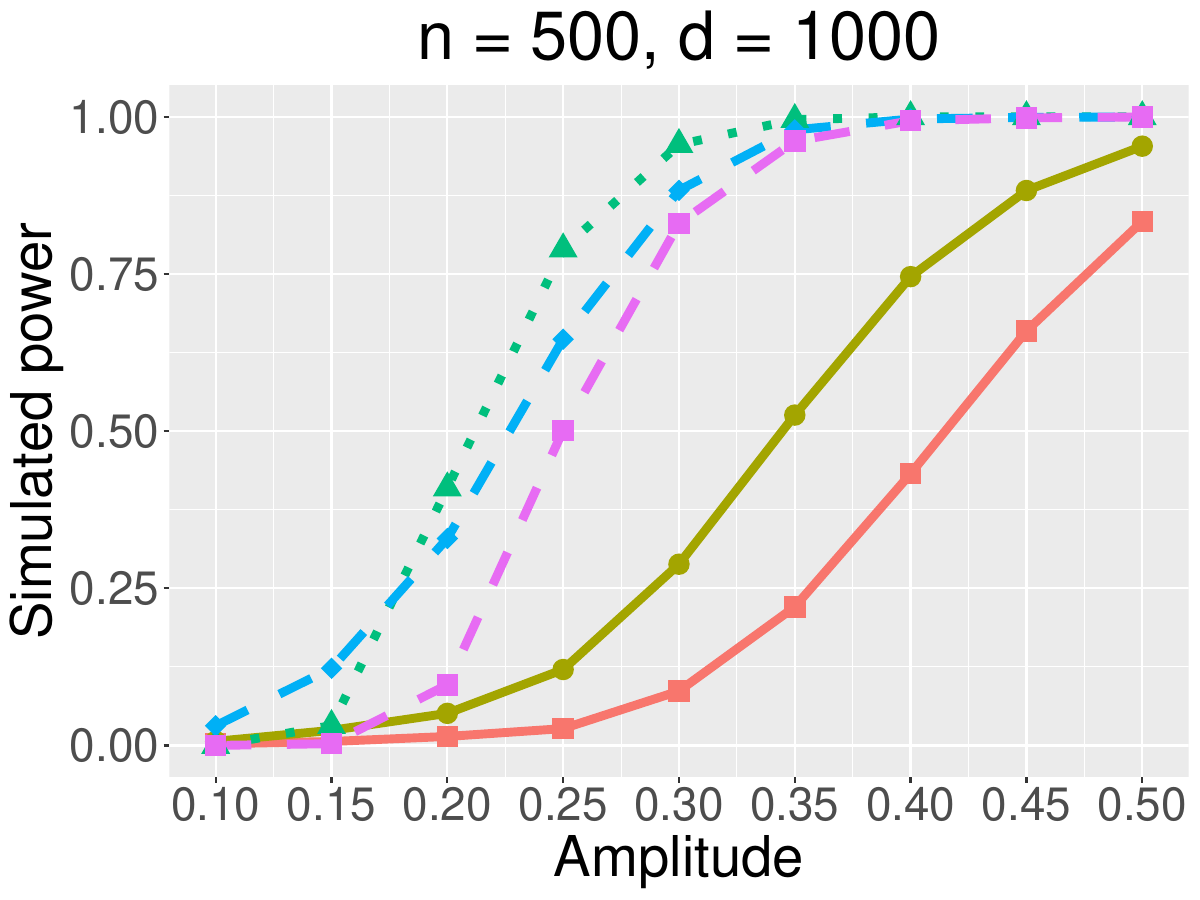}}
	\caption{\small Simulated FDR and power for different combinations of $(n,d)$. The rows of the design matrix were generated from Setting 2. The sparsity level is $k = 0.04d$ and the FDR level is $\alpha = 0.05$. The methods compared are Algorithm 1 (squares and red solid line), Algorithm 2 (circles and yellow solid line), the knockoff-based method of \cite{candes2018panning} (triangles and green dotted line), the Gaussian Mirror method of \cite{Xing2021Controlling} (diamonds and blue dashed line), and the Gaussian Mirror method with FDP+ procedure (squares and purple dashed line).}
    \label{Power-large-0.05}
\end{figure}

\begin{figure}[htbp!]
	\centering
	\subfloat{\includegraphics[width=.27\columnwidth]{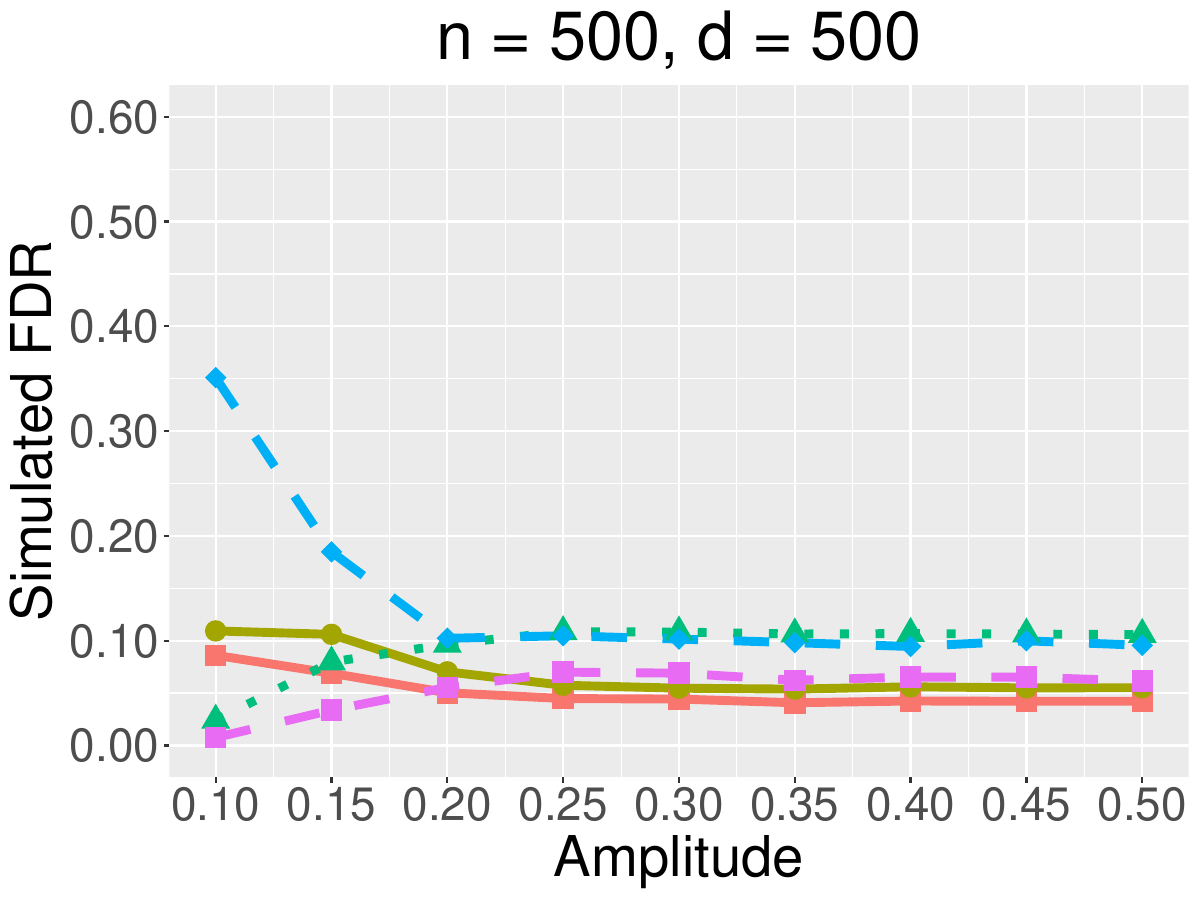}}\hspace{5pt}
	\subfloat{\includegraphics[width=.27\columnwidth]{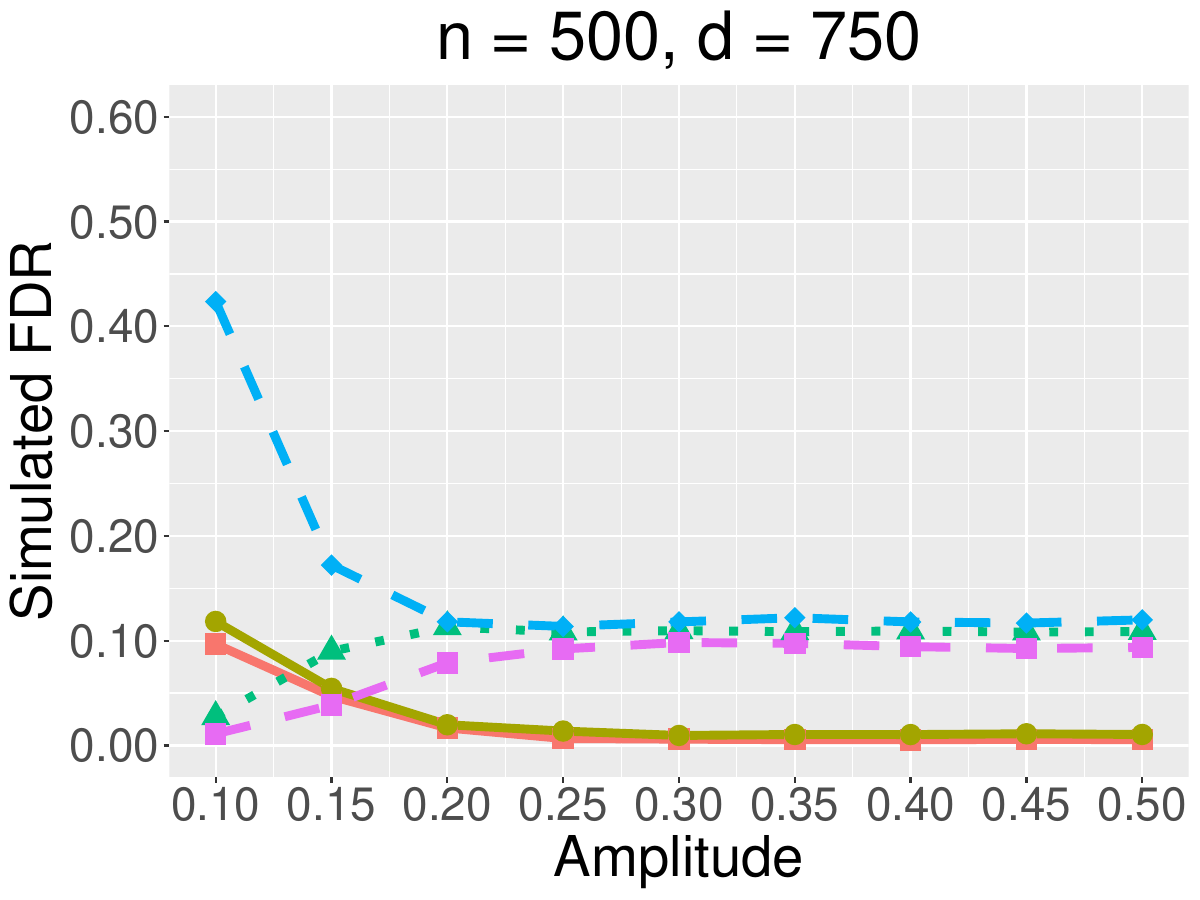}}\hspace{5pt}
	\subfloat{\includegraphics[width=.27\columnwidth]{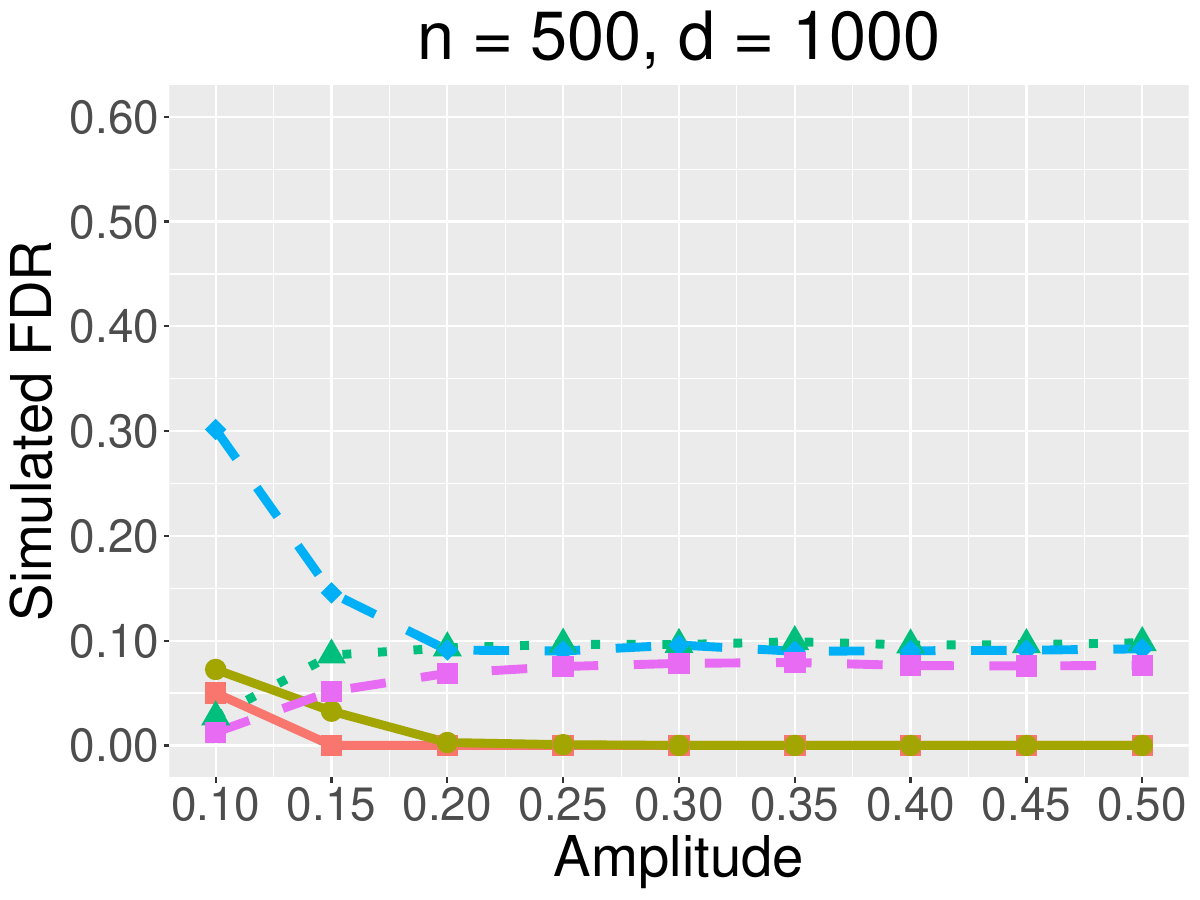}}\\
	\subfloat{\includegraphics[width=.27\columnwidth]{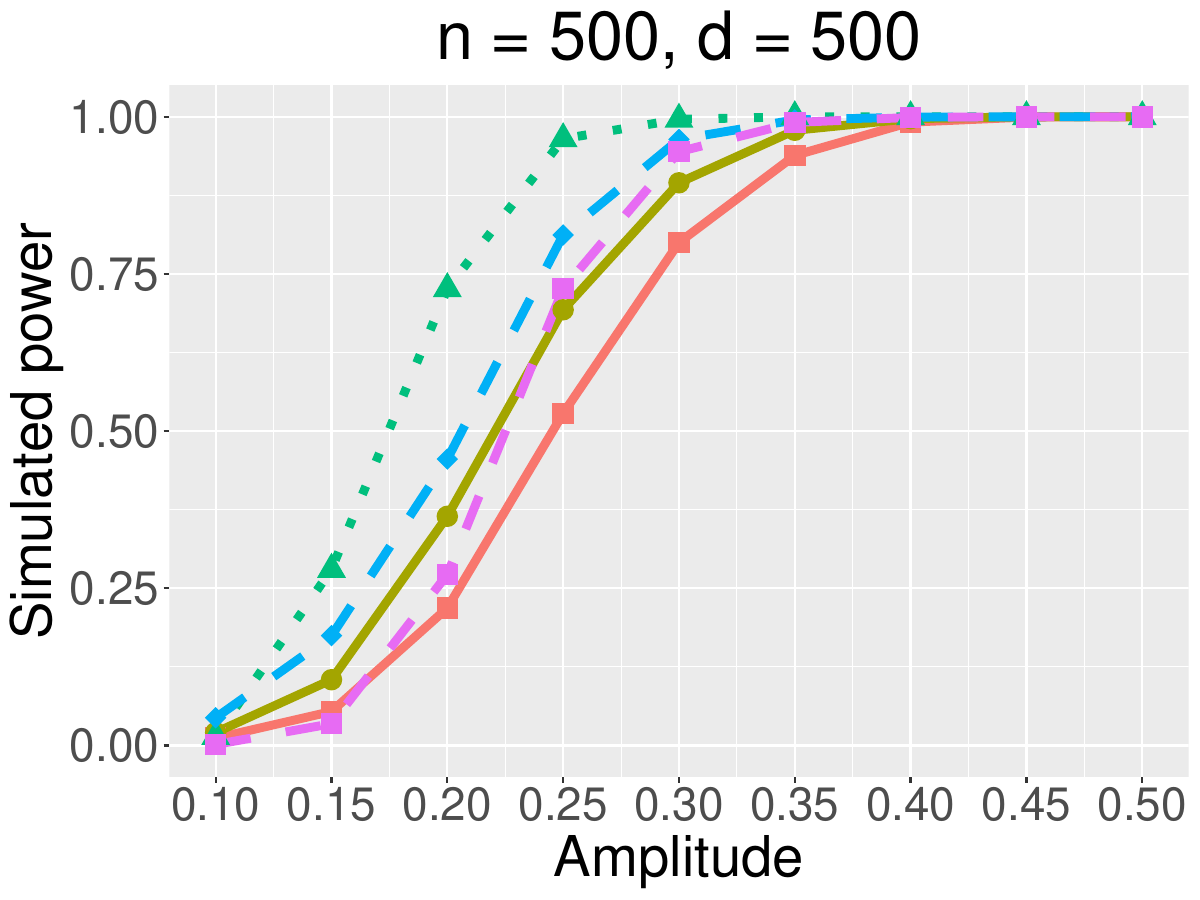}}\hspace{5pt}
    \subfloat{\includegraphics[width=.27\columnwidth]{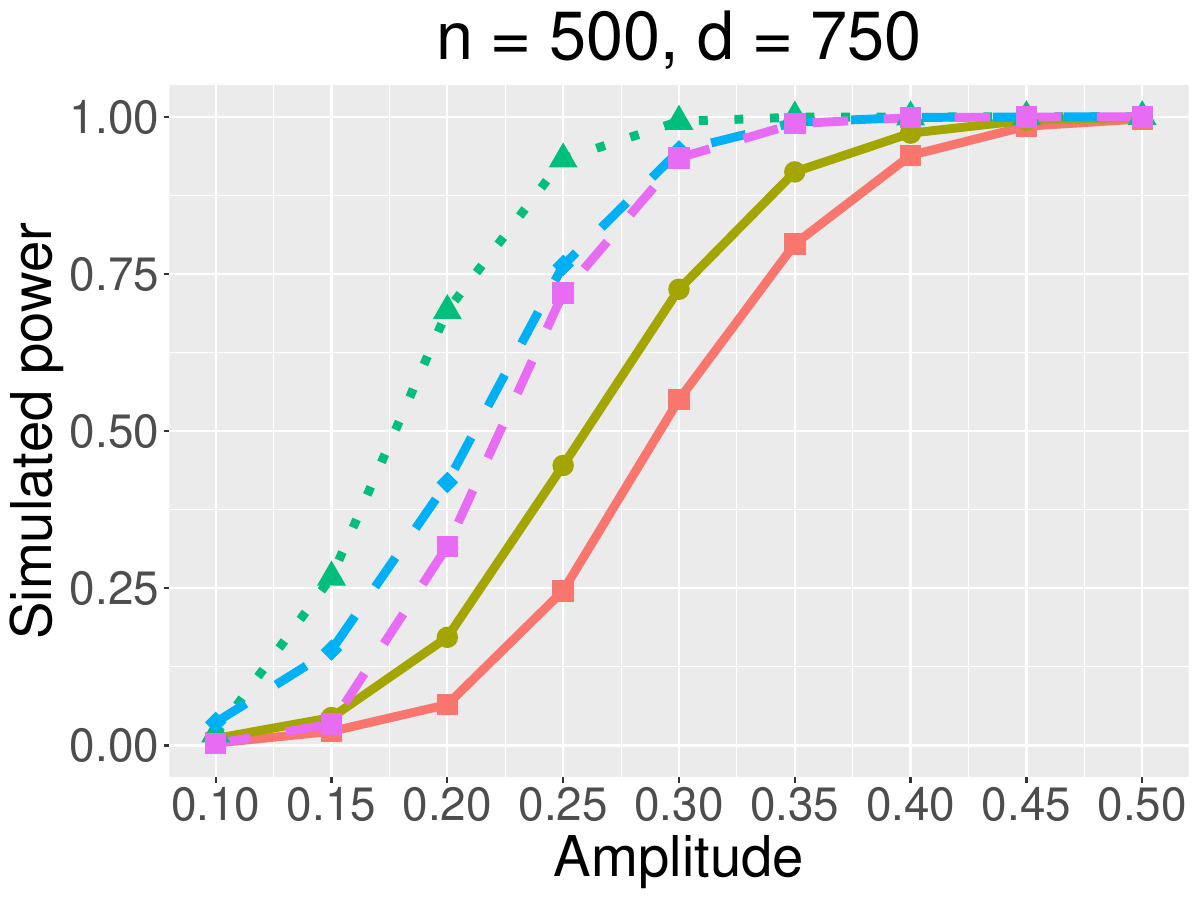}}\hspace{5pt}
    \subfloat{\includegraphics[width=.27\columnwidth]{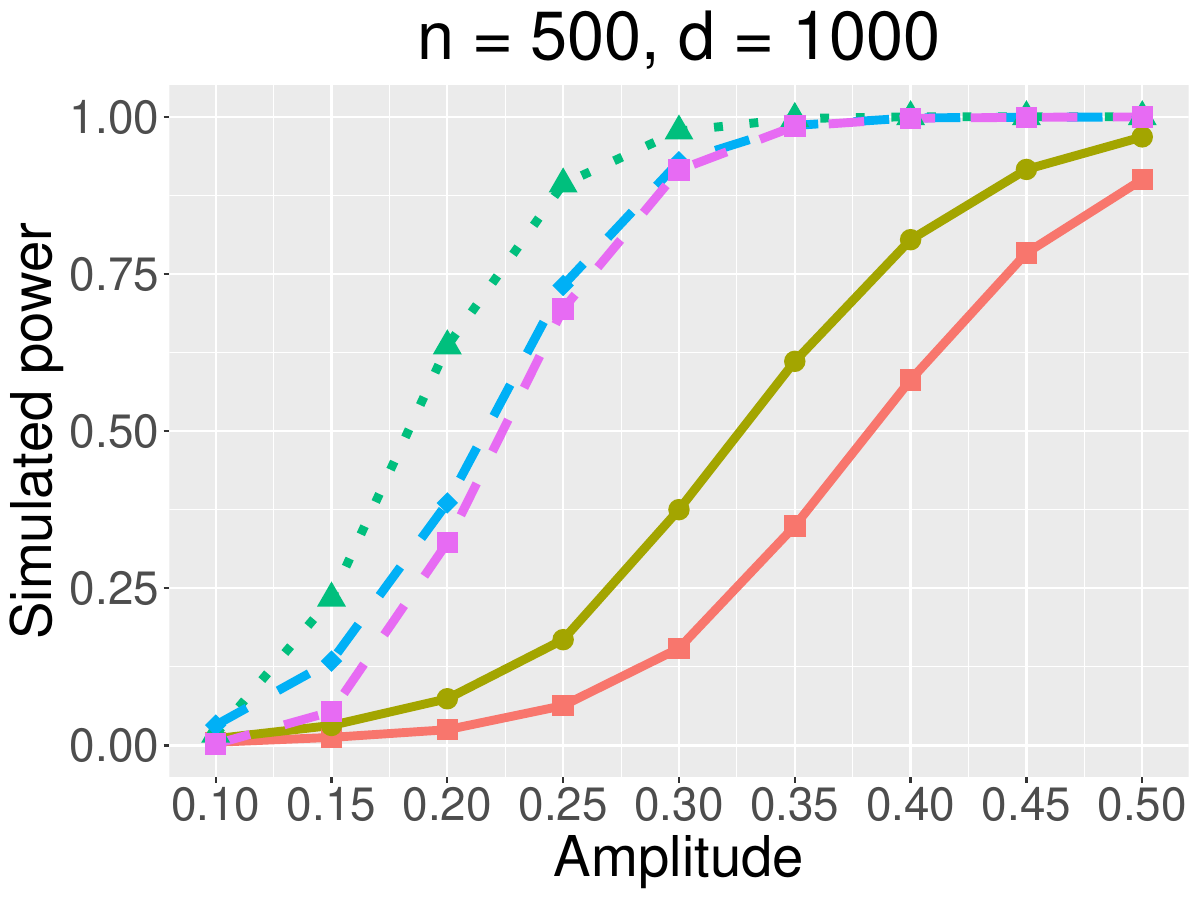}}
	\caption{\small Simulated FDR and power for different combinations of $(n,d)$. The rows of the design matrix were generated from Setting 2. The sparsity level is $k = 0.04d$ and the FDR level is $\alpha = 0.1$. The methods compared are Algorithm 1 (squares and red solid line), Algorithm 2 (circles and yellow solid line), the knockoff-based method of \cite{candes2018panning} (triangles and green dotted line), the Gaussian Mirror method of \cite{Xing2021Controlling} (diamonds and blue dashed line), and the Gaussian Mirror method with FDP+ procedure (squares and purple dashed line).}
    \label{Power-large-0.1}
\end{figure}

\begin{figure}[htbp!]
	\centering
	\subfloat{\includegraphics[width=.27\columnwidth]{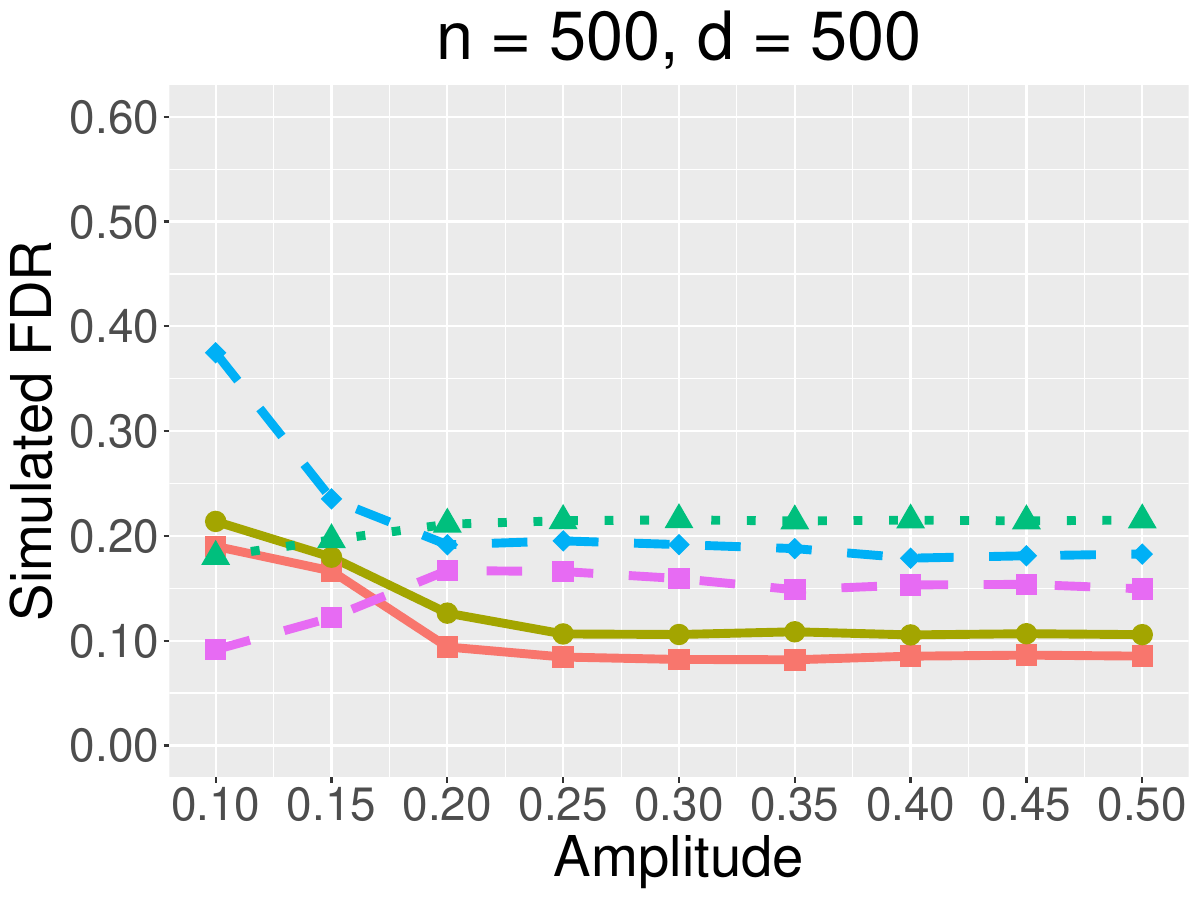}}\hspace{5pt}
	\subfloat{\includegraphics[width=.27\columnwidth]{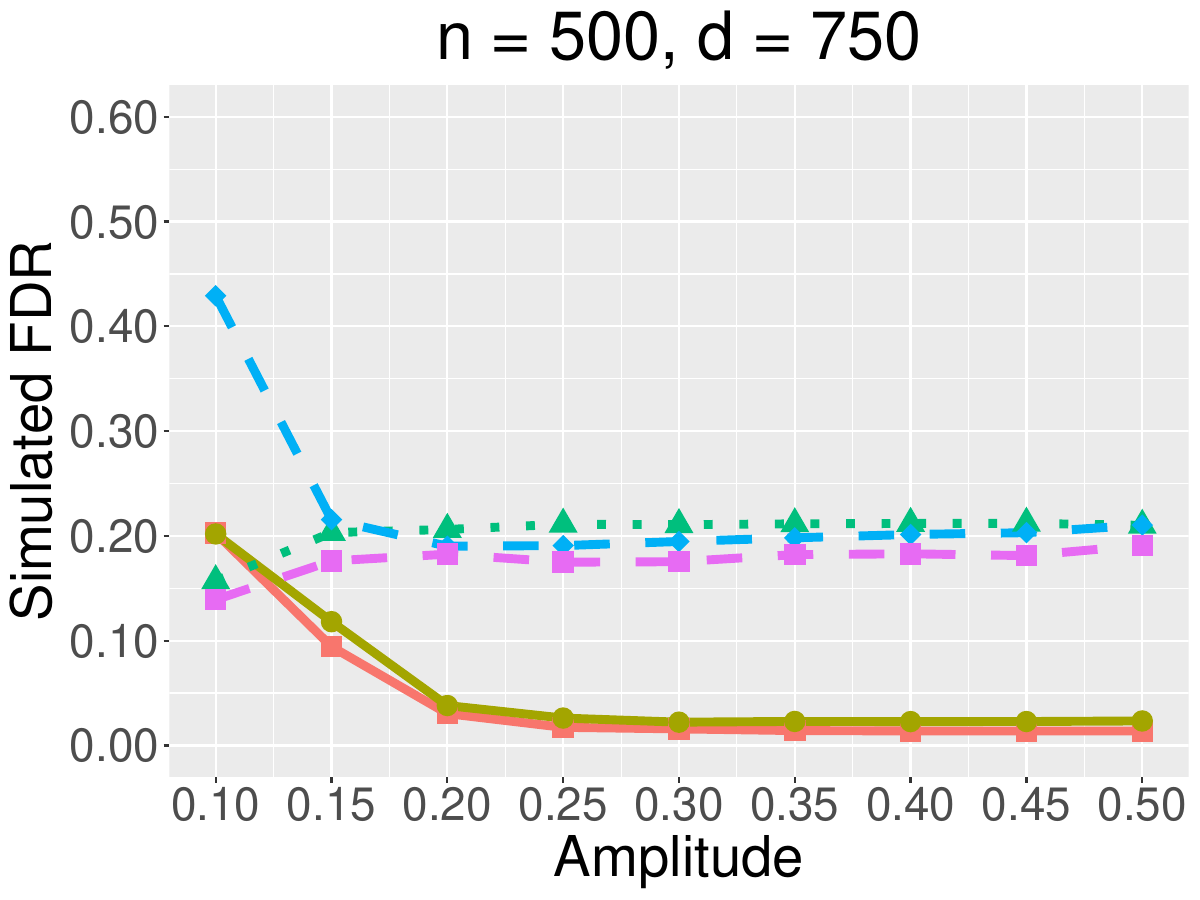}}\hspace{5pt}
	\subfloat{\includegraphics[width=.27\columnwidth]{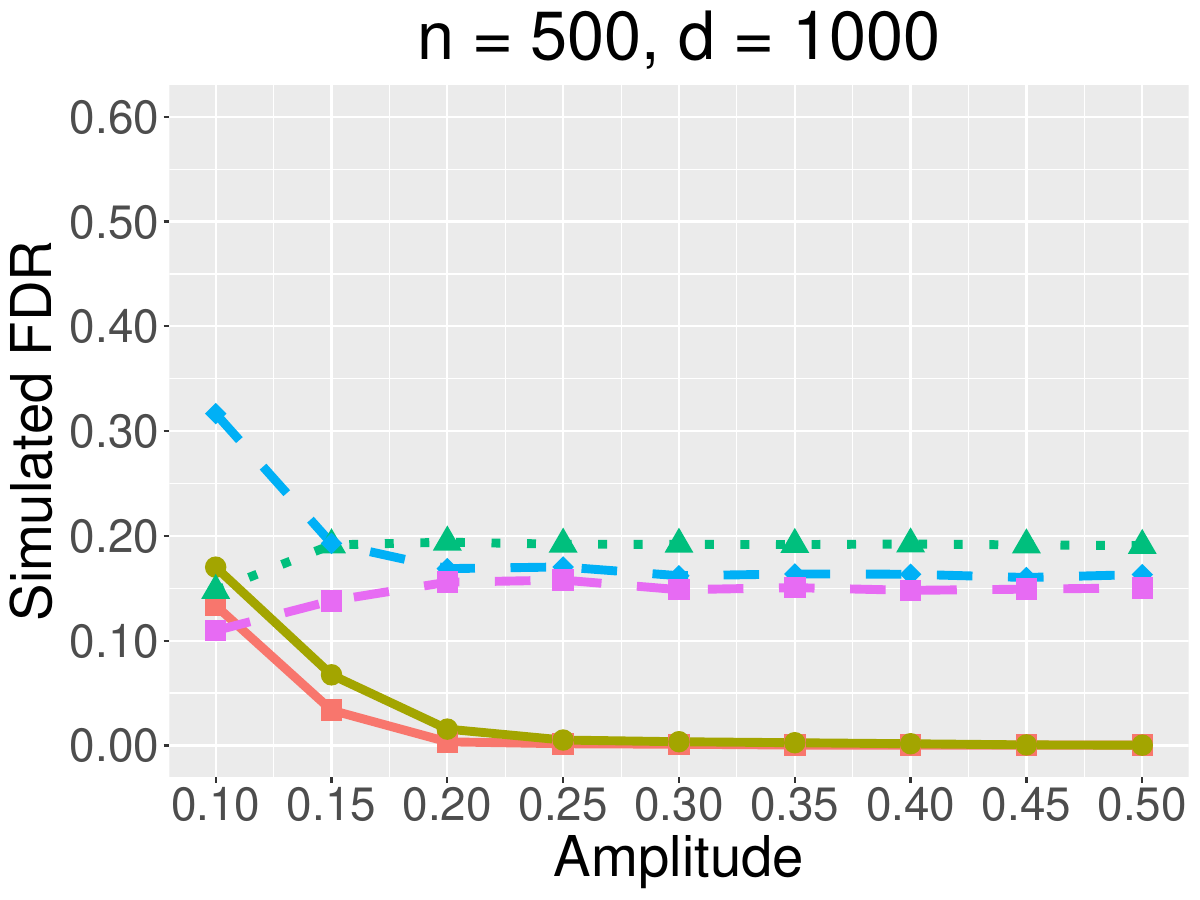}}\\
	\subfloat{\includegraphics[width=.27\columnwidth]{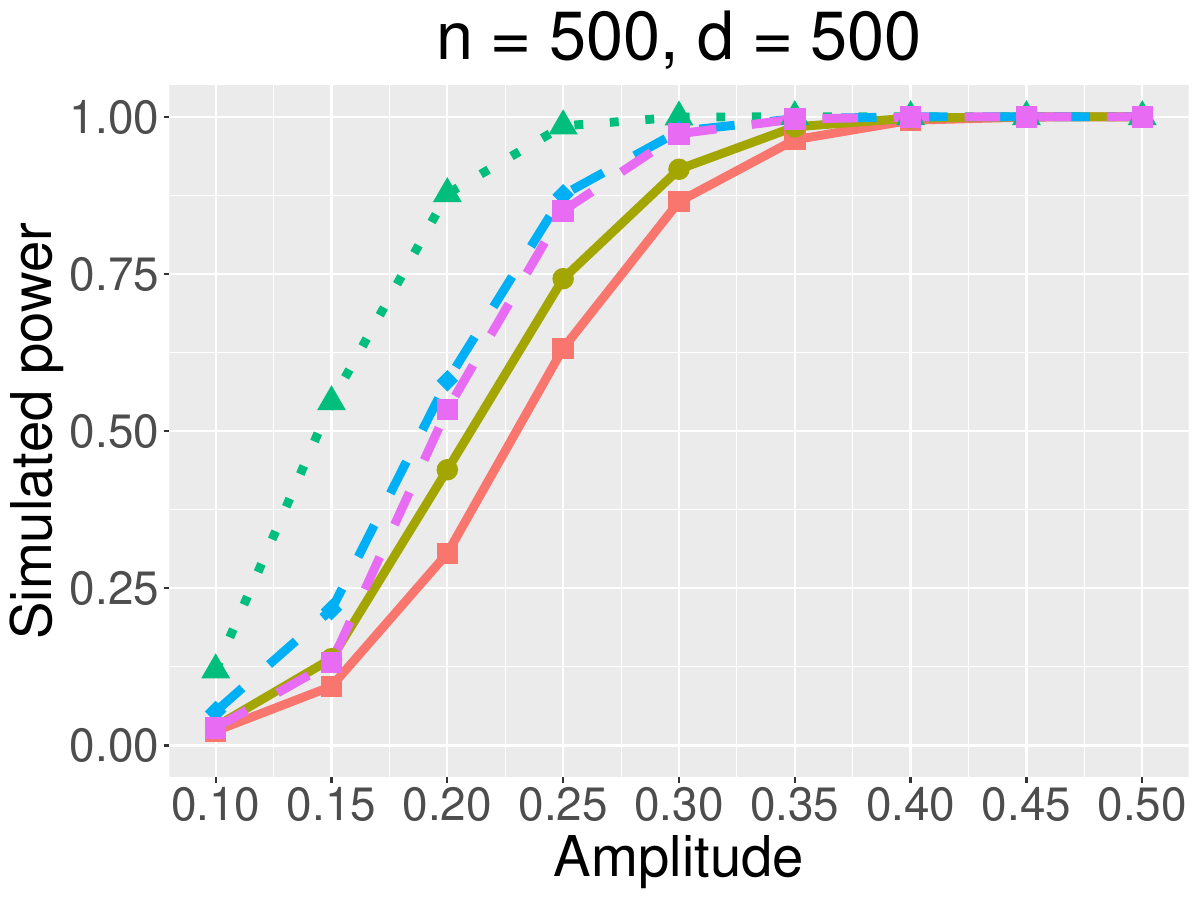}}\hspace{5pt}
    \subfloat{\includegraphics[width=.27\columnwidth]{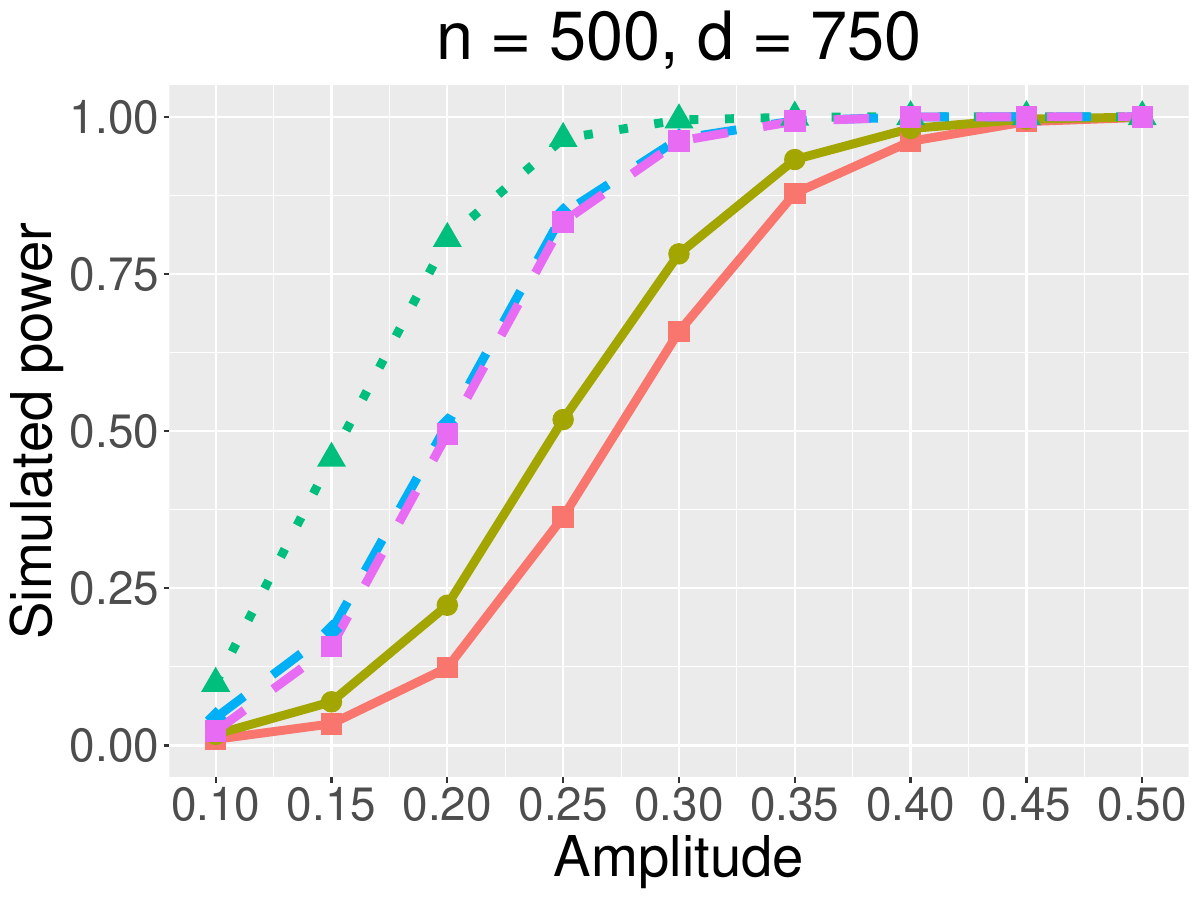}}\hspace{5pt}
    \subfloat{\includegraphics[width=.27\columnwidth]{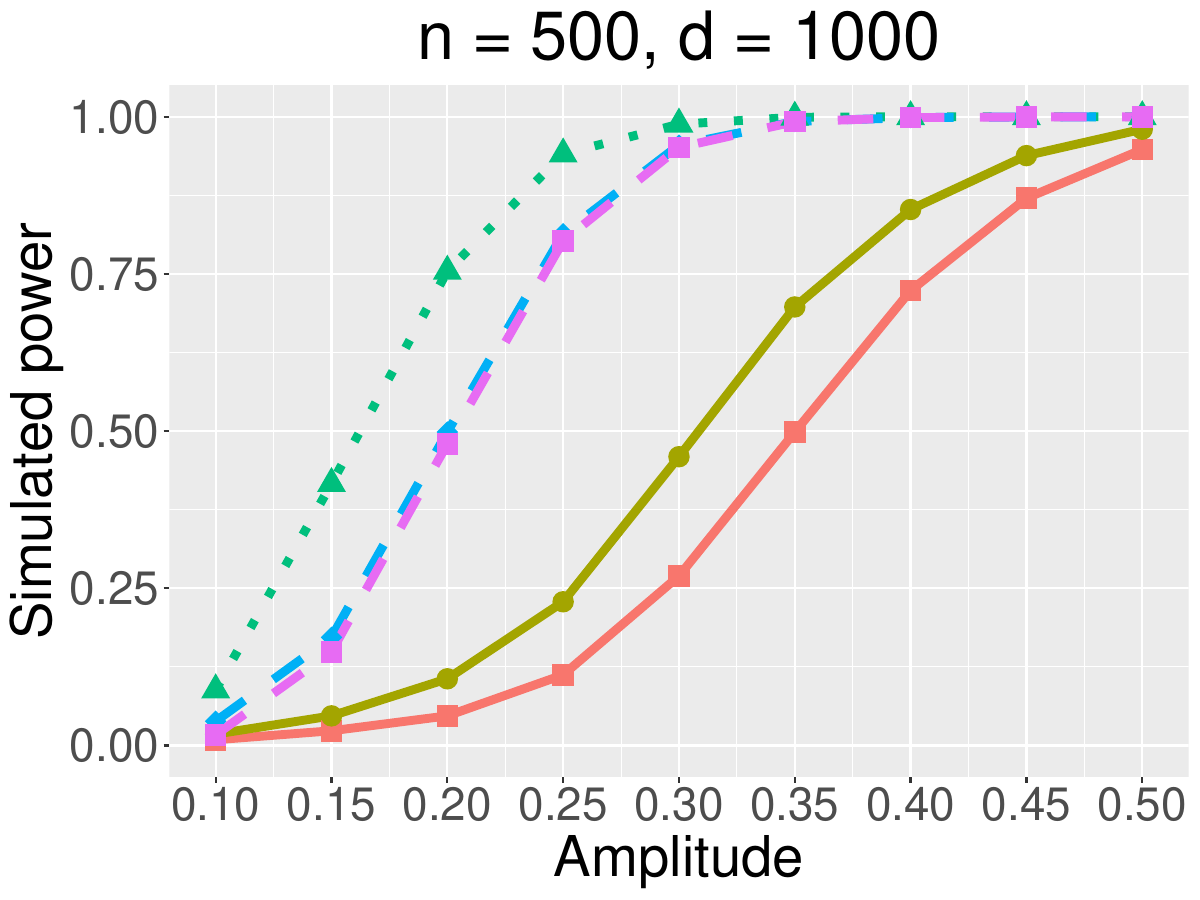}}
	\caption{\small Simulated FDR and power for different combinations of $(n,d)$. The rows of the design matrix were generated from Setting 2. The sparsity level is $k = 0.04d$ and the FDR level is $\alpha = 0.2$. The methods compared are Algorithm 1 (squares and red solid line), Algorithm 2 (circles and yellow solid line), the knockoff-based method of \cite{candes2018panning} (triangles and green dotted line), the Gaussian Mirror method of \cite{Xing2021Controlling} (diamonds and blue dashed line), and the Gaussian Mirror method with FDP+ procedure (squares and purple dashed line).}
    \label{Power-large-0.2}
\end{figure}

All additional simulation results when some null hypotheses are not true under Setting 2 and sparsity level $k = 0.04d$ are reported here. Figures \ref{Power-0.05} and \ref{Power-0.2} present the results with relatively small $n$ and $d$. While Figures \ref{Power-large-0.05}--\ref{Power-large-0.2} present the results with relatively large $n$ and $d$.

\subsection{Additional simulation results with highly correlated variables} \label{gm-extra}
\begin{figure}[htbp!]
	\centering
	\subfloat{\includegraphics[width=.27\columnwidth]{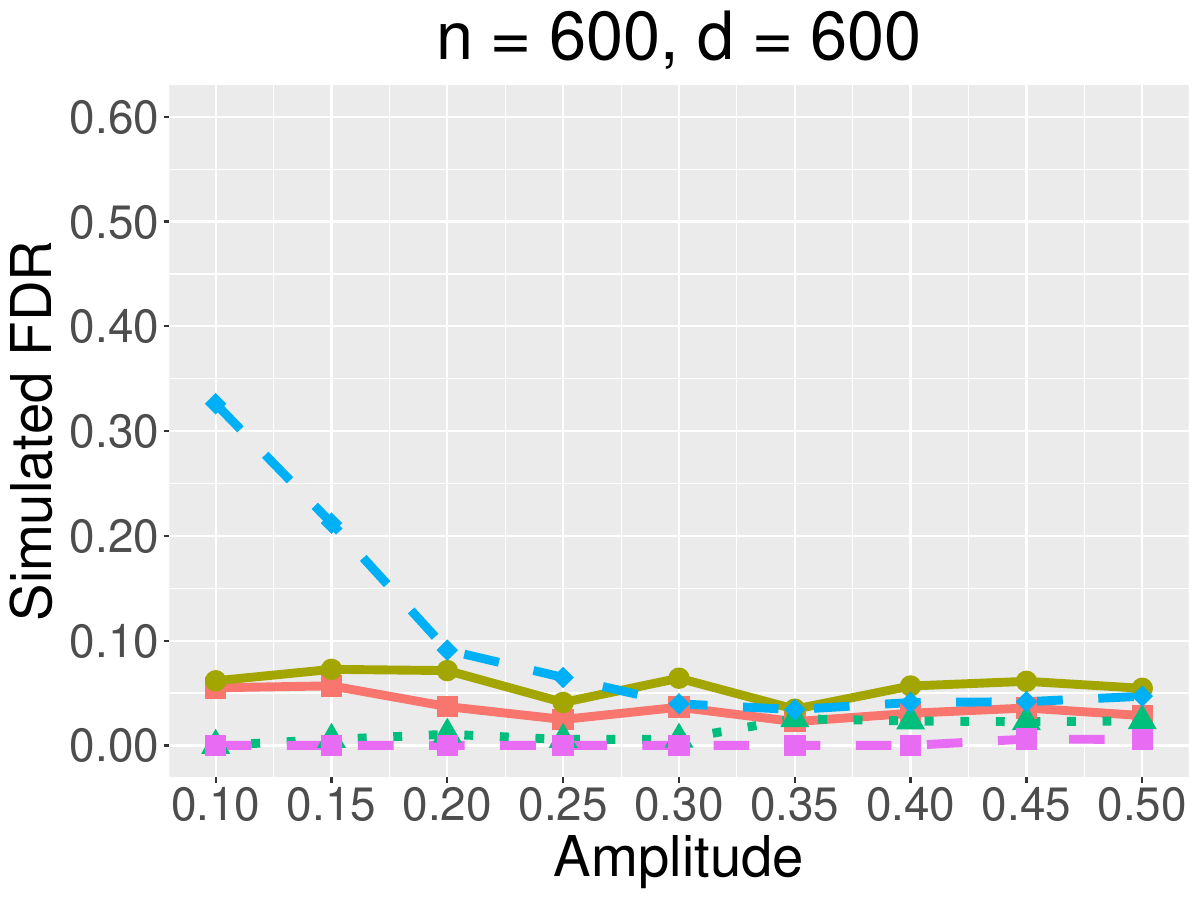}}\hspace{5pt}
	\subfloat{\includegraphics[width=.27\columnwidth]{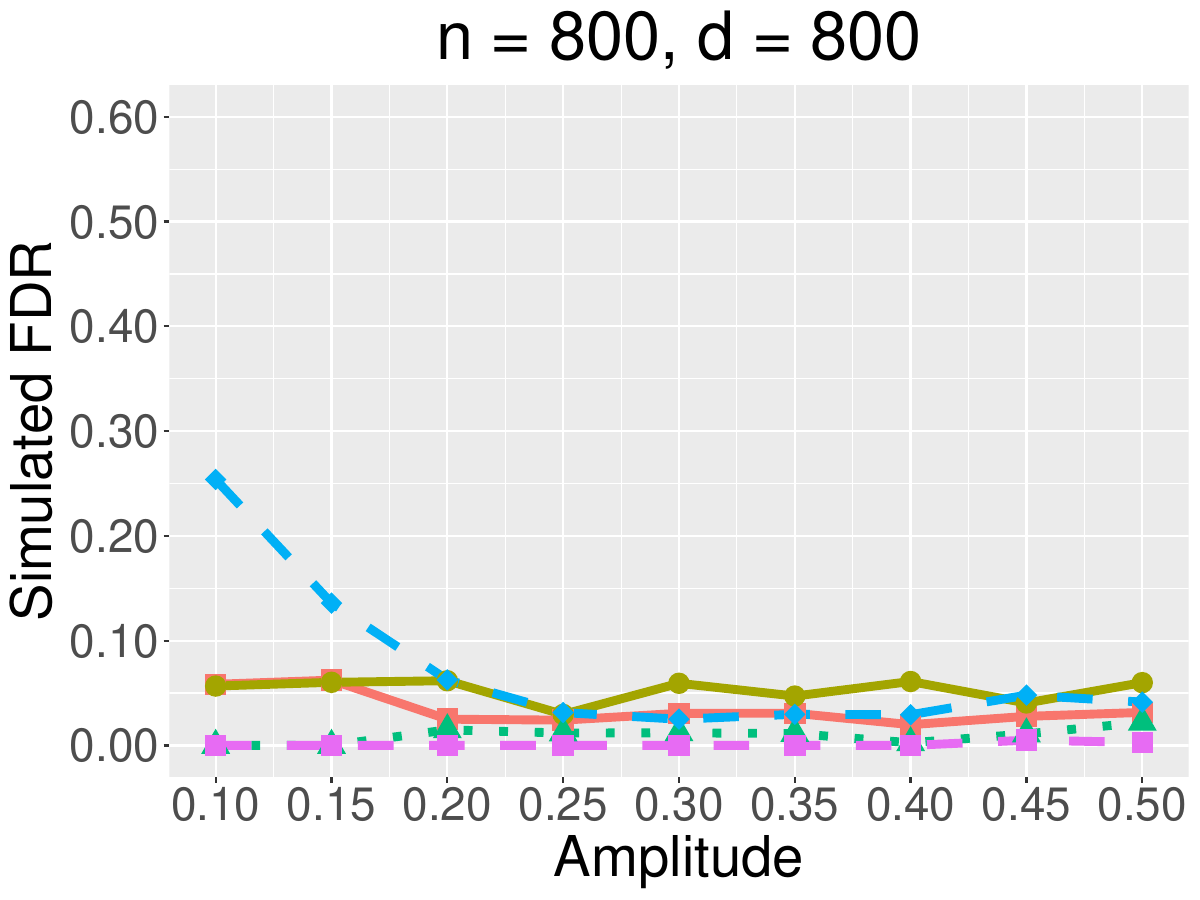}}\hspace{5pt}
	\subfloat{\includegraphics[width=.27\columnwidth]{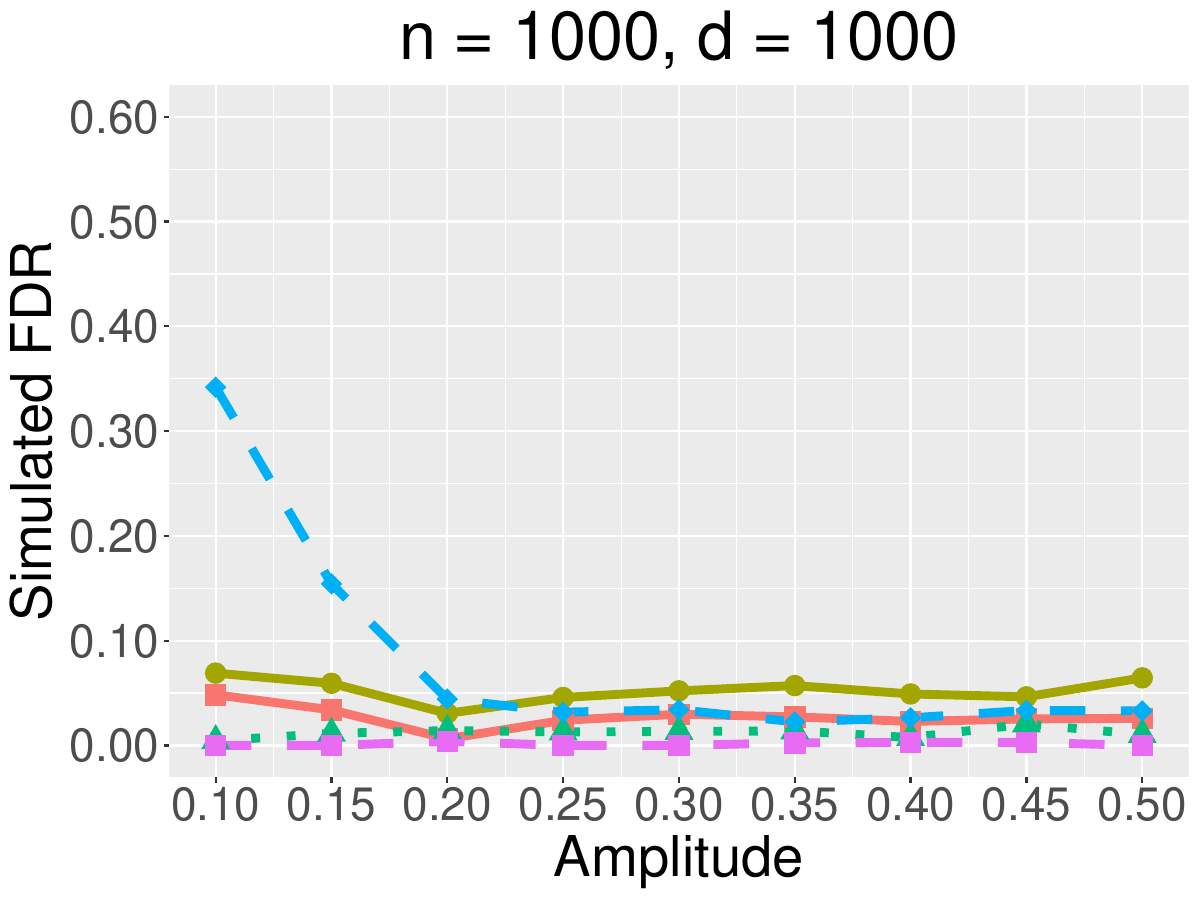}}\\
	\subfloat{\includegraphics[width=.27\columnwidth]{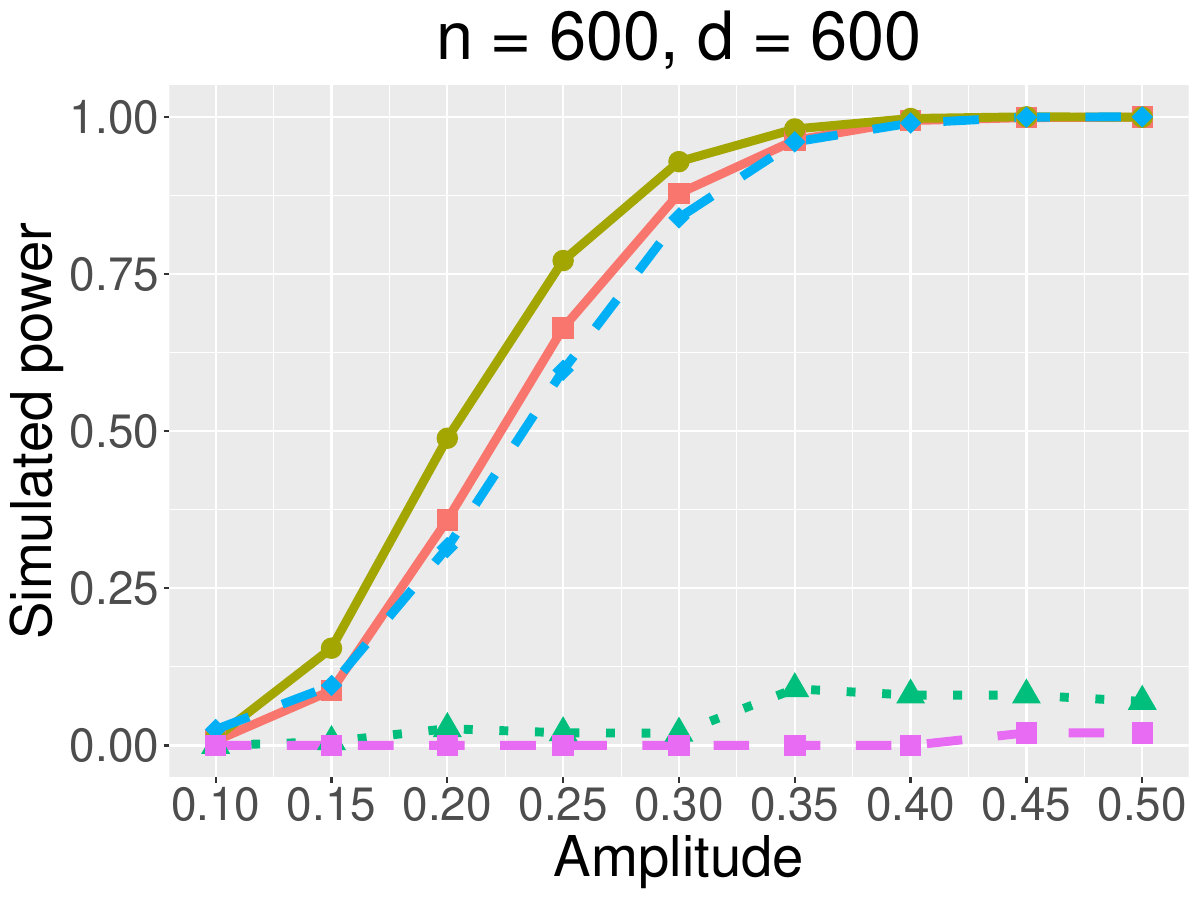}}\hspace{5pt}
    \subfloat{\includegraphics[width=.27\columnwidth]{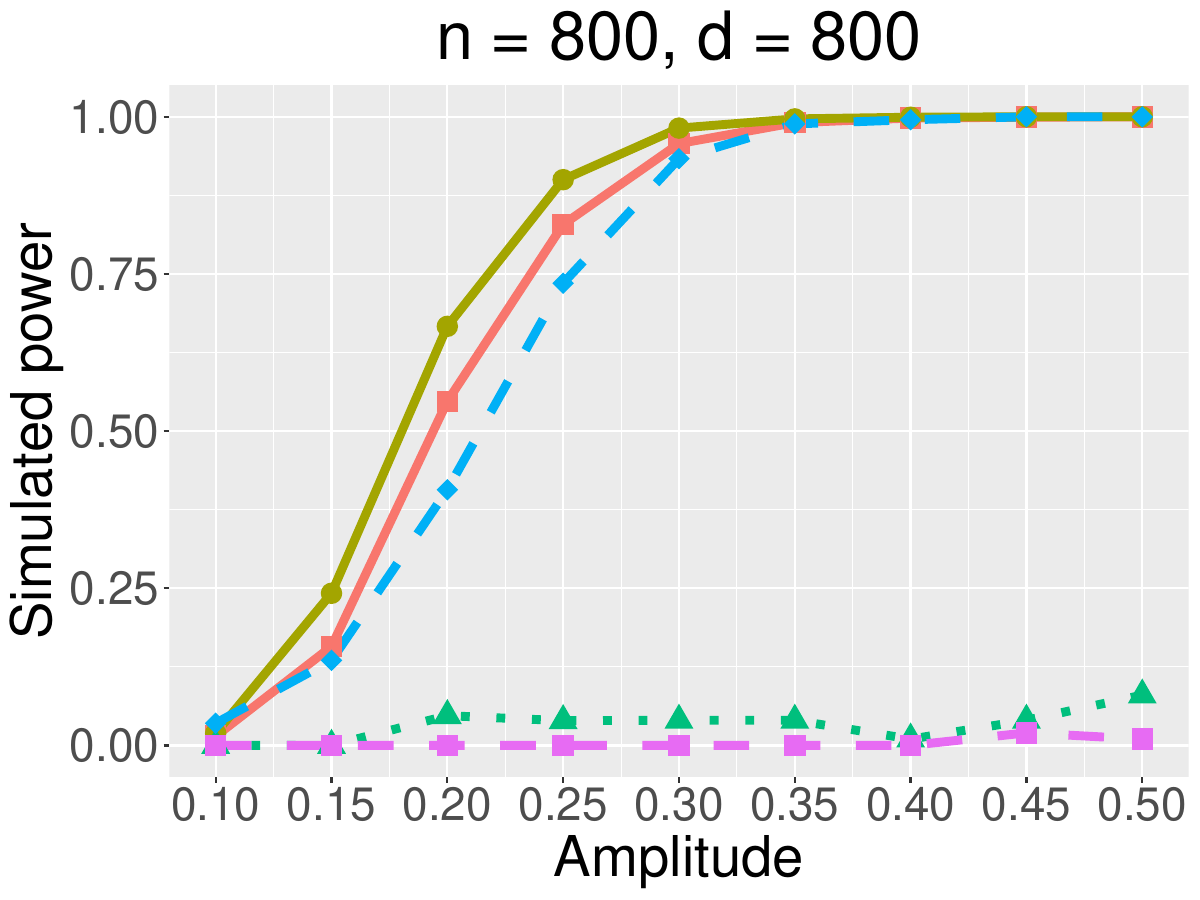}}\hspace{5pt}
    \subfloat{\includegraphics[width=.27\columnwidth]{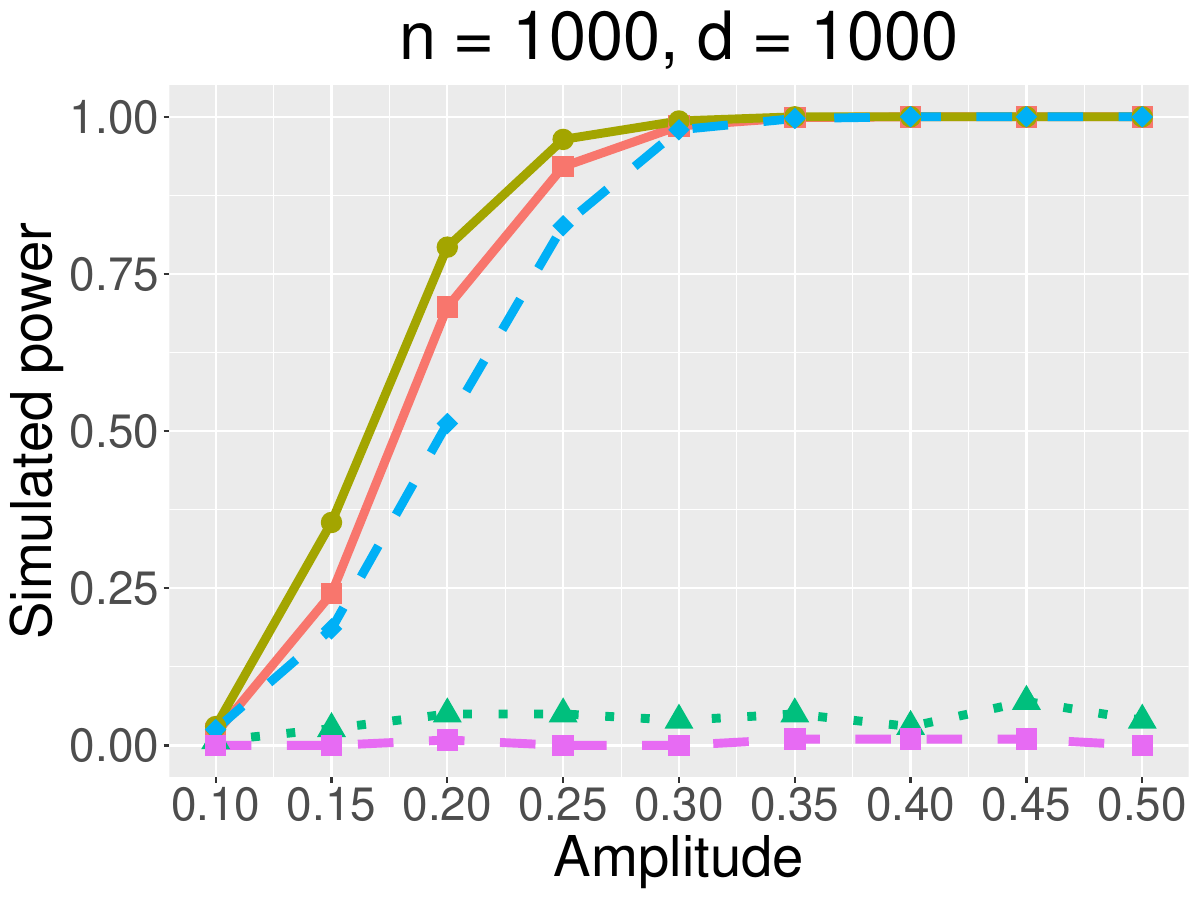}}
	\caption{\small Simulated FDR and power for different combinations of $(n,d)$. The rows of the design matrix were generated from Setting 2 with the AR(1) coefficient being set to 0.67. The sparsity level is $k = 15$ and the FDR level is $\alpha = 0.05$. The methods compared are Algorithm 1 (squares and red solid line), Algorithm 2 (circles and yellow solid line), the knockoff-based method of \cite{candes2018panning} (triangles and green dotted line), the Gaussian Mirror method of \cite{Xing2021Controlling} (diamonds and blue dashed line), and the Gaussian Mirror method with FDP+ procedure (squares and purple dashed line).}
    \label{Highly-correlated-0.05}
\end{figure}

\begin{figure}[htbp!]
	\centering
	\subfloat{\includegraphics[width=.27\columnwidth]{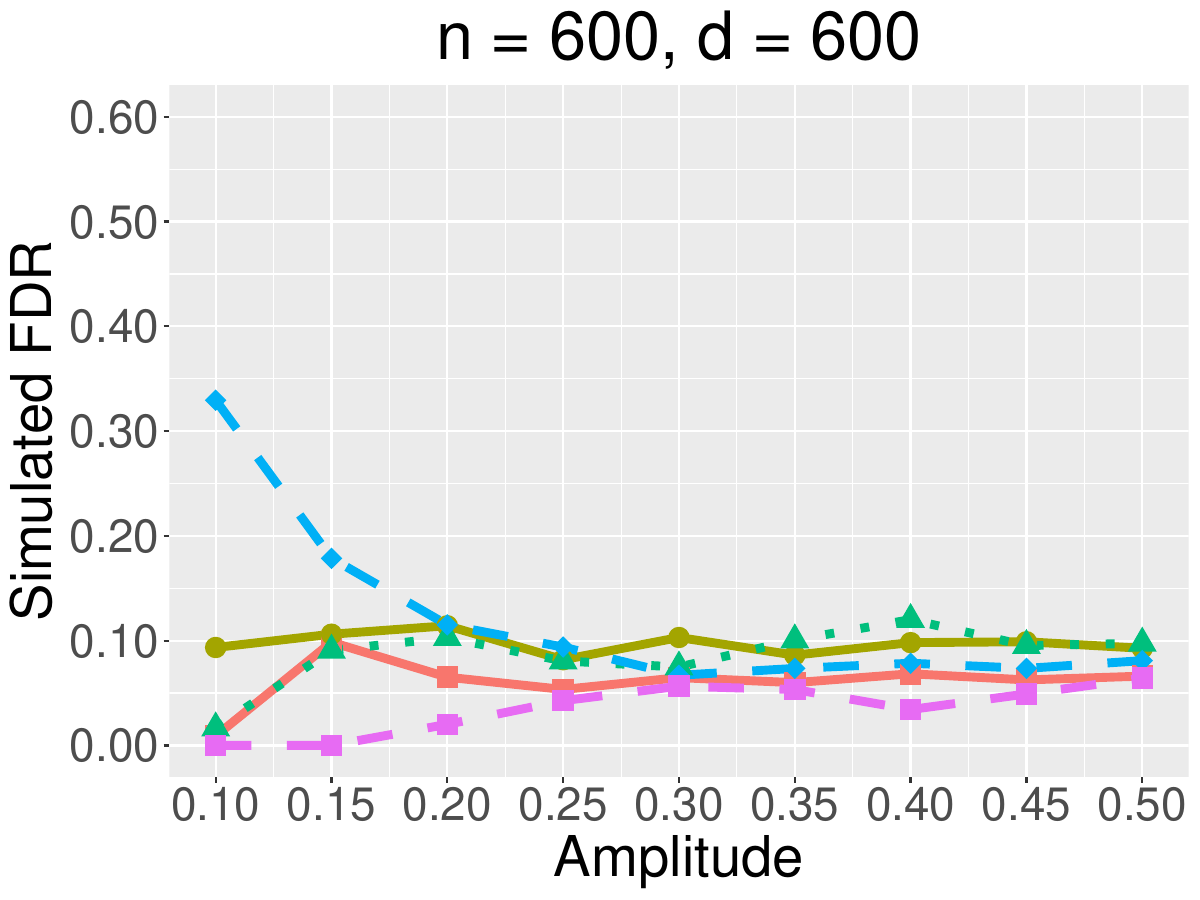}}\hspace{5pt}
	\subfloat{\includegraphics[width=.27\columnwidth]{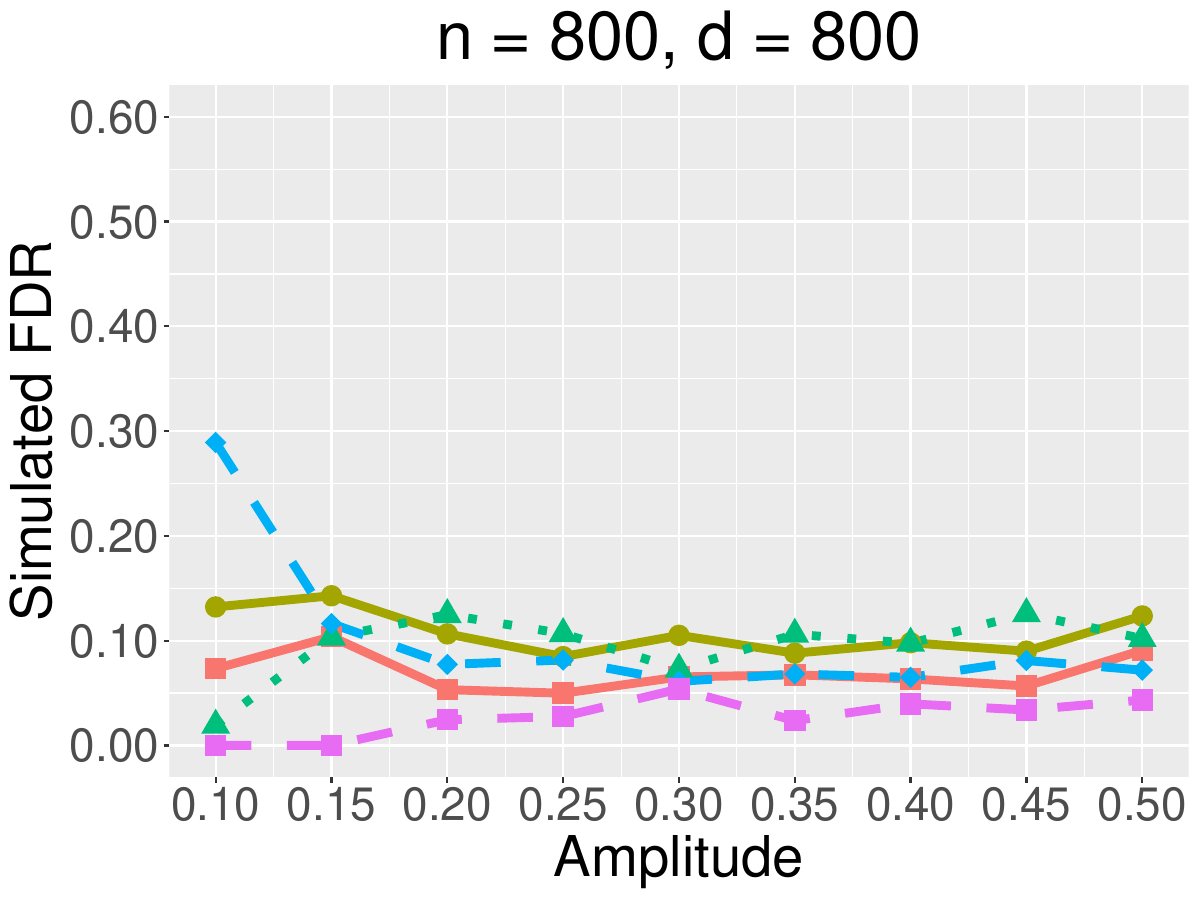}}\hspace{5pt}
	\subfloat{\includegraphics[width=.27\columnwidth]{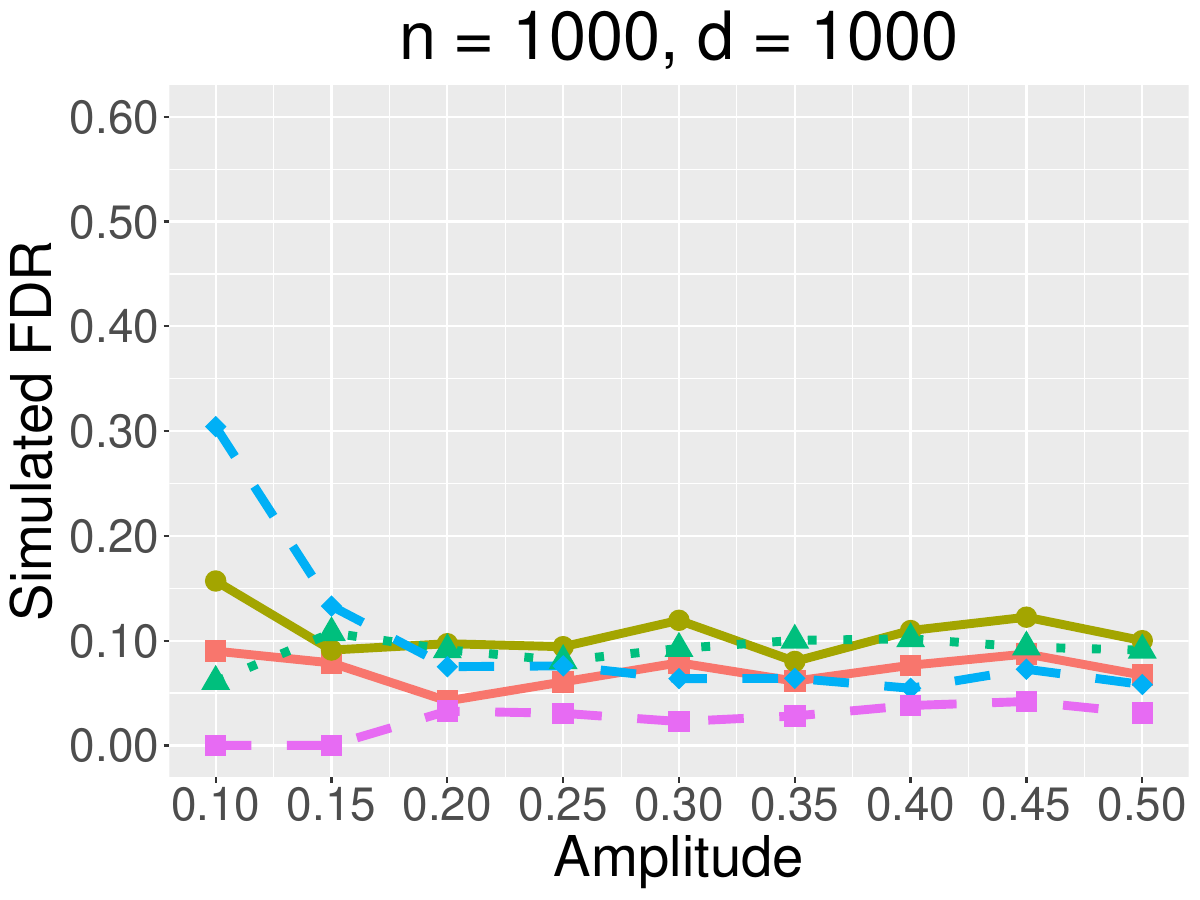}}\\
	\subfloat{\includegraphics[width=.27\columnwidth]{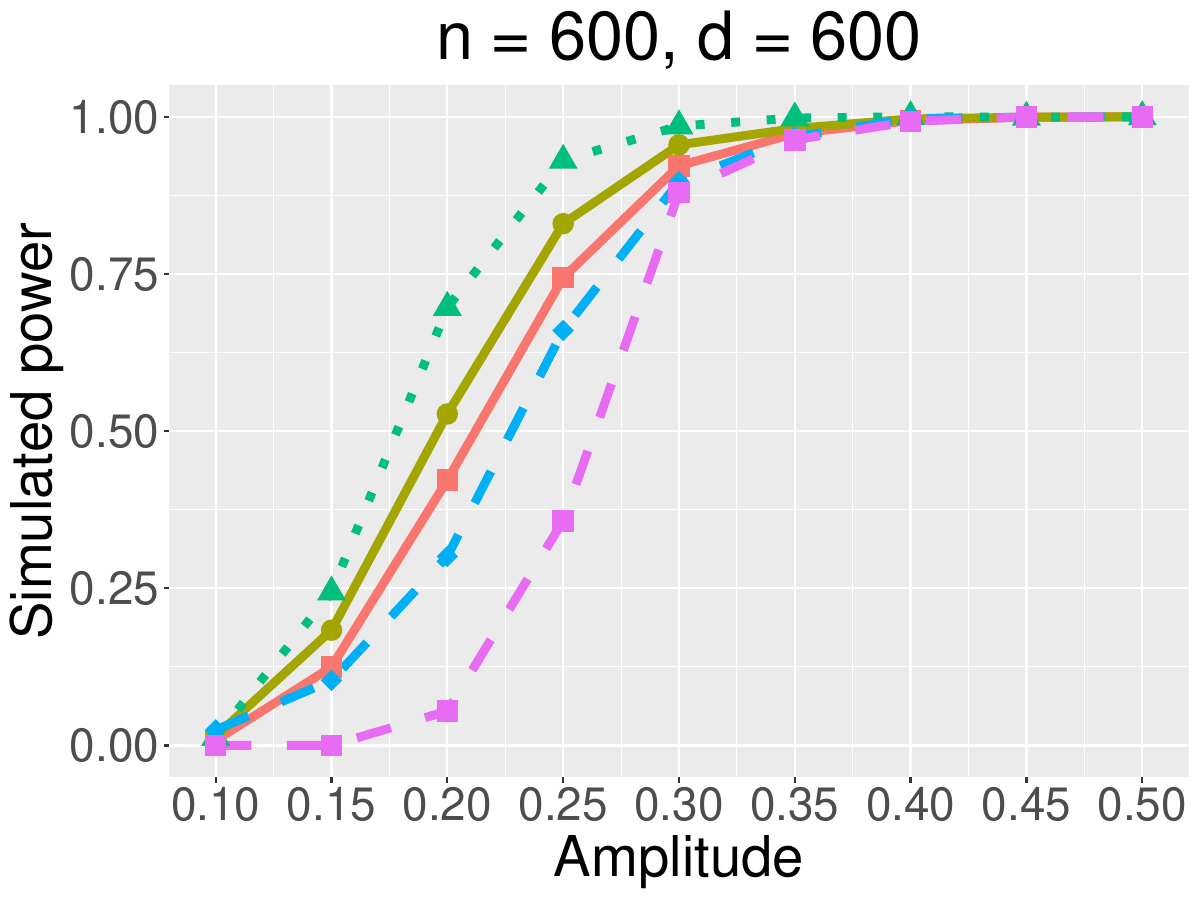}}\hspace{5pt}
    \subfloat{\includegraphics[width=.27\columnwidth]{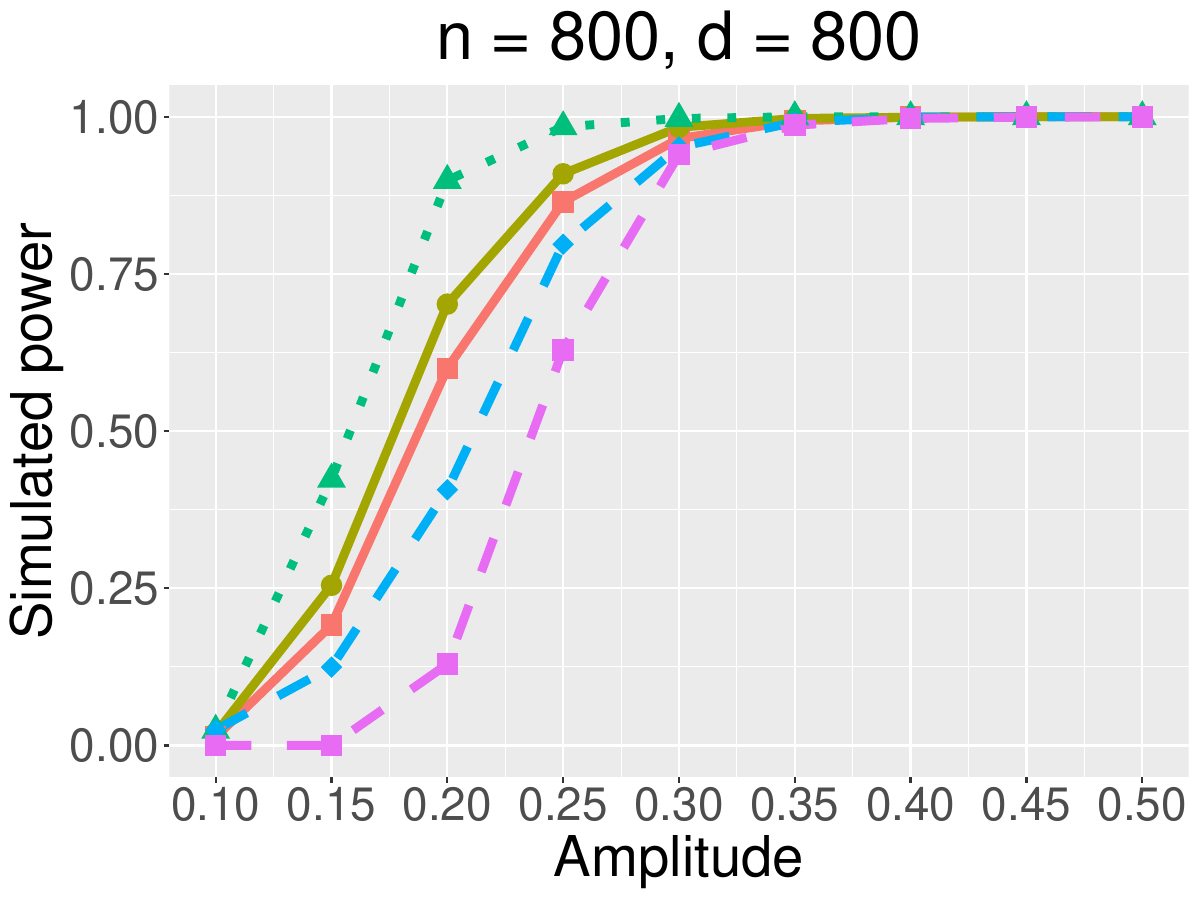}}\hspace{5pt}
    \subfloat{\includegraphics[width=.27\columnwidth]{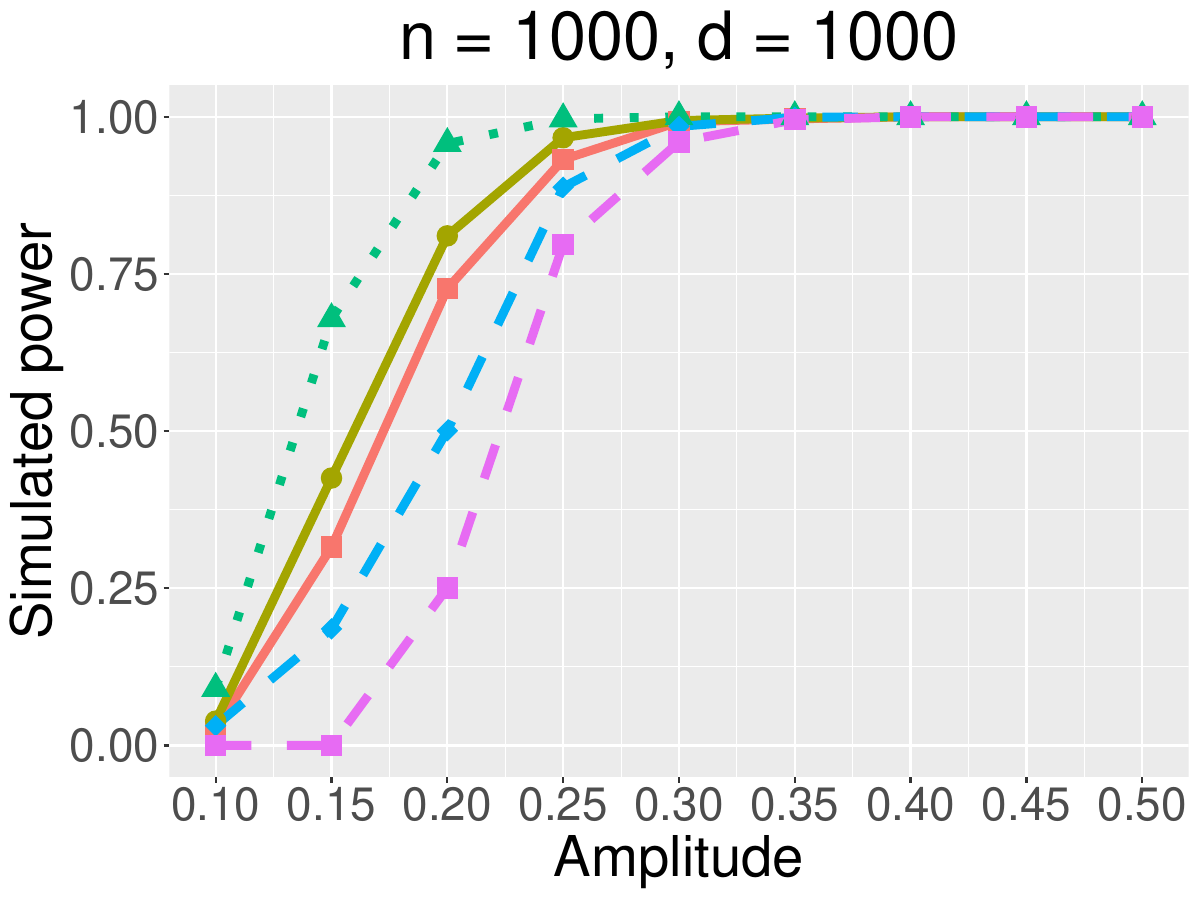}}
	\caption{\small Simulated FDR and power for different combinations of $(n,d)$. The rows of the design matrix were generated from Setting 2 with the AR(1) coefficient being set to 0.67. The sparsity level is $k = 15$ and the FDR level is $\alpha = 0.1$. The methods compared are Algorithm 1 (squares and red solid line), Algorithm 2 (circles and yellow solid line), the knockoff-based method of \cite{candes2018panning} (triangles and green dotted line), the Gaussian Mirror method of \cite{Xing2021Controlling} (diamonds and blue dashed line), and the Gaussian Mirror method with FDP+ procedure (squares and purple dashed line).}
    \label{Highly-correlated-0.1}
\end{figure}

\begin{figure}[htbp!]
	\centering
	\subfloat{\includegraphics[width=.27\columnwidth]{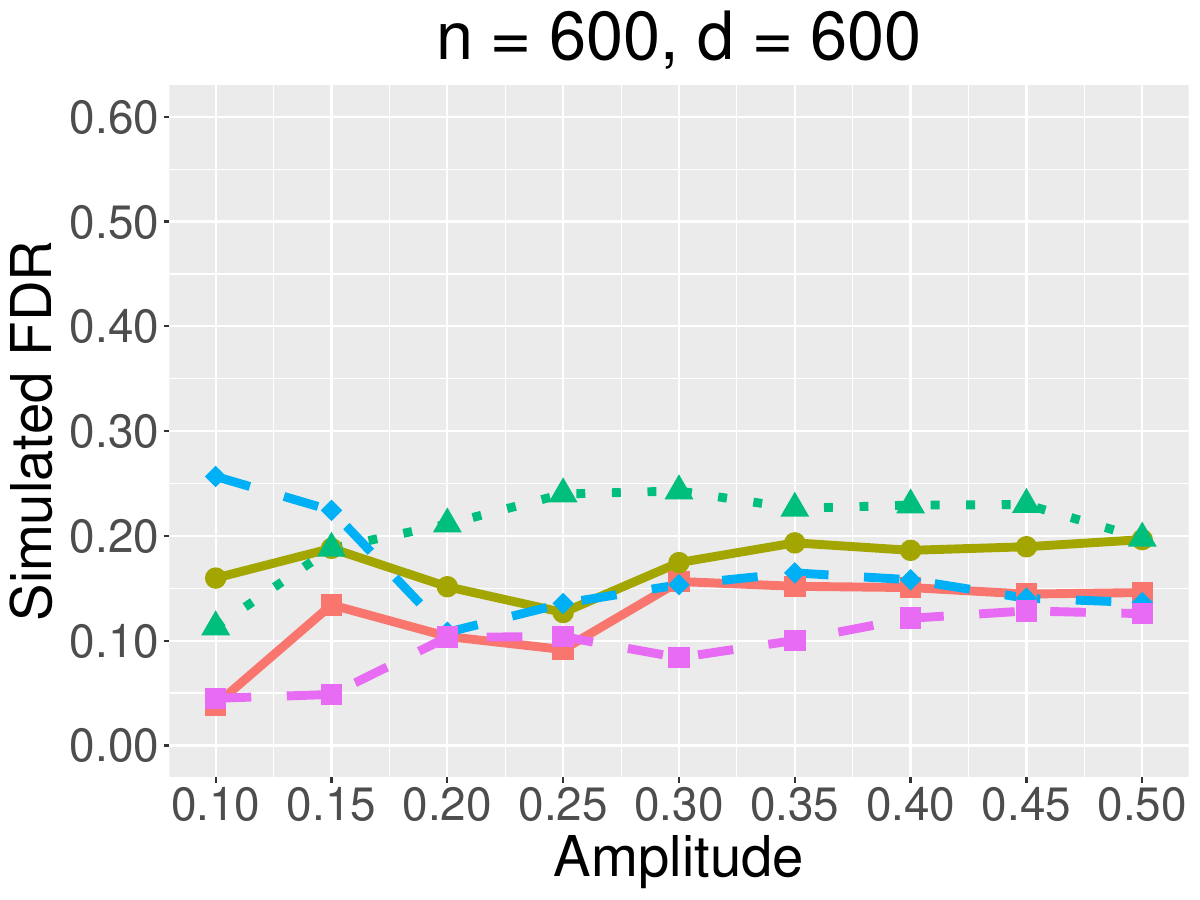}}\hspace{5pt}
	\subfloat{\includegraphics[width=.27\columnwidth]{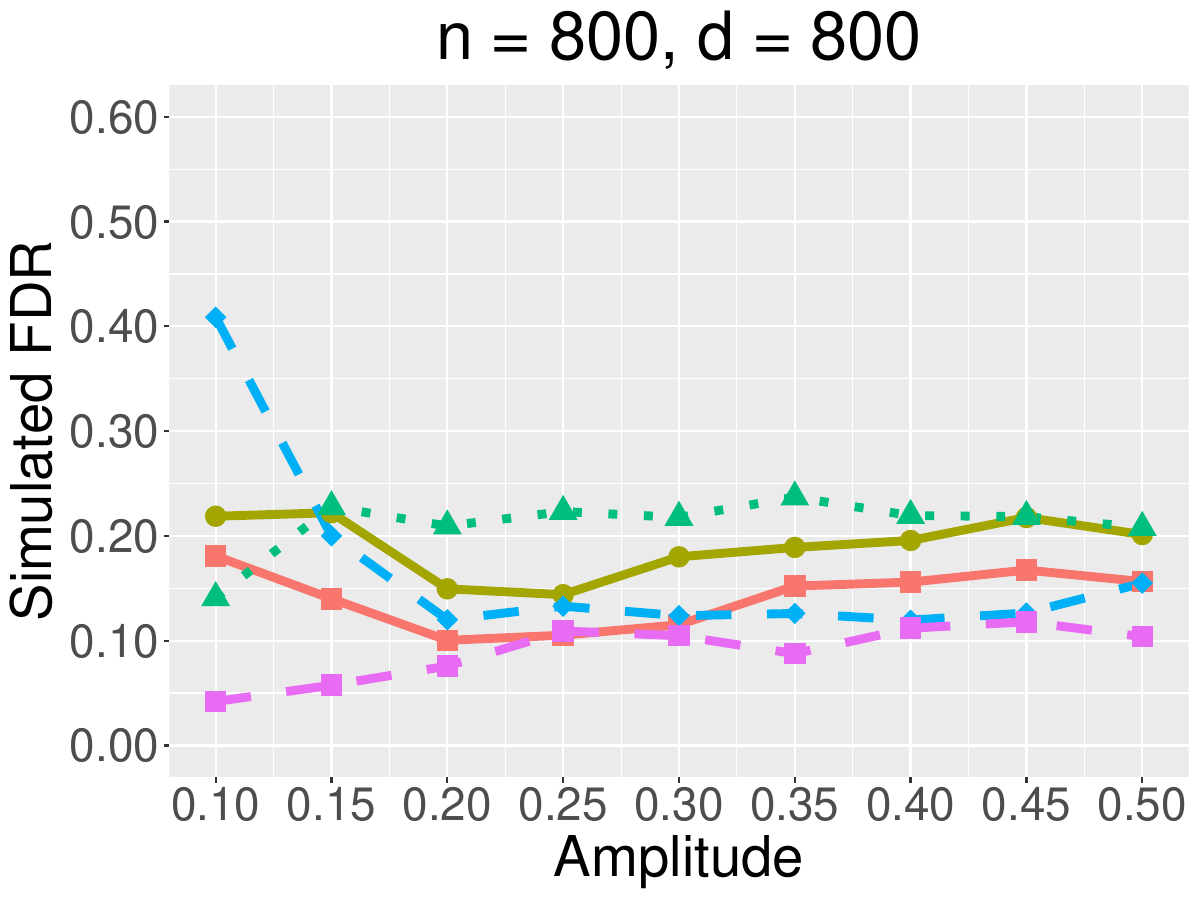}}\hspace{5pt}
	\subfloat{\includegraphics[width=.27\columnwidth]{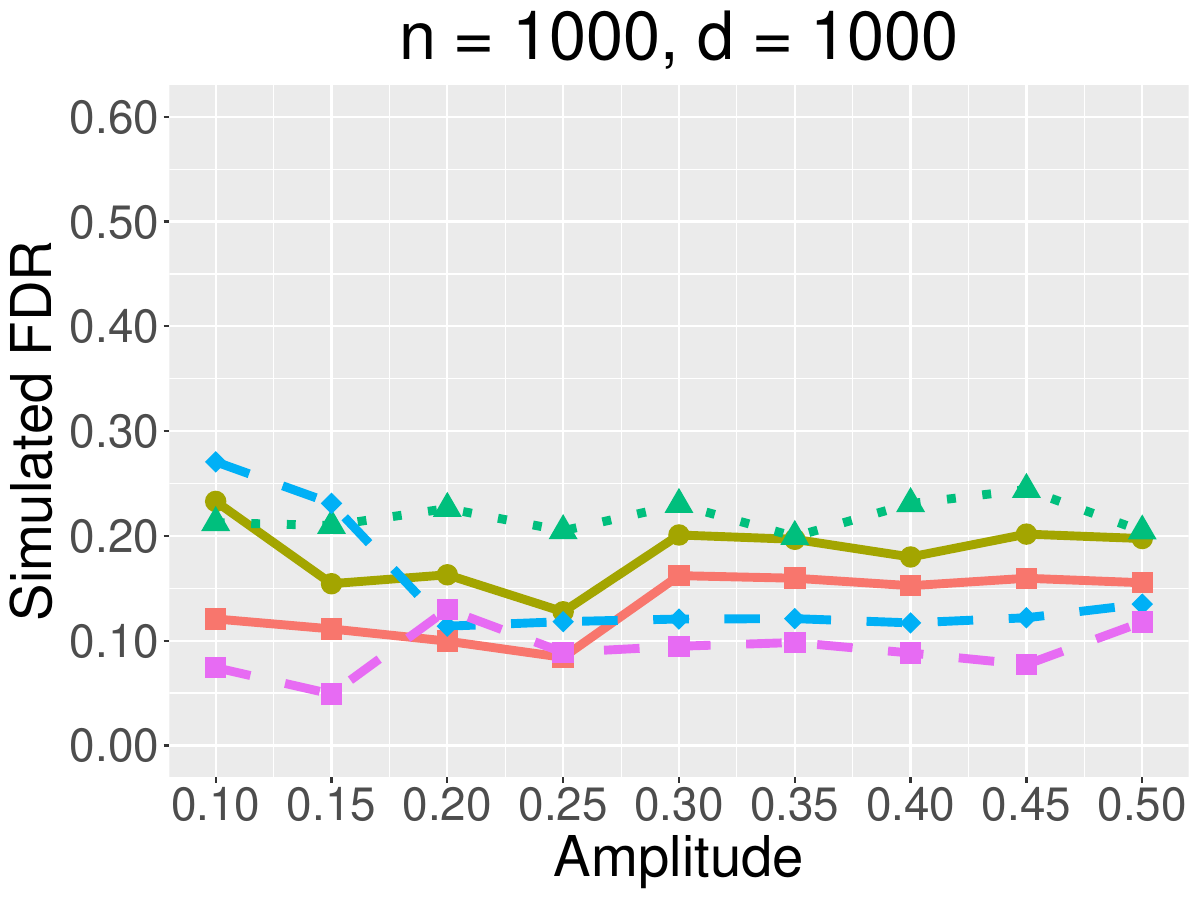}}\\
	\subfloat{\includegraphics[width=.27\columnwidth]{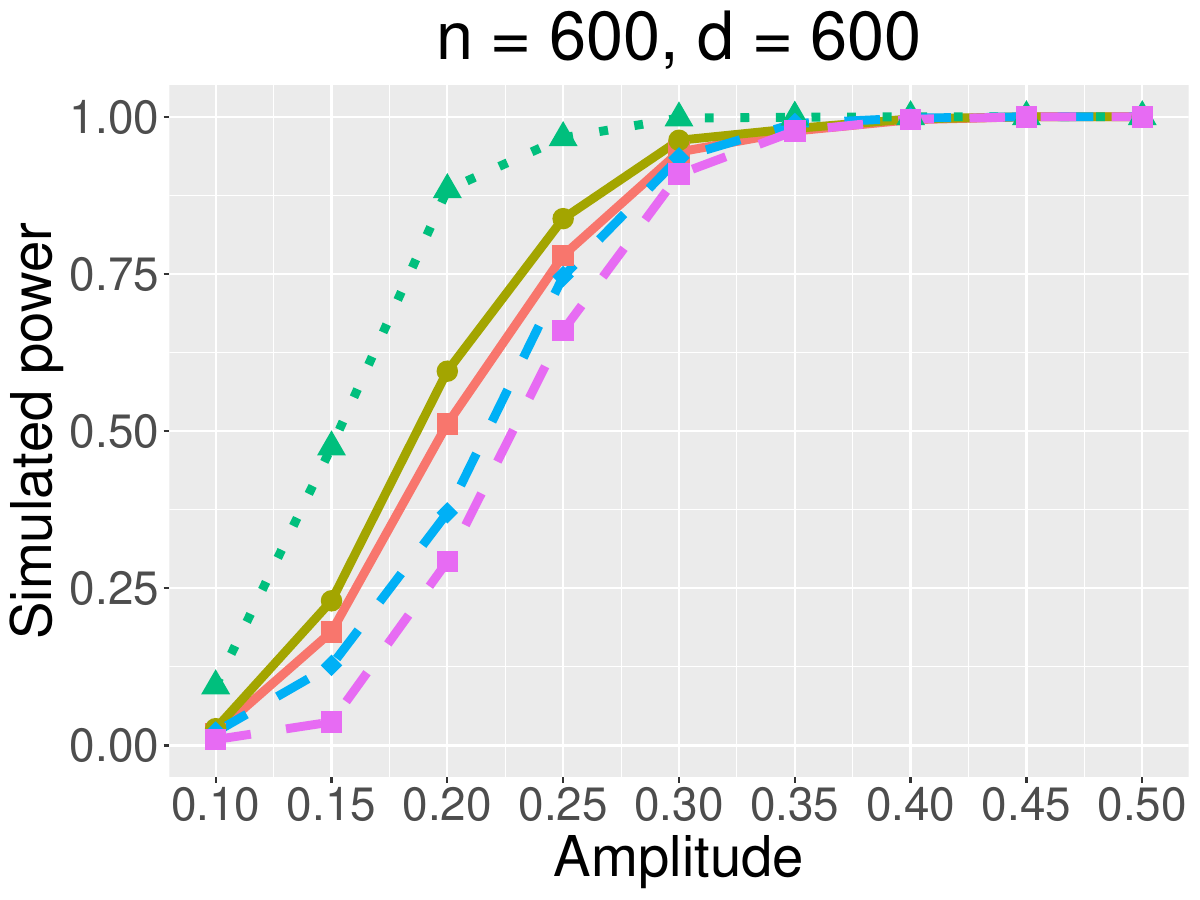}}\hspace{5pt}
    \subfloat{\includegraphics[width=.27\columnwidth]{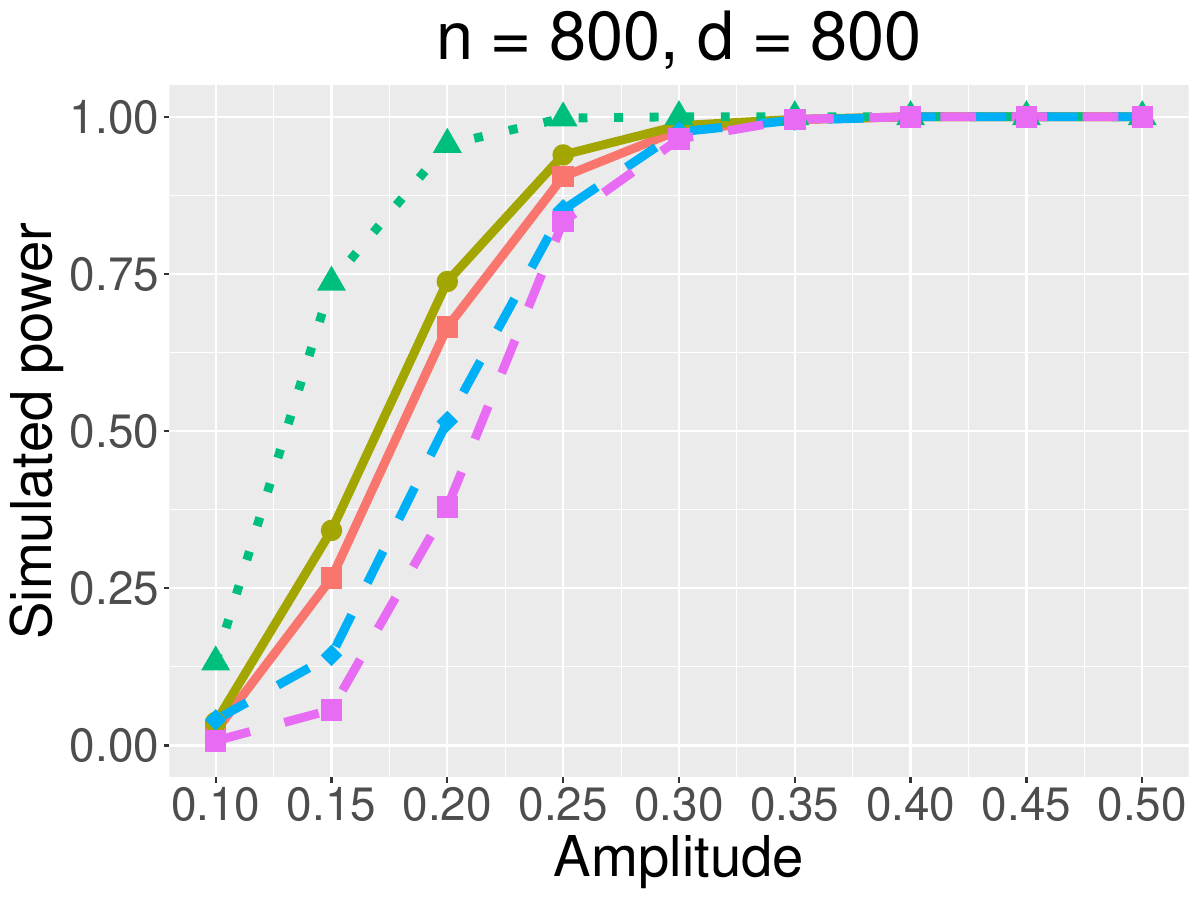}}\hspace{5pt}
    \subfloat{\includegraphics[width=.27\columnwidth]{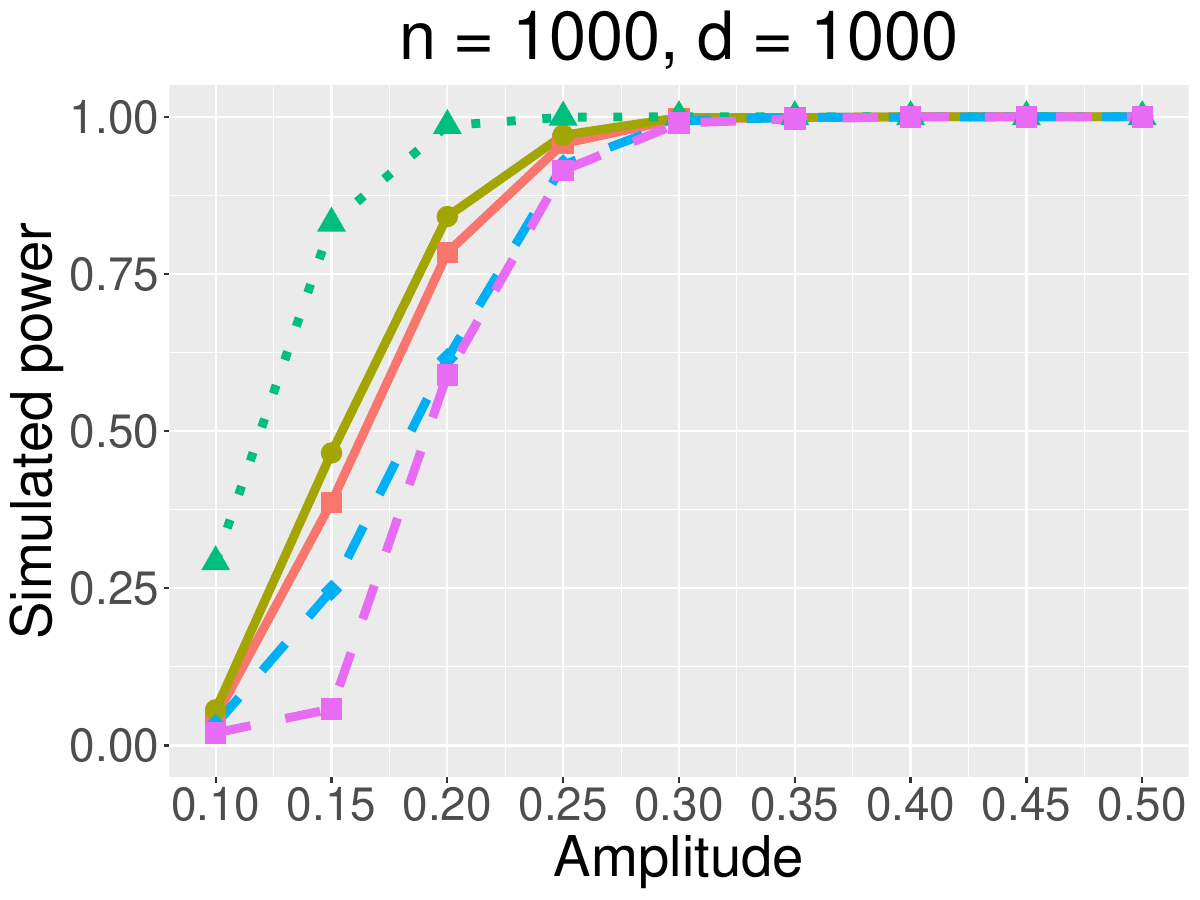}}
	\caption{\small Simulated FDR and power for different combinations of $(n,d)$. The rows of the design matrix were generated from Setting 2 with the AR(1) coefficient being set to 0.67. The sparsity level is $k = 15$ and the FDR level is $\alpha = 0.2$. The methods compared are Algorithm 1 (squares and red solid line), Algorithm 2 (circles and yellow solid line), the knockoff-based method of \cite{candes2018panning} (triangles and green dotted line), the Gaussian Mirror method of \cite{Xing2021Controlling} (diamonds and blue dashed line), and the Gaussian Mirror method with FDP+ procedure (squares and purple dashed line).}
    \label{Highly-correlated-0.2}
\end{figure}

Figures \ref{Highly-correlated-0.05}--\ref{Highly-correlated-0.2} report simulation results with highly correlated variables. The rows of $\bX$ are drawn from a multivariate normal distribution with an autoregressive (AR(1)) dependence structure and correlation coefficient 0.67. The sparsity level is $k = 15$.

\subsection{Additional simulation results of the two-stage approach}\label{simu_twostage}
\begin{figure}[htbp!]
	\centering
	\subfloat{\includegraphics[width=.27\columnwidth]{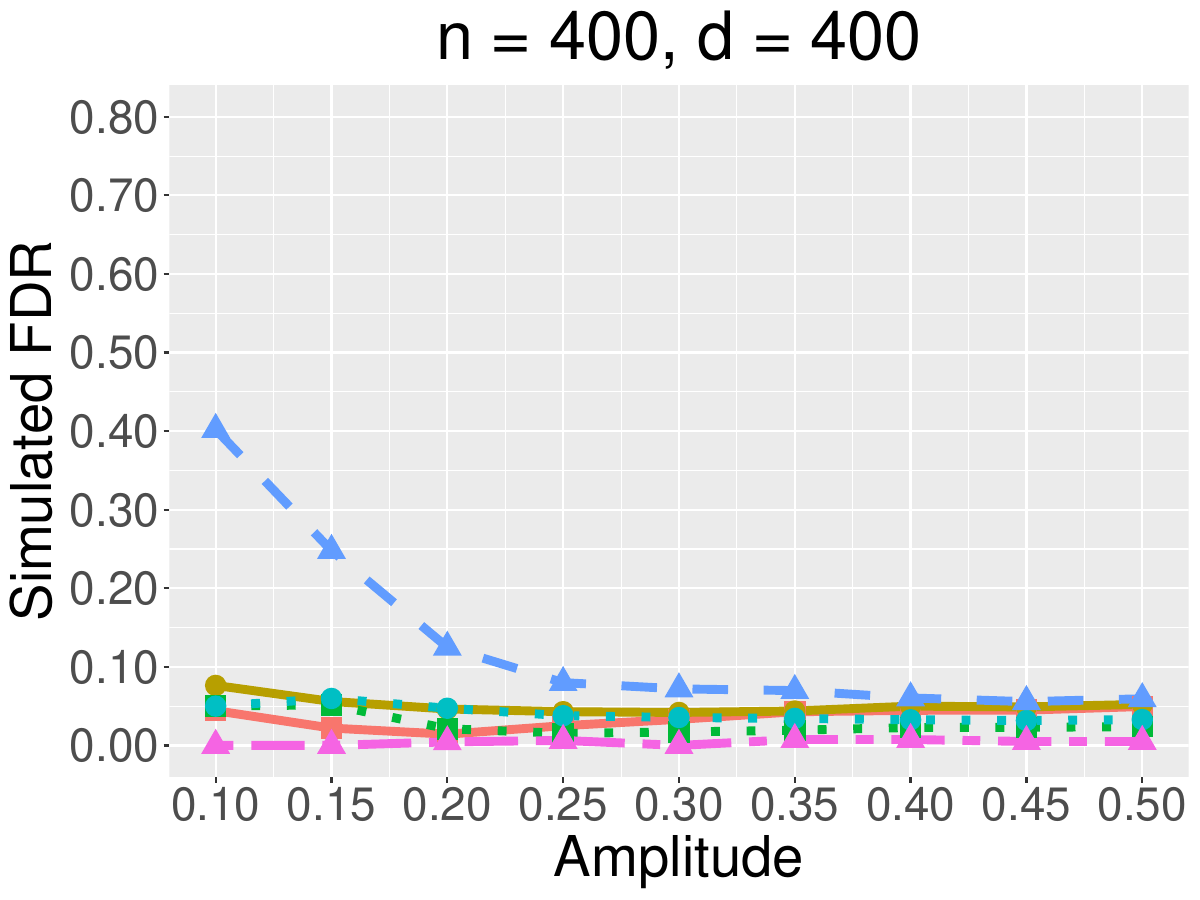}}\hspace{5pt}
	\subfloat{\includegraphics[width=.27\columnwidth]{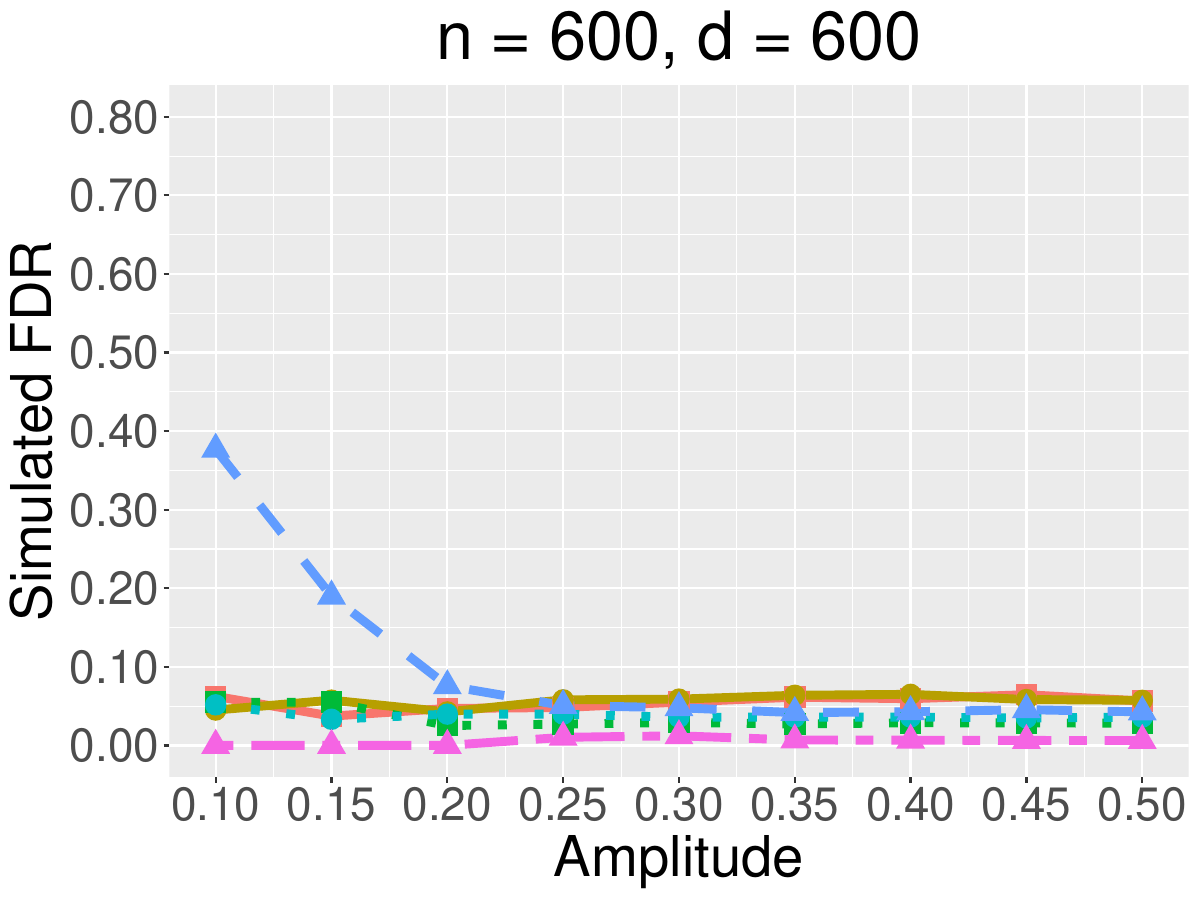}}\hspace{5pt}
	\subfloat{\includegraphics[width=.27\columnwidth]{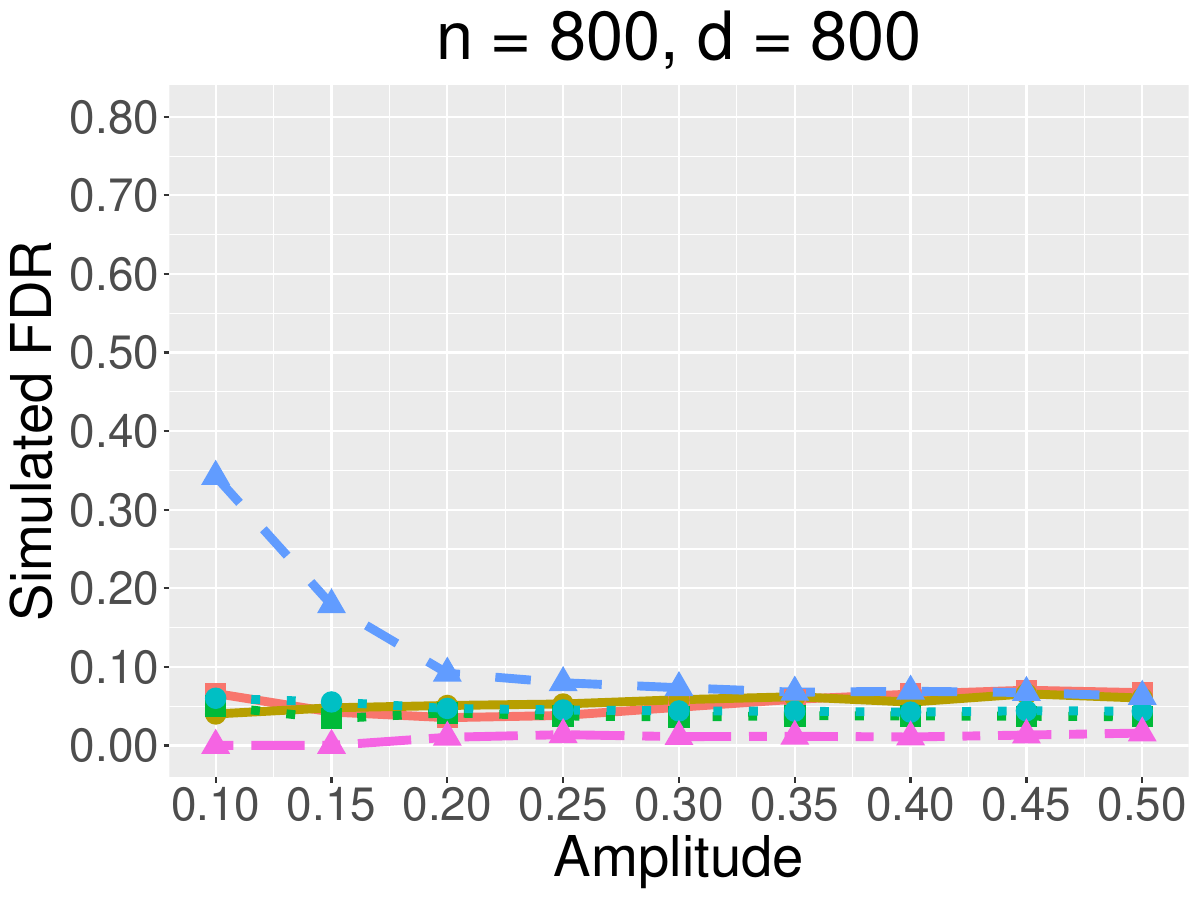}}\\
	\subfloat{\includegraphics[width=.27\columnwidth]{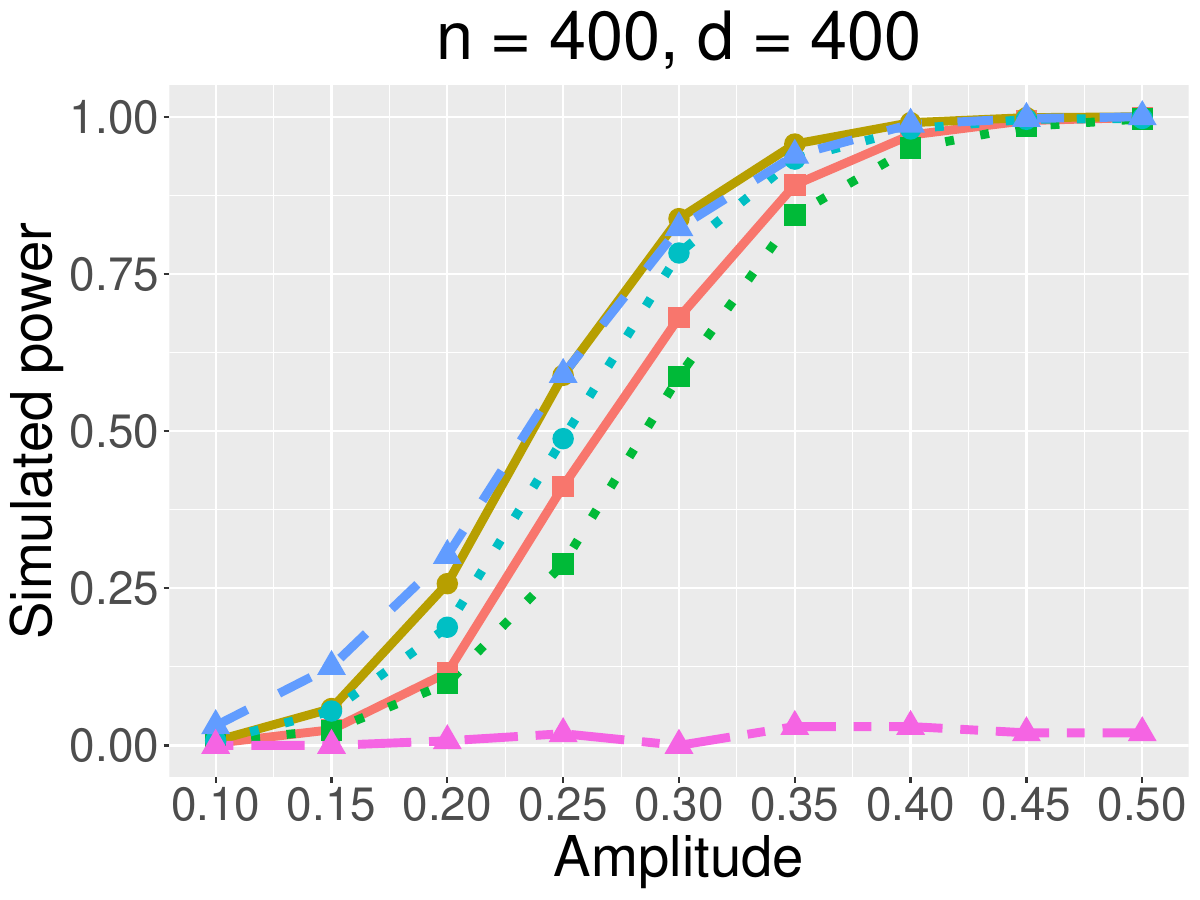}}\hspace{5pt}
    \subfloat{\includegraphics[width=.27\columnwidth]{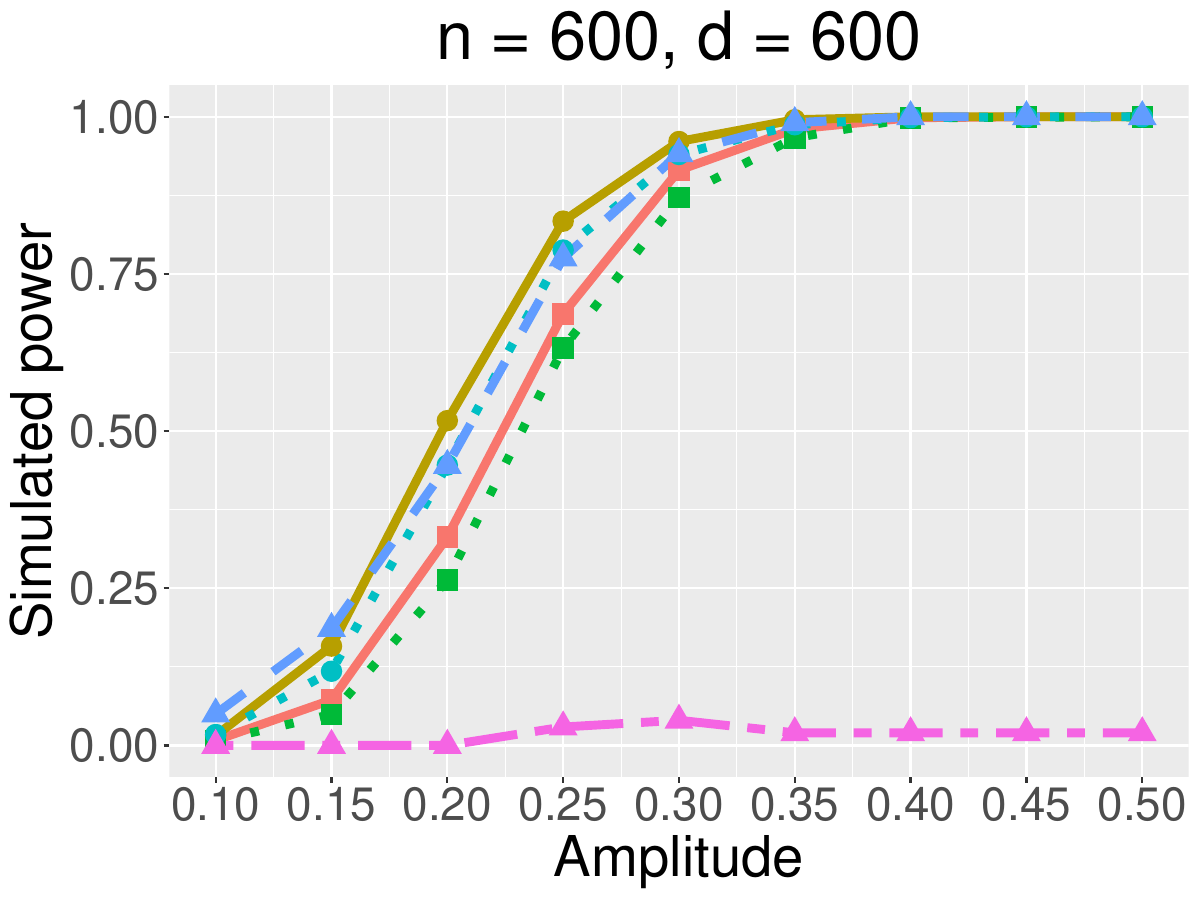}}\hspace{5pt}
    \subfloat{\includegraphics[width=.27\columnwidth]{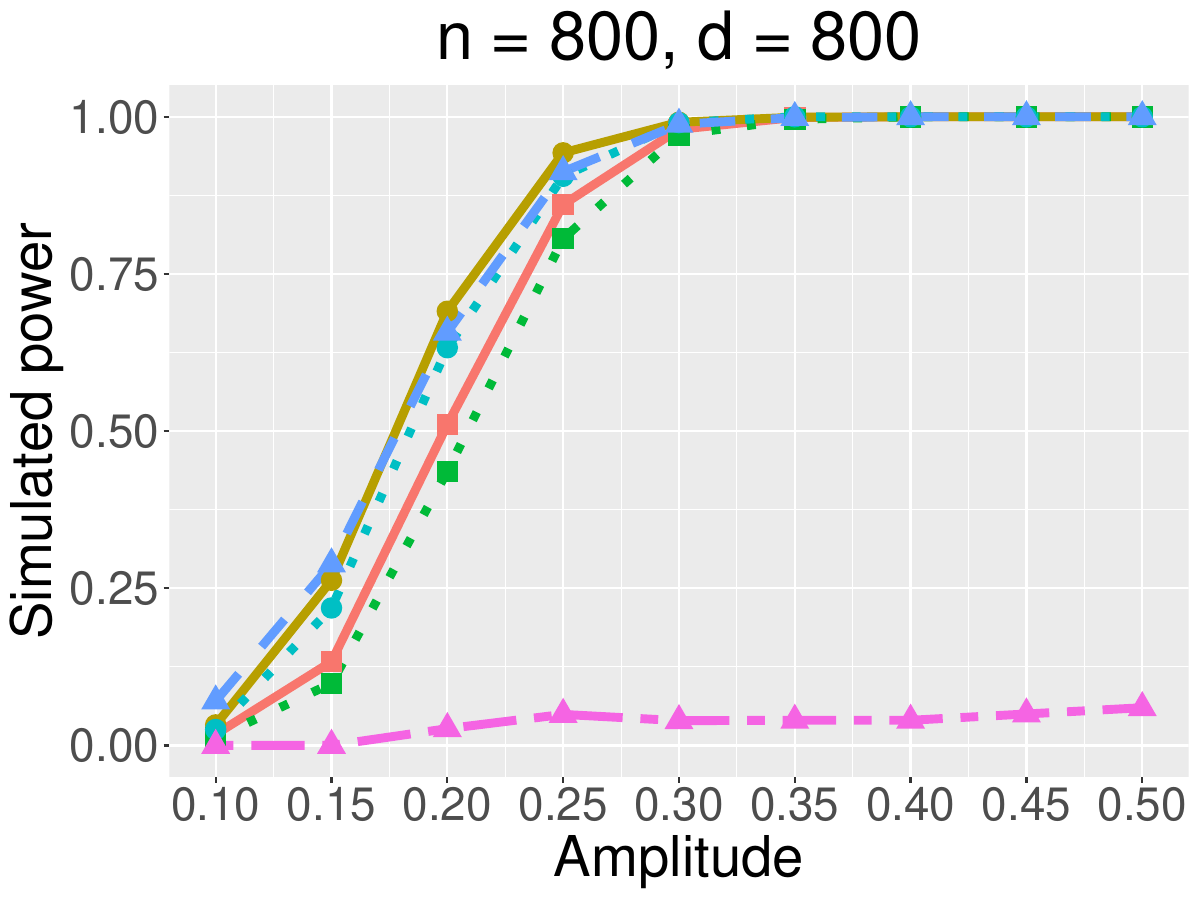}}
	\caption{\small Simulated FDR and power for different combinations of $(n,d)$. The rows of the design matrix were generated from Setting 2. The sparsity level is $k = 15$ and the FDR level is $\alpha = 0.05$. The methods compared are Algorithm 1 (squares and green dotted line), two-stage Algorithm 1 (squares and red solid line), Algorithm 2 (circles and blue dotted line), two-stage Algorithm 2 (circles and yellow solid line), the Gaussian Mirror method of \cite{Xing2021Controlling} (triangles and blue dashed line), and the Gaussian Mirror method with FDP+ procedure (triangles and purple two-dashed line).}
    \label{Two-stage-0.05}
\end{figure}

\begin{figure}[htbp!]
	\centering
	\subfloat{\includegraphics[width=.27\columnwidth]{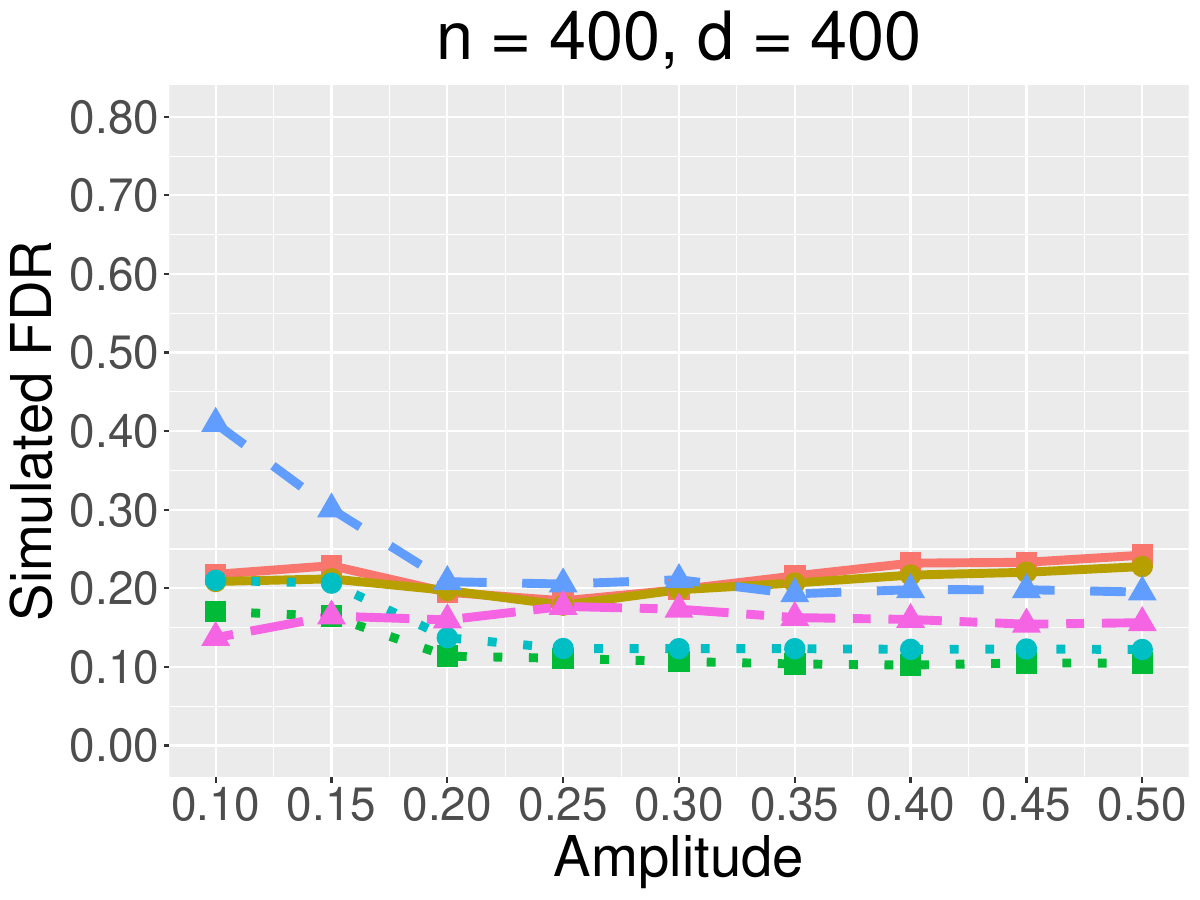}}\hspace{5pt}
	\subfloat{\includegraphics[width=.27\columnwidth]{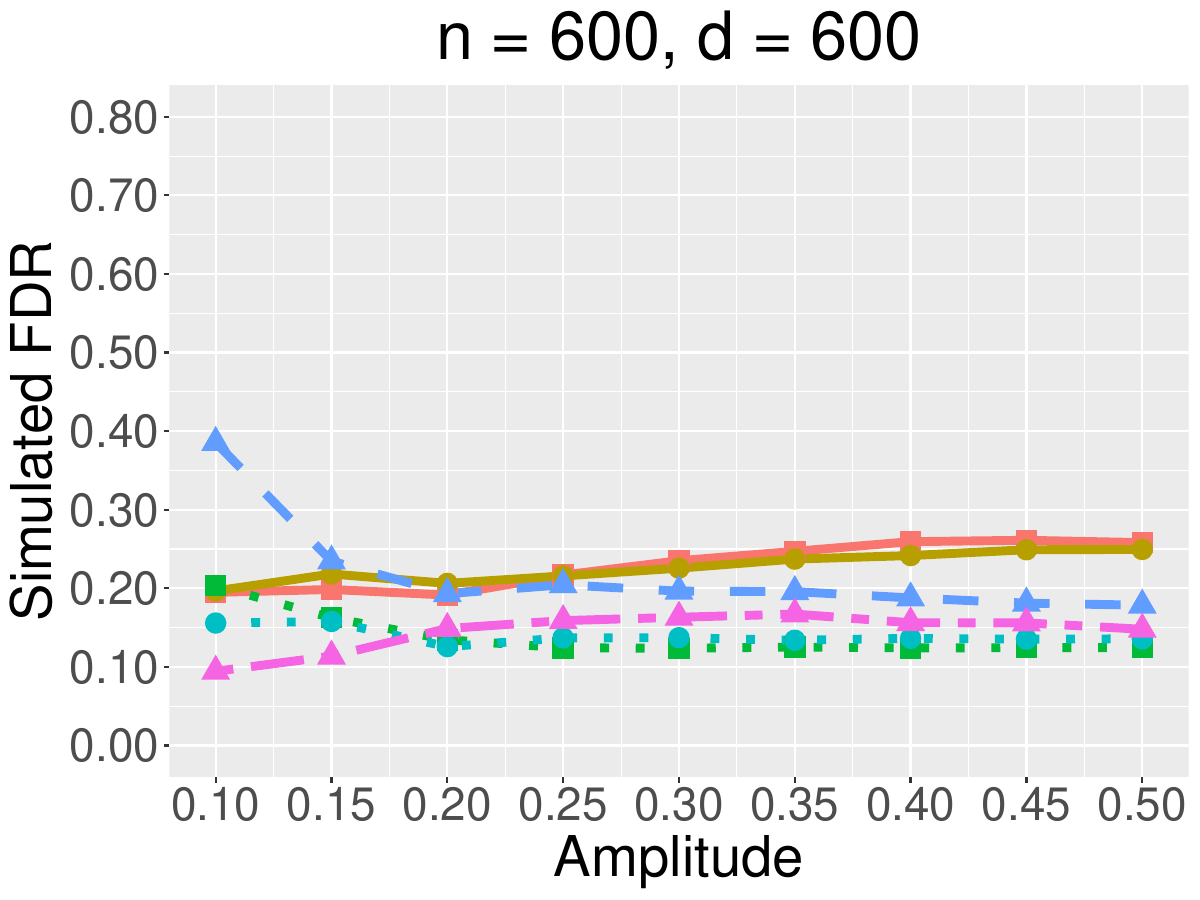}}\hspace{5pt}
	\subfloat{\includegraphics[width=.27\columnwidth]{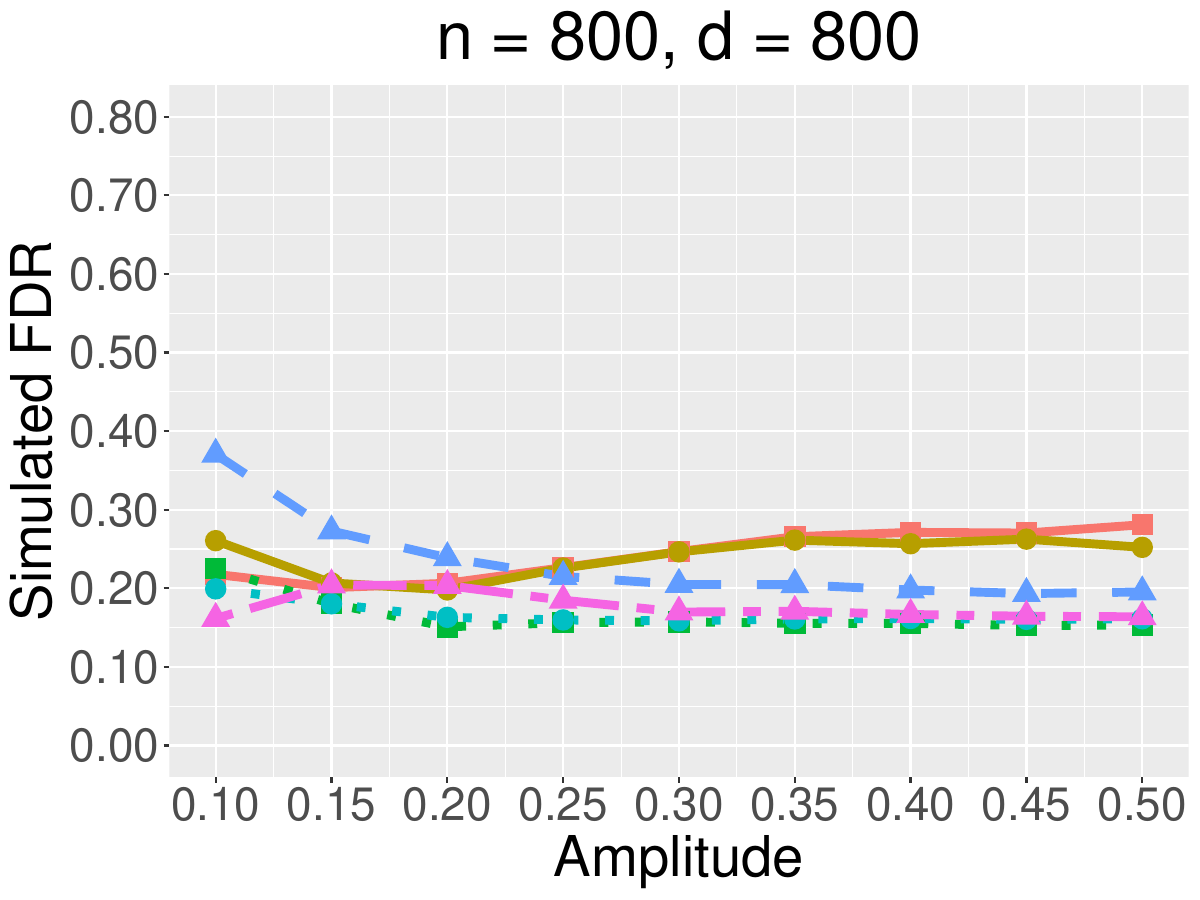}}\\
	\subfloat{\includegraphics[width=.27\columnwidth]{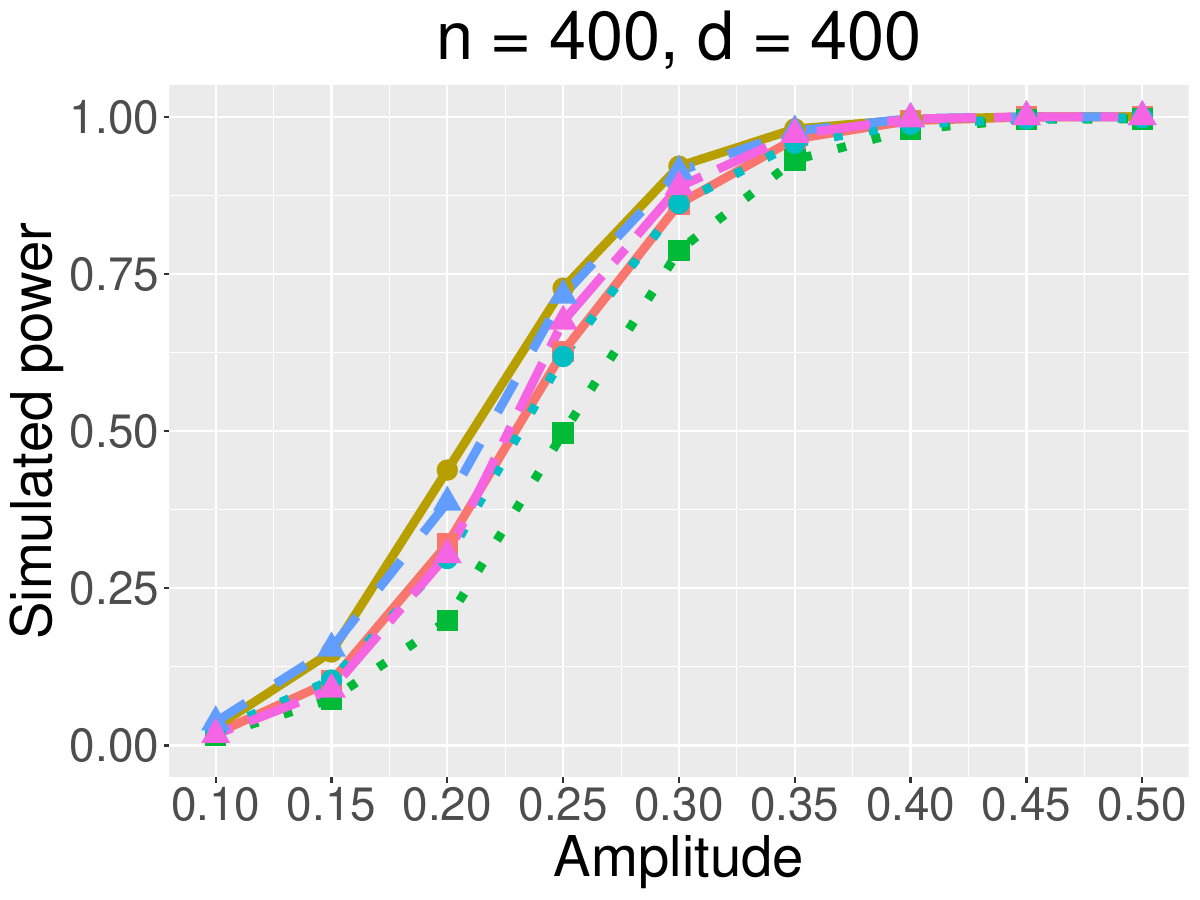}}\hspace{5pt}
    \subfloat{\includegraphics[width=.27\columnwidth]{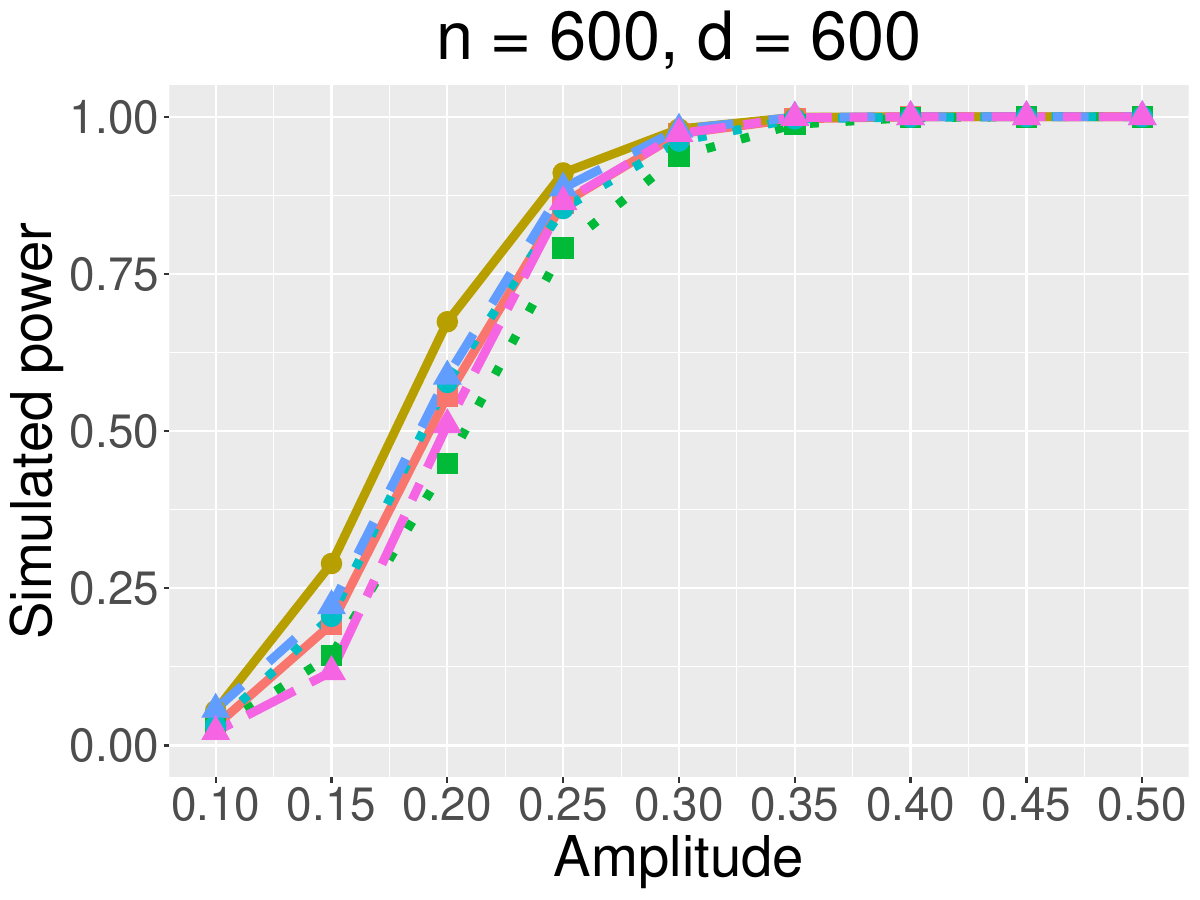}}\hspace{5pt}
    \subfloat{\includegraphics[width=.27\columnwidth]{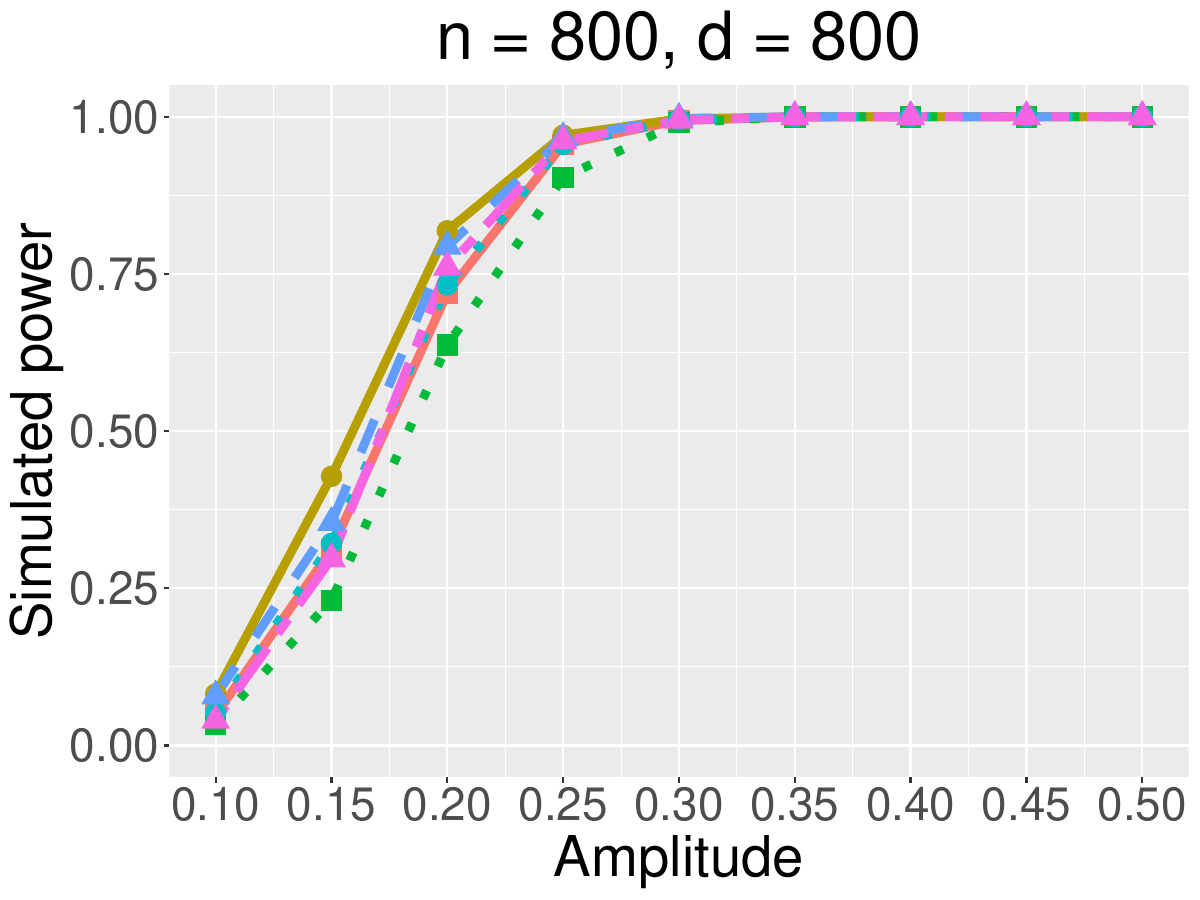}}
	\caption{\small Simulated FDR and power for different combinations of $(n,d)$. The rows of the design matrix were generated from Setting 2. The sparsity level is $k = 15$ and the FDR level is $\alpha = 0.2$. The methods compared are Algorithm 1 (squares and green dotted line), two-stage Algorithm 1 (squares and red solid line), Algorithm 2 (circles and blue dotted line), two-stage Algorithm 2 (circles and yellow solid line), the Gaussian Mirror method of \cite{Xing2021Controlling} (triangles and blue dashed line), and the Gaussian Mirror method with FDP+ procedure (triangles and purple two-dashed line).}
    \label{Two-stage-0.2}
\end{figure}

Figures \ref{Two-stage-0.05} and \ref{Two-stage-0.2} report additional simulation results of the two-stage approach with FDR levels $0.05$ and $0.2$, respectively.

\subsection{Additional simulations  in comparison with the approach of \cite{Sarkar2022Adjusting} } \label{lowd}
\begin{figure}[htbp!]
	\centering
	\subfloat{\includegraphics[width=.27\columnwidth]{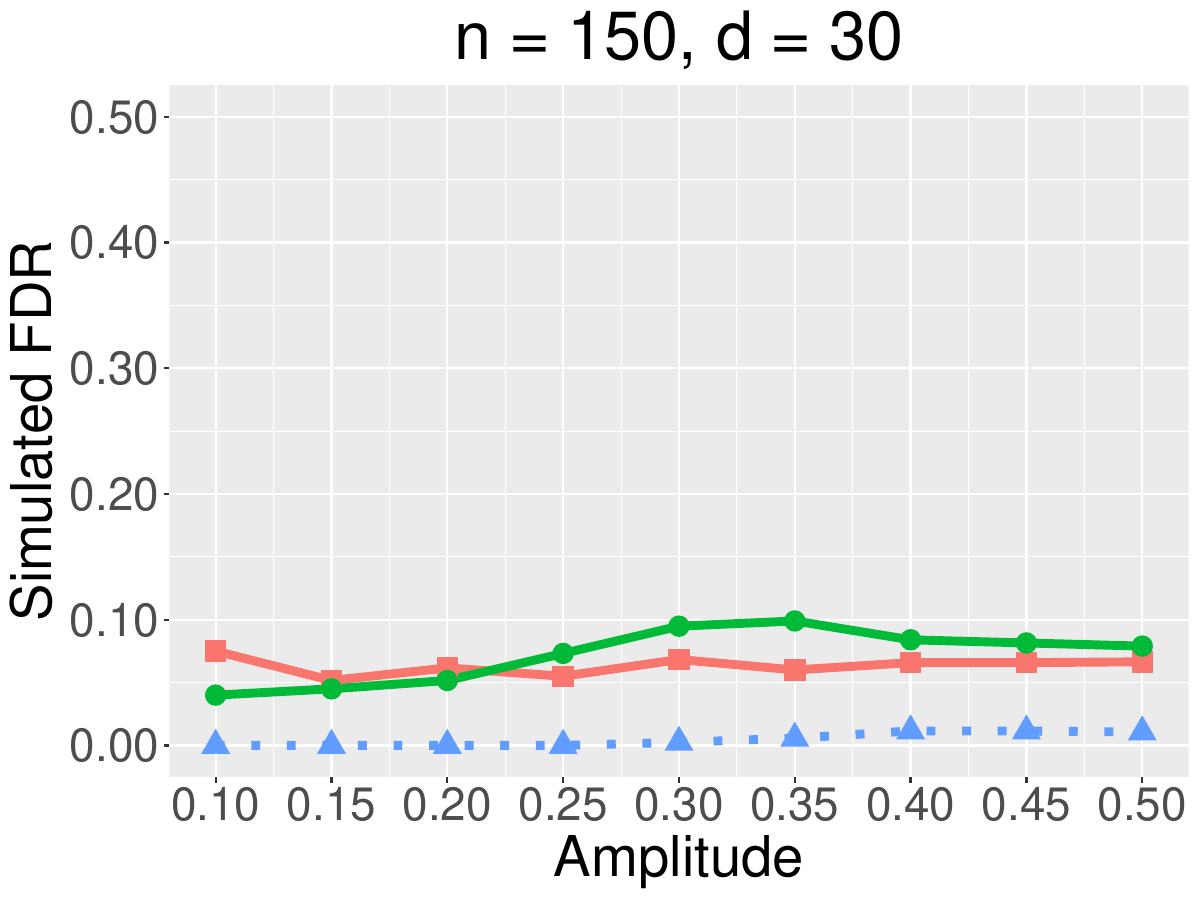}}\hspace{5pt}
	\subfloat{\includegraphics[width=.27\columnwidth]{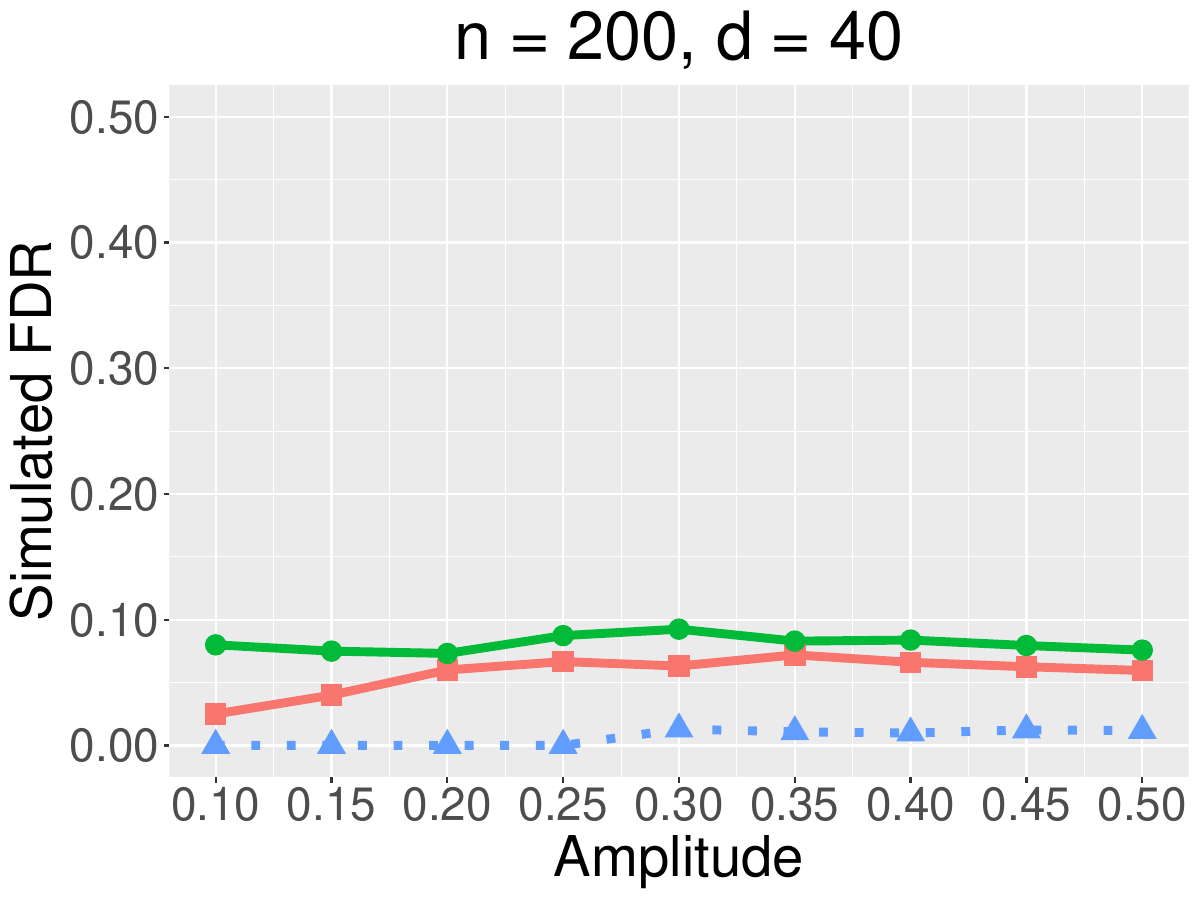}}\hspace{5pt}
	\subfloat{\includegraphics[width=.27\columnwidth]{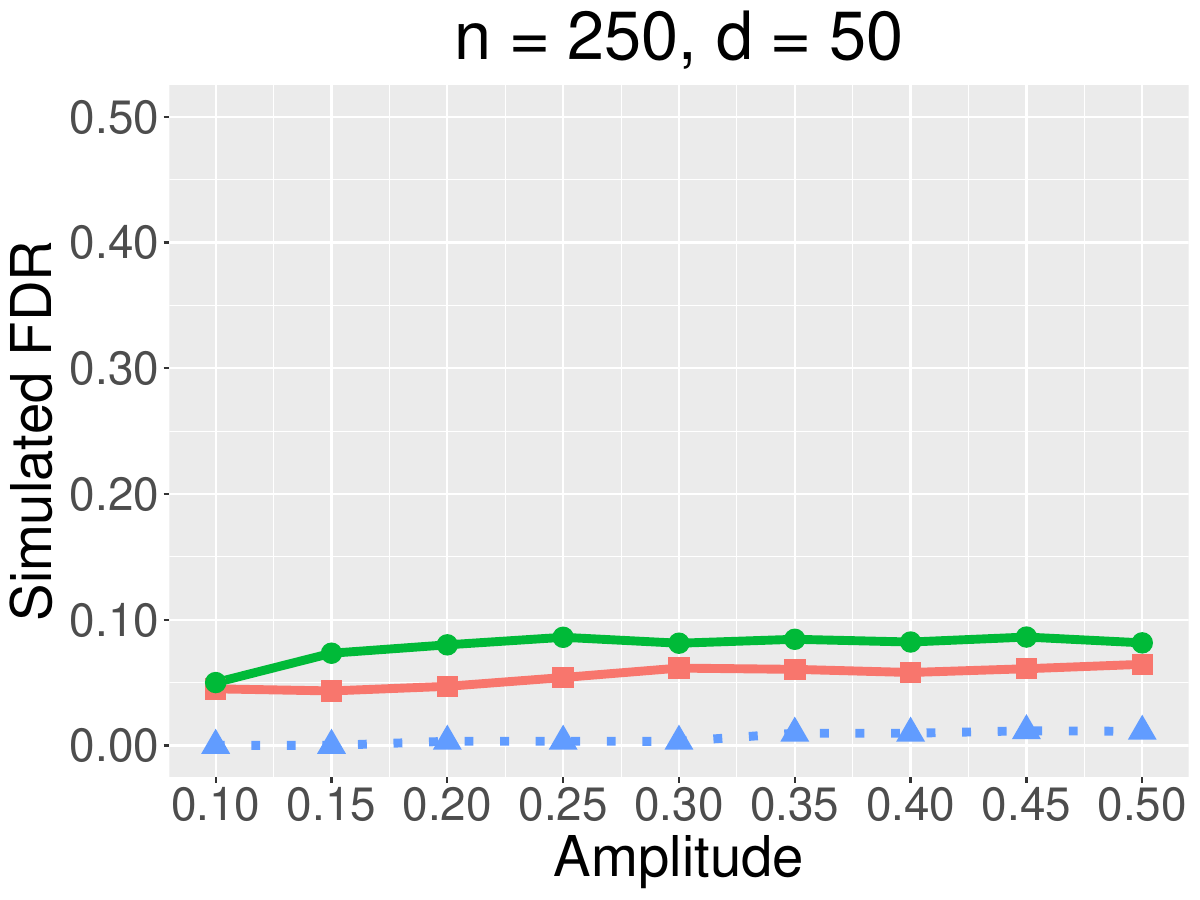}}\\
	\subfloat{\includegraphics[width=.27\columnwidth]{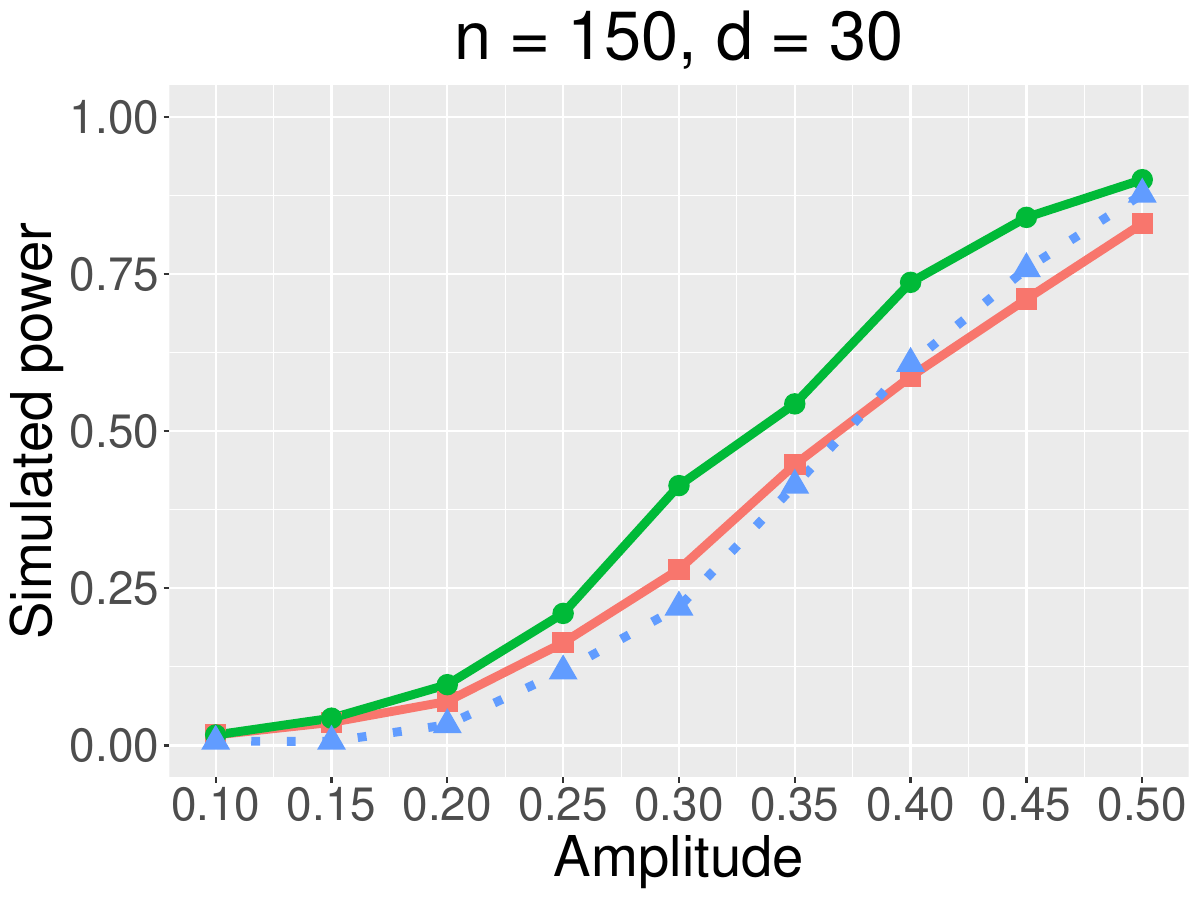}}\hspace{5pt}
    \subfloat{\includegraphics[width=.27\columnwidth]{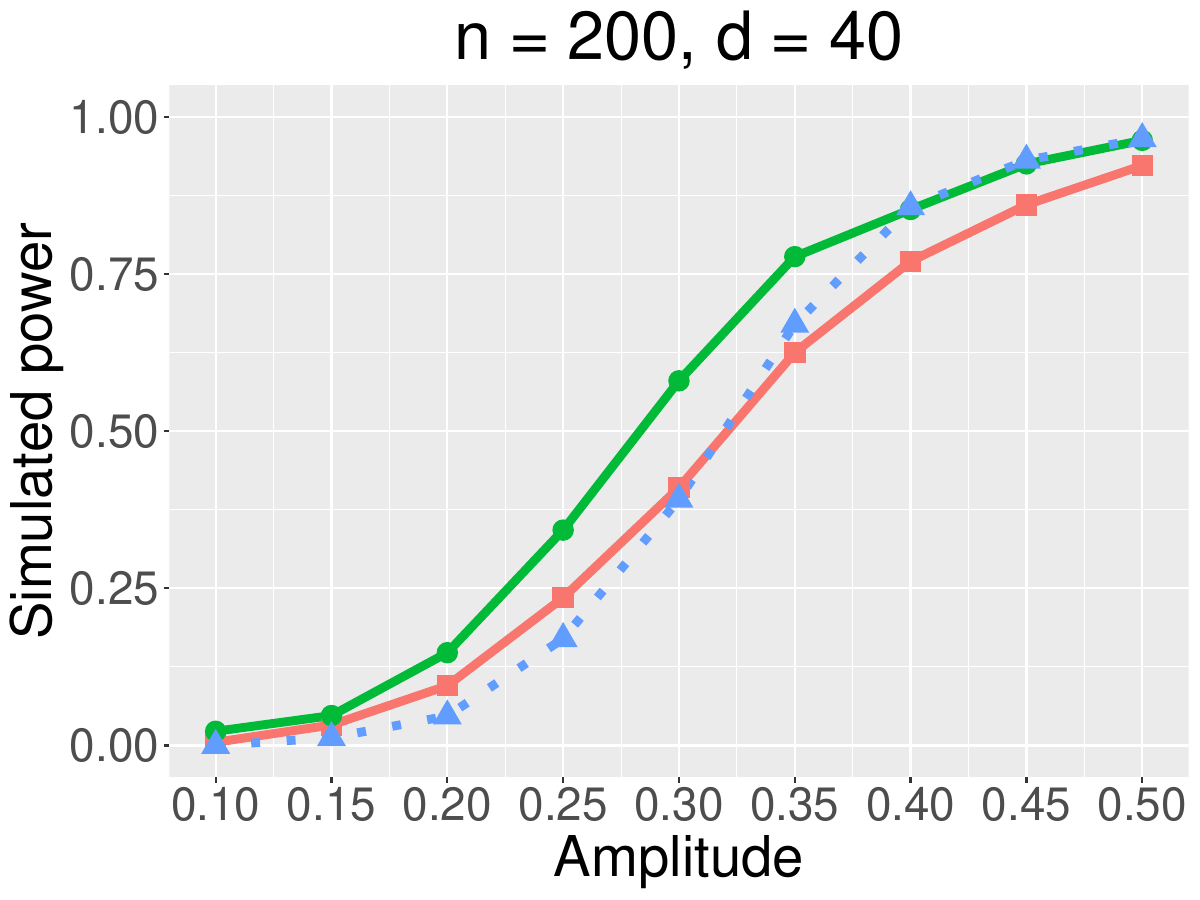}}\hspace{5pt}
    \subfloat{\includegraphics[width=.27\columnwidth]{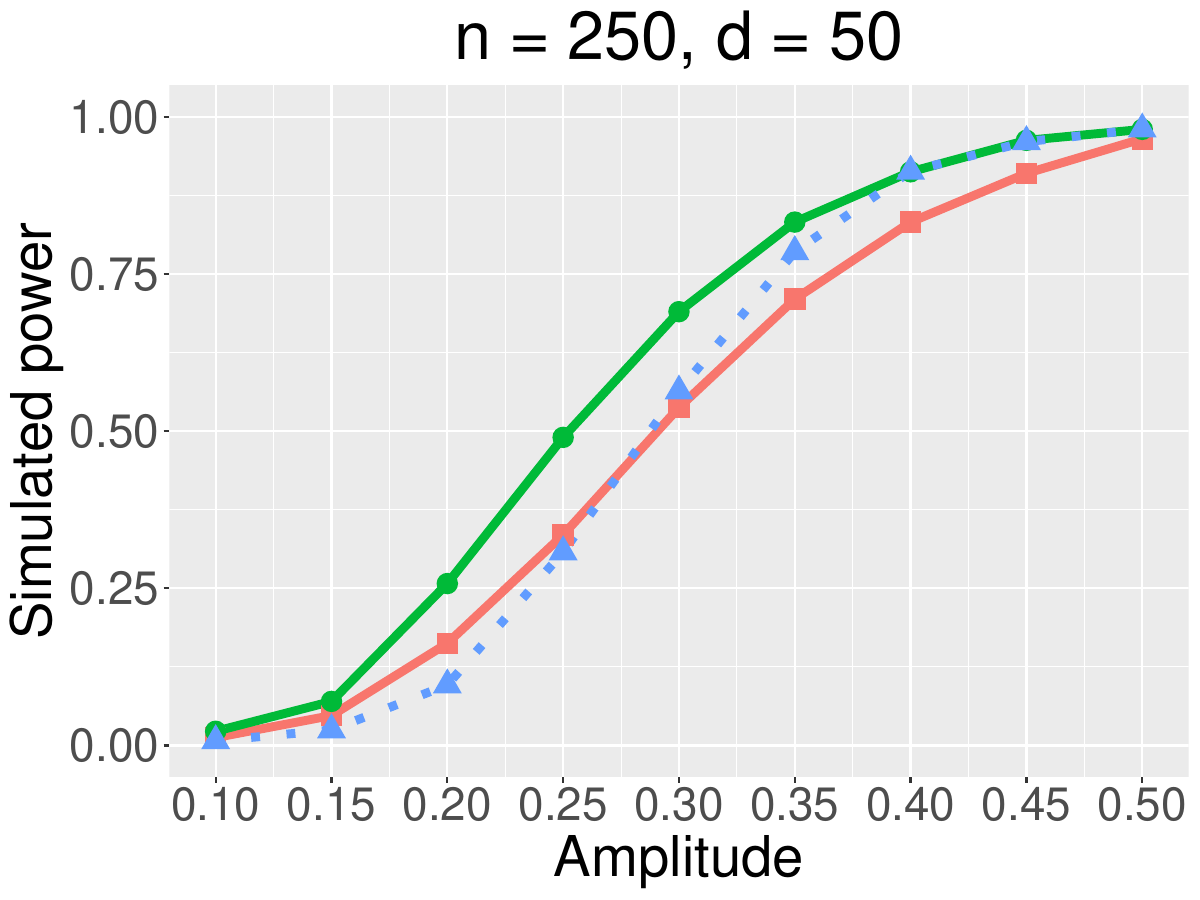}}
	\caption{\small Simulated FDR and power for different combinations of $(n,d)$. The rows of the design matrix were generated from Setting 2. The sparsity level is $k = 0.1d$ and the FDR level is $\alpha = 0.05$. The methods compared are Algorithm 1 (squares and red solid line), Algorithm 2 (circles and green solid line), and the Bonferroni-Benjamini-Hochberg method of \cite{Sarkar2022Adjusting} (triangles and blue dotted line).}
    \label{Low-dimension-0.05}
\end{figure}

\begin{figure}[htbp!]
	\centering
	\subfloat{\includegraphics[width=.27\columnwidth]{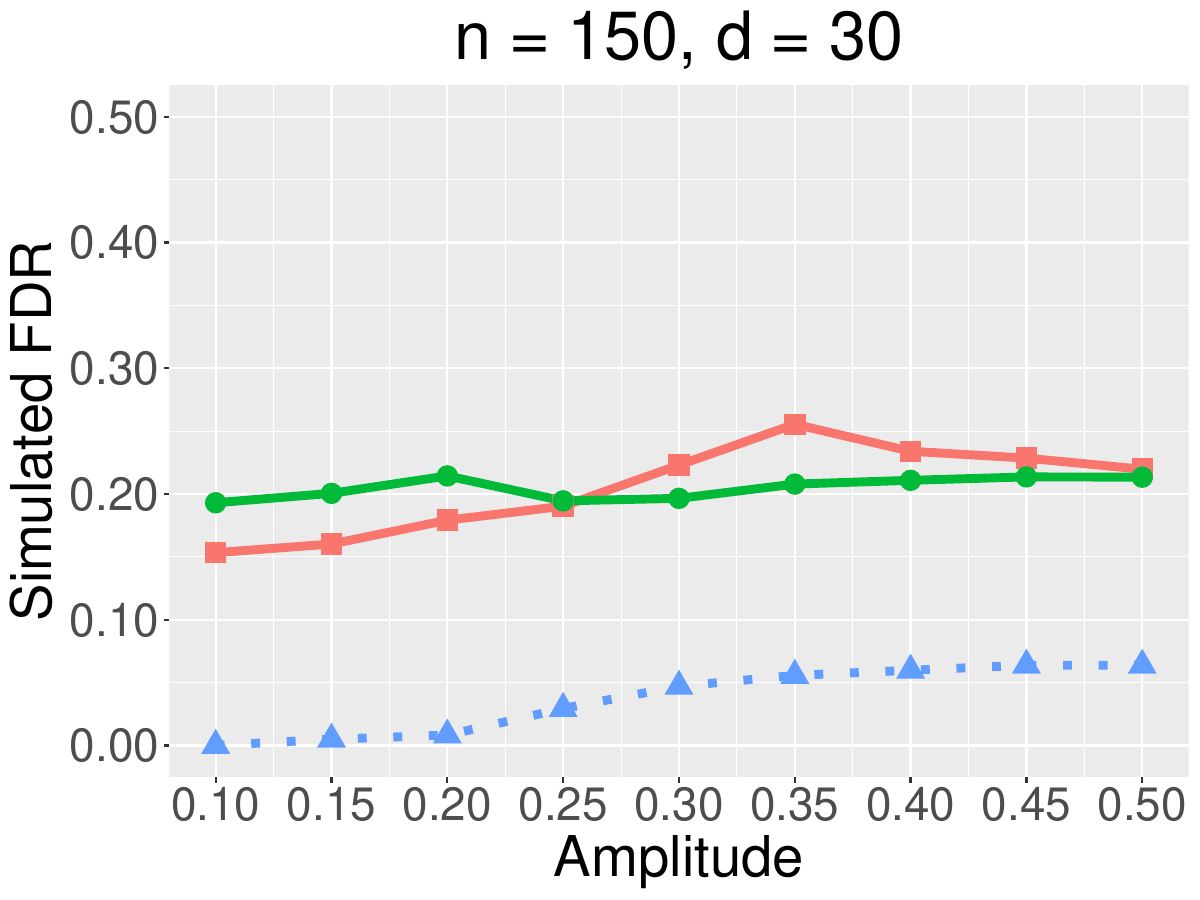}}\hspace{5pt}
	\subfloat{\includegraphics[width=.27\columnwidth]{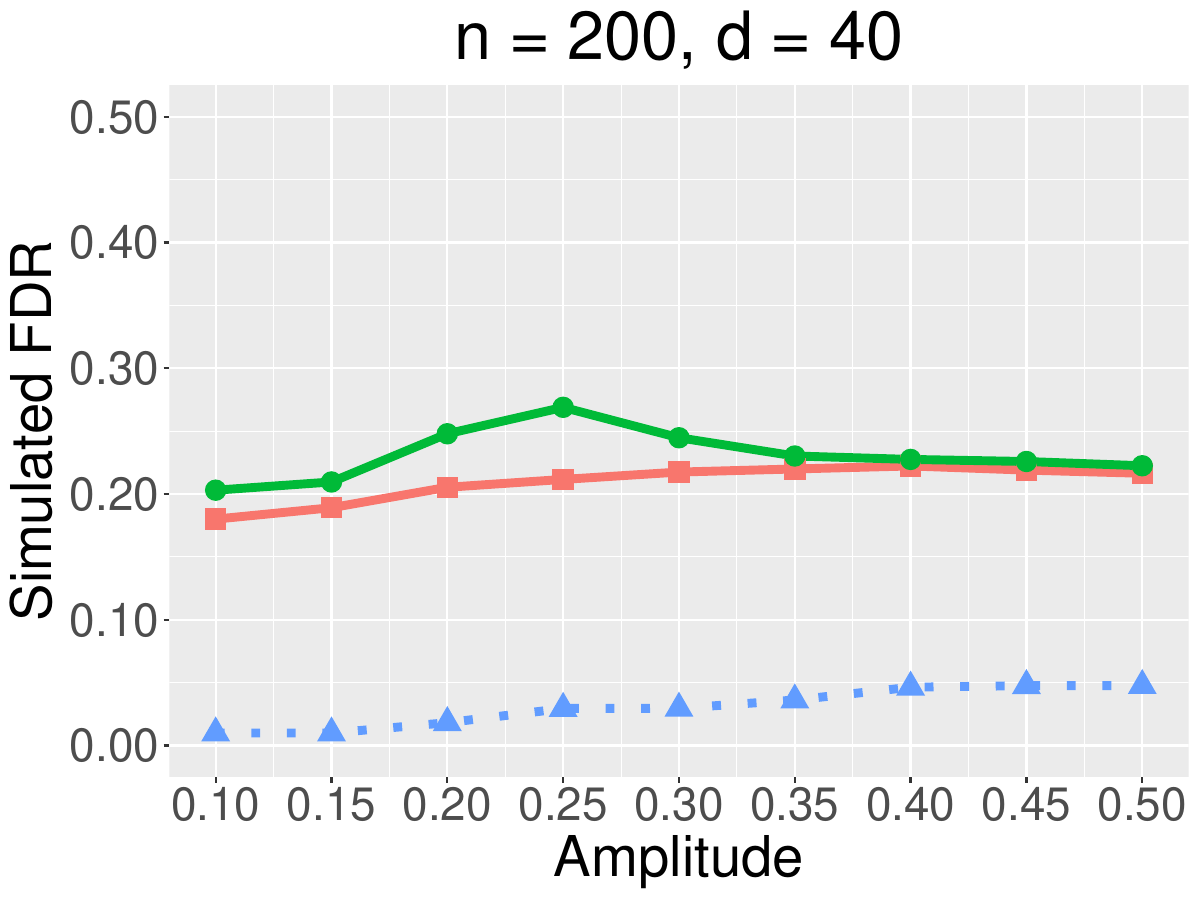}}\hspace{5pt}
	\subfloat{\includegraphics[width=.27\columnwidth]{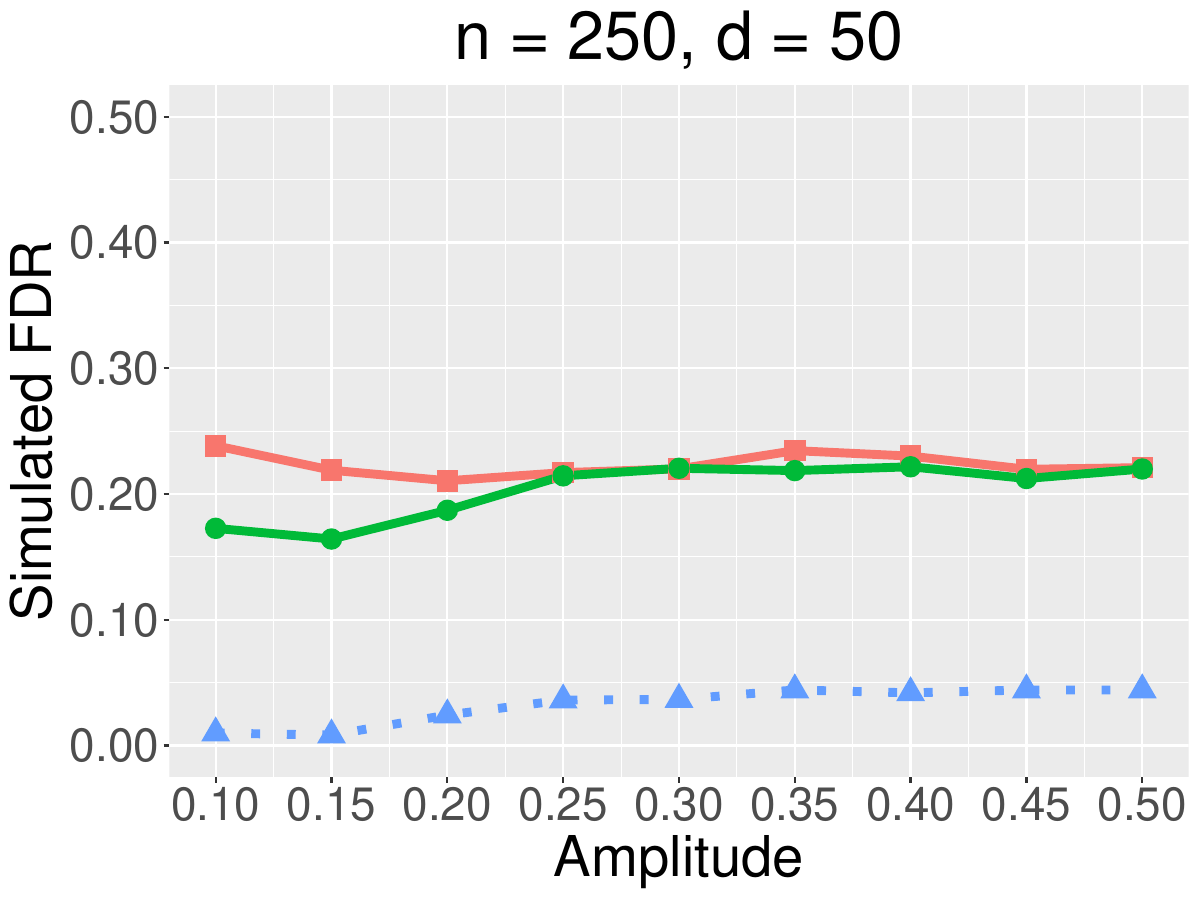}}\\
	\subfloat{\includegraphics[width=.27\columnwidth]{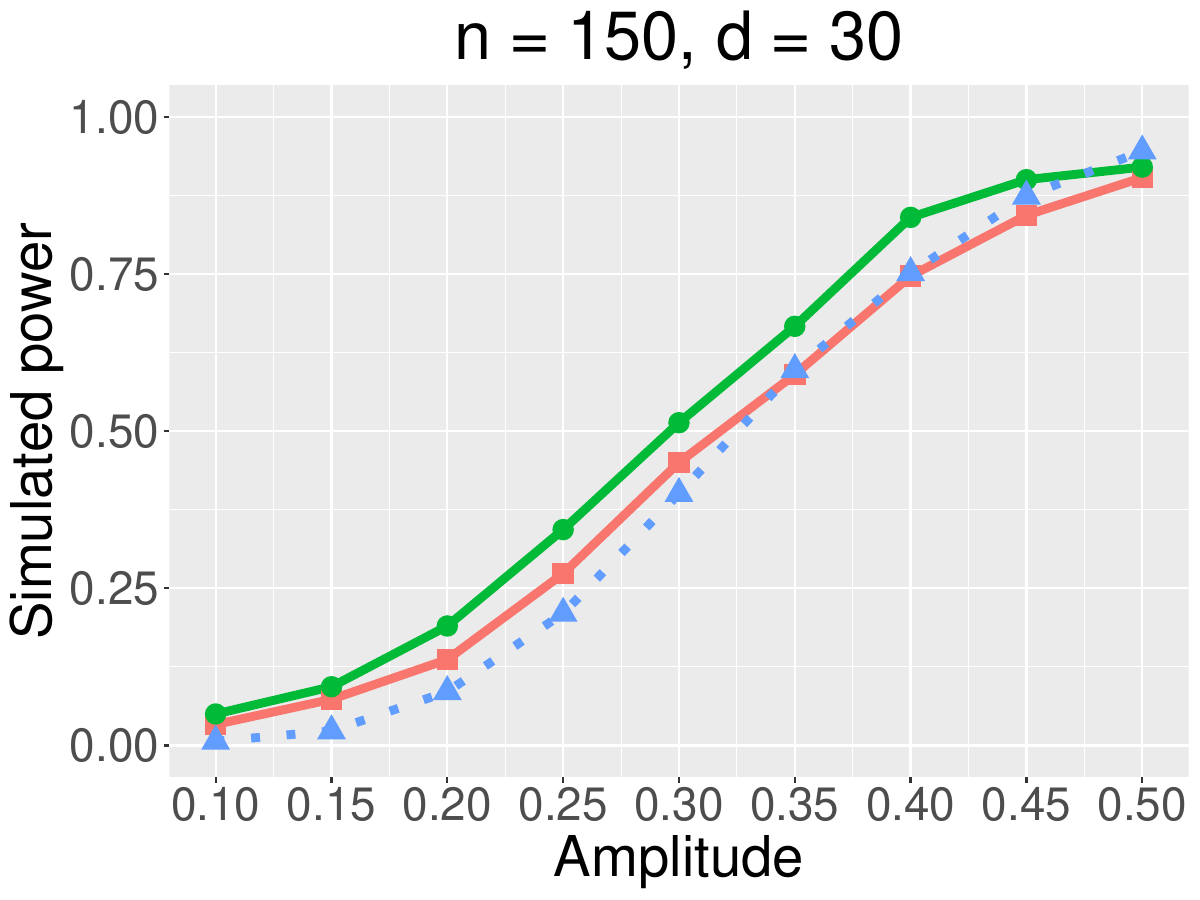}}\hspace{5pt}
    \subfloat{\includegraphics[width=.27\columnwidth]{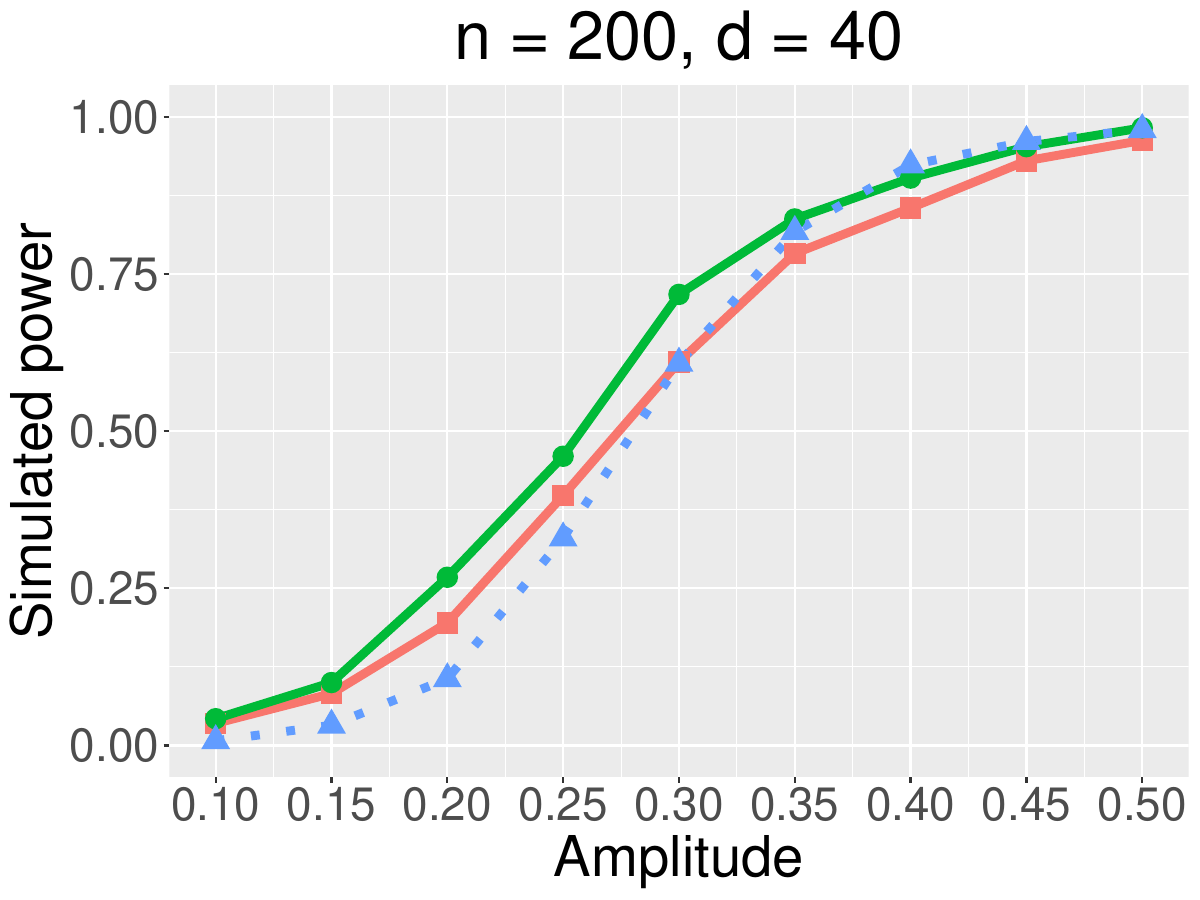}}\hspace{5pt}
    \subfloat{\includegraphics[width=.27\columnwidth]{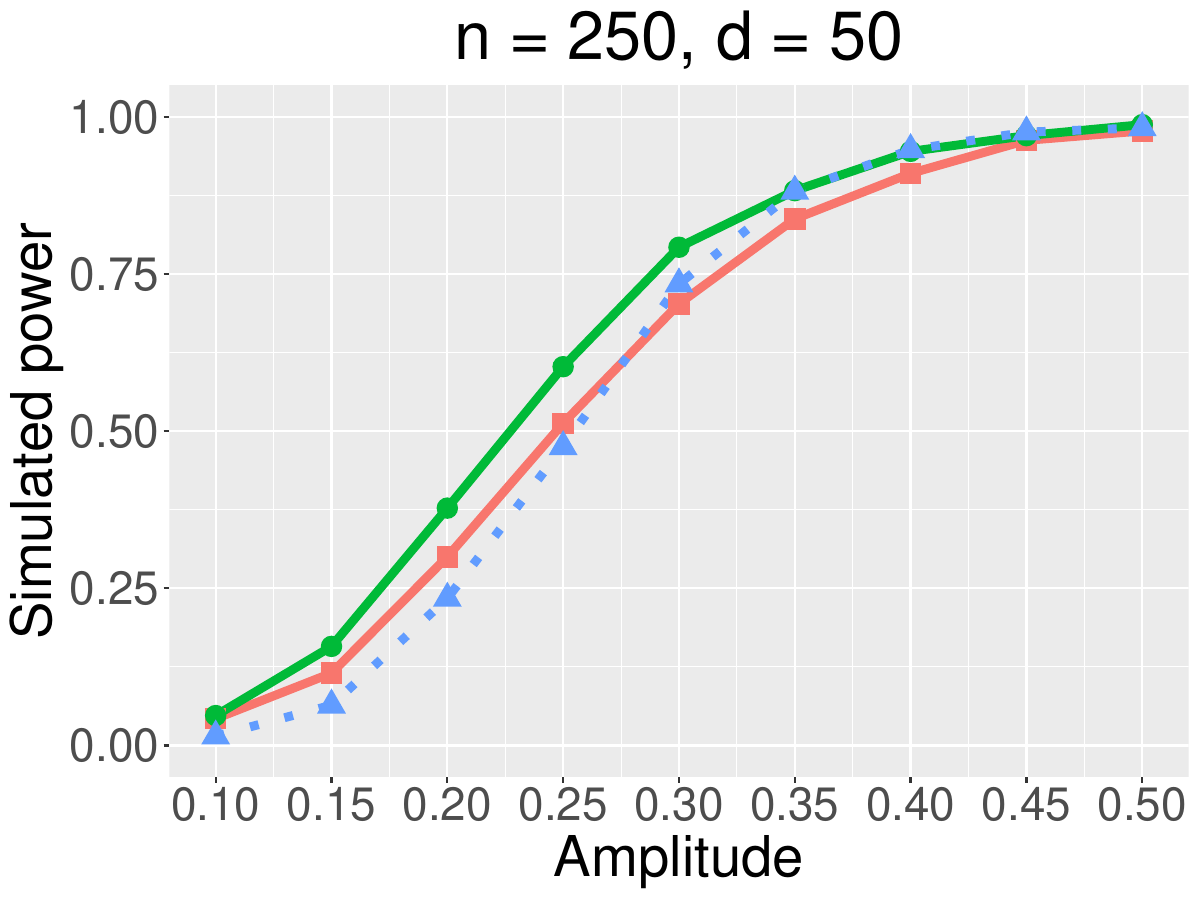}}
	\caption{\small Simulated FDR power for different combinations of $(n,d)$. The rows of the design matrix were generated from Setting 2. The sparsity level is $k = 0.1d$ and the FDR level is $\alpha = 0.2$. The methods compared are Algorithm 1 (squares and red solid line), Algorithm 2 (circles and green solid line), and the Bonferroni-Benjamini-Hochberg method of \cite{Sarkar2022Adjusting} (triangles and blue dotted line).}
    \label{Low-dimension-0.2}
\end{figure}

Figures \ref{Low-dimension-0.05} and \ref{Low-dimension-0.2} report additional simulation results under low dimensional setting with FDR levels $0.05$ and $0.2$, respectively.

\subsection{Additional simulations comparing with debiased Lasso without knockoff variables }\label{simu_pvalue}
In this subsection, we compare the methods proposed in this paper with alternative procedures based on $p$-values. 
Instead of employing model-X knockoffs, we construct $p$-values using the debiased Lasso based solely on the original data, without introducing knockoff variables. 

We begin by describing the debiasing methodology for a linear model using only the original design matrix.

Recall the linear model 
\begin{align*}
\by = \bX\bbeta + \beps\,,
\end{align*}
where \( \by= (y_1,\ldots,y_n)^\top \in \mathbb{R}^n \) is the response vector, \( \beps = (\varepsilon_1,\ldots,\varepsilon_n)^\top \in \mathbb{R}^n \) is the error term with $\mathbb{E}(\varepsilon_i)=0$ and ${\rm Var}(\varepsilon_i)=\sigma^2$, and \( \bbeta \in \mathbb{R}^d \) is the parameter vector. If we rely solely on the original design matrix without constructing knockoff variables, the debiased estimator takes the form
\begin{equation}\label{debiased_origin}
    \hat{\bbeta}^{(\rm bc)} 
    = \hat{\bbeta} 
    + n^{-1} \hat{\bXi}^{\top} \bX^{\top} (\by - \bX \hat{\bbeta})\,.
\end{equation}
Here, \(\hat{\bbeta}\) denotes the Lasso estimator
\begin{equation} \label{Lasso_origin}
    \hat{\bbeta} 
    = \arg\min_{\bbeta \in \mathbb{R}^{d}} 
    \bigg\{ 
    \frac{1}{2n} | \by - \bX\bbeta |_2^2 
    + \varrho_1 | \bbeta |_1 
    \bigg\}\,,
\end{equation}
and \(\hat{\bXi}\) is the decorrelating matrix associated with 
\(\hat{\bSigma} = n^{-1}\bX^{\top} \bX\).
We first construct the debiased Lasso estimator using \eqref{debiased_origin}. With a slight abuse of notation, we estimate the noise level $\hat{\sigma}$ via the scaled Lasso \citepS{sun2012scaled_app} applied to the original data:
\begin{align}\label{scaledLasso_origin}
    \min_{\bbeta \in \mathbb{R}^{d},\, \sigma > 0} 
    \bigg\{ 
    \frac{1}{2\sigma n} | \by - \bX \bbeta |_2^2 
    + \frac{\sigma}{2} 
    + \varrho_3 |\bbeta|_1 
    \bigg\}\,.
\end{align}
Let $\hat{\bOmega} = \hat{\bXi}^\top \hat{\bSigma} \hat{\bXi}$, and denote its diagonal elements by $\{\hat{\Omega}_{j,j}\}_{j=1}^d$. Writing $\hat{\bbeta}^{(\rm bc)} = (\hat{\beta}_1^{(\rm bc)}, \ldots, \hat{\beta}_d^{(\rm bc)})^\top$, we define the corresponding test statistics as
\begin{equation}
    t_j := 
    \frac{n^{1/2}\hat{\beta}_j^{(\rm bc)}}
    {\hat{\sigma}\,\hat{\Omega}_{j,j}^{1/2}}\,,
\end{equation}
and compute the associated $p$-values using the standard normal reference distribution.

The BH procedure and a two-step method that does not use paired statistics, both based solely on the original design matrix, are presented as Algorithm~\ref{method_1_origin} and Algorithm~\ref{method_2_origin}, respectively.

\floatname{algorithm}{Algorithm}
\begin{algorithm}
    \caption{Benjamini-Hochberg with test statistics $\{t_{j}\}_{j=1}^d$}
    \label{method_1_origin}

    \begin{algorithmic}
        \STATE\textbf{Step 1}. Let $\tilde{P}_j=P_j=G(|t_{j}|)$ for each $j\in[d]$, where $G(t)=2\{1-\Phi(t)\}$ and $\Phi(t)$ is the cumulative distribution function of the standard normal distribution.
        \STATE\textbf{Step 2}. Given $\alpha \in(0,1)$, let $\tilde{P}_{(1)} \leq \cdots \leq \tilde{P}_{(d)}$ be the ordered versions of the $\tilde{P}_j$'s, find
        \begin{equation*}
            \tilde{R}=\max\bigg\{i\in[d]: \tilde{P}_{(i)} \leq \frac{i \alpha}{d}\bigg\}\,,
        \end{equation*}
        provided that the maximum exists; otherwise, let $\tilde{R}=0$.
        \STATE\textbf{Step 3}. Reject the null hypotheses corresponding to $\tilde{P}_{(j)}$ with $j \leq \tilde{R}$.
    \end{algorithmic}
\end{algorithm}

\floatname{algorithm}{Algorithm}
\begin{algorithm}
    \caption{Bonferonni-Benjamini-Hochberg with test statistics $\{t_{j}\}_{j=1}^d$ }
    \label{method_2_origin}

    \begin{algorithmic} 
        \STATE\textbf{Step 1}. Given $0<\sqrt{\alpha}<1$, for each $j\in[d]$, let
        \begin{align*}
            \tilde{P}_j= \left\{
        	\begin{aligned}
        		1\,, ~~~&\textrm{if}~P_j>\sqrt{\alpha} \,,\\
        		P_j \,, ~~&\textrm{if}~P_j \leq \sqrt{\alpha} \,,
        	\end{aligned}
        	\right.
        \end{align*}
            where $P_j=G(|t_{j}|)$.
        \STATE\textbf{Step 2}. Let $\tilde{P}_{(1)} \leq \cdots \leq \tilde{P}_{(d)}$ be the ordered versions of the $\tilde{P}_j$'s. Find
        \begin{equation*}
            \tilde{R}=\max\bigg\{i\in[d]: \tilde{P}_{(i)} \leq \frac{i \sqrt{\alpha}}{d}\bigg\}\,,
        \end{equation*}
        provided that the maximum exists; otherwise, let $\tilde{R}=0$.
        \STATE\textbf{Step 3}. Reject the null hypotheses corresponding to $\tilde{P}_{(j)}$ with $j \leq \tilde{R}$.
    \end{algorithmic}
\end{algorithm}

\begin{figure}[htbp!]
	\centering
	\subfloat{\includegraphics[width=.27\columnwidth]{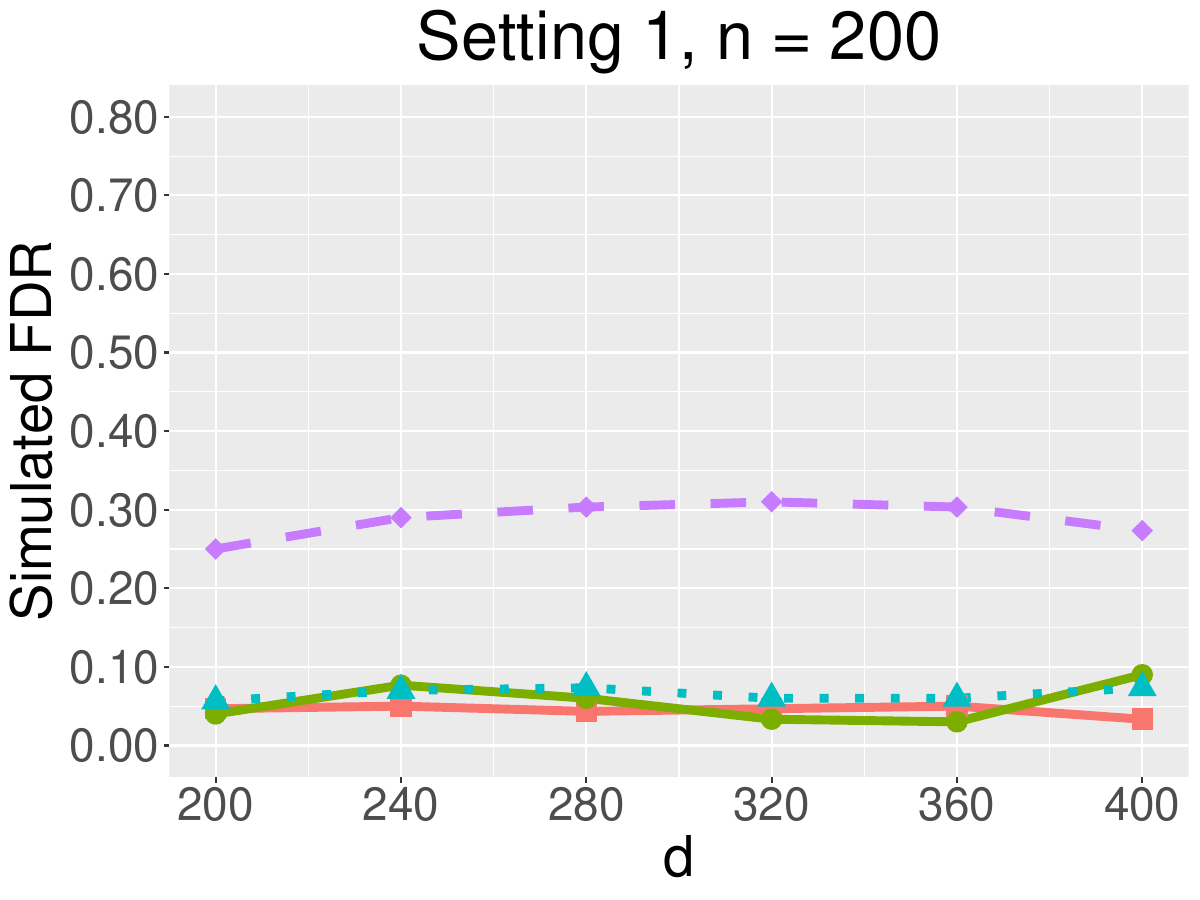}}\hspace{5pt}
	\subfloat{\includegraphics[width=.27\columnwidth]{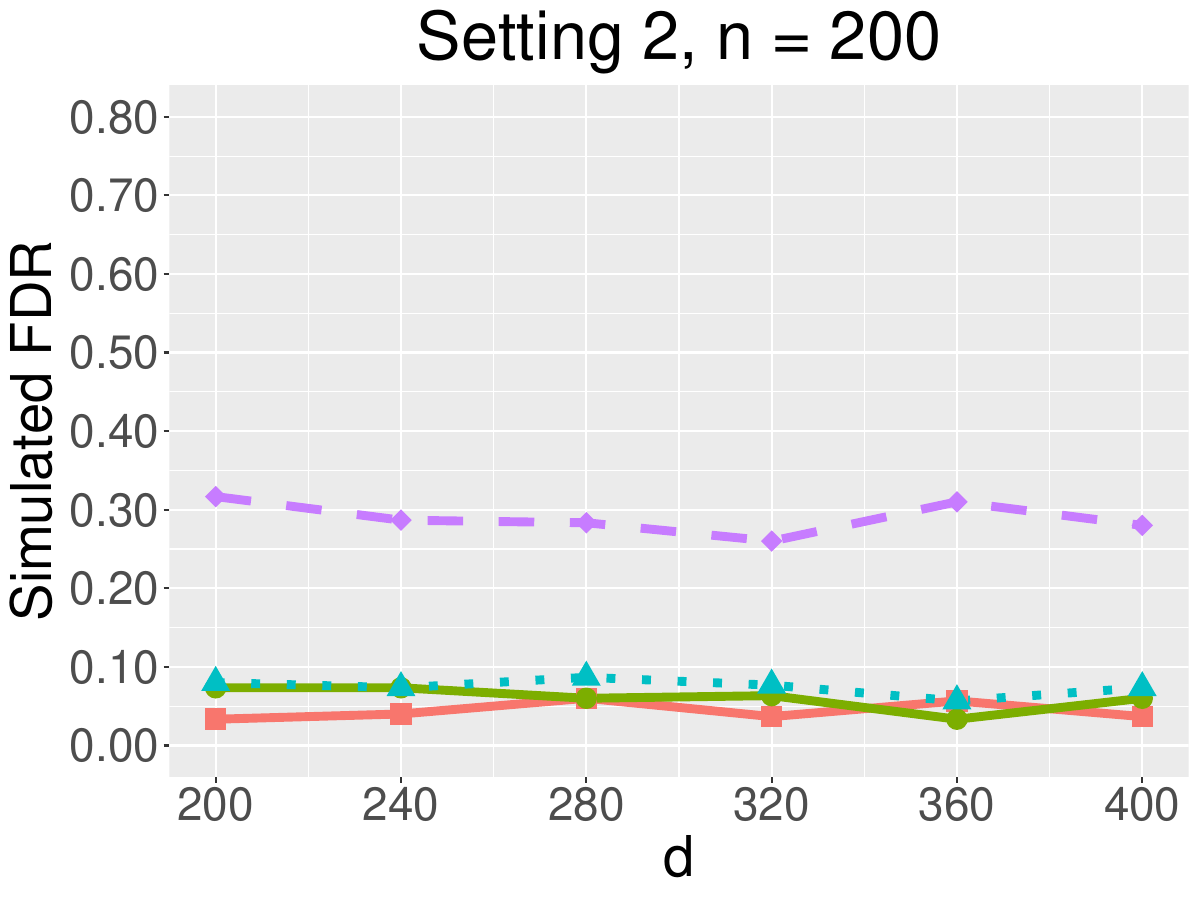}}\hspace{5pt}
	\subfloat{\includegraphics[width=.27\columnwidth]{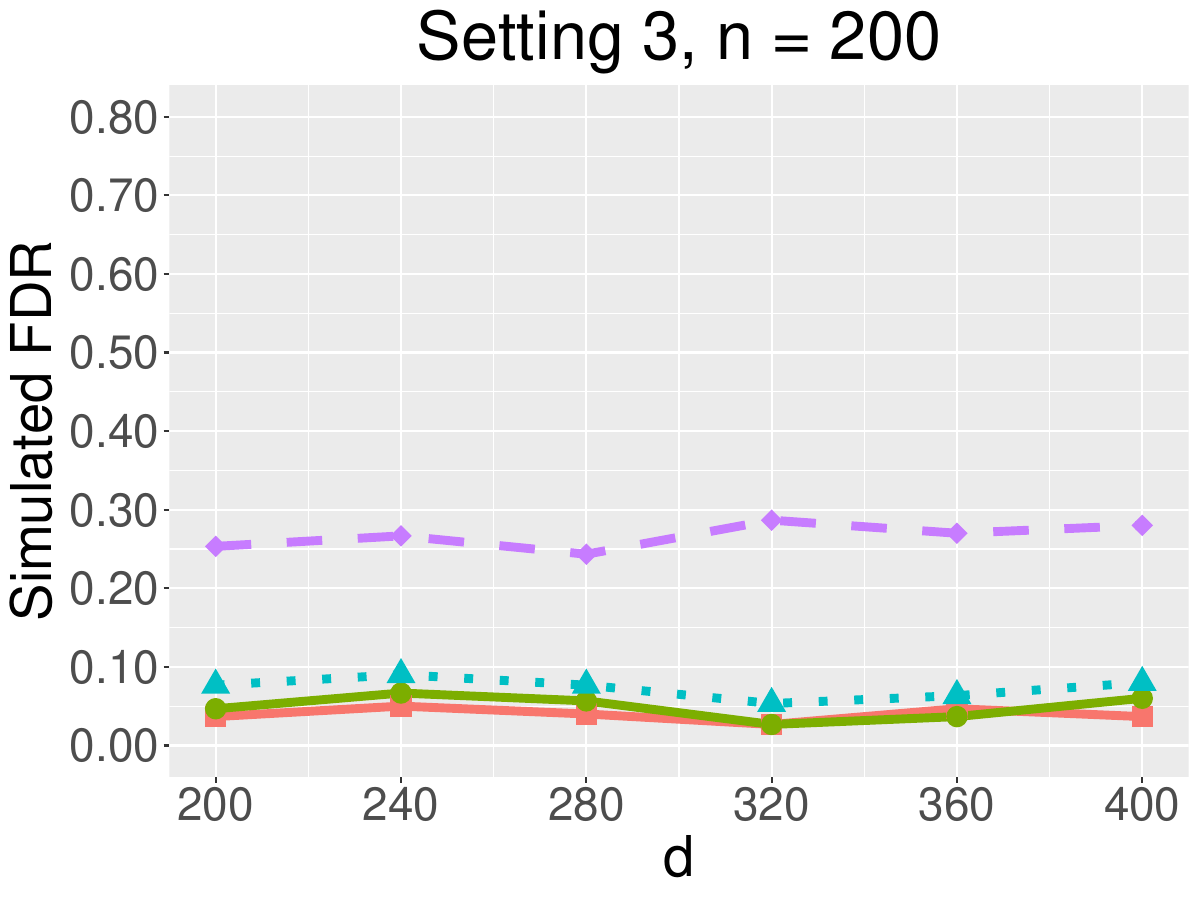}}\\
	\subfloat{\includegraphics[width=.27\columnwidth]{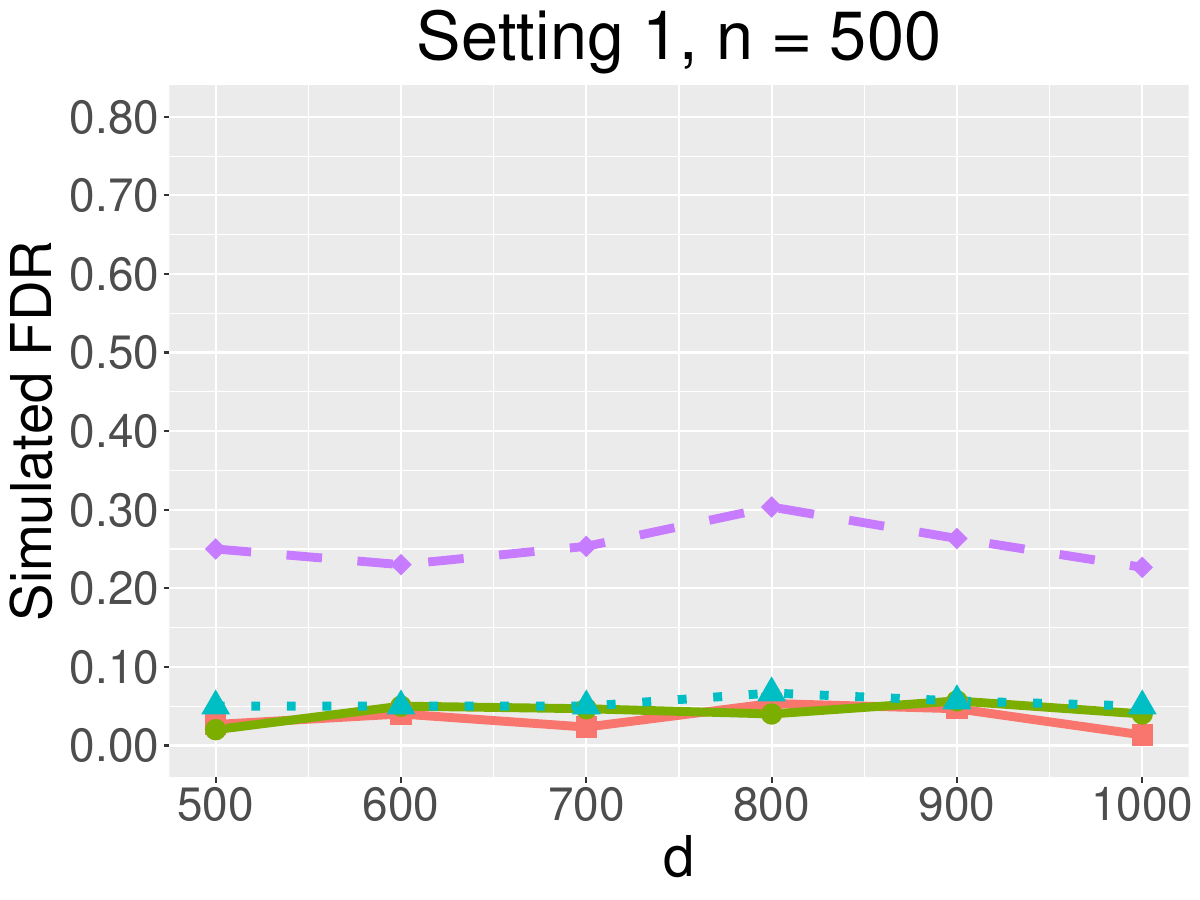}}\hspace{5pt}
    \subfloat{\includegraphics[width=.27\columnwidth]{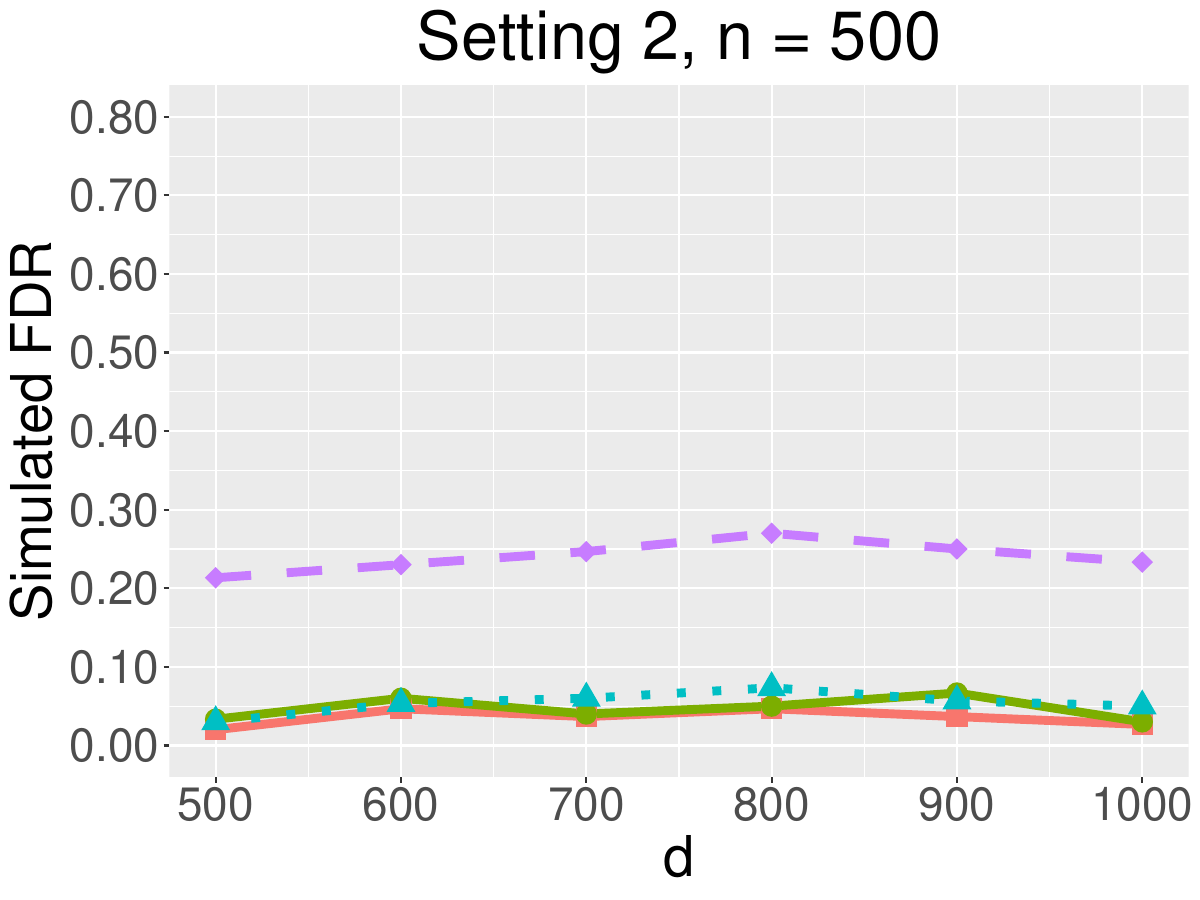}}\hspace{5pt}
    \subfloat{\includegraphics[width=.27\columnwidth]{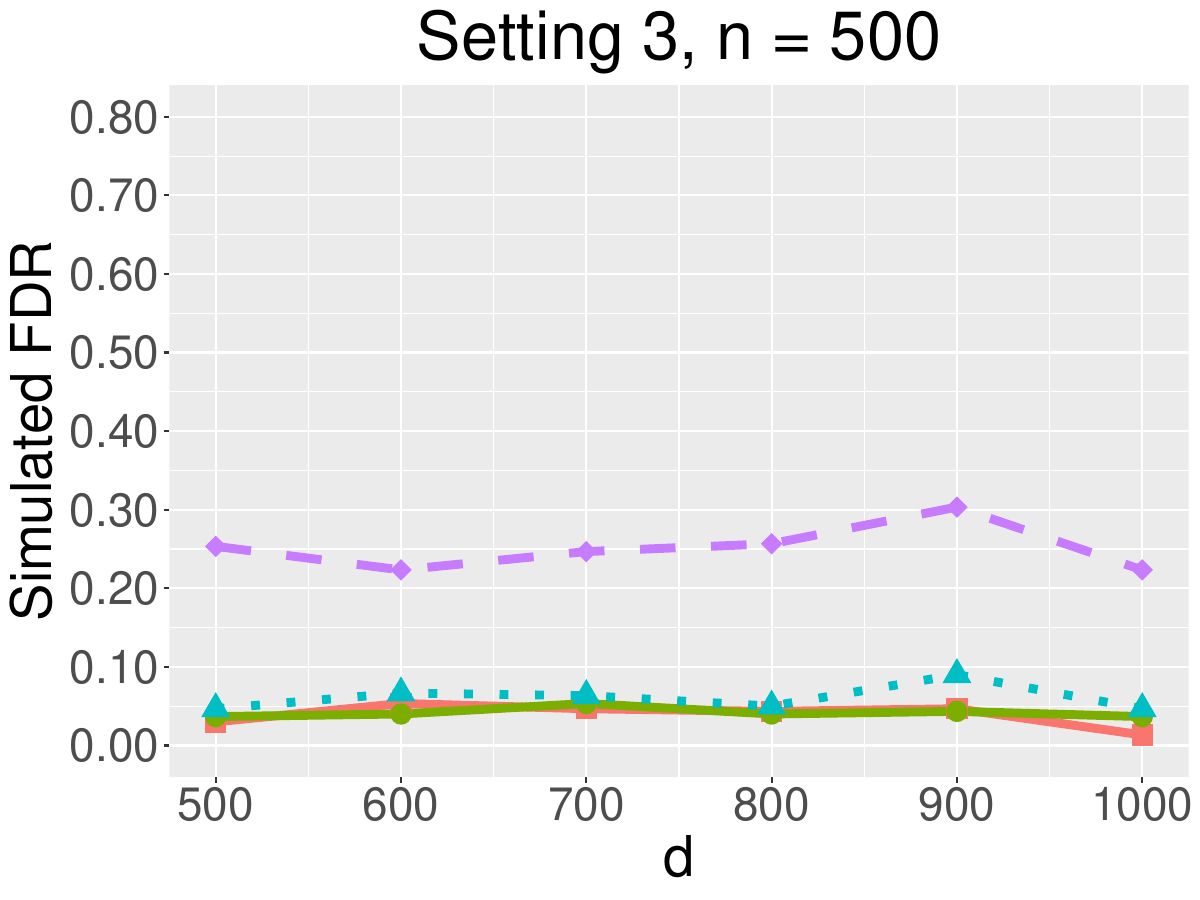}}
	\caption{\small Simulated FDR when all null hypotheses are true, for Setting 1, Setting 2, and Setting 3. The FDR level is $\alpha = 0.05$. The methods compared are Algorithm 1 (squares and red solid line), Algorithm 2 (circles and green solid line), Algorithm A2 (triangles and blue dotted line), and Algorithm A3 (diamonds and purple dashed line).}
    \label{FDR-control-origin-0.05}
\end{figure}

\begin{figure}[htbp!]
	\centering
	\subfloat{\includegraphics[width=.27\columnwidth]{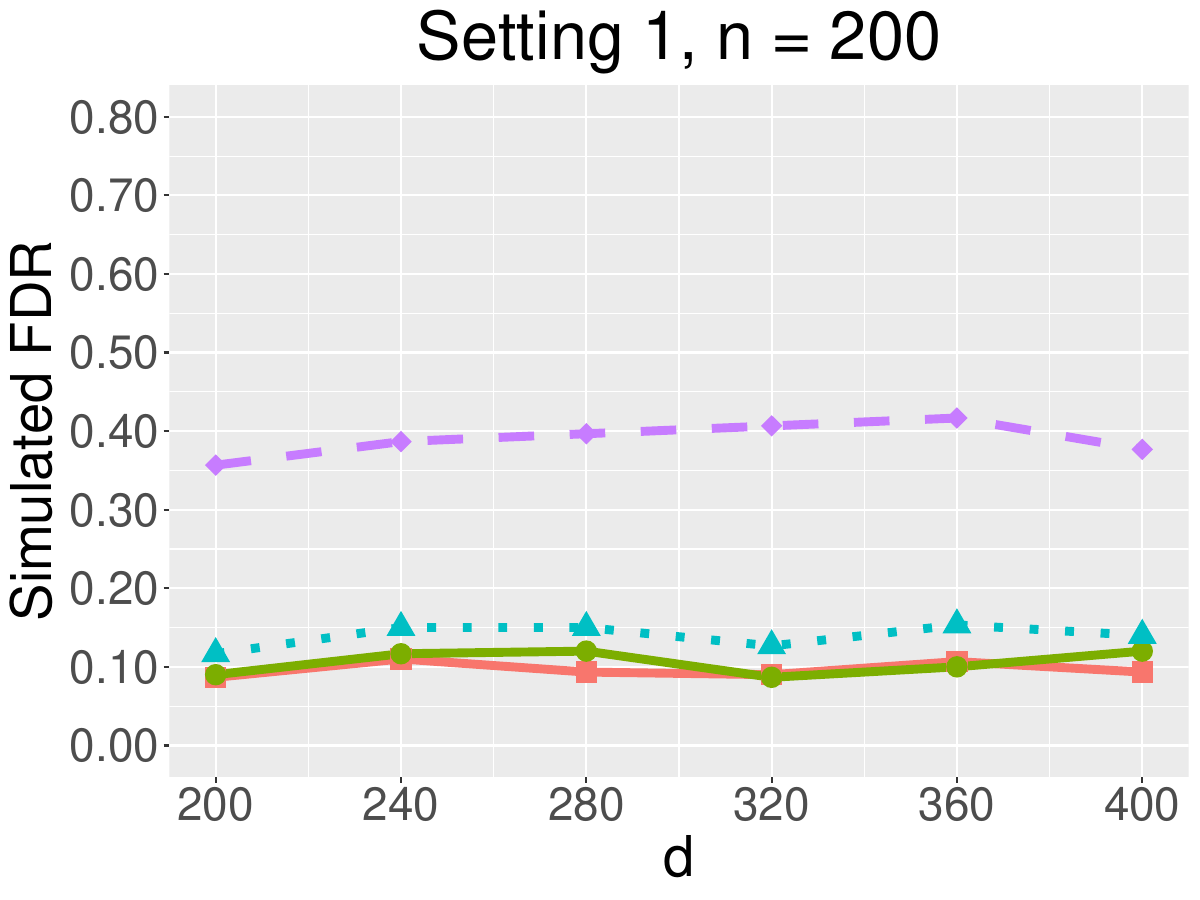}}\hspace{5pt}
	\subfloat{\includegraphics[width=.27\columnwidth]{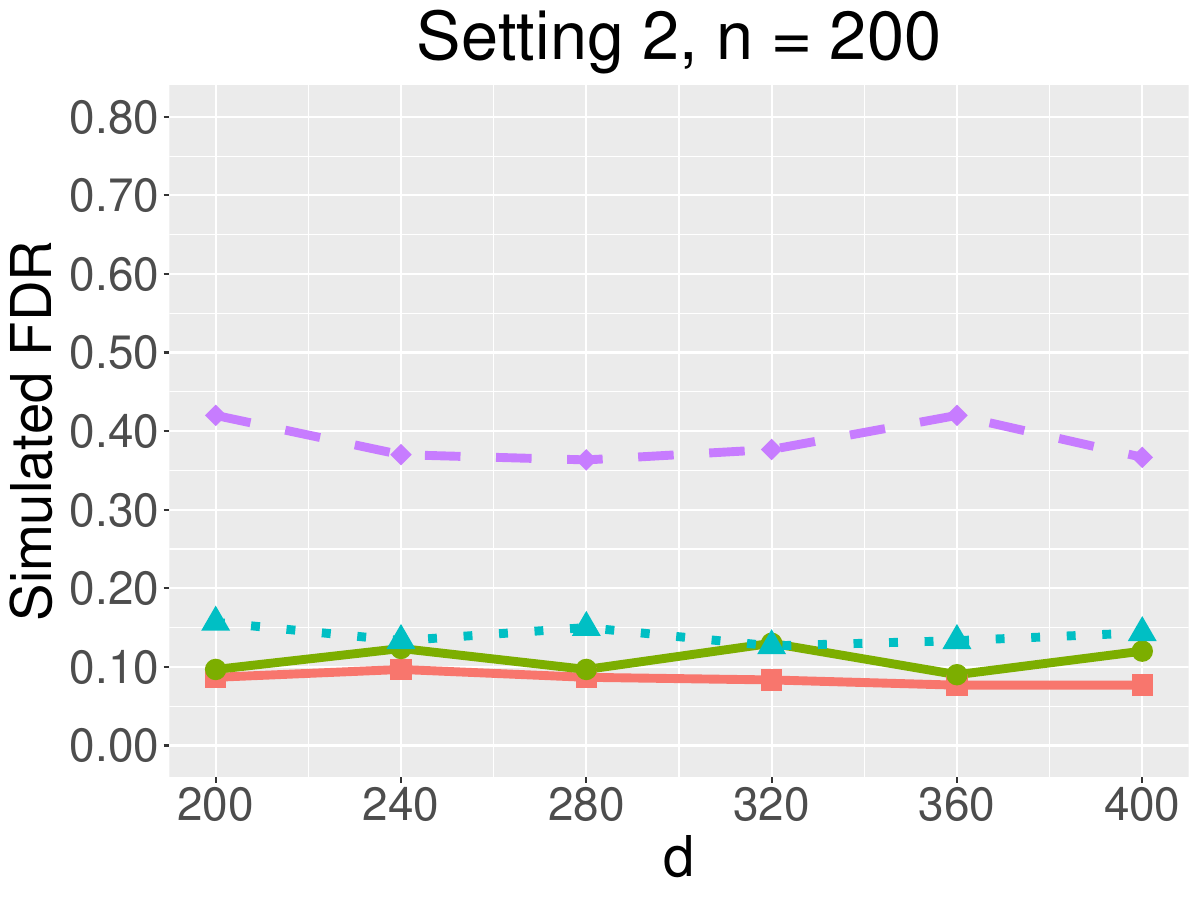}}\hspace{5pt}
	\subfloat{\includegraphics[width=.27\columnwidth]{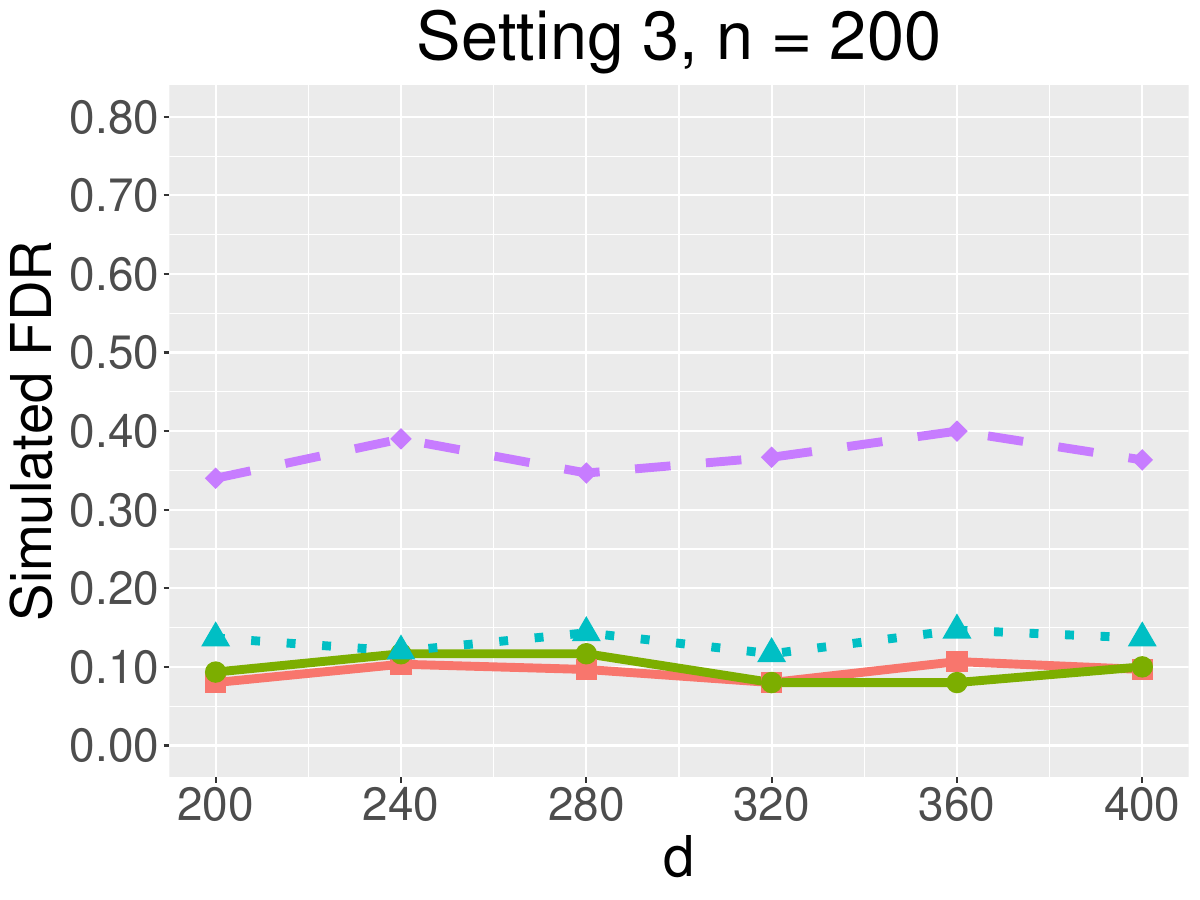}}\\
	\subfloat{\includegraphics[width=.27\columnwidth]{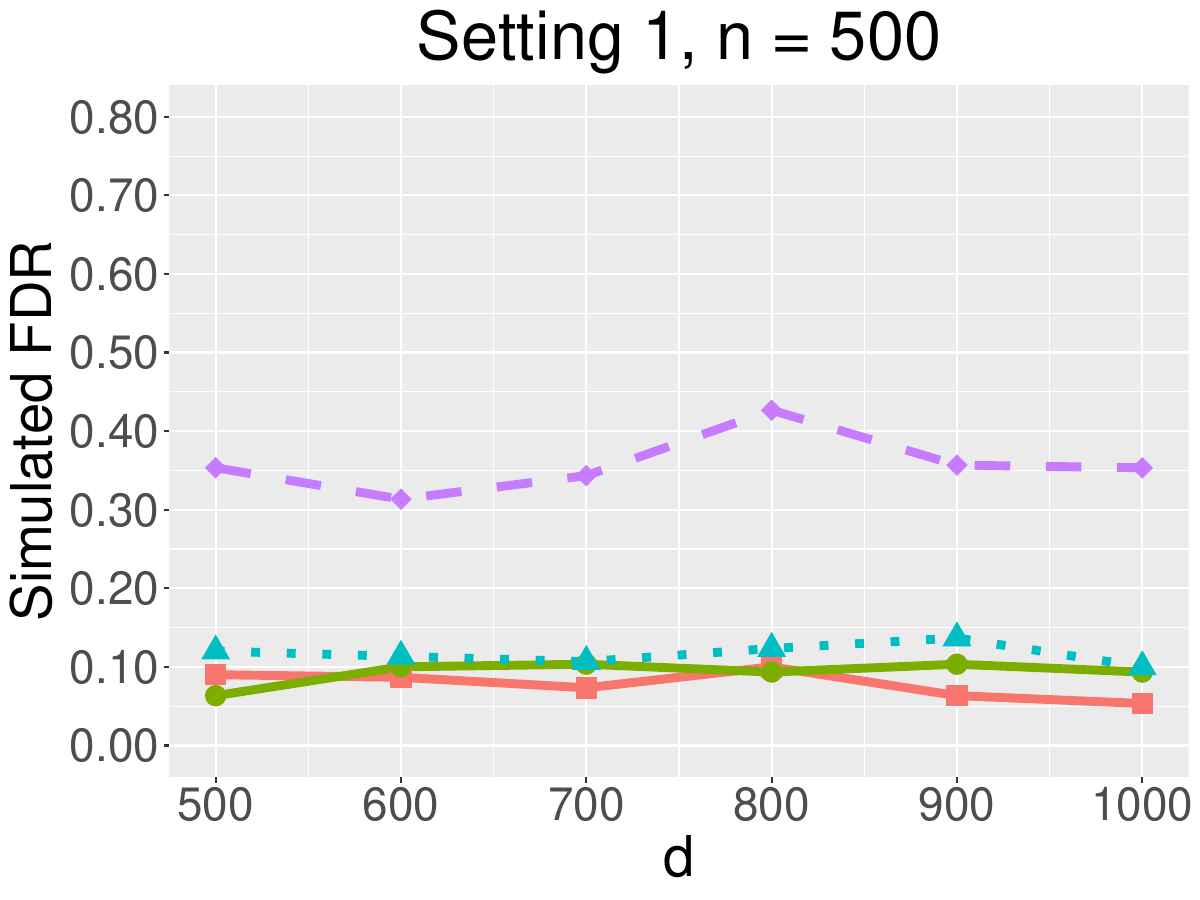}}\hspace{5pt}
    \subfloat{\includegraphics[width=.27\columnwidth]{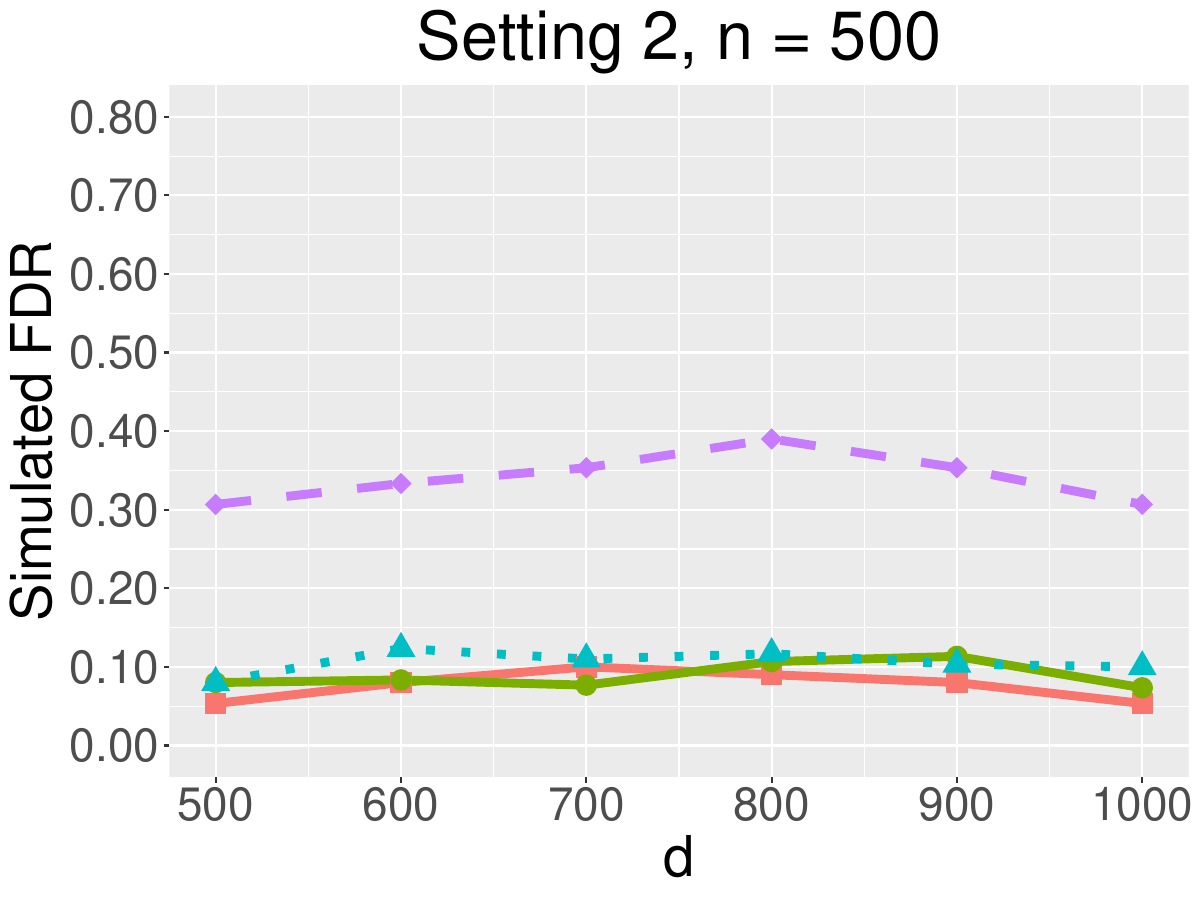}}\hspace{5pt}
    \subfloat{\includegraphics[width=.27\columnwidth]{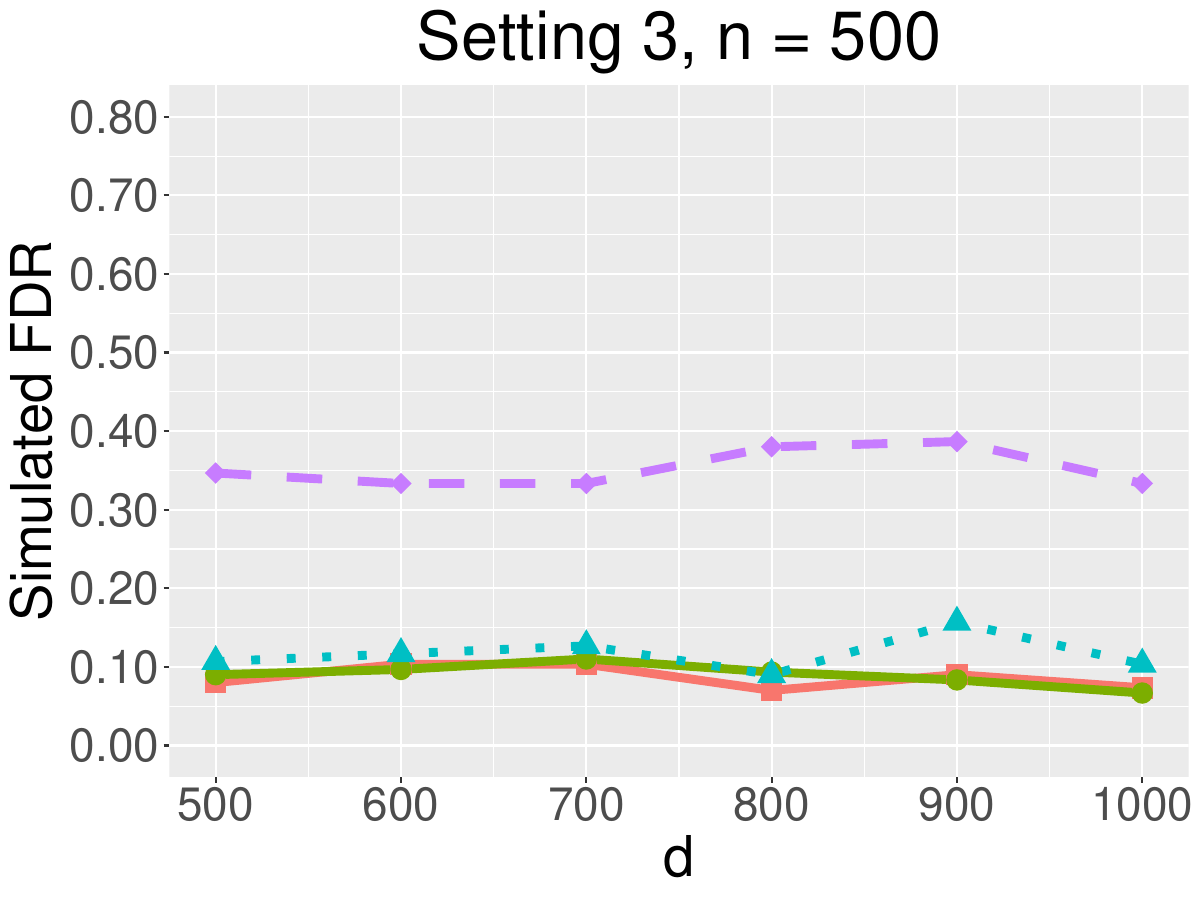}}
	\caption{\small Simulated FDR when all null hypotheses are true, for Setting 1, Setting 2, and Setting 3. The FDR level is $\alpha = 0.1$. The methods compared are Algorithm 1 (squares and red solid line), Algorithm 2 (circles and green solid line), Algorithm A2 (triangles and blue dotted line), and Algorithm A3 (diamonds and purple dashed line).}
    \label{FDR-control-origin-0.1}
\end{figure}

\begin{figure}[htbp!]
	\centering
	\subfloat{\includegraphics[width=.27\columnwidth]{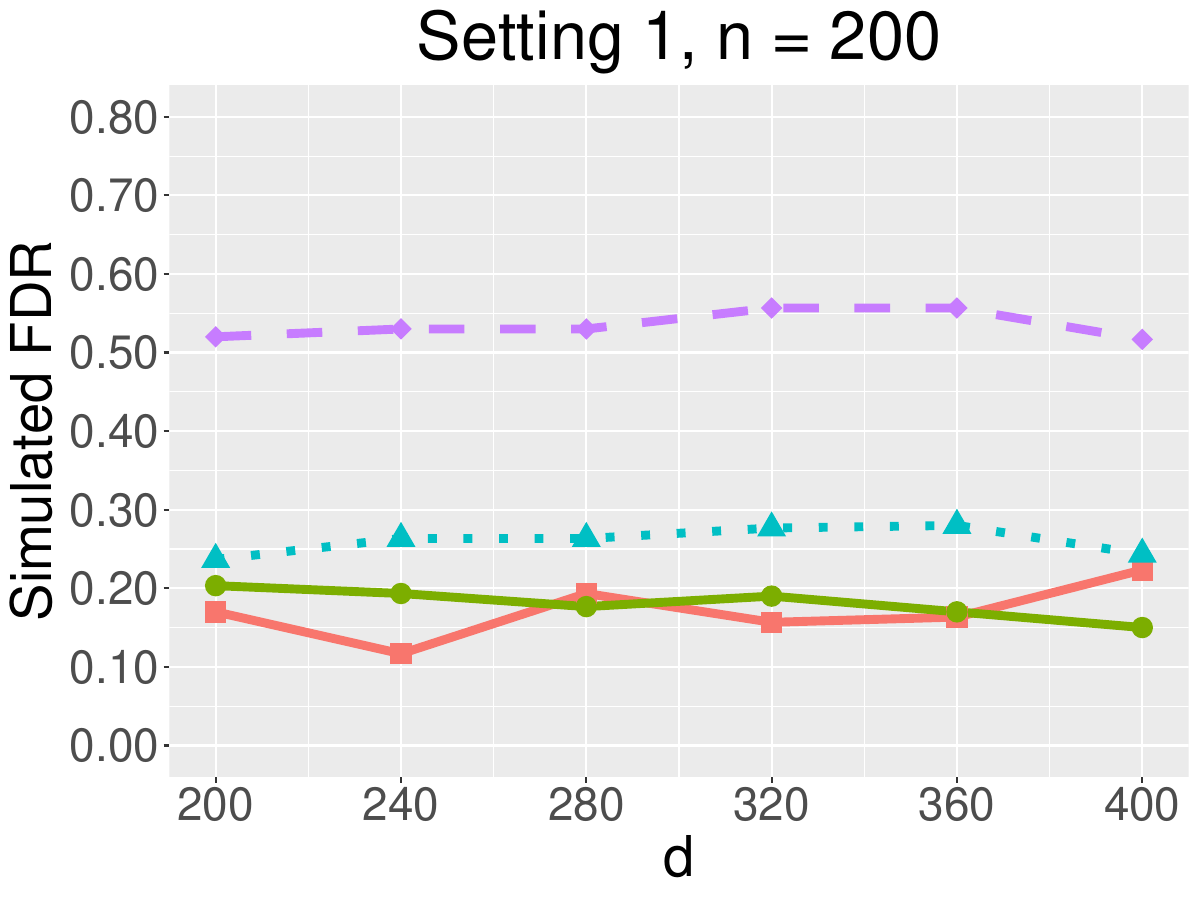}}\hspace{5pt}
	\subfloat{\includegraphics[width=.27\columnwidth]{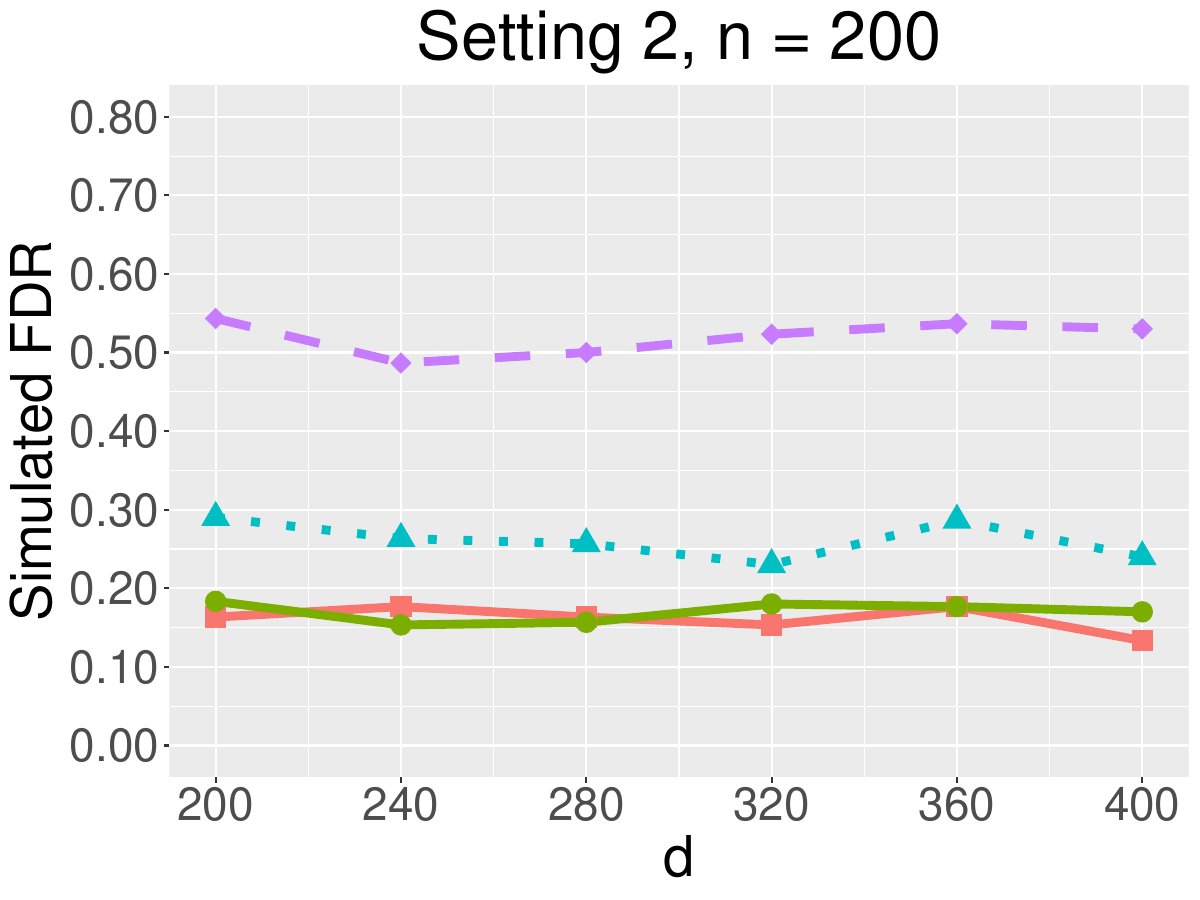}}\hspace{5pt}
	\subfloat{\includegraphics[width=.27\columnwidth]{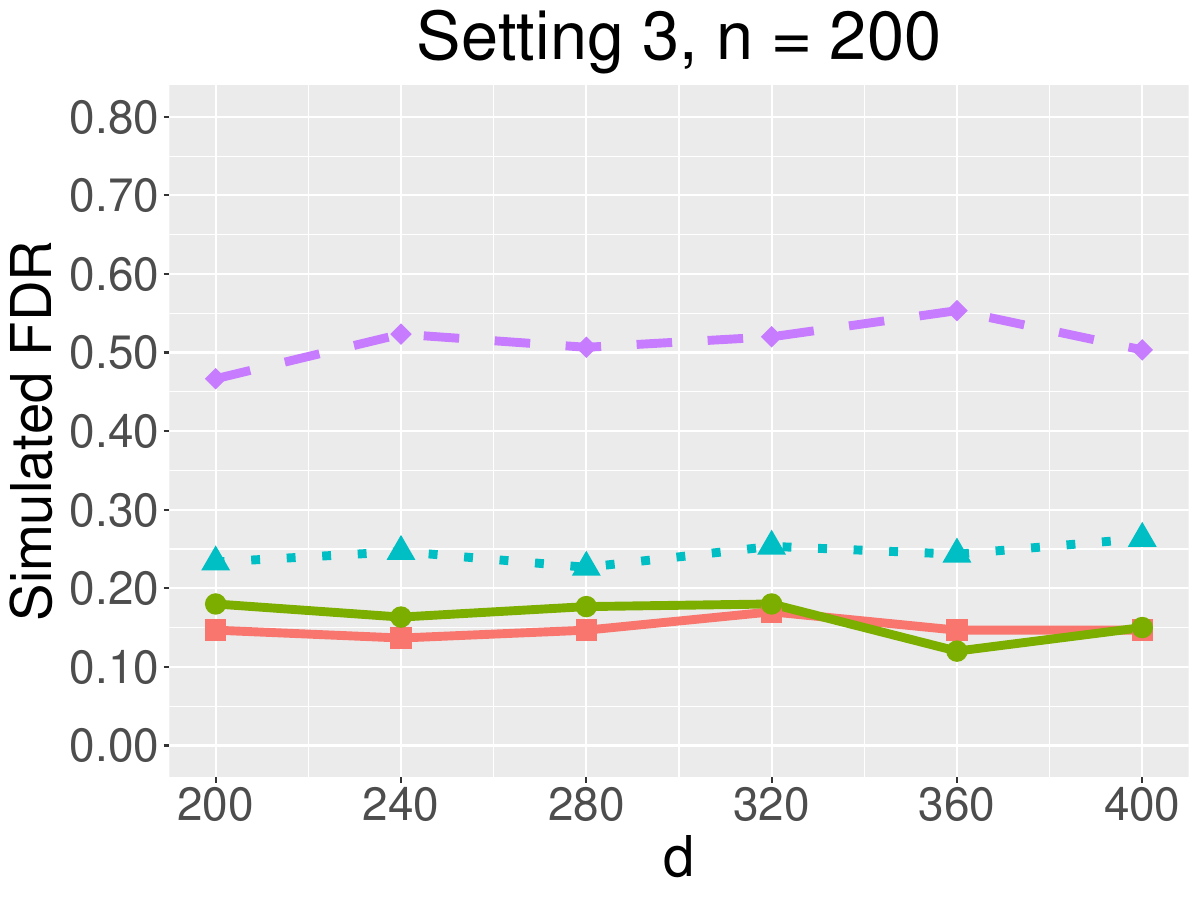}}\\
	\subfloat{\includegraphics[width=.27\columnwidth]{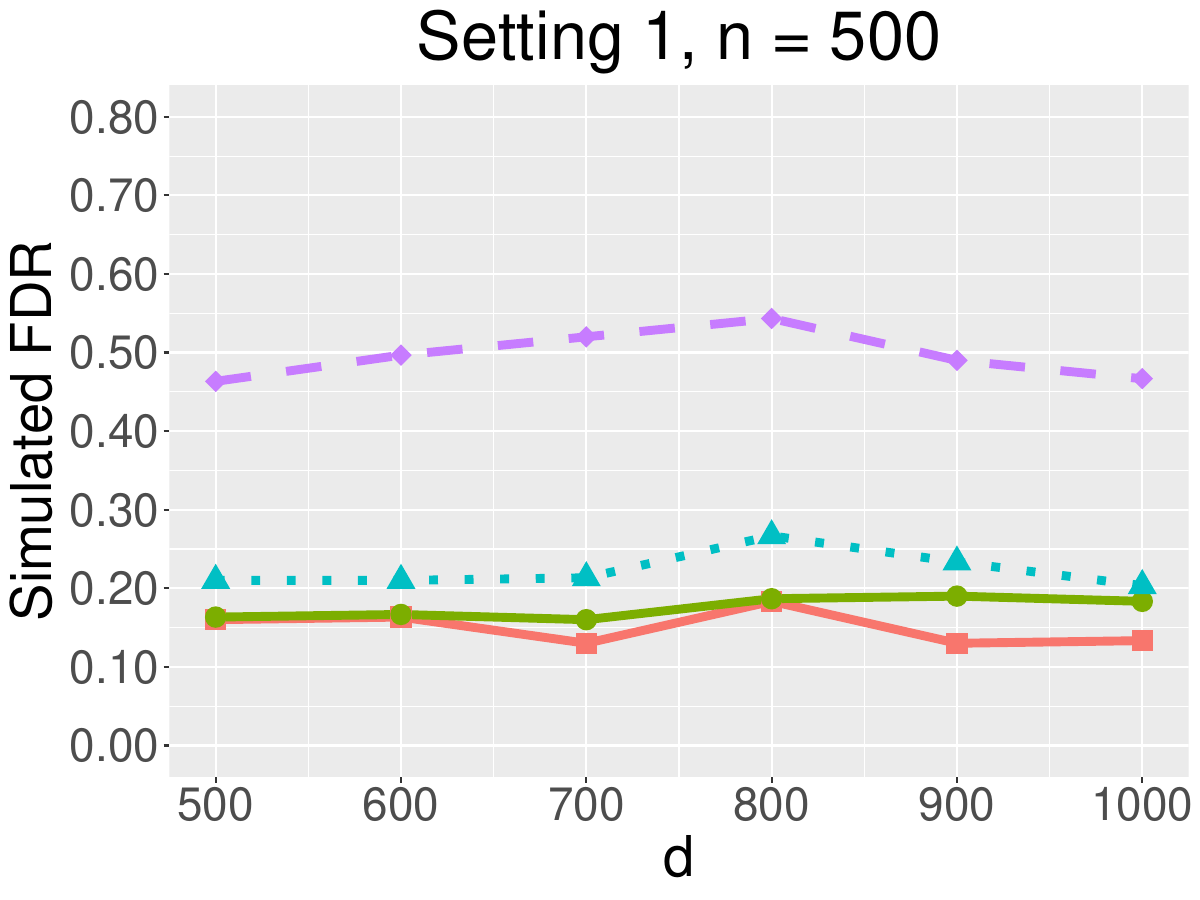}}\hspace{5pt}
    \subfloat{\includegraphics[width=.27\columnwidth]{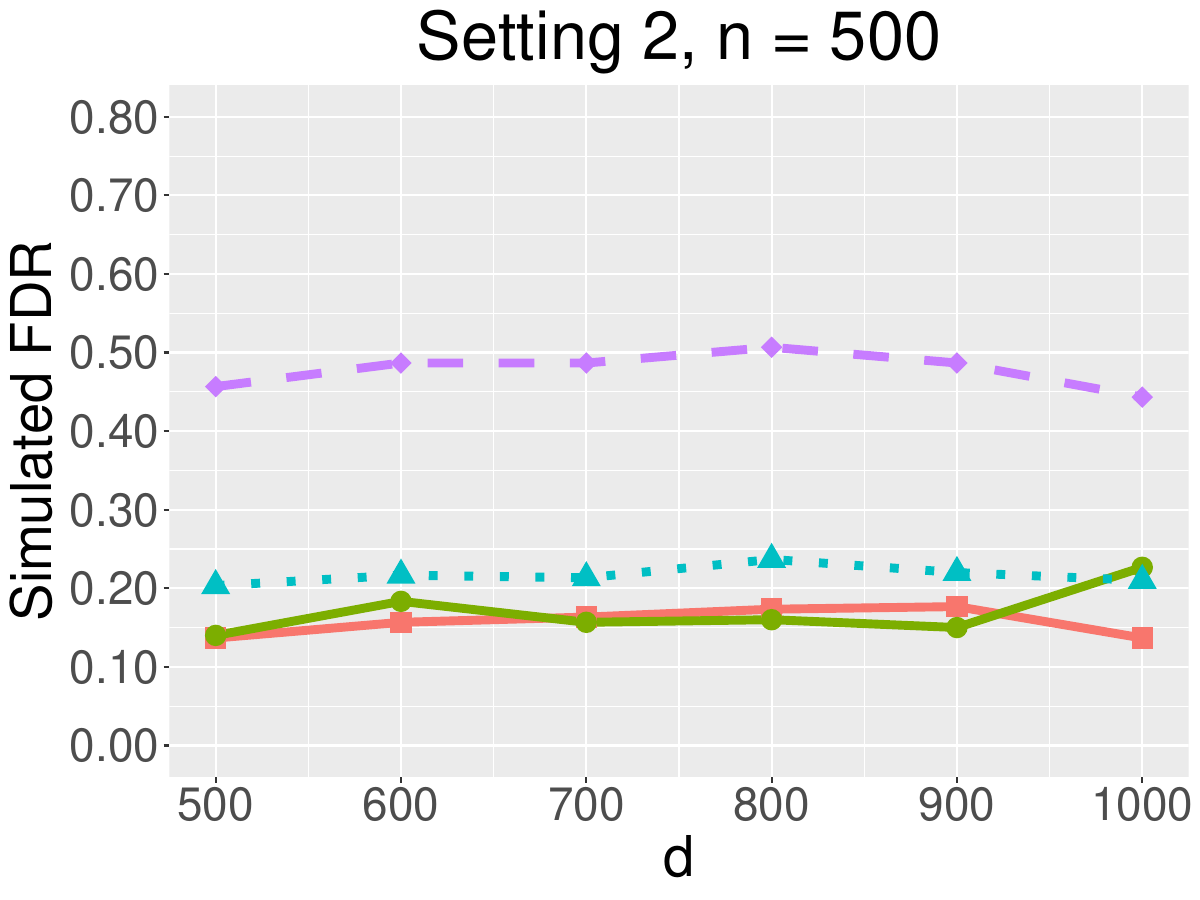}}\hspace{5pt}
    \subfloat{\includegraphics[width=.27\columnwidth]{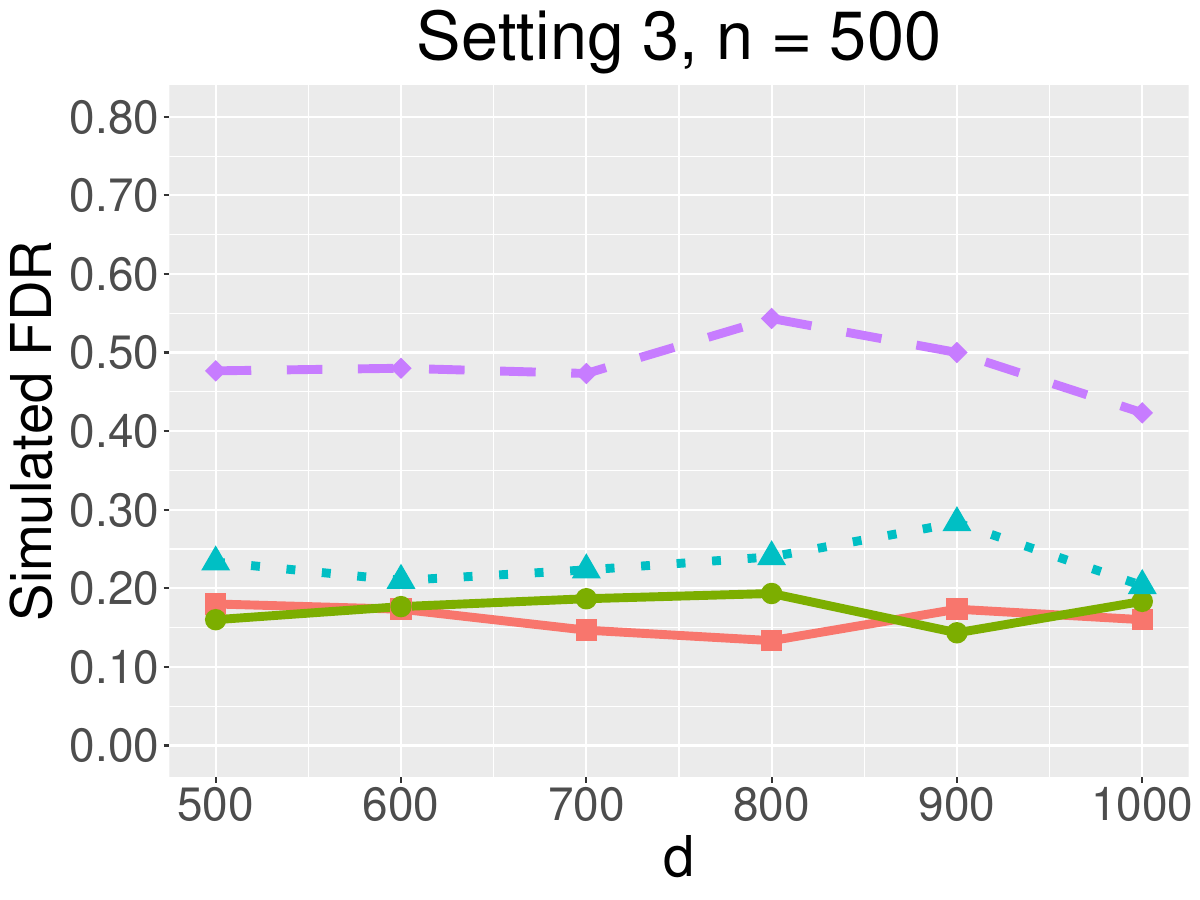}}
	\caption{\small Simulated FDR when all null hypotheses are true, for Setting 1, Setting 2, and Setting 3. The FDR level is $\alpha = 0.2$. The methods compared are Algorithm 1 (squares and red solid line), Algorithm 2 (circles and green solid line), Algorithm A2 (triangles and blue dotted line), and Algorithm A3 (diamonds and purple dashed line).}
    \label{FDR-control-origin-0.2}
\end{figure}

We first examine FDR control under the global null (signal strength equal to zero). The simulation setting is identical to that of Figure~\ref{FDR-control-0.1} in the main text. 
As shown in Figures~\ref{FDR-control-origin-0.05}-\ref{FDR-control-origin-0.2}, Algorithm~\ref{method_1_origin} controls the FDR, with its empirical FDR being comparable to that of our proposed methods. In contrast, Algorithm~\ref{method_2_origin} fails to effectively control the FDR.

\begin{figure}[htbp!]
	\centering
	\subfloat{\includegraphics[width=.27\columnwidth]{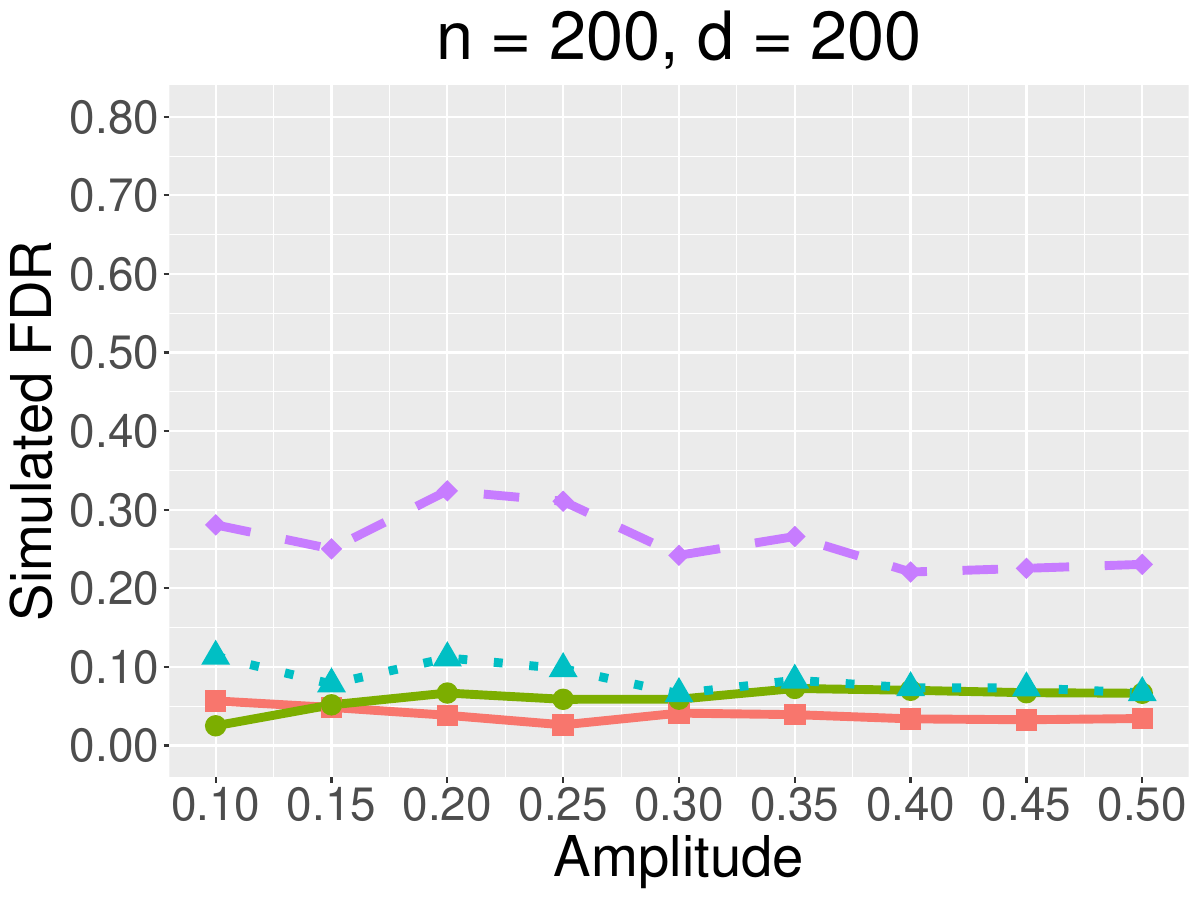}}\hspace{5pt}
	\subfloat{\includegraphics[width=.27\columnwidth]{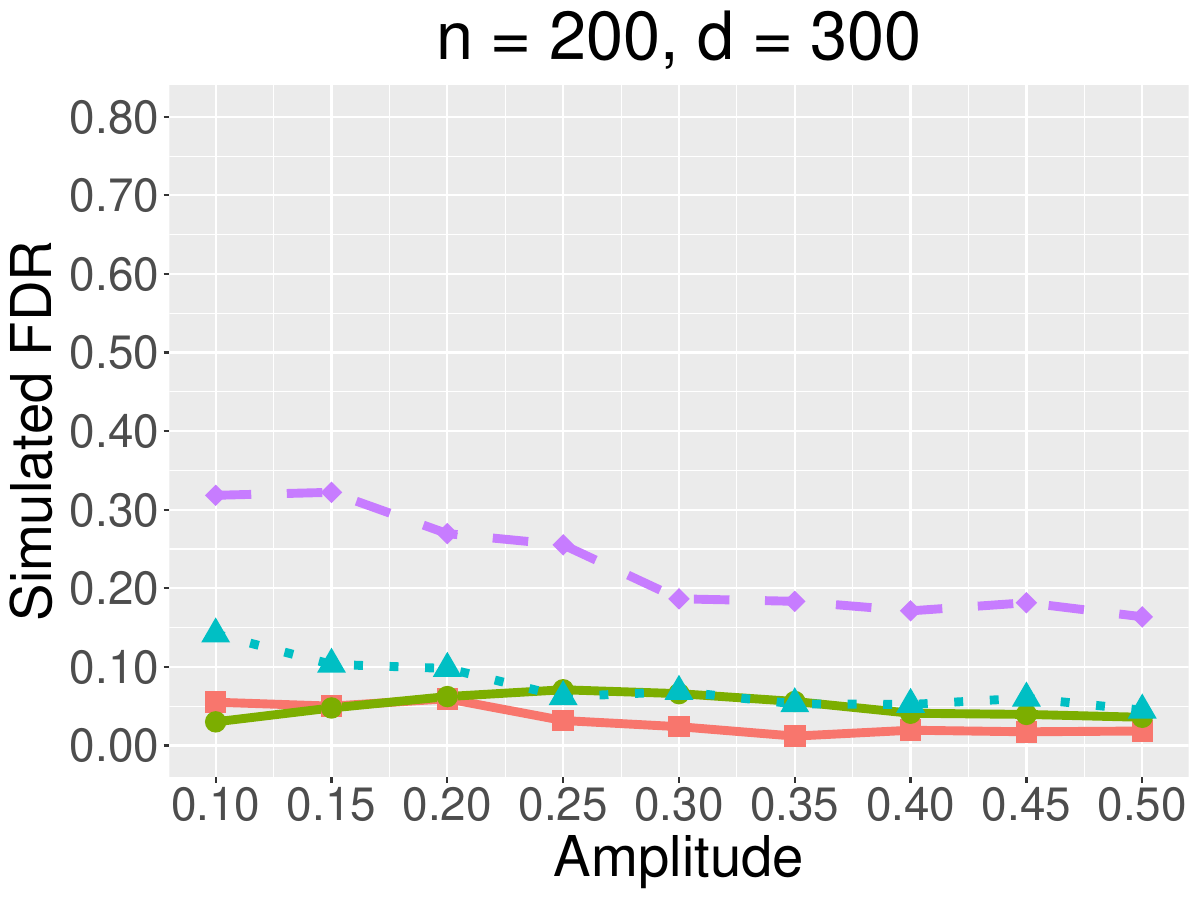}}\hspace{5pt}
	\subfloat{\includegraphics[width=.27\columnwidth]{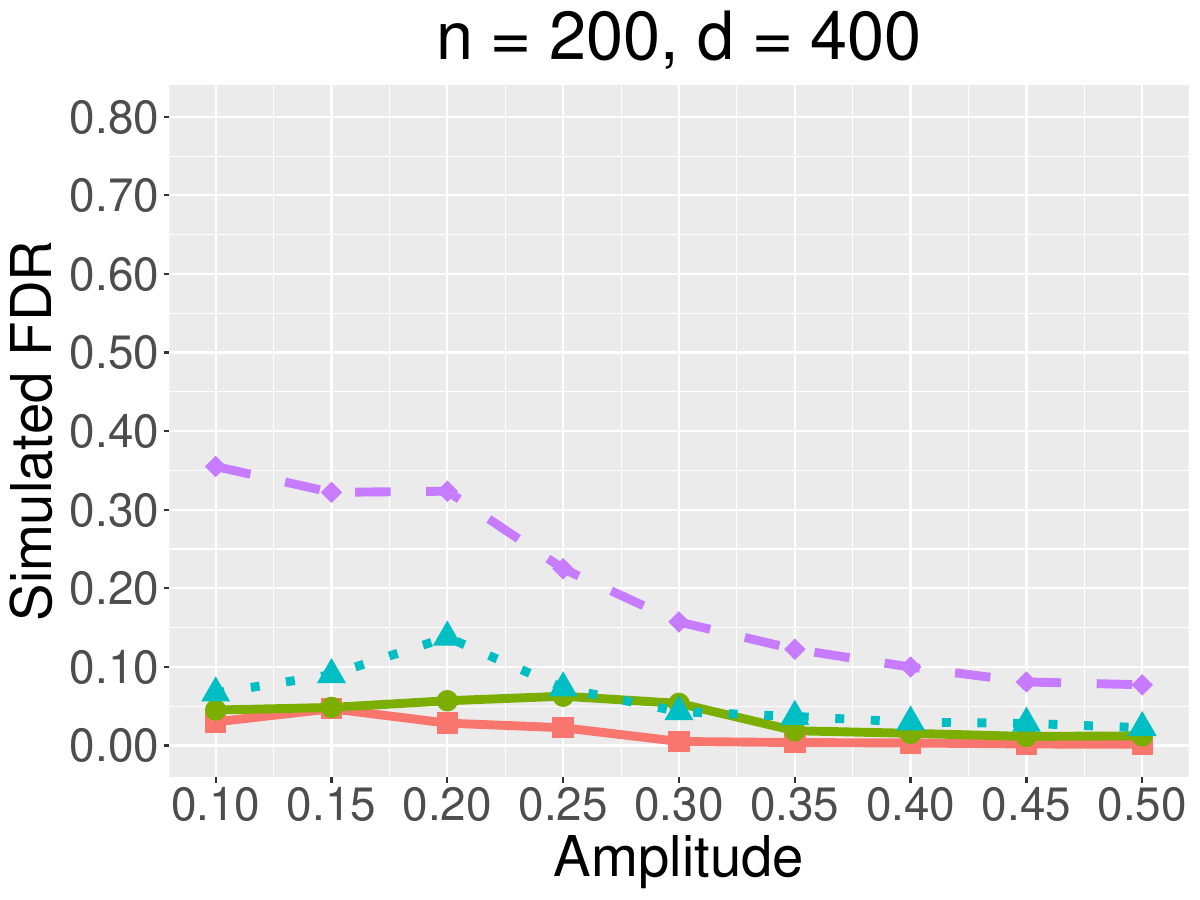}}\\
	\subfloat{\includegraphics[width=.27\columnwidth]{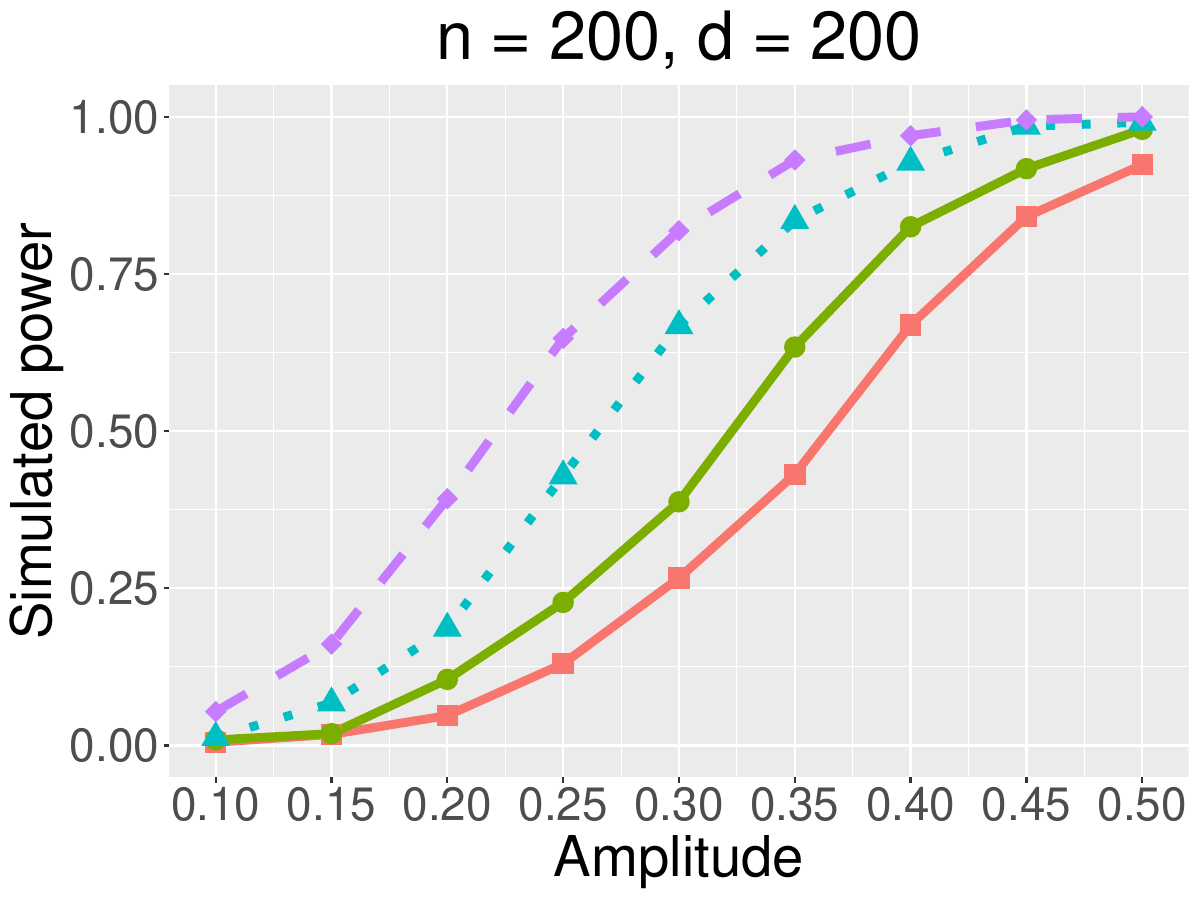}}\hspace{5pt}
    \subfloat{\includegraphics[width=.27\columnwidth]{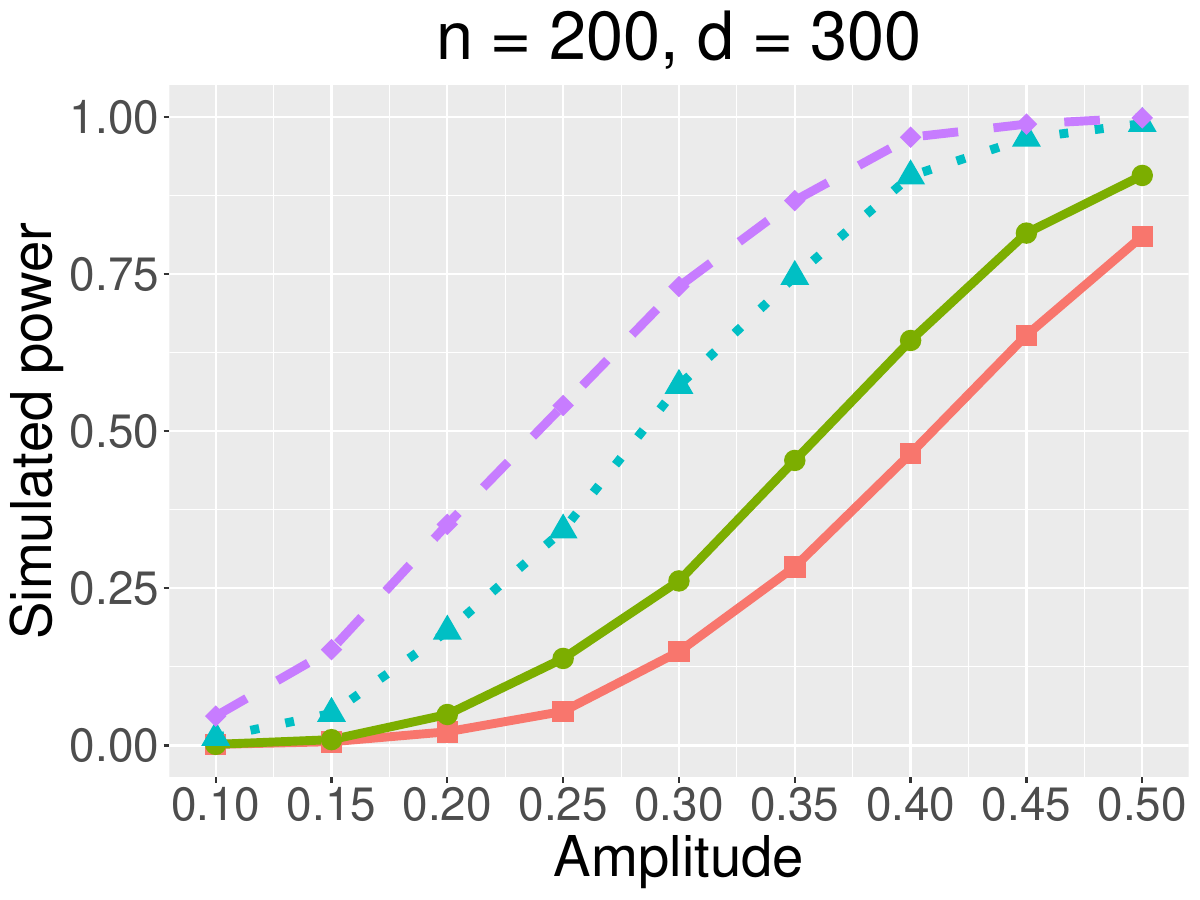}}\hspace{5pt}
    \subfloat{\includegraphics[width=.27\columnwidth]{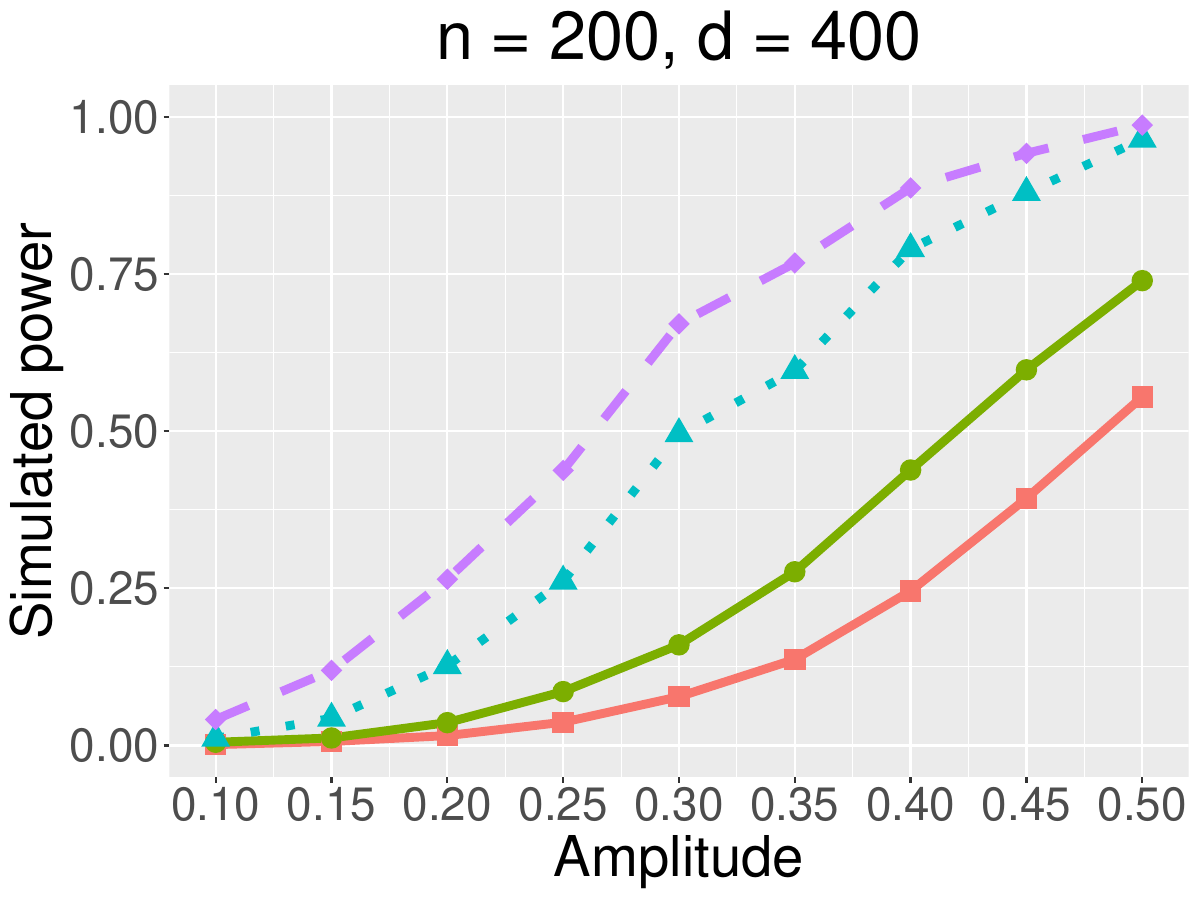}}
	\caption{\small Simulated FDR and power for different combinations of $(n,d)$. The rows of the design matrix were generated from Setting 2. The sparsity level is $k = 0.04d$ and the FDR level is $\alpha = 0.05$. The methods compared are Algorithm 1 (squares and red solid line), Algorithm 2 (circles and green solid line), Algorithm A2 (triangles and blue dotted line), and Algorithm A3 (diamonds and purple dashed line).}
    \label{Power-origin-0.05}
\end{figure}

\begin{figure}[htbp!]
	\centering
	\subfloat{\includegraphics[width=.27\columnwidth]{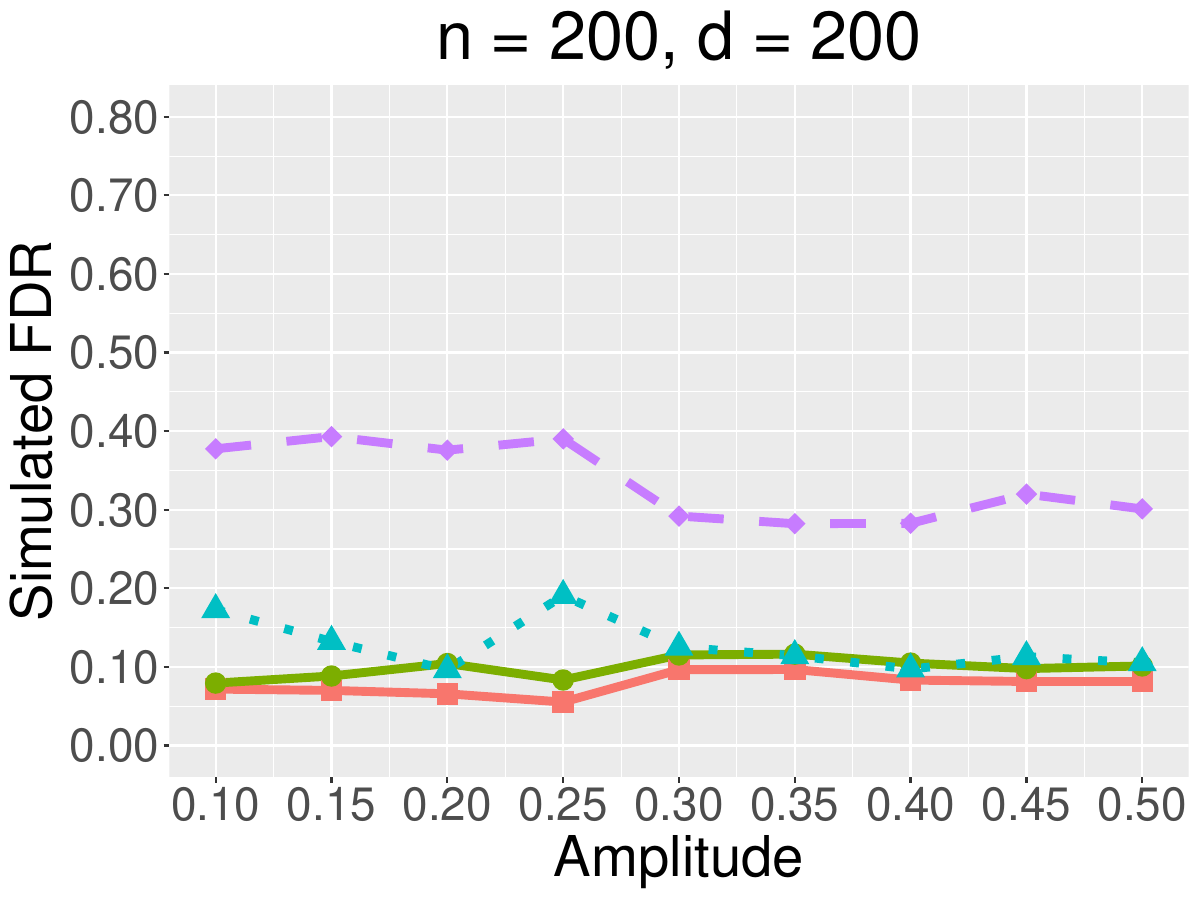}}\hspace{5pt}
	\subfloat{\includegraphics[width=.27\columnwidth]{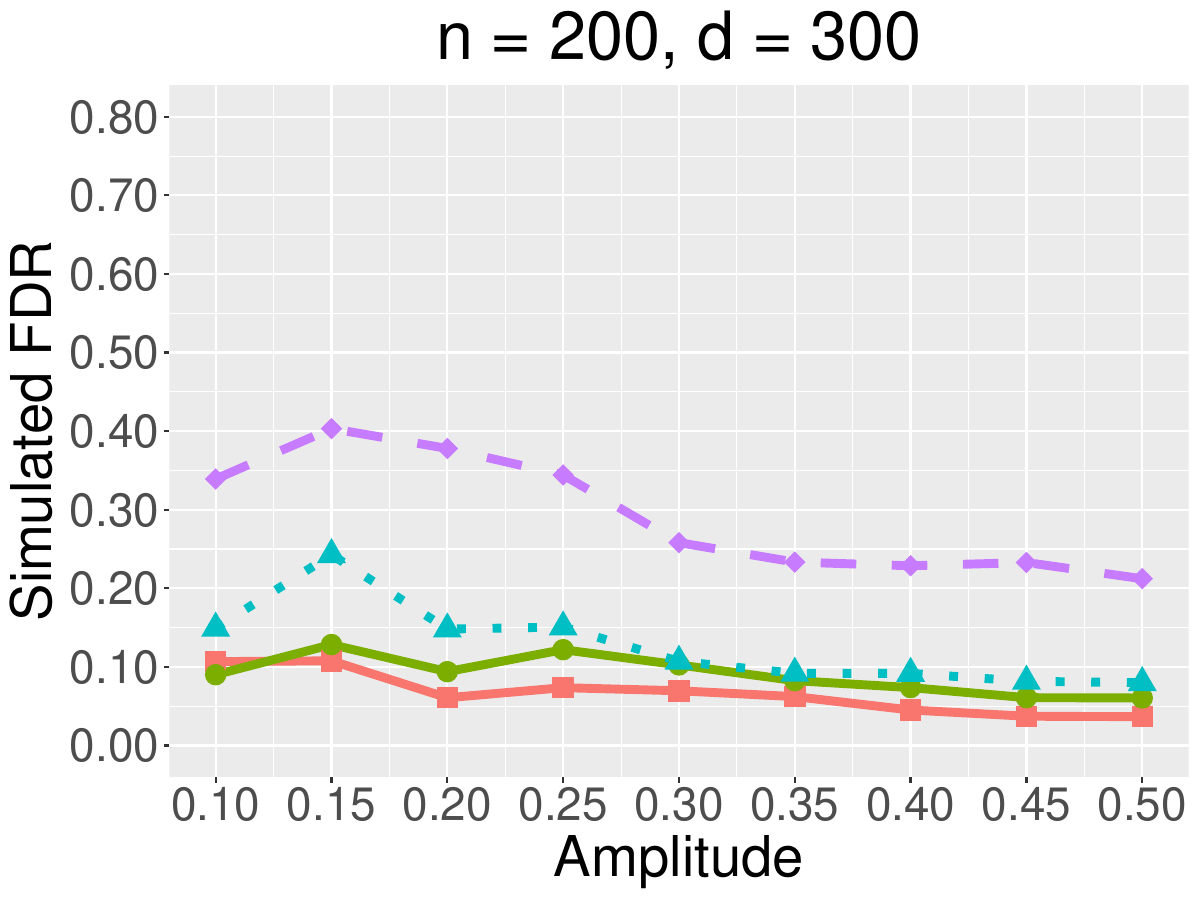}}\hspace{5pt}
	\subfloat{\includegraphics[width=.27\columnwidth]{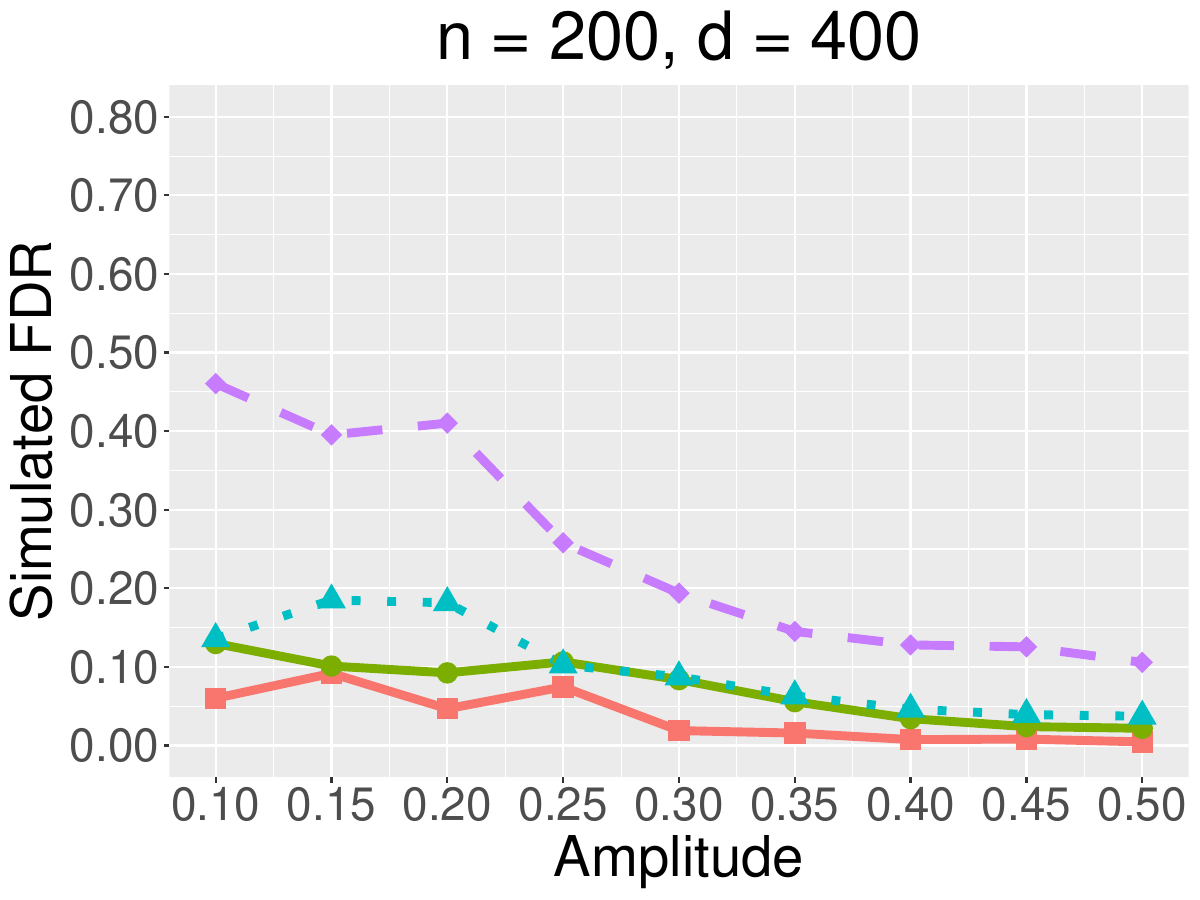}}\\
	\subfloat{\includegraphics[width=.27\columnwidth]{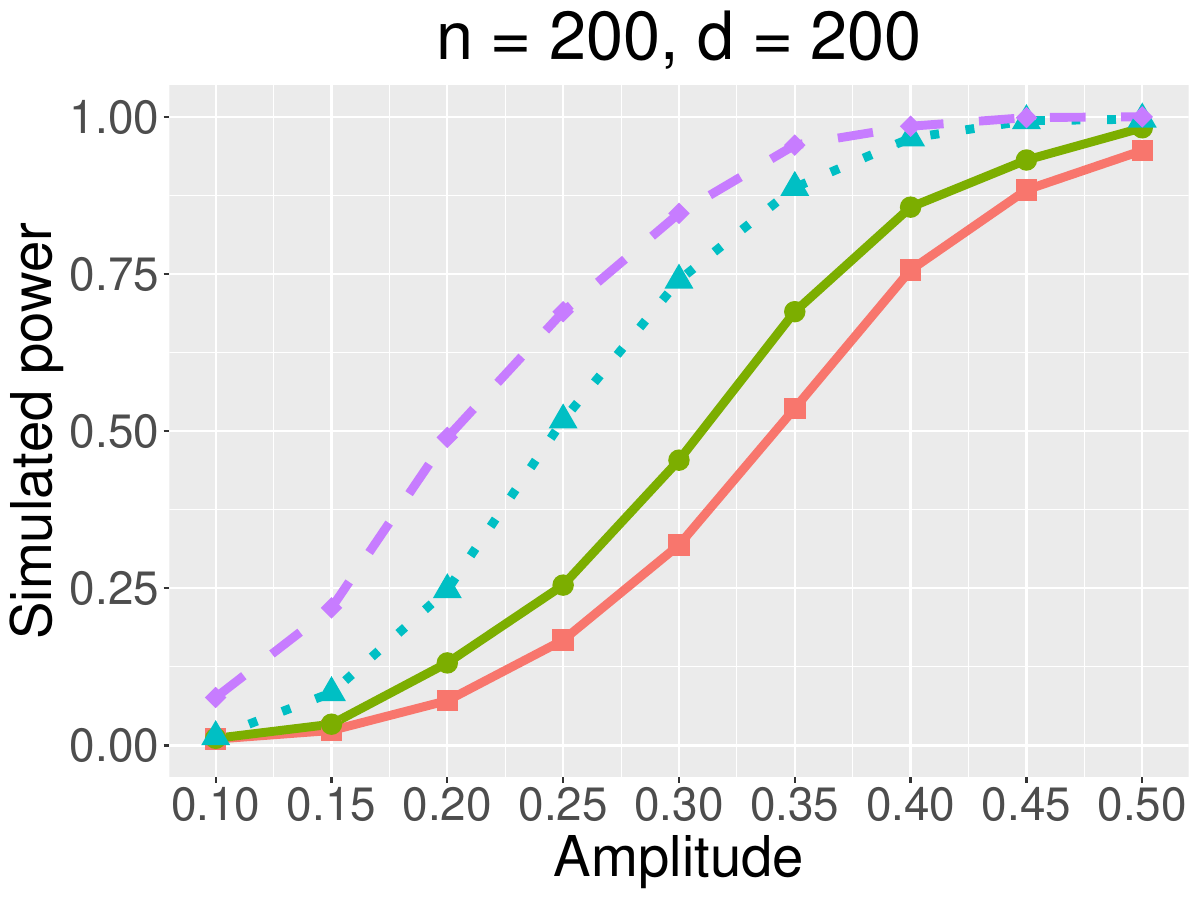}}\hspace{5pt}
    \subfloat{\includegraphics[width=.27\columnwidth]{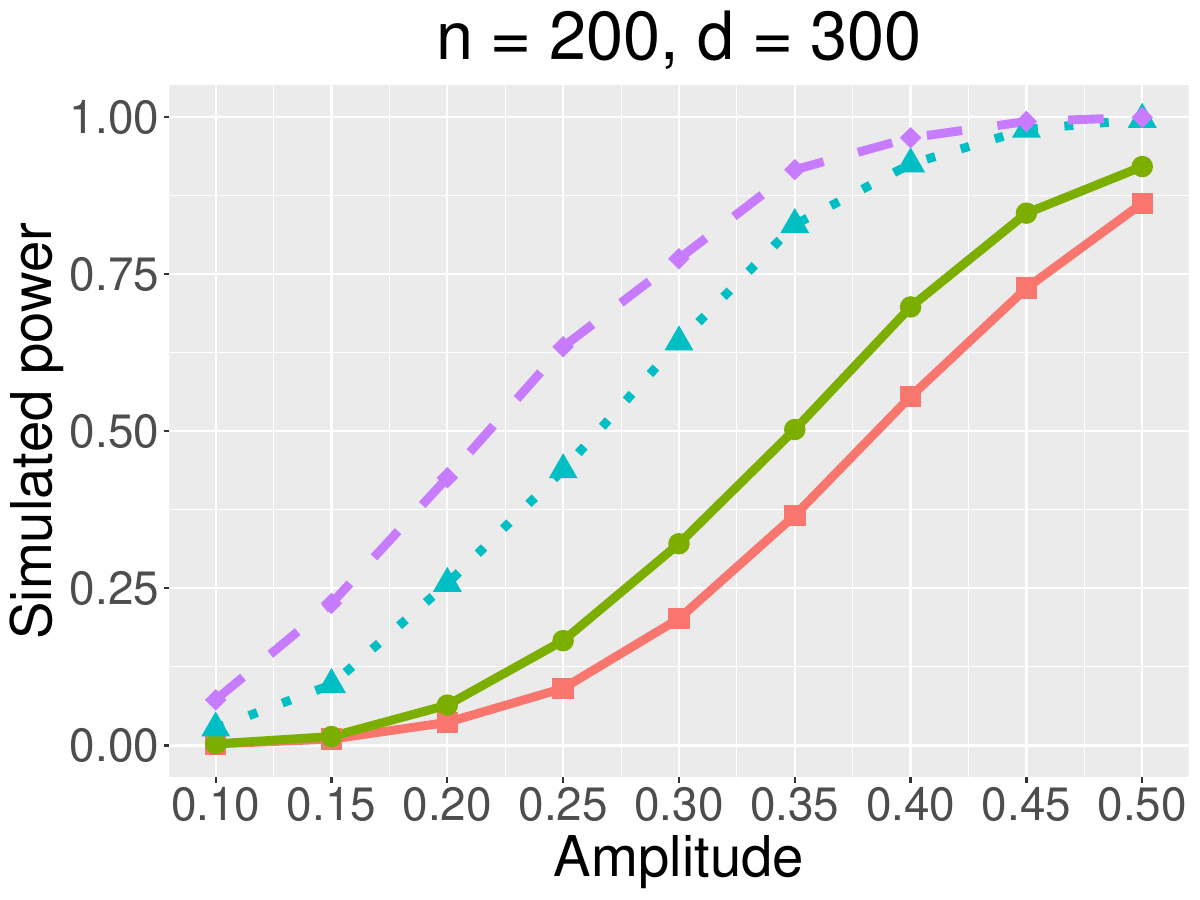}}\hspace{5pt}
    \subfloat{\includegraphics[width=.27\columnwidth]{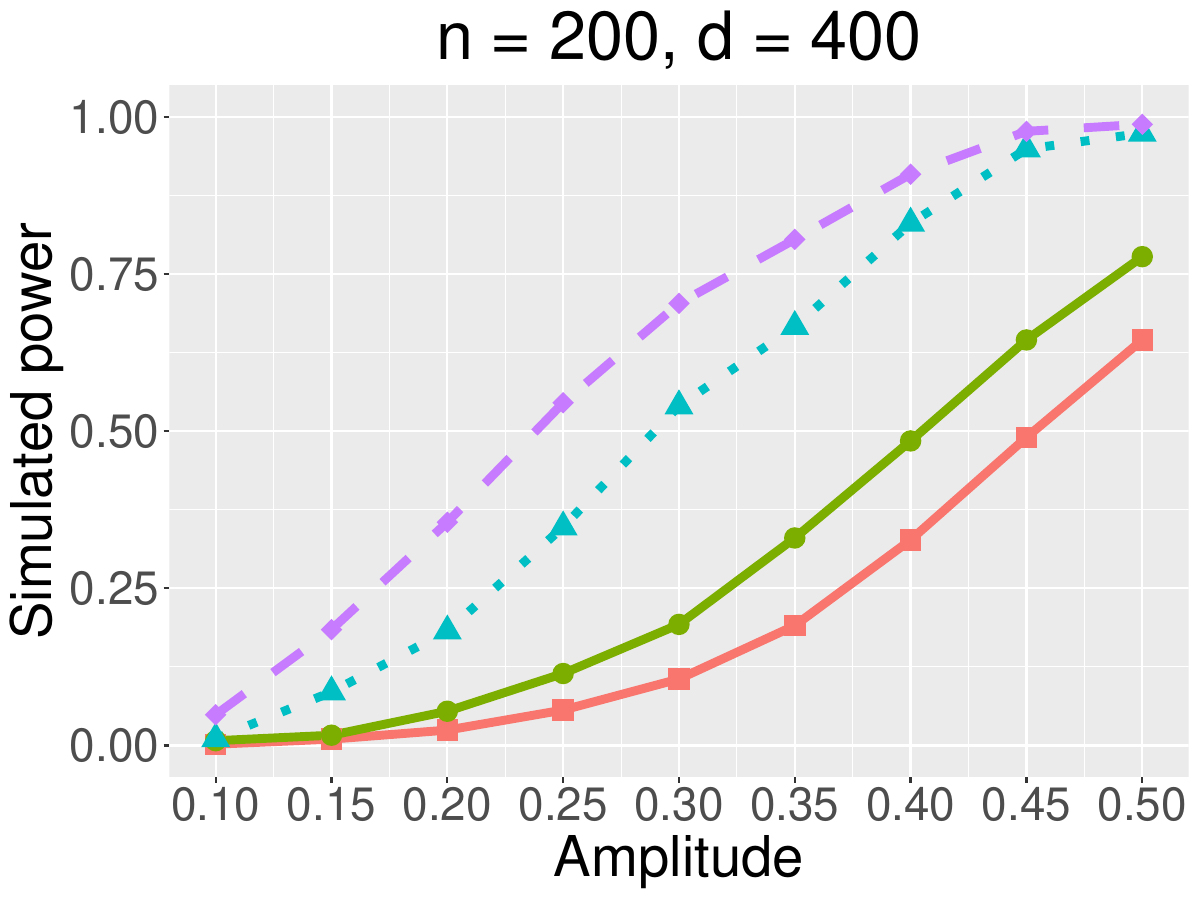}}
	\caption{\small Simulated FDR and power for different combinations of $(n,d)$. The rows of the design matrix were generated from Setting 2. The sparsity level is $k = 0.04d$ and the FDR level is $\alpha = 0.1$. The methods compared are Algorithm 1 (squares and red solid line), Algorithm 2 (circles and green solid line), Algorithm A2 (triangles and blue dotted line), and Algorithm A3 (diamonds and purple dashed line).}
    \label{Power-origin-0.1}
\end{figure}

\begin{figure}[htbp!]
	\centering
	\subfloat{\includegraphics[width=.27\columnwidth]{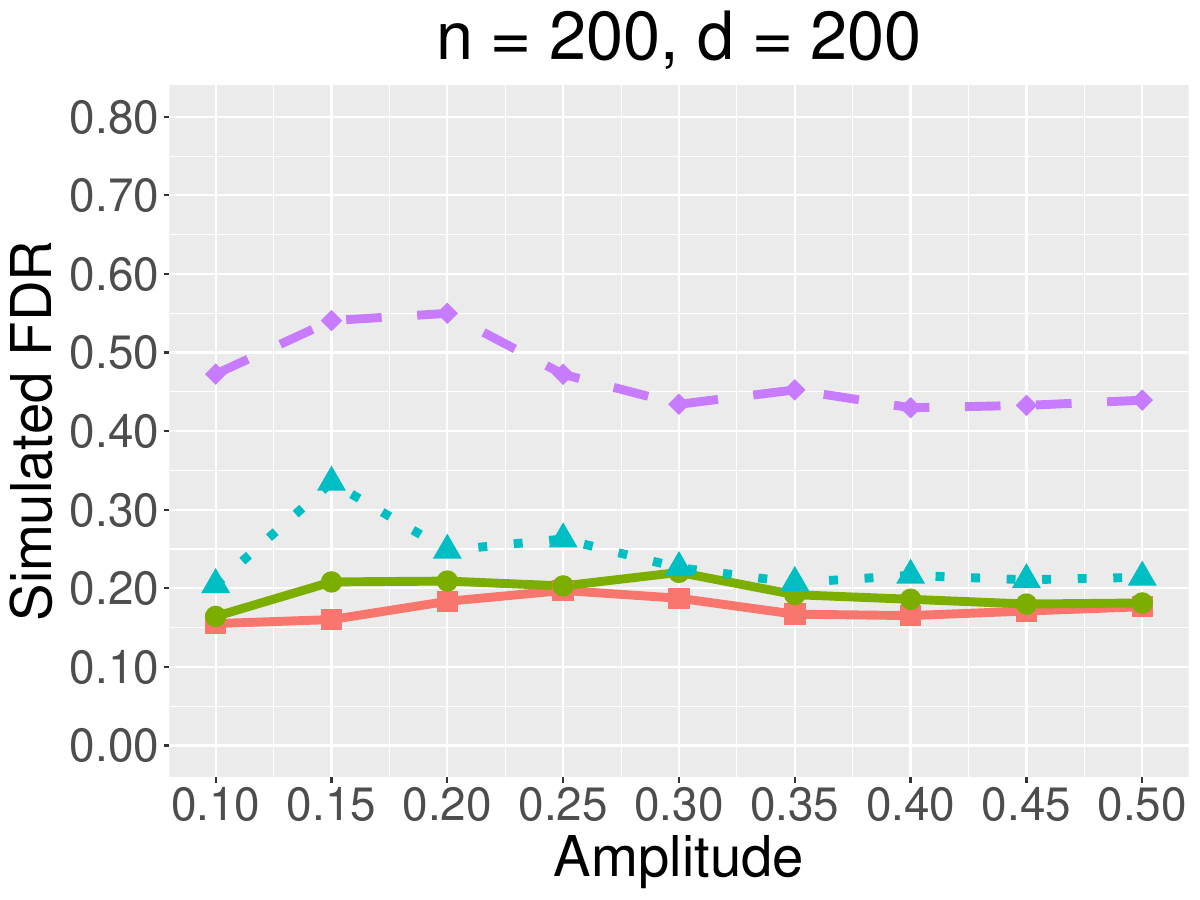}}\hspace{5pt}
	\subfloat{\includegraphics[width=.27\columnwidth]{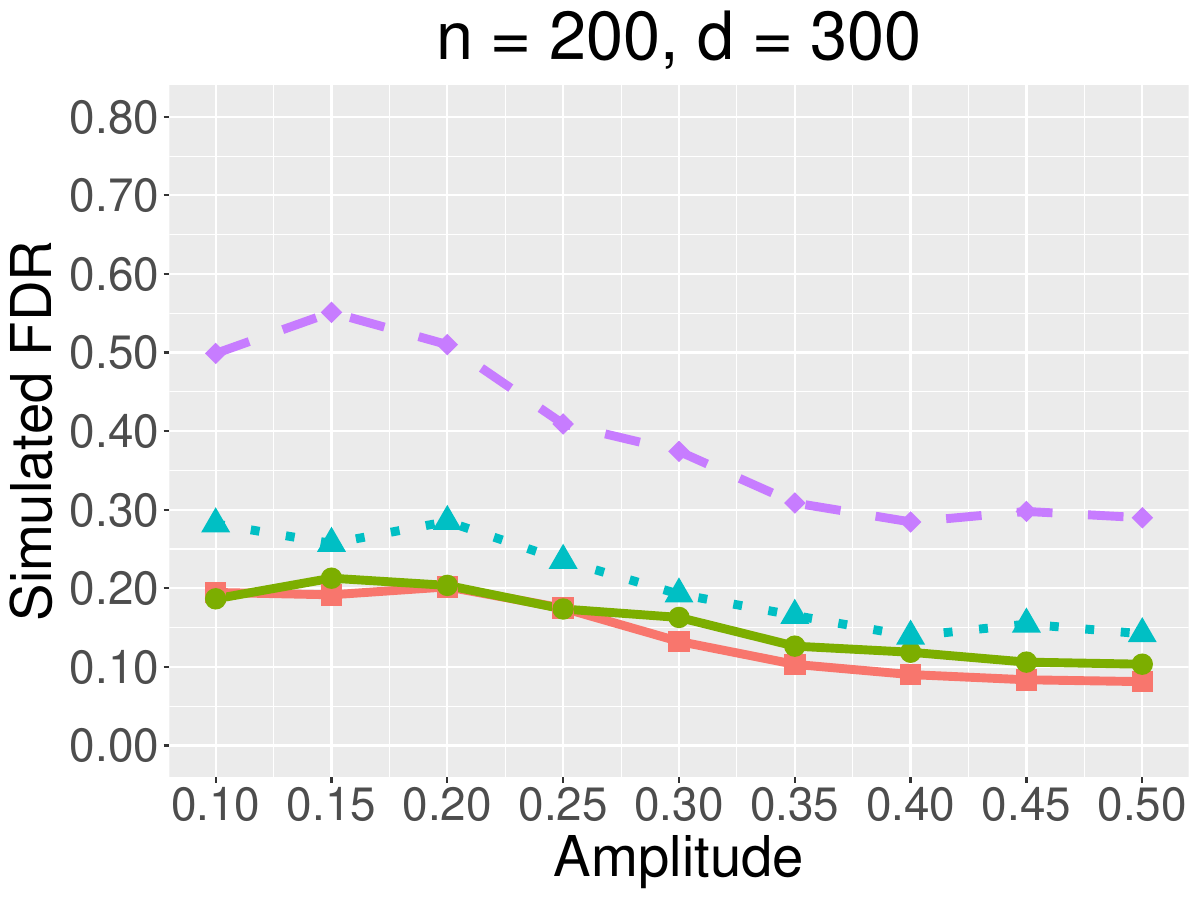}}\hspace{5pt}
	\subfloat{\includegraphics[width=.27\columnwidth]{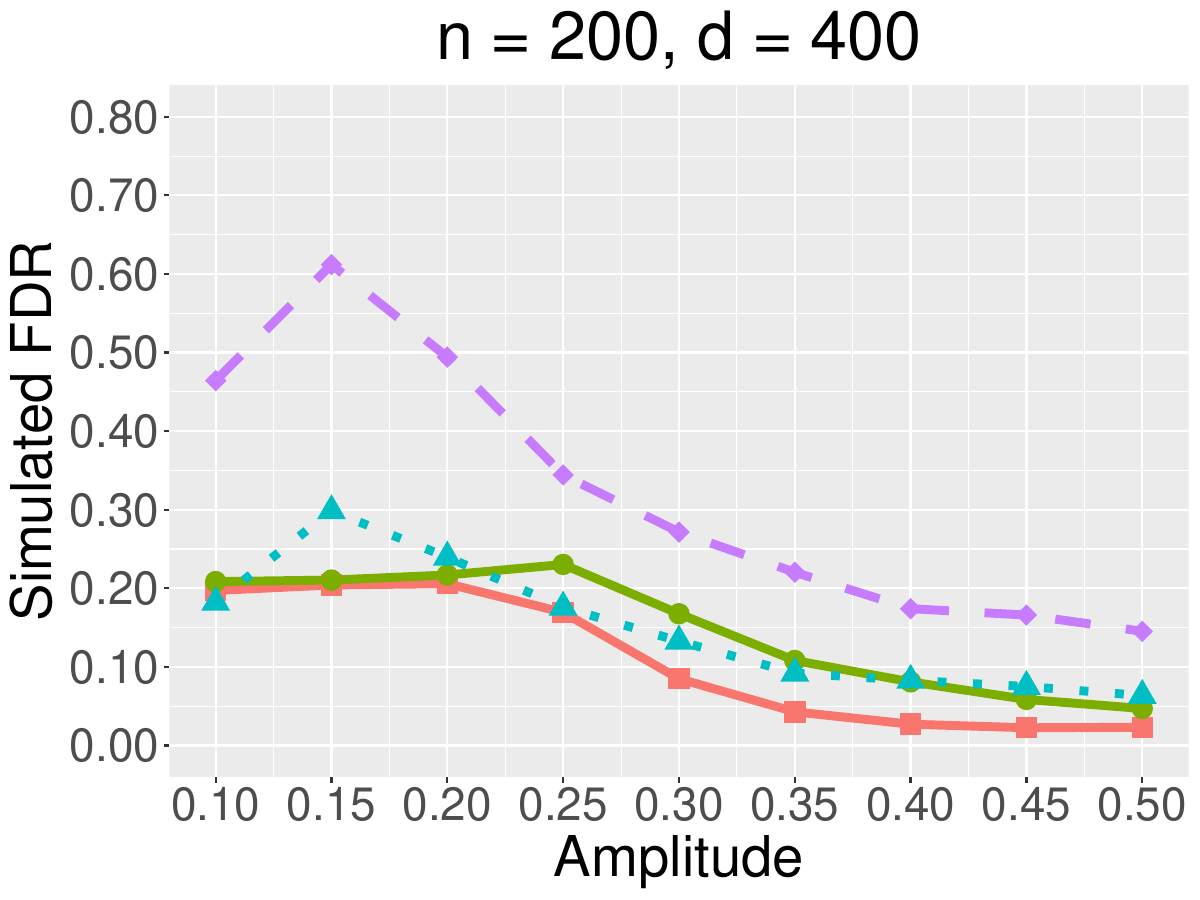}}\\
	\subfloat{\includegraphics[width=.27\columnwidth]{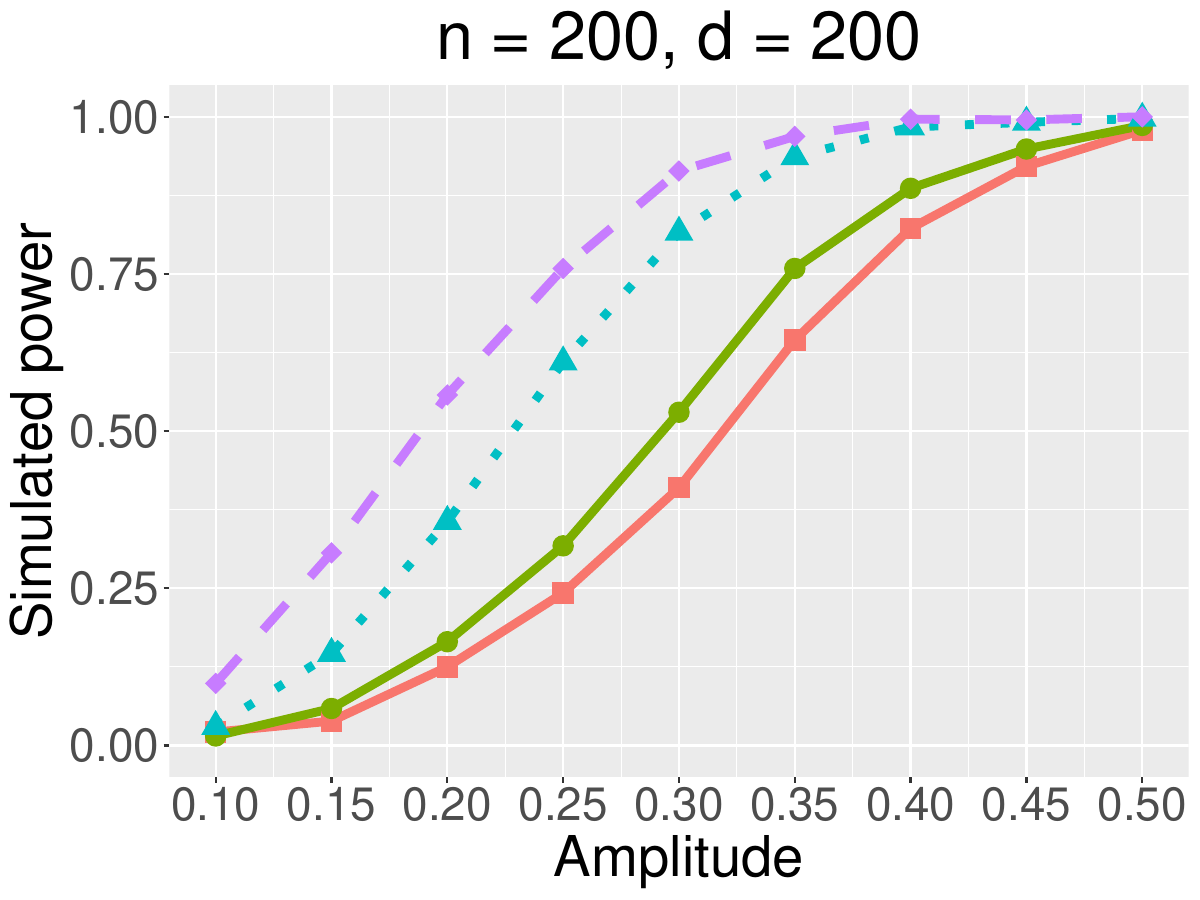}}\hspace{5pt}
    \subfloat{\includegraphics[width=.27\columnwidth]{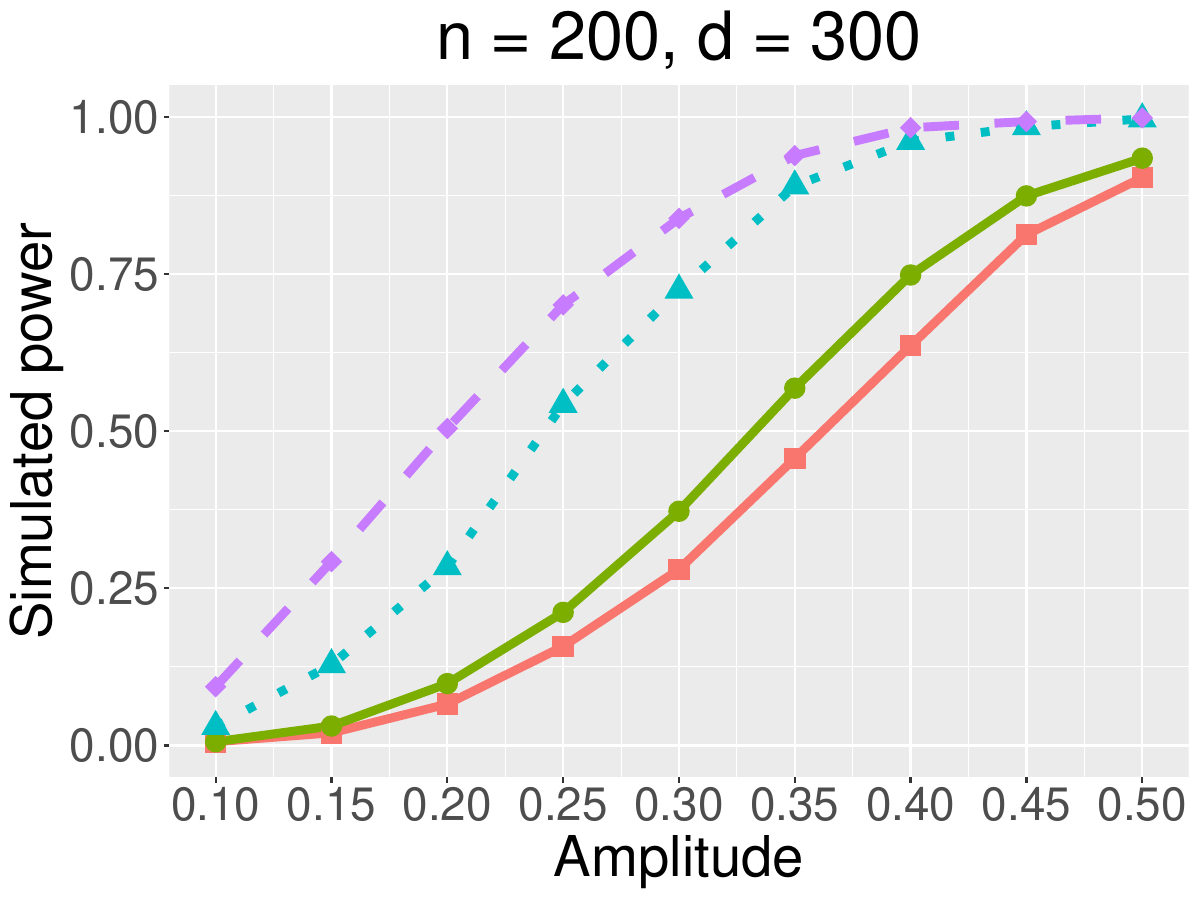}}\hspace{5pt}
    \subfloat{\includegraphics[width=.27\columnwidth]{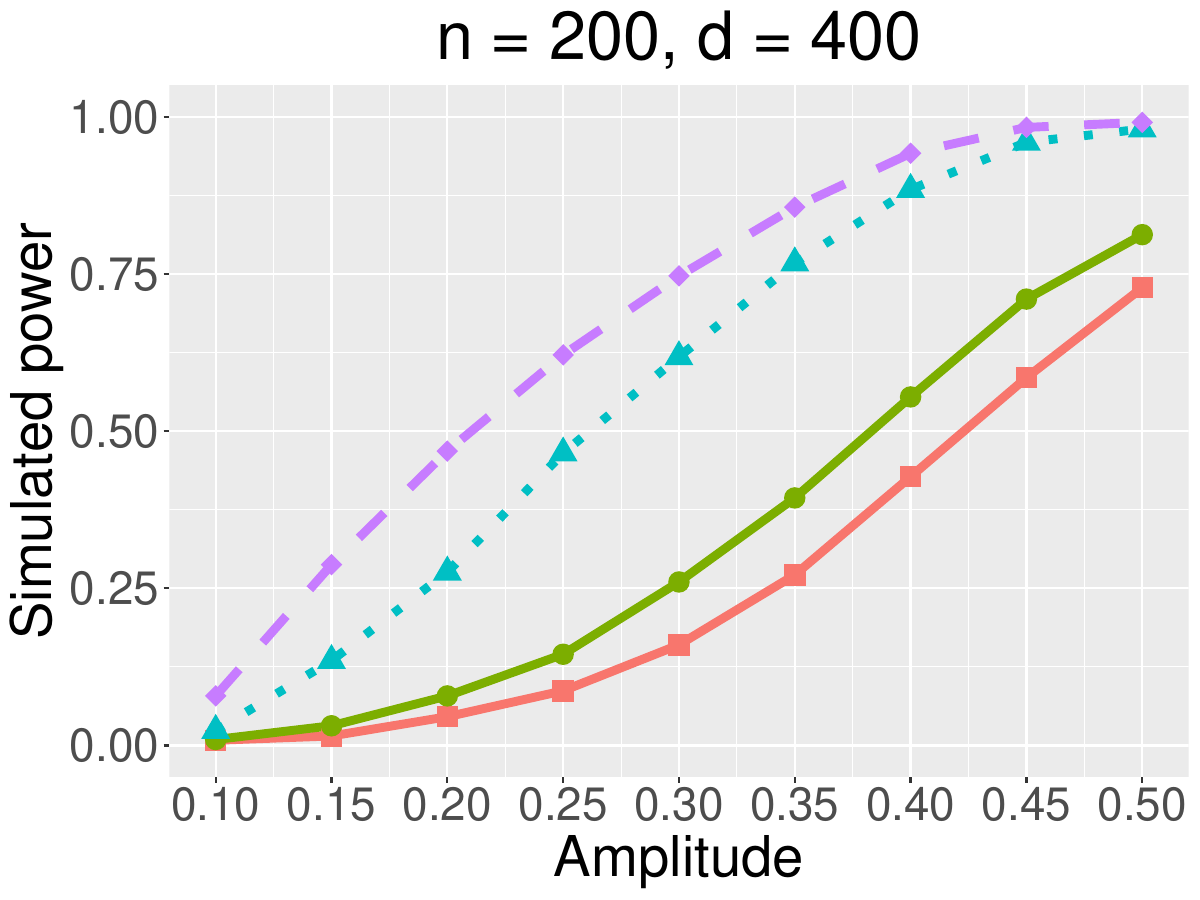}}
	\caption{\small Simulated FDR and power for different combinations of $(n,d)$. The rows of the design matrix were generated from Setting 2. The sparsity level is $k = 0.04d$ and the FDR level is $\alpha = 0.2$. The methods compared are Algorithm 1 (squares and red solid line), Algorithm 2 (circles and green solid line), Algorithm A2 (triangles and blue dotted line), and Algorithm A3 (diamonds and purple dashed line).}
    \label{Power-origin-0.2}
\end{figure}

\begin{figure}[htbp!]
	\centering
	\subfloat{\includegraphics[width=.27\columnwidth]{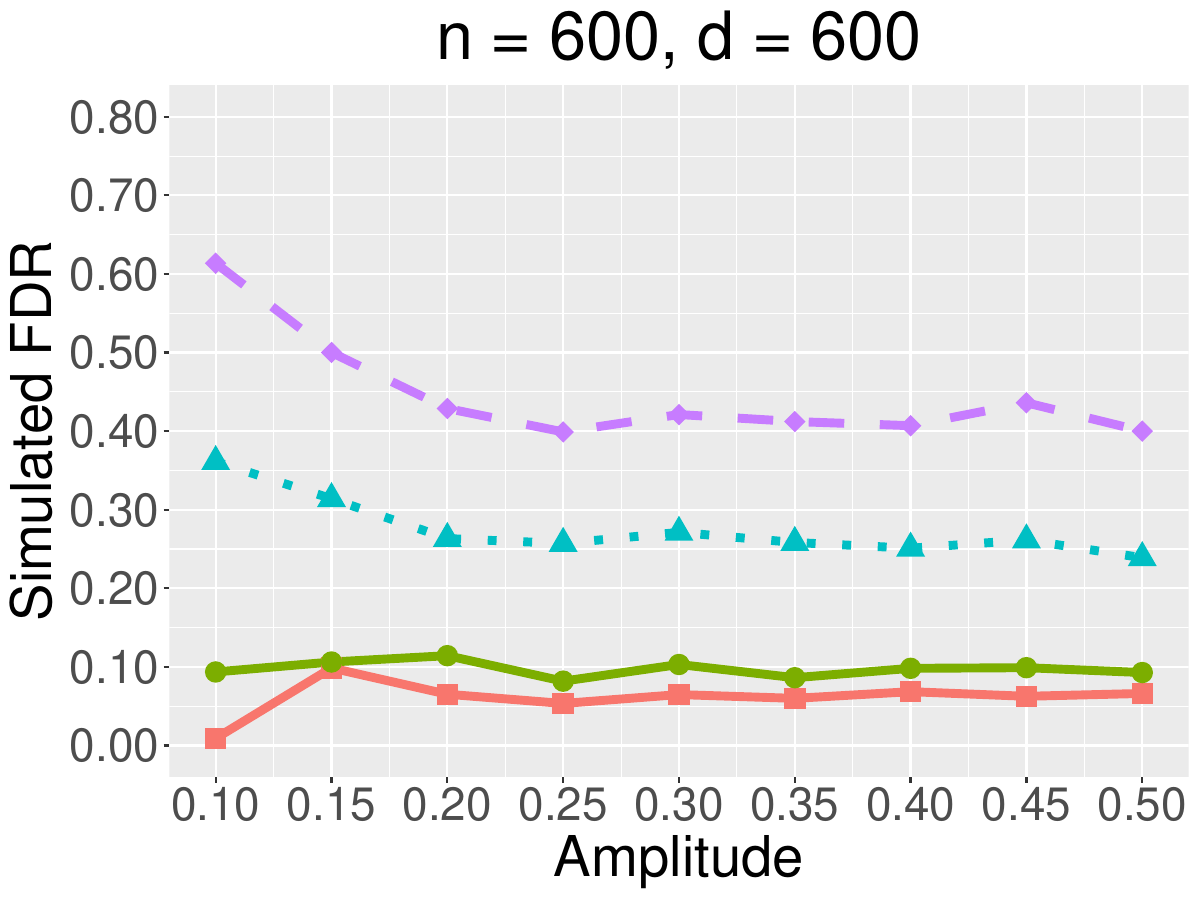}}\hspace{5pt}
	\subfloat{\includegraphics[width=.27\columnwidth]{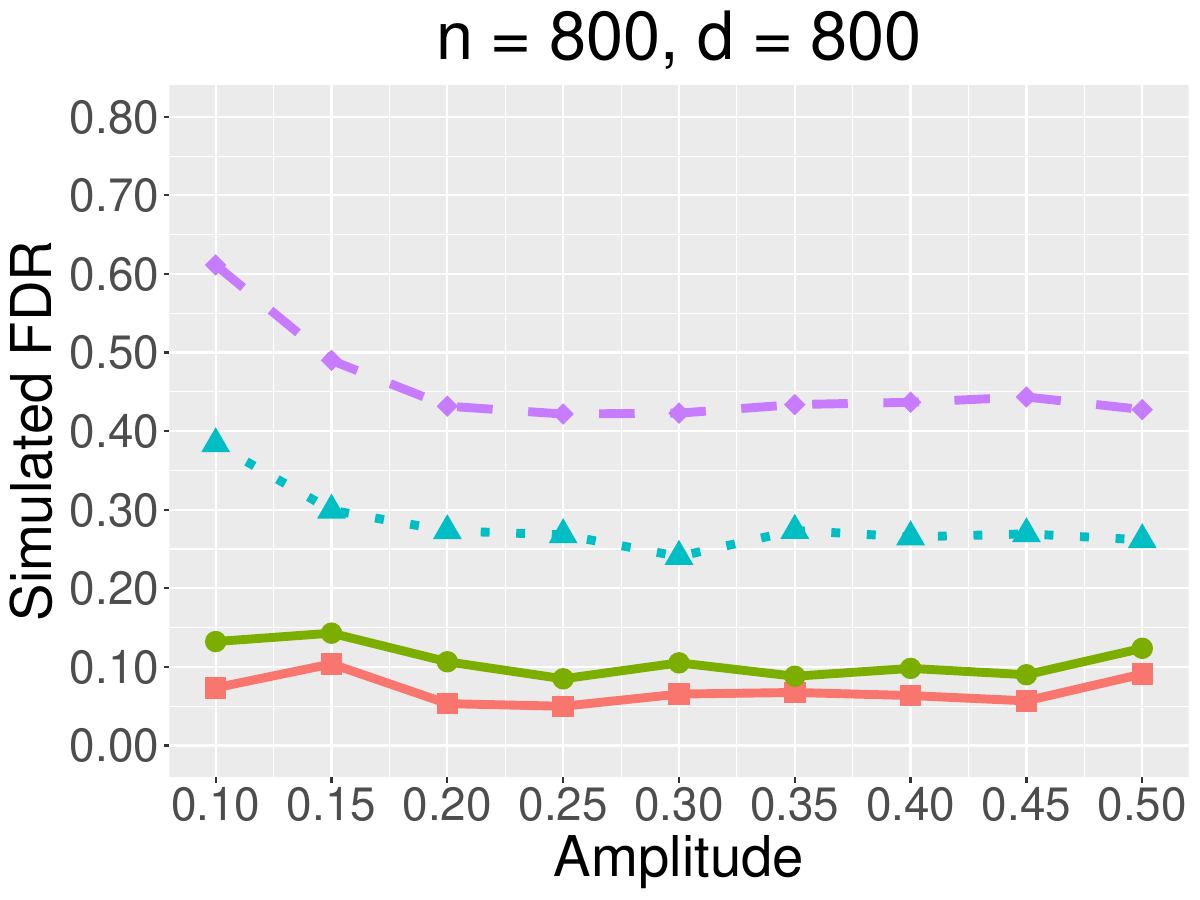}}\hspace{5pt}
	\subfloat{\includegraphics[width=.27\columnwidth]{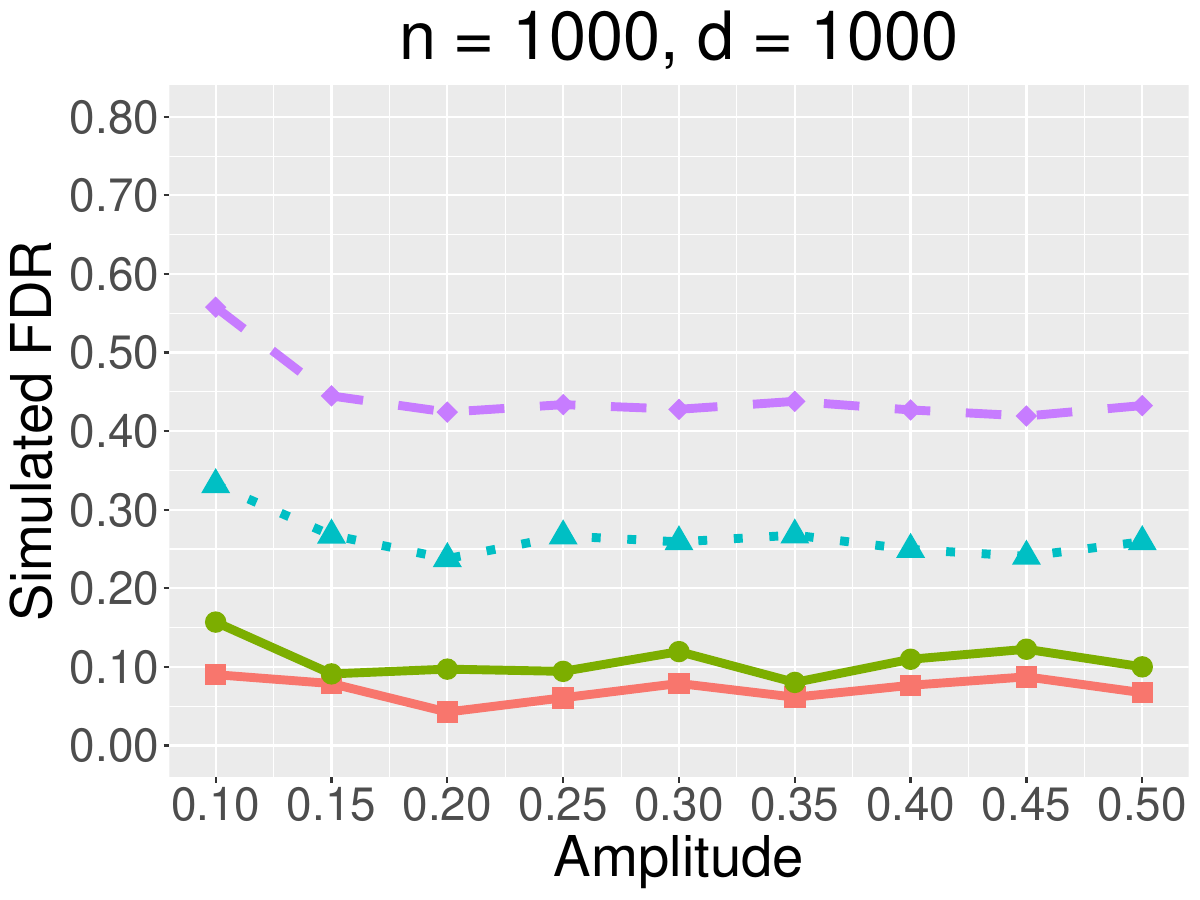}}\\
	\subfloat{\includegraphics[width=.27\columnwidth]{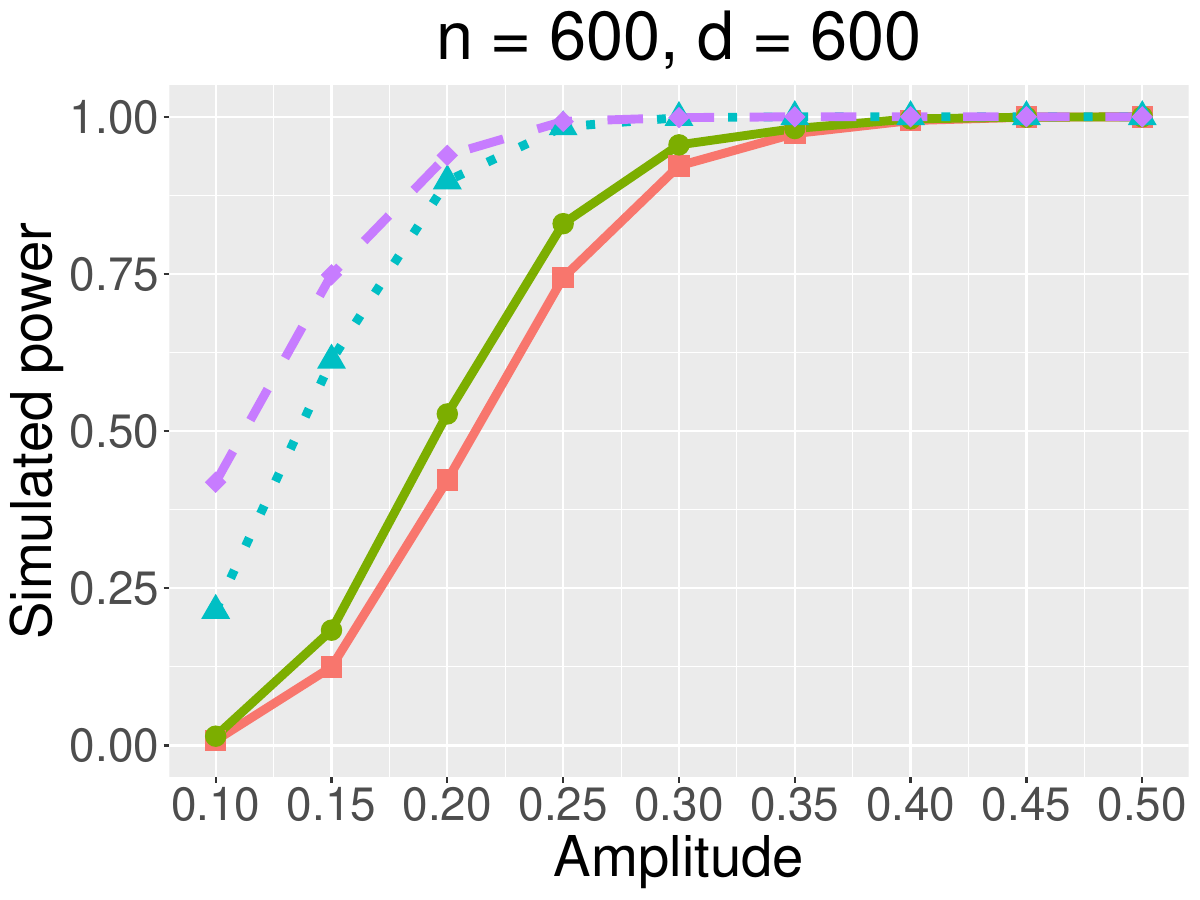}}\hspace{5pt}
    \subfloat{\includegraphics[width=.27\columnwidth]{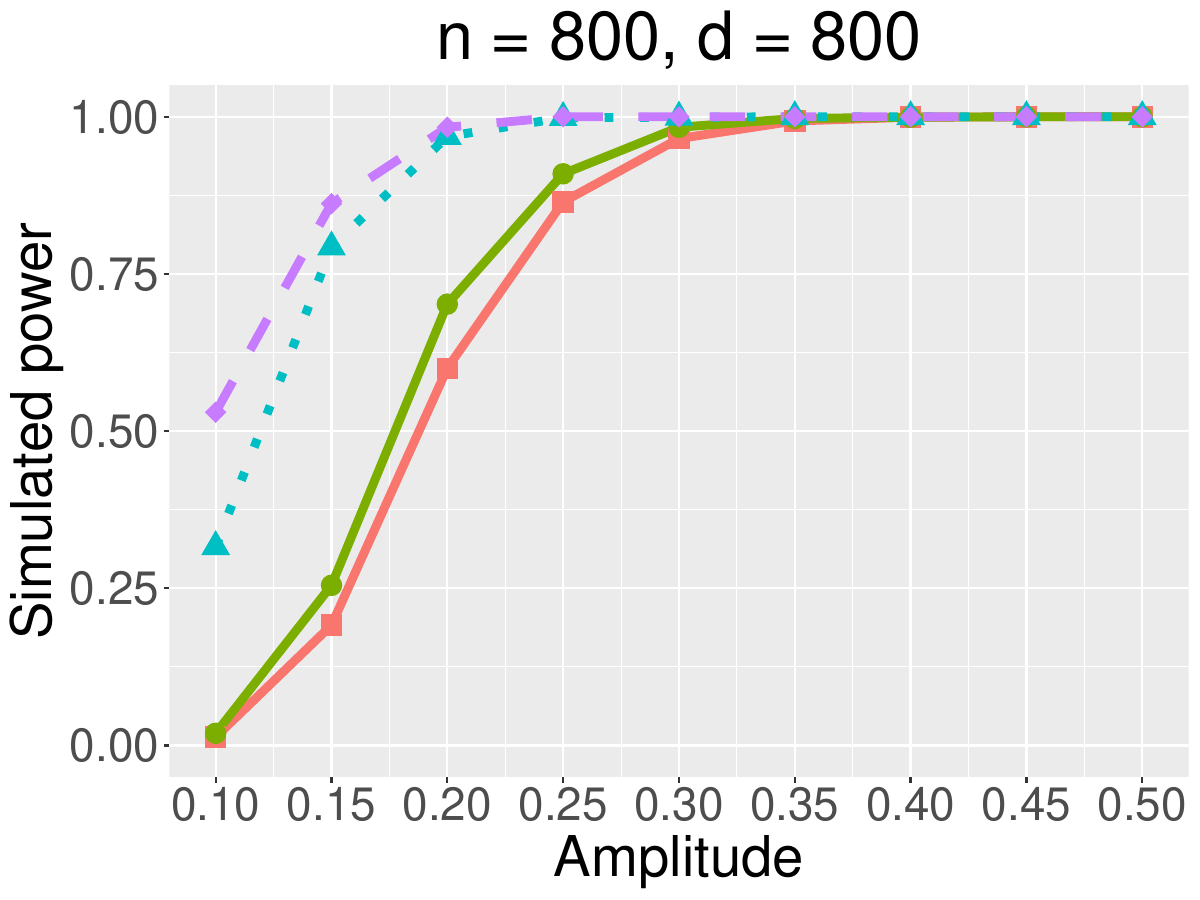}}\hspace{5pt}
    \subfloat{\includegraphics[width=.27\columnwidth]{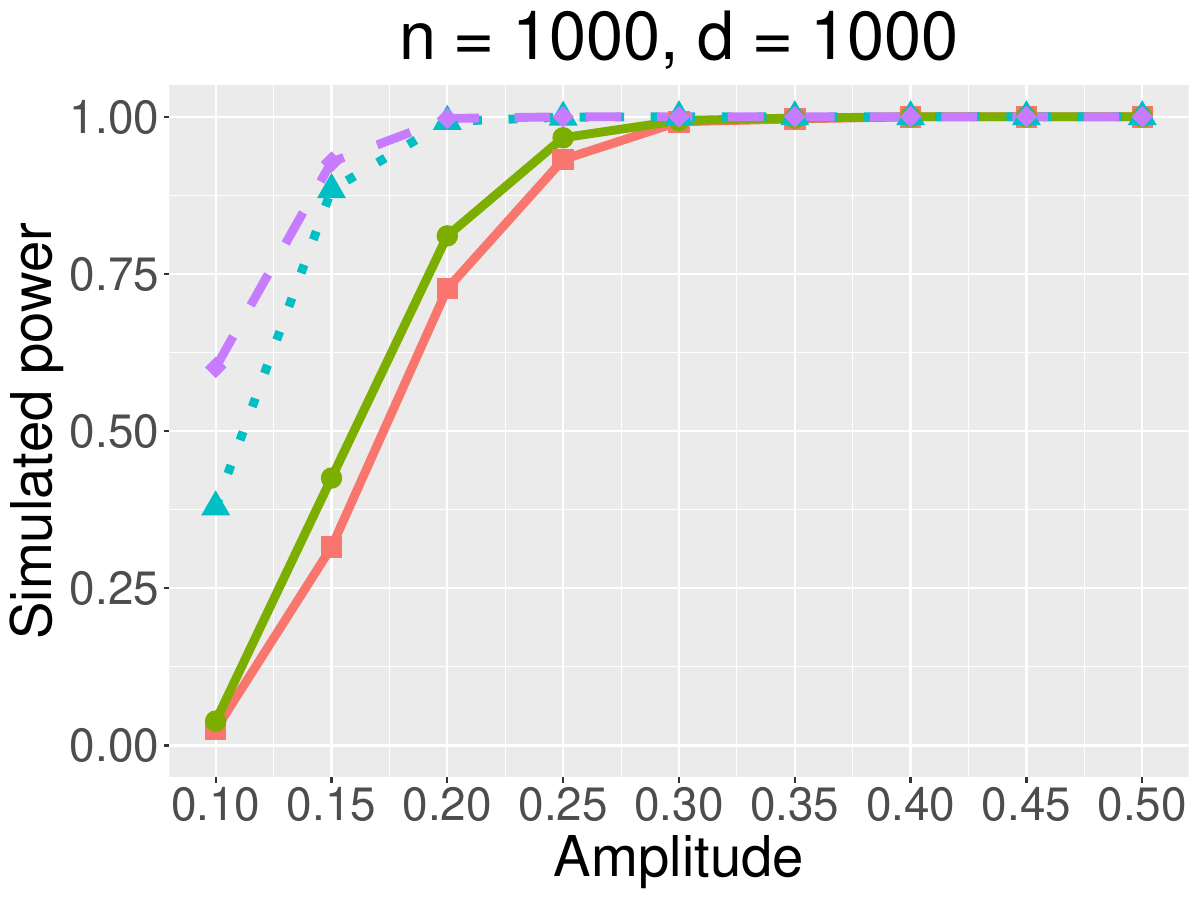}}
	\caption{\small Simulated FDR and power for different combinations of $(n,d)$. The rows of the design matrix were generated from Setting 2 with the AR(1) coefficient being set to 0.67. The sparsity level is $k = 15$ and the FDR level is $\alpha = 0.1$. The methods compared are Algorithm 1 (squares and red solid line), Algorithm 2 (circles and green solid line), Algorithm A2 (triangles and blue dotted line), and Algorithm A3 (diamonds and purple dashed line).}
    \label{Highly-correlated-origin-0.1}
\end{figure}

\begin{figure}[htbp!]
	\centering
	\subfloat{\includegraphics[width=.27\columnwidth]{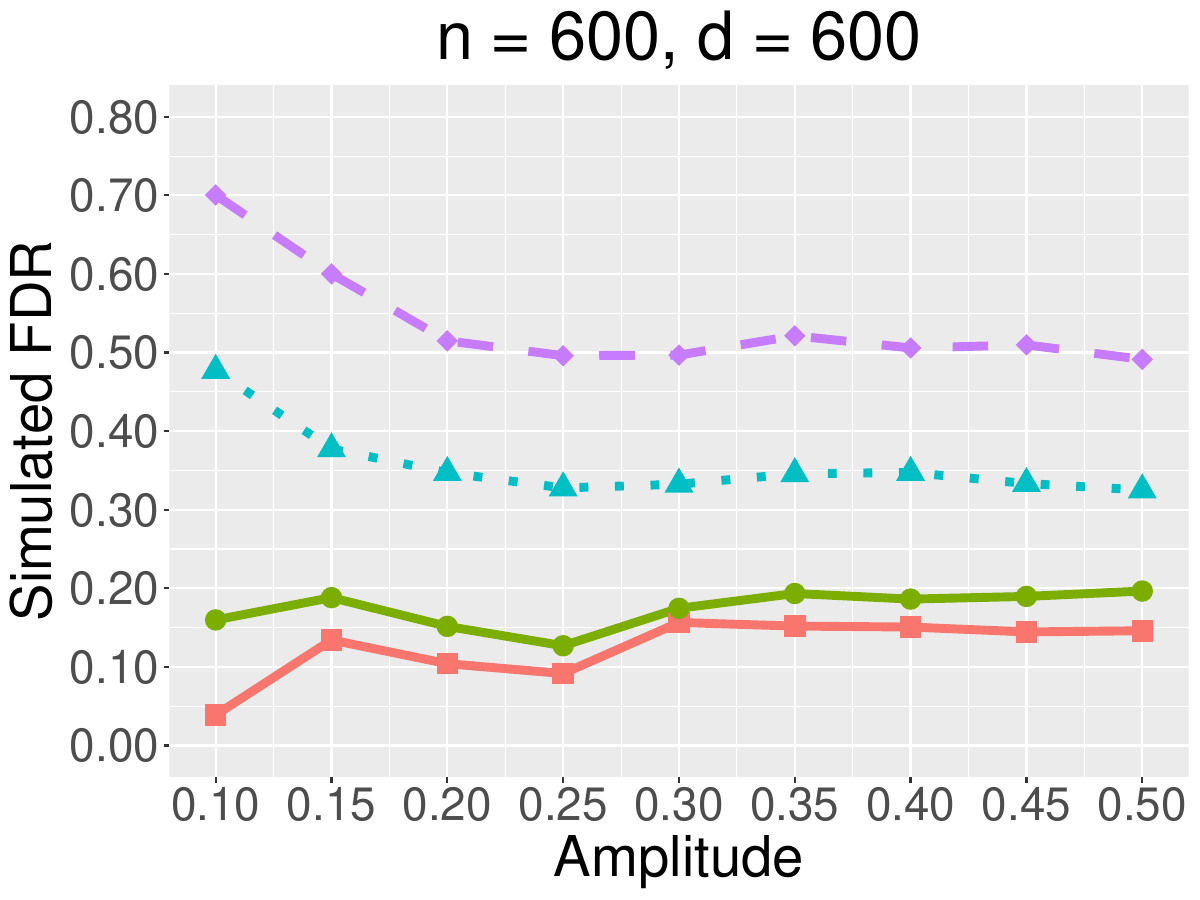}}\hspace{5pt}
	\subfloat{\includegraphics[width=.27\columnwidth]{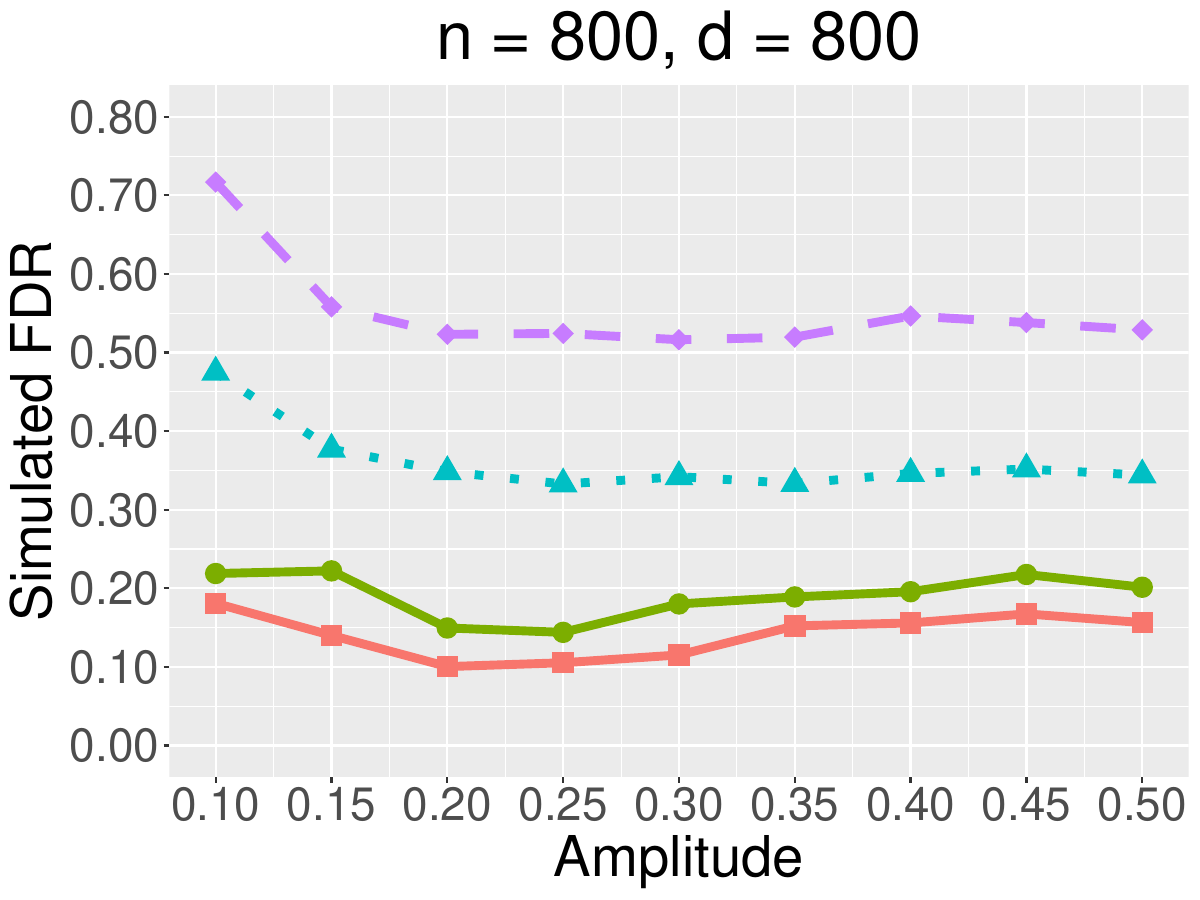}}\hspace{5pt}
	\subfloat{\includegraphics[width=.27\columnwidth]{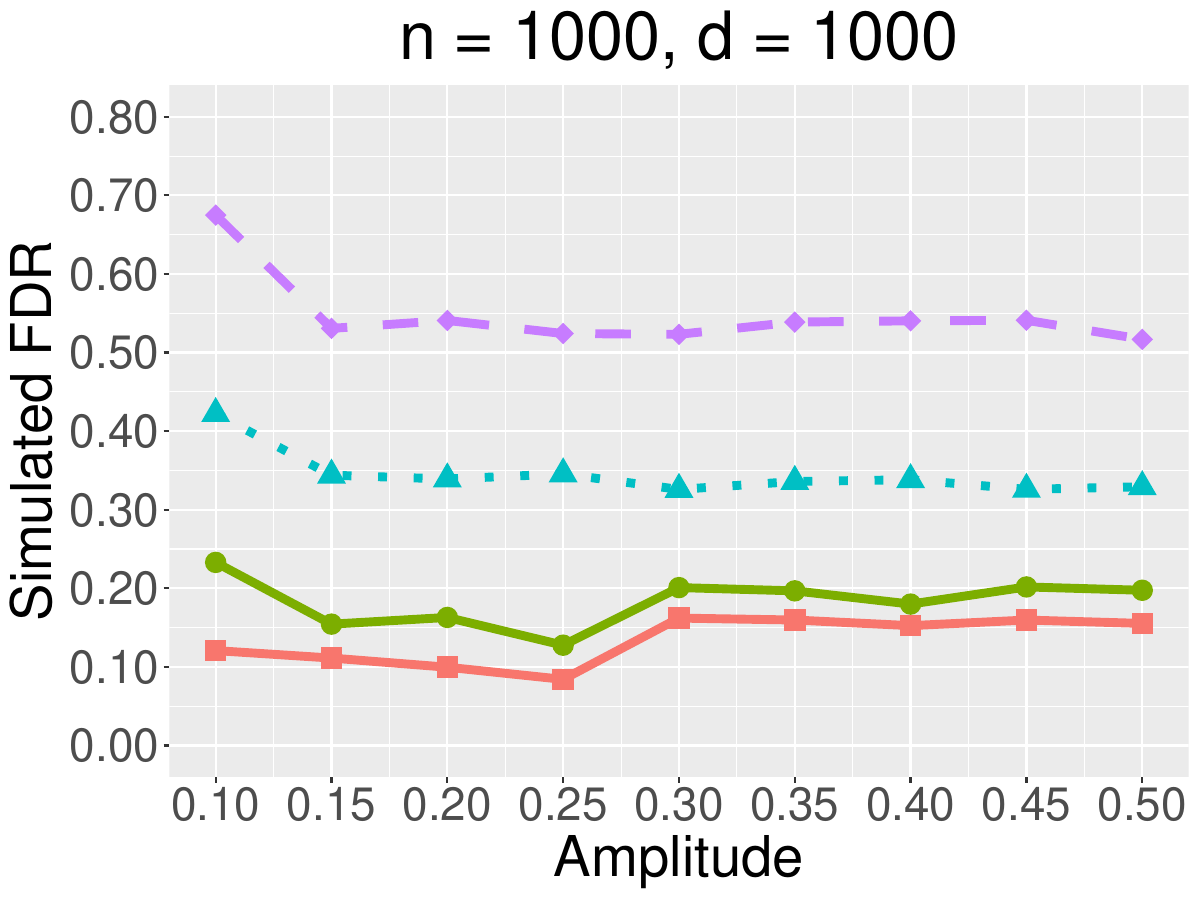}}\\
	\subfloat{\includegraphics[width=.27\columnwidth]{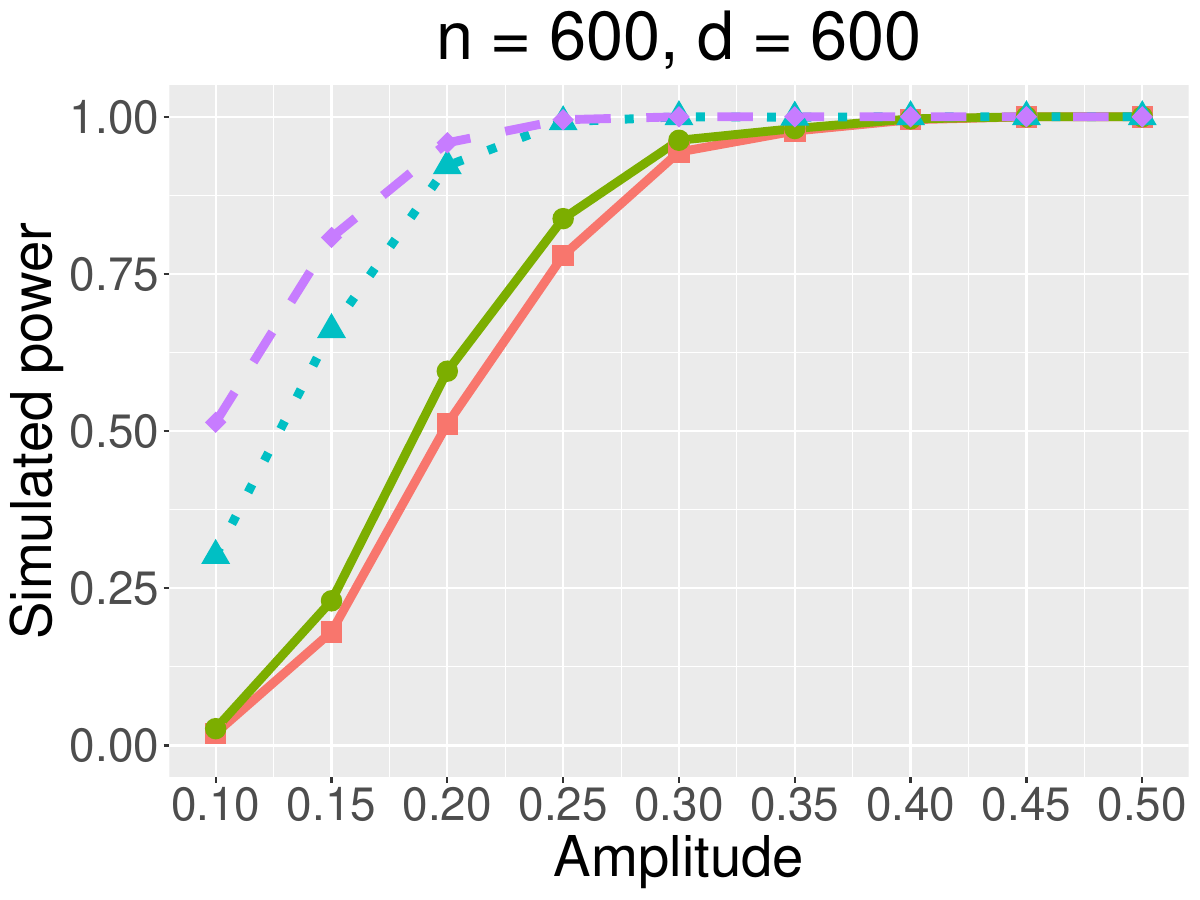}}\hspace{5pt}
    \subfloat{\includegraphics[width=.27\columnwidth]{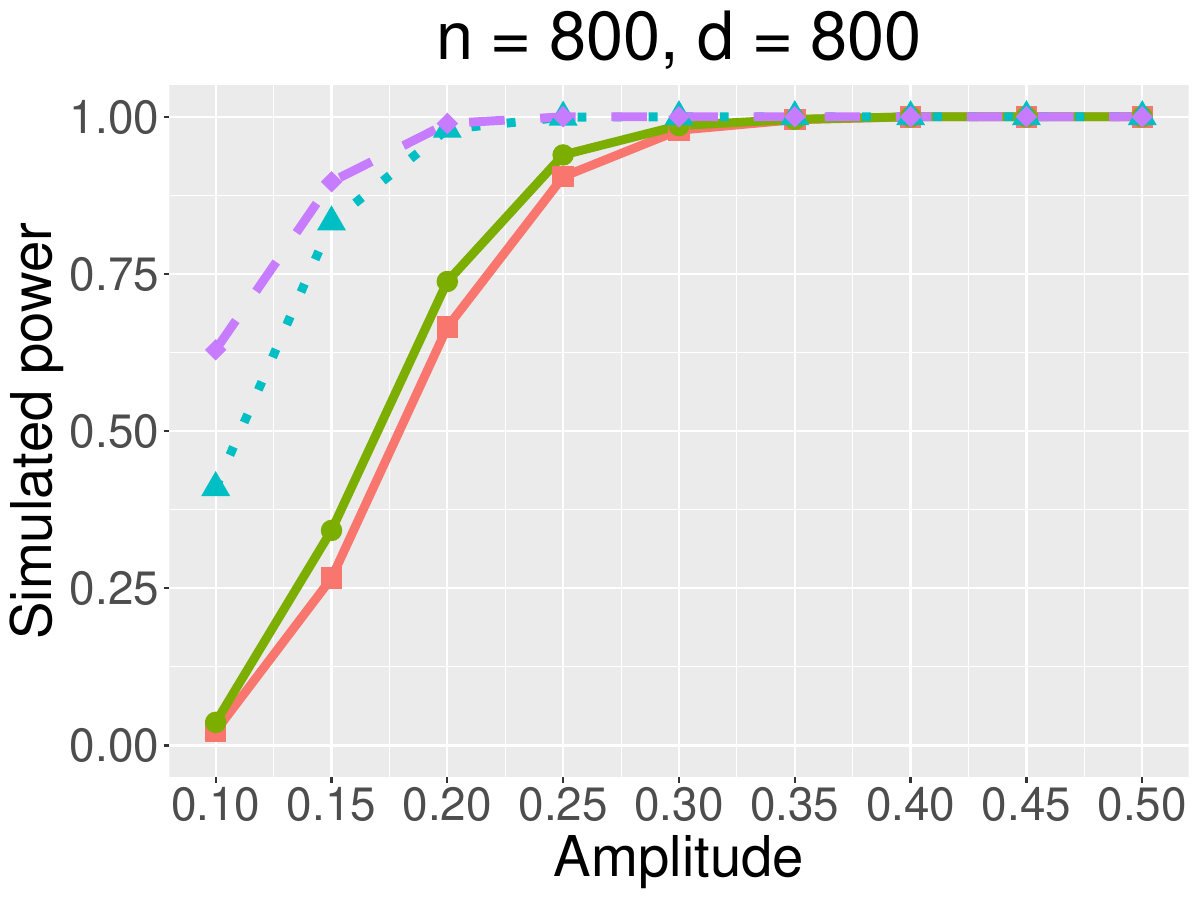}}\hspace{5pt}
    \subfloat{\includegraphics[width=.27\columnwidth]{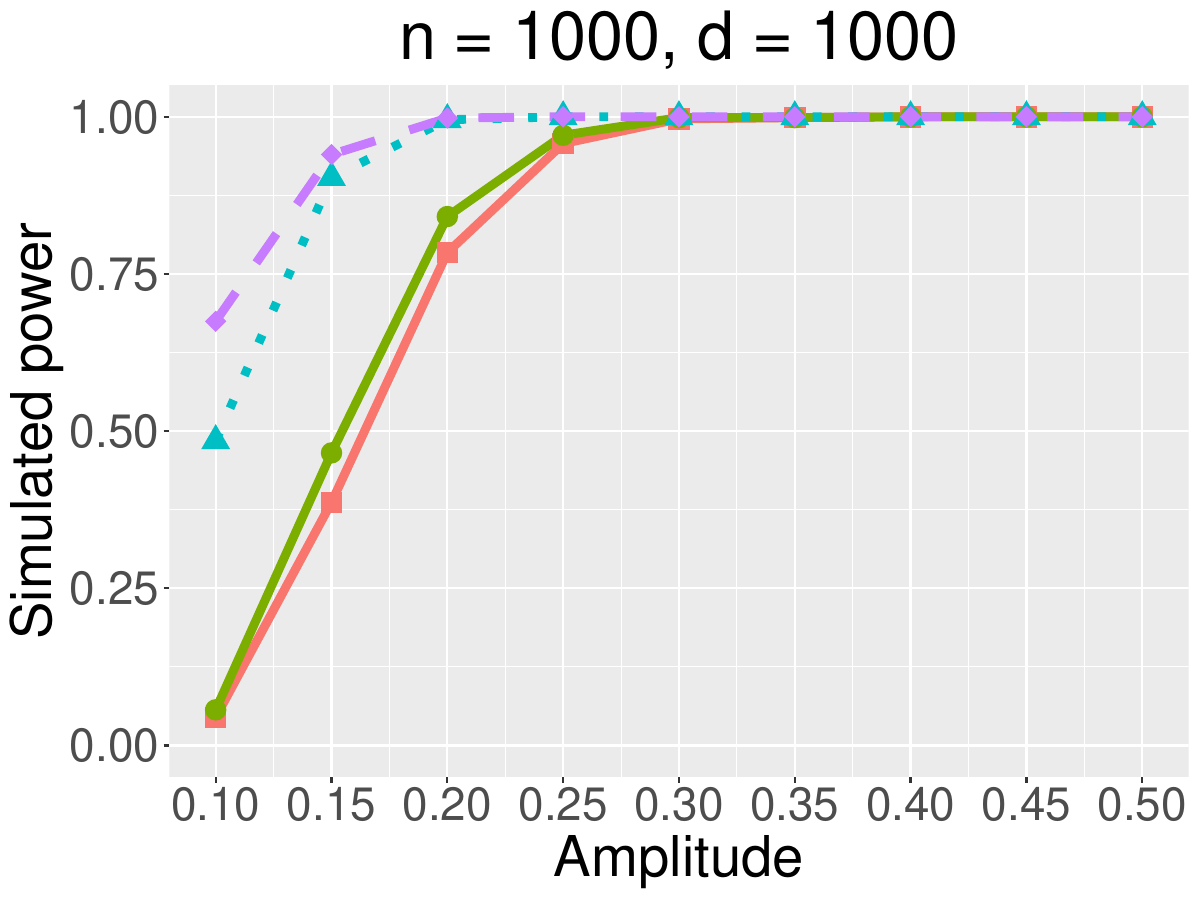}}
	\caption{\small Simulated FDR and power for different combinations of $(n,d)$. The rows of the design matrix were generated from Setting 2 with the AR(1) coefficient being set to 0.67. The sparsity level is $k = 15$ and the FDR level is $\alpha = 0.2$. The methods compared are Algorithm 1 (squares and red solid line), Algorithm 2 (circles and green solid line), Algorithm A2 (triangles and blue dotted line), and Algorithm A3 (diamonds and purple dashed line).}
    \label{Highly-correlated-origin-0.2}
\end{figure}

Next, we investigate FDR control and power as the signal strength increases, using the same setup as in Figure~\ref{Power-0.1} of the main text. Under this setting, the dependence in the design matrix is not particularly strong. From Figures~\ref{Power-origin-0.05}--\ref{Power-origin-0.2}, Algorithm~\ref{method_1_origin} maintains FDR control and achieves higher power than Algorithm~\ref{method_2}. Algorithm~\ref{method_2_origin} exhibits high apparent power; however, this is accompanied by a lack of FDR control, indicating that its performance is not theoretically justified. While Algorithm~\ref{method_1_origin} serves as a valid multiple testing procedure when the test statistics are independent, our empirical results confirm that it faces substantial challenges in controlling the FDR when the test statistics are dependent. In particular, we examine and compare performance under an AR(1) design for the model matrix with a moderately large  coefficient of $0.67$. The setup mirrors that of Figures~\ref{Highly-correlated-0.05}--\ref{Highly-correlated-0.2} in the Supplementary Material. As shown in Figures~\ref{Highly-correlated-origin-0.05}--\ref{Highly-correlated-origin-0.2}, the FDR of Algorithm~\ref{method_1_origin} becomes severely inflated, substantially exceeding the nominal level and the FDR achieved by our proposed methods. 

This demonstrates that the traditional BH procedure, when applied directly to the original debiased statistics without the proposed knockoff-based framework, does not reliably control the FDR. In high-dimensional applications, however, substantial and complex feature dependence is the norm. This highlights the necessity and practical importance of our framework, which remains stable and continues to control the FDR under such general dependence structures.

Furthermore, we examine a two-stage screening strategy in which we first perform variable selection on the original data using the Lasso--following the same initial screening stage as in \citeS{Xing2021Controlling_app}--and then apply the \( p \)-value-based multiple testing procedures (Algorithms~\ref{method_1_origin} and~\ref{method_2_origin}) to the design matrix formed by the selected variables. Under the same setting as in Figure~\ref{Two-stage-0.1} of the main text, both approaches based on the original data consistently yield severely inflated FDP values exceeding $0.8$. Given this severe inflation, we do not present the corresponding figures. These findings suggest that directly combining Lasso screening with the debiased Lasso framework does not provide reliable FDR control in this setting. In contrast, under the same setting, our approach constructs model-X knockoffs after screening and continues to yield asymptotically independent test statistics, which ensures that our approach achieves favorable and reliable performance; see Figure~\ref{Two-stage-0.1} in the main text for details. These additional comparisons further underscore the importance of the knockoff augmentation step for achieving reliable FDR control in high-dimensional settings with strong dependence.

Overall, the additional comparisons demonstrate that procedures based solely on the original design matrix are highly sensitive to feature dependence and may fail to control the FDR, particularly under strong correlations or after variable screening. In contrast, our knockoff-based augmentation consistently achieves stable FDR control across all considered settings while maintaining competitive or improved power. Taken together, these findings reinforce that the dimension augmentation and pairing mechanism are not merely technical refinements, but essential components for reliable high-dimensional multiple testing under complex dependence structures.

\begin{figure}[htbp!]
	\centering
	\subfloat{\includegraphics[width=.27\columnwidth]{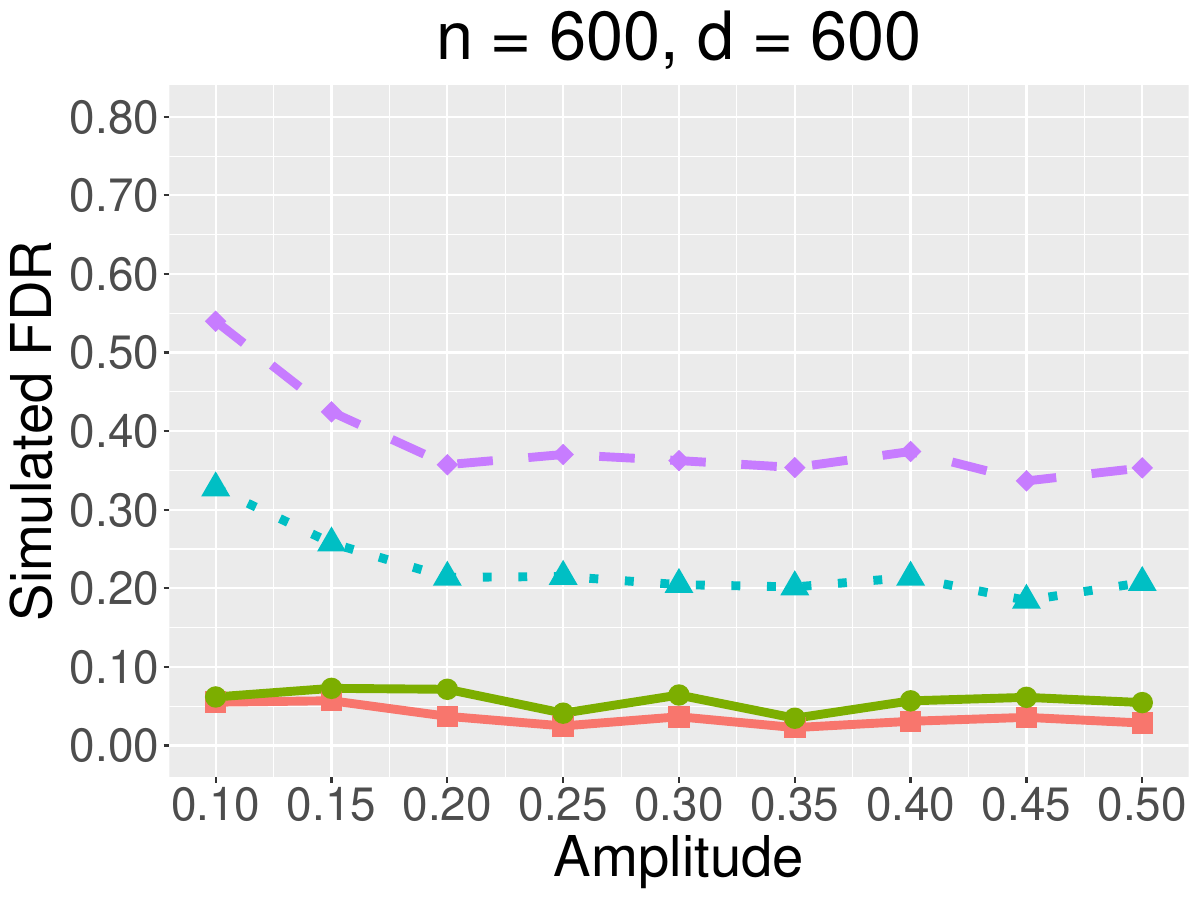}}\hspace{5pt}
	\subfloat{\includegraphics[width=.27\columnwidth]{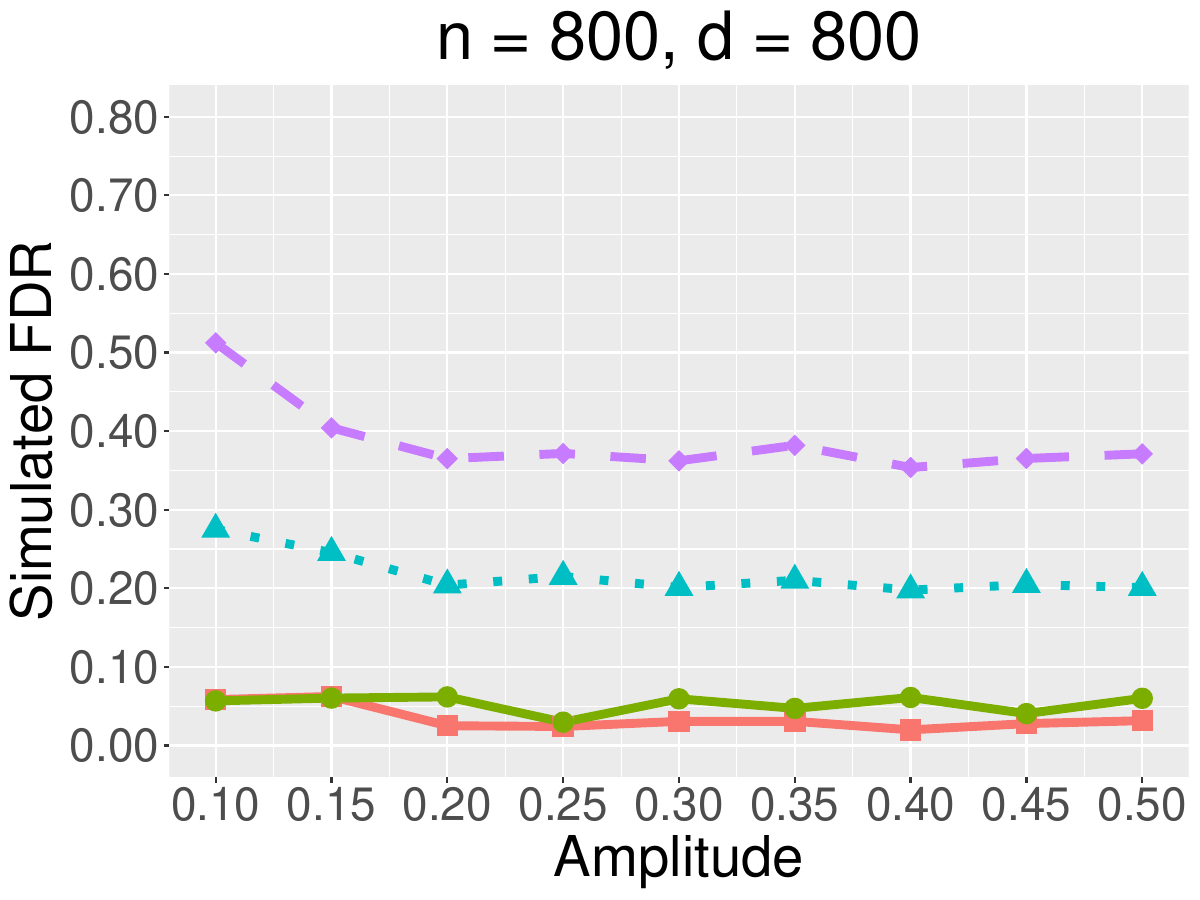}}\hspace{5pt}
	\subfloat{\includegraphics[width=.27\columnwidth]{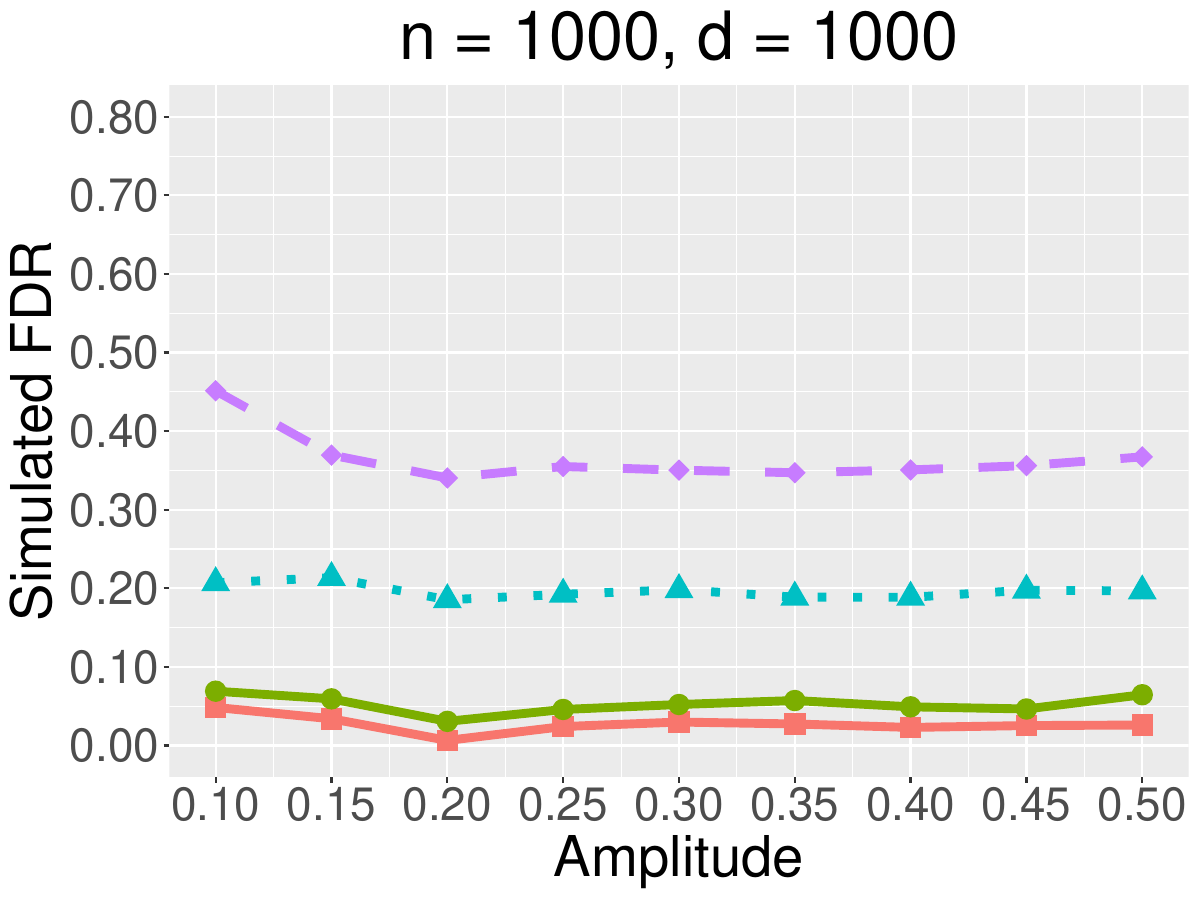}}\\
	\subfloat{\includegraphics[width=.27\columnwidth]{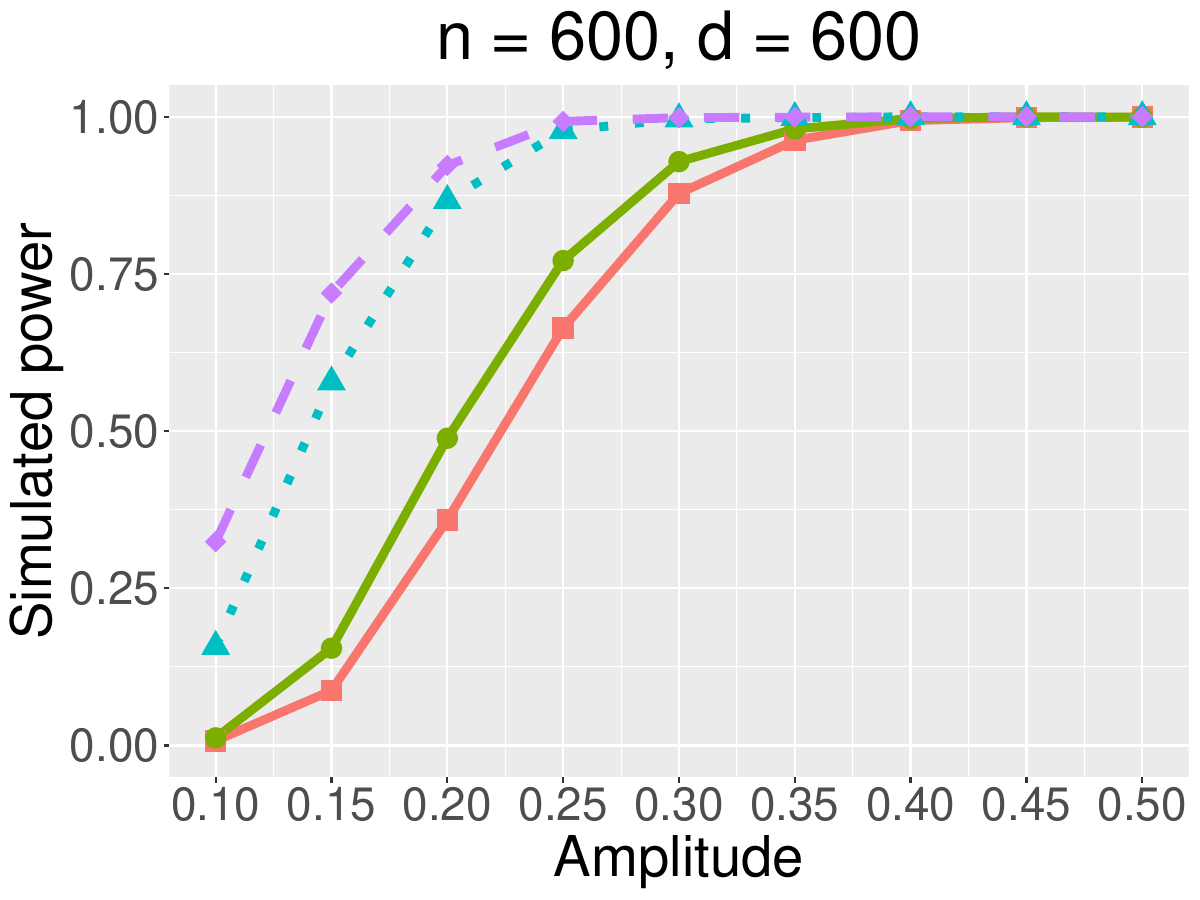}}\hspace{5pt}
    \subfloat{\includegraphics[width=.27\columnwidth]{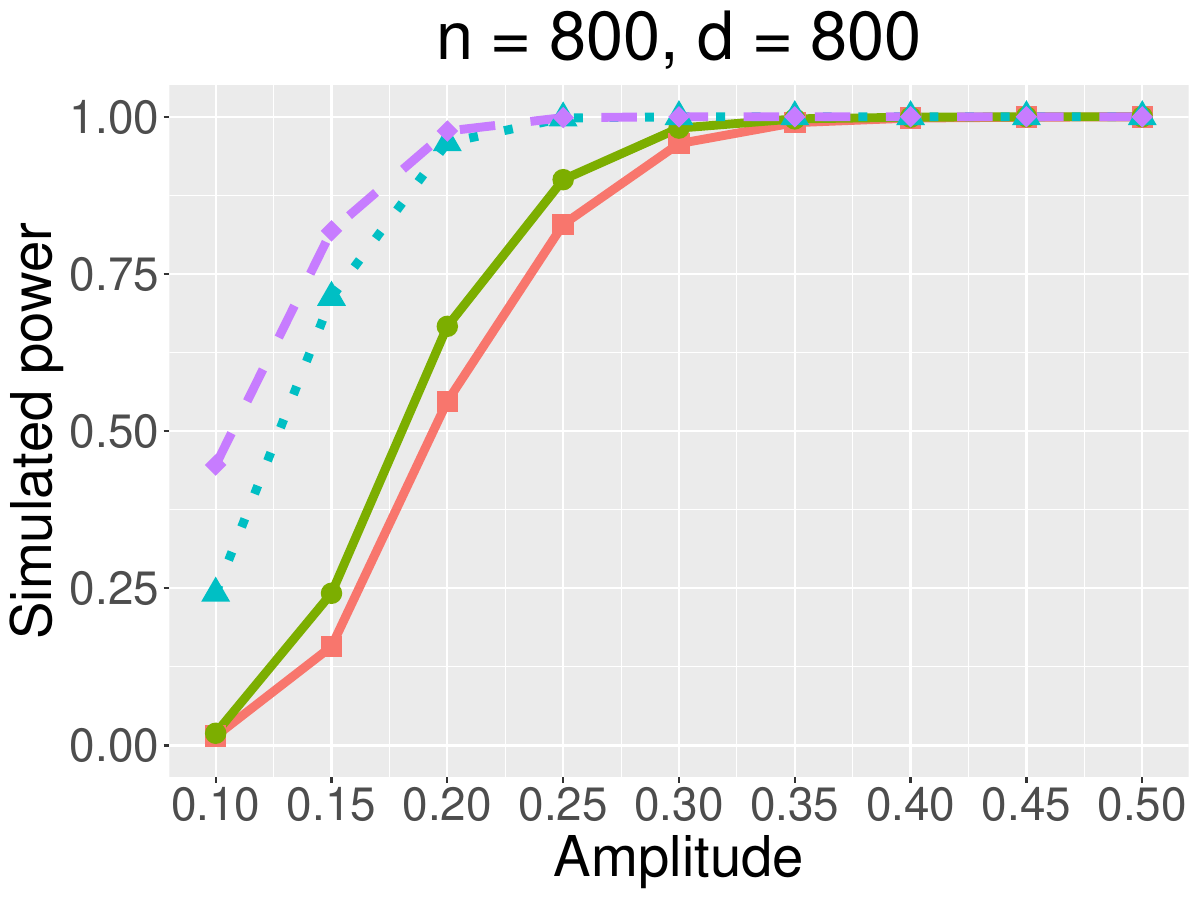}}\hspace{5pt}
    \subfloat{\includegraphics[width=.27\columnwidth]{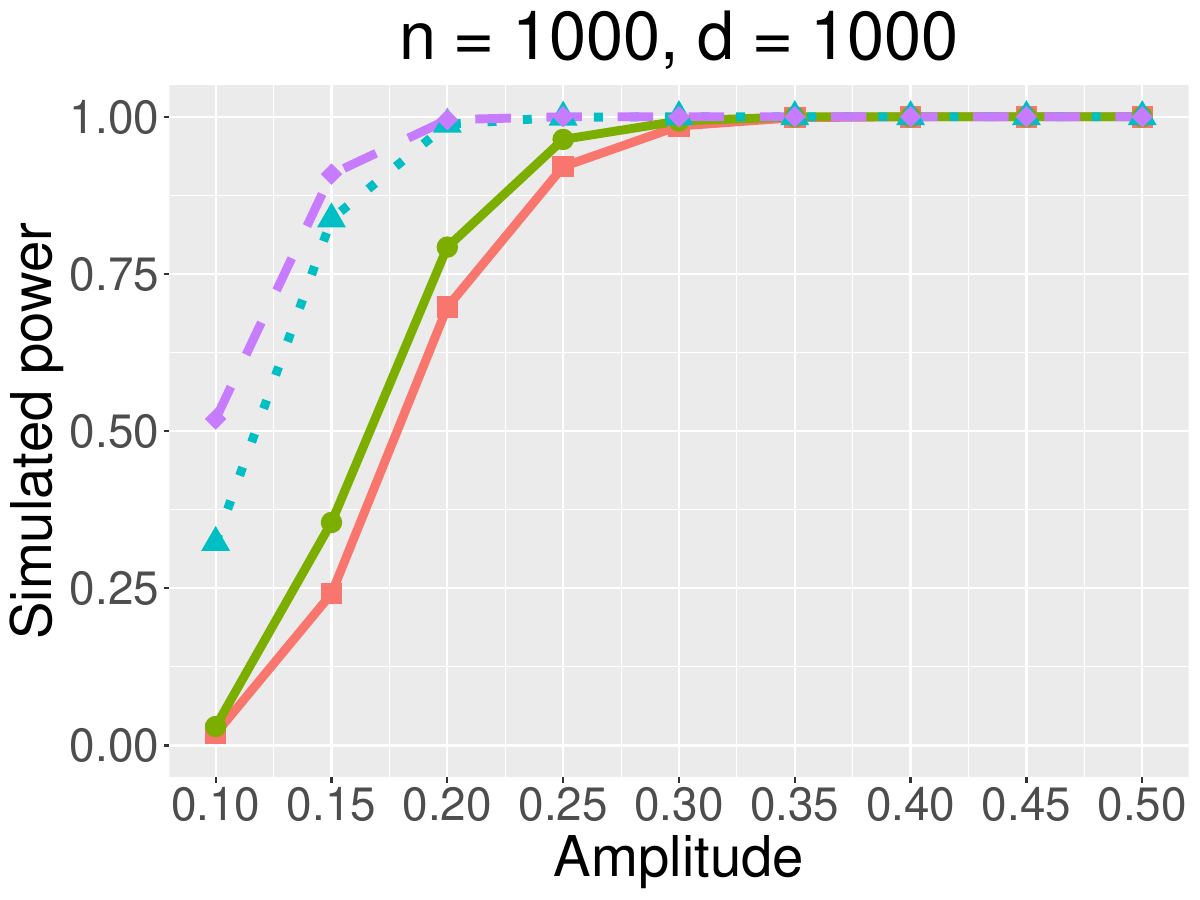}}
	\caption{\small Simulated FDR and power for different combinations of $(n,d)$. The rows of the design matrix were generated from Setting 2 with the AR(1) coefficient being set to 0.67. The sparsity level is $k = 15$ and the FDR level is $\alpha = 0.05$. The methods compared are Algorithm 1 (squares and red solid line), Algorithm 2 (circles and green solid line), Algorithm A2 (triangles and blue dotted line), and Algorithm A3 (diamonds and purple dashed line).}
    \label{Highly-correlated-origin-0.05}
\end{figure}

\subsection{Additional simulation results with non-Gaussian errors}\label{simu_nonGaussian}
In this subsection, we conduct additional simulations in a setting where the error terms follow a non-Gaussian distribution. Specifically, we generate errors from a scaled $t$-distribution with $5$ degrees of freedom (i.e., a Student's $t$-distribution standardized to have unit variance), a setting consistent with the approach used in \citeS{vandeGeer2014on_app}. All other settings remain the same as those in Section~\ref{sec:sim}.

\begin{figure}[htbp!]
	\centering
	\subfloat{\includegraphics[width=.27\columnwidth]{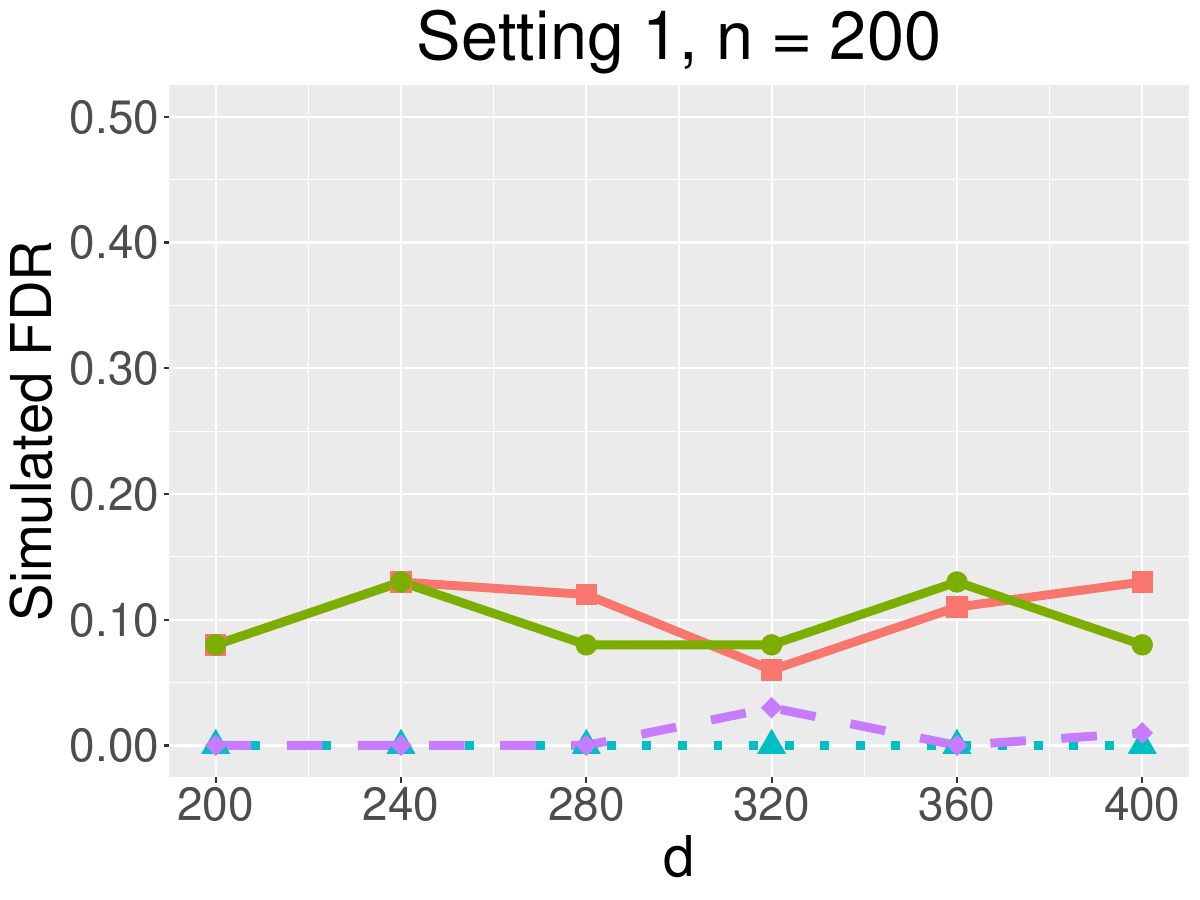}}\hspace{5pt}
	\subfloat{\includegraphics[width=.27\columnwidth]{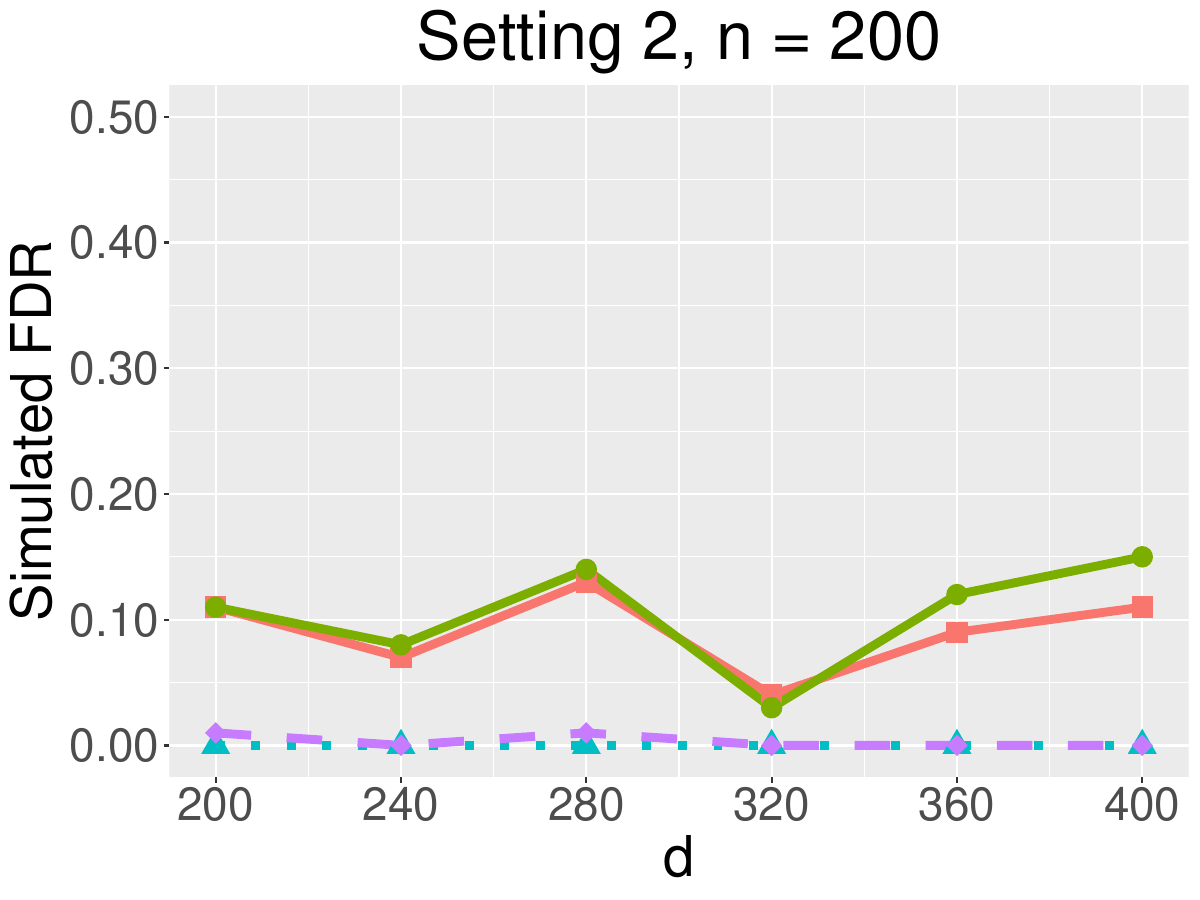}}\hspace{5pt}
	\subfloat{\includegraphics[width=.27\columnwidth]{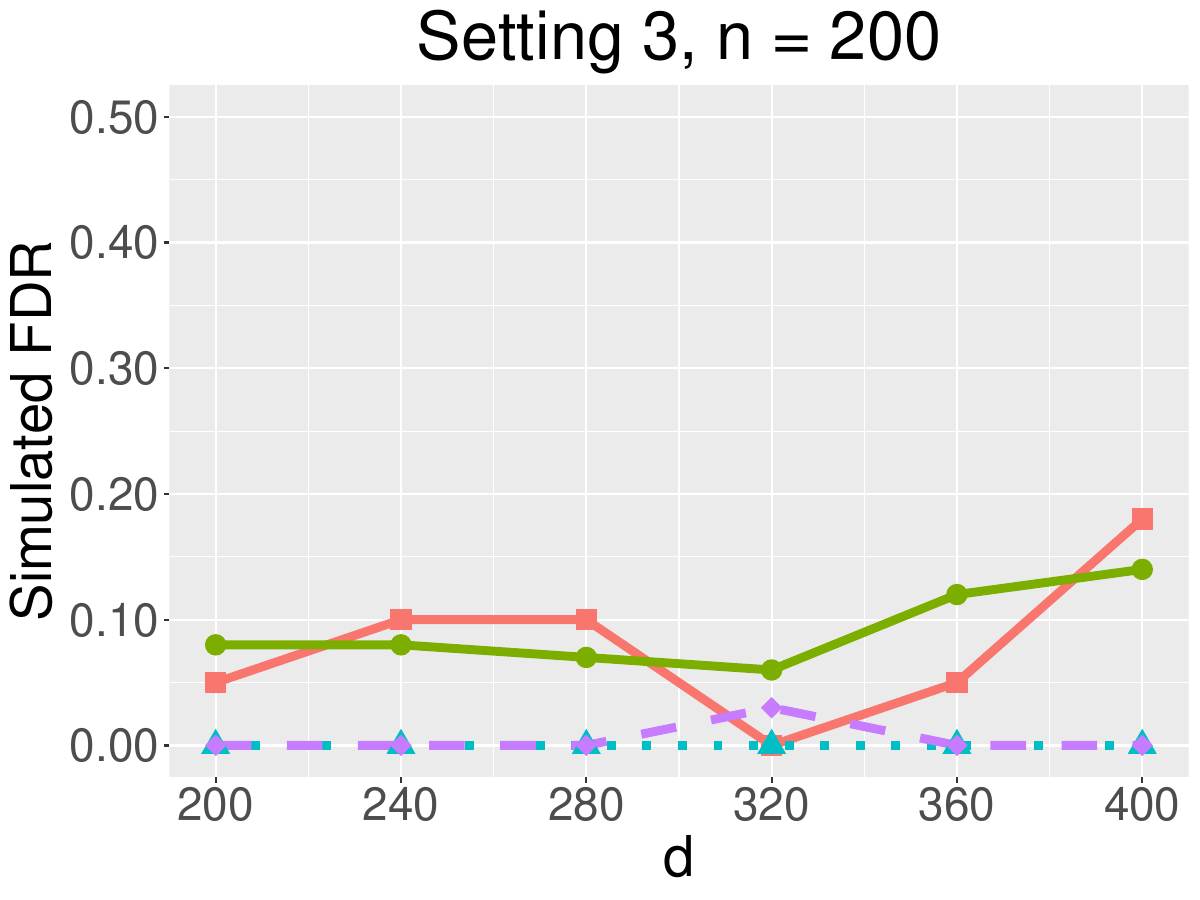}}\\
	\subfloat{\includegraphics[width=.27\columnwidth]{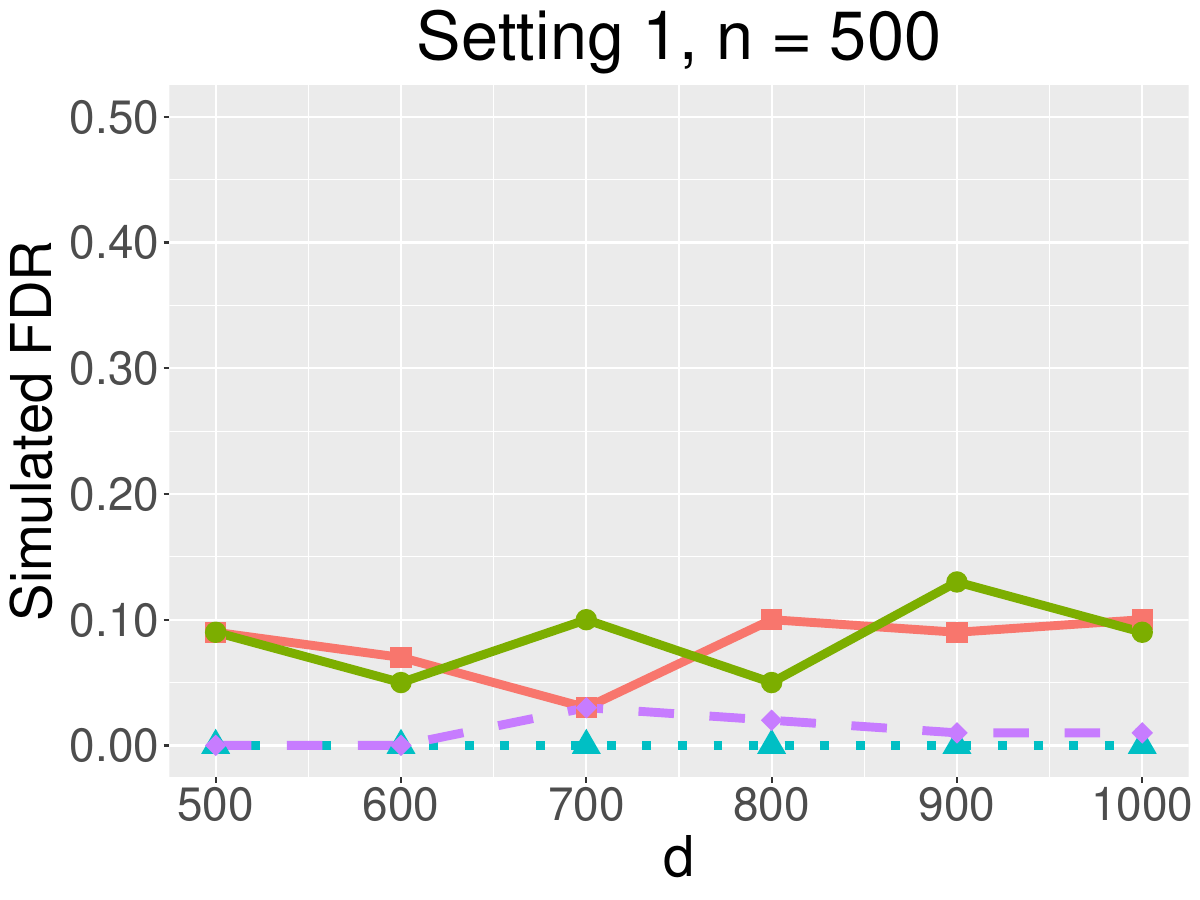}}\hspace{5pt}
    \subfloat{\includegraphics[width=.27\columnwidth]{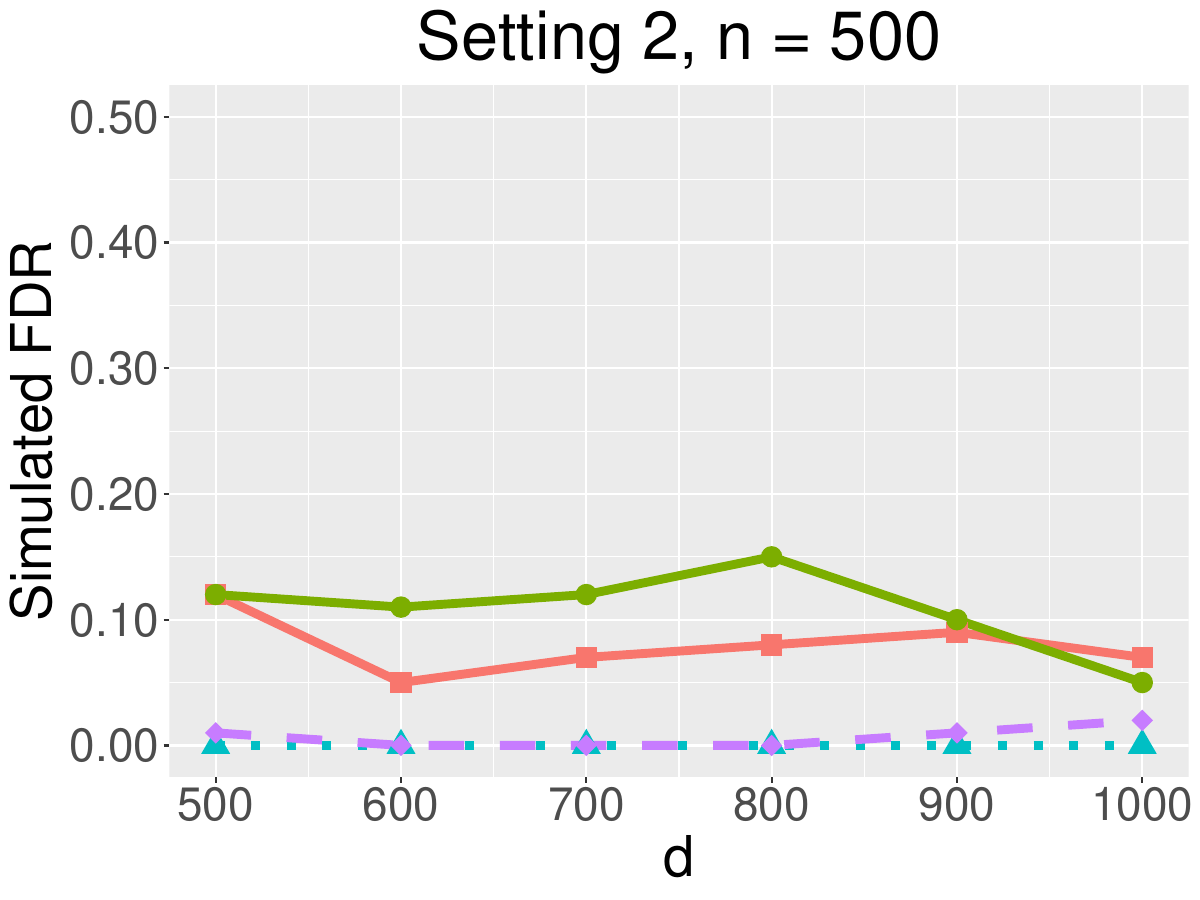}}\hspace{5pt}
    \subfloat{\includegraphics[width=.27\columnwidth]{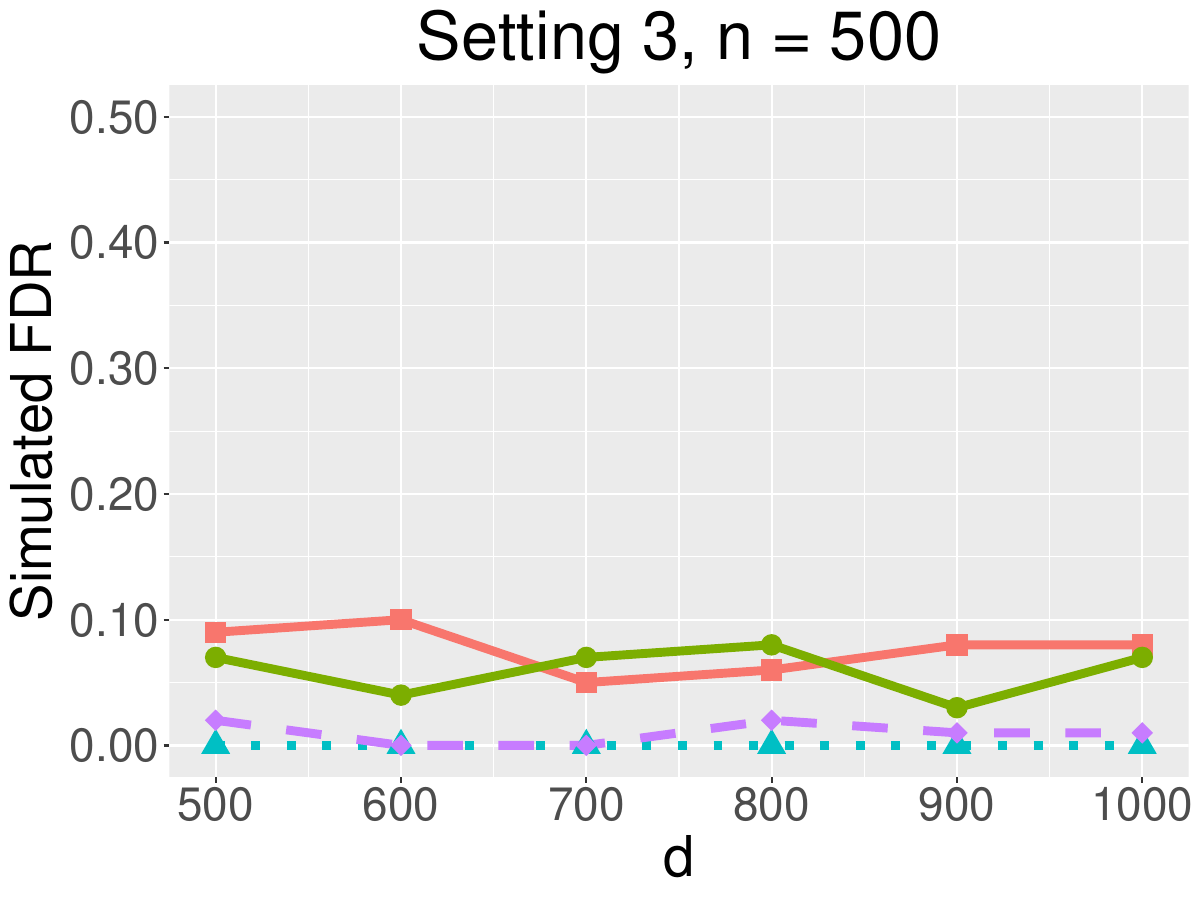}}
	\caption{\small Simulated FDR when all null hypotheses are true, for Setting 1, Setting 2, and Setting 3. The error term is generated from a normalized $t$-distribution with $5$ degrees of freedom. The FDR level is $\alpha = 0.1$. The methods compared are Algorithm 1 (squares and red solid line), Algorithm 2 (circles and green solid line), the knockoff-based method of \cite{candes2018panning} (triangles and blue dotted line), and the Gaussian Mirror method of \cite{Xing2021Controlling} with FDP+ procedure (diamonds and purple dashed line).}
    \label{FDR-control-nongaussian-0.1}
\end{figure}

\begin{figure}[htbp!]
	\centering
	\subfloat{\includegraphics[width=.27\columnwidth]{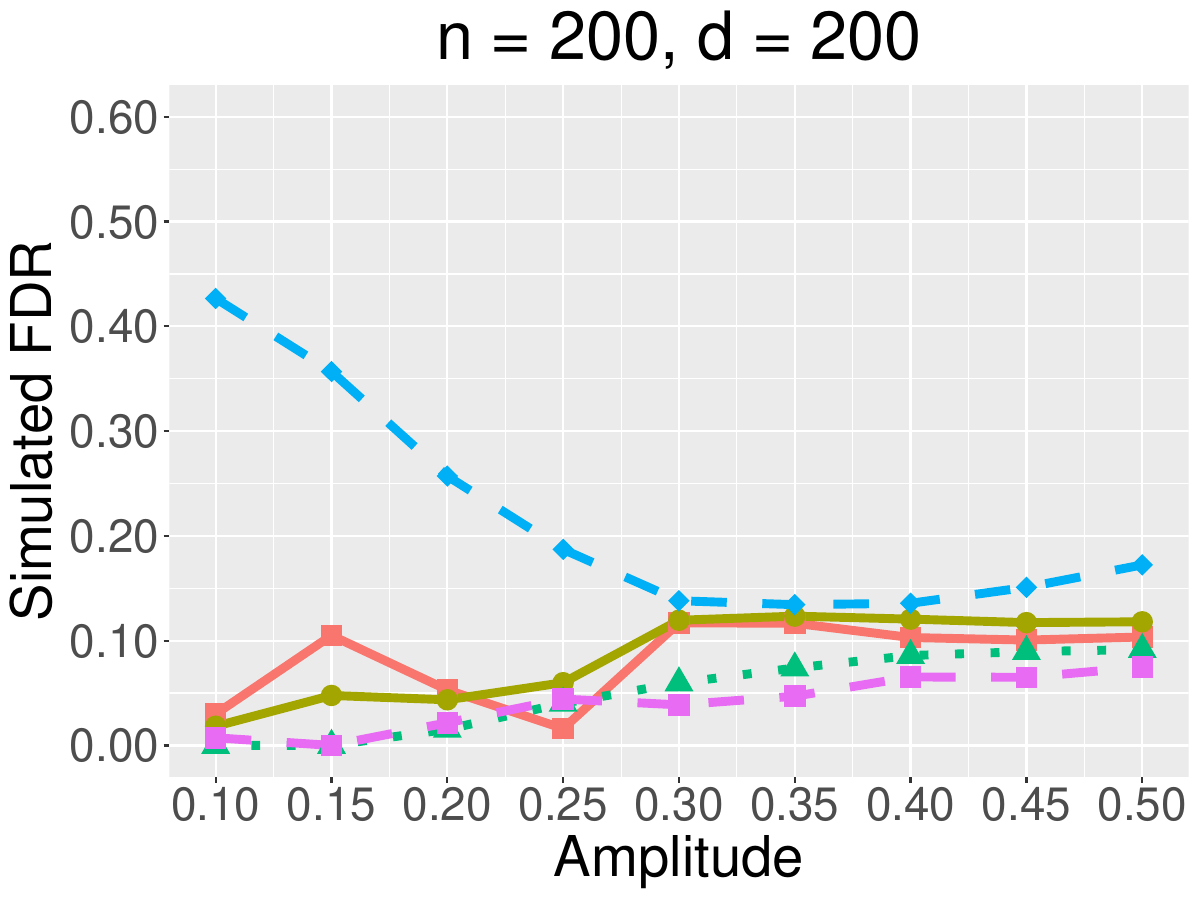}}\hspace{5pt}
	\subfloat{\includegraphics[width=.27\columnwidth]{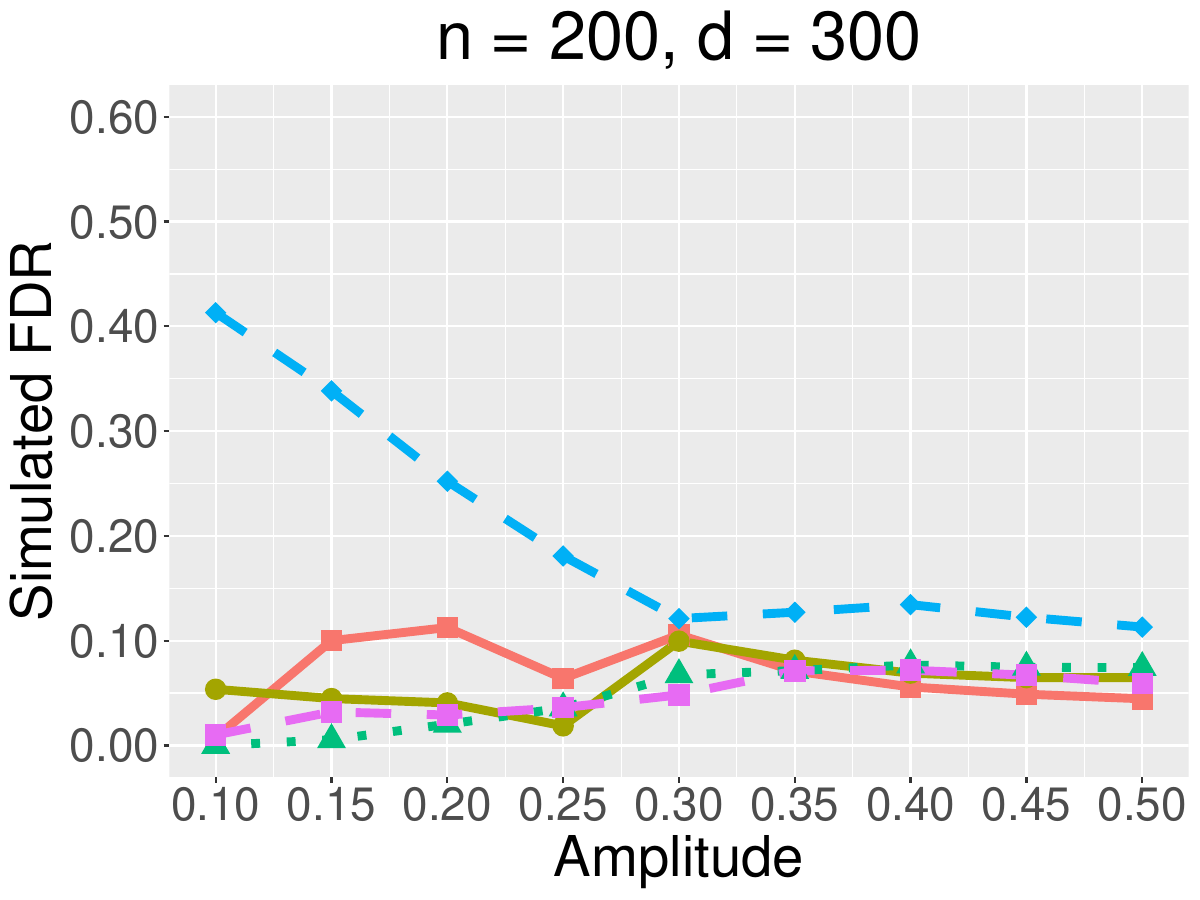}}\hspace{5pt}
	\subfloat{\includegraphics[width=.27\columnwidth]{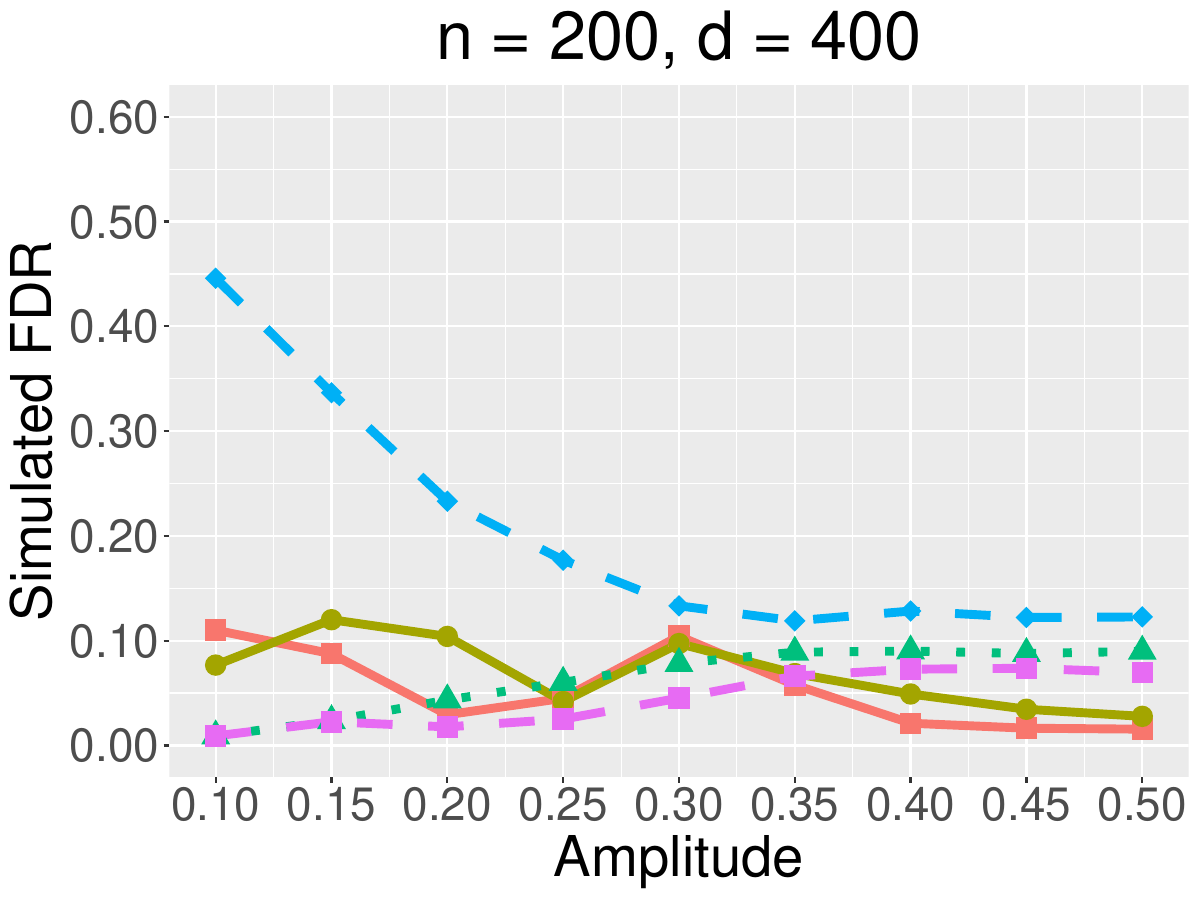}}\\
	\subfloat{\includegraphics[width=.27\columnwidth]{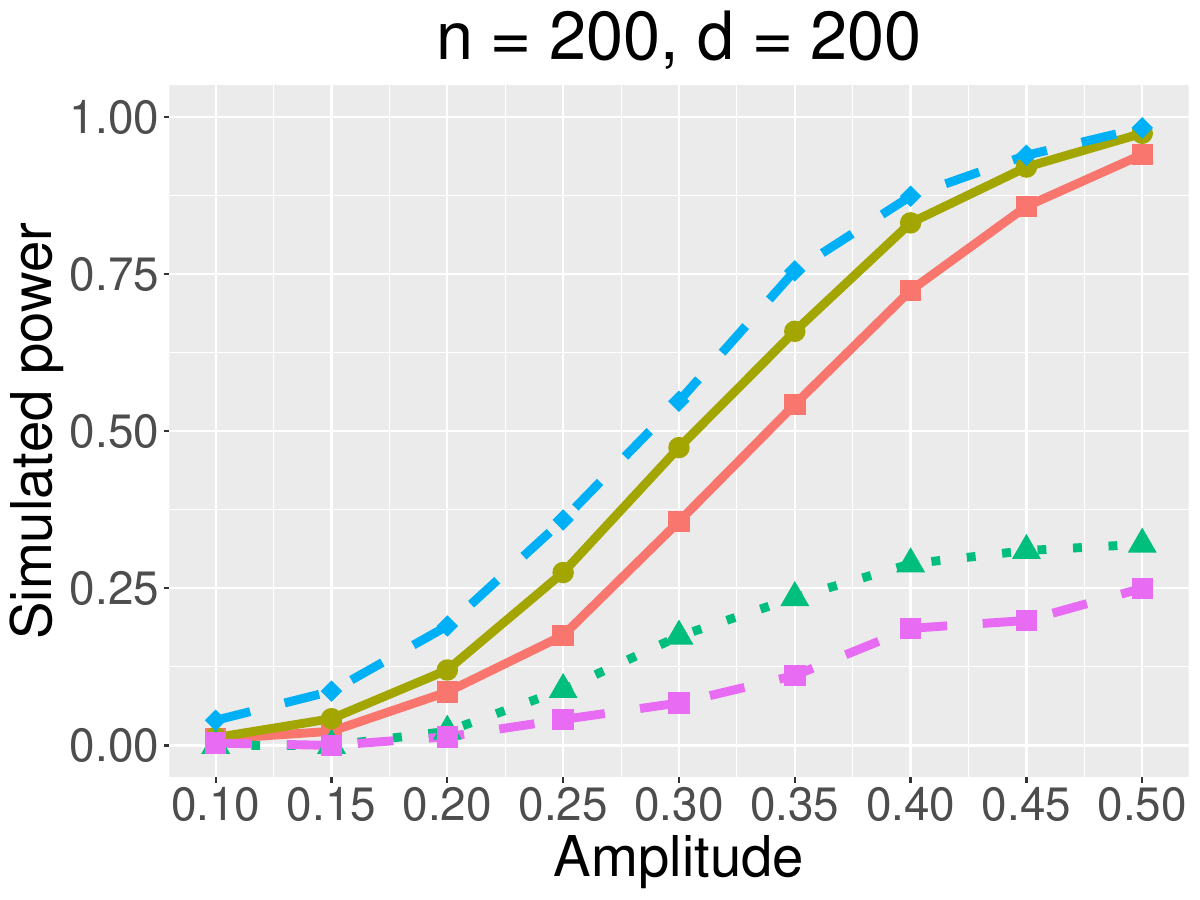}}\hspace{5pt}
    \subfloat{\includegraphics[width=.27\columnwidth]{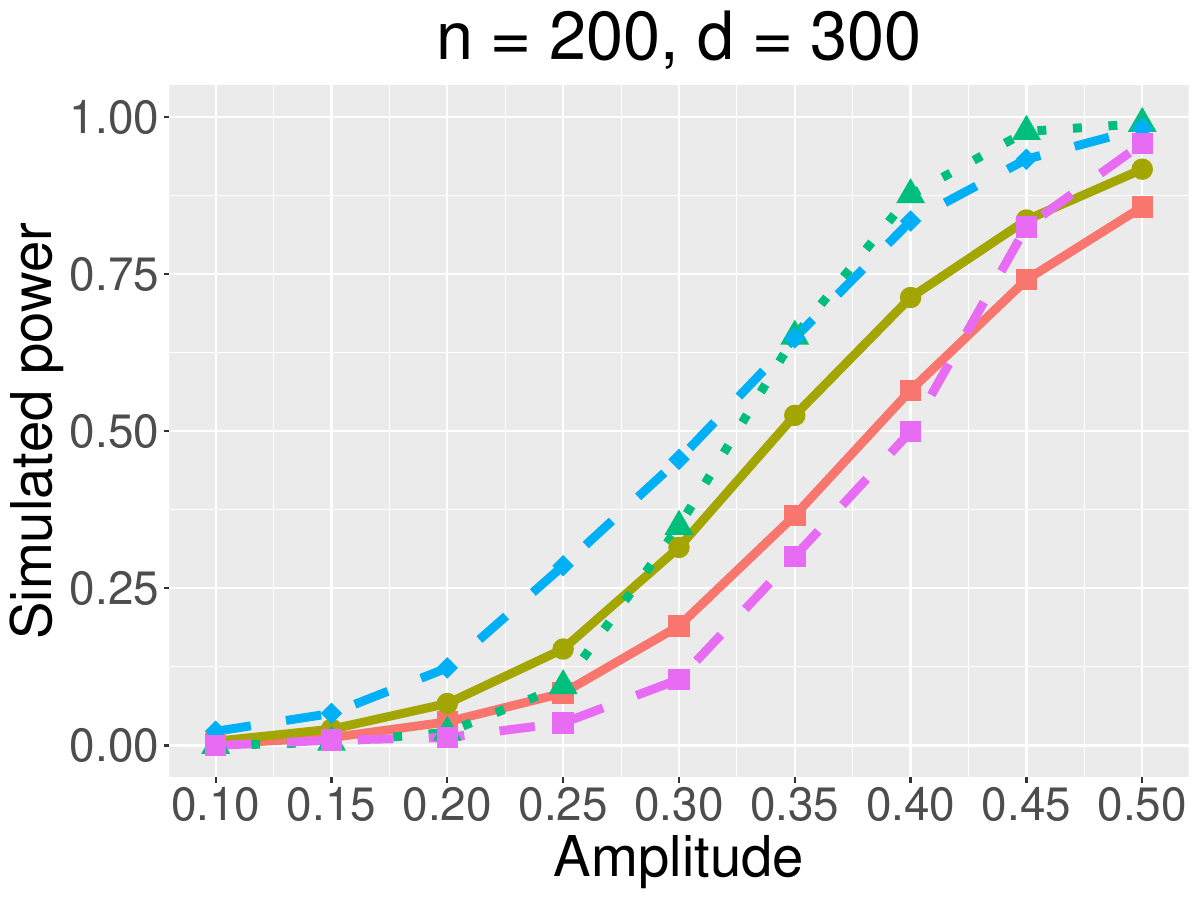}}\hspace{5pt}
    \subfloat{\includegraphics[width=.27\columnwidth]{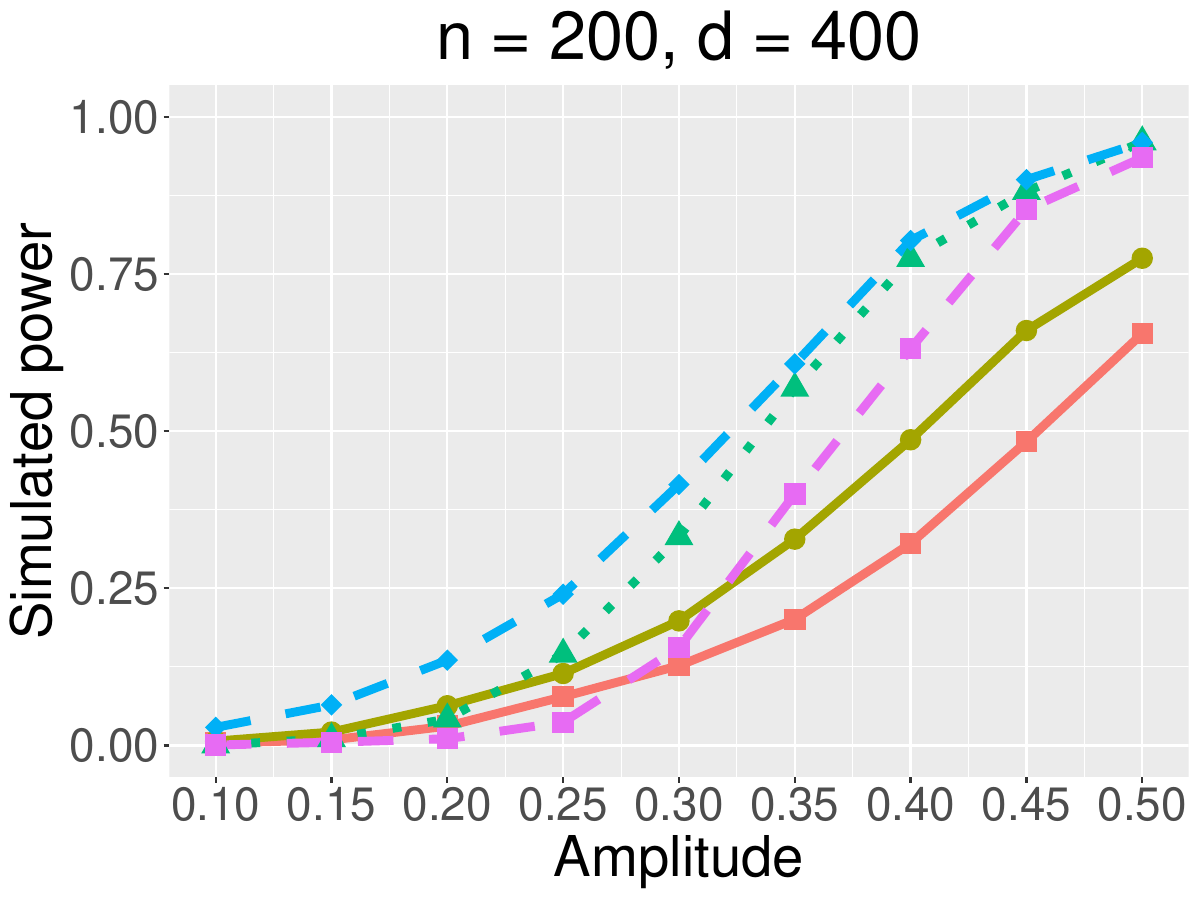}}
	\caption{\small Simulated FDR and power for different combinations of $(n,d)$. The rows of the design matrix were generated from Setting 2. The error term is generated from a normalized $t$-distribution with $5$ degrees of freedom. The sparsity level is $k = 0.04d$ and the FDR level is $\alpha = 0.1$. The methods compared are Algorithm 1 (squares and red solid line), Algorithm 2 (circles and yellow solid line), the knockoff-based method of \cite{candes2018panning} (triangles and green dotted line), the Gaussian Mirror method of \cite{Xing2021Controlling} (diamonds and blue dashed line), and the Gaussian Mirror method with FDP+ procedure (squares and purple dashed line).}
    \label{Power-nongaussian-0.1}
\end{figure}

\begin{figure}[htbp!]
	\centering
	\subfloat{\includegraphics[width=.27\columnwidth]{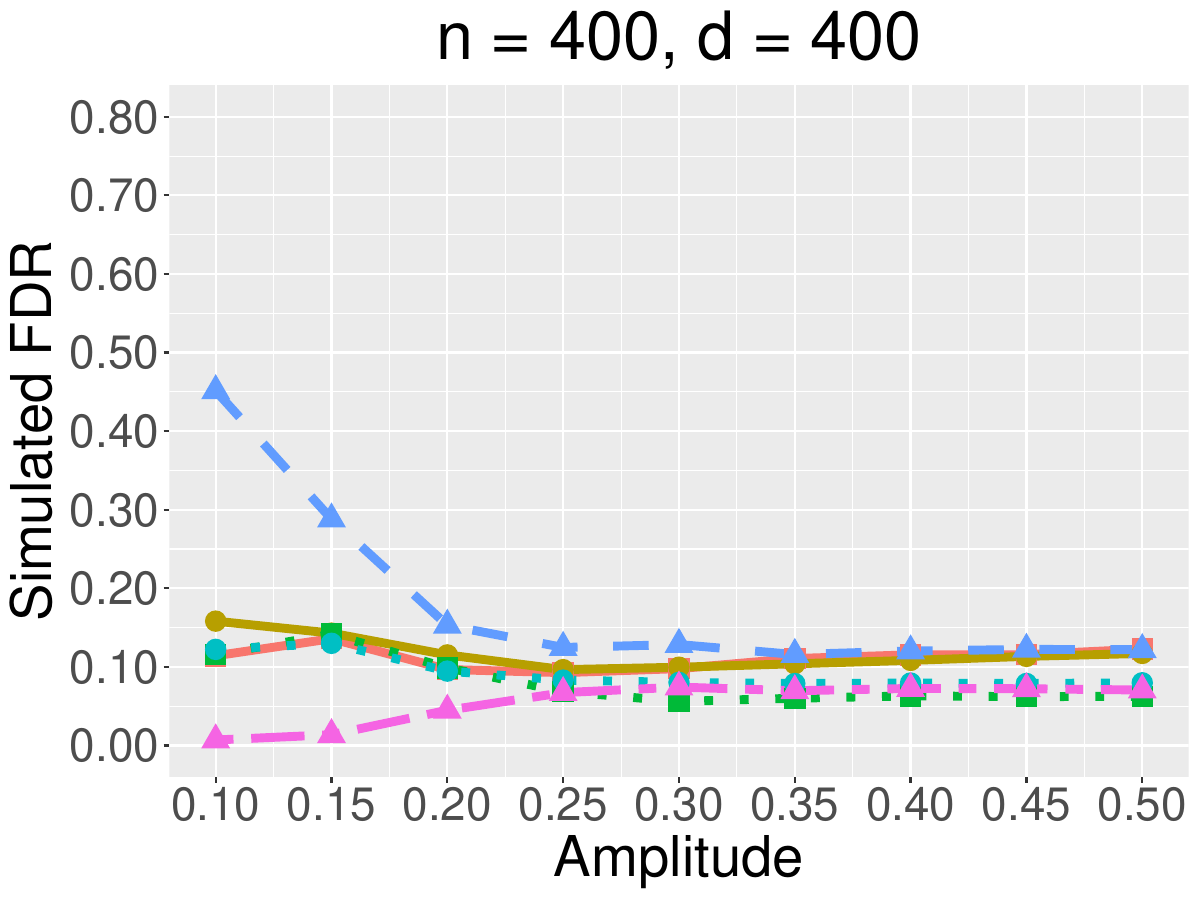}}\hspace{5pt}
	\subfloat{\includegraphics[width=.27\columnwidth]{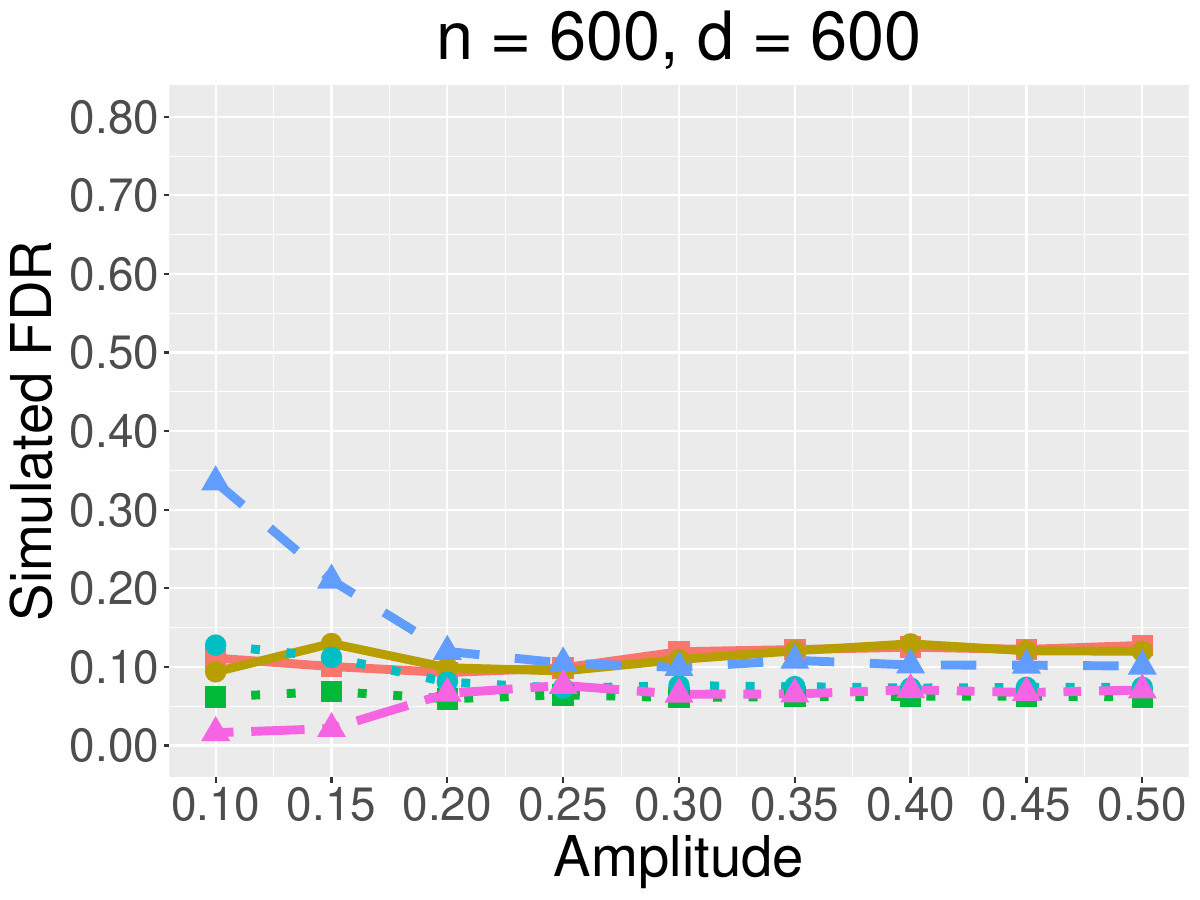}}\hspace{5pt}
	\subfloat{\includegraphics[width=.27\columnwidth]{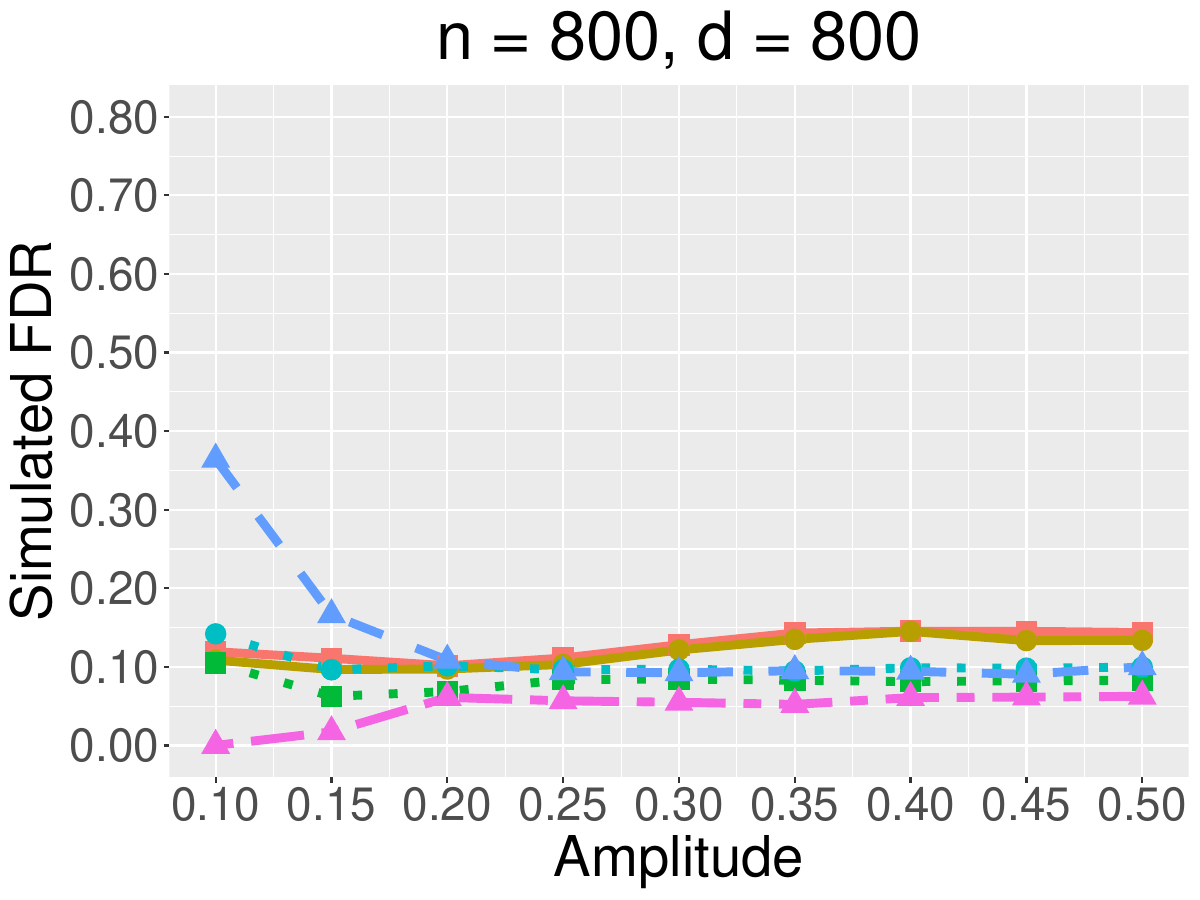}}\\
	\subfloat{\includegraphics[width=.27\columnwidth]{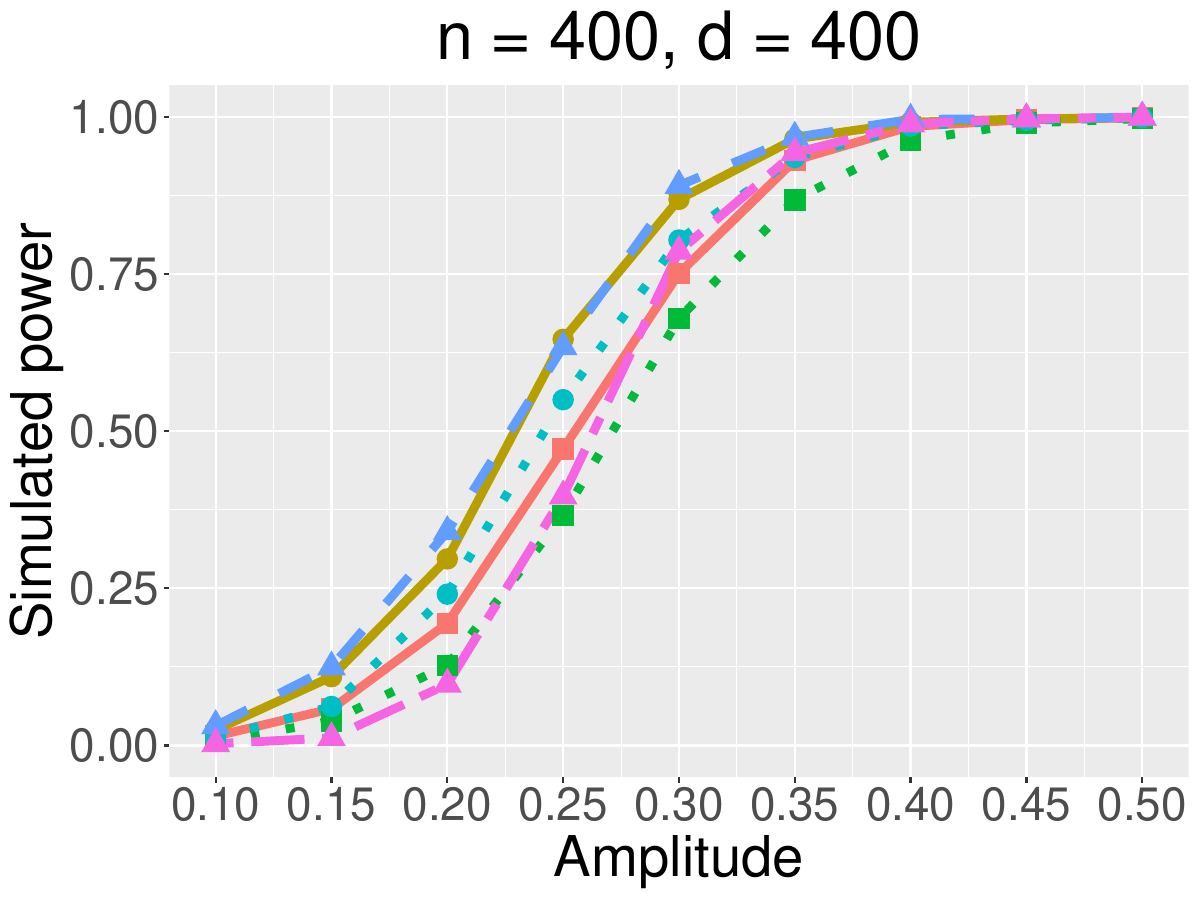}}\hspace{5pt}
    \subfloat{\includegraphics[width=.27\columnwidth]{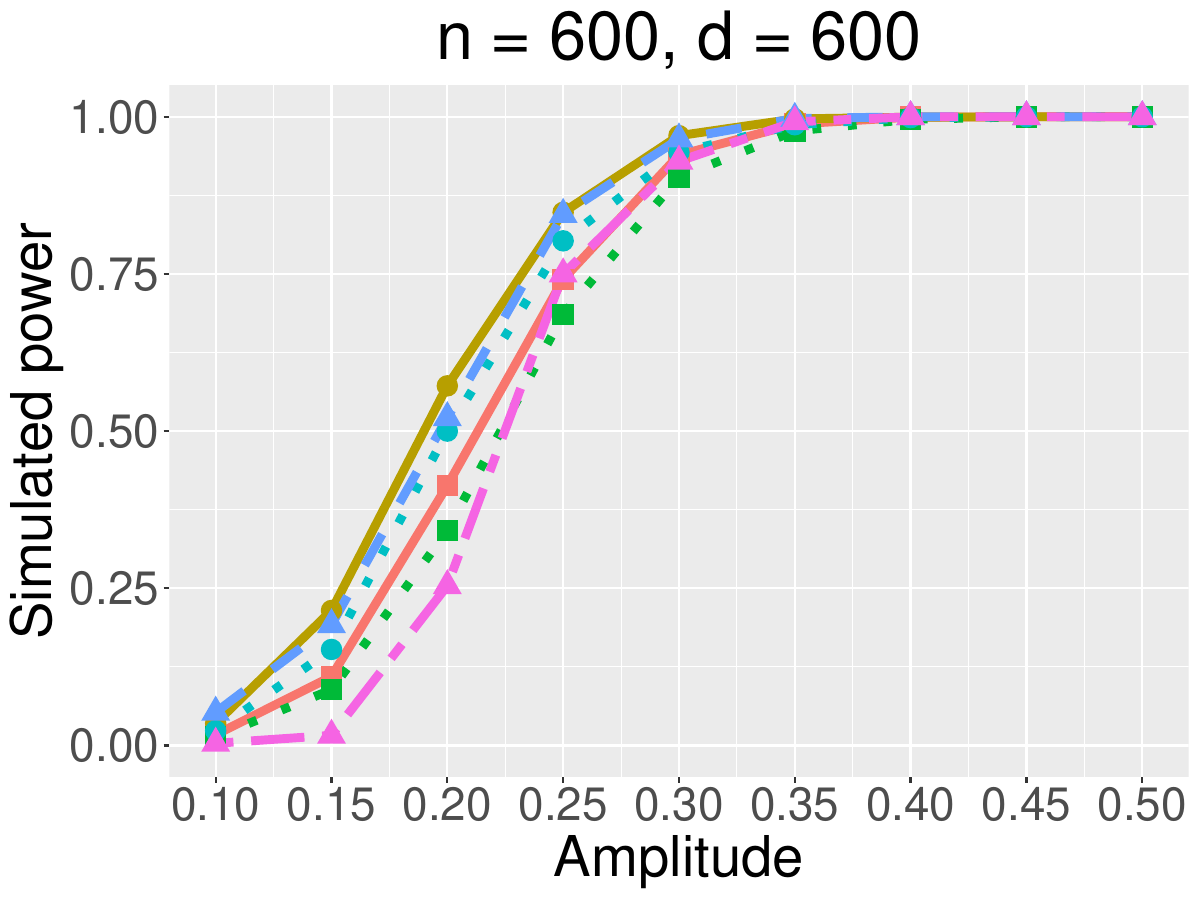}}\hspace{5pt}
    \subfloat{\includegraphics[width=.27\columnwidth]{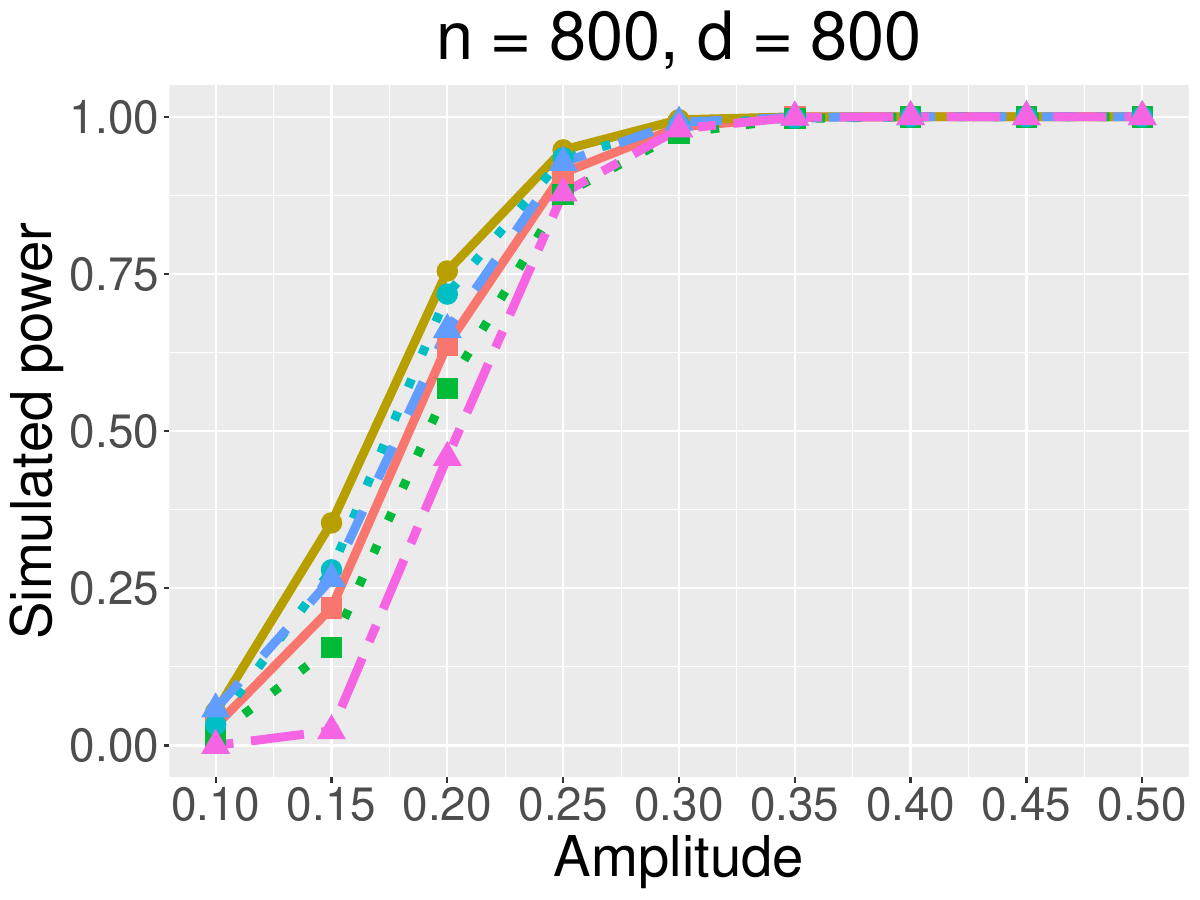}}
	\caption{\small Simulated FDR and power for different combinations of $(n,d)$. The rows of the design matrix were generated from Setting 2. The error term is generated from a normalized $t$-distribution with $5$ degrees of freedom. The sparsity level is $k = 15$ and the FDR level is $\alpha = 0.1$. The methods compared are Algorithm 1 (squares and green dotted line), two-stage Algorithm 1 (squares and red solid line), Algorithm 2 (circles and blue dotted line), two-stage Algorithm 2 (circles and yellow solid line), the Gaussian Mirror method of \cite{Xing2021Controlling} (triangles and blue dashed line), and the Gaussian Mirror method with FDP+ procedure (triangles and purple two-dashed line).}
    \label{Two-stage-nongaussian-0.1}
\end{figure}

Overall, the findings from these simulation studies are consistent with those reported in Section~\ref{sec:sim}. Representative results are presented here. Figure~\ref{FDR-control-nongaussian-0.1} shows that the proposed method remains stable in controlling the empirical FDR. Figure~\ref{Power-nongaussian-0.1} illustrates that our method remains competitive in terms of power. Figure~\ref{Two-stage-nongaussian-0.1} further shows that, after incorporating an additional screening stage, the resulting two-stage procedure can further improve power. These findings confirm the competitive performance of our method across a broad range of settings.

\section{Additional results of the real data analysis}\label{SM3:figure}

\begin{figure}[htbp!]
	\centering
	\subfloat[$n = 767, \ d = 201$]{\includegraphics[width=.31\columnwidth]{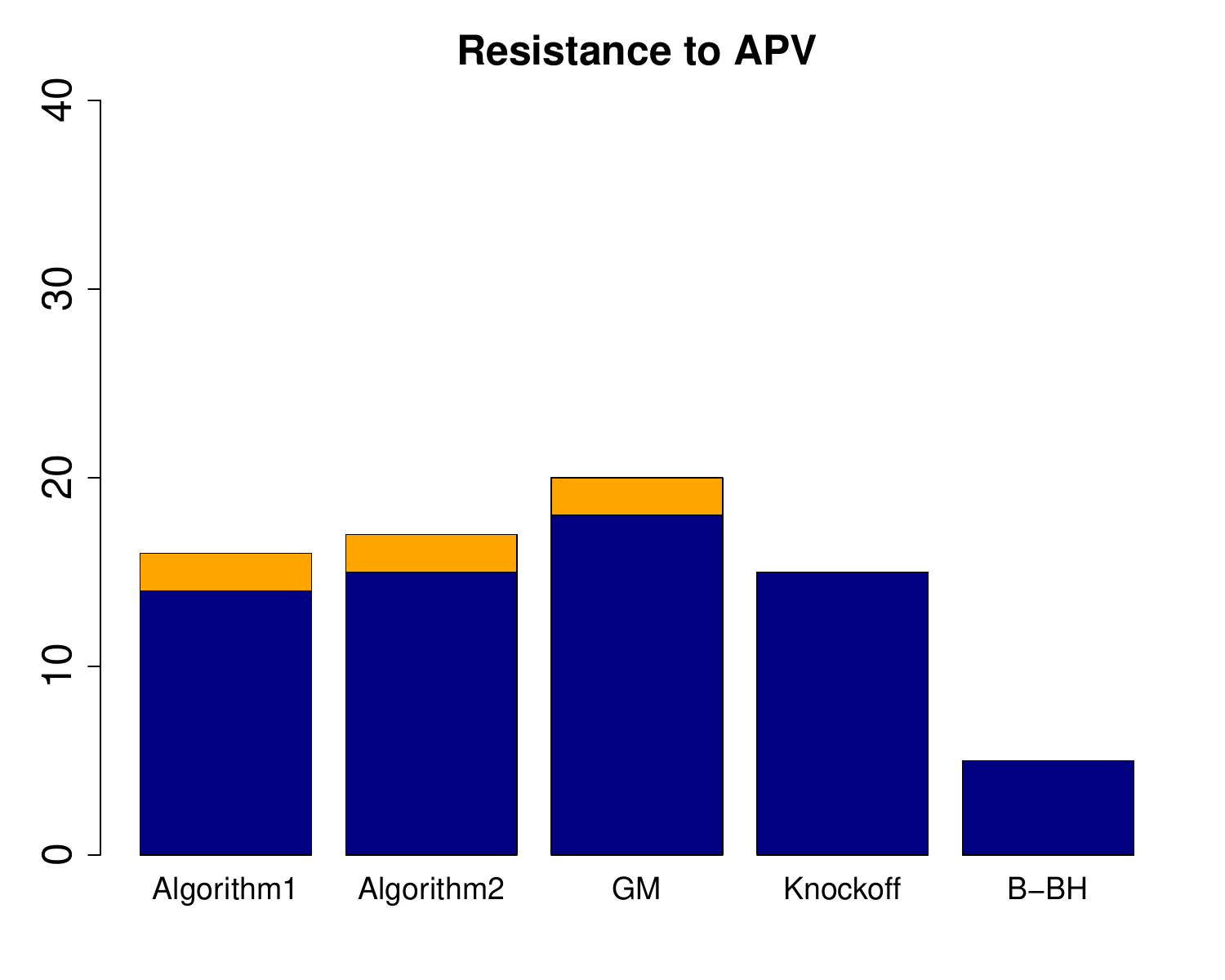}}\hspace{5pt}
	\subfloat[$n = 328, \ d = 147$]{\includegraphics[width=.31\columnwidth]{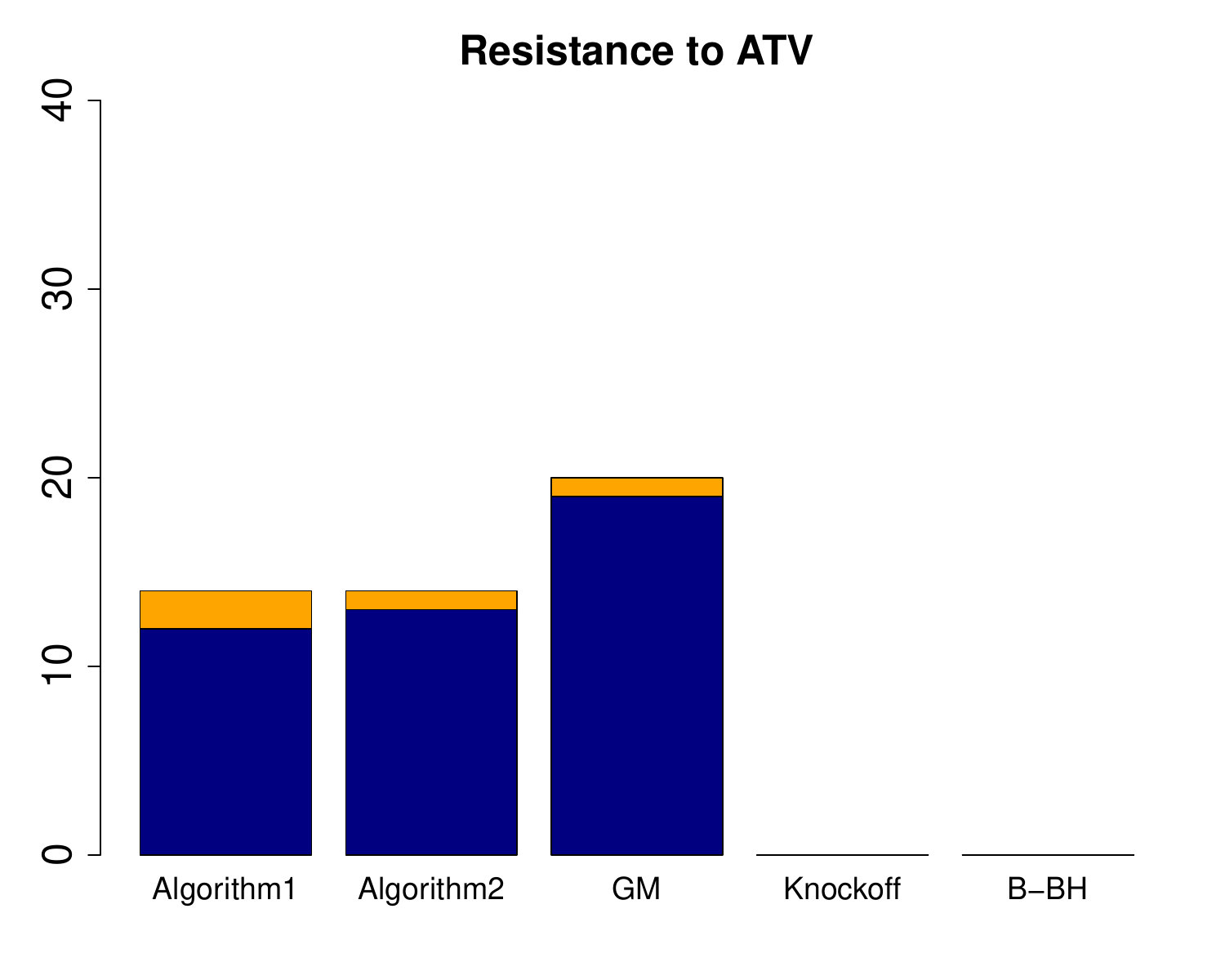}}\hspace{5pt}
	\subfloat[$n = 825, \ d = 206$]{\includegraphics[width=.31\columnwidth]{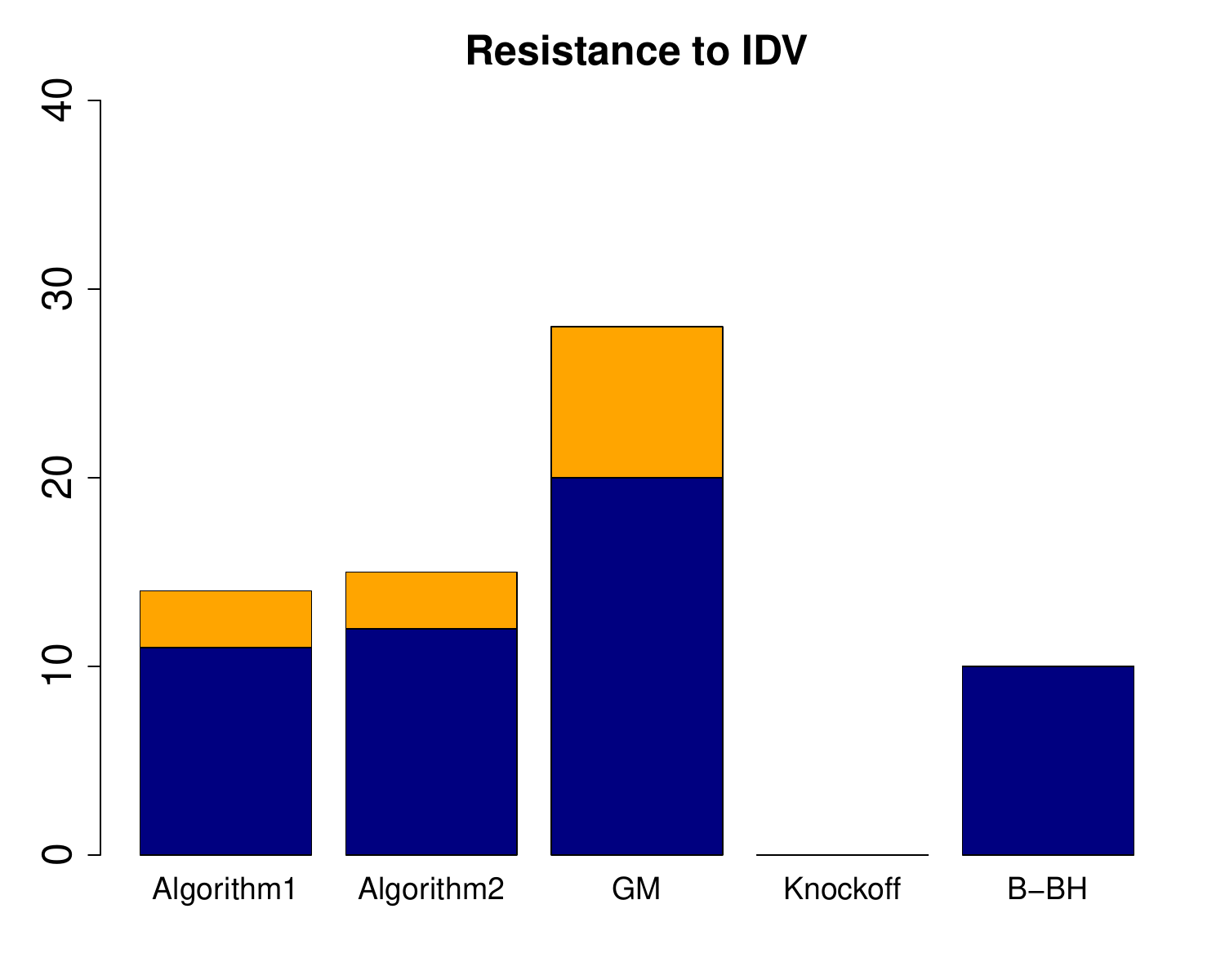}}\\
	\subfloat[$n = 515, \ d = 184$]{\includegraphics[width=.31\columnwidth]{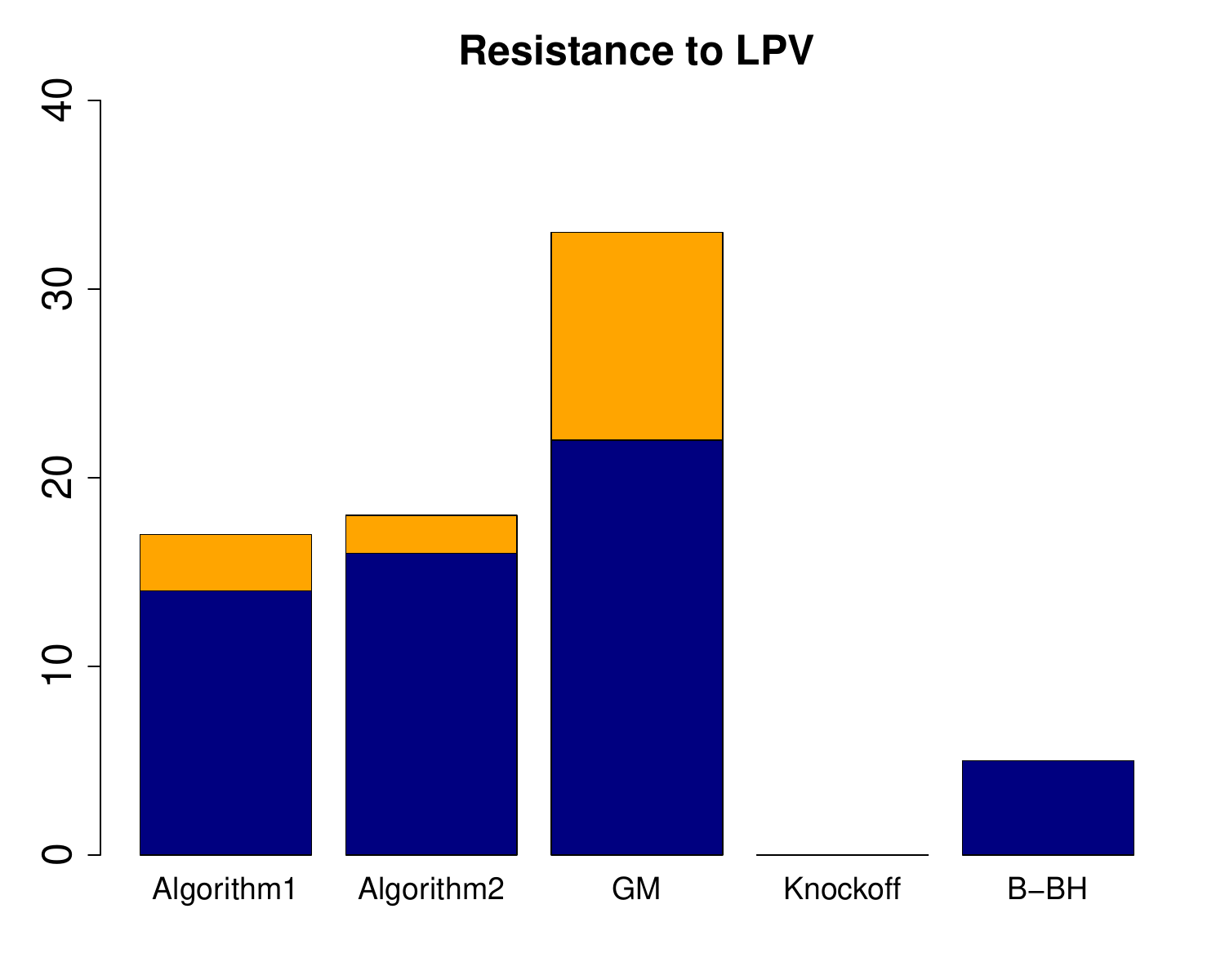}}\hspace{5pt}
    \subfloat[$n = 842, \ d = 207$]{\includegraphics[width=.31\columnwidth]{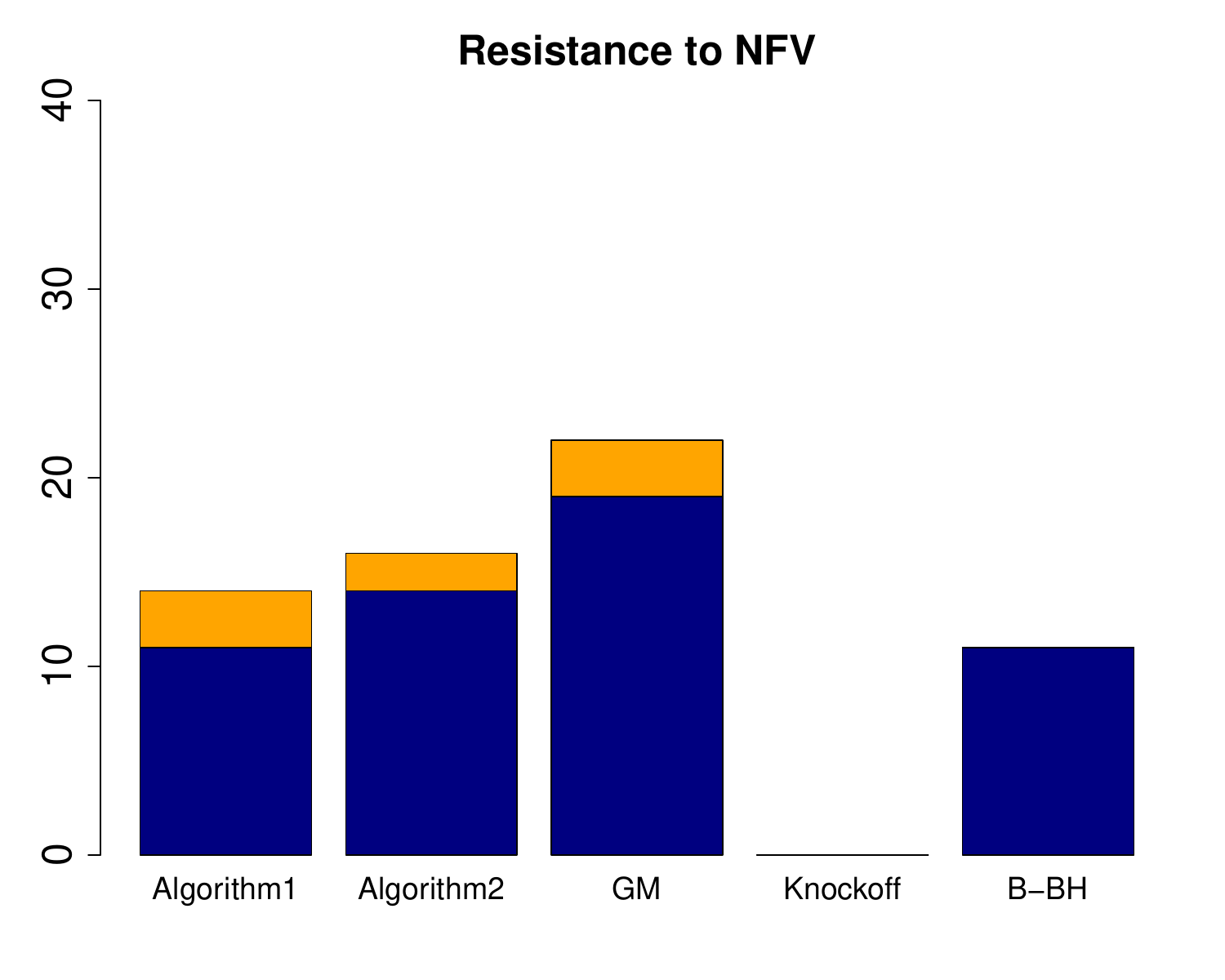}}\hspace{5pt}
    \subfloat[$n = 793, \ d = 205$]{\includegraphics[width=.31\columnwidth]{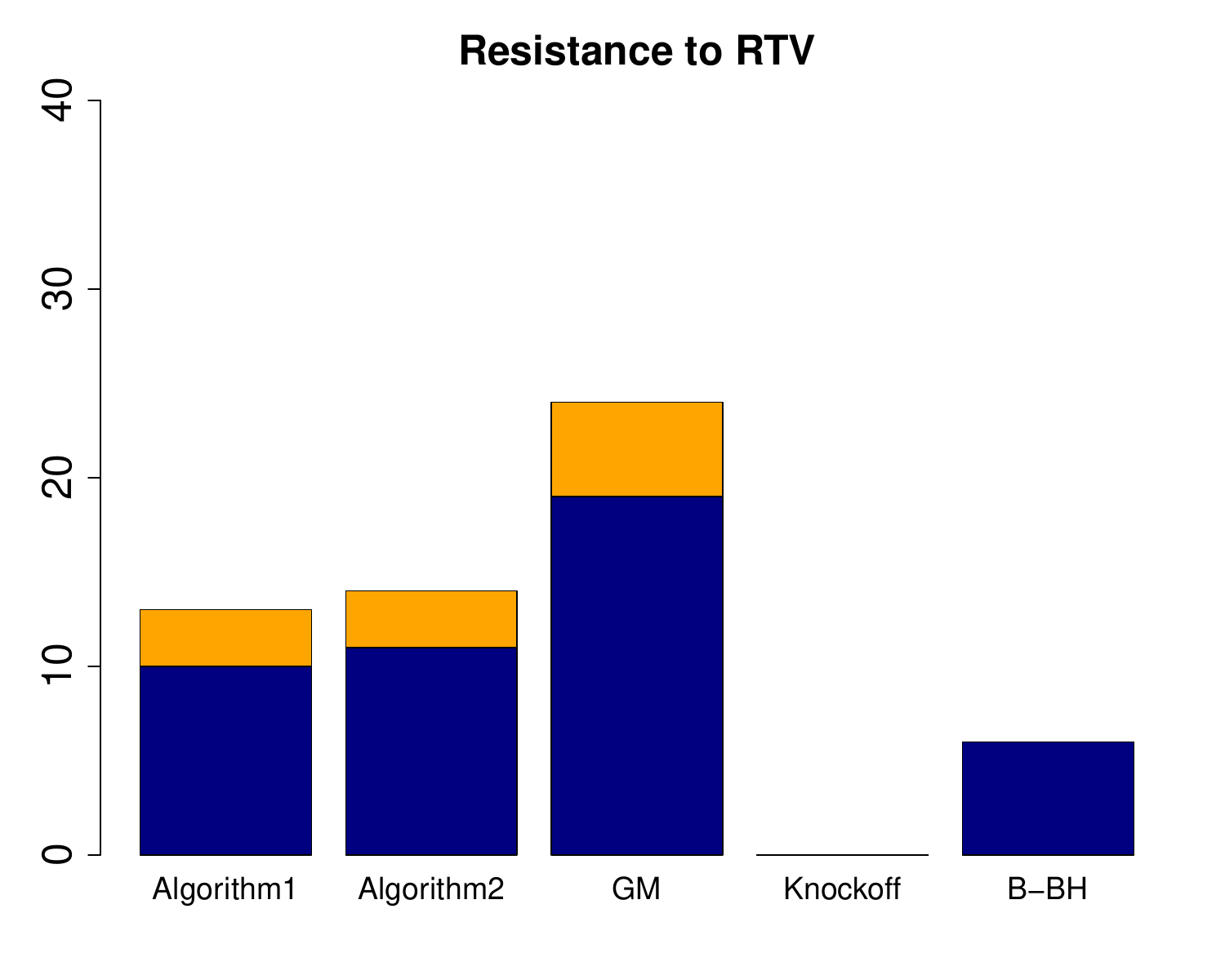}}
	\caption{\small Results of the real data example for $\alpha = 0.05$.  Blue represents the number of discoveries that are in the treatment-selected mutation panels list, and yellow represents the number of discoveries not in the treatment-selected mutation panels list. The total number of HIV-1 protease positions in the treatment-selected mutation panels list is $34$. The methods compared are the proposed Algorithm 1 (Algorithm1), the proposed Algorithm 2 (Algorithm2), the Gaussian Mirror method of \cite{Xing2021Controlling} (GM), the knockoff-based method of \cite{Barber2015Controlling} (Knockoff), and the Bonferroni-Benjamini-Hochberg method of \cite{Sarkar2022Adjusting} (B-BH). }
    \label{fig_realdata_0.05}
\end{figure}

\begin{figure}[htbp!]
	\centering
	\subfloat[$n = 767, \ d = 201$]{\includegraphics[width=.31\columnwidth]{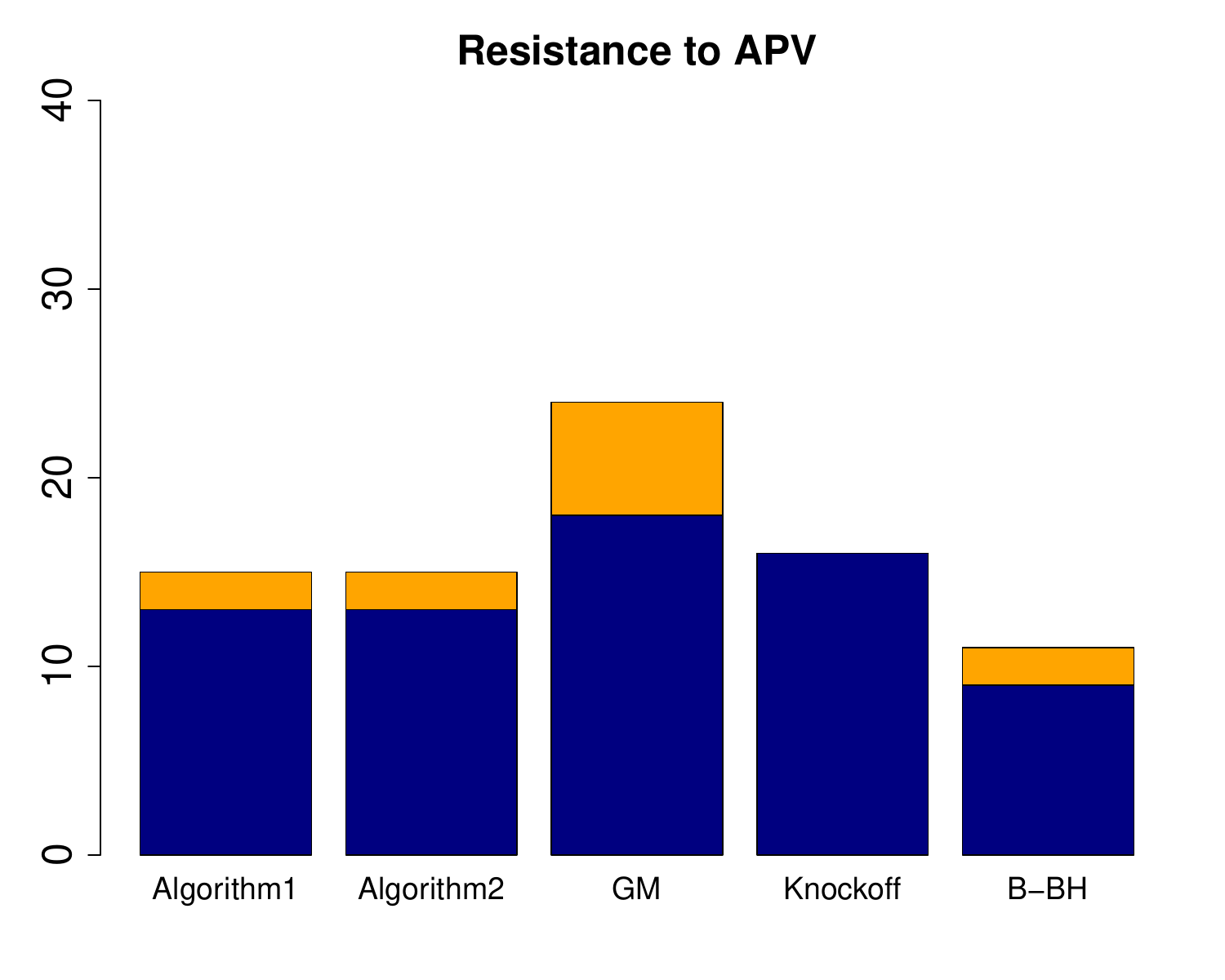}}\hspace{5pt}
	\subfloat[$n = 328, \ d = 147$]{\includegraphics[width=.31\columnwidth]{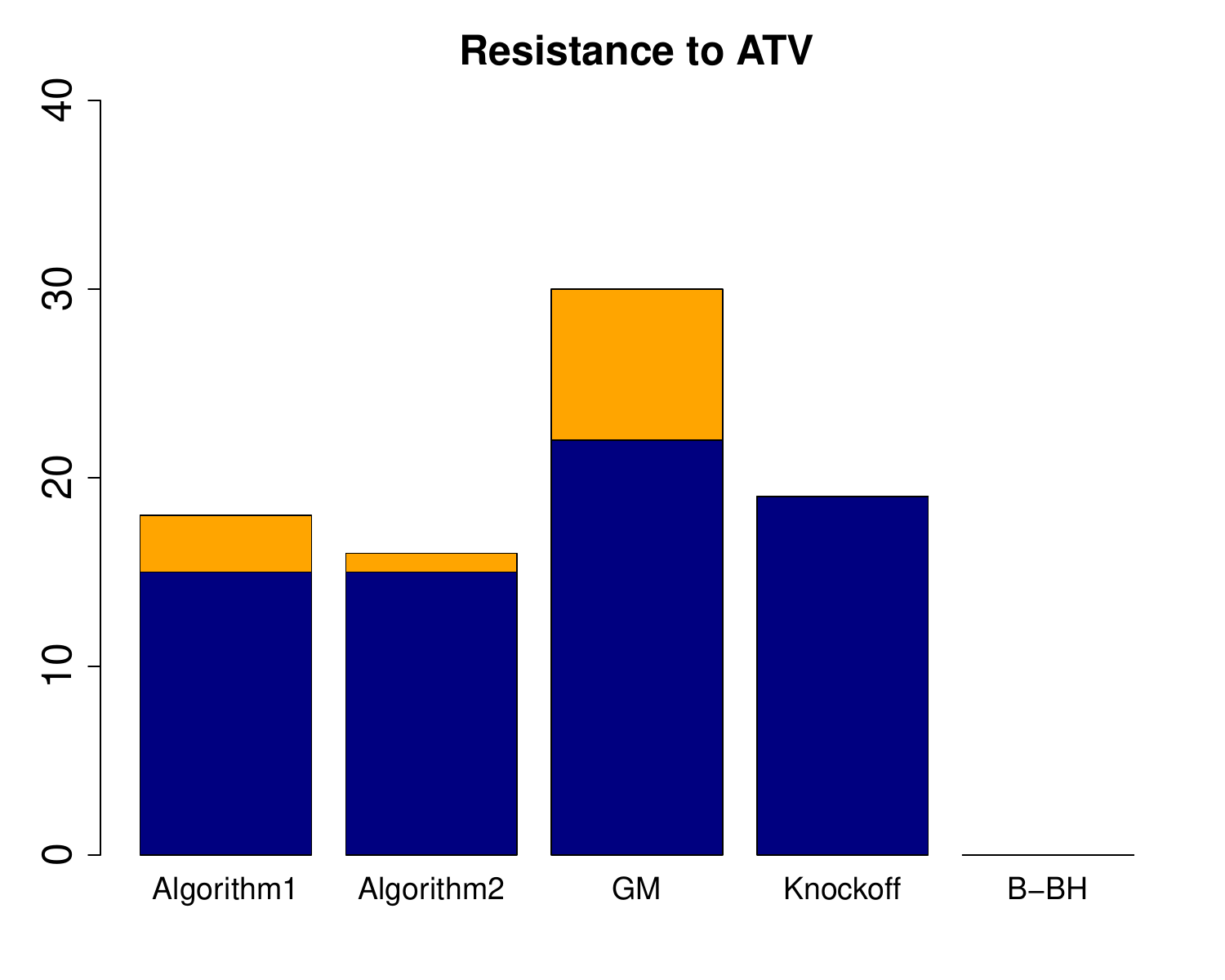}}\hspace{5pt}
	\subfloat[$n = 825, \ d = 206$]{\includegraphics[width=.31\columnwidth]{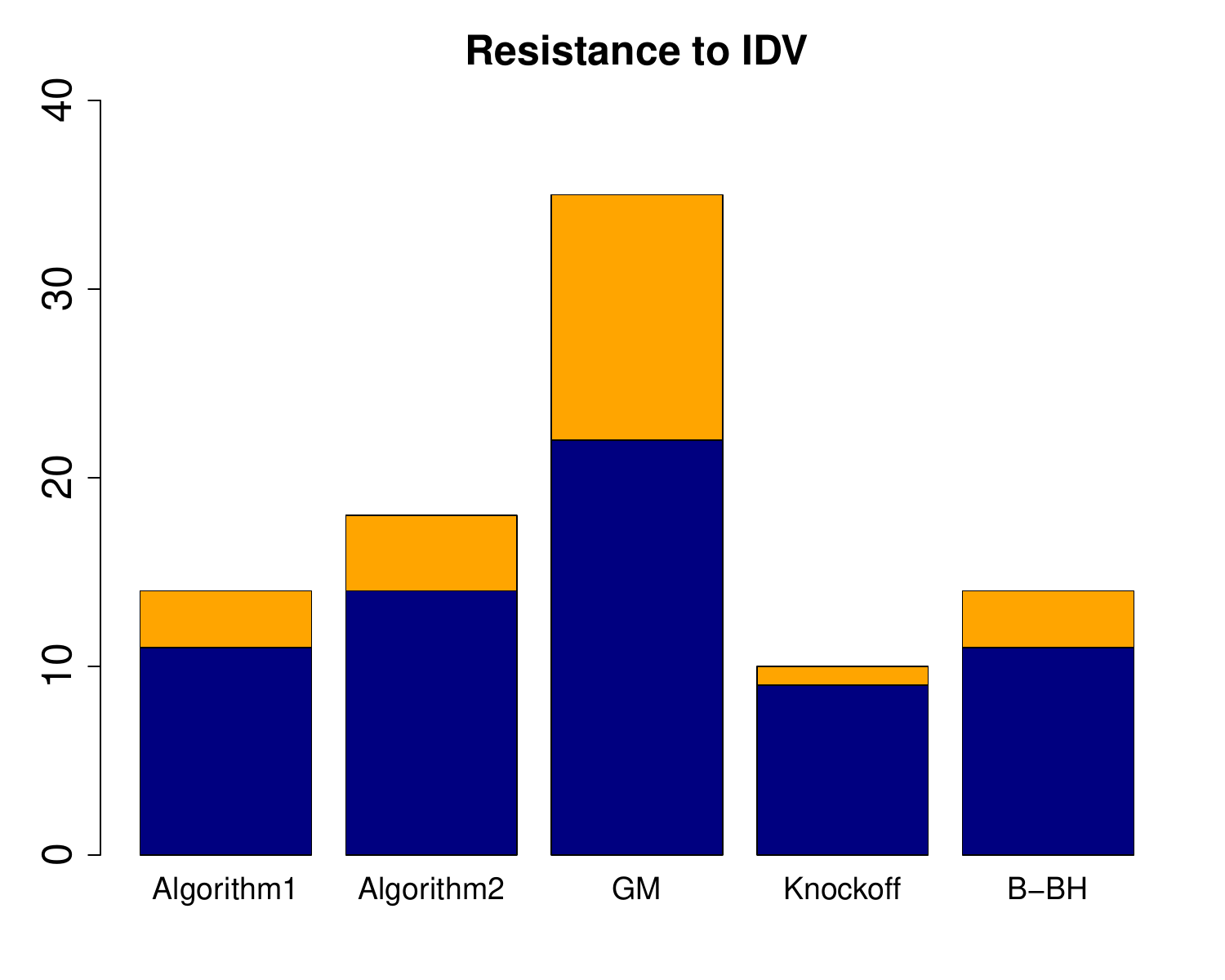}}\\
	\subfloat[$n = 515, \ d = 184$]{\includegraphics[width=.31\columnwidth]{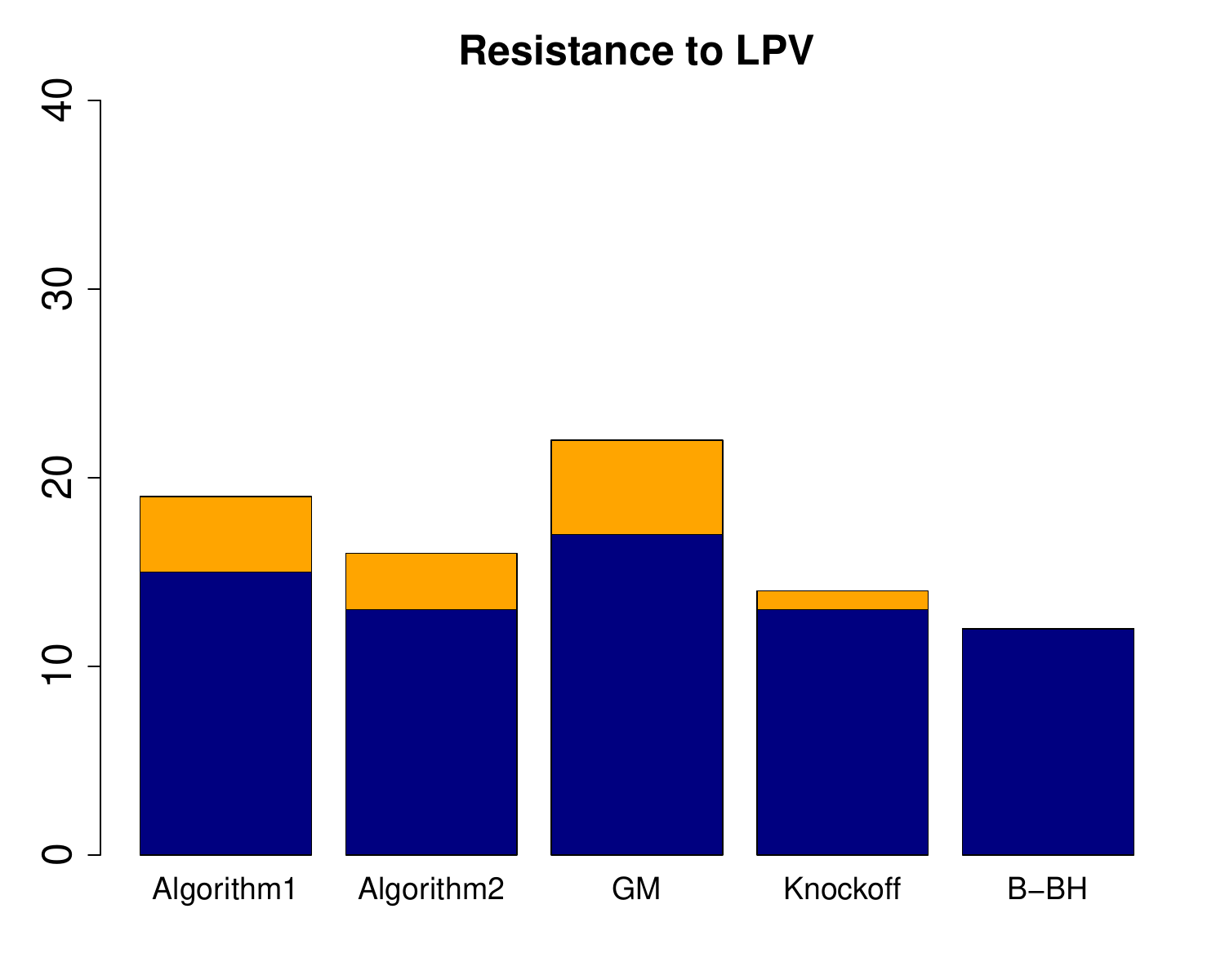}}\hspace{5pt}
    \subfloat[$n = 842, \ d = 207$]{\includegraphics[width=.31\columnwidth]{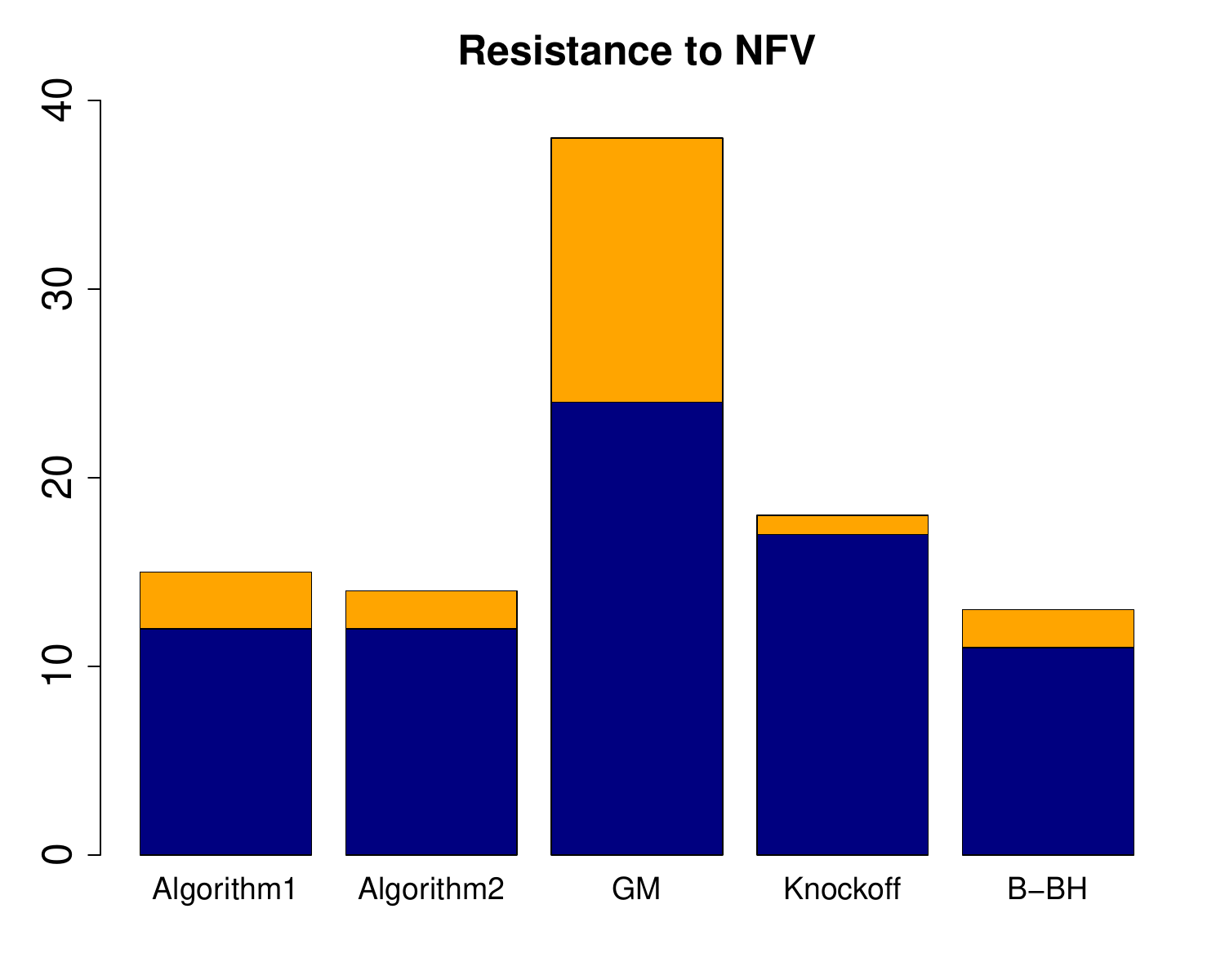}}\hspace{5pt}
    \subfloat[$n = 793, \ d = 205$]{\includegraphics[width=.31\columnwidth]{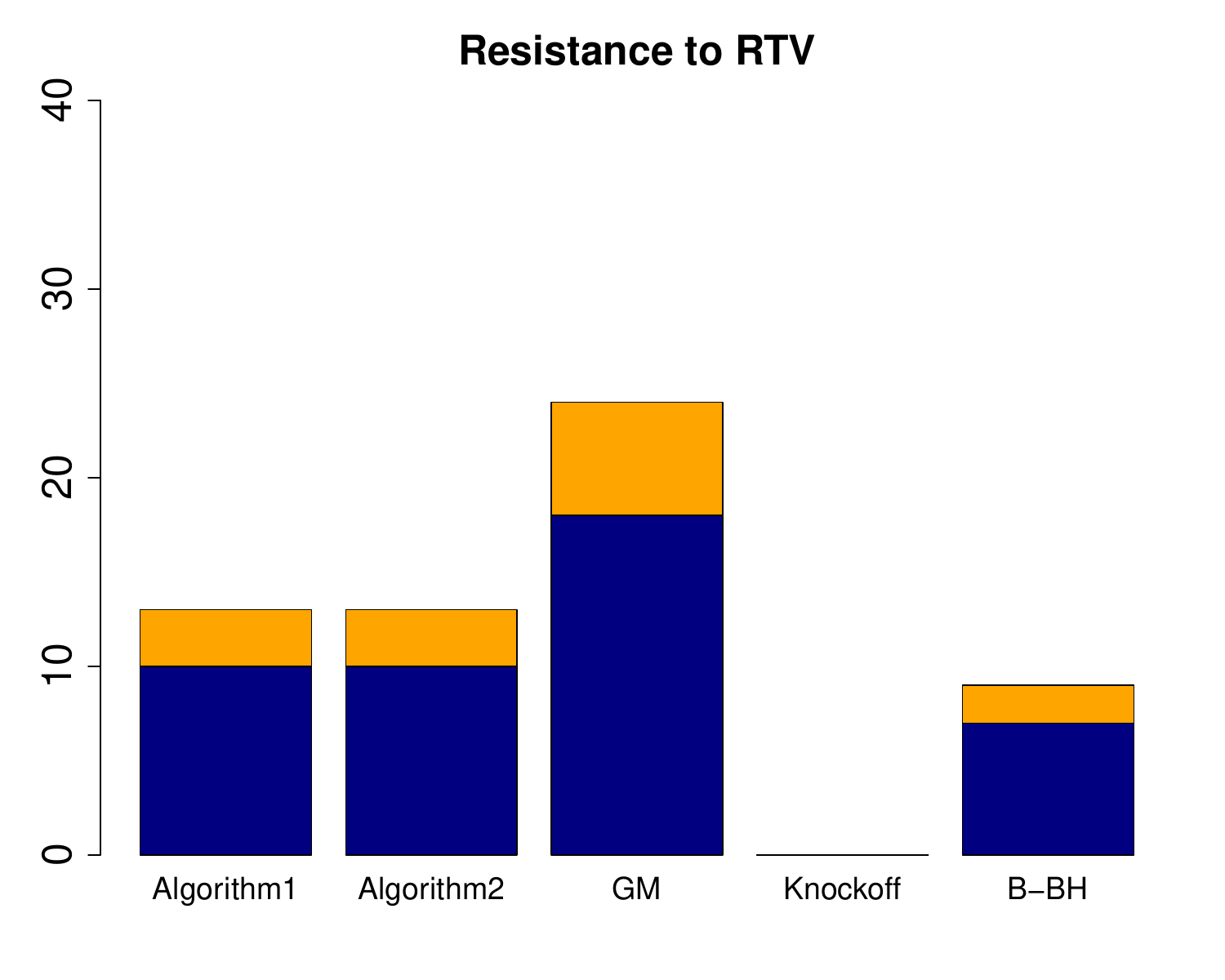}}
	\caption{\small Results of the real data example for $\alpha = 0.1$.  Blue represents the number of discoveries that are in the treatment-selected mutation panels list, and yellow represents the number of discoveries not in the treatment-selected mutation panels list. The total number of HIV-1 protease positions in the treatment-selected mutation panels list is $34$. The methods compared are the proposed Algorithm 1 (Algorithm1), the proposed Algorithm 2 (Algorithm2), the Gaussian Mirror method of \cite{Xing2021Controlling} (GM), the knockoff-based method of \cite{Barber2015Controlling} (Knockoff), and the Bonferroni-Benjamini-Hochberg method of \cite{Sarkar2022Adjusting} (B-BH). }
        \label{fig_realdata_0.1}
\end{figure}

Additional results of the real data analysis are included in Figures \ref{fig_realdata_0.05} and \ref{fig_realdata_0.1}. 

\section{Simulation results for the extended framework in Section~\ref{F:F_0}}\label{simu_ds}

Section~\ref{F:F_0} presents an extension of our framework that accommodates an estimated precision matrix and allows for a broad class of feature distributions in the design matrix. In this subsection, we present numerical experiments demonstrating the finite-sample performance of the method developed in that section.

\begin{figure}[htbp!]
	\centering
	\subfloat{\includegraphics[width=.27\columnwidth]{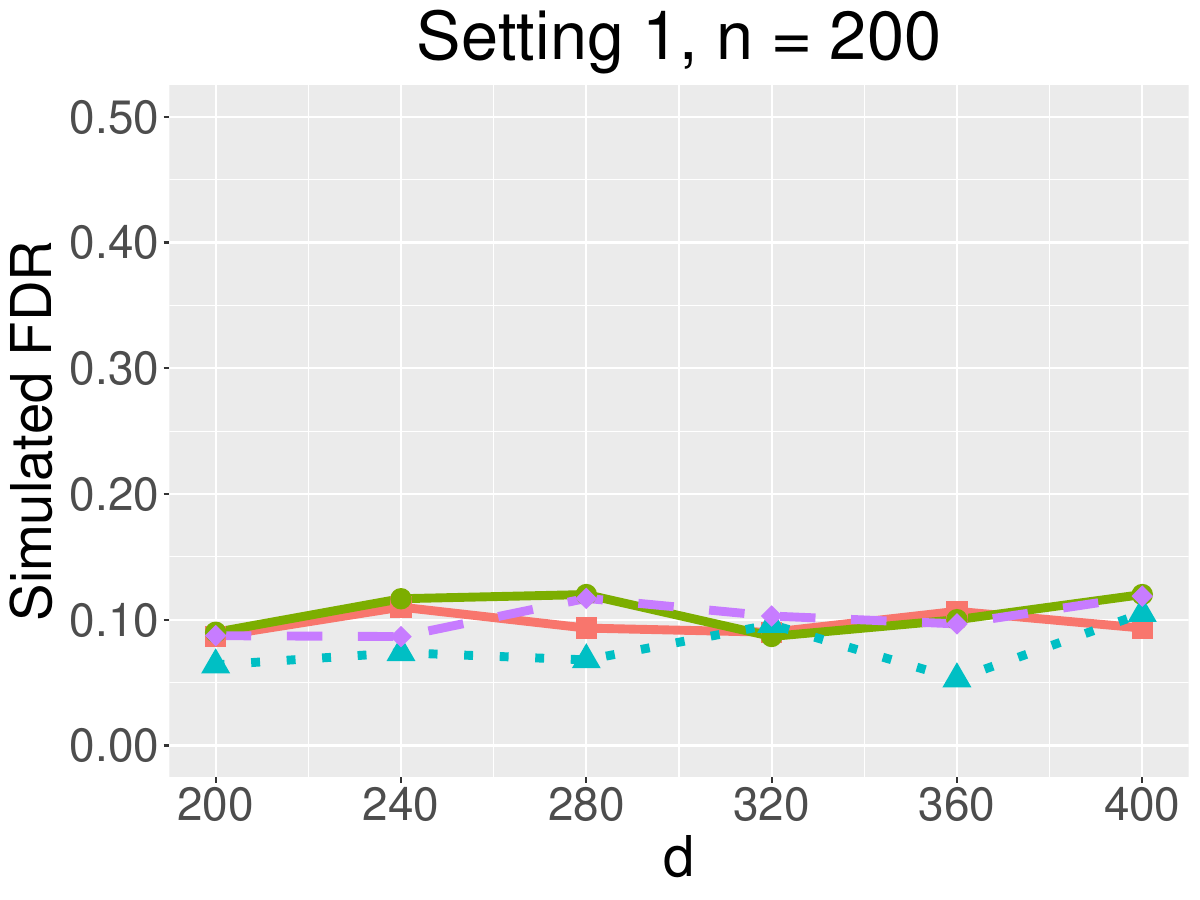}}\hspace{5pt}
	\subfloat{\includegraphics[width=.27\columnwidth]{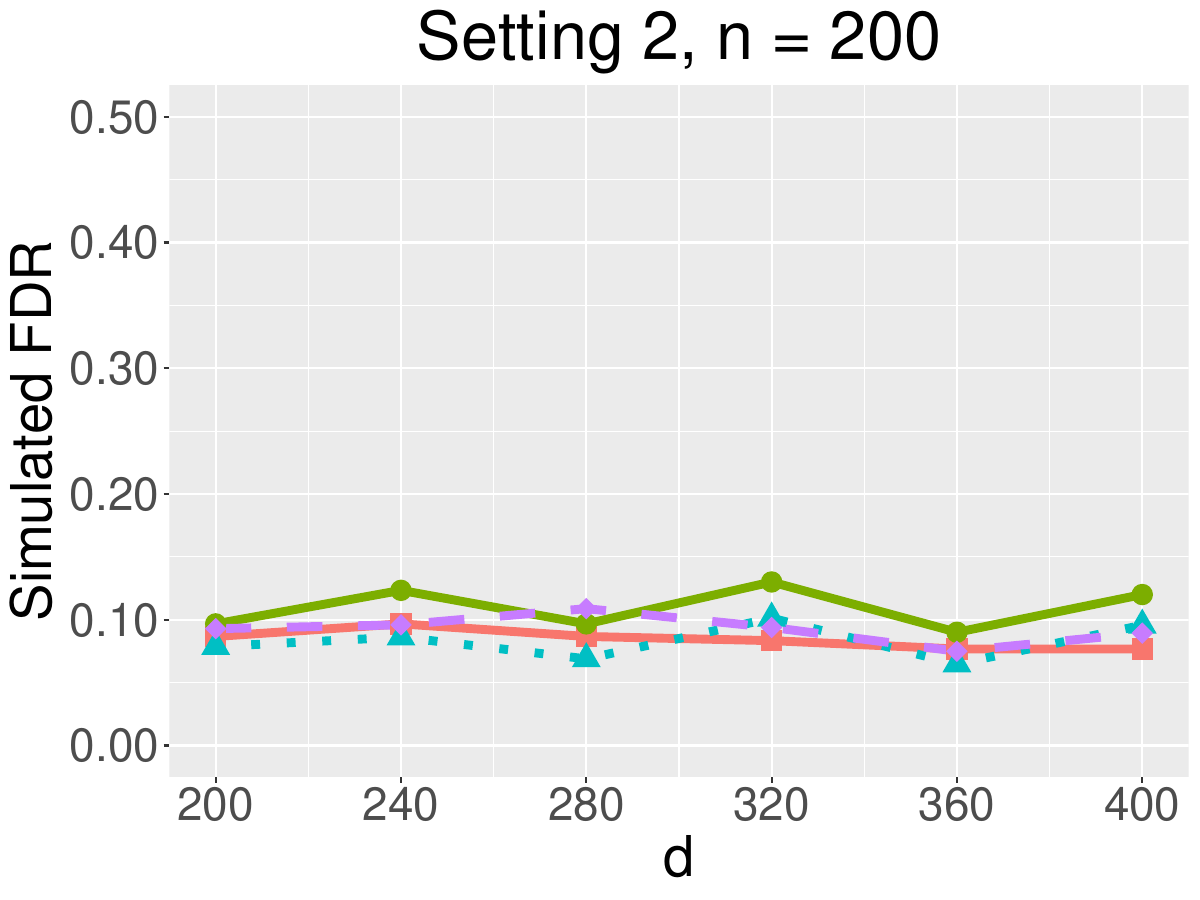}}\hspace{5pt}
	\subfloat{\includegraphics[width=.27\columnwidth]{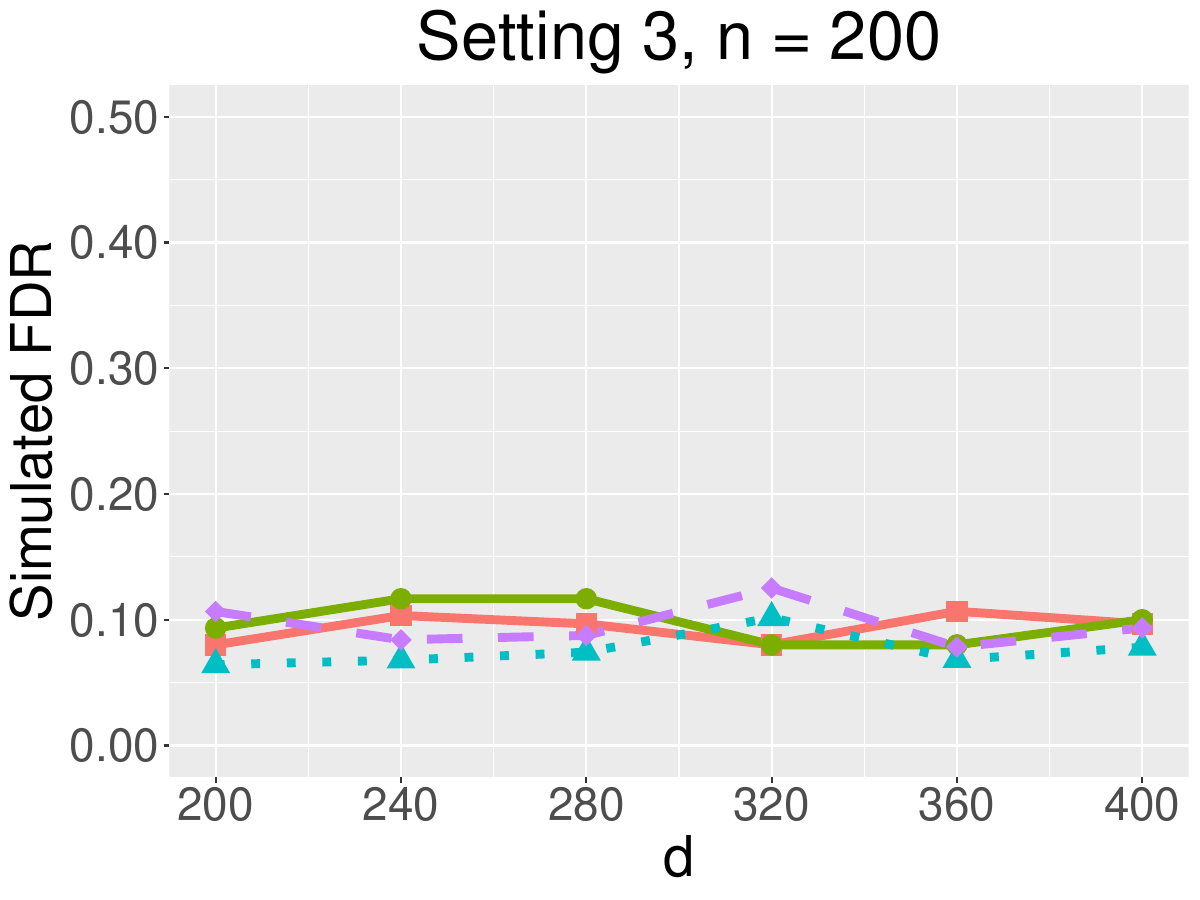}}\\
	\subfloat{\includegraphics[width=.27\columnwidth]{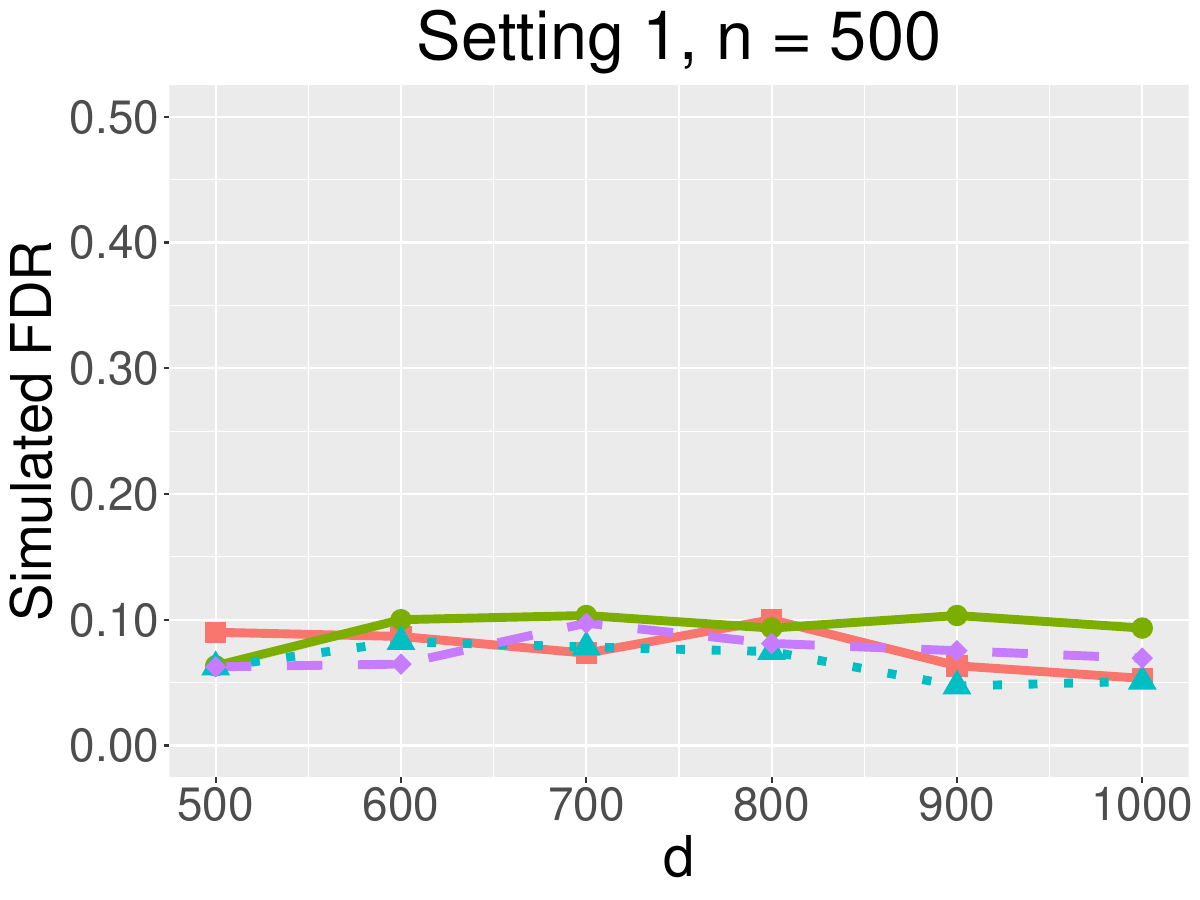}}\hspace{5pt}
    \subfloat{\includegraphics[width=.27\columnwidth]{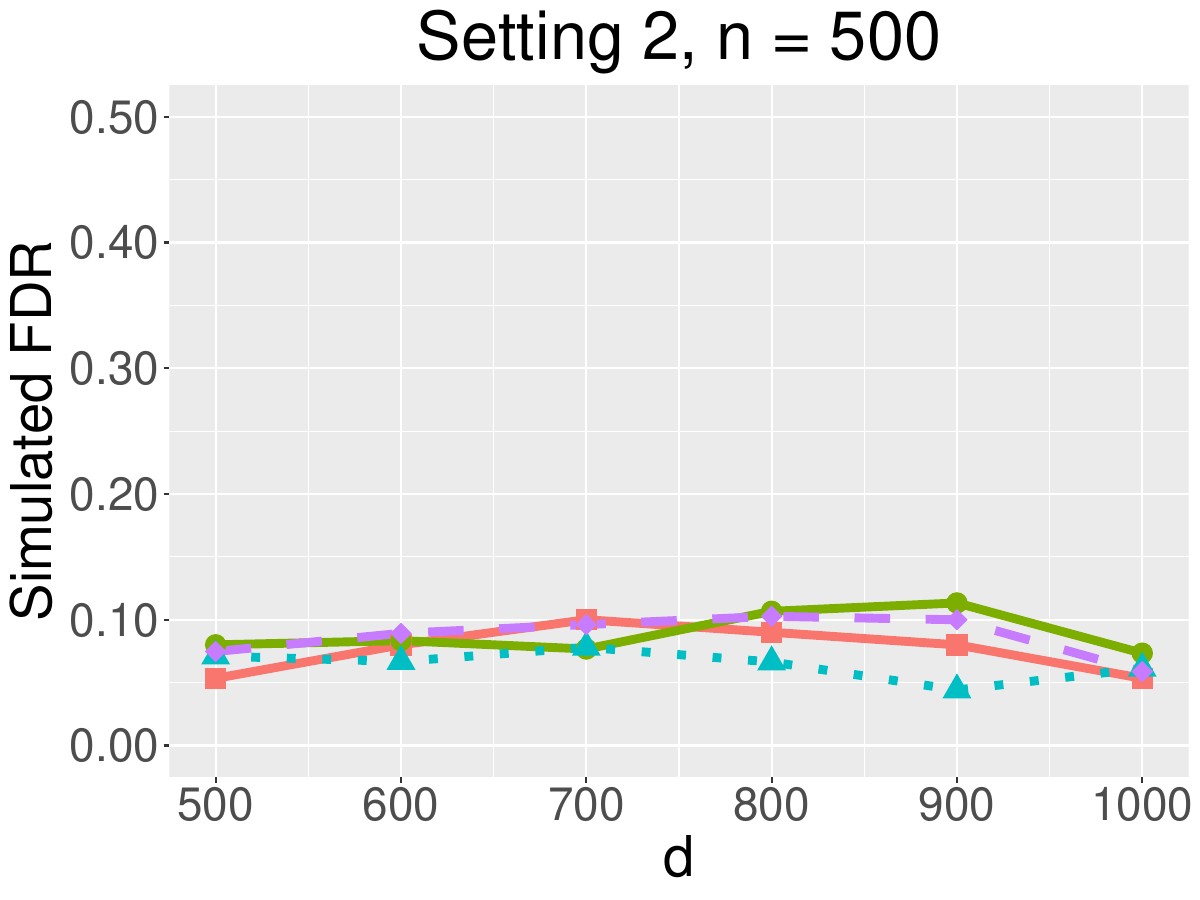}}\hspace{5pt}
    \subfloat{\includegraphics[width=.27\columnwidth]{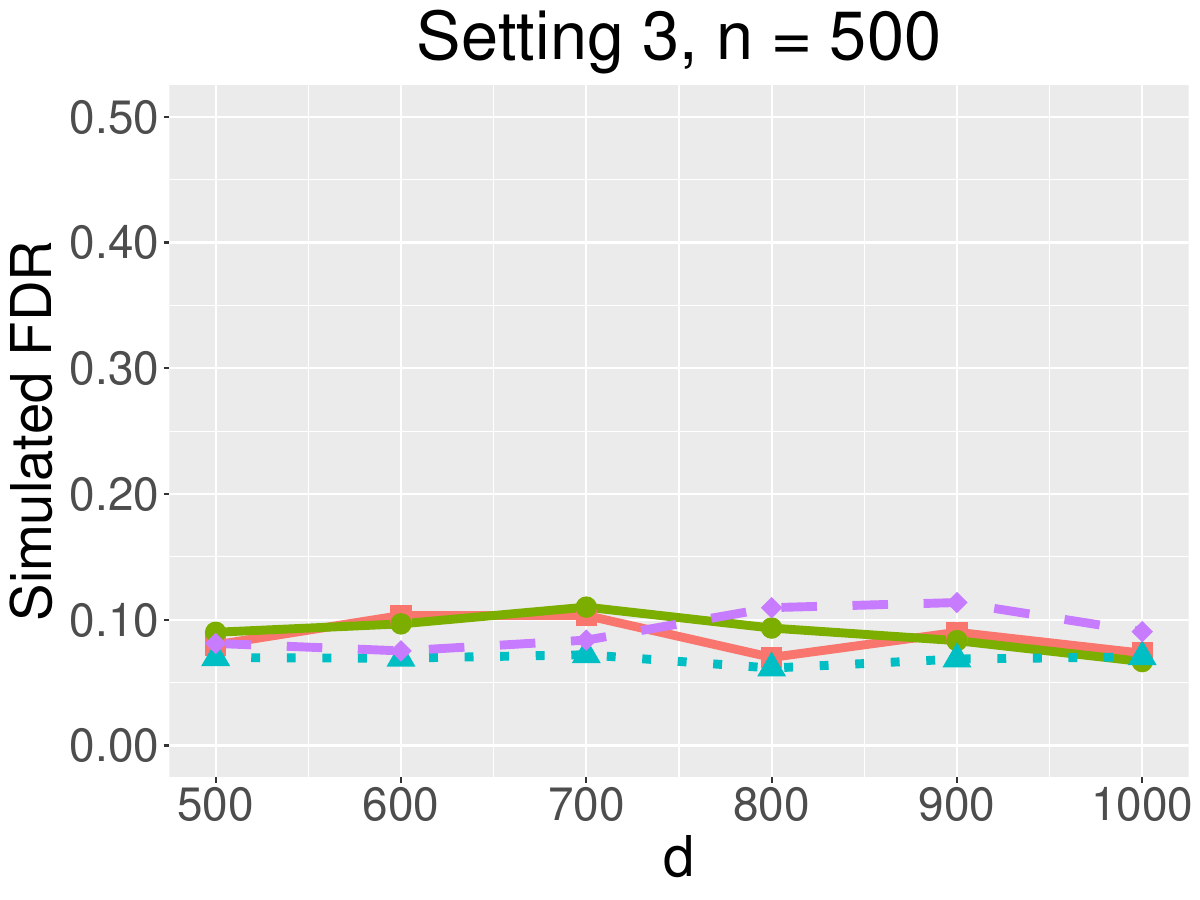}}
	\caption{\small Simulated FDR when all null hypotheses are true, for Setting 1, Setting 2, and Setting 3. The sample sizes of top row and bottom row are $n = 200$ and $n = 500$, respectively. The FDR level is $\alpha = 0.1$. The methods compared are Algorithm 1 without data-splitting (squares and red solid line), Algorithm 2 without data-splitting (circles and green solid line), Algorithm 1 with data-splitting (triangles and blue dotted line), and Algorithm 2 with data-splitting (diamonds and purple dashed line).}
    \label{FDR-control-ds-0.1}
\end{figure}

\begin{figure}[htbp!]
	\centering
	\subfloat{\includegraphics[width=.27\columnwidth]{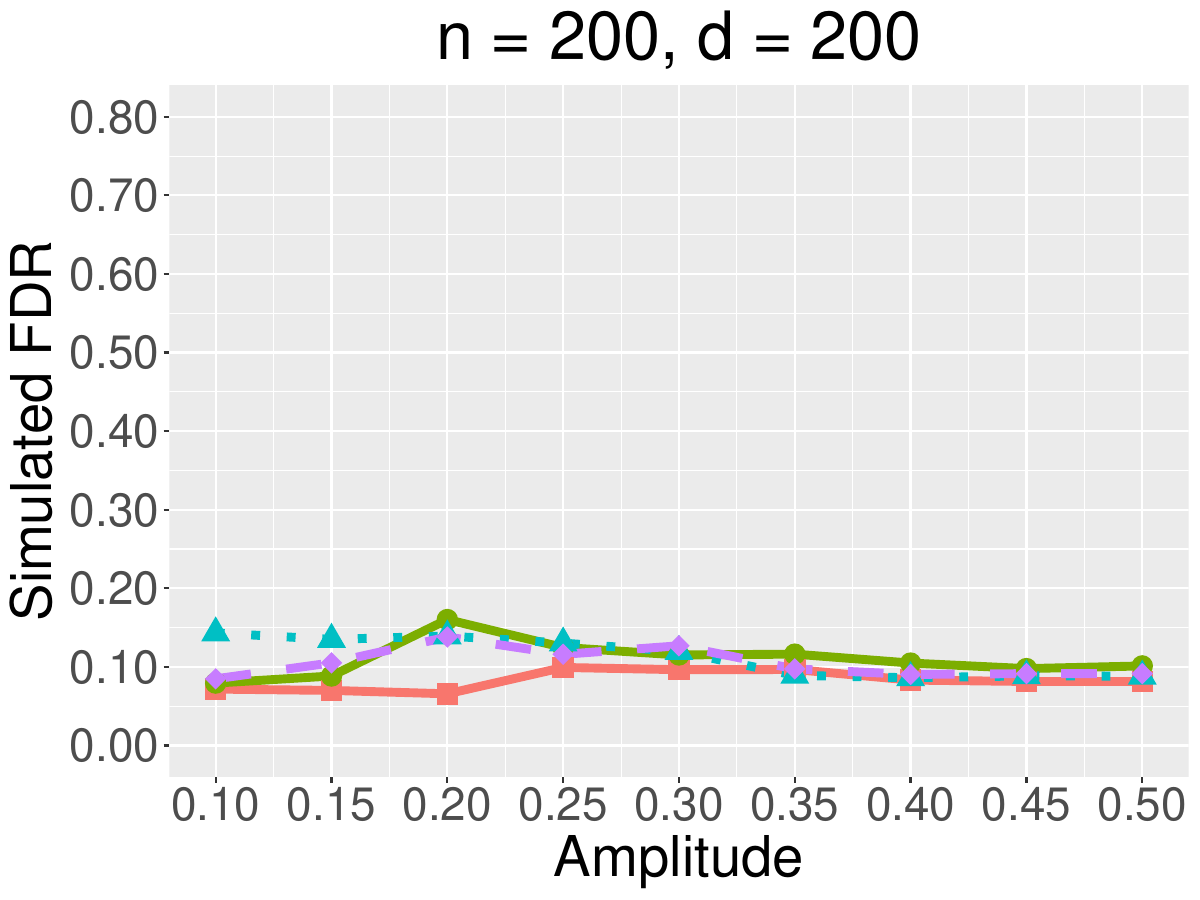}}\hspace{5pt}
	\subfloat{\includegraphics[width=.27\columnwidth]{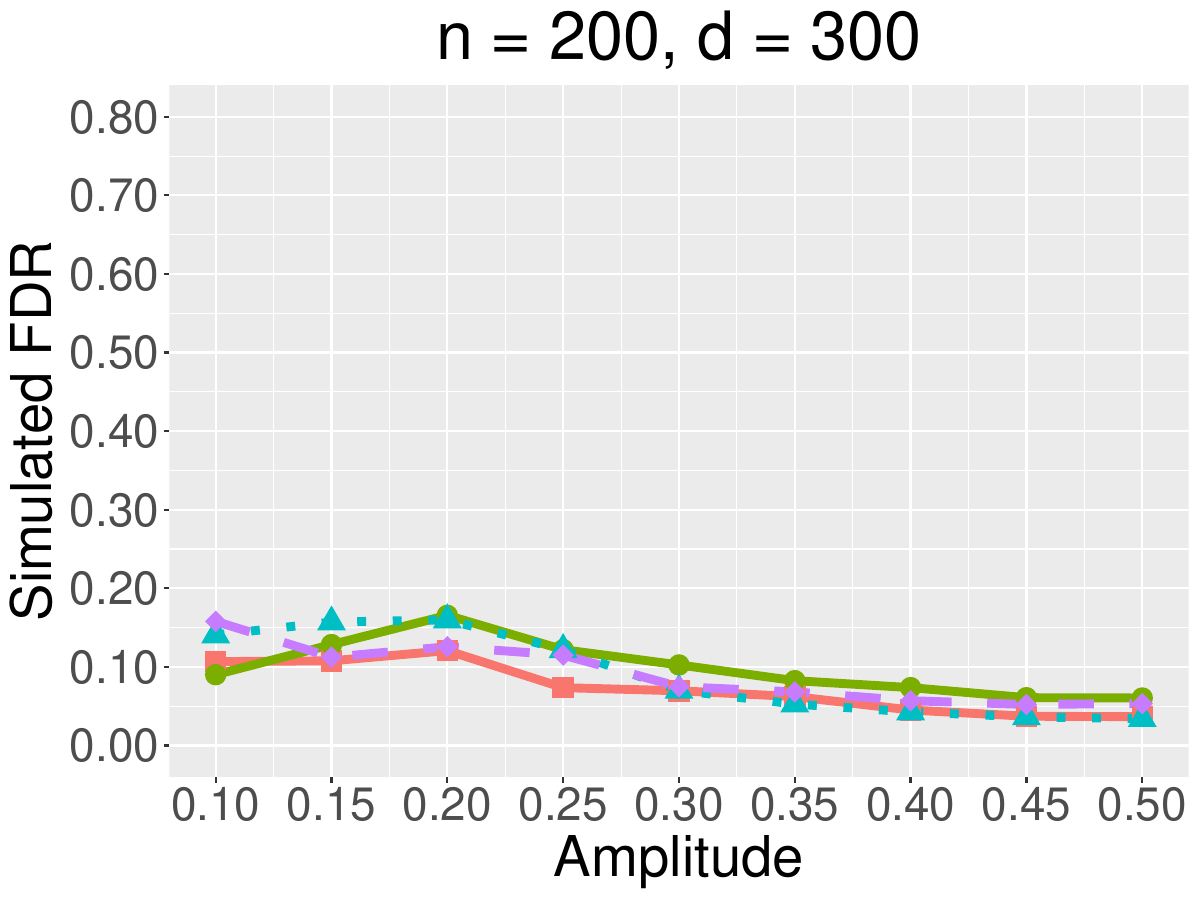}}\hspace{5pt}
	\subfloat{\includegraphics[width=.27\columnwidth]{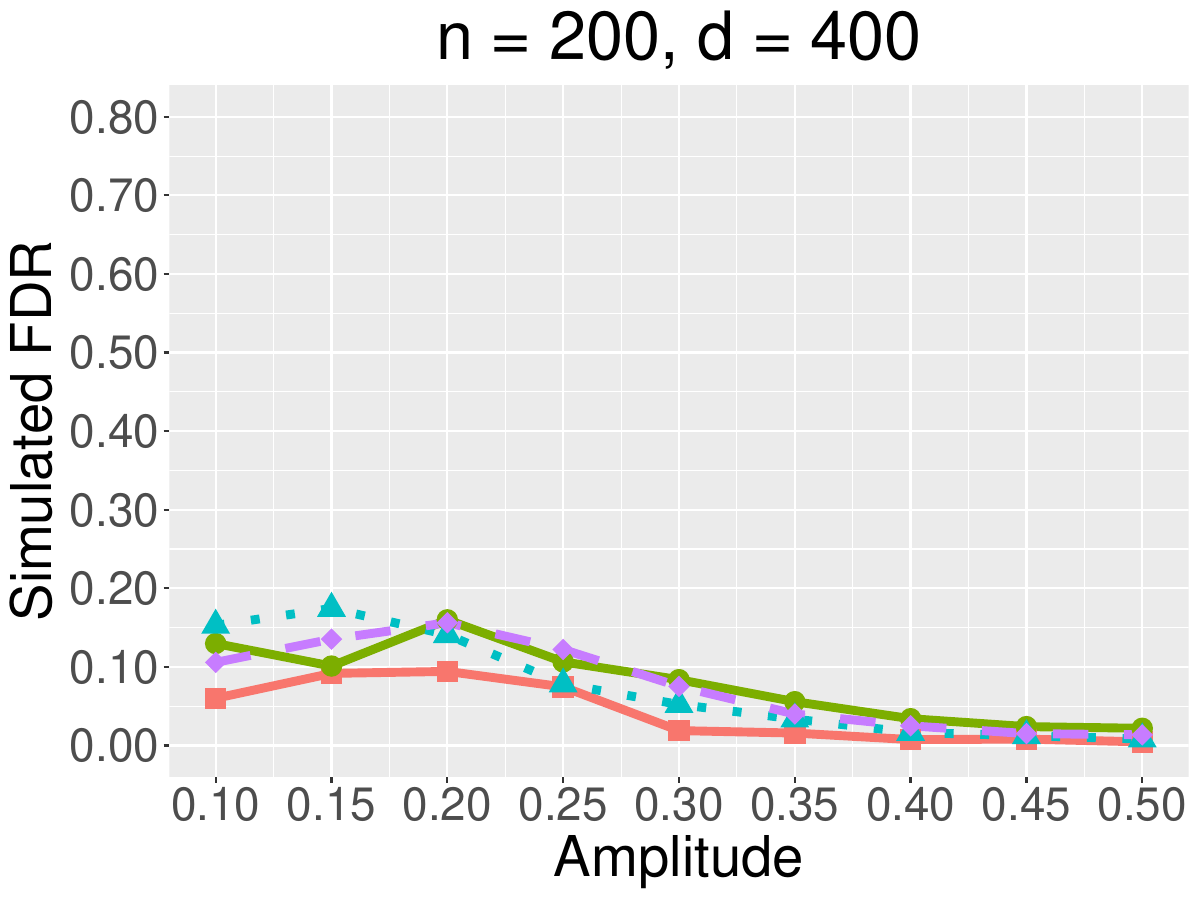}}\\
	\subfloat{\includegraphics[width=.27\columnwidth]{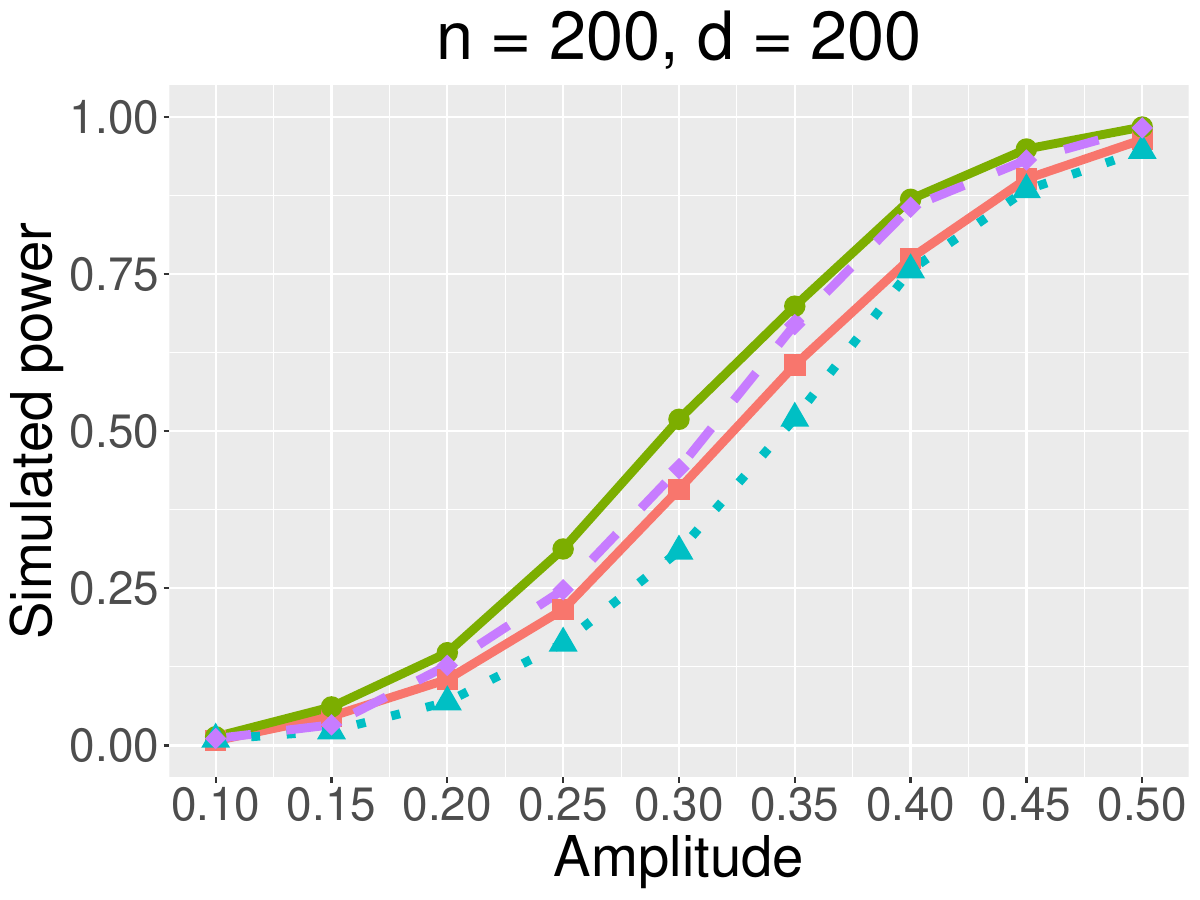}}\hspace{5pt}
    \subfloat{\includegraphics[width=.27\columnwidth]{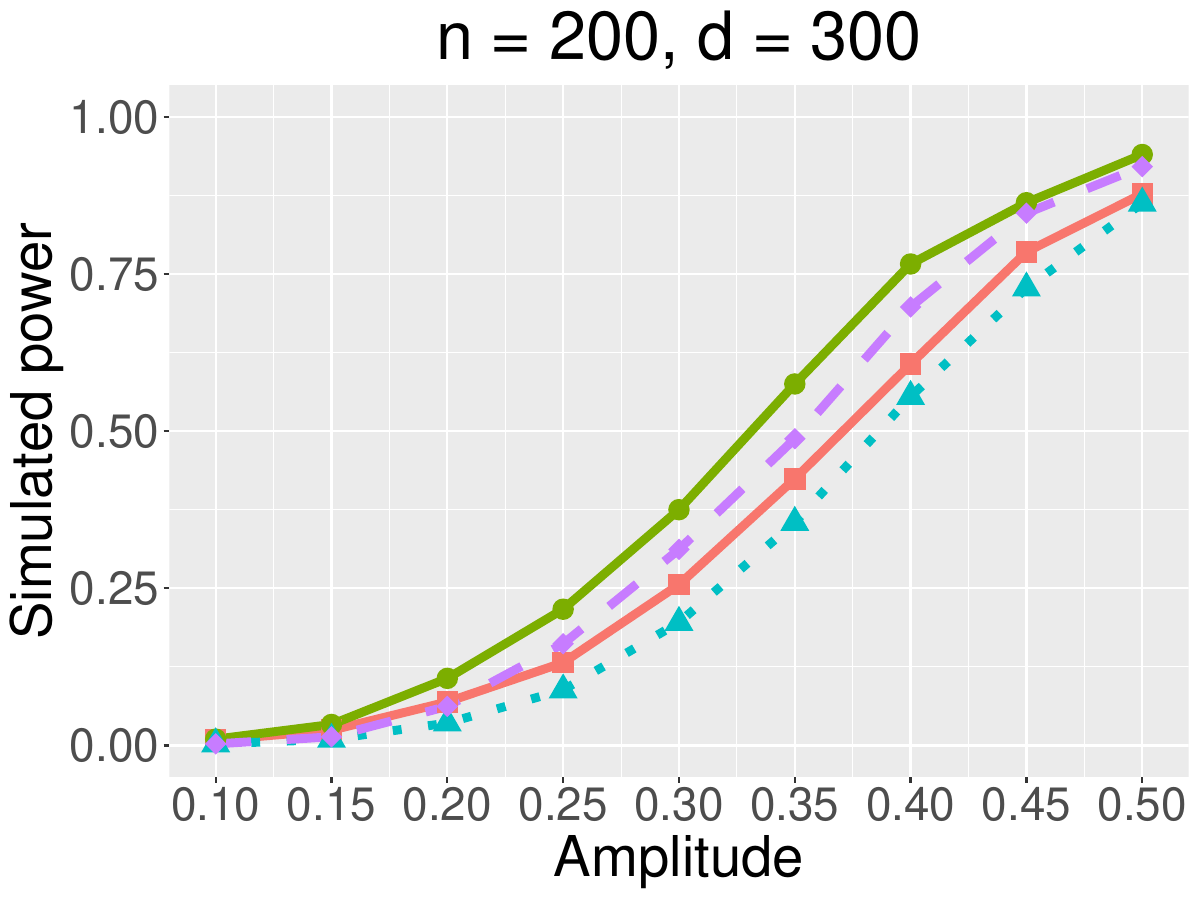}}\hspace{5pt}
    \subfloat{\includegraphics[width=.27\columnwidth]{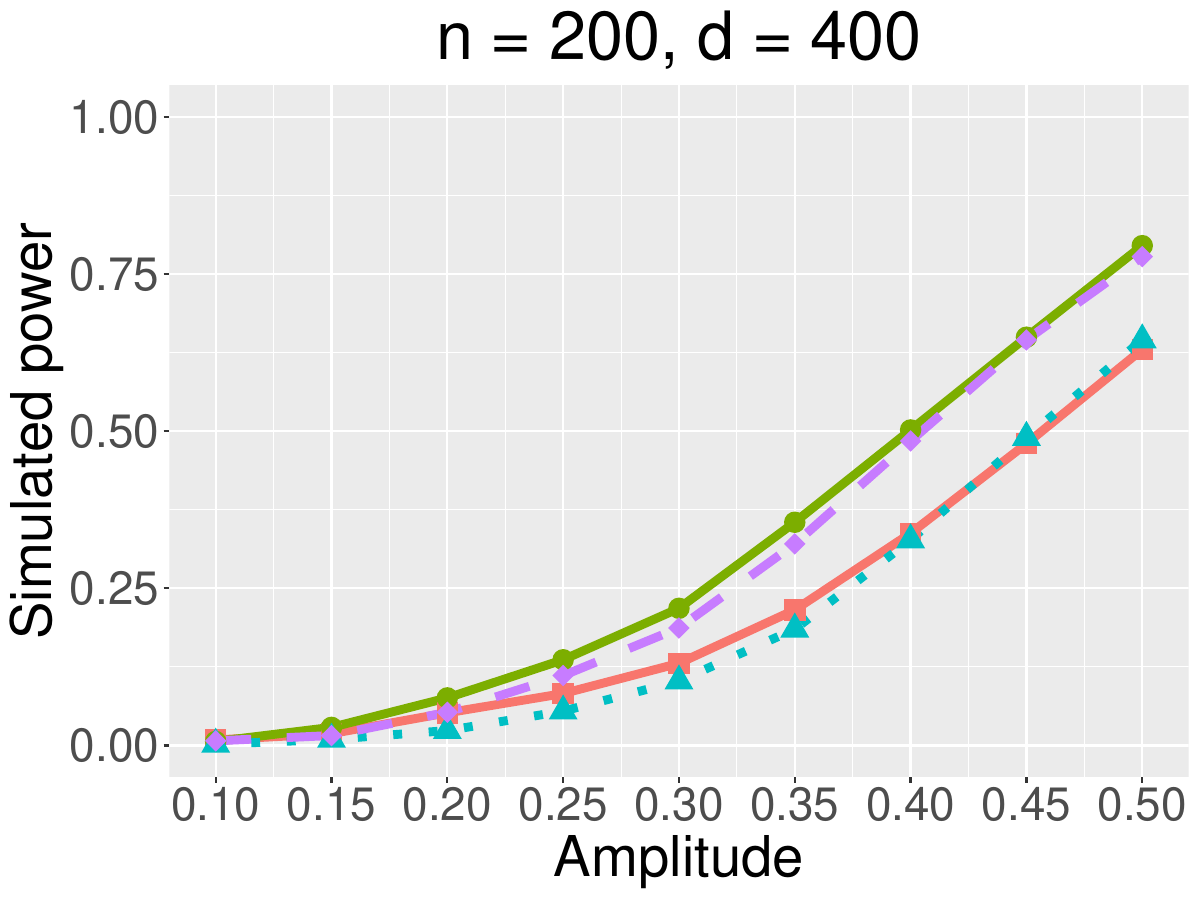}}\\
    \subfloat{\includegraphics[width=.27\columnwidth]{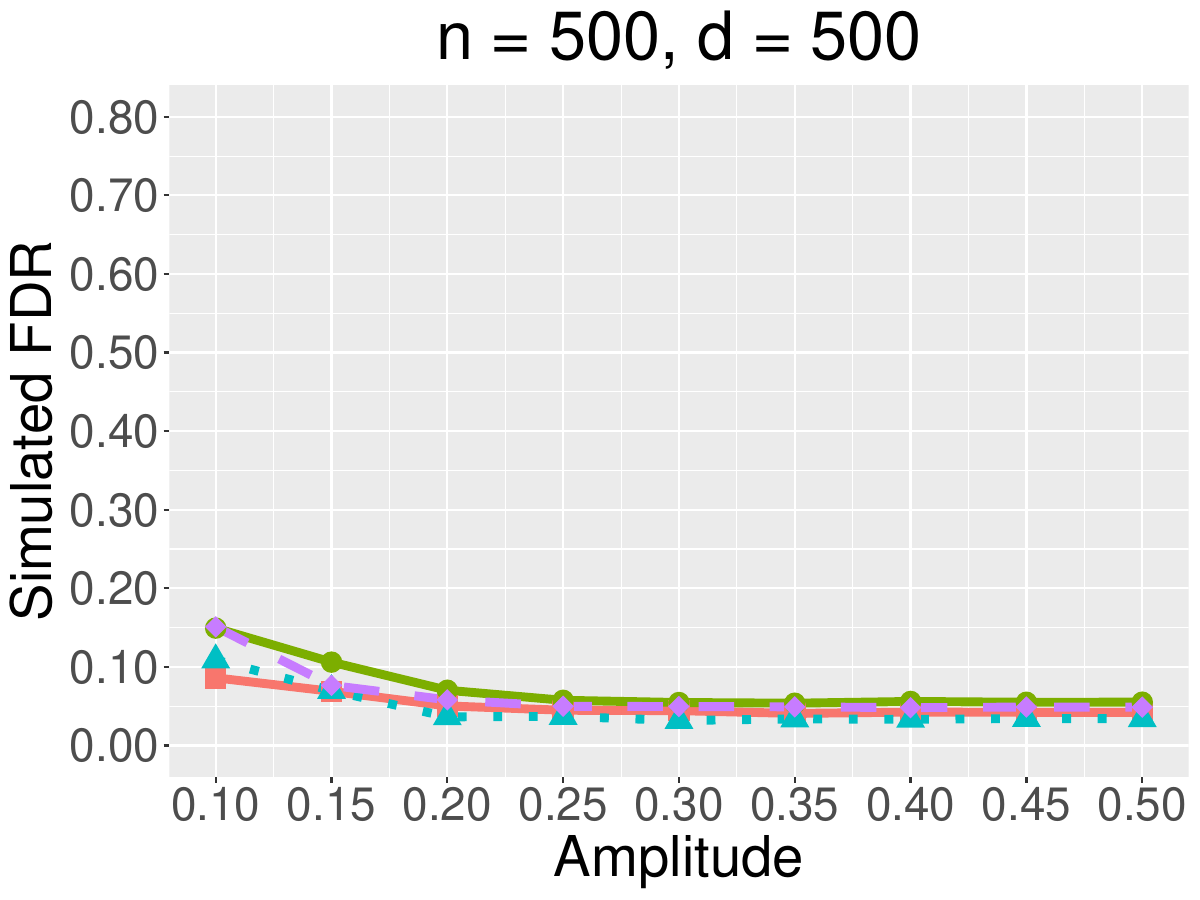}}\hspace{5pt}
	\subfloat{\includegraphics[width=.27\columnwidth]{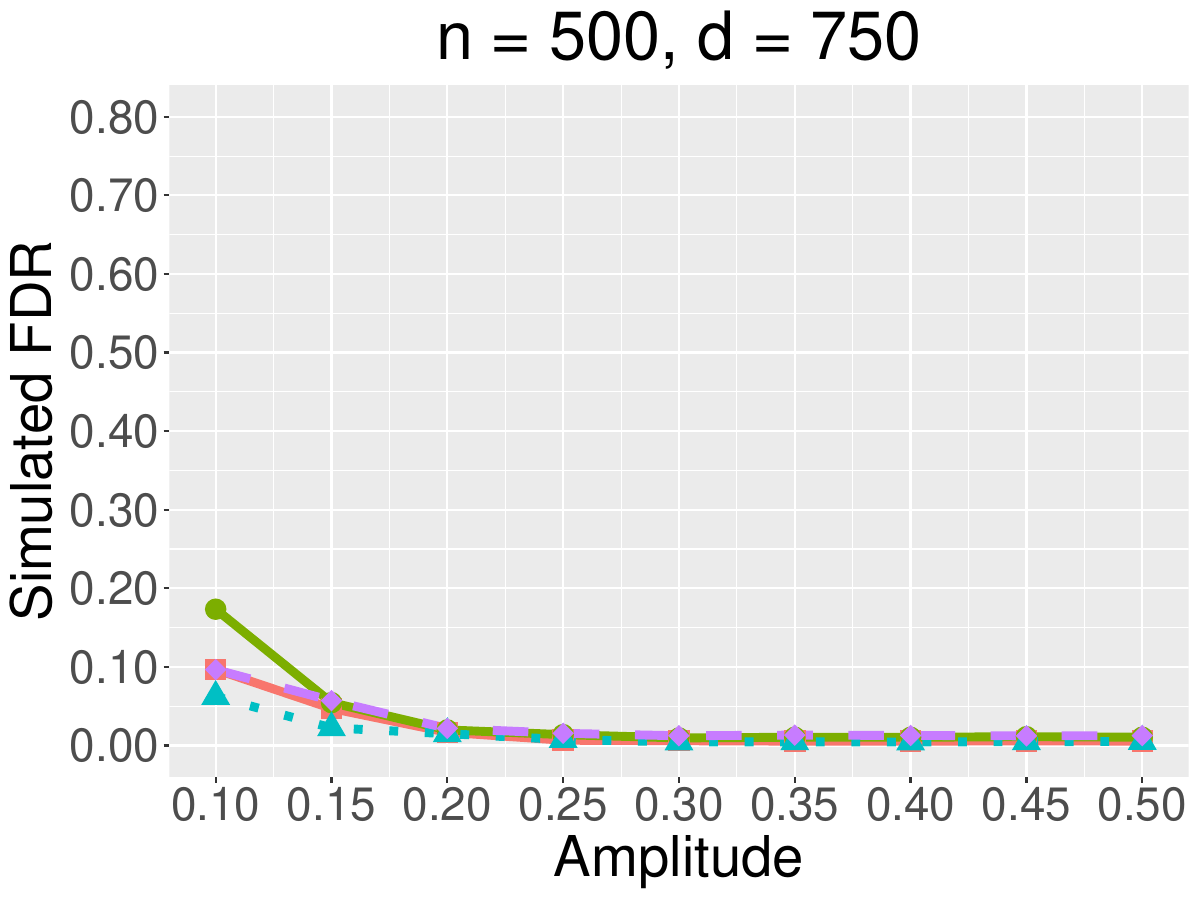}}\hspace{5pt}
	\subfloat{\includegraphics[width=.27\columnwidth]{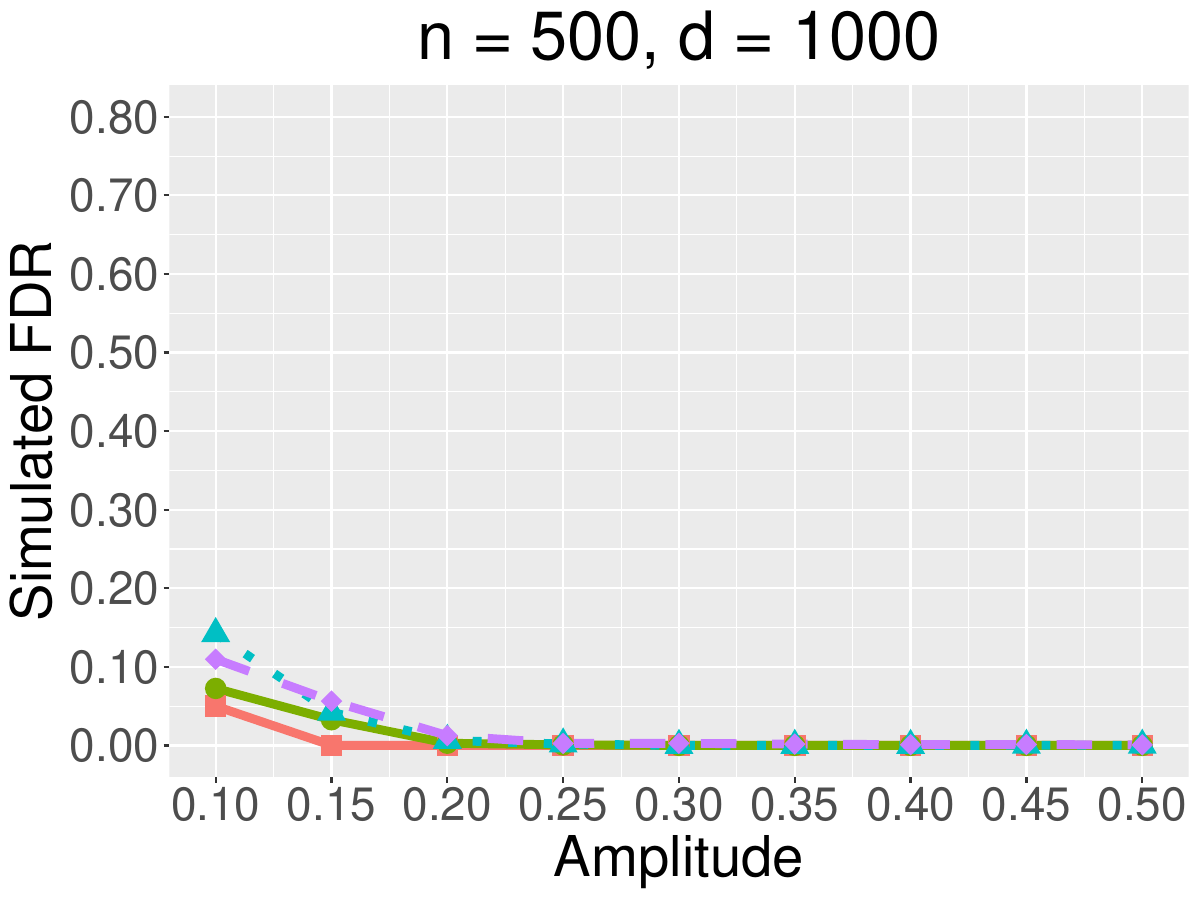}}\\
	\subfloat{\includegraphics[width=.27\columnwidth]{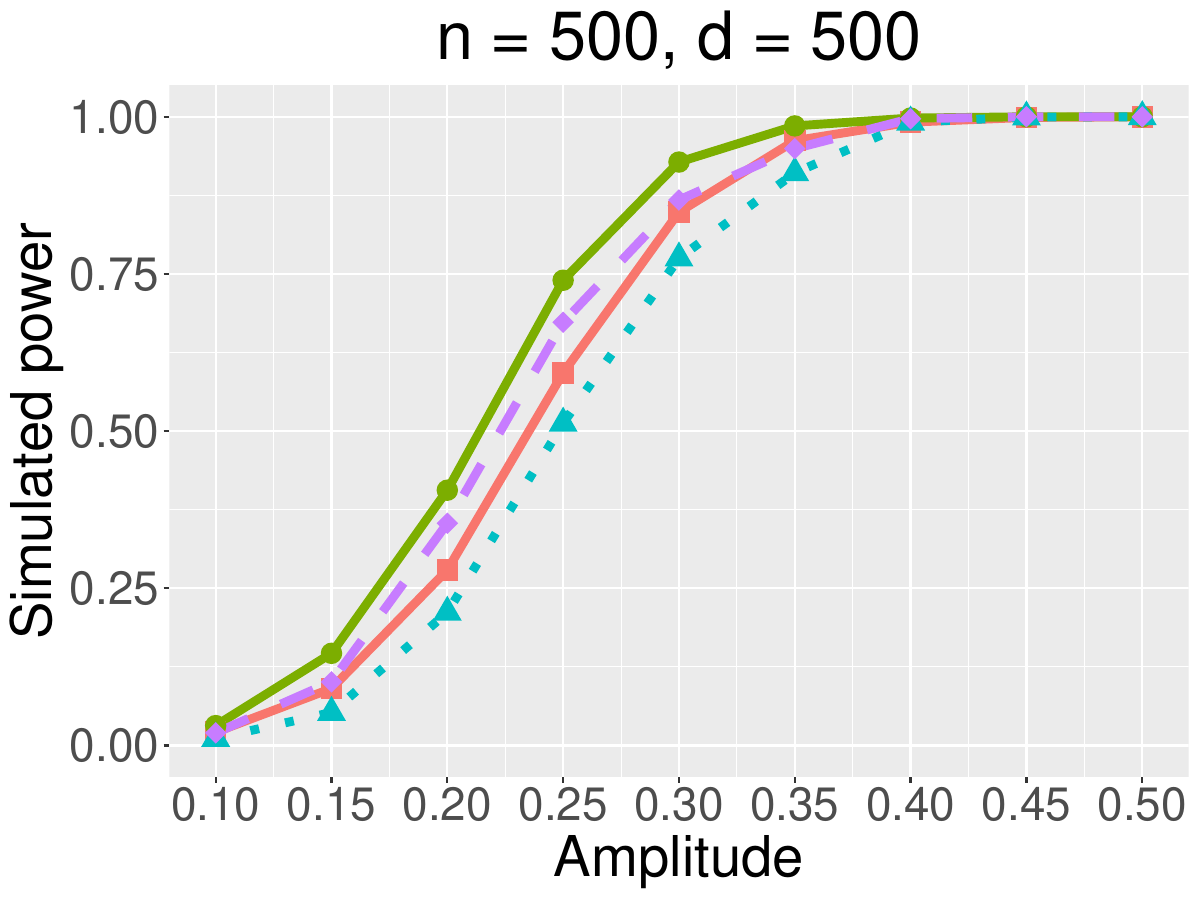}}\hspace{5pt}
    \subfloat{\includegraphics[width=.27\columnwidth]{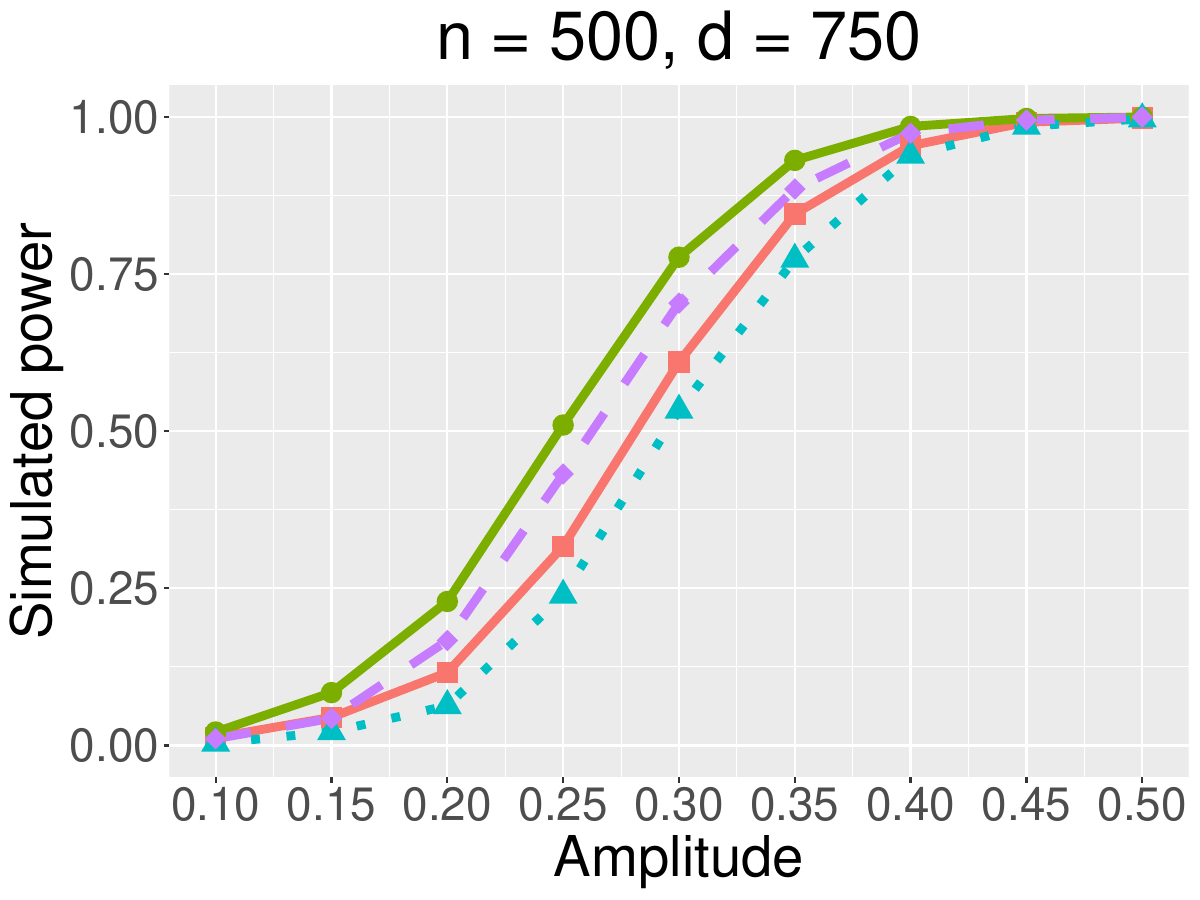}}\hspace{5pt}
    \subfloat{\includegraphics[width=.27\columnwidth]{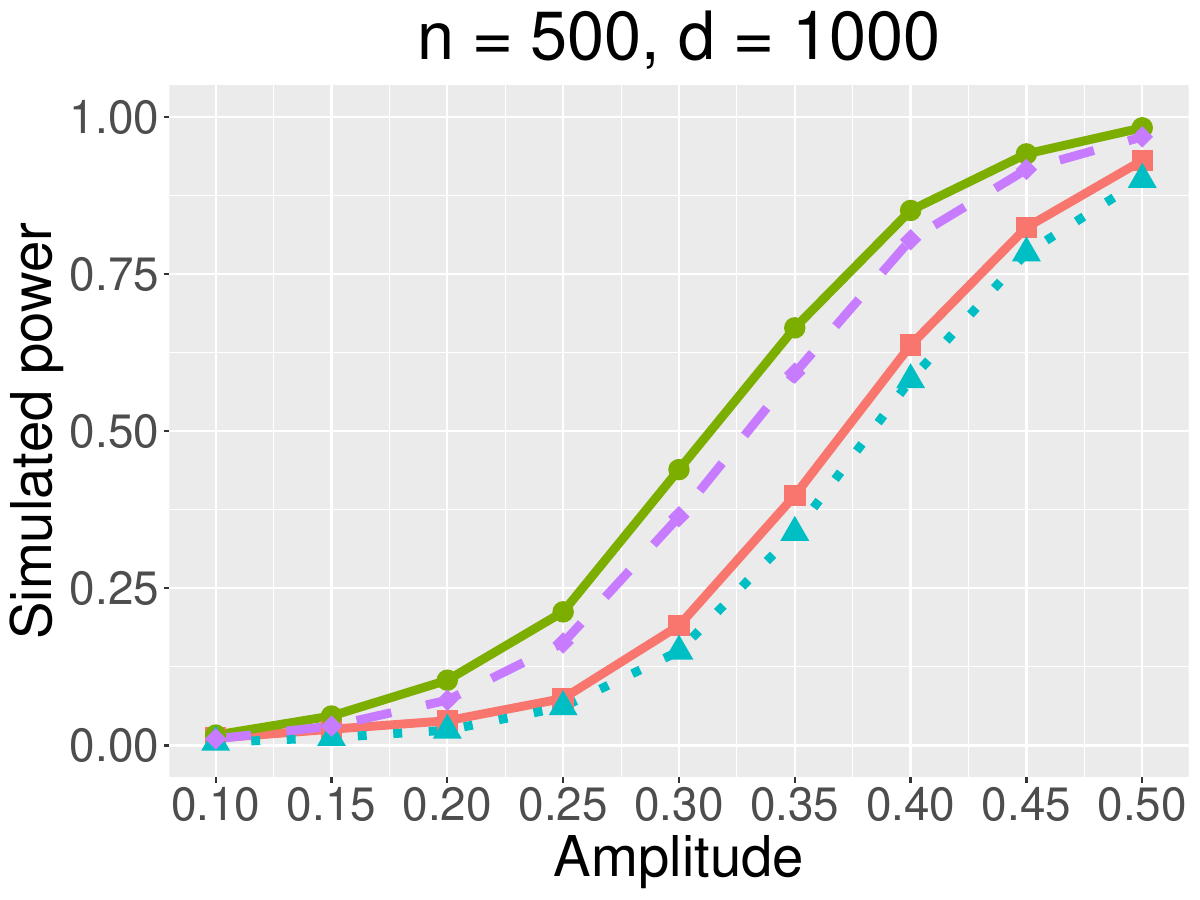}}
	\caption{\small Simulated FDR and power for different combinations of $(n,d)$. The rows of the design matrix were generated from Setting 2. The sparsity level is $k = 0.04d$ and the FDR level is $\alpha = 0.1$. The methods compared are Algorithm 1 without data-splitting (squares and red solid line), Algorithm 2 without data-splitting (circles and green solid line), Algorithm 1 with data-splitting (triangles and blue dotted line), and Algorithm 2 with data-splitting (diamonds and purple dashed line).}
    \label{Power-ds-0.1}
\end{figure}

We set the sample sizes $n_1$ and $n_2$ satisfying $n_1 =\lfloor n/2\rfloor$ and $n_2 = n - n_1$ in data-splitting, and the relevant simulation results are shown in Figures \ref{FDR-control-ds-0.1} and \ref{Power-ds-0.1}. 
Figure \ref{FDR-control-ds-0.1} presents the empirical FDR at $\alpha = 0.1$ when all null hypotheses are true, comparing Algorithms \ref{method_1} and \ref{method_2} both without and with data-splitting across Settings 1--3. The results demonstrate that both Algorithms \ref{method_1} and \ref{method_2} with data-splitting effectively control the FDR at the target level $\alpha = 0.1$, validating the theoretical guarantees established in Section \ref{F:F_0}. Notably, the FDR control performance of Algorithms \ref{method_1} and \ref{method_2} with data-splitting is highly consistent with that of Algorithms \ref{method_1} and \ref{method_2} without data-splitting.

Figure~\ref{Power-ds-0.1} presents the empirical FDR and power under Setting~2 of the main paper for the design matrix, at $\alpha = 0.1$, for $n \in \{200, 500\}$ and selected values of $d$ that are representative of the overall results. As shown in Figure~\ref{Power-ds-0.1}, (i) Algorithms~\ref{method_1} and~\ref{method_2} without data-splitting remain effective in finite samples, and (ii) both algorithms with data-splitting successfully control the FDR, with performance highly consistent with their counterparts without data-splitting. 
When the sample size is $n=200$, the versions without data-splitting achieve higher power. This is likely because the data-splitting strategy uses only part of the data to construct the test statistics, leading to information loss that is more noticeable when the sample size is small. As the sample size increases, the power of Algorithms~\ref{method_1} and~\ref{method_2} with data-splitting becomes increasingly comparable to that of the corresponding methods without data-splitting.
In these scenarios, the numerical results confirm the theoretical guarantees established in Section~\ref{F:F_0} and provide empirical support for the validity and competitiveness of Algorithms~\ref{method_1} and~\ref{method_2} under the extended scope of our framework.

\section{Comparison with selective inference}\label{sec:selective inference}

\subsection{Background on selective inference}

Selective inference considers the setting in which a model is chosen by some data-dependent procedure. Because the same data are used both to select the model and to conduct inference within it, classical inference that ignores the selection step is generally invalid, and selective inference restores validity by explicitly accounting for it \citepS{Berk2013Valid_app}.   A prominent line of work conditions on the selection event: for the Lasso with a fixed penalty, the event that a given model and its sign pattern are selected is a polyhedron in the response space, and, under Gaussian errors with a known noise level, exact finite-sample $p$-values and confidence intervals for the coefficients of the selected submodel follow from the resulting truncated Gaussian laws \citepS{Lee2016Exact_app}.  Related works include sequential procedures such as
forward stepwise regression and least angle regression
\citepS{Tibshirani2016Exact_app}, and the incorporation of unknown error variance via the
square-root Lasso \citepS{Tian2018Selective_app}.  \citetS{Barber2019Knockoff_app} combine a
screening step with the knockoff filter through sample splitting, thus constructing a selective-inference version of the knockoff-filter-based FDR control methodology.

To ensure the validity of inference after selection, data splitting is a
default strategy: one part of the data is used to select the model, and the
other part, independent of the selection, is used to conduct the inference,
which is then valid in the classical sense \citepS{Cox1975_app}. Splitting,
however, incurs an efficiency loss: both the selection and the subsequent inference  are performed with fewer
observations and the associated result is therefore more variable. The
conditional approach described above avoids splitting by using the full data,
but it pays a price of its own. To avoid using the same information twice,
inference proceeds under the conditional distribution of the data given the
selection event. The information carried by the occurrence of the event
itself is therefore forgone \citepS{Fithian2017Optimal_app}. The resulting loss
can be severe: for instance, the expected length of the commonly used
polyhedral confidence interval of \citetS{Lee2016Exact_app} is always infinite
\citepS{KivaranovicLeeb2021_app}.  To improve
efficiency, subsequent work has introduced randomization and data re-use
strategies, including data carving \citepS{Fithian2017Optimal_app}, selection with
a randomized response \citepS{TianTaylor2018Randomized_app}, randomization-based
splitting \citepS{RasinesYoung2023_app}, data fission \citepS{Leiner2025Fission_app},
and approximate maximum-likelihood selective inference
\citepS{PanigrahiTaylor2023_app}.

We remark that the selective inference framework establishes the component-wise
validity of the resulting $p$-values -- each is valid on its own, conditional
on the selection event -- but no FDR control result is established for
multiple testing with these $p$-values. The distinction is consequential:
the validity of the individual $p$-values does not imply valid FDR control
when they are aggregated by the BH procedure. Indeed, the recent negative
results of \citetS{Dobriban2026BH_app} and \citetS{Lei2026BH_app} show that the BH
procedure applied to exactly valid but correlated two-sided Gaussian
$p$-values can fail to control the FDR at its nominal level, and that the
worst-case inflation admits no universal multiplicative bound as the target
level decreases. 

Establishing the validity of FDR control therefore requires
incorporating the joint distribution of the test statistics beyond their marginal
validity -- the requirement that our knockoff-augmented construction is
designed to meet.

\subsection{Distinctions and advantages of the proposed framework}\label{sec:SI-distinctions}

Our framework differs from selective inference in its target problem, its mechanism,
its guarantee, and the role it assigns to the Lasso.

First, the target: selective inference conducts inference
based upon the selected model(s), so that the set of null hypotheses is an
outcome of the selection step, whereas our approach covers all the hypotheses
$H_{0,j}\colon\beta^0_{j}=0$, $j\in[d]$, with one hypothesis for each
original variable; see also Section~1.4 of \citetS{candes2018panning_app}.
This distinction is fundamental: since the set of null hypotheses depends on
the selected model, a true signal falsely excluded in the selection step --
an event that tends to occur in weak-signal scenarios -- can never be
discovered thereafter. Relatedly, \citetS{GoemanSolari2024_app} show that
multiple testing over the full universe of hypotheses is always at least as
powerful as approaches based on selection and conditioning.

Second, the mechanism. As briefly discussed in the previous
subsection, the validity of selective inference relies on data splitting,
which pays a cost in sample size, or on conditioning, which pays a cost in
information that is implicit and can be severe. Knockoff variables
are conceived as controls for the original variables: they are generated
to replicate the dependence structure of the original variables while being
conditionally independent of the response, given the original variables
\citepS{candes2018panning_app}. The knockoff copy $\tilde{\bX}$ is thus external
randomness, generated without access to the response -- the same ingredient
that underlies the randomized strategies reviewed in the previous
subsection. Our knockoff-based construction puts this ingredient to a different use:
through the transformation \eqref{eq:transform}, the augmented design yields
two asymptotically independent sets of statistics. 
Pairing knockoffs
with $p$-values thus complements the model-X
program: \citetS[Section~1.5]{candes2018panning_app} set $p$-values aside because
valid $p$-values with a tractable joint distribution were unavailable in high
dimensions.  Our knockoff-augmented construction supplies that
missing ingredient -- exactly the gap this work fills.
No observation is set aside; our statistics benefit from the
construction alone, and no conditioning is required. 
 The construction's effect is in the dimension: the working problem becomes $2d$-dimensional. The cost paid is thus of
a different nature and is explicit in the increased dimension.

Third, the guarantee: selective $p$-values are component-wise valid, each on
its own and conditional on the selection event. Given such $p$-values, one
can certainly apply the BH procedure to them; the question is whether the
resulting procedure controls the FDR. The answer is not automatic, and there is likely no guarantee: the FDR
guarantee of the BH procedure requires a joint structure among the
$p$-values, such as independence or a suitable form of positive dependence
\citepS{Benjamini1995Controlling_app, BY2001_app}, and conditioning supplies
neither. The validity of the individual $p$-values alone is provably
insufficient: the BH procedure applied to exactly valid but correlated
two-sided Gaussian $p$-values can exceed its nominal FDR level
\citepS{Dobriban2026BH_app}, and the worst-case inflation admits no universal
multiplicative bound \citepS{Lei2026BH_app}.  Rather than relying on component-wise properties
alone, our framework formally derives and exploits the limiting joint
distribution of the constructed paired statistics
$\{(t_{1,j}, t_{2,j})\}_{j=1}^d$: the two sets $\{t_{1,j}\}_{j=1}^d$ and
$\{t_{2,j}\}_{j=1}^d$ are asymptotically independent of each other, and the
statistics $\{t_{2,j}\}_{j=1}^d$ are asymptotically mutually independent.
These joint properties place the BH procedure on a rigorous footing and
deliver asymptotic FDR control for the full fixed family
(Theorems \ref{thm0} and \ref{thm1}), with the pairing further exploited by the
two-step procedure to enhance power (Theorem~\ref{thm_power}).

Fourth, the role of the Lasso. Both frameworks use the Lasso, but they
assign it different roles. In selective inference, the Lasso is the
selection device: its active set and sign pattern determine the selection
event, and hence both the hypotheses to be tested and the conditional law
under which they are tested. The inference is therefore tied to the behavior
of the Lasso as a model selector, and the penalty parameter must be held
fixed for the polyhedral characterization to be exact. In our framework, the
Lasso \eqref{eq:lasso} serves as an initial estimator in the debiasing step
\eqref{de_est}. It supplies a preliminary consistent fit whose bias due to sparse model fitting is
subsequently corrected, and it plays no part in determining which hypotheses
are tested. Accordingly, our development does not rely on sparse support
recovery; indeed, after debiasing, the bias-corrected estimator is no longer
sparse. This difference also has an empirical implication: the performance
of the selective-inference procedures is more sensitive to the dependence in
the design than that of our methods.

We conclude with a remark on our two-step procedure, Algorithm
\ref{method_2}, which screens with the statistics $\{t_{1,j}\}_{j=1}^d$
before applying the BH procedure to the screened statistics
$\{t_{2,j}\}_{j=1}^d$. On the surface, this is a screening-then-testing
structure that selective inference shares. But the key distinction is that
our Algorithm \ref{method_2} remains focused on the same aim throughout:
all the hypotheses $H_{0,j}\colon\beta^0_{j}=0$, $j\in[d]$,
unconditionally. Moreover, the operations differ in nature.
In our procedure, the selection is post-$p$-value construction and is itself
a multiplicity-control device: a classical Bonferroni screening applied to
already-constructed unconditional $p$-values, adaptable to a user-specified
control level. In selective inference, the selection is a model-choice step
that alters the reference distribution: the $p$-values are constructed
post-selection and are valid conditionally and component-wise.
Indeed, the guarantee of FDR control of Algorithm \ref{method_2} is
established in Theorem~\ref{thm1}, and the
pairing is what delivers the power gain of Theorem~\ref{thm_power}.

\subsection{Numerical comparison with selective-inference procedures}

\subsubsection{Data-generating processes and implementation}
We now compare the proposed methods numerically with the selective-inference
procedures reviewed above. 

The simulation setting is the same as that in the
main paper (Section~\ref{sec:sim}): data are generated from the linear model
with independent standard normal errors. The rows of $\bX$ are
independent draws from $\mathcal{N}(\mathbf{0}, \bSigma)$ under three
dependence structures: Setting~1 takes $\bSigma = \bI_d$; Setting~2 uses an
AR(1) structure with correlation coefficient $0.4$; and Setting~3 uses a
block structure with $\Sigma_{i,i} = 1$ and
$\Sigma_{i,j} = 0.2 \cdot I(\lceil i/20 \rceil = \lceil j/20 \rceil)$ for
$i \neq j$. For the power comparisons we consider $n \in \{200, 500\}$ with
$d \in \{n, 1.5n, 2n\}$; a set of $k = 0.04d$ randomly located components of
$\bbeta_0$ receives a common amplitude varied over $[0.1, 0.5]$, and the
remaining components are zero. Under the global null all $d$ coefficients are
zero and $d$ varies over $[n, 2n]$ as displayed. The nominal level is
$\alpha \in \{0.05, 0.1\}$, and every configuration is repeated $300$ times. 

Algorithms \ref{method_1} and \ref{method_2} are implemented exactly as in
the main paper: knockoff copies are
generated by the second-order model-X construction of
\citetS{candes2018panning_app} (function \texttt{create.second\_order} in the R
package \texttt{knockoff}), the Lasso is computed with \texttt{cv.glmnet}
(package \texttt{glmnet}), the decorrelating matrix $\hat{\bTheta}$ follows
\citetS{javanmard2014confidence_app}, and the noise level is estimated by the
scaled Lasso (package \texttt{scalreg}). The implementation of the
selective-inference baselines is described next, with their complete
statements given in Algorithms A4--A7.

\subsubsection{Competing procedures}
We compare Algorithms \ref{method_1} and \ref{method_2} with four procedures
representative of the two
main selective-inference strategies: conditioning (Algorithms A4 and A5) and
sample splitting (Algorithms A6 and A7).

Algorithm A4 implements the post-selection BH procedure built on
\citetS{Lee2016Exact_app}. The Lasso is fitted on the full data with the penalty
parameter $\varrho_{4}^\ast$ chosen by cross-validation (\texttt{cv.glmnet} in the
R package \texttt{glmnet}); for each selected variable, a one-sided
conditional $p$-value is computed from the truncated Gaussian law, with the
test direction determined by the sign with which the variable entered the
active set (\texttt{fixedLassoInf} in the R package
\texttt{selectiveInference}); the BH procedure at level $\alpha$ is then
applied within the selected family.

Algorithm A5 replaces the Lasso by the square-root Lasso, following
\citetS{Tian2018Selective_app}, so that the noise level need not be known. The
regularization parameter is specified in the scale-free form
$\varrho_{5}^\ast = \kappa\, \mathbb{E}\,[\,|\bX^\top \mathbf{g}|_\infty /
|\mathbf{g}|_2\,]$ with $\mathbf{g}\sim\mathcal{N}(\mathbf{0},\mathbf{I}_n)$;
among $\kappa \in \{0.5, 0.6, \ldots, 1\}$, the value $\kappa = 0.8$ gave the
highest power and is the one reported. In practice, we set $\varrho_{5}^\ast$ as $\kappa$ times the 95th percentile of $\,|\bX^\top \mathbf{g}|_\infty /
|\mathbf{g}|_2\,$. Conditional $p$-values are computed
via the truncated-normal approximation (\texttt{TG.pvalue} in
\texttt{selectiveInference}), as in the numerical experiments of
\citetS{Barber2019Knockoff_app}, and aggregated by BH procedure within the selected family.

Algorithm A6 implements the screening-then-knockoff procedure of
\citetS{Barber2019Knockoff_app}: a random $40\%$ of the observations form the
screening set, on which the full Lasso path is fitted; form the screened set consisting of the first
$k_{\max}$ many features entering the path, with $k_{\max}$ set to $25\%$ of the
inference sample size, and record their entry signs. On the inference set, equi-correlated fixed knockoffs are generated
for the screened features (\texttt{create.fixed} in the R package
\texttt{knockoff}), a sign-restricted square-root Lasso is fitted, and the
knockoff filter is applied to the statistics
$W_j = |\hat{\beta}_j| - |\tilde{\beta}_j|$. Both the data-splitting and
data-recycling variants of \citetS{Barber2019Knockoff_app} are included, with
$\kappa = 0.7$ and $\kappa = 0.9$, respectively. 

Algorithm A7 uses the same
screening protocol as Algorithm A6, followed by an ordinary least squares refit on the
inference set with one-sided $t$-tests in the screened directions and BH procedure
within the screened family.

\floatname{algorithm}{Algorithm}
\begin{algorithm}[htbp!]\small
    \renewcommand{\thealgorithm}{A4}
    \caption{Post-selective inference via Lasso \protect\citepS{Lee2016Exact_app} based Benjamini-Hochberg procedure}
    \begin{algorithmic}
        \STATE\textbf{Step 1}. Fit the Lasso with original data,
    \begin{align*}
        \hat{\bbeta}
         = \arg\min_{\bbeta \in \mathbb{R}^{d}}
        \bigg\{
        \frac{1}{n} | \by - \bX\bbeta |_2^2
        + \varrho_{4}^\ast | \bbeta |_1
        \bigg\}\,.
    \end{align*}
      \STATE\textbf{Step 2}. Generate conditional $p$-values $\tilde P_1,\ldots,\tilde P_{|\hat{\mathcal{M}}|}$ for selected variables following \citetS{Lee2016Exact_app}, where $|\hat{\mathcal{M}}|$ is cardinality of the Lasso active set $\hat{\mathcal{M}}$.
        \STATE\textbf{Step 3}. Given $\alpha \in(0,1)$, let $\tilde{P}_{(1)} \leq \cdots \leq \tilde{P}_{(|\hat{\mathcal{M}}|)}$ be the ordered versions of the $\tilde{P}_j$'s, find
        \begin{equation*}
            \tilde{R}=\max\bigg\{i\in[|\hat{\mathcal{M}}|]: \tilde{P}_{(i)} \leq \frac{i \alpha}{|\hat{\mathcal{M}}|}\bigg\}\,,
        \end{equation*}
        provided that the maximum exists; otherwise, let $\tilde{R}=0$.
        \STATE\textbf{Step 4}.
        Select the feature corresponding to $\tilde{P}_{(j)}$ with $j \leq \tilde{R}$ and aggregate them to construct a selection set $\hat{\mathcal{S}}$.
    \end{algorithmic}
\end{algorithm}

\begin{algorithm}[htbp!]\small
    \renewcommand{\thealgorithm}{A5}
    \caption{Post-selection inference via square-root Lasso \protect\citepS{Tian2018Selective_app} based Benjamini-Hochberg procedure}
    \begin{algorithmic}
        \STATE\textbf{Step 1}. Fit the square-root Lasso with original data,
        \begin{align*}
            \tilde{\bbeta}
            = \arg\min_{\bbeta \in \mathbb{R}^{d}}
            \big\{
            | \by - \bX\bbeta |_2
            + \varrho_{5}^\ast | \bbeta |_1
            \big\}\,.
        \end{align*}
        \STATE\textbf{Step 2}. Generate conditional $p$-values $\tilde P_1,\ldots,\tilde P_{|\hat{\mathcal{M}}|}$ for selected variables following \citetS{Tian2018Selective_app}, where $|\hat{\mathcal{M}}|$ is cardinality of the Lasso active set $\hat{\mathcal{M}}$.
        \STATE\textbf{Step 3}. Given $\alpha \in(0,1)$, let $\tilde{P}_{(1)} \leq \cdots \leq \tilde{P}_{(|\hat{\mathcal{M}}|)}$ be the ordered versions of the $\tilde{P}_j$'s, find
        \begin{equation*}
            \tilde{R}=\max\bigg\{i\in[|\hat{\mathcal{M}}|]: \tilde{P}_{(i)} \leq \frac{i \alpha}{|\hat{\mathcal{M}}|}\bigg\}\,,
        \end{equation*}
        provided that the maximum exists; otherwise, let $\tilde{R}=0$.
        \STATE\textbf{Step 4}. Select the feature corresponding to $\tilde{P}_{(j)}$ with $j \leq \tilde{R}$ and aggregate them to construct a selection set $\hat{\mathcal{S}}$.
    \end{algorithmic}
\end{algorithm}

\begin{algorithm}[htbp!]\small
    \renewcommand{\thealgorithm}{A6}
    \caption{Knockoff filter via square-root Lasso \protect\citepS{Barber2019Knockoff_app}}
    \begin{algorithmic}
        \STATE\textbf{Step 1}. Randomly split the full data $(\bX, \by)$ into a screening set $(\bX^{(0)}, \by^{(0)})$ of size $n_0$ and an inference set $(\bX^{(1)}, \by^{(1)})$ of size $n_1 = n - n_0$. Fit the Lasso on the screening set to obtain the active set $\hat{\mathcal{M}} = \{j : \hat{\beta}^{(0)}_j \neq 0\}$ with cardinality $|\hat{\mathcal{M}}|$, where $\hat{\beta}^{(0)}_j$ denotes the $j$-th component of the estimator $\hat{\bbeta}^{(0)}$ given by
\begin{align*}
\hat{\bbeta}^{(0)}
 = \arg\min_{\bbeta \in \mathbb{R}^{d}}
\bigg\{
\frac{1}{n_0} \big| \by^{(0)} - \bX^{(0)}\bbeta \big|_2^2
+ \varrho_{6}^\ast | \bbeta |_1
\bigg\}\,.
\end{align*}
For each feature $j \in \hat{\mathcal{M}}$, record its estimated sign $s_j = \mathrm{sign}(\hat{\beta}^{(0)}_j) \in \{\pm 1\}$ from the screening stage.
        \STATE\textbf{Step 2}. Let $\bX^{(1)}_{\hat{\mathcal{M}}}$ be the submatrix of $\bX^{(1)}$ with columns in $\hat{\mathcal{M}}$. Generate knockoffs $\tilde{\bX}^{(1)}_{\hat{\mathcal{M}}}$ for the inference set. Construct the response vector $\by_{\text{infer}}$ and augmented design matrix $\bX_{\text{aug}}$ based on the chosen procedure:
        \begin{itemize}
            \item \textit{Data Splitting}: Use only the inference set, and let $\by_{\text{infer}} = \by^{(1)}$ and $\bX_{\text{aug}} = \big(\bX^{(1)}_{\hat{\mathcal{M}}}, \tilde{\bX}^{(1)}_{\hat{\mathcal{M}}}\big)$.
            \item \textit{Data Recycling}: Reuse the screening set, and let $\by_{\text{infer}} = \by$ and $\bX_{\text{aug}} = \big(\bX_{\hat{\mathcal{M}}}, \tilde{\bX}_{\hat{\mathcal{M}}}\big)$, where $\bX_{\hat{\mathcal{M}}} = \begin{bmatrix} \bX^{(0)}_{\hat{\mathcal{M}}} \\ \bX^{(1)}_{\hat{\mathcal{M}}} \end{bmatrix}$ and $\tilde{\bX}_{\hat{\mathcal{M}}} = \begin{bmatrix} \bX^{(0)}_{\hat{\mathcal{M}}} \\ \tilde{\bX}^{(1)}_{\hat{\mathcal{M}}} \end{bmatrix}$.
        \end{itemize}
        Solve the sign-restricted square-root Lasso on the formulated data:
        \begin{align*}
            \big(\hat{\bbeta}, \tilde{\bbeta}\big) = \arg\min_{\substack{\bb\in\mathbb{R}^{2|\hat{\mathcal{M}}|} \\ b_j s_j \ge 0,\ \forall j  \\\ b_{j+|\hat{\mathcal{M}}|} s_j \ge 0,\ \forall j }}
            \big\{
            \big| \by_{\text{infer}} - \bX_{\text{aug}}\bb \big|_2
            + \lambda |\bb|_1
            \big\}.
        \end{align*}
        Define the feature importance statistic for each $j \in \hat{\mathcal{M}}$:
        \begin{align*}
            W_j = |\hat{\beta}_j| - |\tilde{\beta}_j|\,,
        \end{align*}
        where $\hat{\beta}_j$ and $\tilde{\beta}_j$ denote the $j$-th components of the estimators $\hat{\bbeta}$ and $\tilde{\bbeta}$, respectively.
        \STATE\textbf{Step 3}. Given $\alpha \in(0,1)$, compute the threshold
        \begin{equation*}
            T = \min\bigg\{ t > 0 : \frac{1 + \#\{j : W_j \le -t\}}{\#\{j : W_j \ge t\}} \le \alpha \bigg\}\,.
        \end{equation*}
        \STATE\textbf{Step 4}. Select the features with $W_j \ge T$ and aggregate them to construct a selection set $\hat{\mathcal{S}}$.
    \end{algorithmic}
\end{algorithm}

\begin{algorithm}[htbp!]\small
    \renewcommand{\thealgorithm}{A7}
    \caption{Ordinary least squares based Benjamini-Hochberg procedure}
    \begin{algorithmic}
        \STATE\textbf{Step 1}. Randomly split the full data $(\bX, \by)$ into a screening set $(\bX^{(0)}, \by^{(0)})$ of size $n_0$ and an inference set $(\bX^{(1)}, \by^{(1)})$ of size $n_1 = n - n_0$. Fit the Lasso on the screening set to obtain the active set $\hat{\mathcal{M}} = \{j : \hat{\beta}^{(0)}_j \neq 0\}$ with cardinality $|\hat{\mathcal{M}}|$, where $\hat{\beta}^{(0)}_j$ denotes the $j$-th component of the estimator $\hat{\bbeta}^{(0)}$ given by
\begin{align*}
\hat{\bbeta}^{(0)}
 = \arg\min_{\bbeta \in \mathbb{R}^{d}}
\bigg\{
\frac{1}{n_0} \big| \by^{(0)} - \bX^{(0)}\bbeta \big|_2^2
+ \varrho_{7}^\ast | \bbeta |_1
\bigg\}\,.
\end{align*}
For each feature $j \in \hat{\mathcal{M}}$, record its estimated sign $s_j = \mathrm{sign}(\hat{\beta}^{(0)}_j) \in \{\pm 1\}$ from the screening stage.
        \STATE\textbf{Step 2}. Let $\bX^{(1)}_{\hat{\mathcal{M}}}$ be the submatrix of $\bX^{(1)}$ with columns in $\hat{\mathcal{M}}$. Compute the OLS estimator
        \begin{align*}
            \hat{\bbeta}_{\mathrm{LS}} = \big(\bX^{(1)\top}_{\hat{\mathcal{M}}} \bX^{(1)}_{\hat{\mathcal{M}}}\big)^{-1} \bX^{(1)\top}_{\hat{\mathcal{M}}} \by^{(1)}\,,
        \end{align*}
        and the variance estimator
        \begin{align*}
            \hat{\sigma}^2 = \frac{\big| \by^{(1)} - \bX^{(1)}_{\hat{\mathcal{M}}} \hat{\bbeta}_{\mathrm{LS}} \big|_2^2}{n_1 - |\hat{\mathcal{M}}|}\,.
        \end{align*}
        For each $j \in \hat{\mathcal{M}}$, define the t-statistic
        \begin{align*}
            t_j = \frac{(\hat{\bbeta}_{\mathrm{LS}})_j}{\hat{\sigma} \cdot \sqrt{\big((\bX^{(1)}_{\hat{\mathcal{M}}})^\top \bX^{(1)}_{\hat{\mathcal{M}}}\big)^{-1}_{jj}}}\,,
        \end{align*}
        and compute the one-sided $p$-value with the test direction determined by $s_j$:
        \begin{align*}
            \tilde P_j =
            \begin{cases}
            1 - T_{n_1-|\hat{\mathcal{M}}|}(t_j), & s_j = +1\,, \\
            ~~~T_{n_1-|\hat{\mathcal{M}}|}(t_j), & s_j = -1\,,
            \end{cases}
        \end{align*}
        where $T_\nu(\cdot)$ denotes the cumulative distribution function of Student's t-distribution with $\nu$ degrees of freedom.
        \STATE\textbf{Step 3}. Given $\alpha \in(0,1)$, let $\tilde P_{(1)} \leq \cdots \leq \tilde P_{(|\hat{\mathcal{M}}|)}$ be the ordered versions of the $\tilde P_j$'s, find
        \begin{equation*}
            \tilde{R}=\max\bigg\{i\in[|\hat{\mathcal{M}}|]: \tilde P_{(i)} \leq \frac{i \alpha}{|\hat{\mathcal{M}}|}\bigg\}\,,
        \end{equation*}
        provided that the maximum exists; otherwise, let $\tilde{R}=0$.
        \STATE\textbf{Step 4}. Select the feature corresponding to $\tilde P_{(j)}$ with $j \leq \tilde{R}$ and aggregate them to construct a selection set $\hat{\mathcal{S}}$.
    \end{algorithmic}
\end{algorithm}

\subsubsection{Evaluation criteria and results}

To fix the scope and ensure a fair comparison, we follow the numerical
experiments of
\citetS{Barber2019Knockoff_app} and evaluate all procedures on a common scale
against the true full model: with $\hat{\mathcal{S}}$
denoting the final selection set,
\begin{align*}
\text{FDR} = \mathbb{E} \Bigg[ \frac{|\hat{\mathcal{S}} \cap \{j : \beta_{j}^0 = 0\}|}{|\hat{\mathcal{S}}| \vee 1} \Bigg]
\quad\text{and}\quad
\text{Power} = \mathbb{E} \Bigg[ \frac{|\hat{\mathcal{S}} \cap \{j : \beta_{j}^0 \neq 0\}|}{|\{j : \beta_{j}^0 \neq 0\}|} \Bigg].
\end{align*}

\begin{figure}[htbp!]
	\centering
	\subfloat{\includegraphics[width=.27\columnwidth]{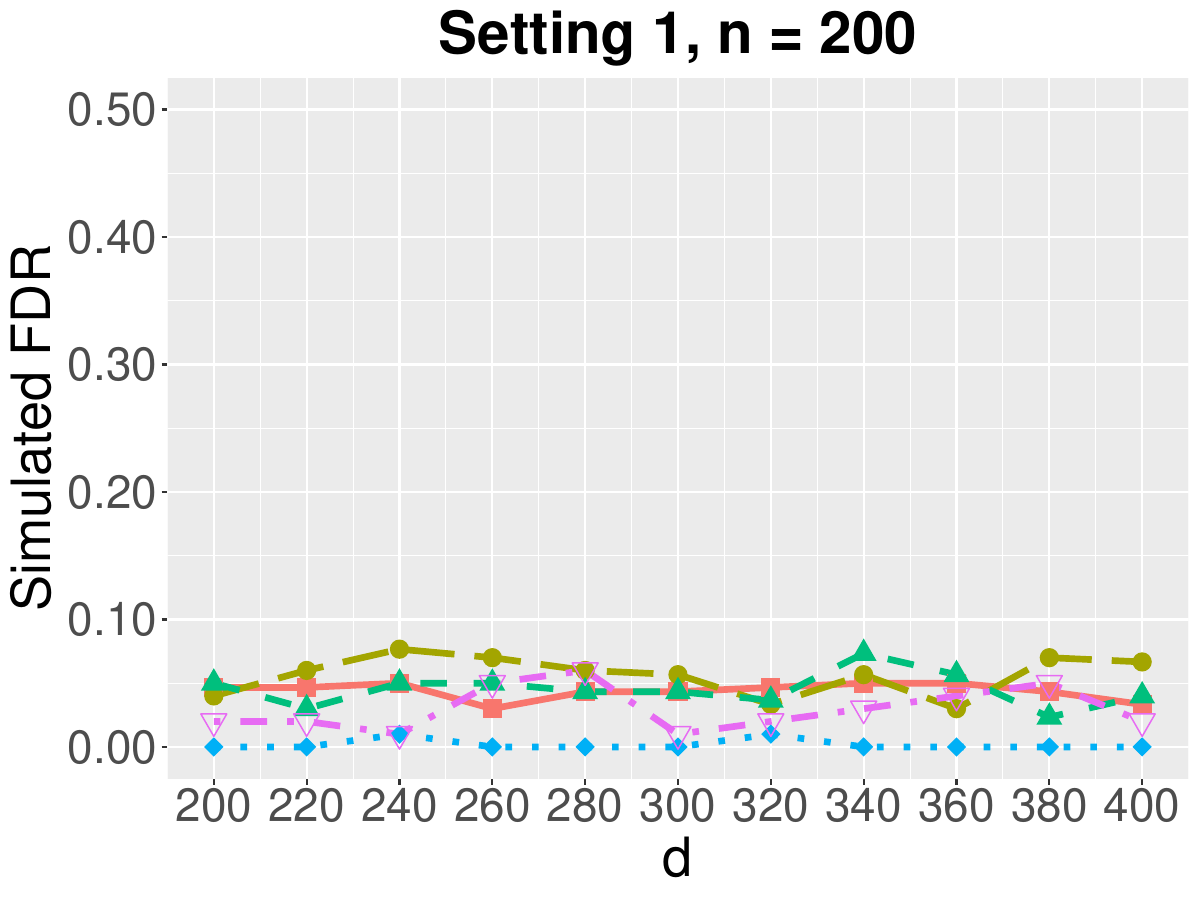}}\hspace{5pt}
	\subfloat{\includegraphics[width=.27\columnwidth]{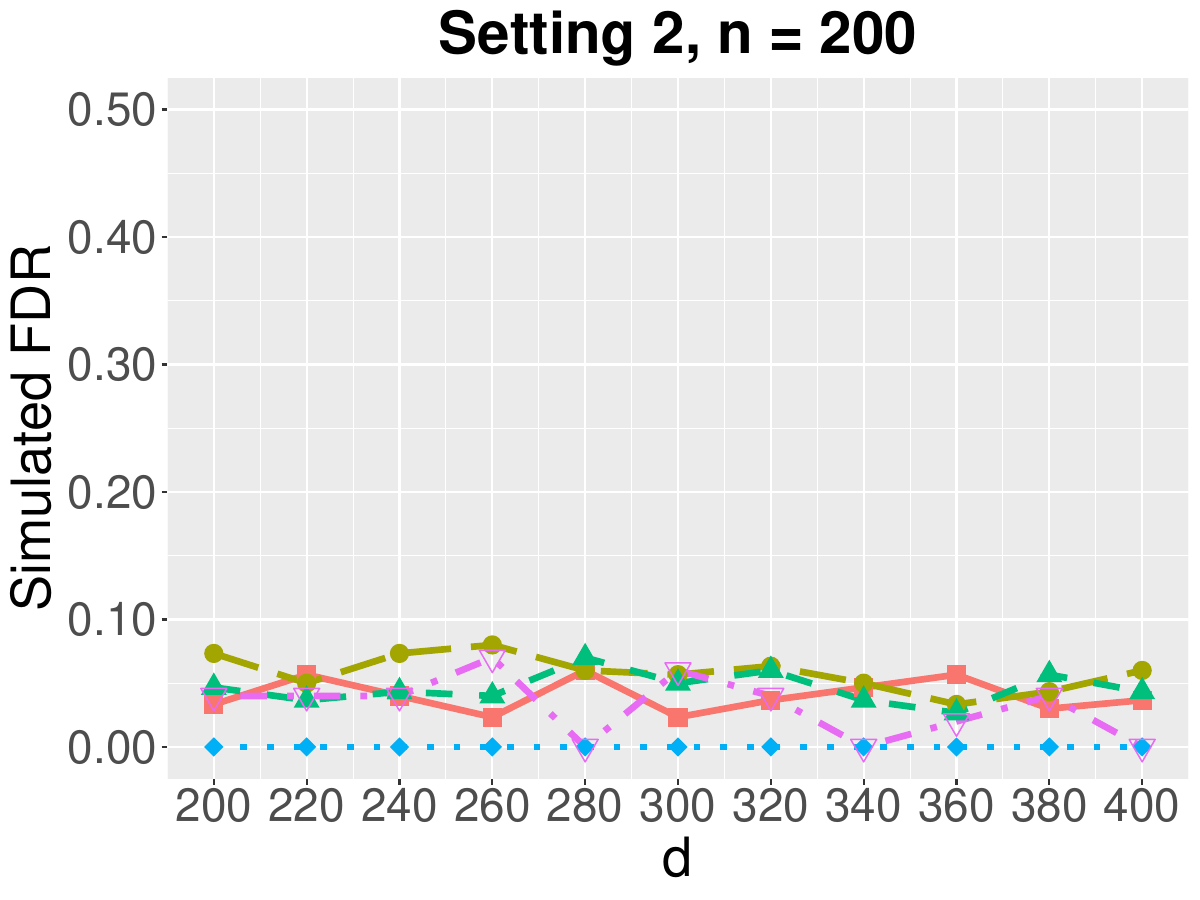}}\hspace{5pt}
	\subfloat{\includegraphics[width=.27\columnwidth]{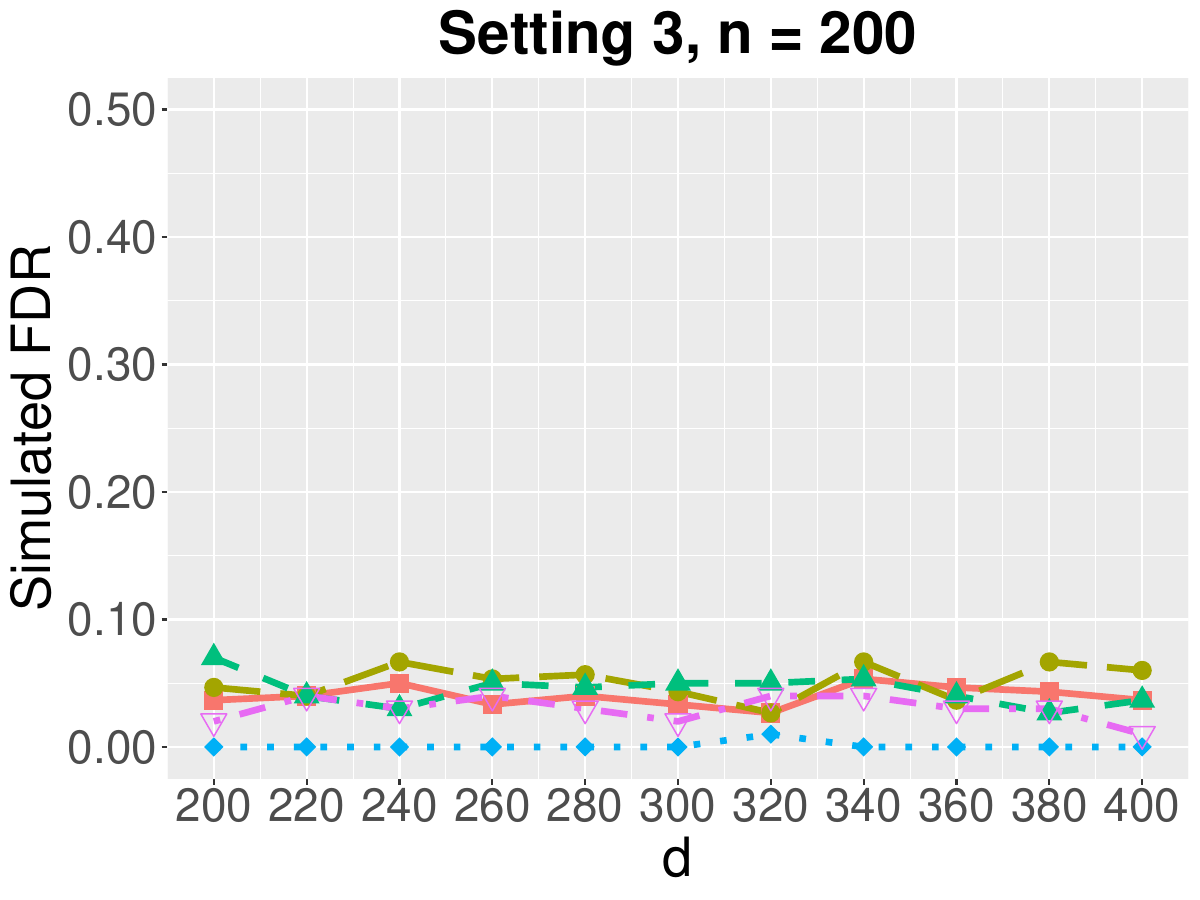}}\\
	\subfloat{\includegraphics[width=.27\columnwidth]{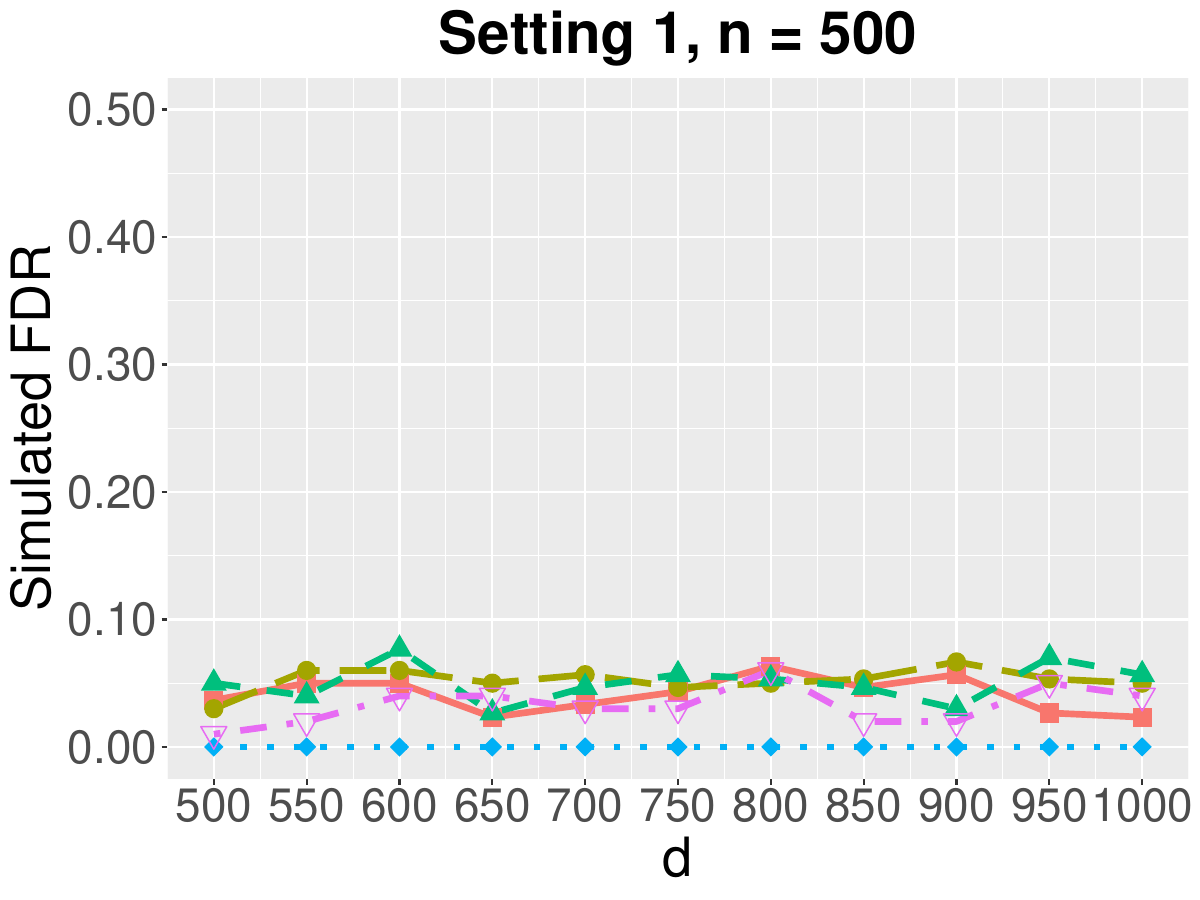}}\hspace{5pt}
    \subfloat{\includegraphics[width=.27\columnwidth]{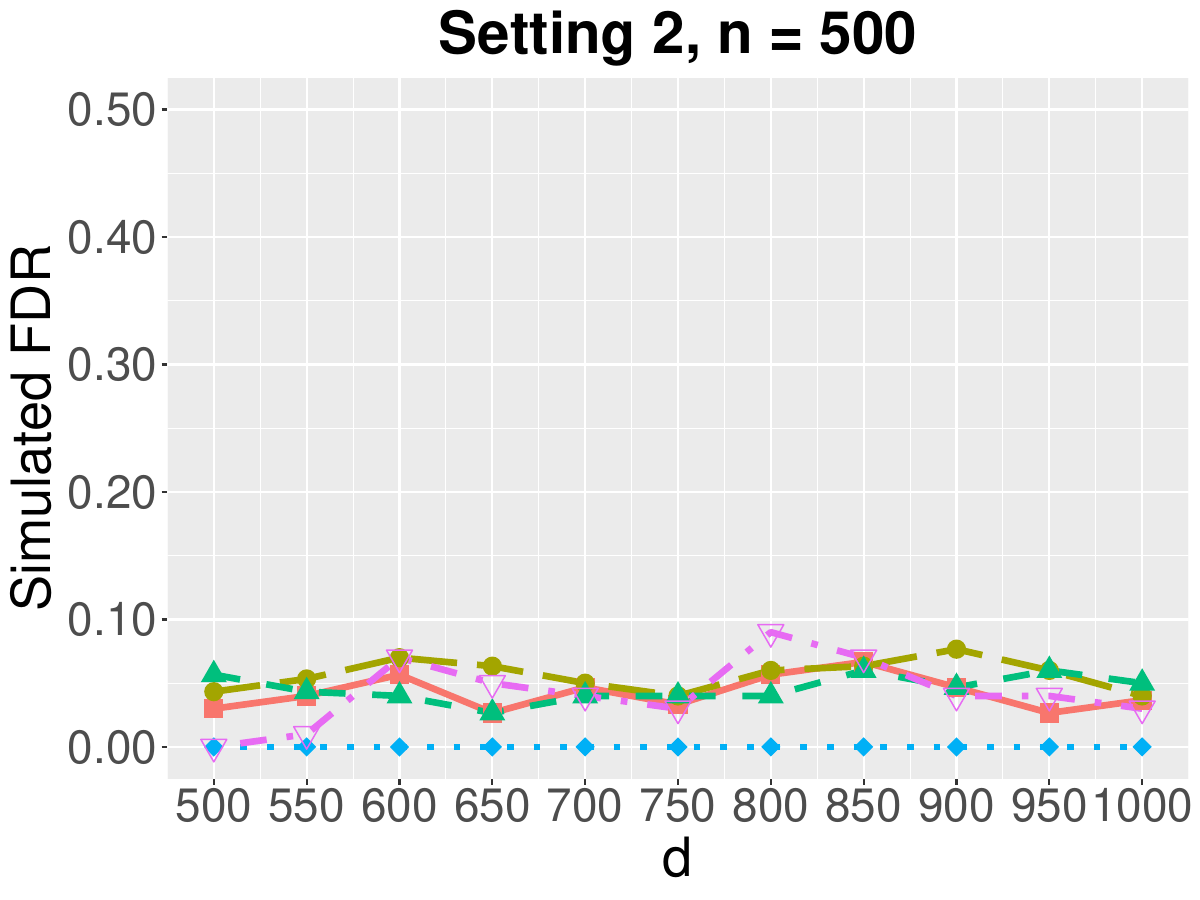}}\hspace{5pt}
    \subfloat{\includegraphics[width=.27\columnwidth]{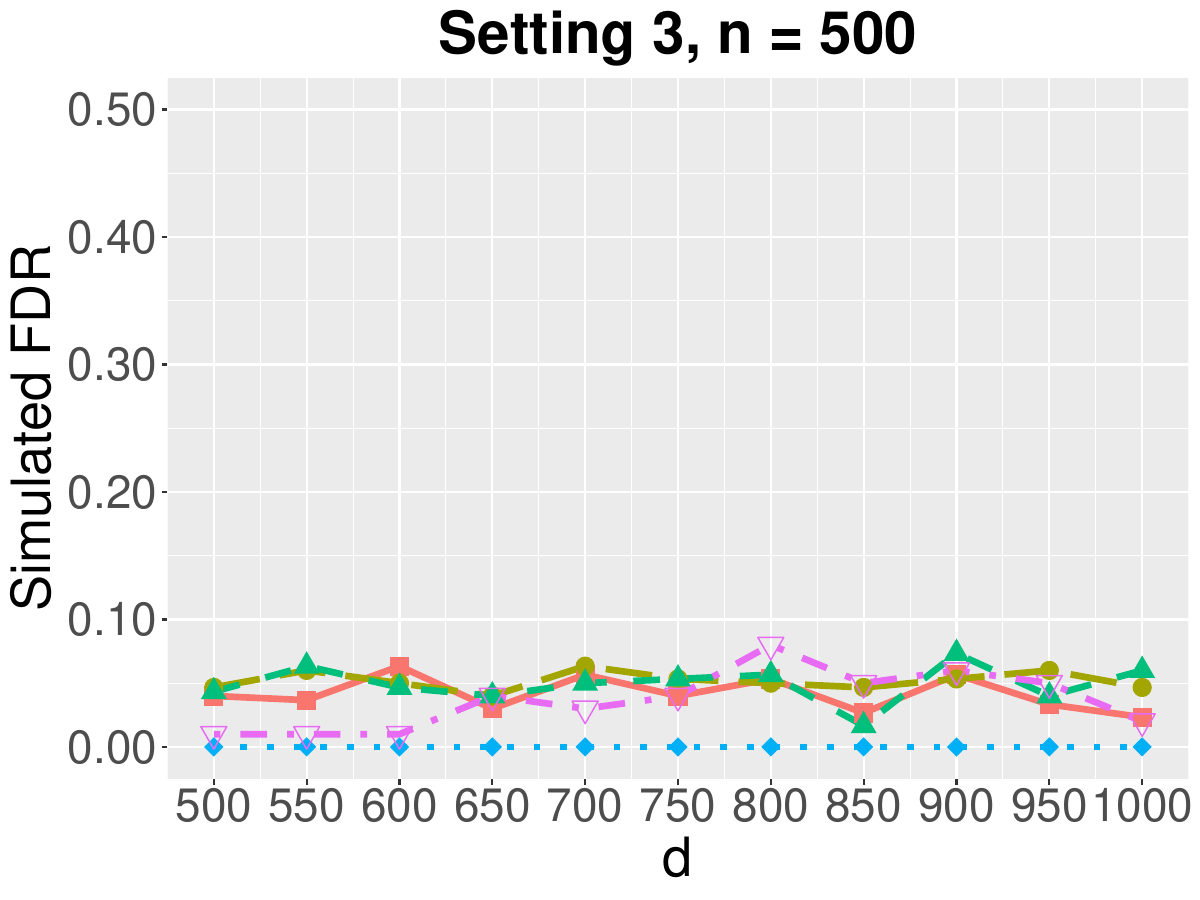}}
	\caption{\small Simulated FDR when all null hypotheses are true, for Setting 1, Setting 2, and Setting 3. The sample sizes of top row and bottom row are $n = 200$ and $n = 500$, respectively. The FDR level is $\alpha = 0.05$. The methods compared are Algorithm 1 (squares and red solid line), Algorithm 2 (circles and yellow long-dashed line), Algorithm A4 (diamonds and blue dotted line), Algorithm A5 (downward triangles and purple dot-dashed line), and Algorithm A7 (upward triangles and green dashed line).}
    \label{FDR-control-SI-005}
\end{figure}

\begin{figure}[htbp!]
	\centering
	\subfloat{\includegraphics[width=.27\columnwidth]{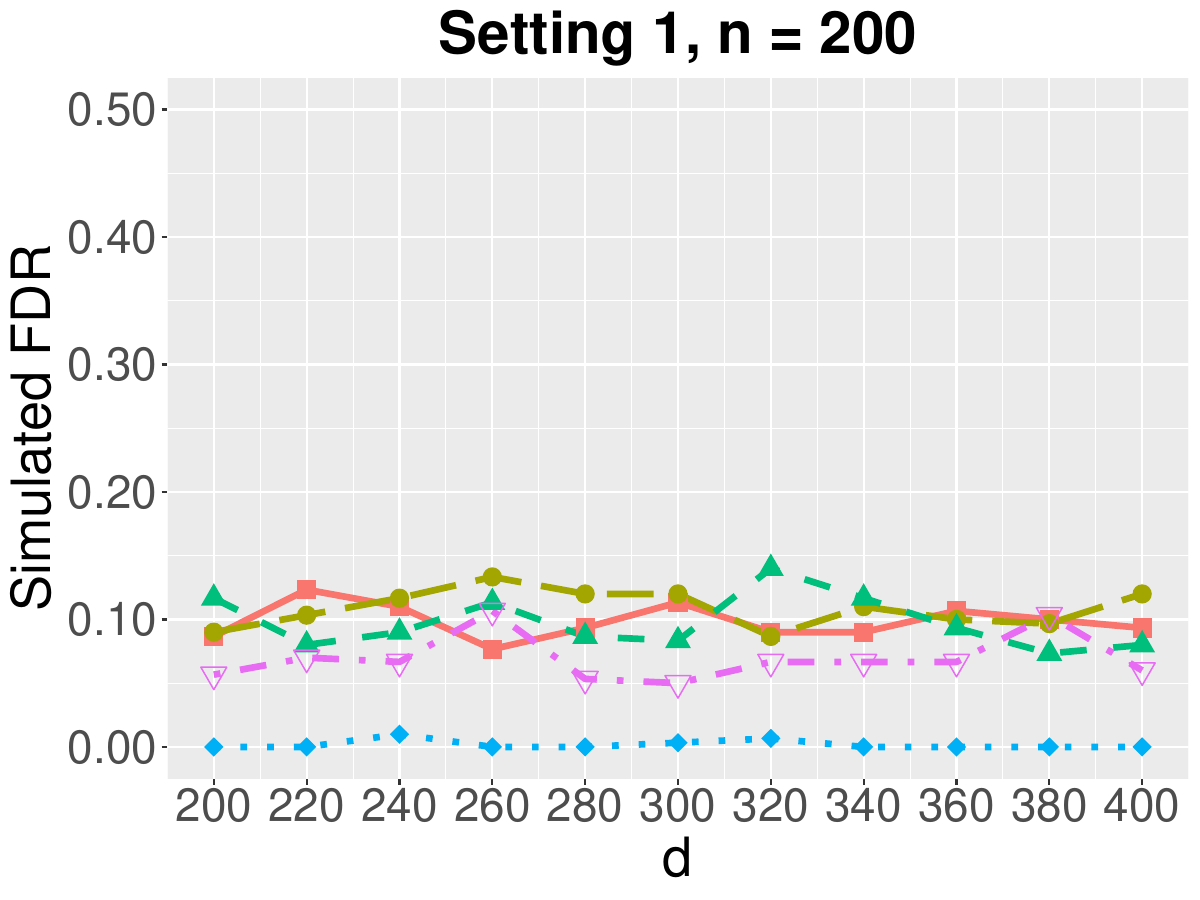}}\hspace{5pt}
	\subfloat{\includegraphics[width=.27\columnwidth]{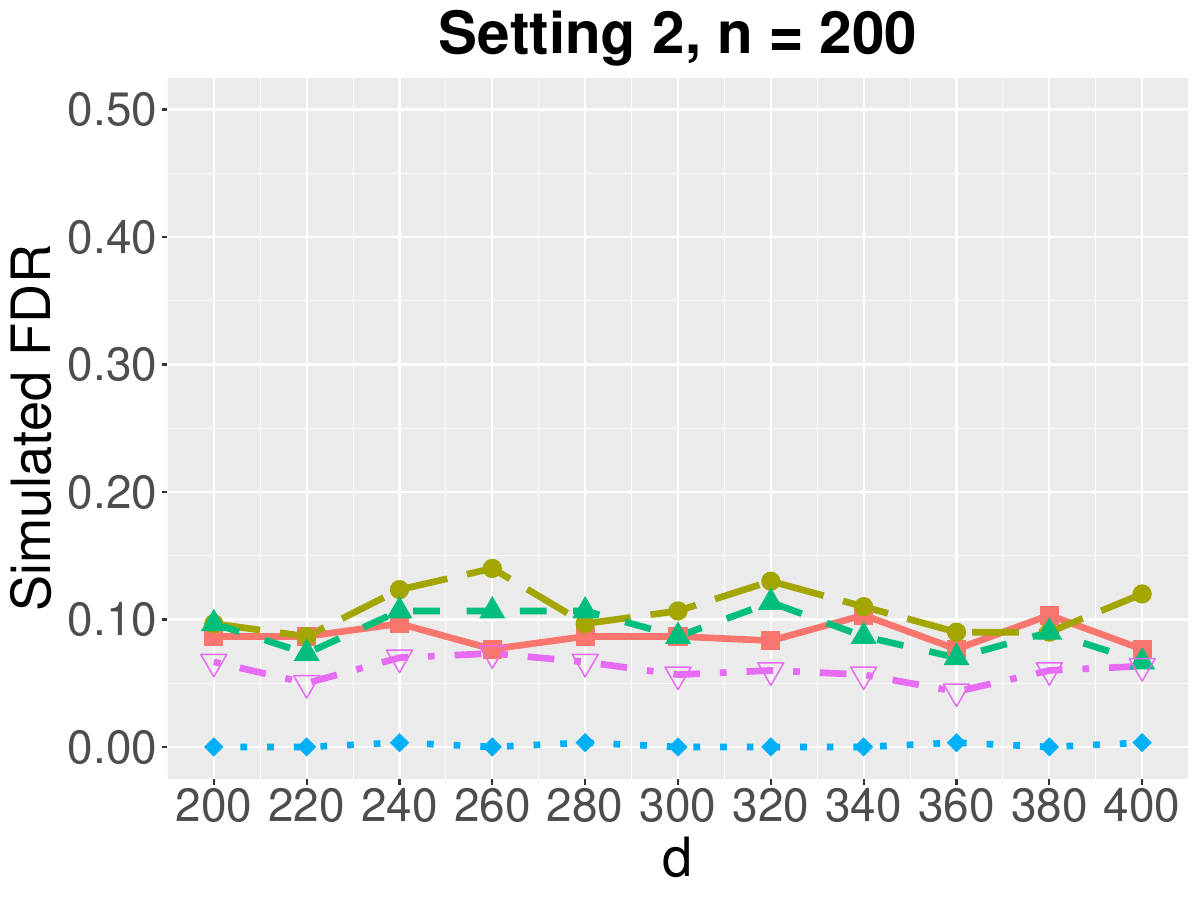}}\hspace{5pt}
	\subfloat{\includegraphics[width=.27\columnwidth]{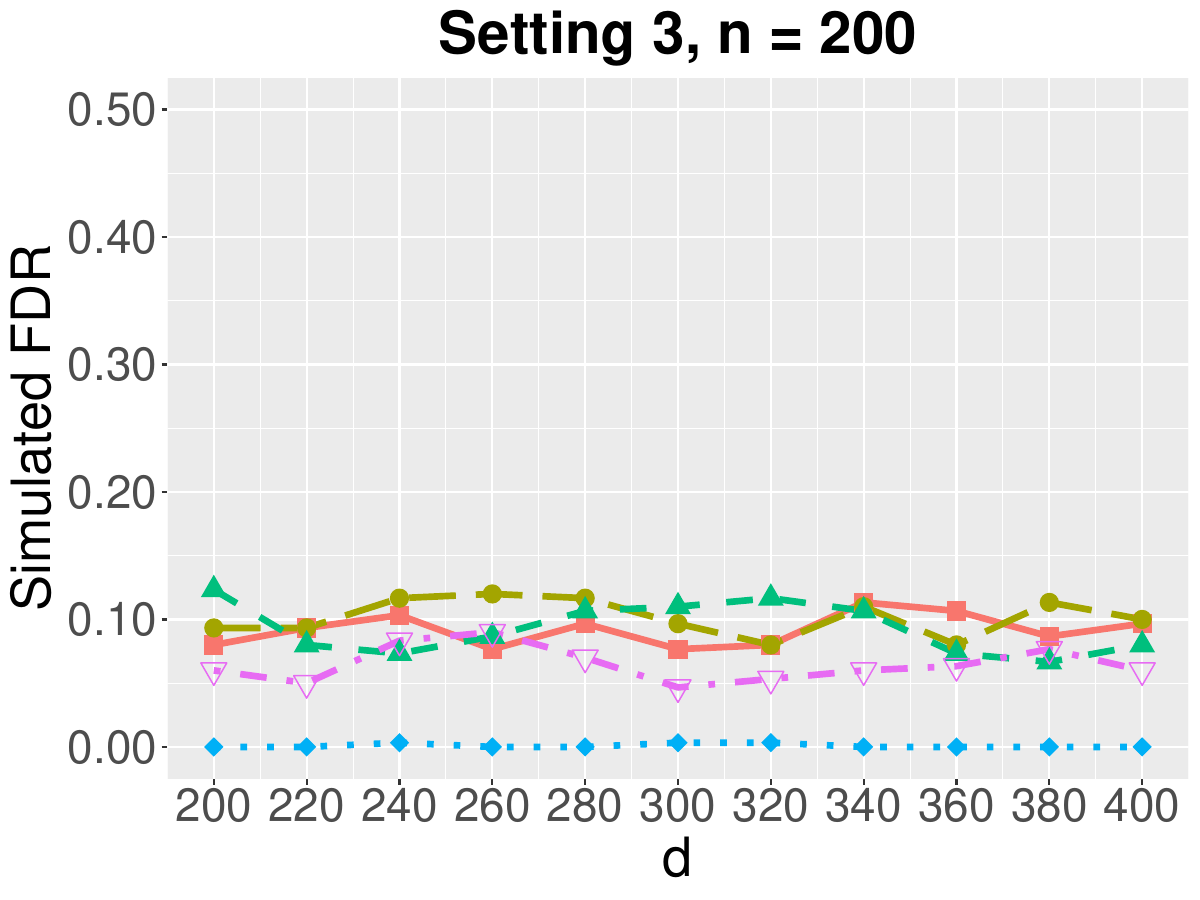}}\\
	\subfloat{\includegraphics[width=.27\columnwidth]{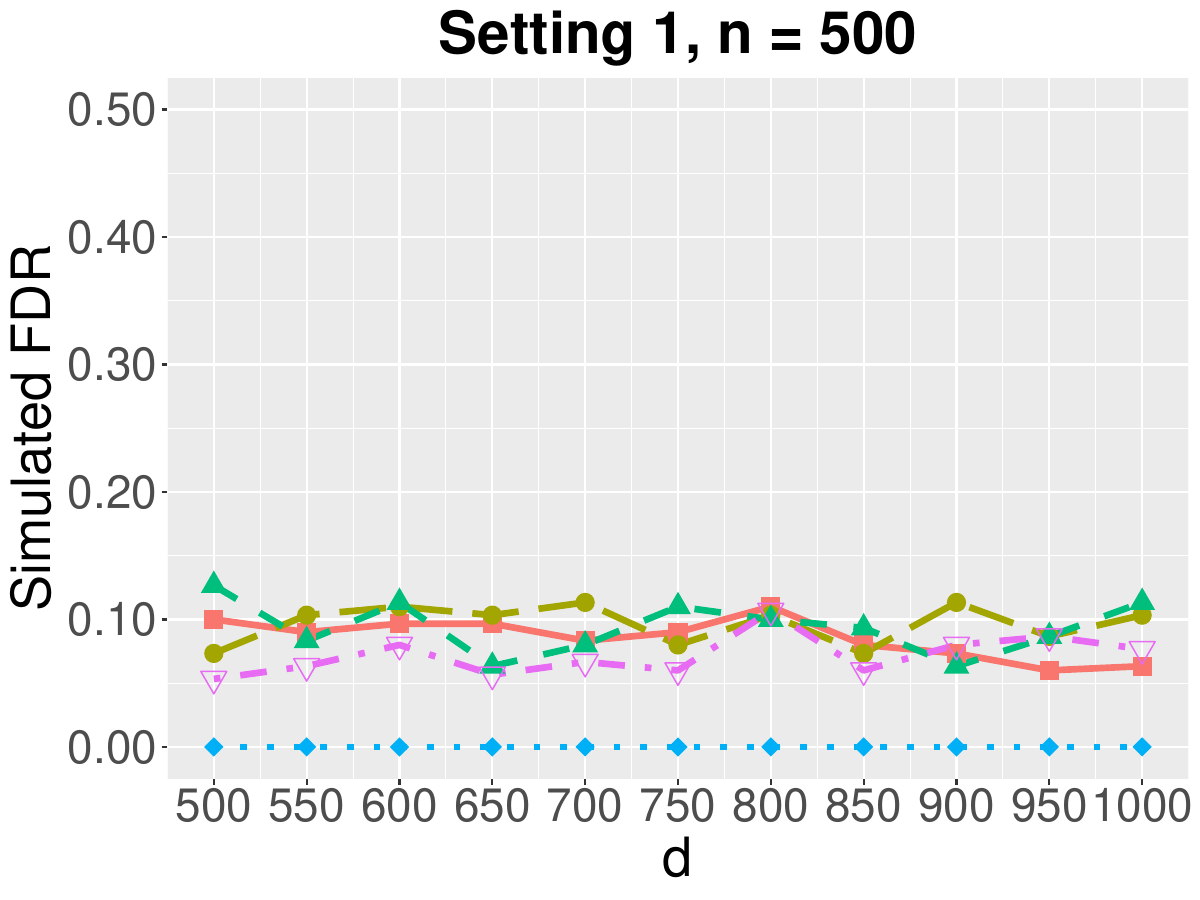}}\hspace{5pt}
    \subfloat{\includegraphics[width=.27\columnwidth]{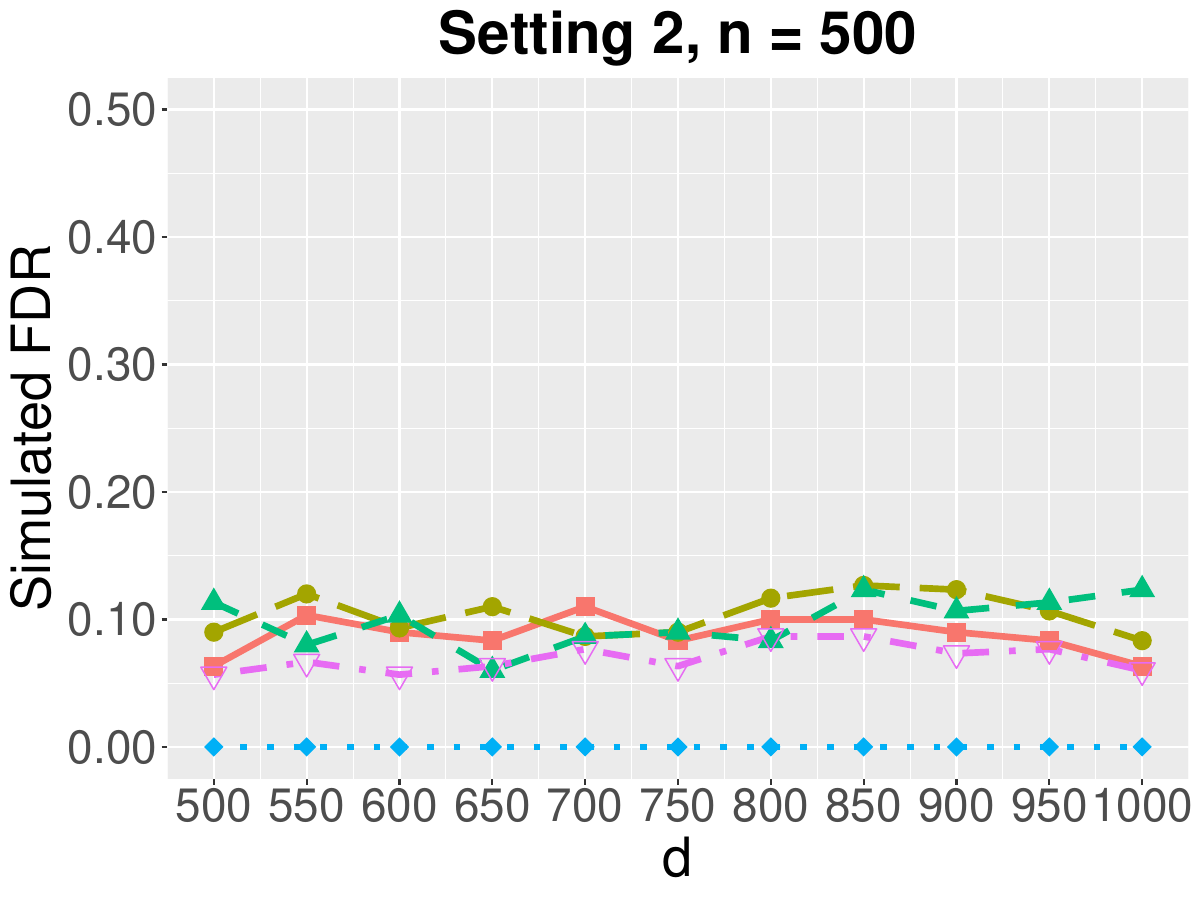}}\hspace{5pt}
    \subfloat{\includegraphics[width=.27\columnwidth]{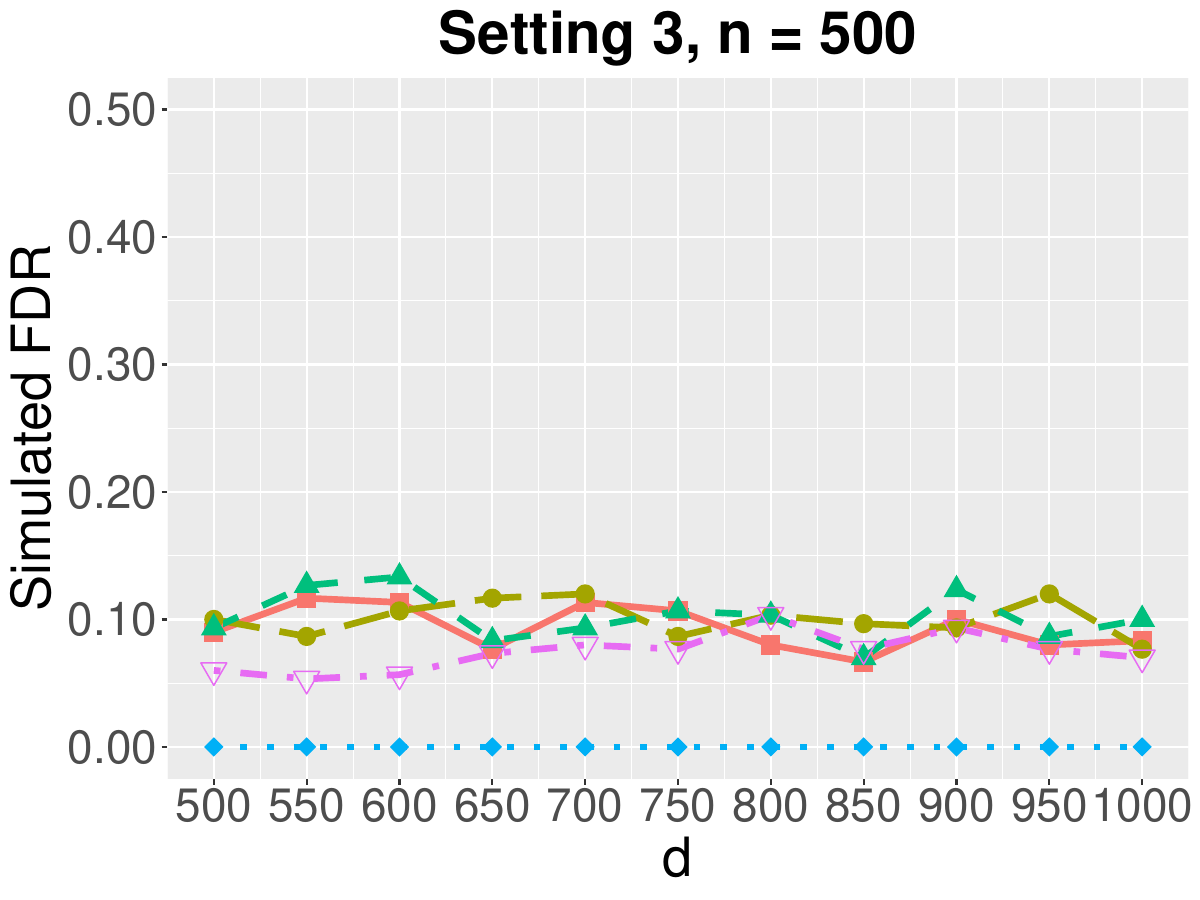}}
	\caption{\small Simulated FDR when all null hypotheses are true, for Setting 1, Setting 2, and Setting 3. The sample sizes of top row and bottom row are $n = 200$ and $n = 500$, respectively. The FDR level is $\alpha = 0.1$. The methods compared are Algorithm 1 (squares and red solid line), Algorithm 2 (circles and yellow long-dashed line), Algorithm A4 (diamonds and blue dotted line), Algorithm A5 (downward triangles and purple dot-dashed line), and Algorithm A7 (upward triangles and green dashed line).}
    \label{FDR-control-SI}
\end{figure}

Under the global null hypothesis
(Figures~\ref{FDR-control-SI-005} and \ref{FDR-control-SI}), Algorithms \ref{method_1} and \ref{method_2},
as well as Algorithms A5 and A7,
keep the empirical FDR close to the nominal level, while Algorithm A4 is markedly
conservative; both variants of Algorithm A6 select no variables under the zero-signal
configuration, so their (identically zero) FDR is omitted from the figure.

\begin{figure}[htbp!]
	\centering
	\subfloat{\includegraphics[width=.27\columnwidth]{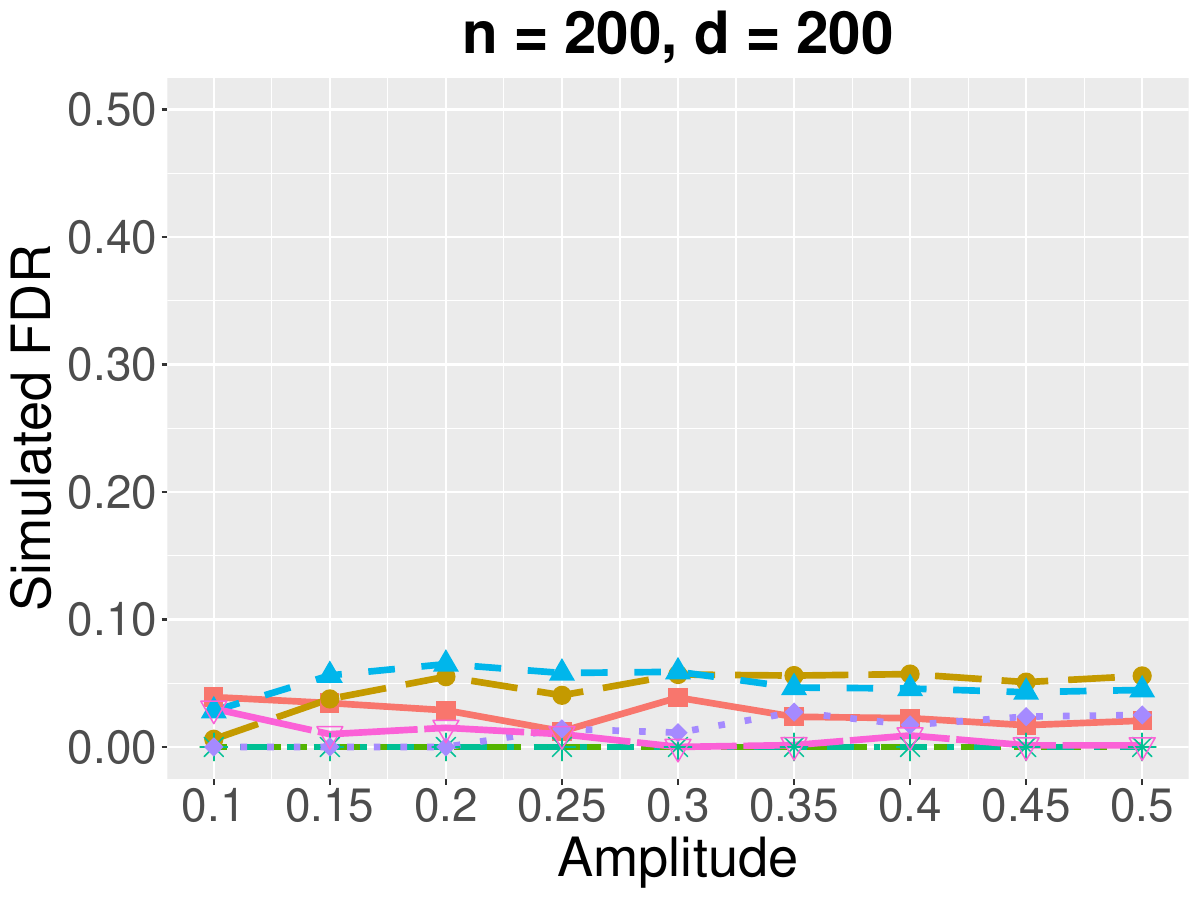}}\hspace{5pt}
	\subfloat{\includegraphics[width=.27\columnwidth]{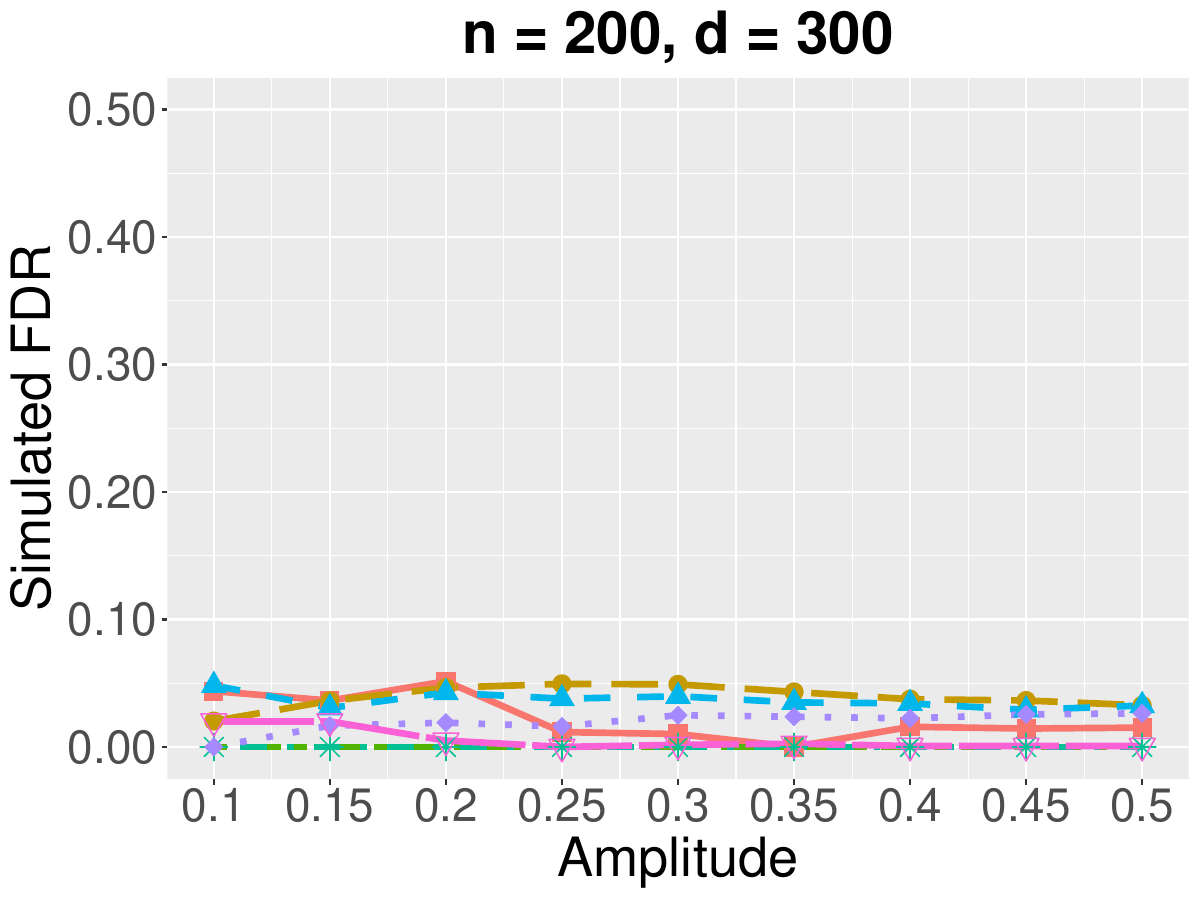}}\hspace{5pt}
	\subfloat{\includegraphics[width=.27\columnwidth]{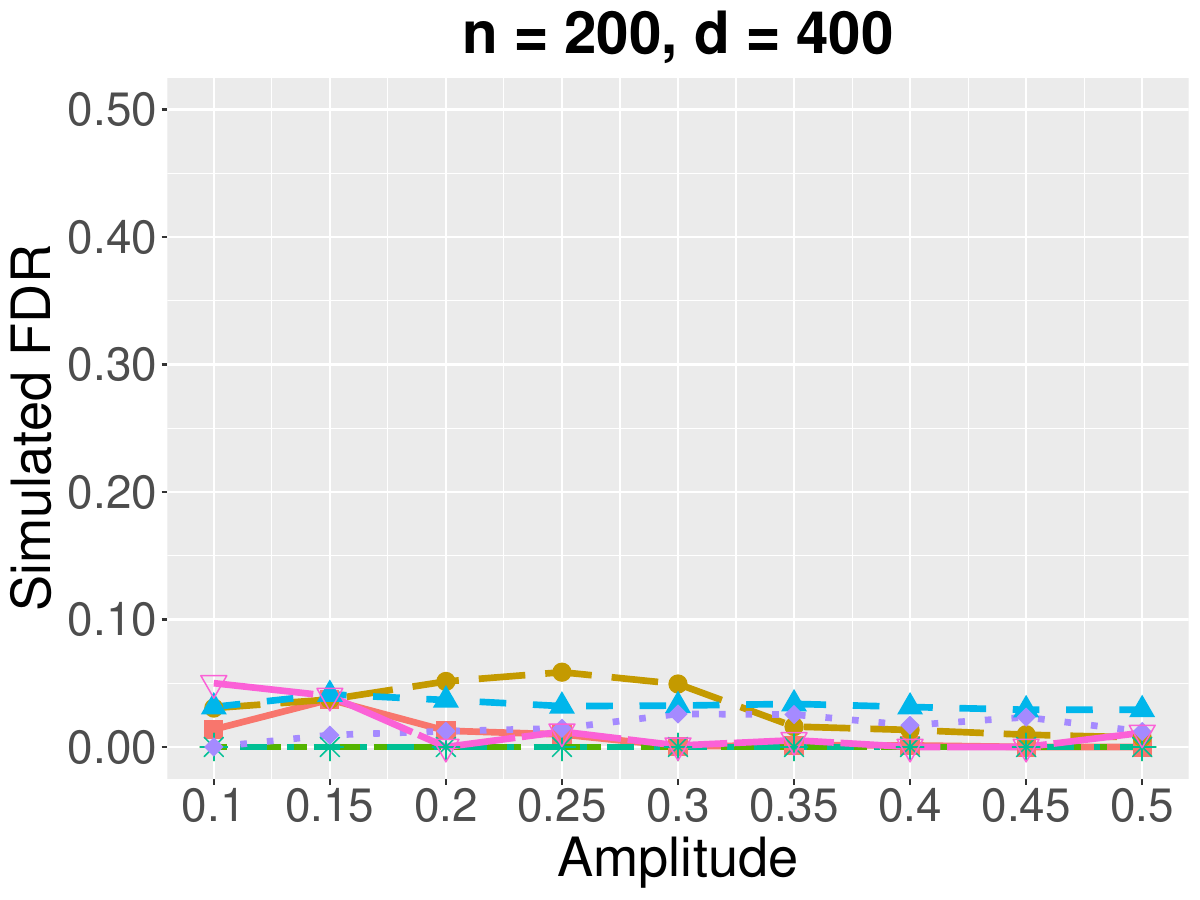}}\\
	\subfloat{\includegraphics[width=.27\columnwidth]{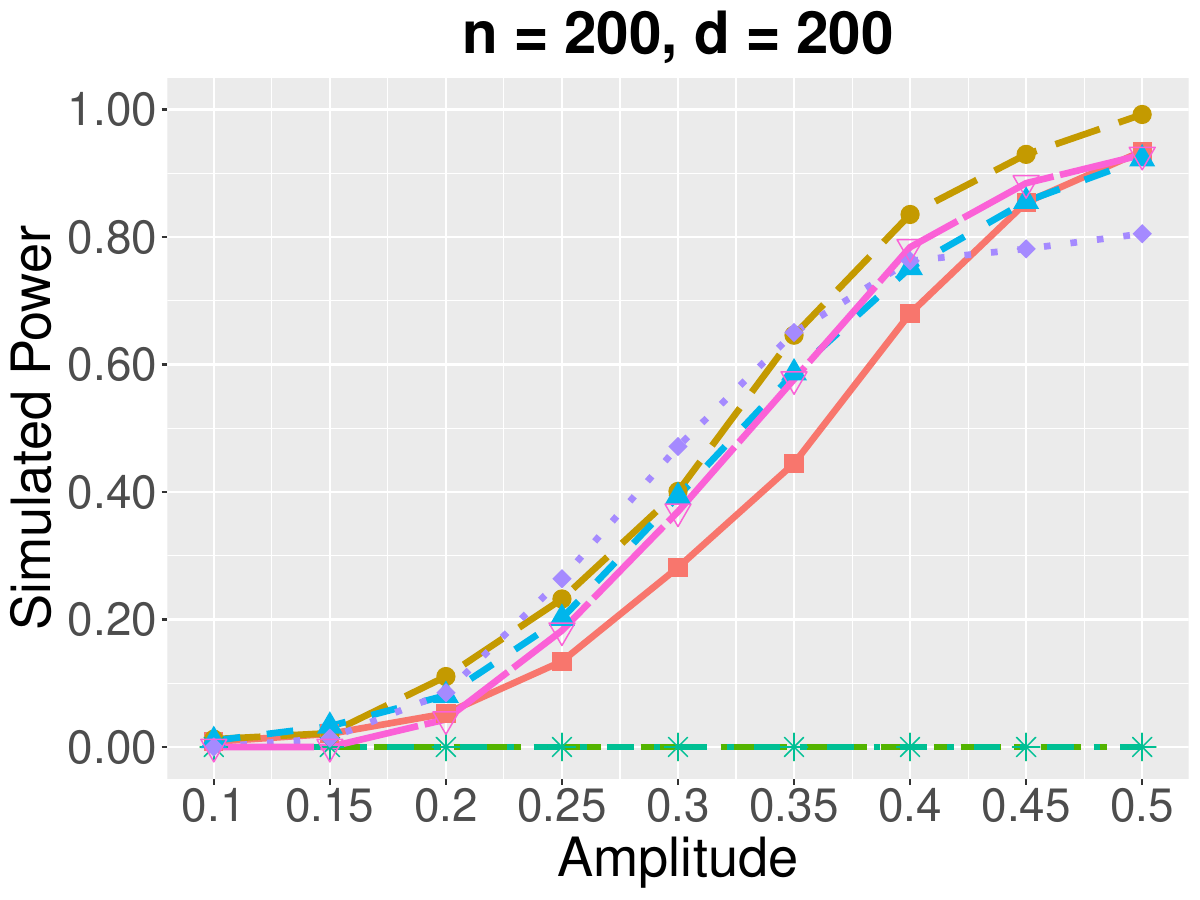}}\hspace{5pt}
    \subfloat{\includegraphics[width=.27\columnwidth]{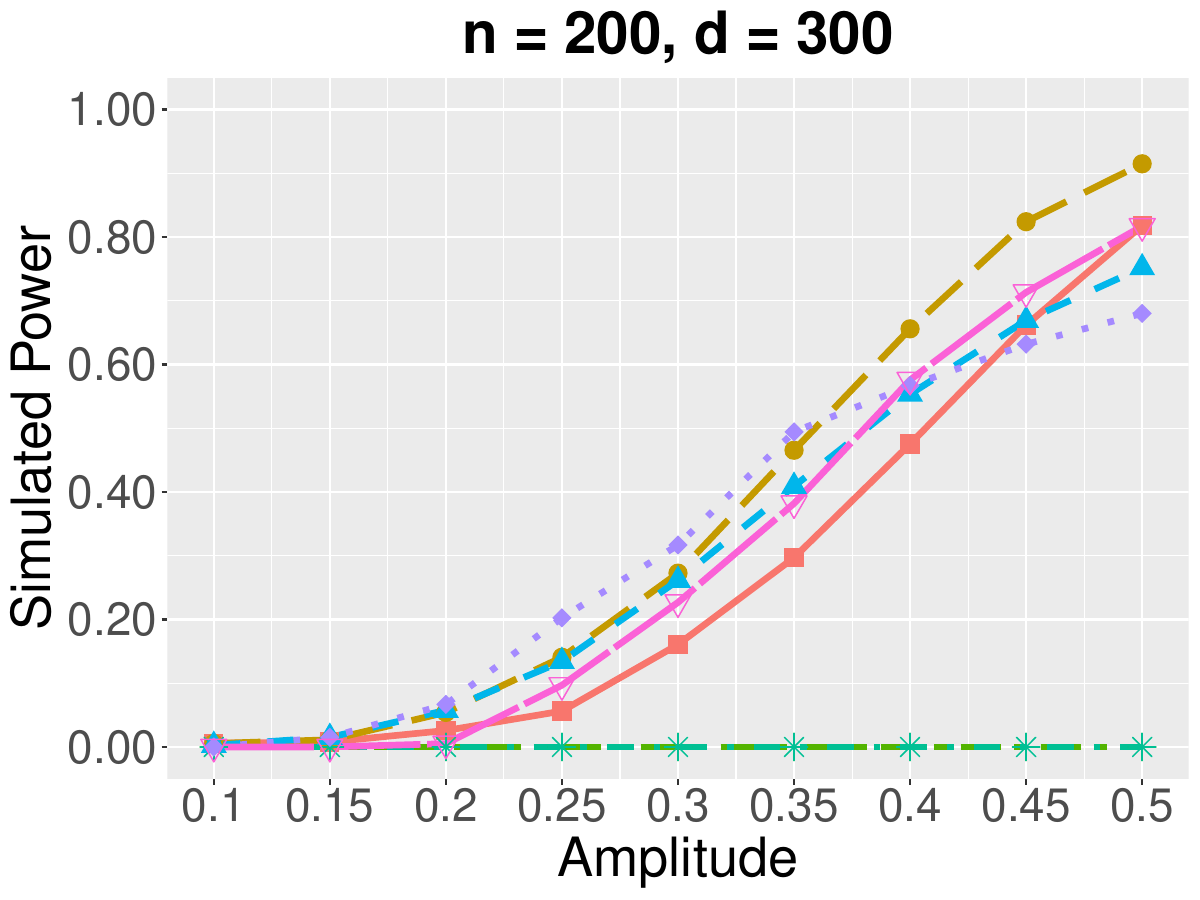}}\hspace{5pt}
    \subfloat{\includegraphics[width=.27\columnwidth]{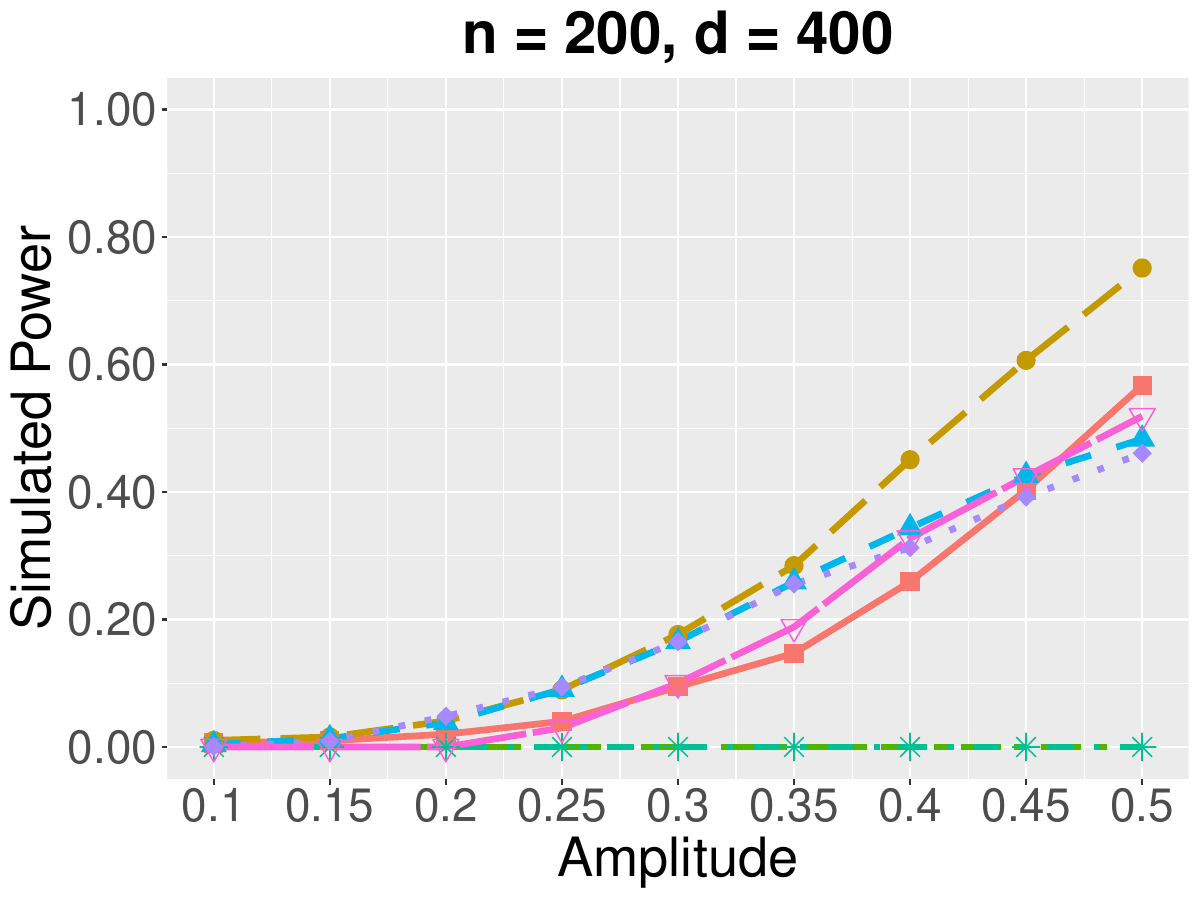}}\\
    \subfloat{\includegraphics[width=.27\columnwidth]{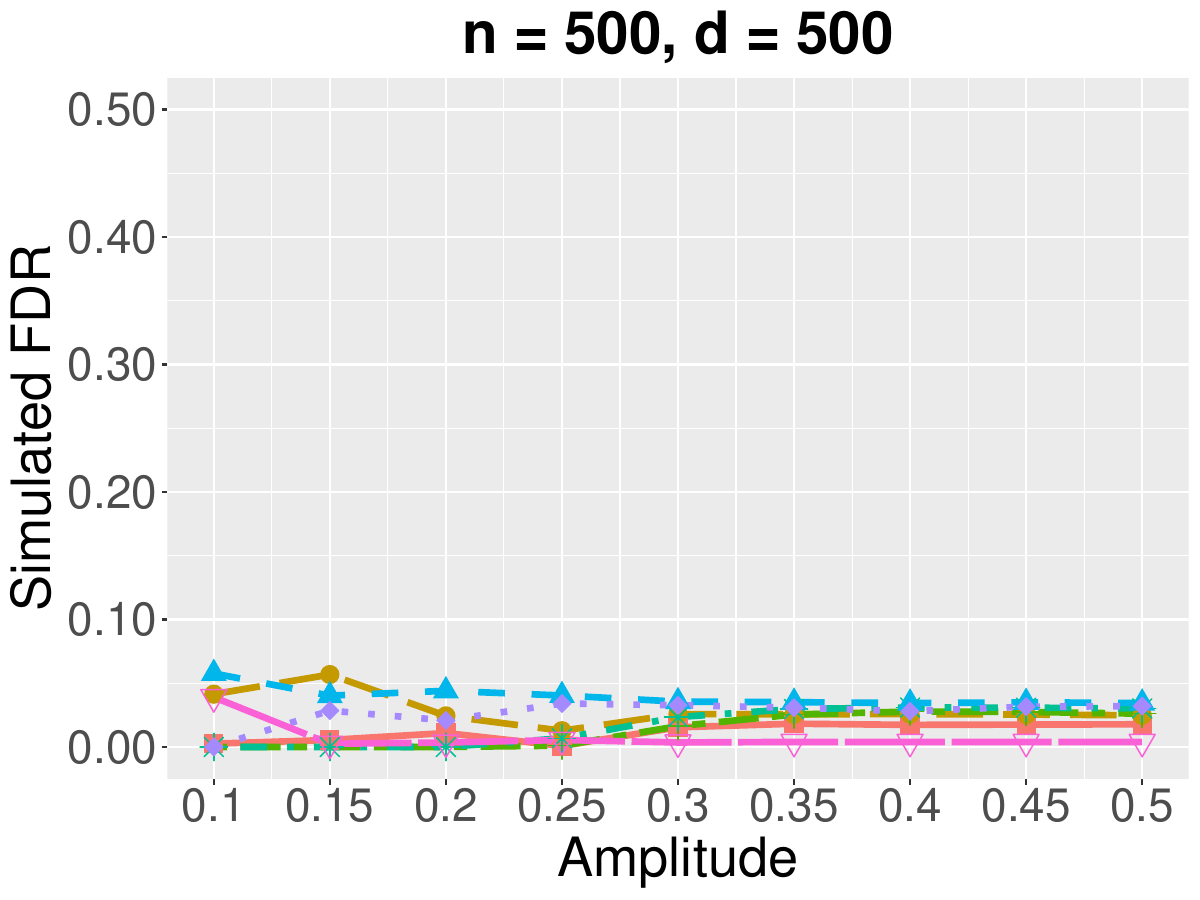}}\hspace{5pt}
	\subfloat{\includegraphics[width=.27\columnwidth]{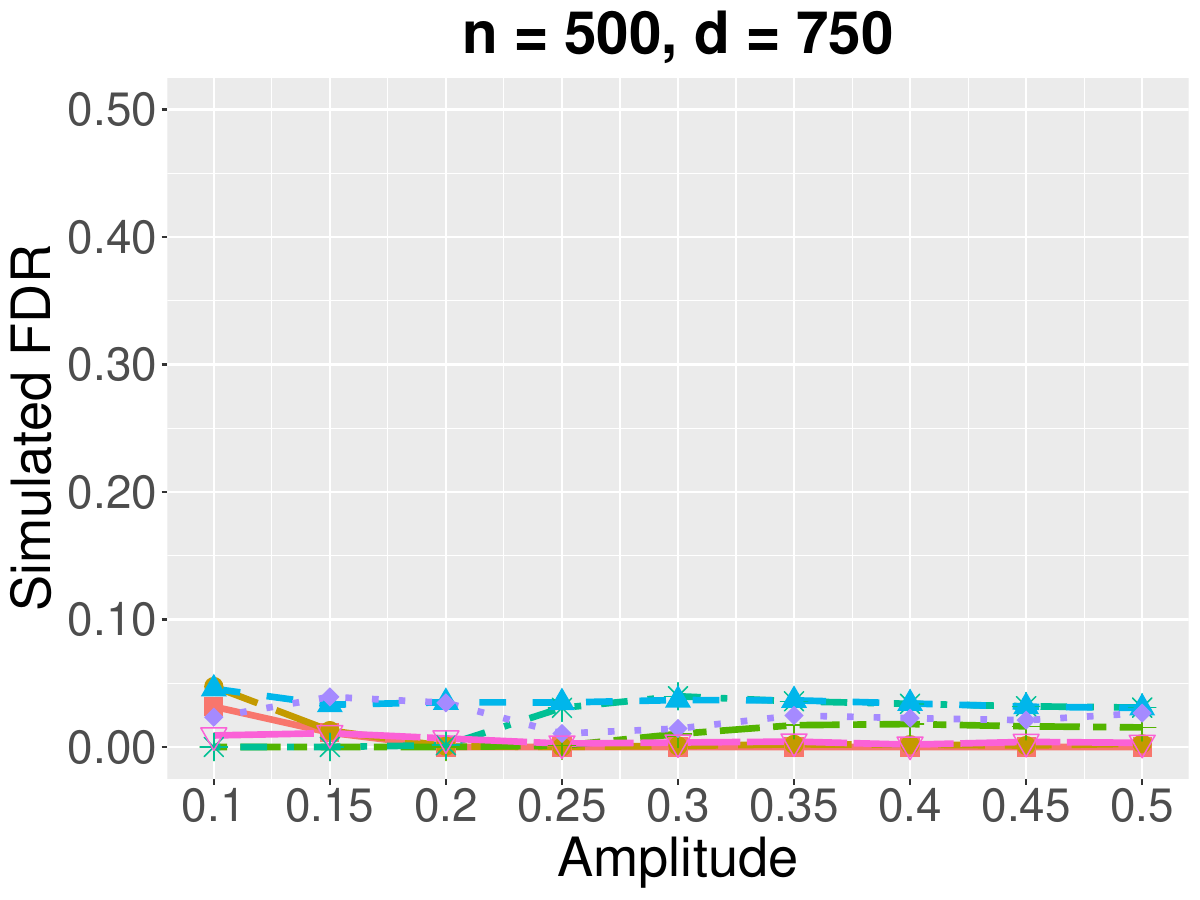}}\hspace{5pt}
	\subfloat{\includegraphics[width=.27\columnwidth]{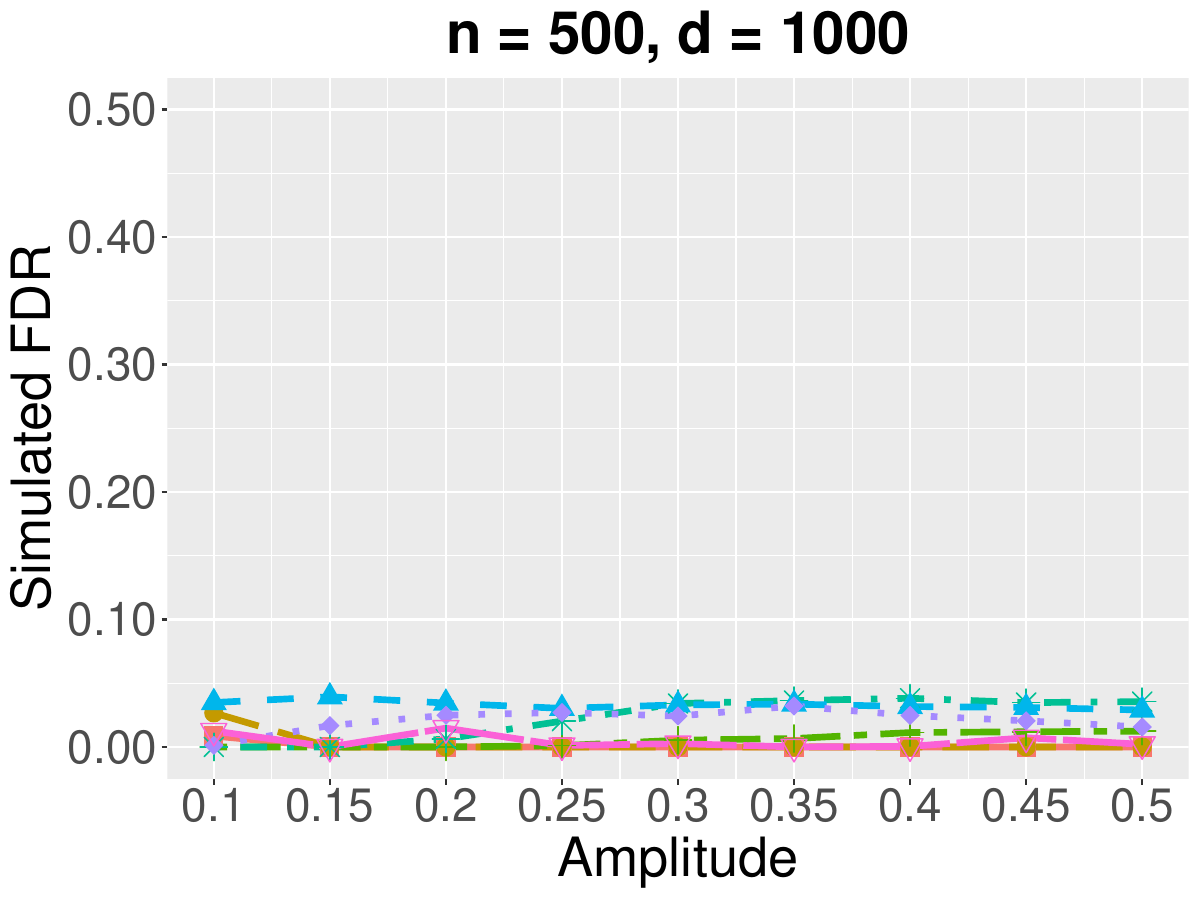}}\\
	\subfloat{\includegraphics[width=.27\columnwidth]{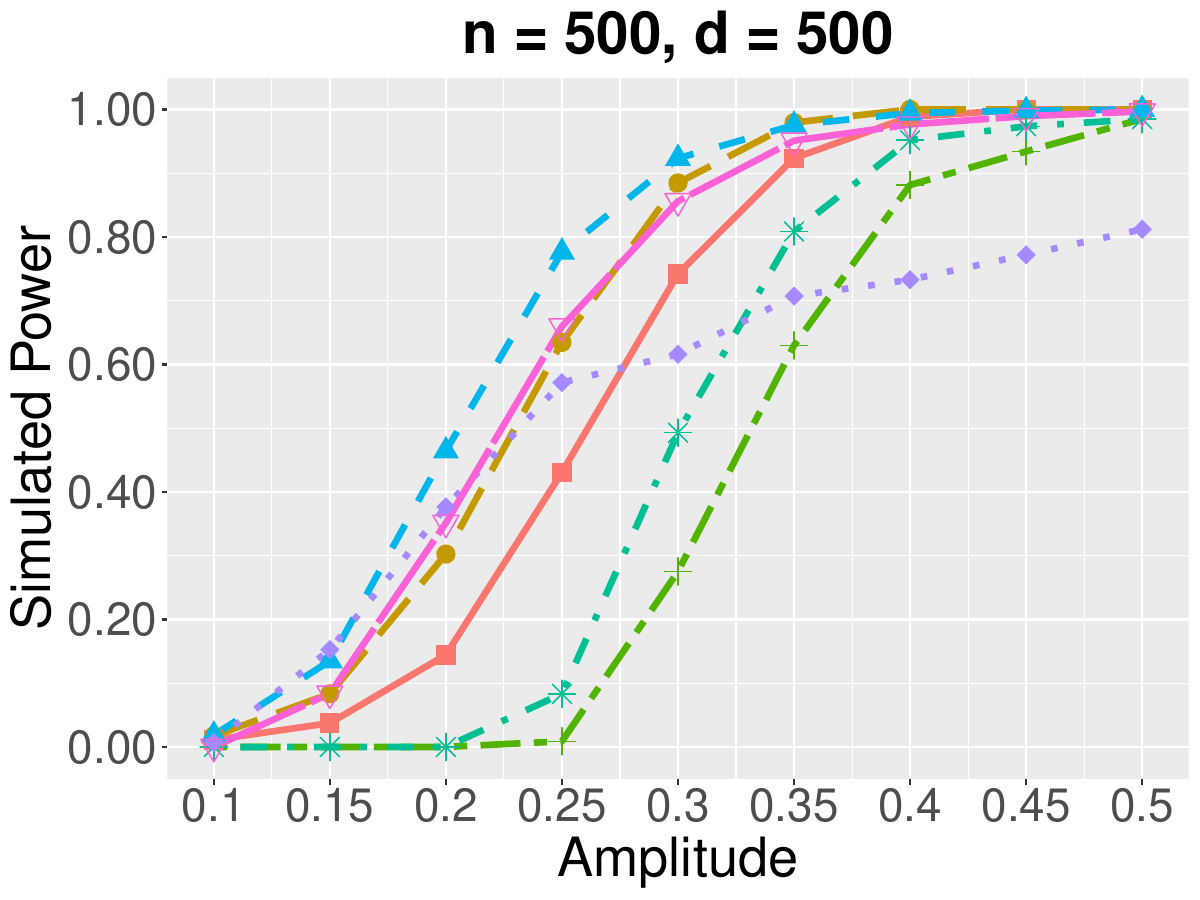}}\hspace{5pt}
    \subfloat{\includegraphics[width=.27\columnwidth]{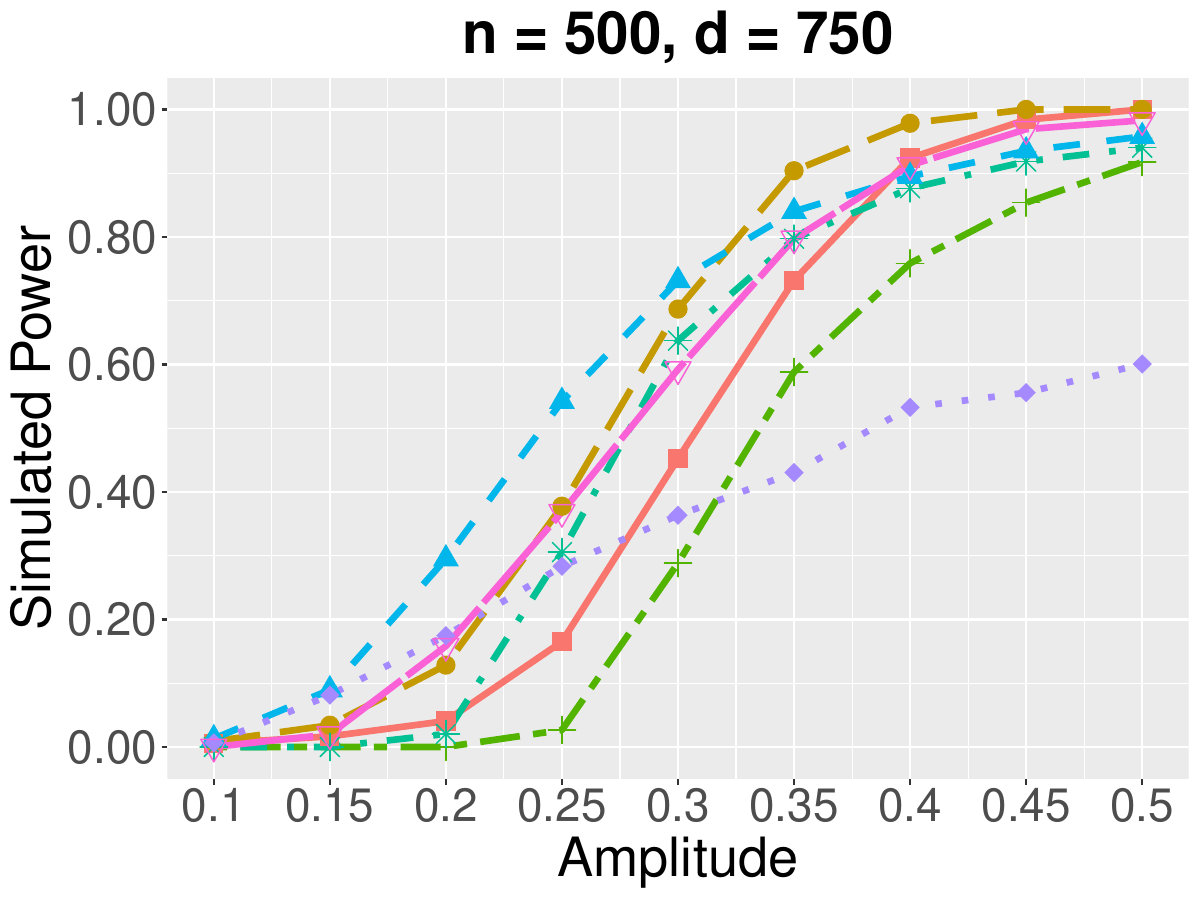}}\hspace{5pt}
    \subfloat{\includegraphics[width=.27\columnwidth]{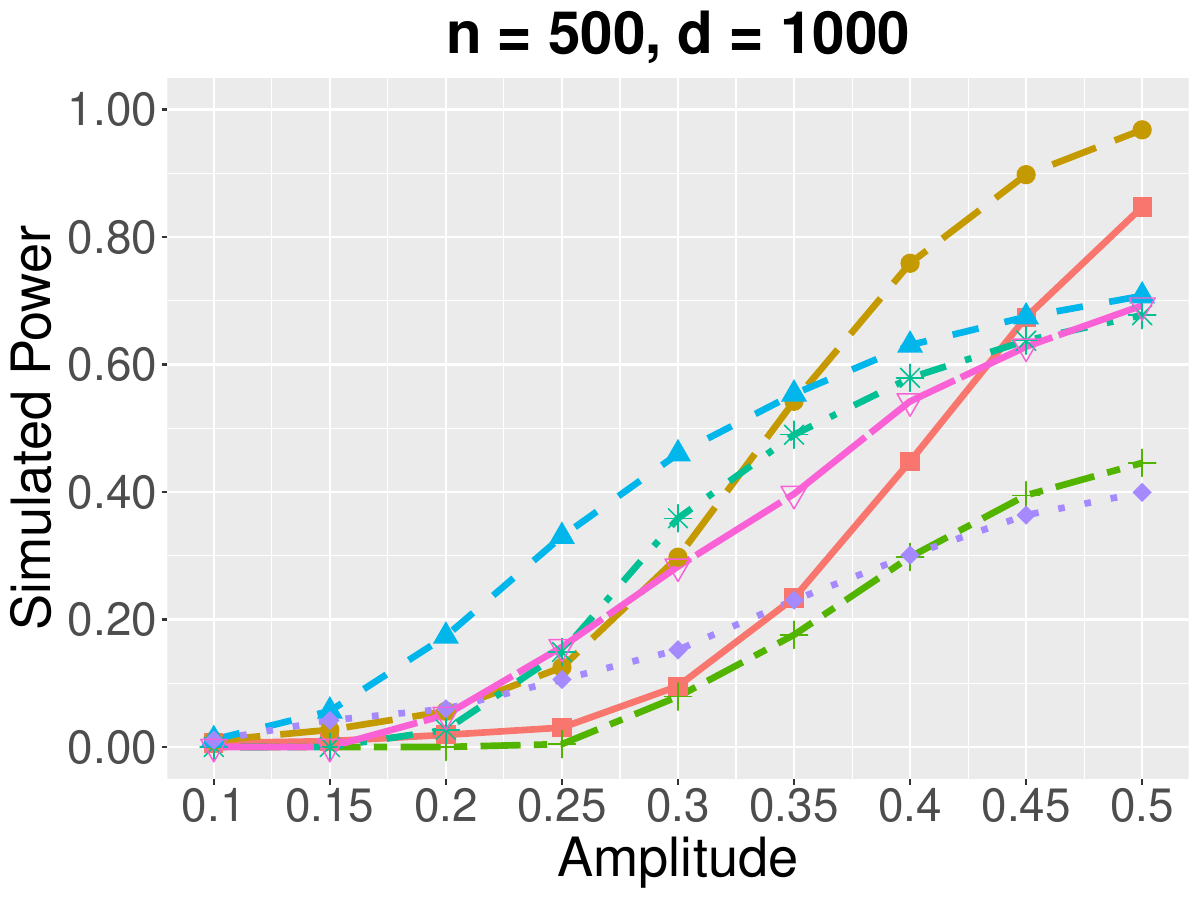}}
	\caption{\small Simulated FDR and power for different combinations of $(n,d)$. The rows of the design matrix were generated from Setting 1. The sparsity level is $k = 0.04d$ and the FDR level is $\alpha = 0.05$. The methods compared are Algorithm 1 (squares and red solid line), Algorithm 2 (circles and yellow long-dashed line), Algorithm A4 (diamonds and purple dotted line), Algorithm A5 (downward triangles and pink dot-dashed line), data splitting version of Algorithm A6 (crosses and dark green two-dashed line), data recycling version of Algorithm A6 (stars and light green dot-dashed line) and Algorithm A7 (upward triangles and blue dashed line).}
    \label{power-SI-indep-005}
\end{figure}

\begin{figure}[htbp!]
	\centering
	\subfloat{\includegraphics[width=.27\columnwidth]{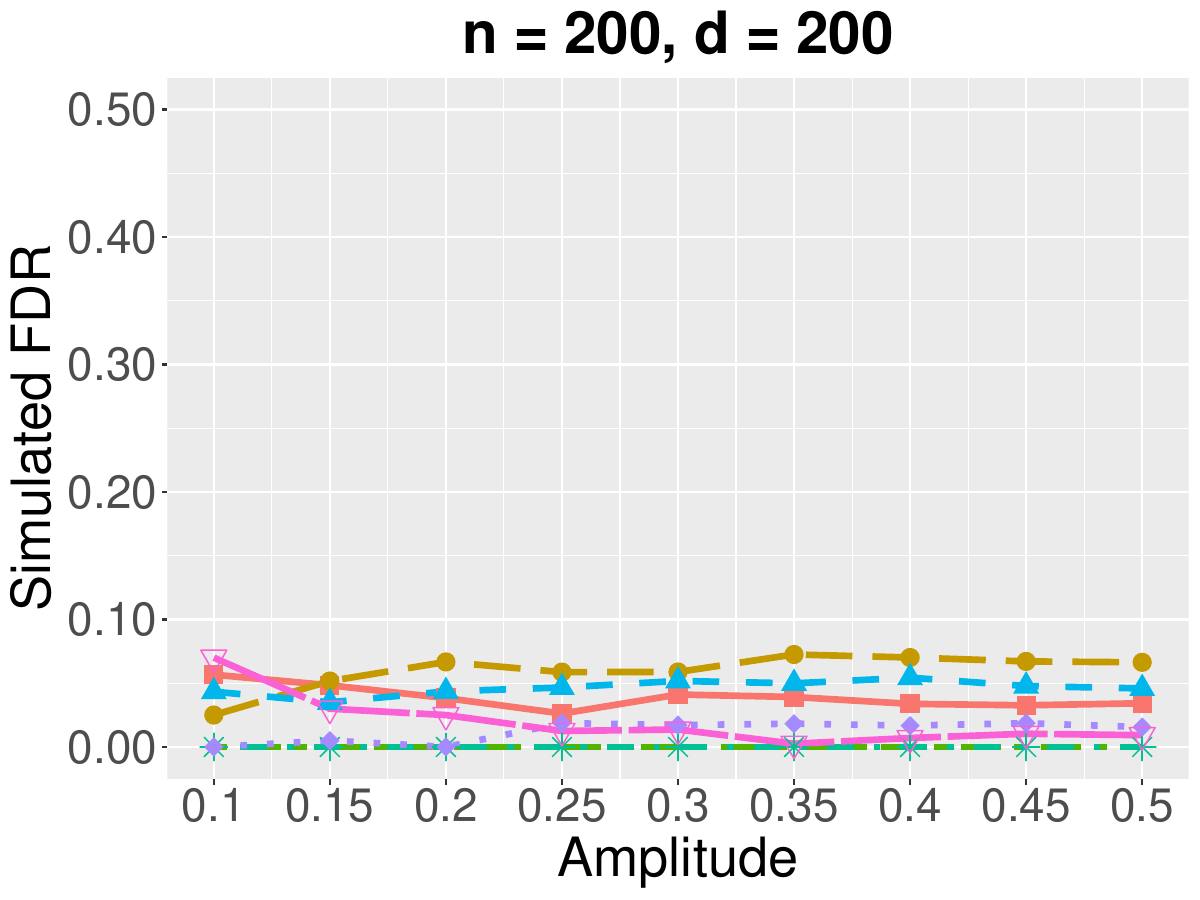}}\hspace{5pt}
	\subfloat{\includegraphics[width=.27\columnwidth]{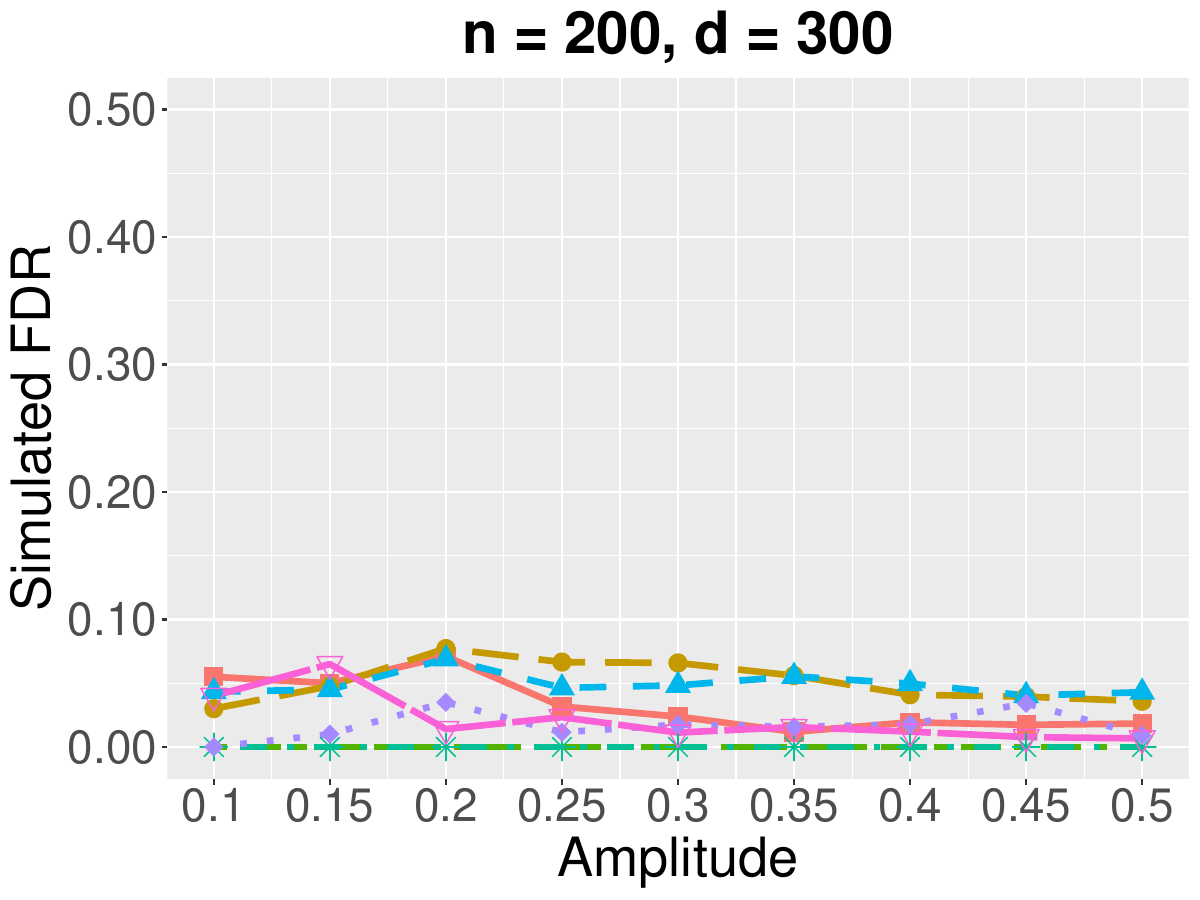}}\hspace{5pt}
	\subfloat{\includegraphics[width=.27\columnwidth]{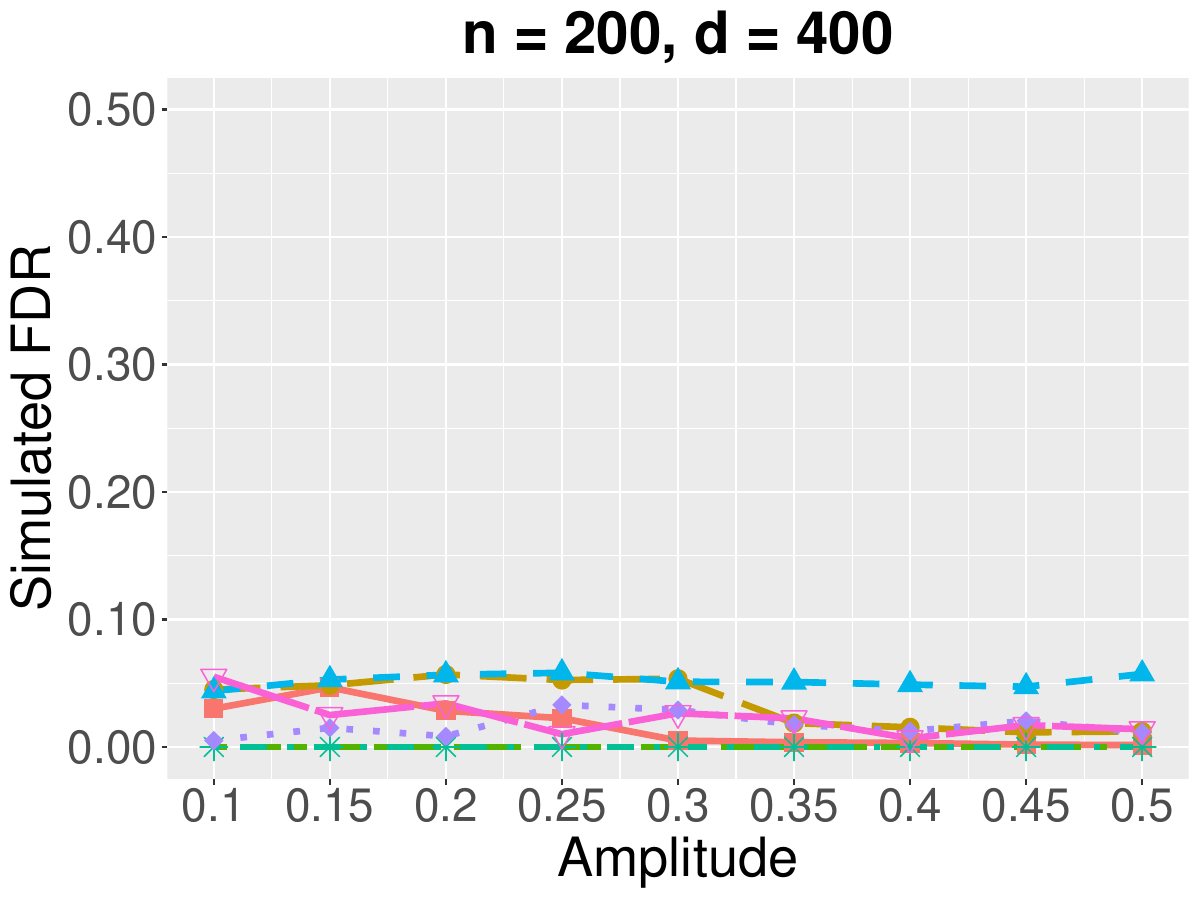}}\\
	\subfloat{\includegraphics[width=.27\columnwidth]{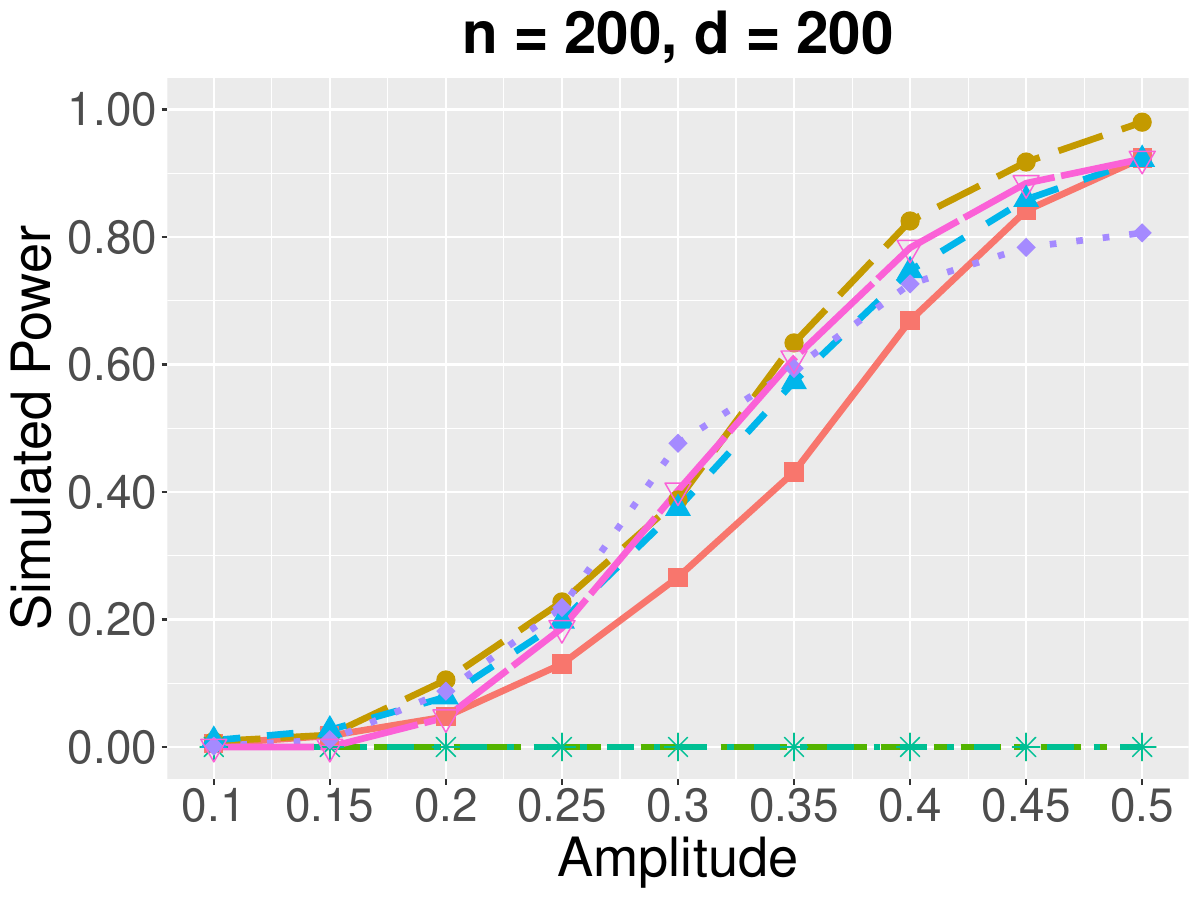}}\hspace{5pt}
    \subfloat{\includegraphics[width=.27\columnwidth]{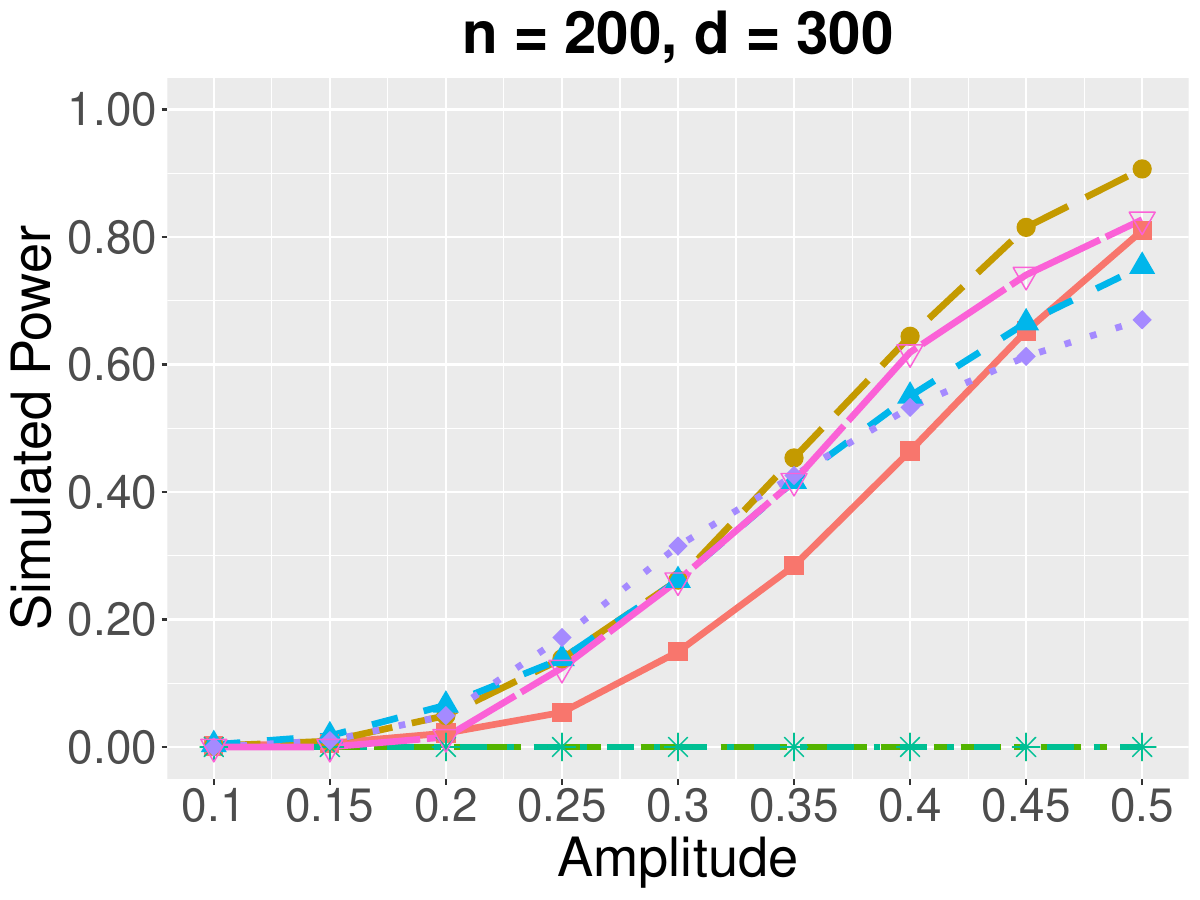}}\hspace{5pt}
    \subfloat{\includegraphics[width=.27\columnwidth]{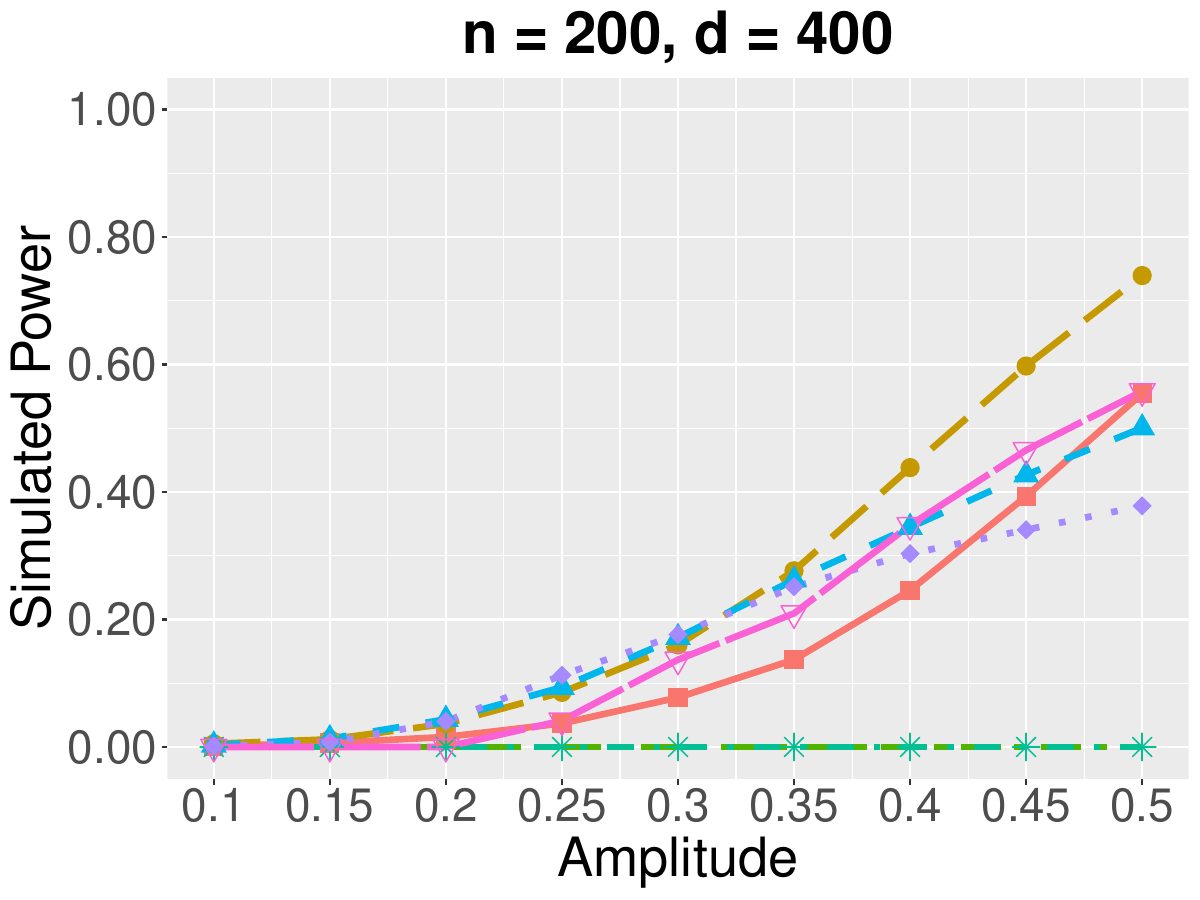}}\\
    \subfloat{\includegraphics[width=.27\columnwidth]{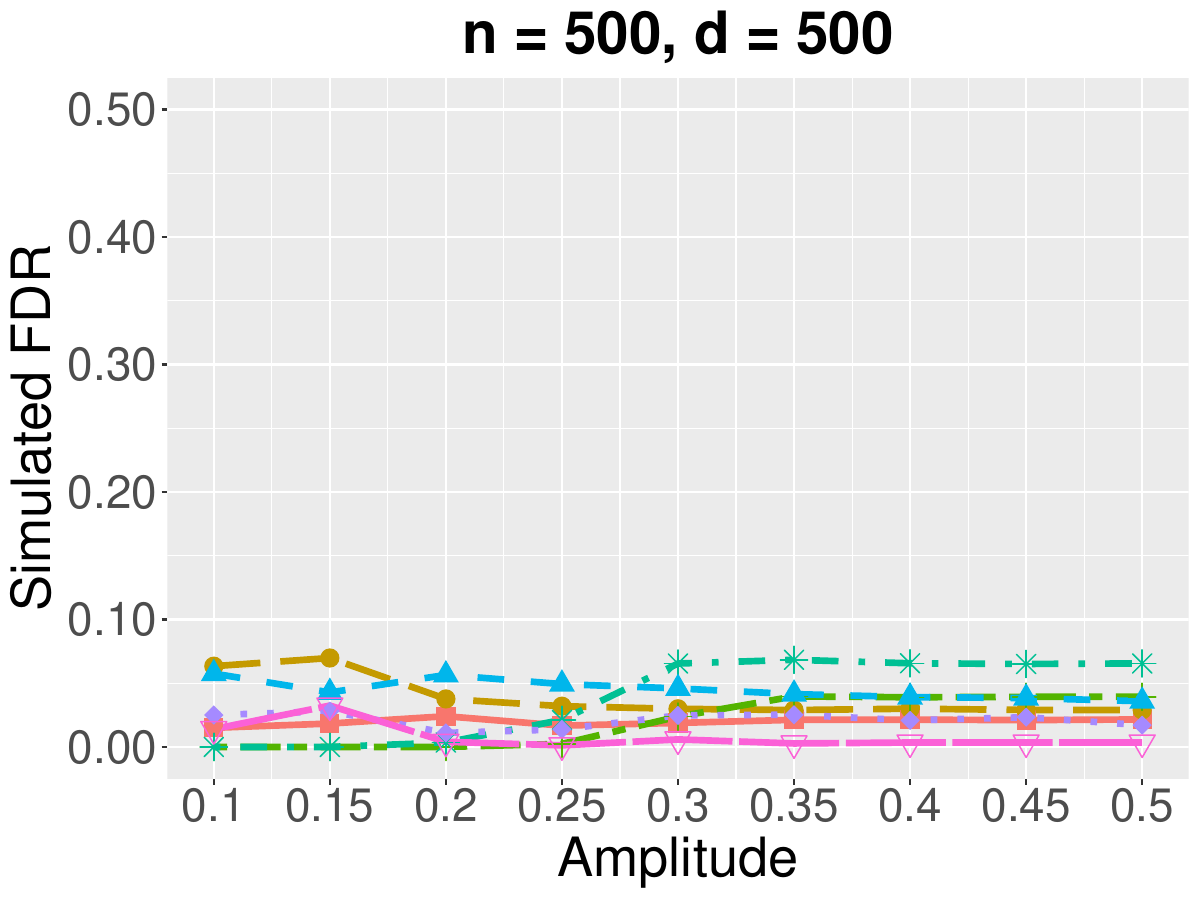}}\hspace{5pt}
	\subfloat{\includegraphics[width=.27\columnwidth]{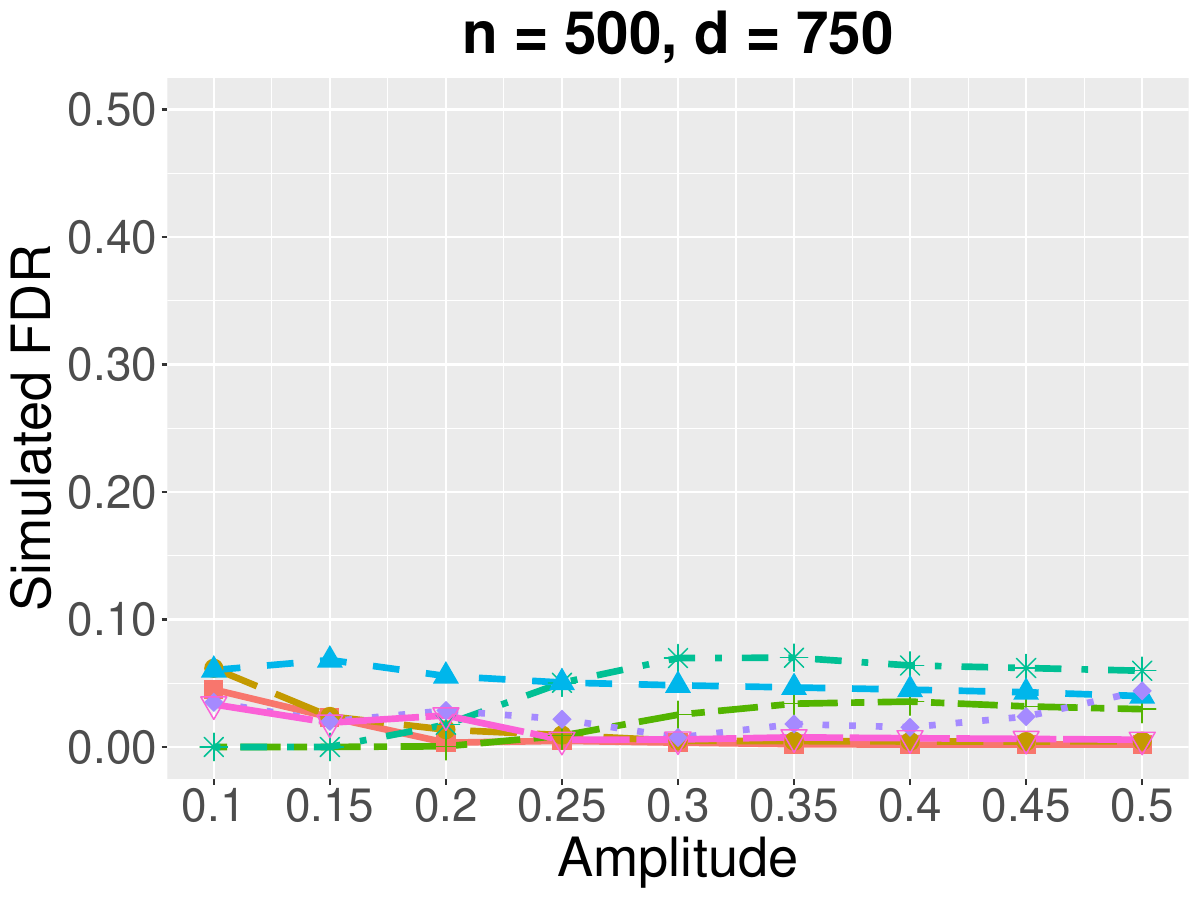}}\hspace{5pt}
	\subfloat{\includegraphics[width=.27\columnwidth]{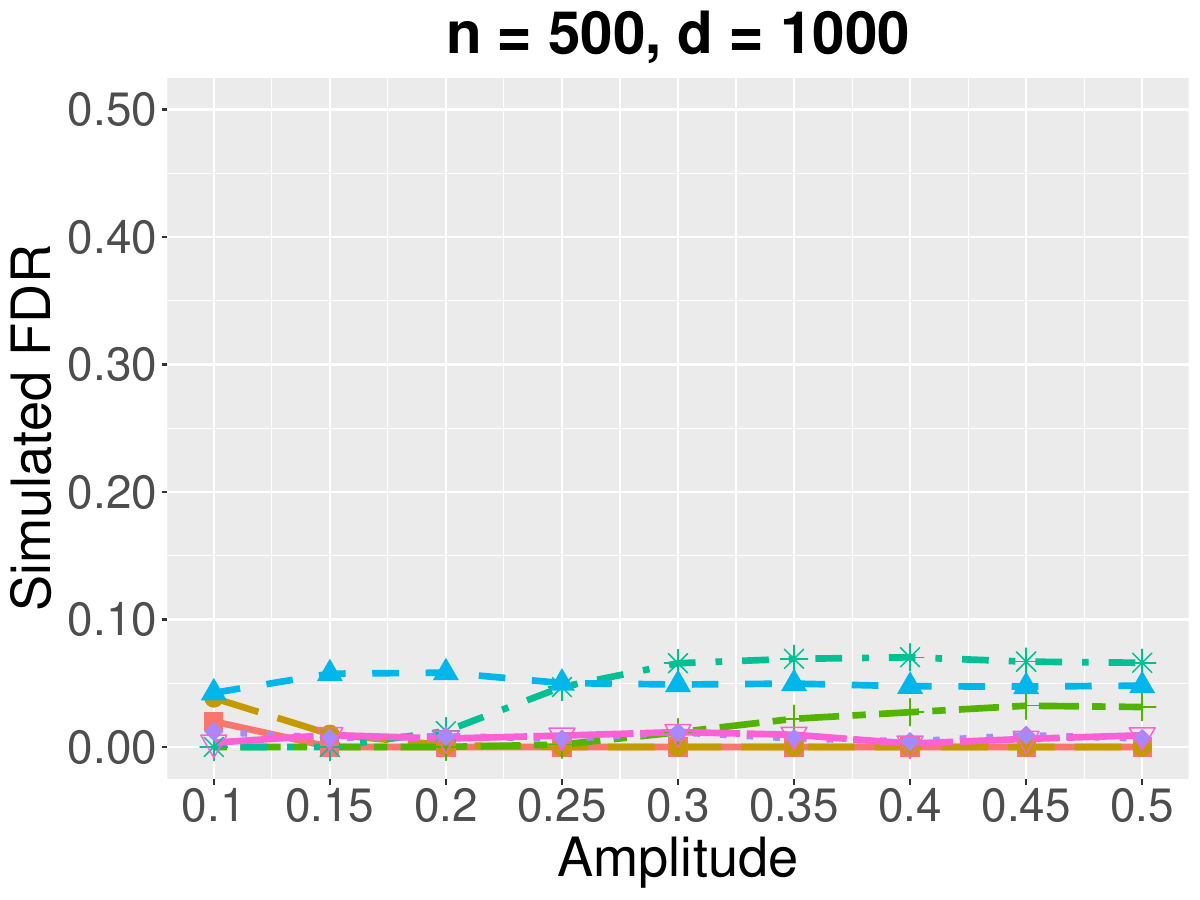}}\\
	\subfloat{\includegraphics[width=.27\columnwidth]{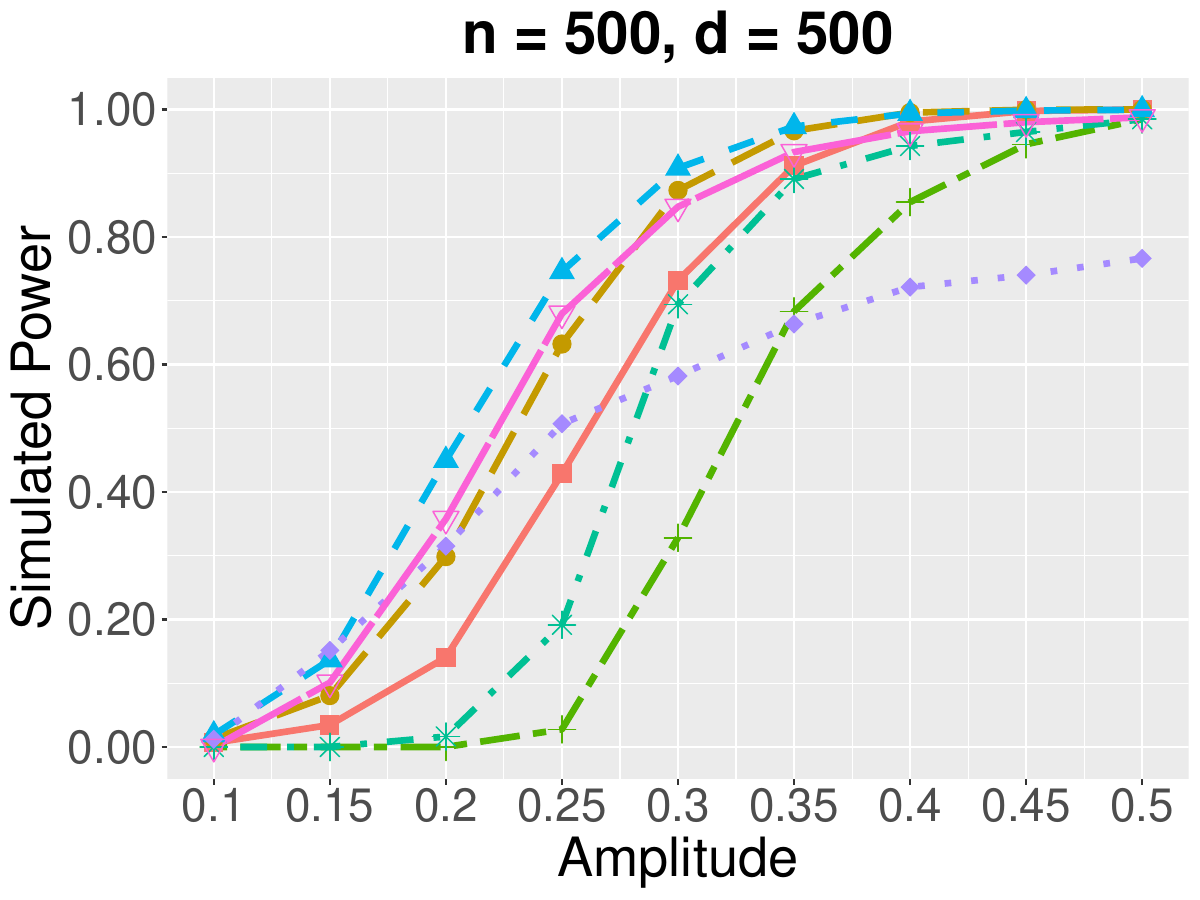}}\hspace{5pt}
    \subfloat{\includegraphics[width=.27\columnwidth]{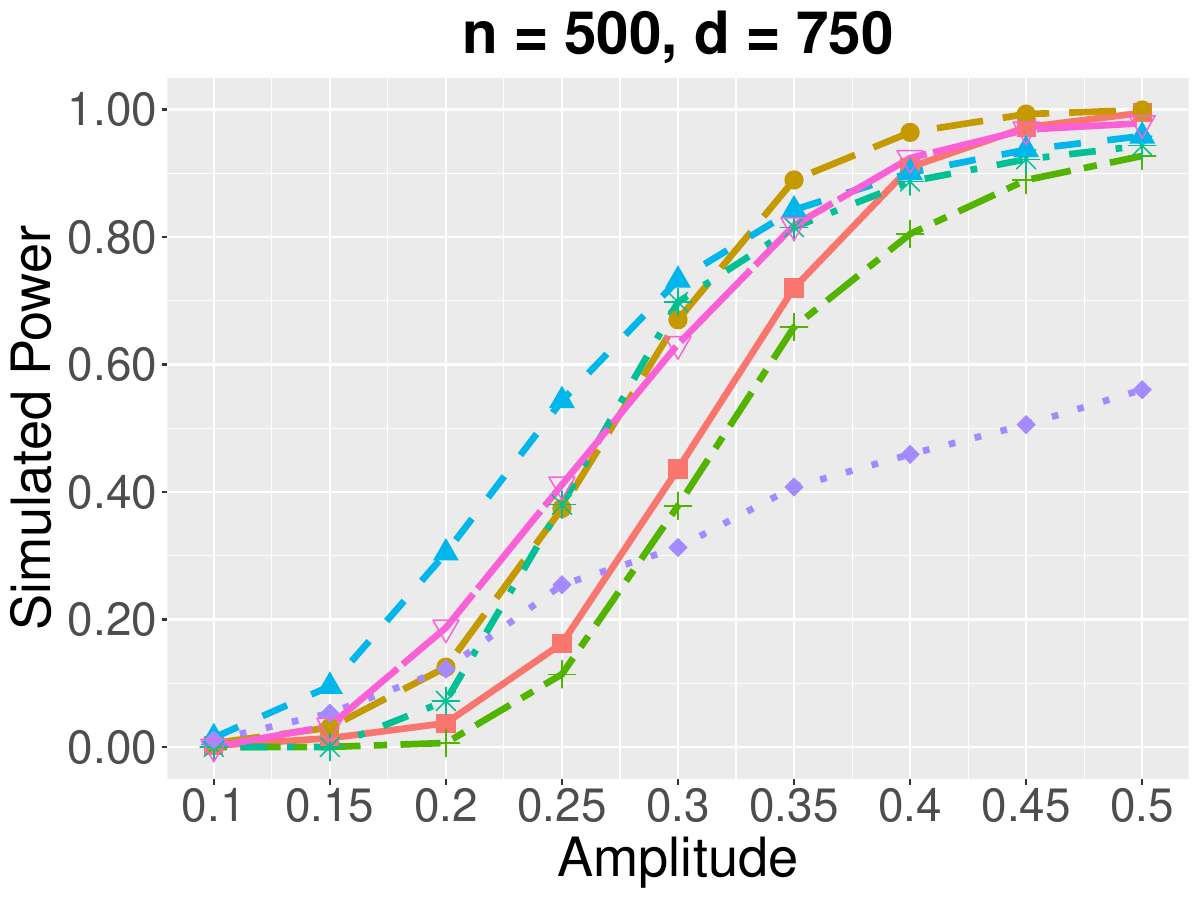}}\hspace{5pt}
    \subfloat{\includegraphics[width=.27\columnwidth]{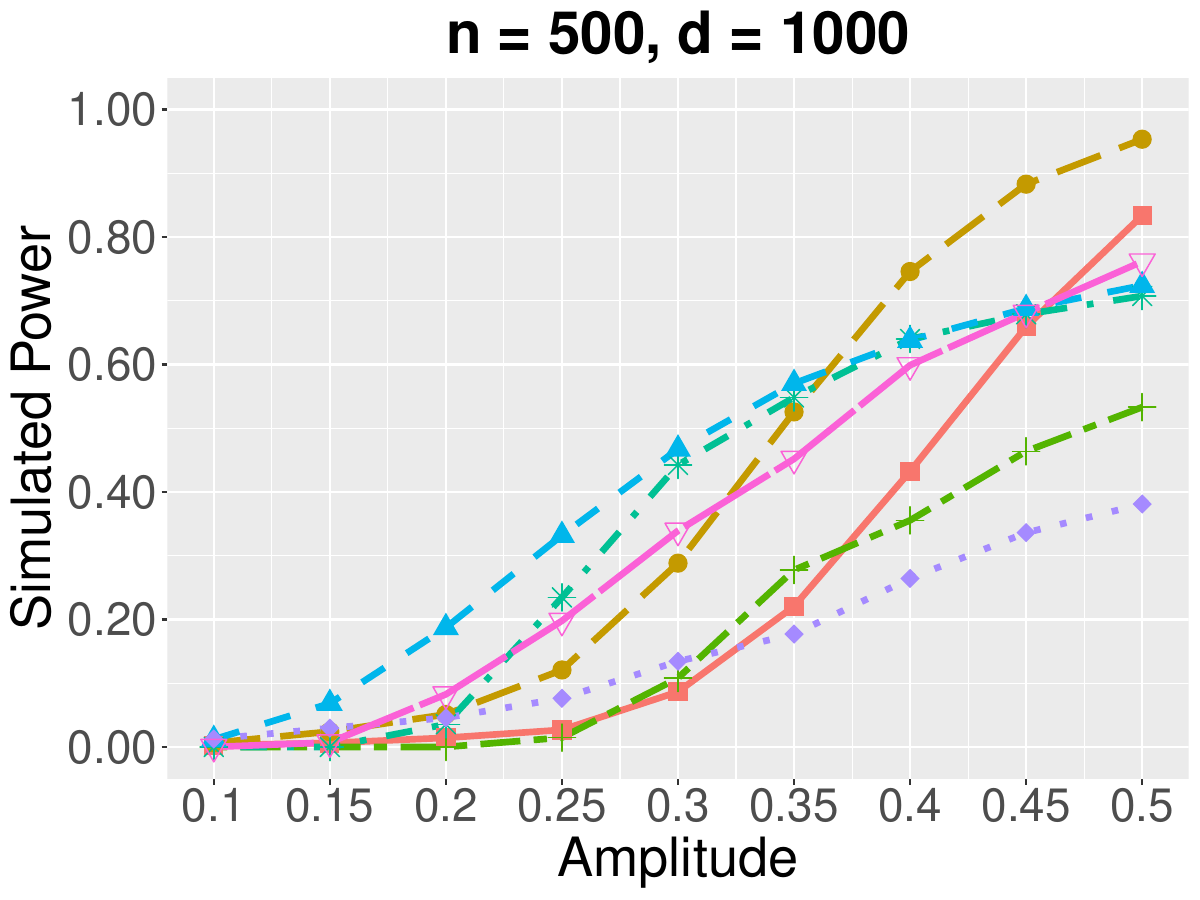}}
	\caption{\small Simulated FDR and power for different combinations of $(n,d)$. The rows of the design matrix were generated from Setting 2. The sparsity level is $k = 0.04d$ and the FDR level is $\alpha = 0.05$. The methods compared are Algorithm 1 (squares and red solid line), Algorithm 2 (circles and yellow long-dashed line), Algorithm A4 (diamonds and purple dotted line), Algorithm A5 (downward triangles and pink dot-dashed line), data splitting version of Algorithm A6 (crosses and dark green two-dashed line), data recycling version of Algorithm A6 (stars and light green dot-dashed line) and Algorithm A7 (upward triangles and blue dashed line).}
    \label{power-SI-ar1-005}
\end{figure}

\begin{figure}[htbp!]
	\centering
	\subfloat{\includegraphics[width=.27\columnwidth]{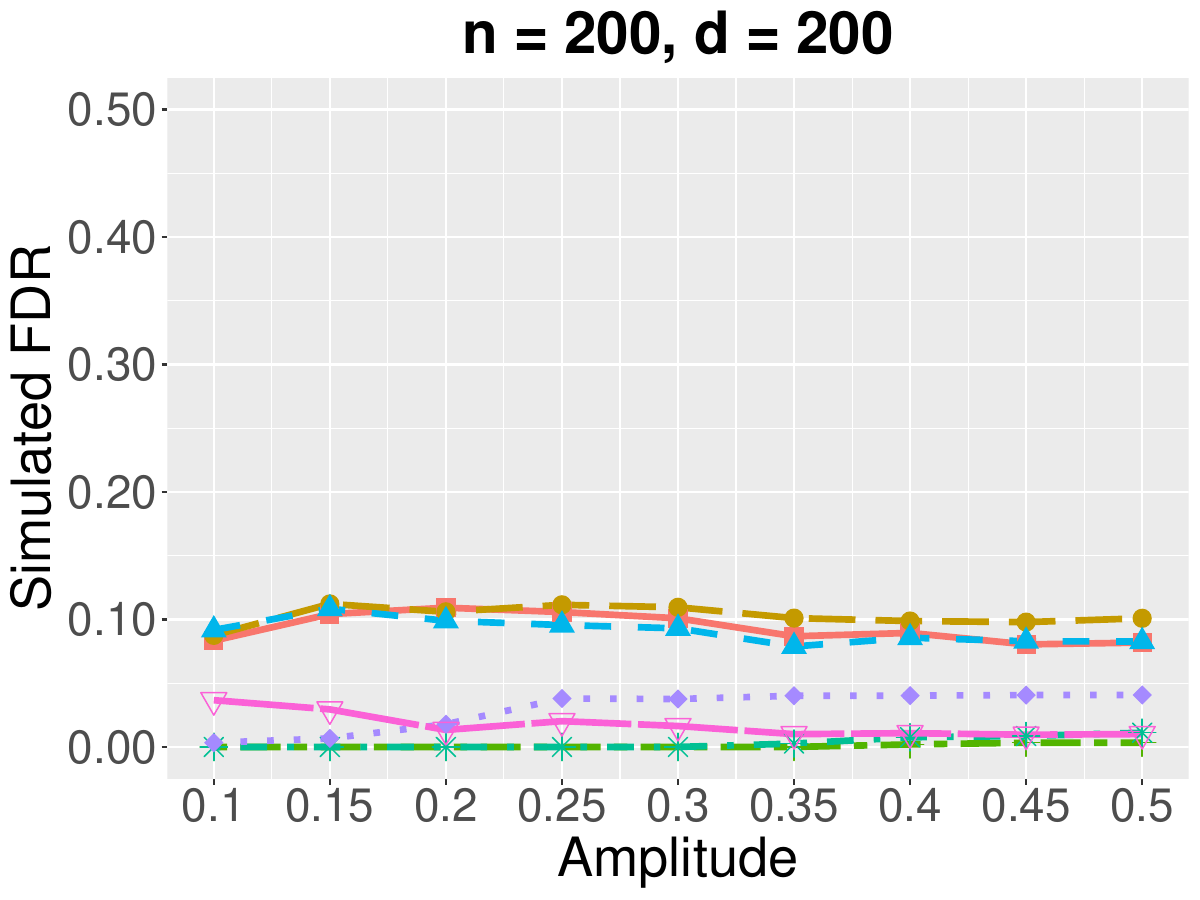}}\hspace{5pt}
	\subfloat{\includegraphics[width=.27\columnwidth]{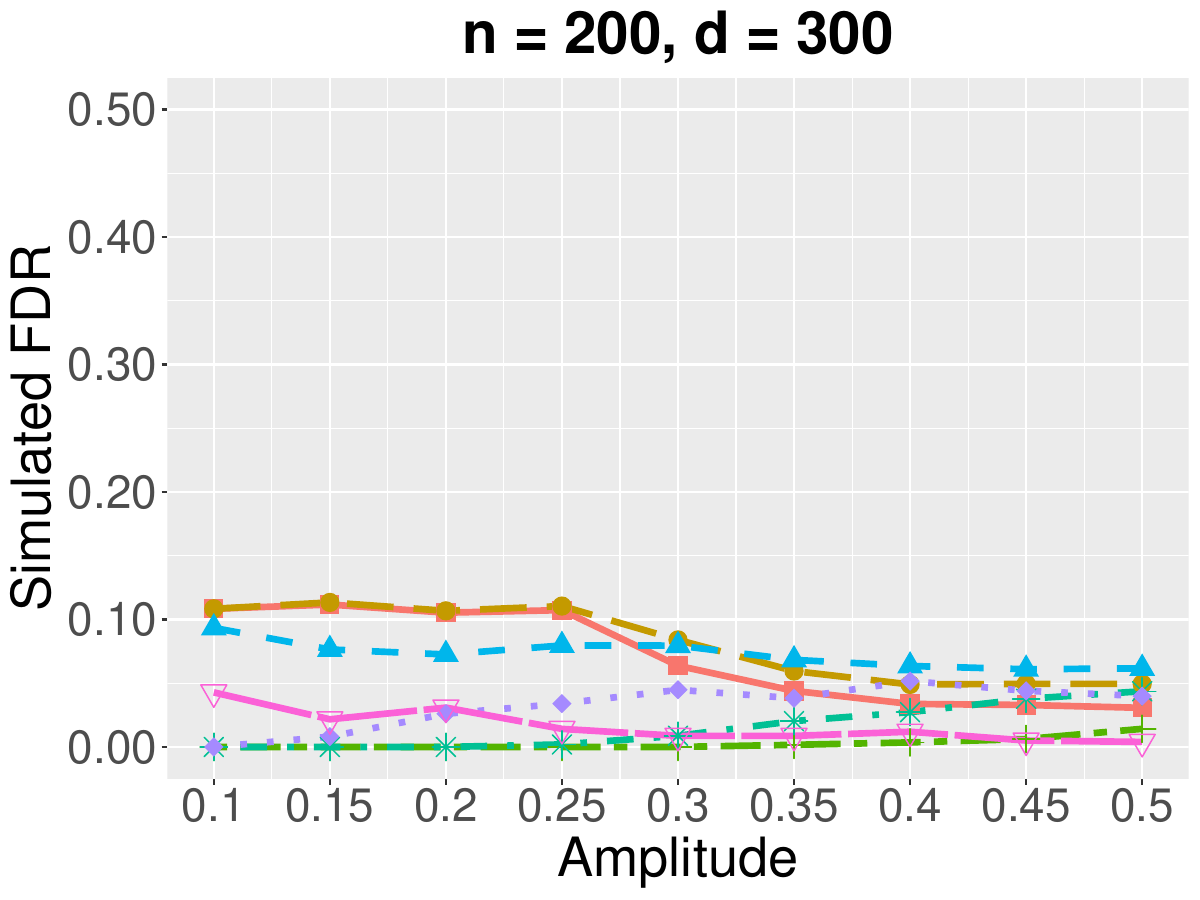}}\hspace{5pt}
	\subfloat{\includegraphics[width=.27\columnwidth]{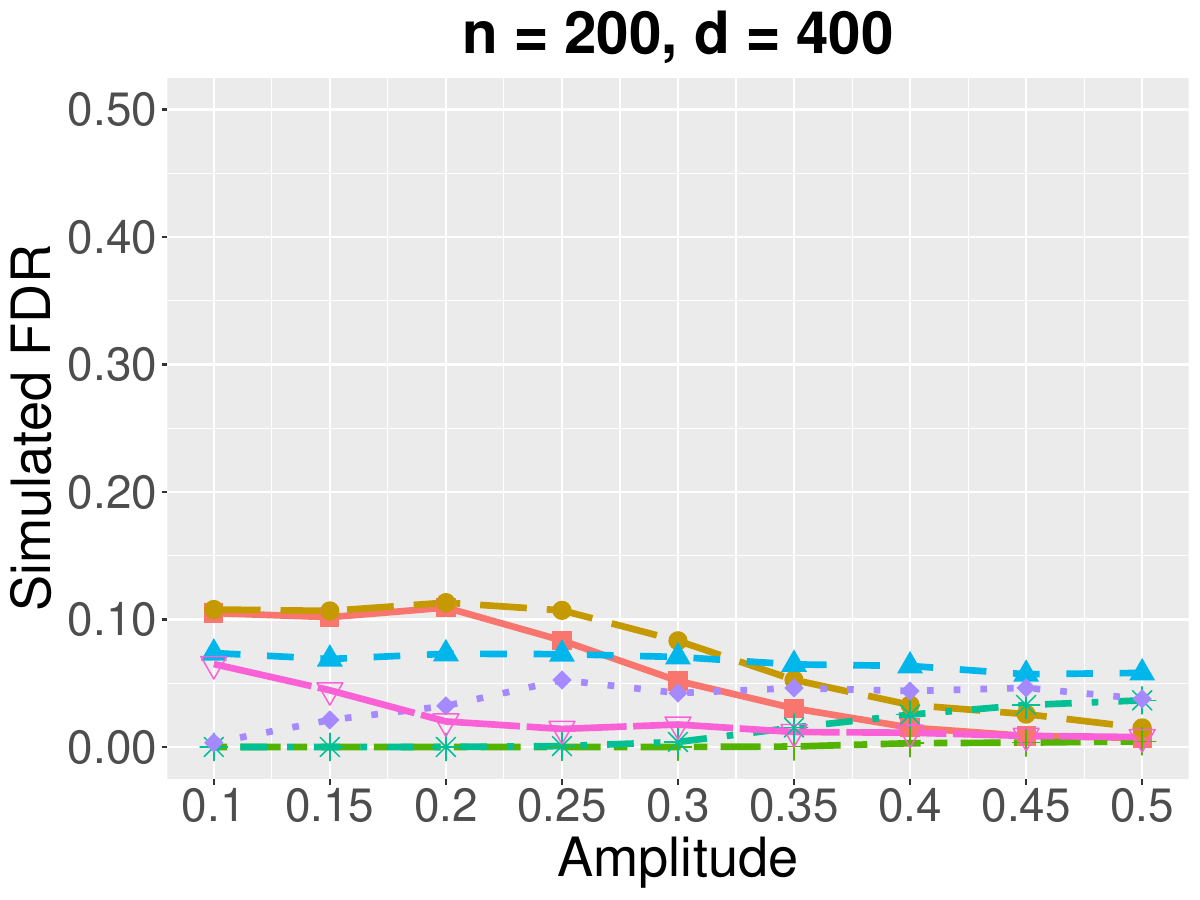}}\\
	\subfloat{\includegraphics[width=.27\columnwidth]{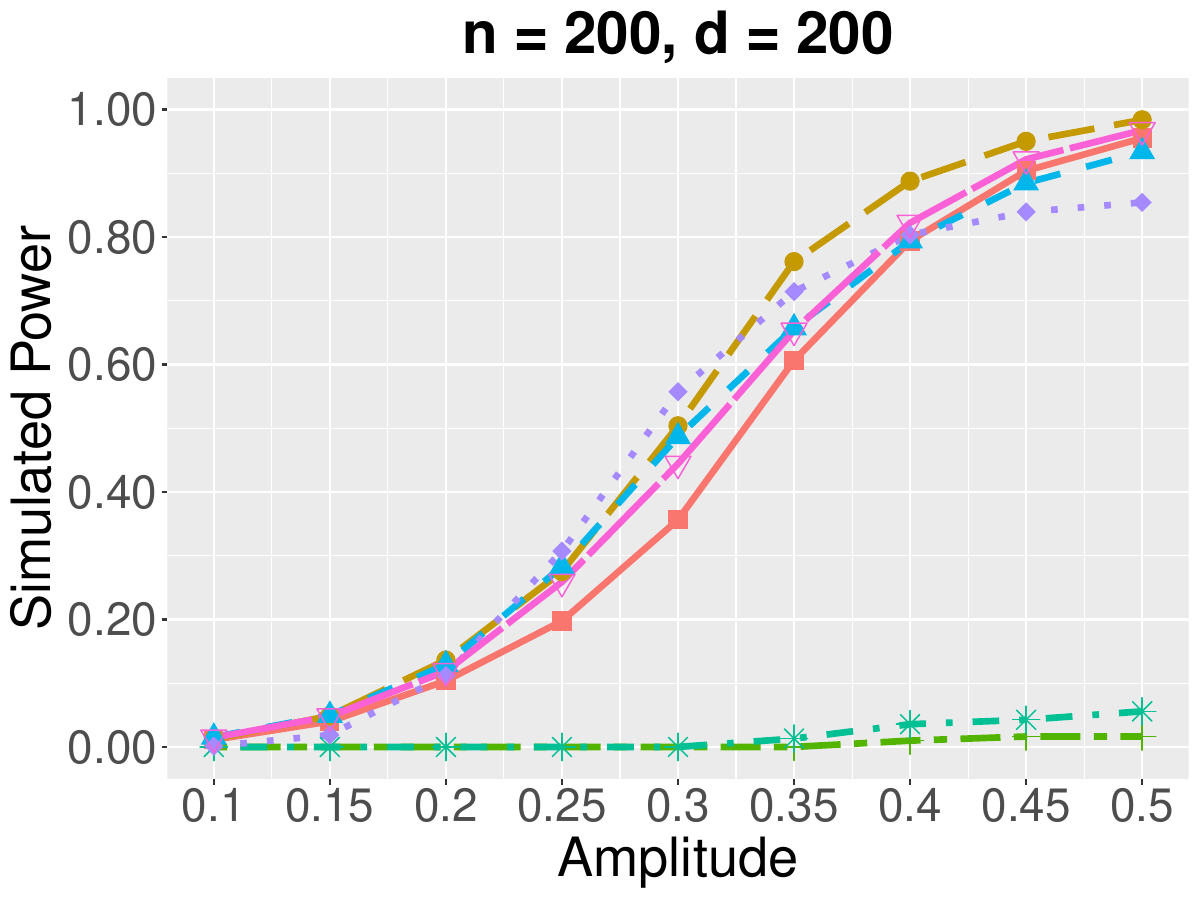}}\hspace{5pt}
    \subfloat{\includegraphics[width=.27\columnwidth]{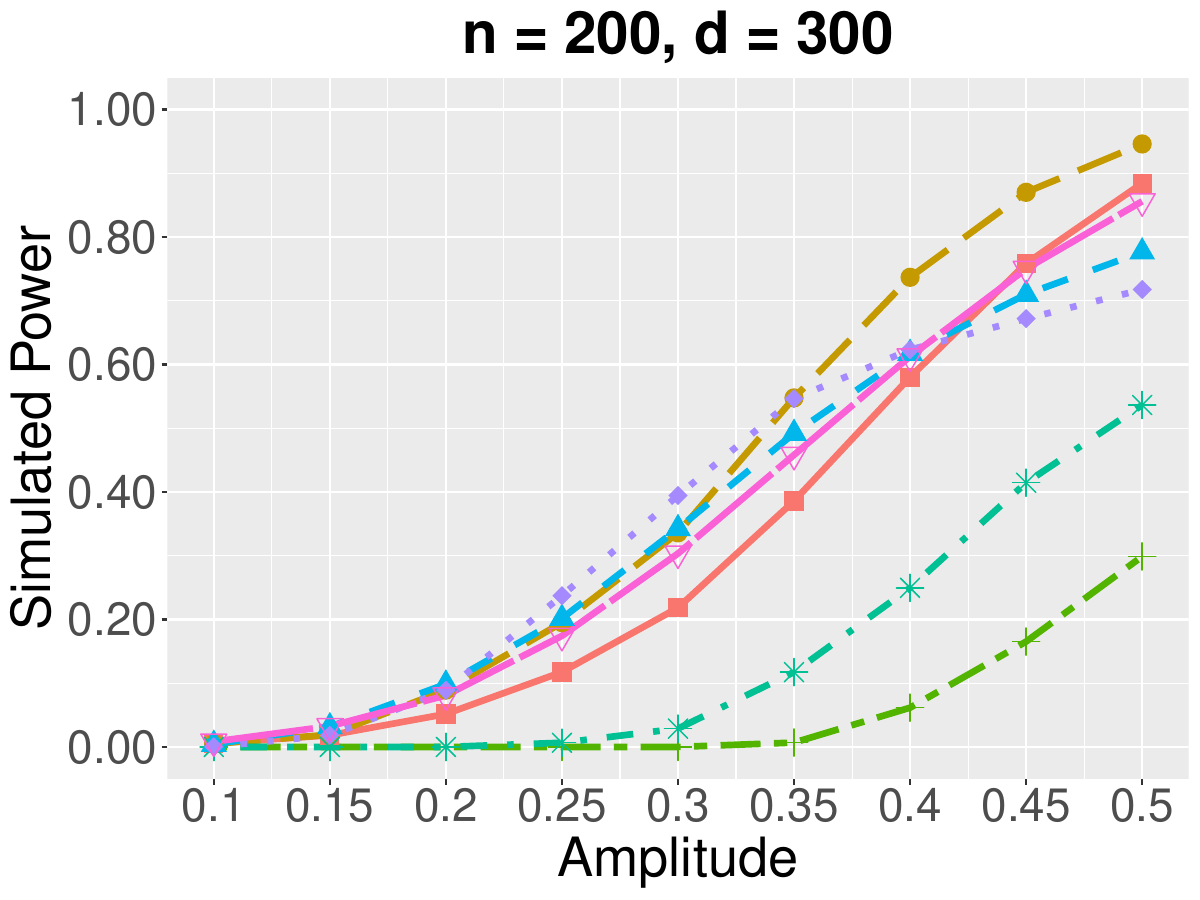}}\hspace{5pt}
    \subfloat{\includegraphics[width=.27\columnwidth]{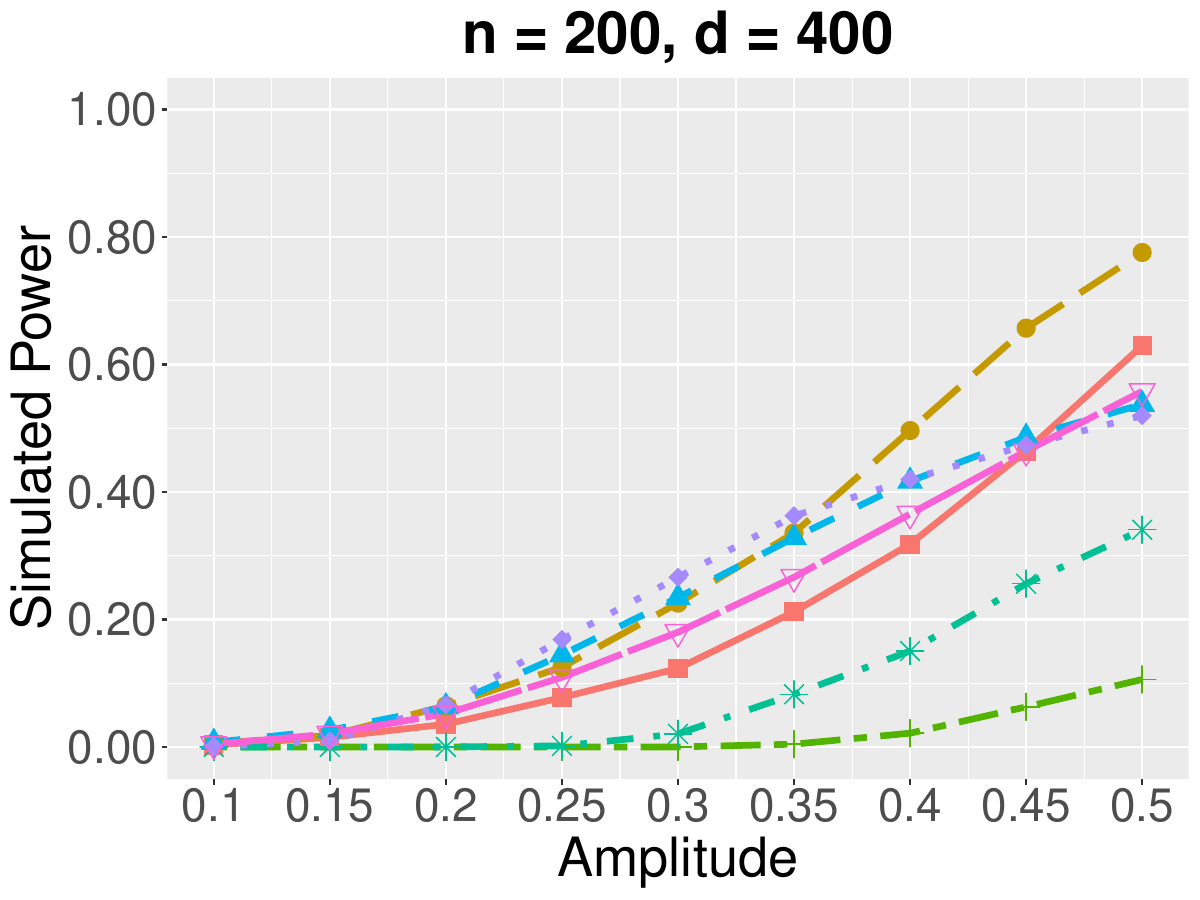}}\\
    \subfloat{\includegraphics[width=.27\columnwidth]{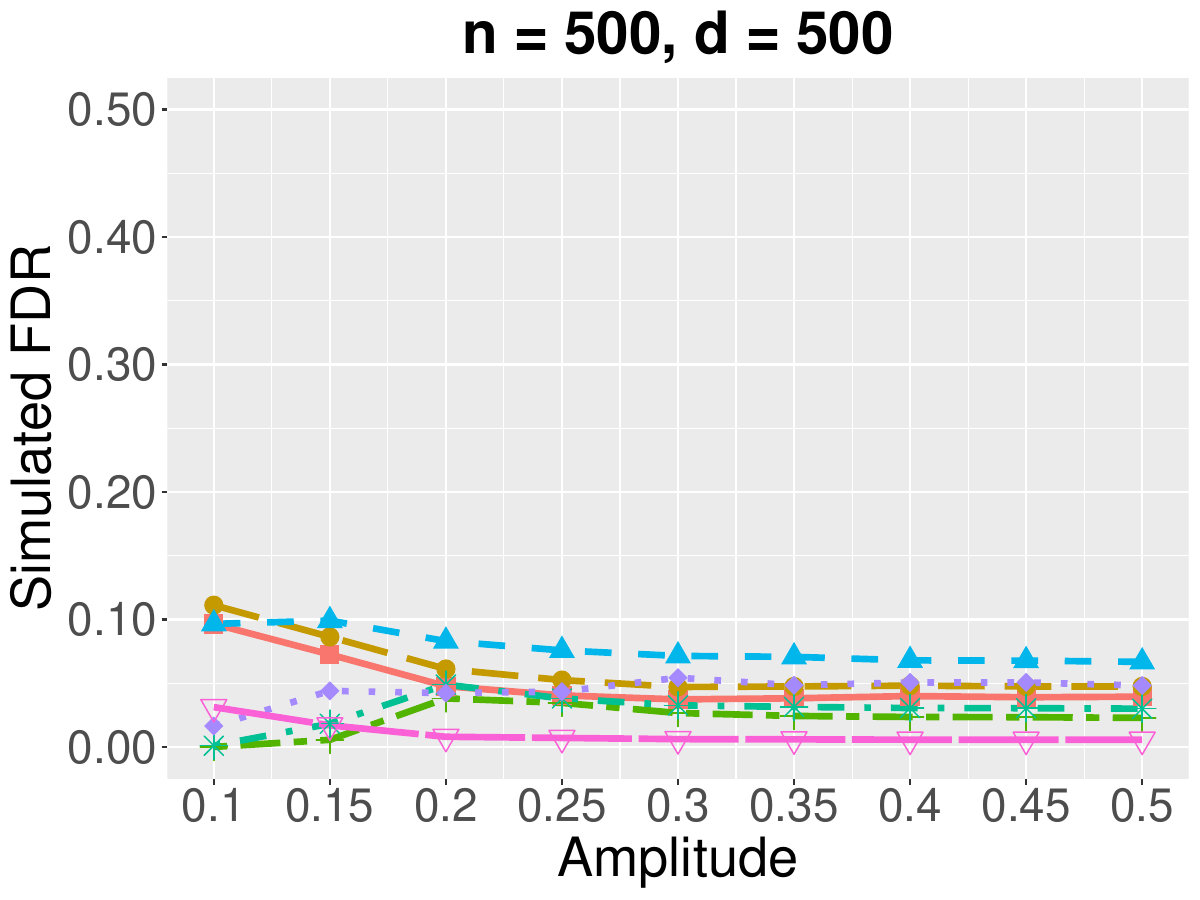}}\hspace{5pt}
	\subfloat{\includegraphics[width=.27\columnwidth]{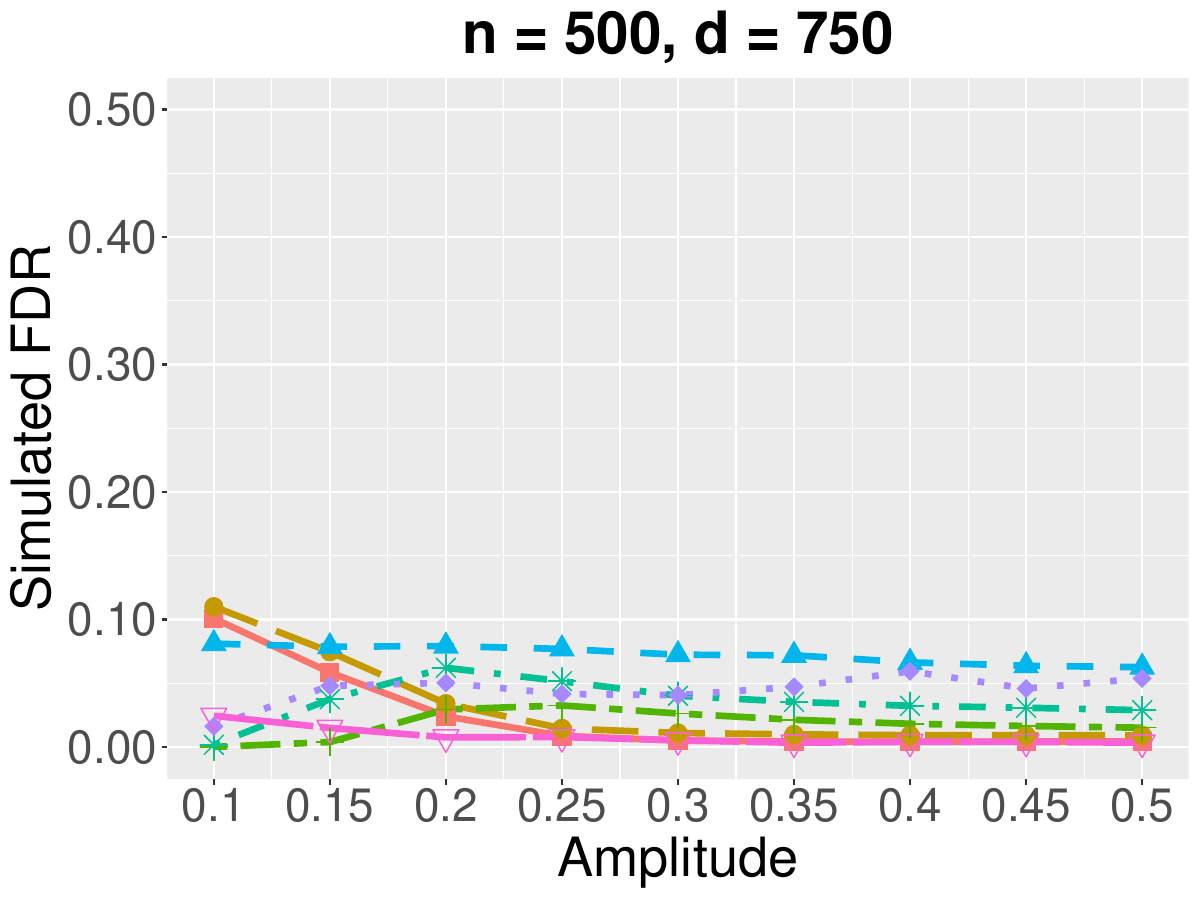}}\hspace{5pt}
	\subfloat{\includegraphics[width=.27\columnwidth]{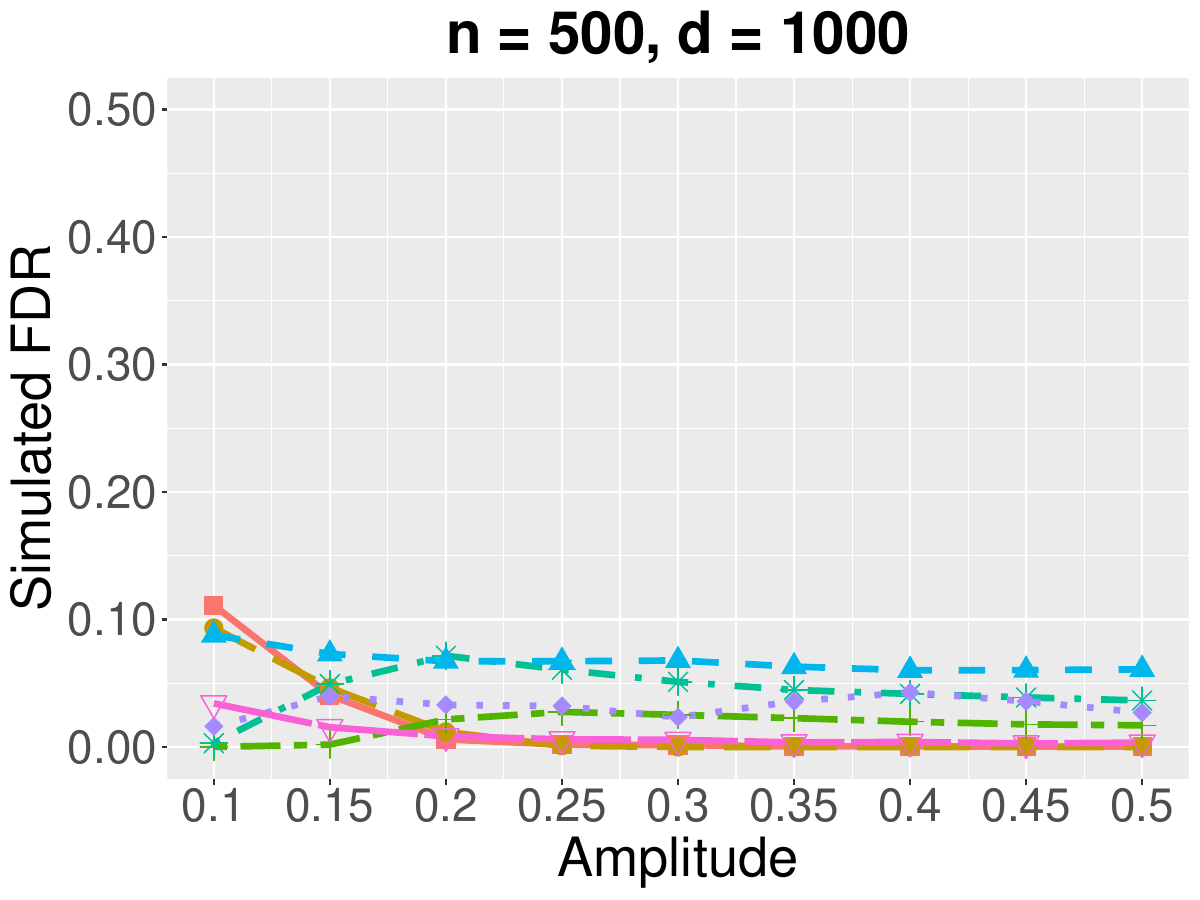}}\\
	\subfloat{\includegraphics[width=.27\columnwidth]{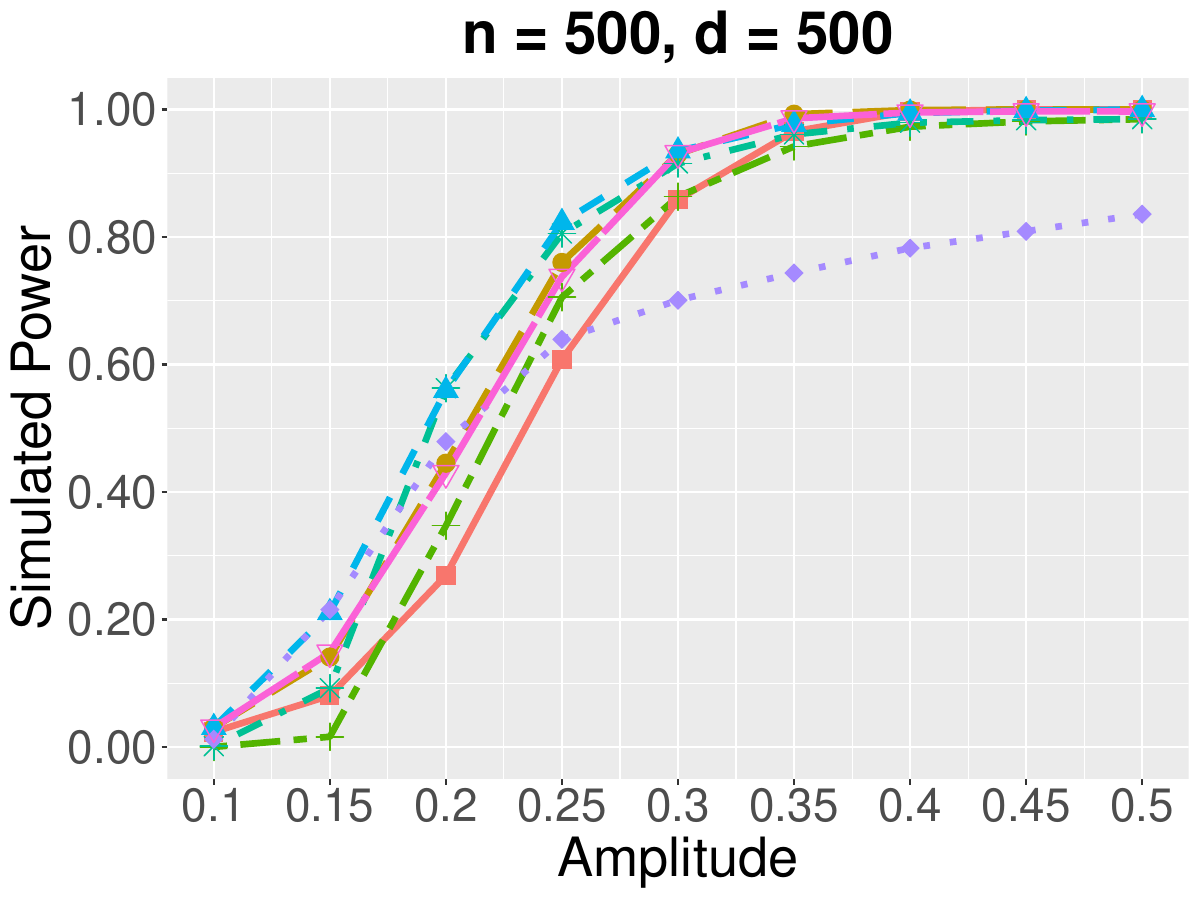}}\hspace{5pt}
    \subfloat{\includegraphics[width=.27\columnwidth]{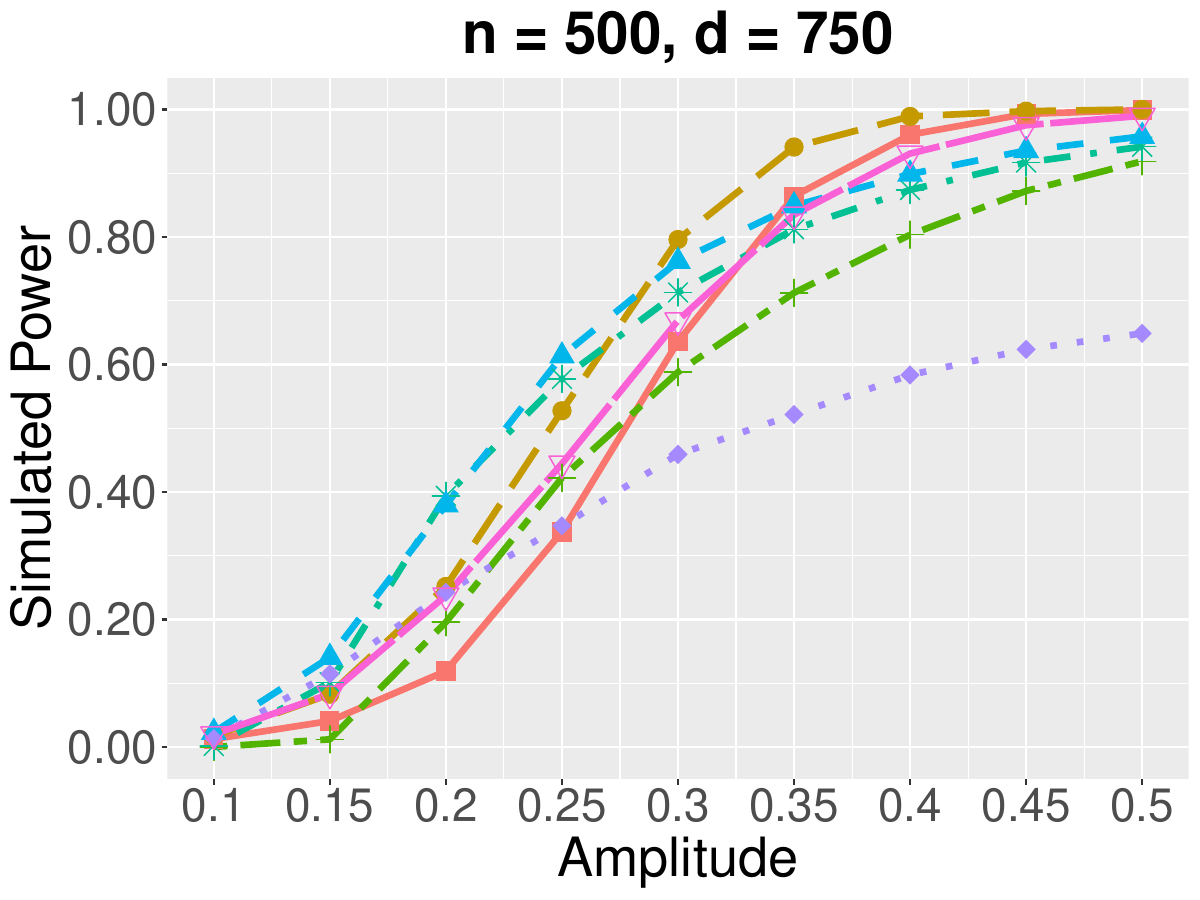}}\hspace{5pt}
    \subfloat{\includegraphics[width=.27\columnwidth]{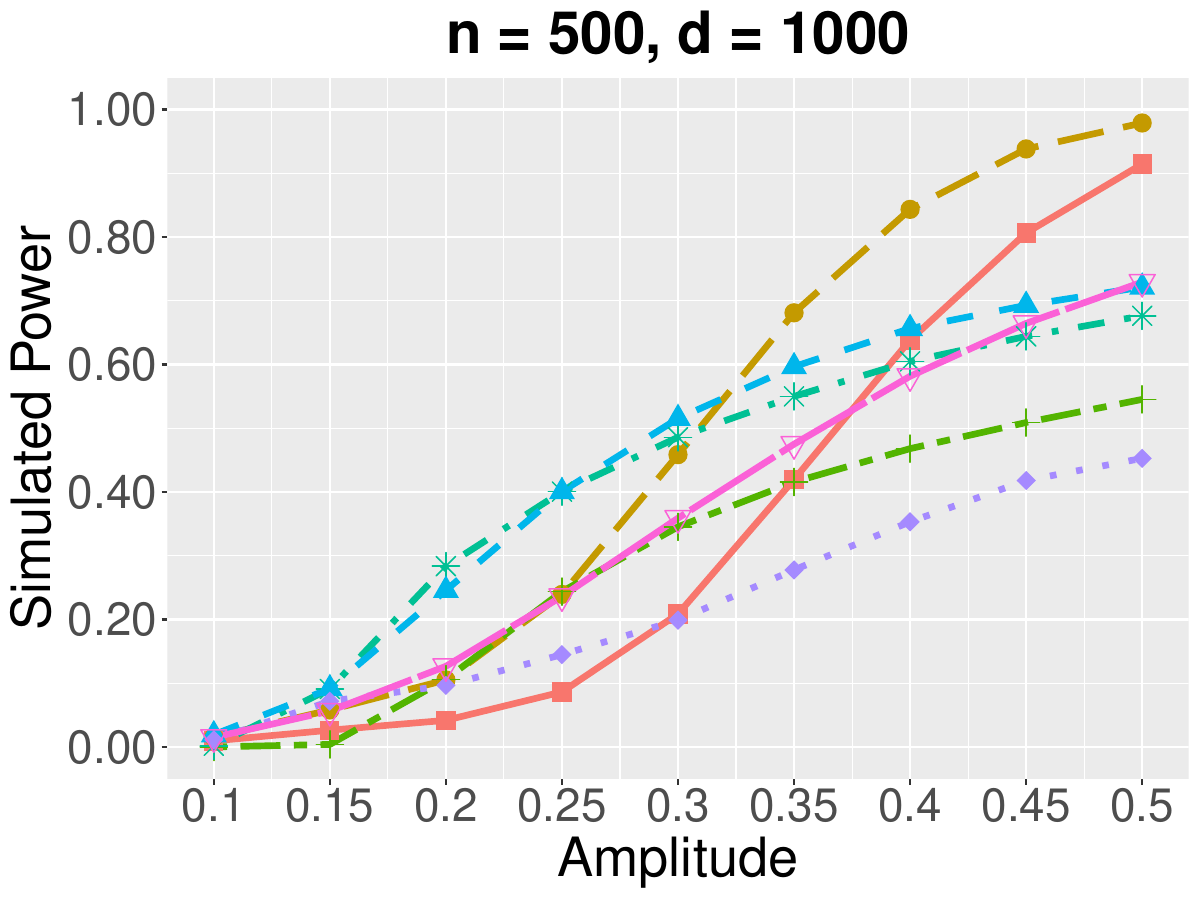}}
	\caption{\small Simulated FDR and power for different combinations of $(n,d)$. The rows of the design matrix were generated from Setting 1. The sparsity level is $k = 0.04d$ and the FDR level is $\alpha = 0.1$. The methods compared are Algorithm 1 (squares and red solid line), Algorithm 2 (circles and yellow long-dashed line), Algorithm A4 (diamonds and purple dotted line), Algorithm A5 (downward triangles and pink dot-dashed line), data splitting version of Algorithm A6 (crosses and dark green two-dashed line), data recycling version of Algorithm A6 (stars and light green dot-dashed line) and Algorithm A7 (upward triangles and blue dashed line).}
    \label{power-SI-indep}
\end{figure}

\begin{figure}[htbp!]
	\centering
	\subfloat{\includegraphics[width=.27\columnwidth]{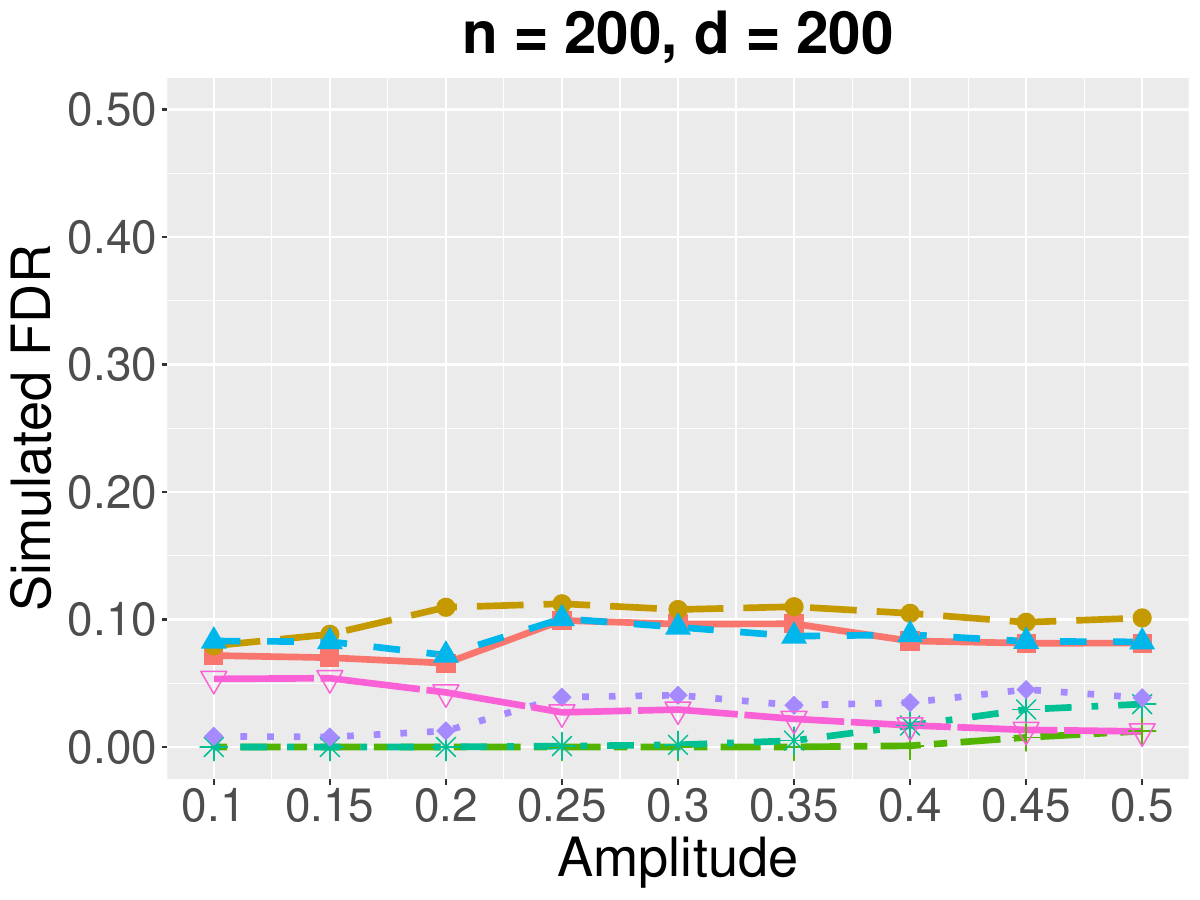}}\hspace{5pt}
	\subfloat{\includegraphics[width=.27\columnwidth]{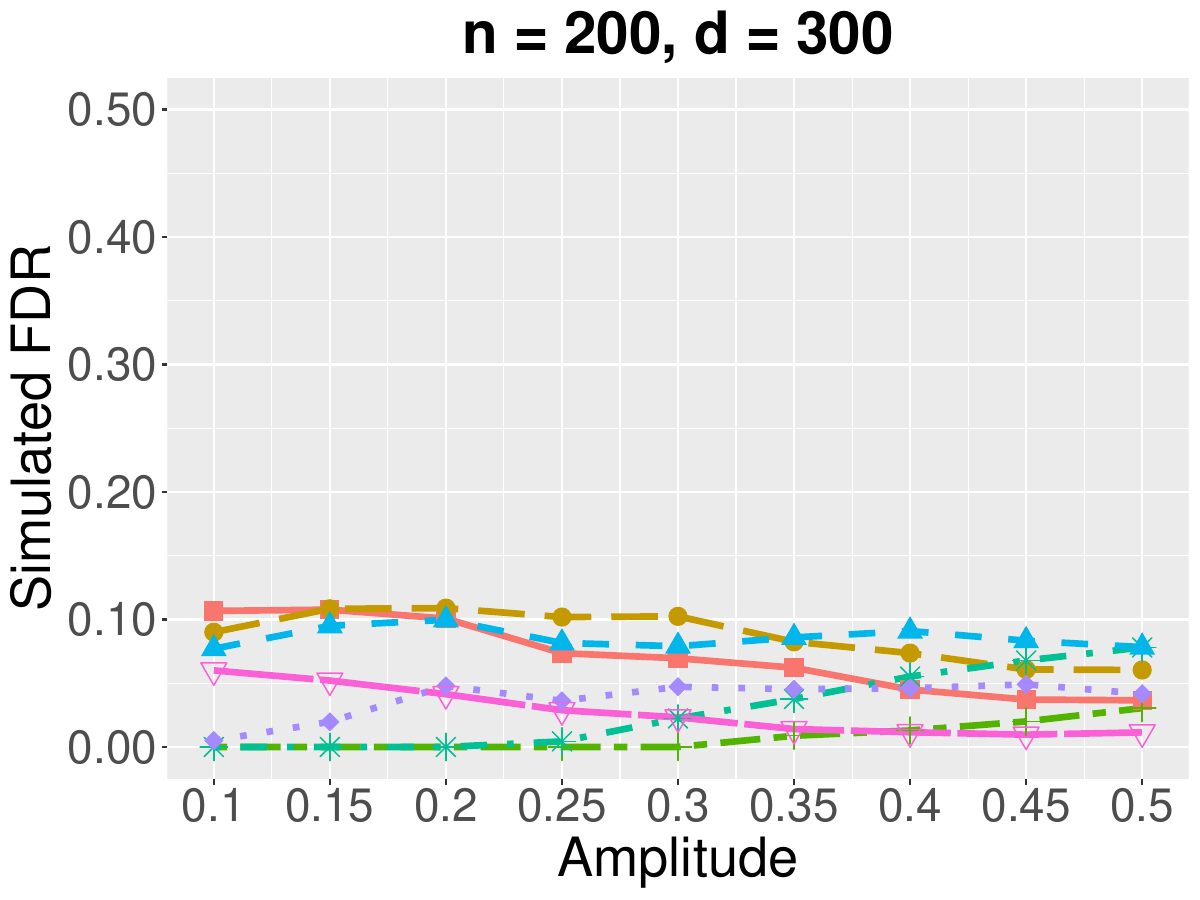}}\hspace{5pt}
	\subfloat{\includegraphics[width=.27\columnwidth]{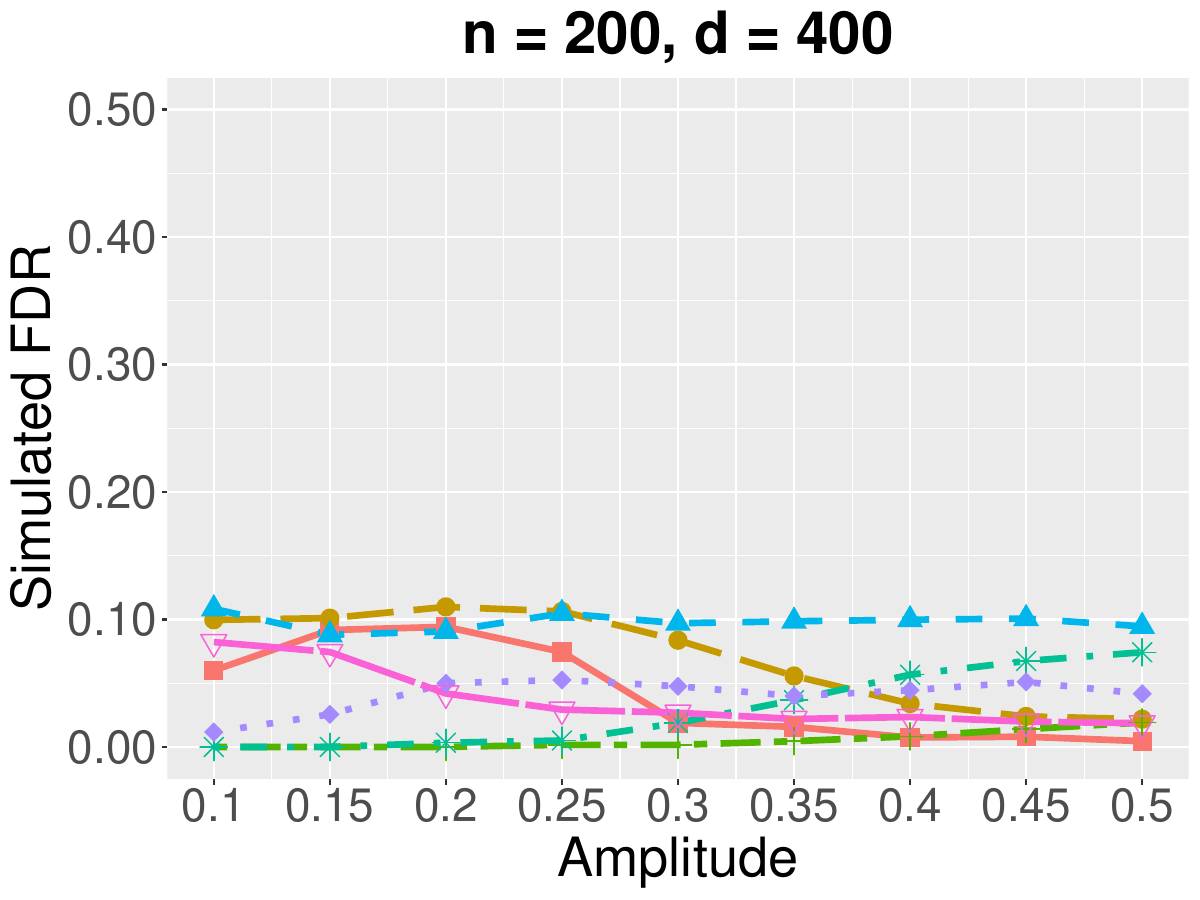}}\\
	\subfloat{\includegraphics[width=.27\columnwidth]{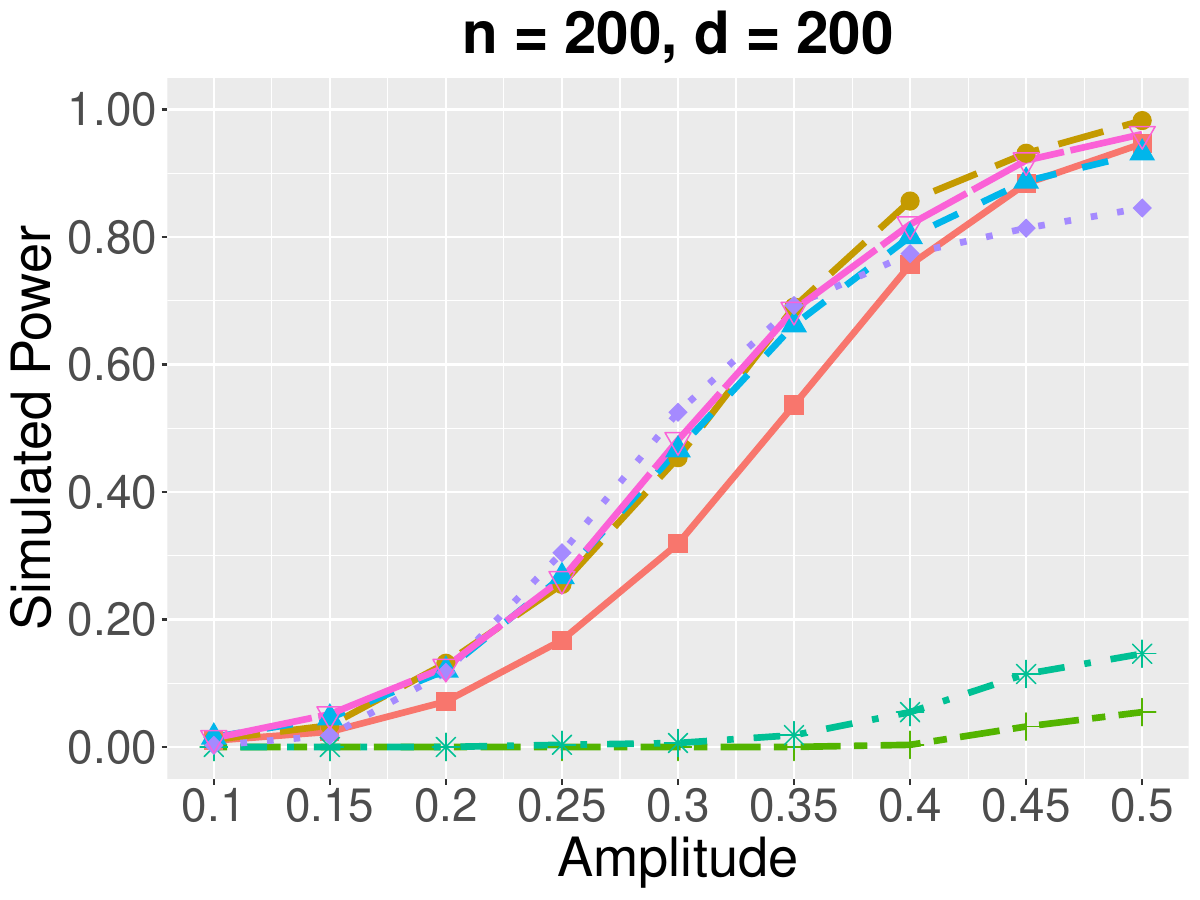}}\hspace{5pt}
    \subfloat{\includegraphics[width=.27\columnwidth]{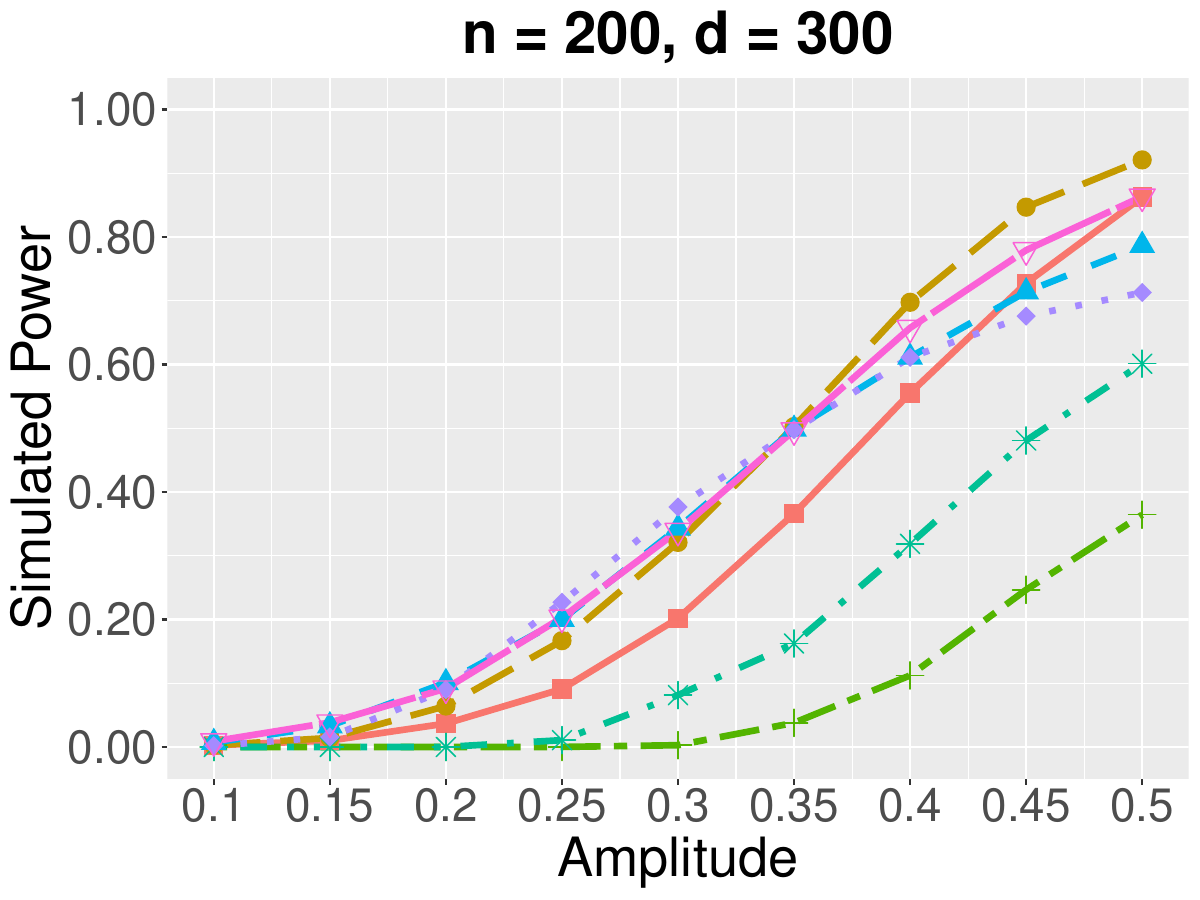}}\hspace{5pt}
    \subfloat{\includegraphics[width=.27\columnwidth]{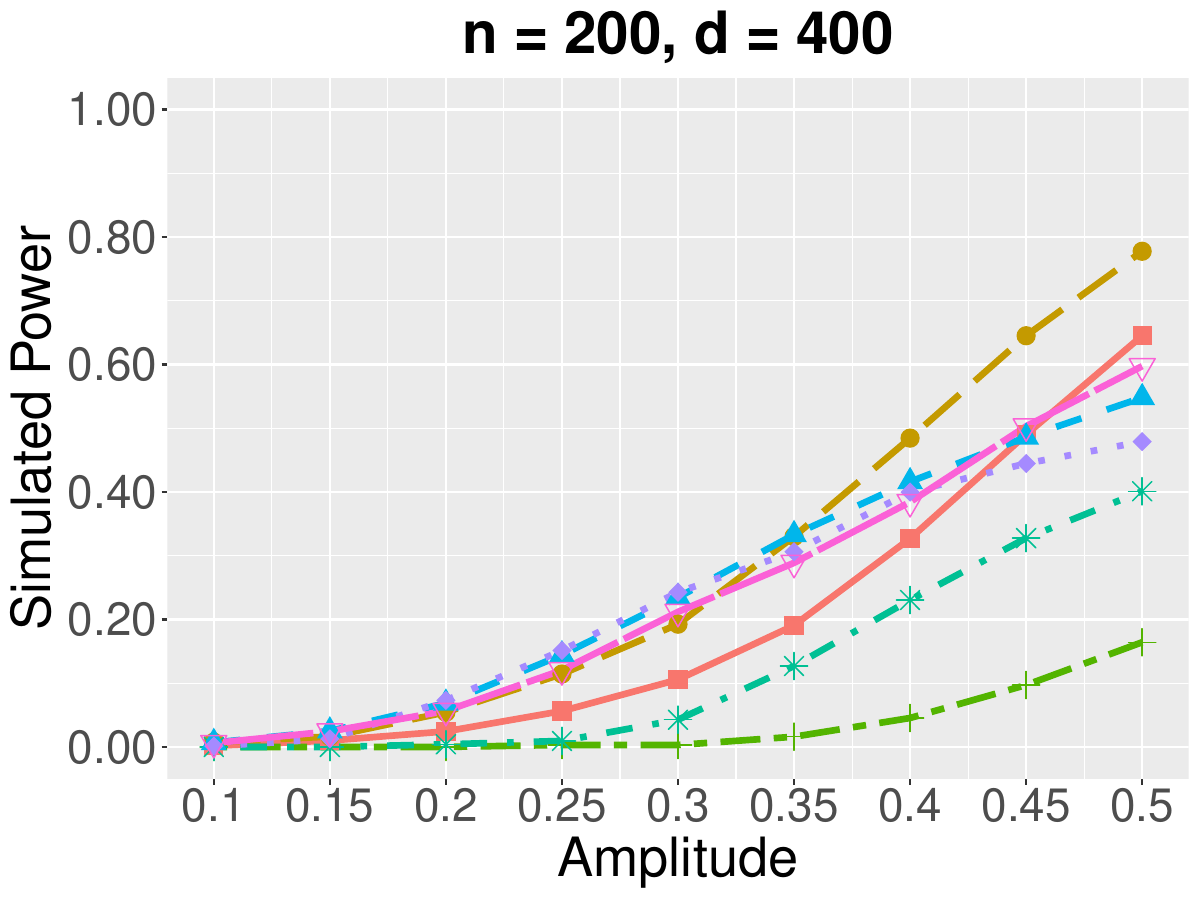}}\\
    \subfloat{\includegraphics[width=.27\columnwidth]{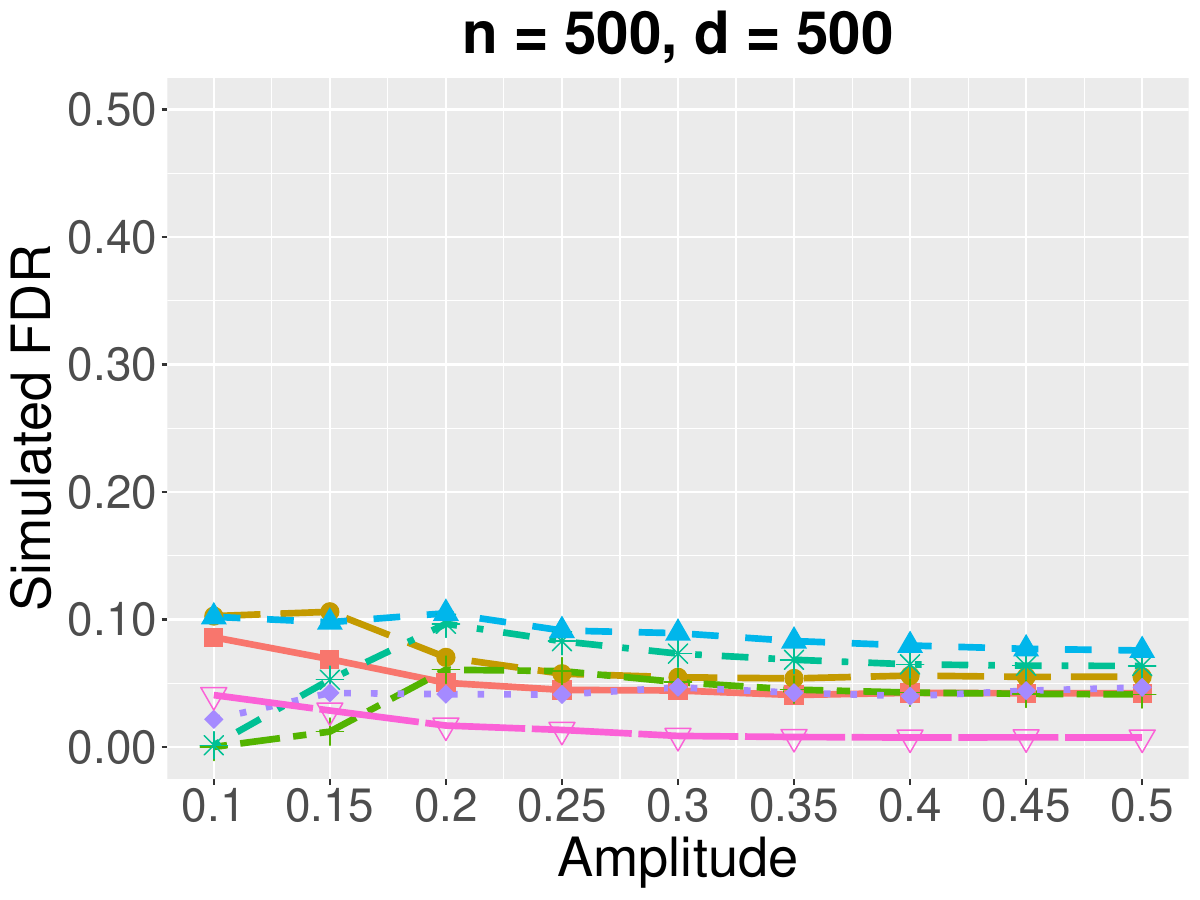}}\hspace{5pt}
	\subfloat{\includegraphics[width=.27\columnwidth]{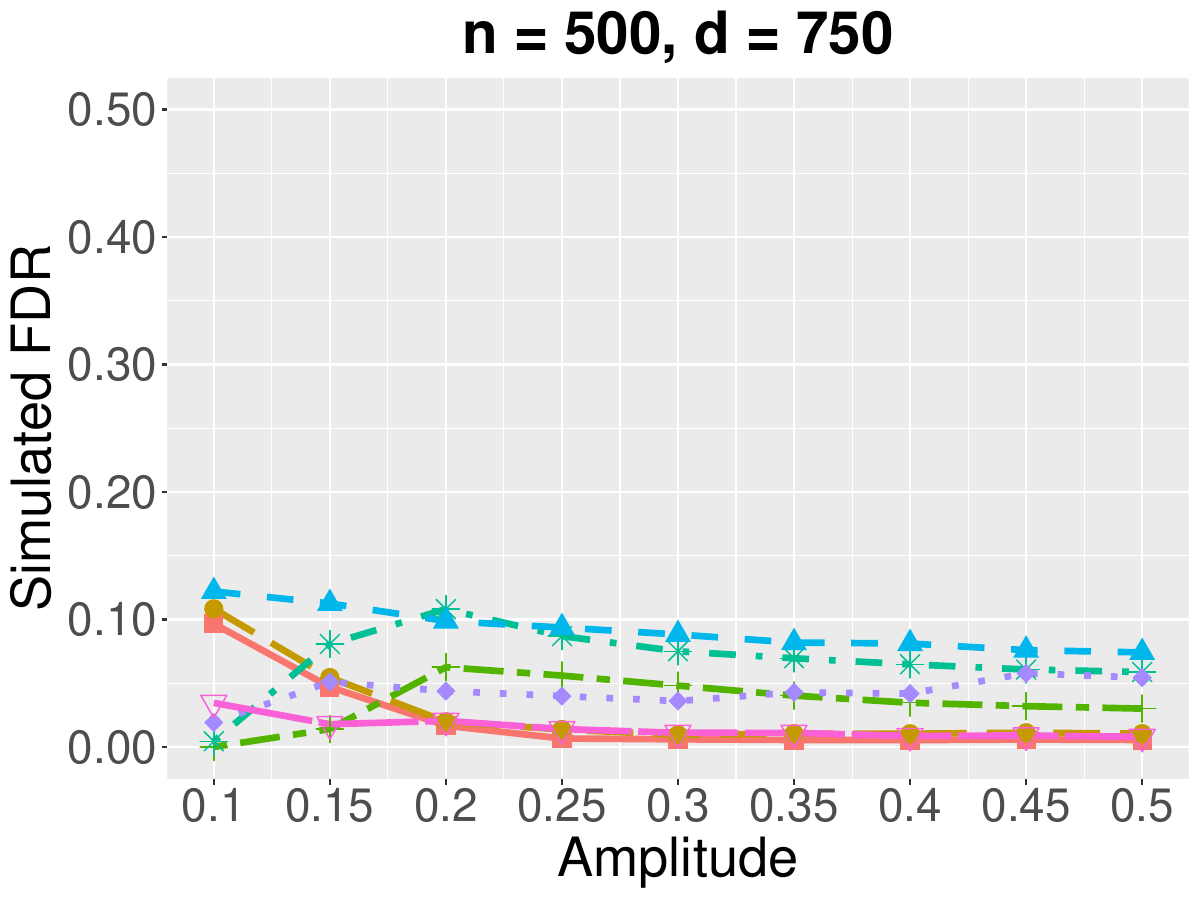}}\hspace{5pt}
	\subfloat{\includegraphics[width=.27\columnwidth]{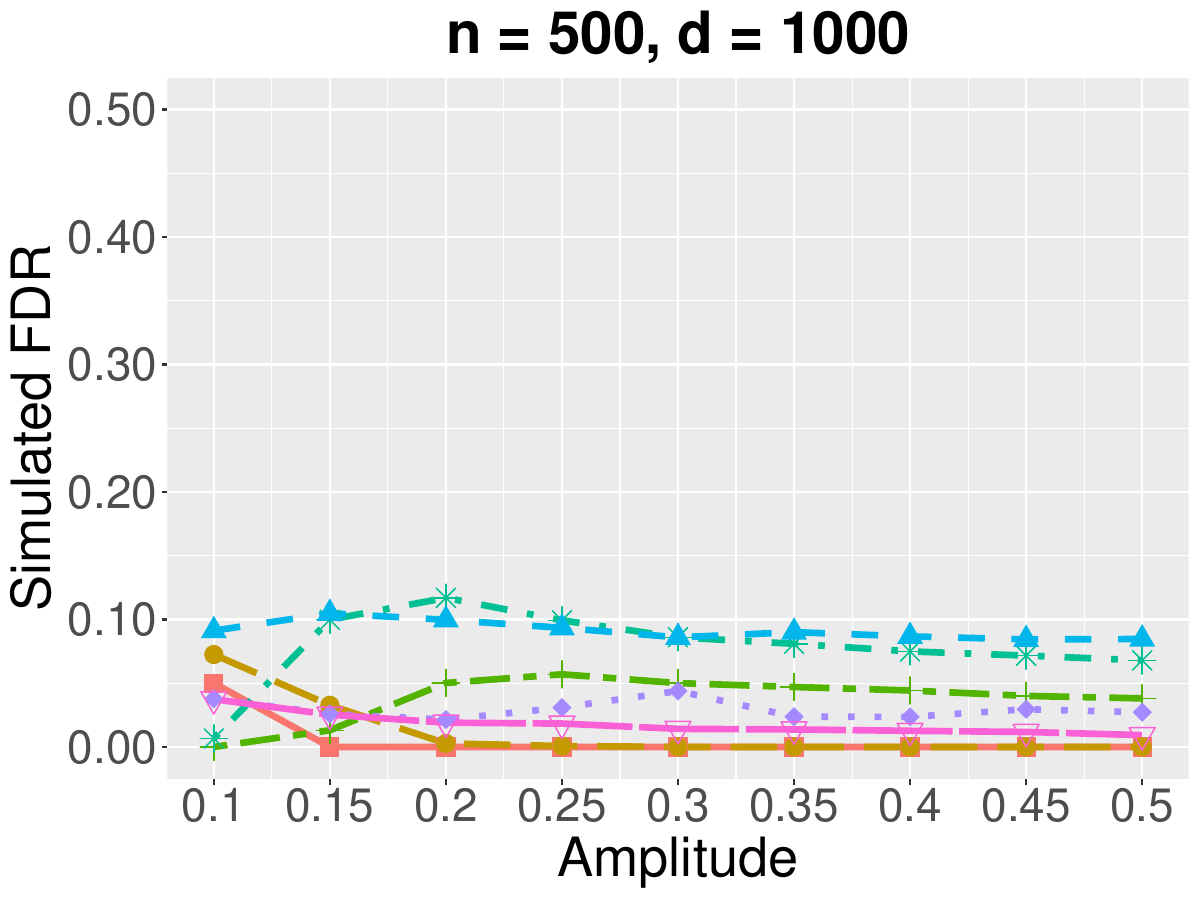}}\\
	\subfloat{\includegraphics[width=.27\columnwidth]{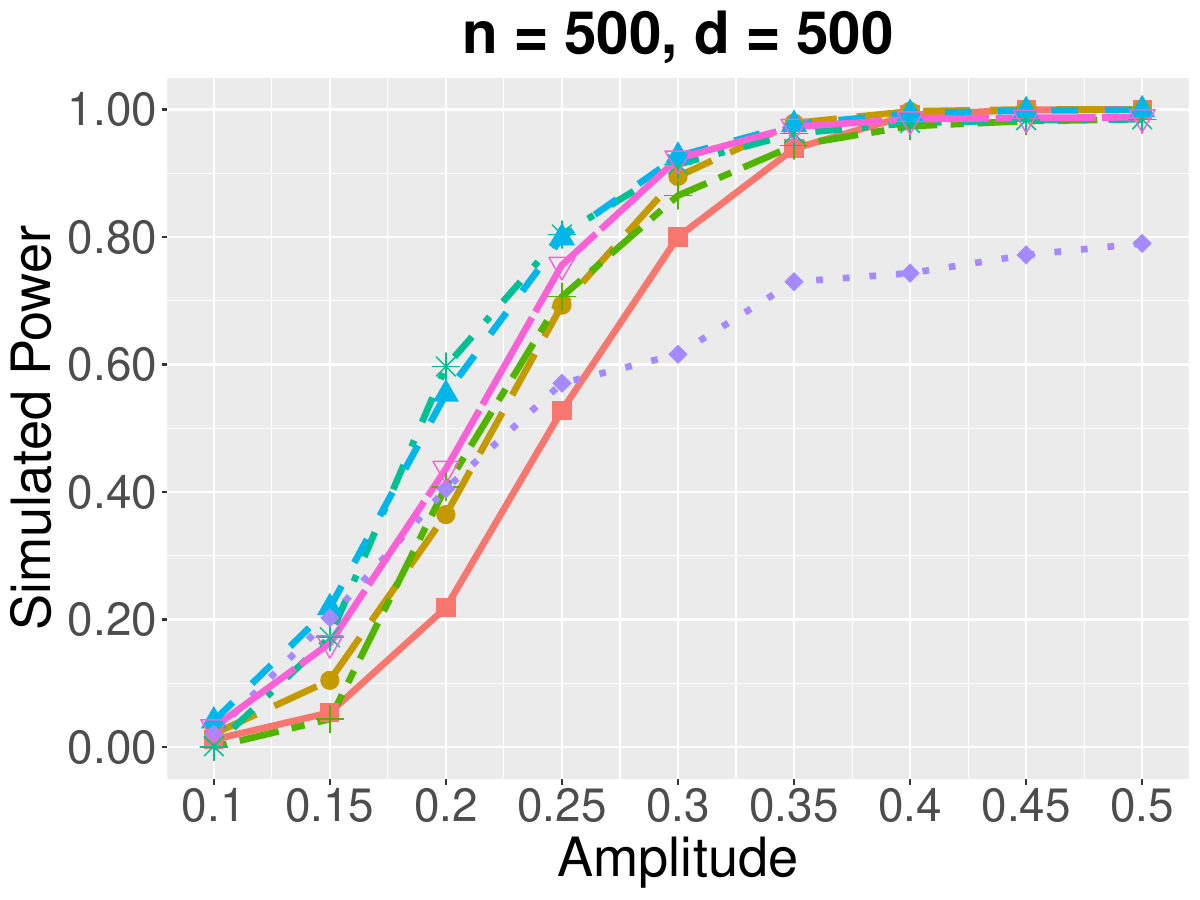}}\hspace{5pt}
    \subfloat{\includegraphics[width=.27\columnwidth]{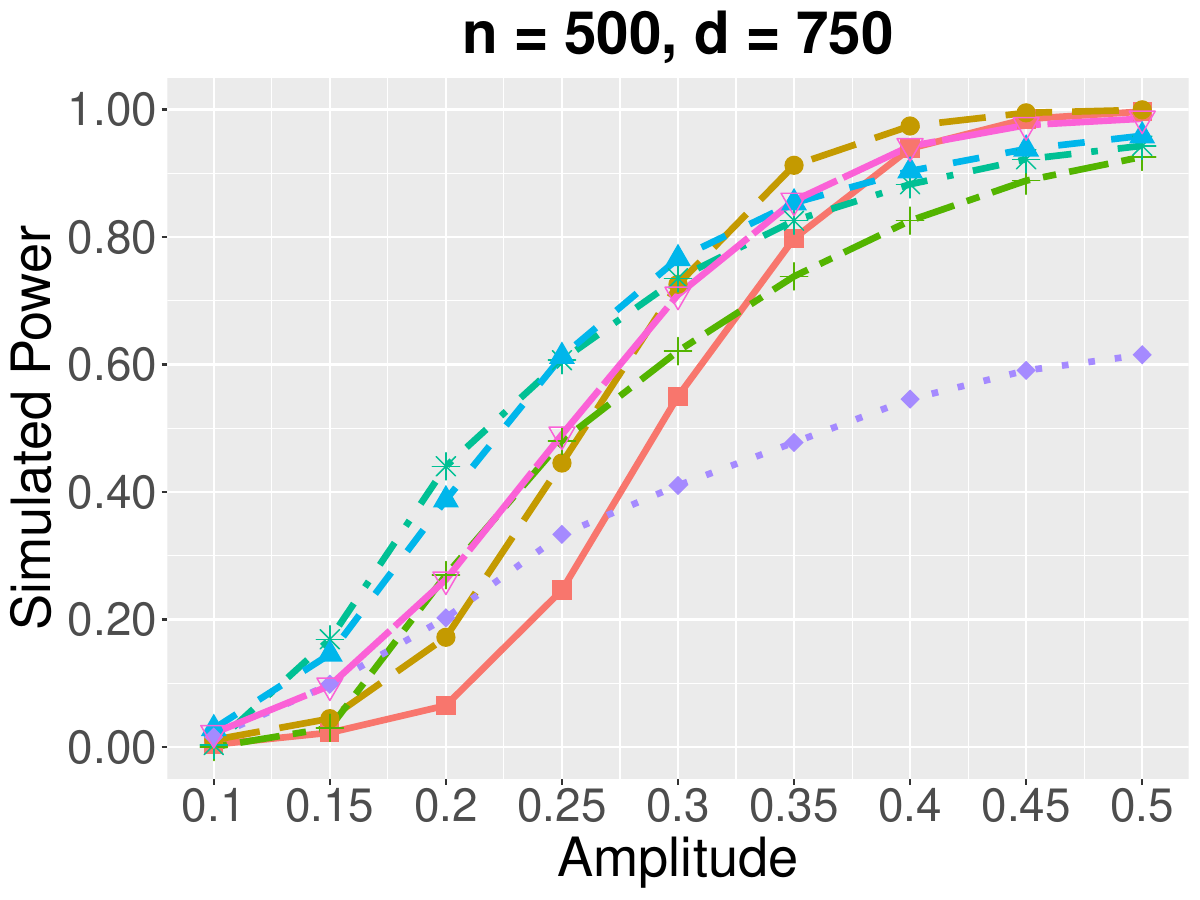}}\hspace{5pt}
    \subfloat{\includegraphics[width=.27\columnwidth]{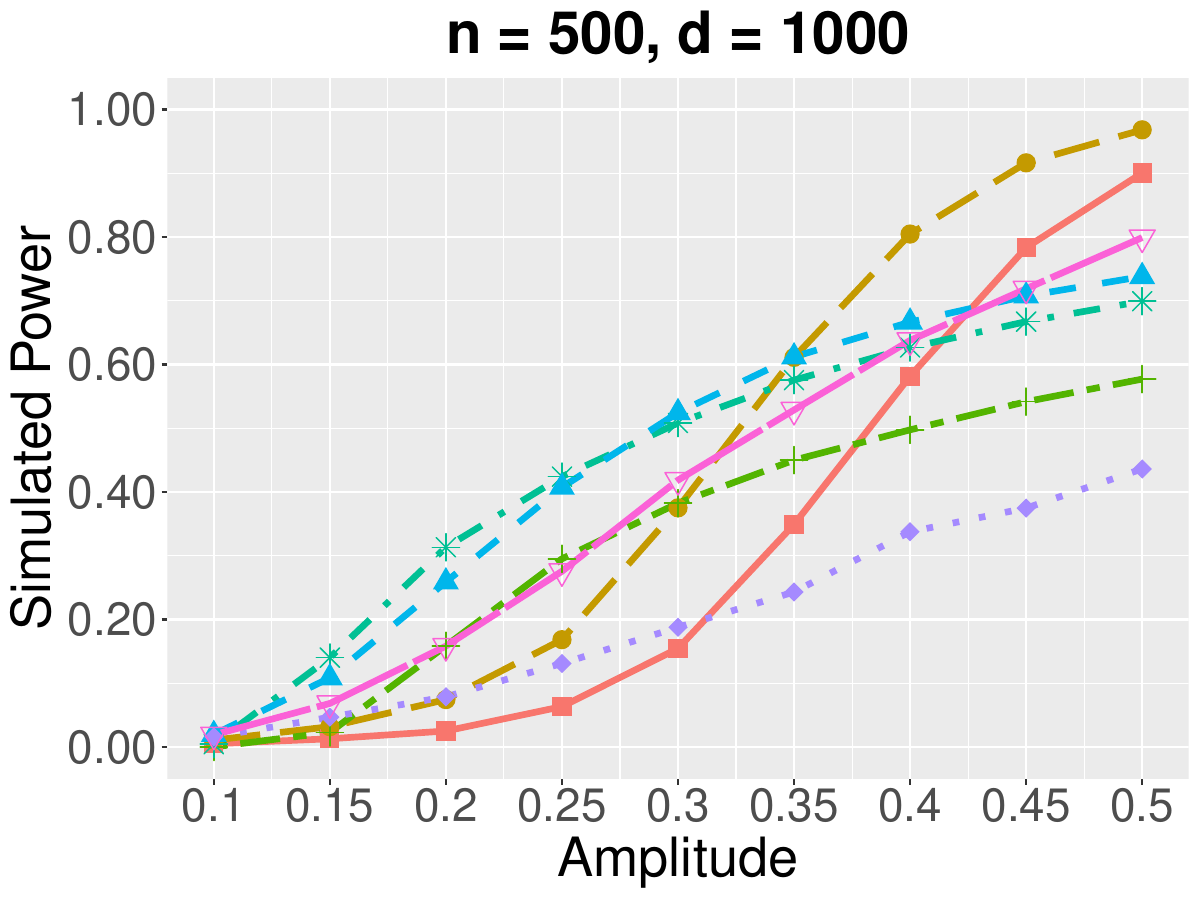}}
	\caption{\small Simulated FDR and power for different combinations of $(n,d)$. The rows of the design matrix were generated from Setting 2. The sparsity level is $k = 0.04d$ and the FDR level is $\alpha = 0.1$. The methods compared are Algorithm 1 (squares and red solid line), Algorithm 2 (circles and yellow long-dashed line), Algorithm A4 (diamonds and purple dotted line), Algorithm A5 (downward triangles and pink dot-dashed line), data splitting version of Algorithm A6 (crosses and dark green two-dashed line), data recycling version of Algorithm A6 (stars and light green dot-dashed line) and Algorithm A7 (upward triangles and blue dashed line).}
    \label{power-SI-ar1}
\end{figure}

Figures~\ref{power-SI-indep-005}--\ref{power-SI-ar1} display results across
signal amplitudes under the independent and AR(1) designs at the two
nominal levels. While our methods keep the empirical full-model FDR below
or close to the nominal level in both settings, the competing methods
behave differently across the two designs: their empirical FDR remains
close to the nominal level when the covariates are independent, whereas under the more correlated design
some of them show visibly inflated empirical FDR, particularly at the lower
level $\alpha = 0.05$ (Figure~\ref{power-SI-ar1-005}). This might be due to
(a) the screening step, whose guarantees concern the selected model rather
than the full model, so that stronger dependence may induce false
selections among correlated nulls; and (b) dependence among the resulting
test statistics that is left unaccounted for in selective inference.

The power
profiles of the methods also differ. From moderate
amplitudes onward, our Algorithm \ref{method_2} generally attains the
highest power, with the power of Algorithm \ref{method_1} improving rapidly
as the amplitude grows. The conditional procedures, Algorithms A4 and A5, are competitive
when $n$ and $d$ are small, and our methods catch up and then outperform
them as $d$ grows: at $(n,d) = (500, 1000)$, the powers of Algorithms A4 and A5 fall
substantially below those of Algorithms \ref{method_1} and \ref{method_2}. As
for the splitting-based procedure, Algorithm A6, in both its data-splitting and
data-recycling variants, is not competitive when $n$ and $d$ are small,
often making no discoveries in settings with weak signals; its
performance picks up as $n$ and $d$ become larger. Our methods also remain
competitive in power compared with Algorithm A7. 

An overarching observation is that the selective-inference methods -- both
those based on data splitting or data re-use and those based on conditional
distributions -- tend to perform better under independence than under the
AR(1) design. This reflects their reliance on the Lasso for selection, whose
performance is known to be affected by dependence in the design; in
contrast, our methods perform stably across the two designs.

The empirical observations mirror the discussion in
Section~\ref{sec:SI-distinctions}. In particular, the power deficit of the
conditional procedures, Algorithms A4 and A5, at larger dimensions is possibly a
finite-sample manifestation of the information cost of conditioning: the
truncated laws discard the information consumed by selection
\citepS{Fithian2017Optimal_app}, and the severity of this loss is reflected in
the expected length of the associated confidence intervals, which can be
infinite \citepS{KivaranovicLeeb2021_app}. The loss also grows with $d$, as the active set
shrinks relative to the true support and the truncation becomes more
aggressive. In our implementation, the truncation-interval computations
became numerically fragile at the larger dimensions.

The comparison with the splitting-based knockoff procedures also confirms
the competitiveness of our methods relative to those based on empirical FDP
estimation: when the FDR level is low and $n$ and $d$ are small, the
knockoff-filter-based procedures do not perform satisfactorily, in line with
our earlier observation in the paper that positions the proposed methodology
as a competitive choice for high-dimensional multiple testing.

\bibliographystyleS{jasa}
\spacingset{0.95}\selectfont
\bibliographyS{mybib_app}

\end{document}